\documentstyle{elsart}
\begin{document}
\input epsf
\input psfig.sty


\newcommand{\aap}       [4]{{\em A\&A\/},                       {\bf #1}, (#4), #2--#3.}
\newcommand{\aaps}      [4]{{\em A\&A Suppl\/},                 {\bf #1}, (#4), #2--#3.}
\newcommand{\aj}        [4]{{\em AJ\/},                         {\bf #1}, (#4), #2--#3.}
\newcommand{\apj}       [4]{{\em ApJ\/},                        {\bf #1}, (#4), #2--#3.}
\newcommand{\apjl}      [4]{{\em ApJ\/},                        {\bf #1}, (#4), #2--#3.}
\newcommand{\apjs}      [4]{{\em ApJS\/},                       {\bf #1}, (#4), #2--#3.}
\newcommand{\araa}      [4]{{\em ARA\&A\/},                     {\bf #1}, (#4), #2--#3.}
\newcommand{\jfm}       [4]{{\em J. Fluid. Mech.\/},            {\bf #1}, (#4), #2--#3.}

\newcommand{\metrica}   [4]{{\em Metrica\/},                    {\bf #1}, (#4), #2--#3.}
\newcommand{\mnras}     [4]{{\em MNRAS\/},                      {\bf #1}, (#4), #2--#3.}
\newcommand{\nat}       [4]{{\em Nature\/},                     {\bf #1}, (#4), #2--#3.}
\newcommand{\newa}      [4]{{\em NewA\/},                       {\bf #1}, (#4), #2--#3.}
\newcommand{\newar}     [4]{{\em NewAR\/},                      {\bf #1}, (#4), #2--#3.}
\newcommand{\pasj}      [4]{{\em Pub. R. Astr. Soc. Jap.\/},    {\bf #1}, (#4), #2--#3.}
\newcommand{\pasp}      [4]{{\em Pub. Astr. Soc. Pac.\/},       {\bf #1}, (#4), #2--#3.}
\newcommand{\pl}        [4]{{\em Phys. Lett.\/},                {\bf #1}, (#4), #2--#3.}
\newcommand{\pr}        [4]{{\em Phys. Rep.\/},                 {\bf #1}, (#4), #2--#3.}
\newcommand{\prd}       [4]{{\em Phys. Rev. D\/},               {\bf #1}, (#4), #2--#3.}

\newcommand{\prdn}      [4]{{\em Phys. Rev. D\/},               {\bf #1}, (#4), #2.}

\newcommand{\prl}       [4]{{\em Phys. Rev. Lett.\/},           {\bf #1}, (#4), #2--#3.}
\newcommand{\rmp}       [4]{{\em Rev. Mod. Phys.\/},            {\bf #1}, (#4), #2--#3.}
\newcommand{\sa}        [4]{{\em Sov. A\/},                     {\bf #1}, (#4), #2--#3.}



\newcommand{\lexp}{\mathop{\langle}}    
\newcommand{\rexp}{\mathop{\rangle}}    
\newcommand{\rexpc}{\mathop{\rangle_c}} 
\def\cum#1{\langle\delta^{#1}\rangle_c} 
\def\mom#1{\langle\delta^{#1}\rangle}   

\def\V{{\cal{V_T}}}     


\font\BF=cmmib10 scaled 1200
\def\de{\delta}
\def\te{\theta}
\def\ds{\delta_s}
\def\dt{\tilde \delta}
\def\dD{[\delta_{\rm D}]}
\def\del{\nabla}
\def\knl{k_{n\ell}}
\def\pleg{{\mathsf P}}

\def\vf{{\hbox{\BF f}}}
\def\vk{{\hbox{\BF k}}}
\def\vl{{\hbox{\BF l}}}
\def\vx{{\hbox{\BF x}}}
\def\vy{{\hbox{\BF y}}}
\def\vr{{\hbox{\BF r}}}
\def\vs{{\hbox{\BF s}}}
\def\vq{{\hbox{\BF q}}}
\def\vp{{\hbox{\BF p}}}
\def\vv{{\hbox{\BF v}}}
\def\vV{{\hbox{\BF V}}}
\def\vX{{\hbox{\BF X}}}
\def\vY{{\hbox{\BF Y}}}
\def\vw{{\hbox{\BF w}}}
\def\uz{{\hbox{$v_z$}}}
\def\uu{{\hbox{\BF v}}}
\def\vu{{\hbox{\BF u}}}
\def\tk{\hat k}
\def\tvk{{\hat{\k}}}
\def\ttheta{\hat \theta}
\def\tphi{\hat \varphi}
\def\tev{{\hbox{\BF $\theta$}}} 
\def\tF{\hat F}                 
\def\tvx{{\bf \hat x}}             
\def\tvy{{\bf \hat y}}          
\def\tvf{{\bf \hat f}}
\def\tvg{{\bf \hat g}}
\def\tvs{{\bf \hat s}} 
\def\tvn{{\bf \hat n}}
\def\tg{{\hat g}}
\def\vPsi{{\bf \Psi}}
\def\wW{{\bf W}}
\def\vf{{\bf f}}
\def\tf{{\hat f}}  
\def\tx{\hat x}                 
\def\ty{\hat y}                 
\def\tP{\hat P}                 
\def\bn{\bar N}                 
\def\tB{\hat B}                 
\def\tA{\hat{A}}                
\def\txi{\hat{\xi}}             
\def\tw{\hat{w}}             
\def\xiav{\bar{\xi}}            
\def\dim{\cal D}                

\def\barN{\bar N}               
\def\calD{\cal{D}}              
\def\B{{\cal{B}}}               
\def\facmomcor{W}               
\def\tfacmomcor{W}

\newcommand{\nbar}{\overline{n}}
\newcommand{\Nbar}{\overline{N}}
\newcommand{\wbar}{\overline{\omega}}
\newcommand{\rhobar}{\overline{\rho}}
\newcommand{\tet}{\hbox{\twelveBF $\theta$}}
\newcommand{\xibar}{\overline{\xi}}

\def\Or{{\cal O}}
\def\vf{{\bf f}}
\def\vF{{\bf F}}
\def\vS{{\bf S}}
\def\vM{{\bf M}}
\def\vN{{\bf N}}
\def\vs{{\bf s}}
\def\vn{{\bf n}}
\def\va{{\bf a}}
\def\vb{{\bf b}}
\def\vC{{\bf C}}
\def\vx{{\bf x}}
\def\vy{{\bf y}}
\def\vq{{\bf q}}
\def\vu{{\bf u}}
\def\vr{{\bf r}}
\def\vp{{\bf p}}
\def\vv{{\bf v}}
\def\vw{{\bf w}}
\def\vk{{\bf k}}
\def\vg{{\bf g}}
\def\ii{{\rm i}}
\def\d{{\rm d}}
\def\vW{{\bf W}}
\def\vL{{\bf L}}
\def\vA{{\bf A}}
\def\vC{{\bf C}}
\def\vQ{{\bf Q}}
\def\vT{{\bf T}}
\def\dr{\partial}
\def\grad{\nabla}
\def\rhob{\overline{\rho}}
\def\gradx{\nabla_{\hbox{\rm x}}\,}
\def\gradq{\nabla_{\hbox{\rm q}}\,}
\def\disp{\displaystyle}
\def\dta{\delta}
\def\mg{\big<}
\def\md{\big>}
\def\mA{{\cal A}}
\def\mC{{\cal C}}
\def\mF{{\cal F}}
\def\mH{{\cal H}}
\def\mL{{\cal L}}
\def\mM{{\cal M}}
\def\mH{{\cal H}}
\def\mI{{\cal I}}
\def\mP{{\cal P}}
\def\mQ{{\cal Q}}
\def\mR{{\cal R}}
\def\mS{{\cal S}}
\def\mG{{\cal G}}
\def\mGd{{\cal G}_{\delta}}
\def\mGv{{\cal G}_{\theta}}
\def\mGdd{{\cal G}_{\delta}^{2D}}
\def\mGt{{\cal G}_{\theta}}
\def\WTHt{W\left(\vert\vk_1+\vk_2+\vk_3\vert\ R\right)}
\def\gm{\gamma}
\def\ort{\bot}
\def\xib{\overline{\xi}}

\def\bce{\begin{center}}
\def\ece{\end{center}}
\def\ben{\begin{enumerate}}
\def\een{\end{enumerate}}
\def\ul{\underline}
\def\ni{\noindent}
\def\nn{\nonumber}
\def\bs{\bigskip}
\def\ms{\medskip}
\def\wt{\widetilde}

\def\brr{\begin{array}}
\def\err{\end{array}}
\def\dsp{\displaystyle}
\def\beq#1{\begin{equation}#1\end{equation}}
\def\be{\begin{equation}}
\def\ee{\end{equation}}
\def\bea{\begin{eqnarray}}
\def\eea{\end{eqnarray}}
\def\ba{\begin{eqnarray}}
\def\ea{\end{eqnarray}}

\newcommand{\eq}{{Eq.~}}

\def\Mpc{{\,h^{-1}\,{\rm Mpc}}}
\def\la{\mathrel{\mathpalette\fun <}}
\def\ga{\mathrel{\mathpalette\fun >}}
\def\fun#1#2{\lower3.6pt\vbox{\baselineskip0pt\lineskip.9pt
        \ialign{$\mathsurround=0pt#1\hfill##\hfil$\crcr#2\crcr\sim\crcr}}}
\def\etal{{\rm et~al. }}
\def\spose#1{\hbox to 0pt{#1\hss}}


\begin{frontmatter}

\title{Large-Scale Structure of the Universe and \\ 
Cosmological Perturbation Theory}

\author[francis]{F.~Bernardeau}, 
\author[stephane]{S.~Colombi},
\author[enrique]{E.~Gazta\~naga},
\author[roman]{R.~Scoccimarro} 

\address[francis]{Service de Physique Th\'eorique, C.E. de Saclay,
F-91191 \\ Gif-sur-Yvette Cedex, France} 

\address[stephane]{Institut d'Astrophysique de Paris, CNRS, 98 bis
Boulevard Arago, \\ F-75014 Paris, France} 

\address[enrique]{Instituto   Nacional  de   Astrof\'{\i}sica,   \'Optica  y
Electr\'onica (INAOE), \\ Luis  Enrique Erro 1, Tonantzintla, Cholula, 78840
Puebla, Mexico  \\ Institut d'Estudis Espacials de  Catalunya, ICE/CSIC, 
\\ Edf. nexus-201 - c/ Gran Capit\`a 2-4, 08034 Barcelona, Spain }

\address[roman]{Department of Physics, New York University, 4 Washington
Place, \\ New York, NY 10003, USA \\
Institute for Advanced Study, School of Natural
Sciences, Einstein Drive, \\ Princeton, NJ 08540, USA }

\date{1 December 2001}

\begin{abstract}
We review the formalism and applications of non-linear perturbation
theory (PT) to understanding the large-scale structure of the
Universe.  We first discuss the dynamics of gravitational instability,
from the linear to the non-linear regime.  This includes Eulerian and
Lagrangian PT, non-linear approximations, and a brief description of
numerical simulation techniques.  We then cover the basic statistical
tools used in cosmology to describe cosmic fields, such as
correlations functions in real and Fourier space, probability
distribution functions, cumulants and generating functions.  In
subsequent sections we review the use of PT to make quantitative
predictions about these statistics according to initial conditions,
including effects of possible non Gaussianity of the primordial
fields.  Results are illustrated by detailed comparisons of PT
predictions with numerical simulations.  The last sections deal with
applications to observations.  First we review in detail practical
estimators of statistics in galaxy catalogs and related errors,
including traditional approaches and more recent developments.  Then,
we consider the effects of the bias between the galaxy distribution
and the matter distribution, the treatment of redshift distortions in
three-dimensional surveys and of projection effects in angular
catalogs, and some applications to weak gravitational lensing.  We
finally review the current observational situation regarding
statistics in galaxy catalogs and what the future generation of galaxy
surveys promises to deliver.
\end{abstract}
\end{frontmatter}

\parskip=2pt
\newpage
\setcounter{tocdepth}{5}
\tableofcontents

\clearpage 

\section{\bf Introduction and Notation}

Understanding the large scale structure of the Universe is one of the
main goals of cosmology. In the last two decades it has become widely
accepted that gravitational instability plays a central role in giving
rise to the remarkable structures seen in galaxy surveys. Extracting
the wealth of information contained in galaxy clustering to learn
about cosmology thus requires a quantitative understanding of the
dynamics of gravitational instability and application of sophisticated
statistical tools that can best be used to test theoretical models
against observations.

In this work we review the use of non-linear cosmological perturbation
theory (hereafter PT) to accomplish this goal. The usefulness of PT in
interpreting results from galaxy surveys is based on the fact that in
the gravitational instability scenario density fluctuations become
small enough at large scales (the so-called ``weakly non-linear
regime'') that a perturbative approach suffices to understand their
evolution. Since early developments in the 80's, PT has gone through a
period of rapid evolution in the last decade which gave rise to
numerous useful results. Given the imminent completion of
next-generation large-scale galaxy surveys ideal for applications of
PT, it seems timely to provide a comprehensive review of the subject.

The purpose of this review is twofold: 

1) To summarize the most important theoretical results, which are
sometimes rather technical and appeared somewhat scattered in the
literature with often fluctuating notation, in a clear, consistent and
unified fashion. We tried in particular to unveil approximations that
might have been overlooked in the original papers, and to highlight
the outstanding theoretical issues that remain to be addressed.

2) To present the state of the art observational knowledge of galaxy
clustering with particular emphasis in constraints derived from
higher-order statistics on galaxy biasing and primordial
non-Gaussianity, and give a rigorous basis for the confrontation of
theoretical results with observational data from upcoming galaxy
catalogues.

We assume throughout this review that the universe satisfies the
standard homogeneous and isotropic big bang model.  The framework of
gravitational instability, in which PT is based, assumes that gravity
is the only agent at large scales responsible for the formation of
structures in a universe with density fluctuations dominated by dark
matter.  This assumption is in very good agreement with observations
of galaxy clustering, in particular, as we discuss in detail here,
from higher-order statistics which are sensitive to the detailed
structure of the dynamics responsible for large-scale
structures\footnote{As opposed to just properties of the linearized
equations of motion, which can be mimicked by nongravitational
theories of structure formation in some cases~\cite{BWDO94}.}.  The
non-gravitational effects associated with galaxy formation may alter
the distribution of luminous matter compared to that of the underlying
dark matter, in particular at small scales: such `galaxy biasing' can
be probed with the techniques reviewed in this work.

Inevitably, we had to make some decisions in the choice of topics to
be covered. Our presentation is definitely focused on the density
field, with much less coverage on peculiar velocities.  This choice is
in particular motivated by the comparatively still preliminary stage
of cosmic velocity fields, at least from an observational point of
view (see however~\cite{StWi95,CSW00} for a review). On the other
hand, note that since velocity field results are often obtained by
identical techniques to those used for the density field, we mention
some of these results but without giving them their due importance.

In order to fully characterize the density field, we choose to follow
the traditional approach of using statistical methods, in particular,
$N$-point correlation functions~\cite{Peebles80}. Alternative methods
include morphological descriptors such as Minkowski functionals (of
which the genus is perhaps the most widely known), percolation
analysis, etc. Unlike correlation functions, however, these other
statistics are not as directly linked to dynamics as correlation
functions, and thus are not as easy to predict from theoretical
models. Furthermore, applications of PT to make predictions of these
quantities is still in its infancy (see e.g~\cite{Matsubara00} and
references therein for recent work).

Given that PT is an approximate method to solve the dynamics of
gravitational clustering, it is desirable to test the validity of the
results with other techniques. In particular, we resort to numerical
simulations, which involve different approximations in solving the
equations of motion that are not restricted to the weakly non-linear
regime. There is a strong and healthy interplay between PT and
$N$-body simulations which we extensively illustrate throughout this
review. At large scales PT can be used to test quantitatively for
spurious effects in numerical simulations (e.g. finite volume effects,
transients from initial conditions), whereas at smaller, non-linear
scales $N$-body simulations can be used to investigate the regime of
validity of PT predictions.

Although reviewing the current understanding of clustering at small
scales is beyond the scope of this review, we have also included a
discussion of the predictions of non-linear clustering amplitudes
because connections between PT and strongly non-linear behavior have
been suggested in the literature. We also include a discussion about
stable clustering at small scales which, when coupled with
self-similarity, leads to a connection between the large and
small-scale scaling behavior of correlations functions.

This review is structured so that different chapters can be read
independently, although there are inevitable relations. Chapter~2
deals with the basic equations of motion and their solution in PT,
including a brief summary of numerical simulations. Chapter~3 is a
review of the basics of statistics; we have made it as succinct as
possible to swiftly introduce the reader to the core of the
review. For a more in-depth treatment we refer the reader
to~\cite{KeSt94,Bertschinger92}. The next two chapters represent the
main theoretical results; Chapter~4 deals with $N$-point
functions, whereas Chapter~5 reviews results for the smoothed
one-point moments and PDF's. These two chapters heavily rely on
material covered in Chapters~2 and~3.

In Chapter~6 we describe in detail the standard theory of estimators
and errors for application to galaxy surveys, with particular
attention to the issue of cosmic bias and errors of estimators of the
two-point correlation function, power spectrum, and higher-order
moments such as the skewness. Chapter~7 deals with theoretical issues
related to surveys, such as redshift distortions, projection effects,
galaxy biasing and weak gravitational lensing. Chapter~8 presents the
current observational status of galaxy clustering, including future
prospects in upcoming surveys, with particular emphasis on
higher-order statistics. Chapter~9 contains our conclusions and
outlook. A number of appendices extend the material in the main text
for those interested in carrying out detailed calculations. Finally,
to help the reader, Tables~\ref{tab:abbrev}--\ref{fb:tabStats} list
the main abbreviations and notations used for various cosmological
variables, fields and statistics.

\newpage
\begin{table}[p]
\caption{Abbreviations}
\vspace{.1 cm}
\begin{tabular}{cp{12cm }}
\hline
PT                &  Perturbation Theory;\\
2LPT              &  Second Order Lagrangian Perturbation Theory;\\
EPT               &  Extended Perturbation Theory;\\
HEPT              &  HyperExtended Perturbation Theory;\\
ZA                &  Zel'dovich Approximation;\\
SC                &  Spherical Collapse;\\
CDM               &  Cold Dark Matter (model);\\
SCDM              &  Standard CDM model;\\
$\Lambda$CDM      &  Flat CDM model with a cosmological constant;\\
PDF               &  Probability Distribution Function;\\
CPDF               &  Count Probability Distribution Function.\\
\hline
\end{tabular}
\label{tab:abbrev}
\end{table}

\begin{table}[p]
\caption{Notation for Various Cosmological Variables} 
\vspace{.1 cm}
\begin{tabular}{cp{12cm }}
\hline
$\Omega_m$                   &  The total matter density in units of critical density;\\
$\Omega_\Lambda$             &  The reduced cosmological constant;\\
$\Omega_{tot}$               &  The total energy density of the universe
                                in units of critical density, $\Omega_{tot}=\Omega_m+\Omega_\Lambda$;\\
$H$                          &  The Hubble constant;\\
$h$                          &  The Hubble constant at present time,
                                in units of $100$ km/s/Mpc, $h \equiv H_0/100$;\\
$a$                          &  The scale factor;\\
$\tau$                       &  The conformal time, $d\tau=dt/a$;\\
${\cal H}$                   &  The conformal expansion rate, ${\cal
                                 H}=aH$;\\
$D_1$                        &  The linear growth factor;\\
$D_n$                        &  The $n$-th order growth factor;\\
$f(\Omega_m,\Omega_\Lambda)$ &  The logarithmic derivative of (the
                                fastest growing mode of) the linear growth
                                factor with respect to $a$: 
                                $f(\Omega_m,\Omega_\Lambda)\equiv d \ln D_1/d\ln a$.\\
\hline
\end{tabular}
\label{tab:cosmopar}
\end{table}

\begin{table}[p]
\caption{Notation for the Cosmic Fields}
\vspace{.1 cm}
\begin{tabular}{cp{12cm }}
\hline
${\tilde X}$                    & The Fourier transform of field $X$;\\
& $\tilde{X}(\vk)=(2\pi)^{-3} \int \d^3\vx \,{\rm e}^{-\ii\vk\cdot\vx}\,X(\vx)$
(except in Sect.~\ref{sec:sec3}) \\ 
$\vx$                           & The comoving position in real space;\\
$\rho(\vx)$                     & The local cosmic density;\\
$\delta(\vx)$                   & The local density contrast, $\delta=\rho/\rhobar-1$;\\
$\Phi(\vx)$                     & The gravitational potential;\\
$\vu(\vx)$                      & The local peculiar velocity field;\\
$\theta(\vx)$                   & The local velocity divergence in units of ${\cal H}=aH$;\\
$F_p(\vk_1,\dots,\vk_p)$        & The $p^{\rm th}$ order density field kernel;\\
$G_p(\vk_1,\dots,\vk_p)$        & The $p^{\rm th}$ order velocity divergence field kernel;\\
$\psi(\vq)$                     & The Lagrangian displacement field;\\ 
$J(\vq)$                        & The Jacobian of the Lagrangian-Eulerian mapping.\\
\hline
\end{tabular}
\label{fb:tabFields}
\end{table}

{ }
\begin{table}[p]
\caption{Notation for Statistical Quantities}
\vspace{.1 cm} 
\begin{tabular}{cp{12cm }}
\hline
$P(k)$                          & The density power spectrum;\\
$\Delta(k)$                     & The dimensionless power, $\Delta=4\pi k^3 P(k)$;\\
$B(k_1,k_2,k_3)$                & The bispectrum;\\
$P_N(\vk_1,\ldots,\vk_N)$       & The $N$-point polyspectrum;\\
$P_N$                           & The count-in-cell probability distribution function;\\
$p(\delta)\d\delta$             & The cosmic density probability distribution function;\\
$F_k$                           & The factorial moment of order $k$;\\
$\xi_2(\vx_1,\vx_2)\equiv \xi_{12}\equiv \xi$   & The two-point
                                  correlation function, $\xi_2(\vx_1,\vx_2)=\langle \delta(\vx_1)
                                  \delta(\vx_2) \rangle=\langle \delta(\vx_1)
                                  \delta(\vx_2) \rangle_c$;\\
$\sigma^2\equiv \overline{\xi}\equiv \overline{\xi}_2$  & The
                                  cell-average $two-$point correlation function;\\    
$\sigma_8$                      & The value of the (linearly extrapolated) $\sigma$ in a sphere of $8\,h^{-1}$ Mpc radius;\\
$\Gamma$                        & Shape parameter of the linear power-spectrum, $\Gamma \simeq \Omega_m h$;\\
\hline
\end{tabular}
\label{fb:tabStats}
\end{table}

\begin{table}
\noindent Table~\ref{fb:tabStats} (continued)
\vspace{.1 cm} 

\begin{tabular}{cp{12cm }}
\hline
$\xi_N(\vx_1,\dots,\vx_N)$      & The $N$-point correlation functions 
                                  $\xi_N(\vx_1,\dots,\vx_N)=\langle
                                  \delta(\vx_1)\ldots\delta(\vx_N) \rangle_c$;\\
$w_N(\theta_1,\ldots,\theta_N)$ & The angular $N$-point correlation functions;\\
$\overline{\xi}_N$              & The cell-averaged $N-$point
                                  correlation functions $\overline{\xi}_N=\langle \delta_R^N \rangle_c$;\\
$\overline{w}_N$                & The cell-averaged angular $N-$point correlation functions;\\ 
$S_p$                           & The density normalized cumulants, 
                                  $S_p=\langle\delta_R^p\rangle_c/\langle \delta_R^2 \rangle^{p-1}
                                  =\overline{\xi}_p/\overline{\xi}^{p-1}$;\\
$S_3$, $S_4$                    & The (reduced) skewness/kurtosis;\\
$s_p$                           & The projected density normalized cumulants;\\
$Q\equiv Q_3$, ${\tilde Q}\equiv{\tilde Q}_3$  & The three-point hierarchical amplitude
                                  in real/Fourier space;\\
$Q_N$, $\tilde{Q}_N$            & The $N$-point hierarchical amplitude in real/Fourier space; 
                                  $Q_N$ can also stand for $S_N/N^{N-2}$ (Chap.~\ref{sec:chapter7});\\
$q_N$, $\tilde{q}_N$            & The projected $N$-point hierarchical
                                  amplitude in real/Fourier space;
                                  $q_N$ can also stand for $s_N/N^{N-2}$ (Chap.~\ref{sec:chapter7});\\
$T_p$                           & The velocity divergence normalized cumulants;\\
$C_{pq}$                        & The~two-point~density~normalized~cumulants, $C_{pq}=
                                  \langle\delta_1^p \delta_2^q \rangle_c/(\xi_{12}\mom2^{p+q-2})$;\\
$\varphi(y)$                    & The one-point cumulant generating function, 
                                  $\varphi(y)=\sum_pS_p\,(-y)^p/p!$;\\
$\nu_p$, $\mu_p$                & The density/velocity field vertices;\\
$\mGd(\tau)\equiv\mGd^L(\tau) $, $\mGv(\tau)\equiv\mGv^L(\tau) $ 
                                & The vertex generating function for
                                  the density/velocity field,
                                  $\mGd(\tau)\equiv \sum_{p\geq 1}
                                  \nu_p (-\tau)^p/p!$, and  
                                  $\mGv(\tau)\equiv -f(\Omega_m,\Omega_\Lambda)\sum_{p\geq 1}
                                  \nu_p (-\tau)^p/p!$;\\
$\langle X \rangle$             & The ensemble average of statistic $X$;\\
${\hat X}$                      & The estimator of statistic $X$;\\
$\Upsilon(\hat X)\d{\hat X}$    & The cosmic distribution function of
                                  estimator ${\hat X}$;\\
$\Delta X$                      & The cosmic error on estimator ${\hat X}$.\\
\hline
\end{tabular}
\end{table}

\clearpage 
                                                                    
\section{\bf Dynamics of Gravitational Instability}
\label{chapter2}

The most natural explanation for the large-scale structures seen in
galaxy surveys (e.g. superclusters, walls, and filaments) is that they
are the result of gravitational amplification of small primordial
fluctuations due to the gravitational interaction of collisionless
cold dark matter (CDM) particles in an expanding
universe~\cite{Peebles82,BFPR84,DEFW85,DEFW92}. Throughout this review
we will assume this framework and discuss how PT can be used to
understand the physics of gravitational instability and test this
hypothesis against observations.

Although the nature of dark matter has not yet been identified, all
candidates for CDM particles are extremely light compared to the mass
scale of typical galaxies, with expected number densities of at least
$10^{50}$ particles per Mpc$^3$~\cite{KoTu93}. In this limit where the
number of particles $N \gg 1$, discreteness effects such as two-body
relaxation (important e.g. in globular clusters) are negligible, and
collisionless dark matter\footnote{There has been recently a renewed
interest in studying {\em collisional} dark
matter~\cite{SpSt00,YSWT00,DSSW01}, which may help solve some problems
with collisionless CDM at small scales, of order few kpc. }  obeys the
Vlasov equation for the distribution function in phase space,
Eq.~(\ref{vlasov}) below. This is the master equation from which all
subsequent calculations of gravitational instability are derived.

Since CDM particles are non-relativistic, at scales much smaller than
the Hubble radius the equations of motion reduce essentially to those
of Newtonian gravity\footnote{A detailed treatment of relativistic
linear perturbation theory of gravitational instability can be found
in~\cite{Bardeen80,MFB92,LiLy00}.}. The expansion of the universe
simply calls for a redefinition of the variable used to describe the
position and momentum of particles, and a redefinition of the
gravitational potential. For a detailed discussion of the Newtonian
limit from general relativity see e.g.~\cite{Peebles80}. We will
simply motivate the results without giving a derivation.

\subsection{The Vlasov Equation}
\label{sec:vlasov}

Let's consider a set of particles of a mass $m$ that interact only
gravitationally in an expanding universe.  The equation of motion for
a particle of velocity $\vv$ at position $\vr$ is thus,

\begin{equation}
{\d\vv\over \d t}= G\,m\sum_{i}{\vr_i-\vr\over
\vert\vr_i-\vr\vert^3}
\end{equation}
where the summation is made over all other particles at position
$r_i$.

In the limit of a large number of particles, this equation can be
rewritten in terms of a smooth gravitational potential due to the
particle distribution,
\begin{equation}
{\d\vv\over \d t}= -{\partial\phi\over\partial\vr}
\label{fb:partmotion}
\end{equation}
where $\phi$ is the Newtonian potential induced by the local mass
density $\rho(\vr)$,
\begin{equation}
\phi(\vr)=G\int\d^3\vr'{\rho(\vr')\over
\vert\vr'-\vr\vert}.
\end{equation}

In the context of gravitational instabilities in an expanding universe
we have to consider the departures from the homogeneous Hubble
expansion.  Positions of particles are described by their {\em
comoving} coordinates $\vx$ such that the physical coordinates are
$\vr=a(\tau)\,\vx$ where $a$ is the cosmological scale factor.  We
choose to describe the equations of motion in terms of the conformal
time $\tau$ related to cosmic time by $\d t=a(\tau) \d\tau$.  The
equations of motion that follow are valid in an arbitrary homogeneous
and isotropic background Universe, which evolves according to
Friedmann equations:

\begin{equation}
{{\partial{\cal H}(\tau)} \over{\partial \tau}} = - {\Omega_m(\tau)
\over 2} {\cal H}^2(\tau) + {\Lambda \over 3} a^2(\tau) \equiv
\left( \Omega_\Lambda(\tau) - \frac{\Omega_m(\tau)}{2} \right) 
{\cal H}^2(\tau) 
\label{friedmann1}
\end{equation}
\begin{equation}
(\Omega_{\rm tot}(\tau) -1) {\cal H}^2(\tau) = k
\label{friedmann2},
\end{equation}
where $ {\cal H}\equiv {\d\ln a /{\d\tau}}=H a$ is the conformal
expansion rate, $H$ is the Hubble constant, $\Omega_m$ is the ratio of
matter density to critical density, $\Lambda$ is the cosmological
constant and $k=-1,0,1$ for $\Omega_{\rm tot}<1$, $\Omega_{\rm tot}=1$ and
$\Omega_{\rm tot}>1$ respectively ($\Omega_{\rm tot} \equiv \Omega_m +
\Omega_\Lambda $). Note that $\Omega_m$ and $\Omega_{\Lambda}$ are 
time dependent.

We then define the density contrast $\delta(\vx)$ by,
\begin{equation}
\rho(\vx,\tau) \equiv \bar{\rho}(\tau)\ \left[1+\de(\vx,\tau)\right],
\label{rho}
\end{equation} 
the peculiar velocity $\vu$ with
\begin{equation}
\vv(\vx,\tau) \equiv {\cal H} \vx + \vu(\vx,\tau),
\label{vel}
\end{equation}
and the cosmological gravitational potential $\Phi$ with
\begin{equation}
\phi(\vx,\tau)\equiv - \frac{1}{2}
\frac{\partial {\cal H}}{\partial \tau} x^2 + \Phi(\vx,\tau),
\label{pot}
\end{equation}
so that the latter is sourced only by density fluctuations, as
expected; indeed the Poisson equation reads,
\begin{equation}
\nabla^2 \Phi(\vx,\tau) = {3\over 2} \Omega_m(\tau)\ {\cal H}^2(\tau)\ 
\delta{(\vx,\tau)}. \label{poisson}
\end{equation}
In the following we will only use comoving coordinates as the spatial
variable so that all space derivatives should be understood as done
with respect to $\vx$.

The equation of motion Eq.~(\ref{fb:partmotion}) then reads
\begin{equation}
{\d\vp\over\d \tau}=-a\,m \nabla \Phi(\vx)
\end{equation}
with
\begin{equation}
\vp=a\,m\,\vu.
\end{equation}
Let us now define the particle number density in phase space by
$f(\vx,\vp,\tau)$; phase-space conservation implies the Vlasov
equation,

\begin{equation}
{\d f\over \d \tau}= {\dr f\over\dr \tau} +{\vp\over
m\,a} \cdot \nabla f
-a\,m \nabla \Phi \cdot {\dr f\over\dr\vp}=0
\label{vlasov}
\end{equation}
Needless to say, this equation is very difficult to solve, being a
non-linear partial differential equation involving seven variables.
The non-linearity is induced by the fact that the potential $\Phi$
depends through Poisson equation on the integral of the distribution
function over momentum (which gives the density field, see
Eq.~(\ref{rhof}) below).

\subsection{Eulerian Dynamics}

In practice however we are usually not interested in solving the full
phase-space dynamics, but rather the evolution of the 
spatial distribution. This
can be conveniently obtained by taking momentum moments of the
distribution function. The zeroth order moment
simply relates the phase space density to the local mass density
field,
\begin{equation}
\label{rhof}
\int \d^3\vp\ f(\vx,\vp,\tau) \equiv \rho(\vx,\tau).
\end{equation}
The next order moments,
\begin{eqnarray}
\int \d^3\vp\ \frac{\vp}{a\,m}\
f(\vx,\vp,\tau) &\equiv& \rho(\vx,\tau) \vu(\vx,\tau) \\ & & \nonumber  \\
\int \d^3\vp\ \frac{p_i p_j}{a^2\,m^2}\ f(\vx,\vp,\tau) &\equiv& 
\rho(\vx,\tau) \vu_i(\vx,\tau) \vu_j(\vx,\tau) +
\sigma_{ij}(\vx,\tau),
\label{vmoments}
\end{eqnarray}
define the {\em peculiar velocity flow} $\vu(\vx,\tau)$ and the {\em
stress tensor} $\sigma_{ij}(\vx,\tau)$. The equation for these fields
follow from taking moments of the Vlasov equation. The zeroth moment
gives the continuity equation,
\begin{equation}
{\partial \delta(\vx,\tau) \over{\partial \tau}} + \nabla \cdot \left\{
\left[1+\de(\vx,\tau)\right] \vu(\vx,\tau) \right\} = 0
\label{continuity},
\end{equation}
which describes conservation of mass. Taking the first moment of
Eq.~(\ref{vlasov}) and subtracting $\vu(\vx,\tau)$ times the continuity
equation we obtain the Euler equation,
\begin{eqnarray}
{\partial \vu(\vx,\tau) \over{\partial \tau}} + {{\cal H}(\tau)}\
\vu(\vx,\tau) + \vu(\vx,\tau) \cdot \nabla \vu(\vx,\tau)
&=&\nonumber\\
&&\hspace{-2cm}-\nabla \Phi(\vx,\tau) -{1\over\rho}\nabla_j\left(\rho\,\sigma_{ij}\right)
\label{euler},
\end{eqnarray}
which describes conservation of momentum. Note that the continuity
equation couples the zeroth ($\rho$) to the first moment ($\vu$) of the
distribution function, the Euler equation couples the first moment
($\vu$) to the second moment ($\sigma_{ij}$), and so on. 
However, having integrated out the phase-space information,
we are here in a more familiar ground, and we have reasonable
phenomenological models to close the hierarchy by postulating an
ansatz for the stress tensor $\sigma_{ij}$, i.e. the {\em equation of
state} of the cosmological fluid. For example, standard fluid
dynamics~\cite{LaLi87} gives $\sigma_{ij}=-p \de_{ij}+ \eta
(\nabla_{i}u_{j} +\nabla_{j}u_{i}-\frac{2}{3} \de_{ij}\nabla\cdot\vu) +
\zeta \de_{ij}\nabla\cdot\vu$, where $p$ denotes the pressure and $\eta$
and $\zeta$ are viscosity coefficients.

The equation of state basically relies on the assumption that
cosmological structure formation is driven by matter with negligible
velocity dispersion or pressure, as for example cold dark matter
(CDM). Note that from its definition, Eq.~(\ref{vmoments}), the stress
tensor characterizes the deviation of particle motions from a single
coherent flow (single stream), for which the first term will be the
dominant contribution. Therefore, it is a good approximation to set
$\sigma_{ij}\approx 0 $, at least in the first stages of gravitational
instability when structures did not have time to collapse and
virialize. As time goes on, this approximation will break down at
progressively larger scales, but we will see that at present times at
the scales relevant to large-scale structure, a great deal can be
explored and understood using this simple approximation. In
particular, the breakdown of $\sigma_{ij}\approx 0 $ describes the
generation of velocity dispersion (or even anisotropic pressure) due
to multiple streams, generically known as {\em shell crossing}. We
will discuss this issue further below.

We now turn to a systematic investigation of the solutions of
Eqs.~(\ref{poisson},\ref{continuity},\ref{euler}) for vanishing stress
tensor.

\subsection{Eulerian Linear Perturbation Theory}
\label{eulin}

At large scales, where we expect the Universe to become smooth, the
fluctuation fields in Eqs.~(\ref{rho}-\ref{pot}) can be assumed to be
small compared to the homogeneous contribution described by the first
terms. Therefore, it follows that we can linearize
Eqs.~(\ref{poisson},\ref{continuity},\ref{euler}) to obtain the
equations of motion in the {\em linear regime}

\begin{equation}
{\partial \delta(\vx,\tau) \over{\partial \tau}} + \theta(\vx,\tau) = 0 
\label{linearC},
\end{equation}
\begin{equation}
{\partial \vu(\vx,\tau) \over{\partial \tau}} + {{\cal H}(\tau)}\
\vu(\vx,\tau) = - \nabla
\Phi(\vx,\tau)
\label{linearE},
\end{equation}

where $\theta(\vx,\tau) \equiv \nabla \cdot \vu(\vx,\tau)$ is the
divergence of the velocity field. These equations are now
straightforward to solve. The velocity field, as any vector field,
can be completely described by its divergence $\theta(\vx,\tau)$ and
its vorticity $\vw(\vx,\tau) \equiv \nabla \times \vu(\vx,\tau)$ ,
whose equations of motion follow from Eq.~(\ref{linearE})

\begin{equation}
        {\partial \theta(\vx,\tau) \over{\partial \tau}} + {{\cal H}(\tau)}\
\theta(\vx,\tau) + {3\over 2} \Omega_m(\tau) {\cal H}^2(\tau)
\delta{(\vx,\tau)} =0
        \label{linearV},
\end{equation}
\begin{equation}
        {\partial \vw(\vx,\tau) \over{\partial \tau}} + {{\cal
H}(\tau)}\ \vw(\vx,\tau) = 0 \label{linearW}.
\end{equation}

The vorticity evolution readily follows from Eq.~(\ref{linearW}), $\vw
(\tau) \propto a^{-1}$, i.e. in the linear regime any initial
vorticity decays away due to the expansion of the Universe. The
density contrast evolution follows by taking the time derivative of
Eq.~(\ref{linearV}) and replacing in Eq.~(\ref{linearC}),

\begin{equation}
{\d^{2} D_{1}(\tau) \over{\d \tau^{2}}} + {{\cal 
H}(\tau)}\ \frac{\d D_{1}(\tau)}{\d \tau} = 
{3\over 2} \Omega_m(\tau) {\cal H}^2(\tau) D_{1}(\tau),
        \label{D1}
\end{equation}

where we wrote $\delta(\vx,\tau) = D_{1}(\tau) \delta(\vx,0)$, with
$D_{1}(\tau) $ the {\em linear growth factor}. This equation,
together with the Friedmann equations,
Eqs.~(\ref{friedmann1}-\ref{friedmann2}), determines the growth of
density perturbations in the linear regime as a function of cosmology.
Since it is a second-order differential equation, it has two
independent solutions, let's denote the fastest growing mode
$D_{1}^{(+)}(\tau)$ and the slowest one $D_{1}^{(-)}(\tau)$. The
evolution of the density is then

\begin{equation}
\de(\vx,\tau)= D_{1}^{(+)}(\tau) A(\vx) + D_{1}^{(-)}(\tau) B(\vx),
\label{dlinear}
\end{equation}

where $A(\vx)$ and $B(\vx)$ are two arbitrary functions of position
describing the initial density field configuration, whereas the
velocity divergence [using Eq.~(\ref{linearC})] is given by 

\begin{equation}
\theta(\vx,\tau)= -{\cal H}(\tau) \left[f(\Omega_m,\Omega_\Lambda) A(\vx)
+ g(\Omega_m,\Omega_\Lambda) B(\vx) \right], 
\label{tlinear}
\end{equation}

\begin{equation}
f(\Omega_m,\Omega_\Lambda)\equiv \frac{\d \ln D_{1}^{(+)}}{\d \ln a}=
\frac{1}{{\cal H}} \frac{\d \ln D_{1}^{(+)}}{\d \tau} \qquad
g(\Omega_m,\Omega_\Lambda)=\frac{1}{{\cal H}} \frac{\d \ln D_{1}^{(-)}}{\d
\tau} \label{f1}.
\end{equation}

The most important cases are

\begin{enumerate}
\item When $\Omega_m=1$, $\Omega_{\Lambda}=0$, we have the simple
solution

\begin{equation} D_{1}^{(+)}= a, \qquad \qquad
D_{1}^{(-)}=a^{-3/2}, \qquad \qquad f(1,0)=1,
\label{D1EdS}
\end{equation}
thus density fluctuations grow as the scale factor.

\item When $\Omega_m<1$, $\Omega_{\Lambda}=0$ we have ($x \equiv
1/\Omega_m -1$)~\cite{PeGr75}

\begin{equation}
D_{1}^{(+)}=1+\frac{3}{x}+3\sqrt{\frac{1+x}{x^{3}}}\ln
\left[ \sqrt{1+x}-\sqrt{x} \right] \qquad D_{1}^{(-)}=
\sqrt{\frac{1+x}{x^{3}}}, \label{D1open}
\end{equation}
and the logarithmic derivative can be approximated by~\cite{Peebles76}
\begin{equation} 
f(\Omega_m,0) \approx \Omega_m^{3/5}. \label{f1O}
\end{equation}
        
As $\Omega_m\rightarrow 0$ ($x \gg 1$), $D_{1}^{(+)}\rightarrow 1$ and 
$D_{1}^{(-)}\rightarrow x^{-1}$ and perturbations cease to grow.

\item In the case where there is only matter and vacuum energy, the
linear growth factor admits the integral representation~\cite{Heath77}
as a function of $\Omega_m$ and $\Omega_\Lambda$

\begin{equation}
D_1^{(+)} = H(a)\ \frac{5 \Omega_m}{2}\ \int_0^a \frac{\d a}{a^3
H(a)},
\label{D1int}
\end{equation}
where $H(a)= \sqrt{\Omega_m a^{-3}+ (1-\Omega_m-\Omega_\Lambda) a^{-2}
+ \Omega_\Lambda}$. In general, it is not possible to solve
analytically for $D_{1}^{(+)}$ (unlike $D_{1}^{(-)}$, see
\cite{Heath77}), but can be approximated by~\cite{LLPR91,CPT92}

\begin{equation}
D_{1}^{(+)}\approx \left( \frac{5}{2} \right) \frac{a
 \Omega_m}{\Omega_m^{4/7} -\Omega_\Lambda +
 (1+\Omega_m/2)(1+\Omega_\Lambda/70)},
\label{D1L}
\end{equation}
\begin{equation}
 D_{1}^{(-)}= \frac{{\cal H}}{a},
\label{D1Lm}
\end{equation}
\begin{equation}
f(\Omega_m,\Omega_\Lambda) \approx \frac{1}{[ 1-(\Omega_0 + 
\Omega_\Lambda^0 -1) a + \Omega_\Lambda^0 a^3]^{0.6}},
\label{f1La}
\end{equation}
where $\Omega_\Lambda^0\equiv \Omega_\Lambda(a=1)$. When
$\Omega_m+\Omega_{\Lambda}=1$, we have

\begin{equation}
f(\Omega_m,1-\Omega_m) \approx \Omega_m^{5/9}.
\label{f1L}
\end{equation}
        
\end{enumerate}

Due to Eq.~(\ref{D1Lm}) and Eq.~(\ref{friedmann1}),
$g(\Omega_m,\Omega_\Lambda) = \Omega_m-\Omega_\Lambda/2-1$ holds for 
arbitrary $\Omega_m$ and $\Omega_\Lambda$.

\subsection{Eulerian Non-Linear Perturbation Theory}

We will now consider the evolution of density and velocity fields 
beyond the linear approximation. To do so, we shall first make 
a {\em self-consistent} approximation, that is, we will characterize 
the velocity field by its divergence, and neglect the vorticity 
degrees of freedom. This can be justified as follows. {}From 
Eq.~(\ref{euler}) we can write the vorticity equation of motion

\begin{equation}
\frac{\partial \vw(\vx,\tau)}{\partial \tau} + {\cal H}(\tau) 
\vw(\vx,\tau) - \nabla \times \left[ \vu(\vx,\tau) \times \vw(\vx,\tau) 
\right] = \nabla \times \left(\frac{1}{\rho} \nabla \cdot
\vec{\sigma} \right) 
\label{vorticity},
\end{equation} 

where we have temporarily restored the stress tensor contribution
($\sigma_{ij}$) to the conservation of momentum. We see that if
$\sigma_{ij} \approx 0$, as in the case of a pressureless perfect
fluid, if the primordial vorticity vanishes, it remains zero at all
times. On the other hand, if the initial vorticity is non-zero, we
saw in the previous section that in the linear regime vorticity decays
due to the expansion of the Universe; however, it can be amplified
non-linearly through the third term in Eq.~(\ref{vorticity}). In what
follows, we shall assume that the initial vorticity vanishes, thus
Eq.~(\ref{vorticity}) together with the equation of state $\sigma_{ij}
\approx 0$ guarantees that vorticity remains zero throughout the
evolution. We must note, however, that this assumption is
self-consistent only as long as the condition $\sigma_{ij} \approx 0$
remains valid; in particular, multi-streaming and shocks can generate
vorticity (see for instance~\cite{PiBe99}).  This is indeed expected
to happen at small enough scales.  We will come back to this point in
order to interpret the breakdown of perturbation theory at small
scales.

The assumption of perturbation theory is that it is possible to expand
the density and velocity fields about the linear solutions,
effectively treating the variance of the linear fluctuations as a
small parameter (and assuming no vorticity in the velocity
field). Linear solutions correspond to simple (time dependent)
scalings of the {\em initial} density field; thus we can write

\be
\delta(\vx,t)=\sum_{n=1}^{\infty}
\delta^{(n)}(\vx,t),\ \ \ 
\theta(\vx,t)=\sum_{n=1}^{\infty}\theta^{(n)}(\vx,t), 
\label{deltathetaPT}
\ee
where $\delta^{(1)}$ and $\theta^{(1)}$ are linear in the initial
density field, $\delta^{(2)}$ and $\theta^{(2)}$ are quadratic in the
initial density field, etc.

\subsubsection{The Equations of Motion in the Fourier Representation}

At large scales, when fluctuations are small, linear perturbation
theory provides an adequate description of cosmological fields. In
this regime, different Fourier modes evolve independently conserving
the primordial statistics. Therefore, it is natural to Fourier
transform Eqs.~(\ref{poisson},\ref{continuity},\ref{euler}) and work
in Fourier space.  Our convention for the Fourier transform of a field
$A(\vx,\tau)$ is:

\begin{equation}
\tilde{A}(\vk,\tau)=\int {\d^3\vx \over{(2 \pi)^3}} \exp (-i
\vk \cdot \vx)\ A(\vx,\tau)
\label{fourier}. 
\end{equation}

When non-linear terms in the perturbation series are taken into
account, the equations of motion in Fourier space show the coupling
between different Fourier modes characteristic of non-linear
theories. Taking the divergence of Equation~(\ref{euler}) and Fourier
transforming the resulting equations of motion we get:

\begin{equation}
{\partial \tilde{\delta}(\vk,\tau) \over{\partial \tau}} +
\tilde{\theta}(\vk,\tau) = - \int \d^3\vk_1 \d^3\vk_2
\delta_D(\vk - \vk_{12}) \alpha(\vk_1, \vk_2)
\tilde{\theta}(\vk_1,\tau) \tilde{\delta}(\vk_2,\tau) \label{ddtdelta},
\end{equation}
\begin{eqnarray}
& & {\partial \tilde{\theta}(\vk,\tau) \over{\partial \tau}} +
{\cal H}(\tau)\ \tilde{\theta}(\vk,\tau) + {3\over 2}
\Omega_m {\cal H}^2(\tau) \tilde{\delta}(\vk,\tau) = - \int
\d^3\vk_1 \d^3\vk_2 \delta_D(\vk - \vk_{12}) \nonumber \\
& & \qquad \qquad \qquad \qquad \qquad \qquad \qquad \qquad
\times \beta(\vk_1, \vk_2) 
\tilde{\theta}(\vk_1,\tau) \tilde{\theta}(\vk_2,\tau) 
\label{ddttheta},
\end{eqnarray}

($\delta_D$ denotes the three-dimensional Dirac delta distribution)
where the functions

\begin{equation}
\alpha(\vk_1, \vk_2) \equiv {\vk_{12} \cdot \vk_1
\over{ k_1^2}}, \ \ \ \ \ \beta(\vk_1, \vk_2) \equiv
{k_{12}^2 (\vk_1 \cdot \vk_2 )\over{2 k_1^2 k_2^2}} \label{albe}
\end{equation}

encode the non-linearity of the evolution (mode coupling) and come 
from the non-linear terms in the continuity 
equation~(\ref{continuity}) and the Euler equation~(\ref{euler}) 
respectively. {}From equations~(\ref{ddtdelta})-(\ref{ddttheta}) we see 
that the evolution of $\tilde{\delta}(\vk,\tau)$ and $ 
\tilde{\theta}(\vk,\tau)$ is determined by the mode coupling of the 
fields at all pairs of wave-vectors $\vk_1$ and $\vk_2$ whose sum is
$\vk$, as required by translation invariance in a spatially
homogeneous Universe.

\subsubsection{General Solutions in Einstein-de~Sitter Cosmology}
\label{subsec:eds}

Let's first consider an Einstein-de Sitter Universe, for which
$\Omega_m =1$ and $\Omega_\Lambda =0$. In this case the Friedmann
equation, Eq.~(\ref{friedmann1}), implies $a(\tau) \propto \tau^2$,
${\cal H}(\tau)=2/\tau$, and scaling out an overall factor of ${\cal
H}$ from the velocity field brings
Eqs.~(\ref{ddtdelta}-\ref{ddttheta}) into homogeneous form in $\tau$
or, equivalently, in $a(\tau)$.  As a consequence, these equations can
formally be solved with the following perturbative
expansion~\cite{GGRW86,JaBe94,MSS92},
\begin{equation}
\tilde{\delta}(\vk,\tau) = \sum_{n=1}^{\infty} a^n(\tau)
\delta_n(\vk),\ \ \ \ \ \tilde{\theta}(\vk,\tau) = - 
{\cal H}(\tau) \sum_{n=1}^{\infty} a^n(\tau) \theta_n(\vk)
\label{ptansatz},
\end{equation}
where only the fastest growing mode is taken into account.  Remarkably
it implies that the PT expansions defined in Eq.~(\ref{deltathetaPT})
are actually expansions with respect to the linear density field with
time independent coefficients. At small $a$ the series are dominated
by their first term, and since $\theta_1(\vk) =
\delta_1(\vk) $ from the continuity equation, $\delta_1(\vk)$
completely characterizes the linear fluctuations.

The equations of motion, Eqs.~(\ref{ddtdelta}-\ref{ddttheta})
determine $\delta_n(\vk) $ and $\theta_n(\vk)$ in terms of the linear
fluctuations to be:
\label{solu}
\begin{equation}
\delta_n(\vk) = \int \d^3\vq_1 \ldots \int \d^3\vq_n\, \delta_D(\vk -
\vq_{1\ldots n}) F_n(\vq_1, \ldots ,\vq_n) \delta_1(\vq_1) \ldots
\delta_1(\vq_n) \label{ec:deltan},
\end{equation}
\begin{equation}
\theta_n(\vk) = \int \d^3\vq_1 \ldots \int \d^3\vq_n\, \delta_D(\vk -
\vq_{1\ldots n}) G_n(\vq_1, \ldots ,\vq_n) \delta_1(\vq_1) \ldots
\delta_1(\vq_n) \label{ec:thetan},
\end{equation}

where $F_n$ and $G_n$ are homogeneous functions
of the wave vectors \{$\vq_1, \ldots ,\vq_n $\} with degree
zero. They are constructed from the fundamental mode coupling
functions $\alpha(\vk_1, \vk_2)$ and $\beta(\vk_1,
\vk_2)$ according to the recursion relations ($n \geq 2$,
see~\cite{GGRW86,JaBe94} for a derivation):

\begin{eqnarray}
F_n(\vq_1, \ldots ,\vq_n) &=& \sum_{m=1}^{n-1} { G_m(\vq_1, \ldots ,\vq_m)
 \over{(2n+3)(n-1)}} \Bigl[(2n+1) \alpha(\vk_1,\vk_2) F_{n-m}(\vq_{m+1},
 \ldots ,\vq_n) \nonumber \\ & & +2 \beta(\vk_1, \vk_2)
 G_{n-m}(\vq_{m+1}, \ldots ,\vq_n) \Bigr] \label{Fn},
\end{eqnarray}
\begin{eqnarray}
G_n(\vq_1, \ldots ,\vq_n) &=& \sum_{m=1}^{n-1} { G_m(\vq_1, \ldots ,\vq_m)
\over{(2n+3)(n-1)}} \Bigl[3 \alpha(\vk_1,\vk_2) F_{n-m}(\vq_{m+1}, \ldots
,\vq_n) \nonumber \\ & & +2n \beta(\vk_1, \vk_2) G_{n-m}(\vq_{m+1},
\ldots ,\vq_n) \Bigr] \label{Gn},
\end{eqnarray}

(where $\vk_1 \equiv \vq_1 + \ldots + \vq_m$, $\vk_2 \equiv \vq_{m+1} + 
\ldots + \vq_n$, $\vk \equiv \vk_1 +\vk_2$, and $F_1= G_1 \equiv 1$) 

For $n=2$ we have:
\begin{equation}
F_2(\vq_1,\vq_2) = \frac{5}{7} + \frac{1}{2} \frac{\vq_1 \cdot
 \vq_2}{q_1 q_2} (\frac{q_1}{q_2} + \frac{q_2}{q_1}) + \frac{2}{7}
 \frac{(\vq_1 \cdot \vq_2)^2}{q_1^2 q_2^2} \label{F2},
\end{equation}
\begin{equation}
G_2(\vq_1,\vq_2) = \frac{3}{7} + \frac{1}{2} \frac{\vq_1 \cdot
\vq_2}{q_1 q_2} (\frac{q_1}{q_2} + \frac{q_2}{q_1}) + \frac{4}{7}
\frac{(\vq_1 \cdot \vq_2)^2}{q_1^2 q_2^2} \label{G2}.
\end{equation}

Explicit expressions for the kernels $F_3$ and $F_4$ are 
given in~\cite{GGRW86}. Note that 
the symmetrized kernels, $F_n^{(s)}$ (obtained by a summation of $F_n$
with all possible permutations of the variables),
have the following properties~\cite{GGRW86,Wise88}:

\begin{enumerate}
\item As $\vk = \vq_1 + \dots + \vq_n$ goes to zero, but the individual
$\vq_i$ do not, $F_n^{(s)} \propto k^2$. This is a consequence of
momentum conservation in center of mass coordinates.

\item As some of the arguments of $F_n^{(s)}$ get large
but the total sum $\vk = \vq_1 + \dots +\vq_n$ stays fixed, the kernels
vanish in inverse square law. That is, for $p \gg q_i$, we have:
        
\begin{equation} 
F_n^{(s)}(\vq_1, \dots ,\vq_{n-2},\vp,-\vp) \propto k^2/p^2
\label{inversesq}, 
\end{equation}
        
and similarly for $G_n^{(s)}$.

\item If one of the arguments $\vq_i$ of $F_n^{(s)}$ or $G_n^{(s)}$
goes to zero, there is an infrared divergence of the form $\vq_i
/q_i^2$. This comes from the infrared behavior of the mode coupling
functions $\alpha(\vk_1,\vk_2)$ and $\beta(\vk_1, \vk_2)$. There are no
infrared divergences as partial sums of several wavevectors go to
zero.
\end{enumerate}

A simple application of the recursion relations is to derive the 
corresponding recursion relation for {\em vertices} $\nu_{n}$ and 
$\mu_{n}$ which correspond to the spherical average of the PT kernels:

\begin{equation}
        \nu_{n} \equiv n! \int \frac{\d\Omega_{1}}{4\pi}
        \ldots \frac{\d\Omega_{n}}{4\pi}
        F_{n}(\vk_{1},\ldots,\vk_{n}) \label{nu_n},
\end{equation}
\begin{equation}
        \mu_{n} \equiv n! \int \frac{\d\Omega_{1}}{4\pi}
        \ldots \frac{\d\Omega_{n}}{4\pi}
        G_{n}(\vk_{1},\ldots,\vk_{n}) \label{mu_n}.
\end{equation}

Since the kernels $F_n$ and $G_n$ depend only on the ratios $k_i/k_j$,
the vertices depend a priori on these quantities as well. Considering
the equations (\ref{Fn}, \ref{Gn}), one can see that the angle
integrations can be done recursively: it is possible to integrate
first on the angle between the vectors $\vk_1=\vq_1+\dots+
\vq_m$ and $\vk_2=\vq_{m+1}+\dots+\vq_n$, which amounts to replace
$\alpha(\vk_1,\vk_2)$ and $\beta(\vk_1,\vk_2)$ by their angular
averages $\overline{\alpha}=1$ and $\overline{\beta}=1/3$.
As a result we have,
\begin{equation}
        \nu_{n} = \sum_{m=1}^{n-1} {n \choose m}
        \frac{\mu_{m}}{(2n+3)(n-1)} \left[ (2n+1) \nu_{n-m} +
        \frac{2}{3} \mu_{n-m} \right],
\label{fb:nunrec}
\end{equation}
\begin{equation}
        \mu_{n} = \sum_{m=1}^{n-1} {n \choose m}
        \frac{\mu_{m}}{(2n+3)(n-1)} \left[ 3 \nu_{n-m} + \frac{2}{3} n
        \mu_{n-m} \right],
\label{fb:munrec}
\end{equation}
and the vertices are thus pure numbers, e.g.: 
\be
\nu_1=\mu_1= 1 ~;~
\nu_2={34\over 21} ~;~ \nu_3={682\over189} 
 ~;~ \mu_2=- {26\over 21} ~;~ \mu_3={142\over 63} 
\label{nuq}
\ee
This recursion relation plays a central role for the derivation of
many results in PT~\cite{Bernardeau92b}.

In particular, it can be shown that it is directly related to the
spherical collapse dynamics~\cite{Bernardeau92b,FoGa98a}. In this case
the initial density field is such that it has a spherical symmetry
around $\vx=0$. As a consequence the Fourier transform of the linear
density field $\delta_1(\vk)$ depends only on the norm of $\vk$, and
this property remains valid at any stage of the dynamics. Then the
central density for such initial conditions, $\delta_{\rm sc}$, can be
written (assuming $\Omega_m=1$ for definiteness) 
\be
\delta_{\rm sc}(a)=\sum_n a^n\int\d^3\vq_1\dots\int\d^3\vq_n
F_n(\vq_1,\dots,\vq_n)\delta_1(\vert\vq_1\vert)\dots\delta_1(\vert\vq_n\vert).
\ee
Performing first the integration over the angles of the wave vectors,
one recovers,
\be
\delta_{\rm sc}(a )=\sum_n {\nu_n\over n!}a^n\,\epsilon^n
\ee
with $\epsilon=\int\d^3\vq\,\delta_1(\vert\vq\vert)$.
Similarly the central velocity divergence for the spherical
collapse is expanded in terms of the $\mu_n$ parameters.
The angular averages of the PT kernels are thus directly related
to the spherical collapse dynamics. This result is valid for any
cosmological model.

\subsubsection{Cosmology Dependence of Non-Linear Growth Factors}
\label{sec:codeno}

In general the PT expansion is more complicated because the solutions
at each order become non-separable 
functions of $\tau$ and $\vk$~\cite{BJCP92,BCHJ95,Bernardeau94c,CLMM95}.
In particular the growing mode at order $n$ does not scale as 
$D_{1}^n(\tau)$ (or $a^{n}(\tau)$ as in Eq.~(\ref{ptansatz})). 

However, using the recursion relations, we can easily find the full
dependence on cosmological parameters for the vertices, that is, the
dependence that arises in the spherical collapse approximation. The PT
kernels can then be constructed order by order in terms of these
solutions~\cite{Bernardeau94c}. In the spherical model, we can write

\begin{equation}
        \de(\tau)= \sum_{n=1}^{\infty} \frac{\nu_{n}(\tau)}{n!}\ [D_{1}(\tau)\ 
        \epsilon]^{n},
        \label{dSC}
\end{equation}
\begin{equation}
        \te(\tau)= -{\cal H}(\tau) f(\Omega_m,\Omega_{\Lambda}) 
        \sum_{n=1}^{\infty} \frac{\mu_{n}(\tau)}{n!}\ [D_{1}(\tau)\ 
        \epsilon]^{n}.
        \label{tSC}
\end{equation}

{}From the Fourier space equations of motion,
Eqs.~(\ref{ddtdelta}-\ref{ddttheta}), and taking into account that the
spherical averages of $\alpha$ and $\beta$ can be taken at once, one
gets,
\begin{equation}
{\d \nu_n\over \d \log D_1}+ n \nu_{n} -\mu_{n} =
\sum_{m=1}^{n-1} {n \choose m} \nu_{n-m} \mu_{m},
\label{nunt}
\end{equation}
\begin{equation}
{\d \mu_n\over \d \log D_1}+ n \mu_{n} + 
\left(\frac{3\Omega_m}{2f^{2}}-1\right)\mu_{n} -\frac{3\Omega_m}{2f^{2}} 
\nu_{n}=\frac{1}{3} \sum_{m=1}^{n-1} {n \choose m} \mu_{n-m} \mu_{m},
\label{munt}
\end{equation}
noting that ${\d \log D_1}={{\cal H} f\d \tau}$.  This hierarchy of
differential equations must then be solved numerically at each
order. The results for $n=2,3$ show that indeed the dependence of the
vertices on cosmological parameters is a few percent effect at
most~\cite{Bernardeau94c,FoGa98b}.

For a perfect fluid with a equation of state  $p=\eta \rho$
 we have ~\cite{GaLo01}
\be
\nu_2 = {\frac{2\,\left( 17 + 48\,\eta + 27\,\eta^2 \right) }
{3\,\left( 1 + \eta \right) \,\left( 7 + 15\,\eta \right) }}. 
\ee
for an Einstein-de Sitter Universe. Of course, this
reduces to Eq.~(\ref{nuq}) as $\eta \rightarrow 0$.
For the Brans-Dicke Cosmology~\cite{BrDi61}, with a 
coupling $\omega$ to gravity:
\be
\nu_2 = {{34\omega + 56}\over{21 \omega + 36}},
\ee
which reduces to the standard result $\nu_2=34/21$ in the
limit $\omega \rightarrow \infty $ (see~\cite{GaLo01} for details
and results for $\nu_4$). Even in these extreme cosmologies, the
possible variations of $\nu_2$ are quite small given the observational
constraints on $\eta$ and $\omega$~\cite{GaLo01}.

\subsubsection{Approximate Solutions in Arbitrary Cosmology}

This quite remarkable result is asking for an explanation.  It is
indeed possible to show that a simple approximation to the equations
of motion for general $\Omega_m$ and $\Omega_\Lambda$ leads to
separable solutions to arbitrary order in PT and the same recursion
relations as in the Einstein-de Sitter case~\cite{SCFFHM98}. All the
information on the dependence of the PT solutions on the cosmological
parameters $\Omega_m$ and $\Omega_\Lambda$ is then encoded in the
linear growth factor, $D_1(\tau)$.

In linear PT, the growing-mode solution to the equations of motion
(\ref{ddtdelta}) and (\ref{ddttheta}) reads
\begin{eqnarray}
\delta(\vk,\tau) &=& D_1(\tau) \delta_1(\vk), \\
\theta(\vk,\tau) &=& - {\cal H}(\tau) f(\Omega_m,\Omega_\Lambda) D_1(\tau) 
\delta_1(\vk),
\end{eqnarray} 

where $D_1(\tau)$ is linear growing mode. As mentioned before, we look
for separable solutions of the form (compare with Eq.~(\ref{ptansatz})
)

\begin{eqnarray}
\delta(\vk,\tau) &=& \sum_{n=1}^{\infty} D_n(\tau) \delta_n(\vk), \\
\theta(\vk,\tau) &=& - {\cal H}(\tau) f(\Omega_m,\Omega_\Lambda)
\sum_{n=1}^{\infty} E_n(\tau) \theta_n(\vk),
\end{eqnarray}

{}From the equations of motion (\ref{ddtdelta}) and (\ref{ddttheta}) we
get for the $n^{\rm th}$ order solutions,

\begin{eqnarray}
\frac{\d D_n}{\d\log D_1} \de_n - E_n \te_n &=& \int \d^3k_1 \d^3k_2
\delta_D(\vk-\vk_{12}) \alpha(\vk,\vk_1)\nonumber \\ & & \qquad \qquad
\times \sum_{m=1}^{n-1} D_{n-m}E_m \te_m(\vk_1) \de_{n-m}(\vk_2),
\label{dn}
\end{eqnarray}

\begin{eqnarray}
& & \frac{\d E_n}{\d\log D_1} \te_n + \Big(\frac{3\
\Omega_m}{2f^2}-1\Big) E_n \te_n - \frac{3\ \Omega_m}{2f^2} D_n \de_n =
\nonumber \\ & & \int \d^3\vk_1 \d^3\vk_2 \delta_D(\vk-\vk_{12})
\beta(\vk,\vk_1,\vk_2) \sum_{m=1}^{n-1} E_{n-m}E_m \te_m(\vk_1)
\te_{n-m}(\vk_2).
\label{tn}
\end{eqnarray}
By simple inspection, we see that if $f(\Omega_m,\Omega_\Lambda) =
\Omega_m^{1/2}$, then the system of equations becomes indeed separable,
with $D_n = E_n = (D_1)^n$. In fact, the recursion relations then
reduce to the standard $\Omega_m=1$, $\Omega_\Lambda=0$ case, shown in
equations (\ref{Fn}) and (\ref{Gn}). Then $\Omega_m/f^2=1$ leads to
separability of the PT solutions to any order, generalizing what has
been noted before in the case of second order PT~\cite{MaFr91}. {}From
Section~\ref{eulin}, the approximation $f(\Omega_m,\Omega_\Lambda) \approx
\Omega_m^{1/2}$ is actually very good in practice. As a result, for
example, as we review in the next section, the exact solution for the
$\Omega_\Lambda=0$ case gives $D_2/(D_1)^2 =1+3/17 (\Omega_m^{-2/63}-1)$,
extremely insensitive to $\Omega_m$, even more than what the
approximation $f(\Omega_m,\Omega_\Lambda) = \Omega_m^{3/5}\approx
\Omega_m^{1/2}$ would suggest, since for most of the time evolution 
$\Omega_m$ and $\Omega_\Lambda$ are close to their Einstein-de Sitter values.

\subsubsection{The Density and Velocity Fields up to Third Order}

The computations of the local density field can be done order by order
for any cosmological model.  We give here their explicit expression
up to third order.  The detailed calculations can be found in
\cite{Bernardeau94c}.  Different approaches have been used in the
literature to do such calculations~\cite{Buchert94,CLMM95,BCHJ95}. 
The direct calculation appears to be the most secure, if not the rapid
or most instructive.

The time dependence of the solutions can be written as a function of 
$D_1(\tau)$, $\nu_2(\tau)$, $\nu_3(\tau)$ and an auxiliary function
$\lambda_3(\tau)$ which satisfies,
\be
{\d^2(\lambda_3\,D_1^3)\over\d\tau^2}+\mH\,{\d(\lambda_3 \,D_1^3)\over\d\tau}-
{3\over2}\mH^2\,\Omega_m\,\lambda_3\,D_1^3={3\over2}\mH^2\,\Omega_m\,D_1^3,
\ee
with $\lambda_3\sim9/10$ when $\tau\to 0$.
The geometrical dependences can all be expressed in terms of
the two functions $\alpha(\vq_1,\vq_j)$ [see Eq.~(\ref{albe})] and 
\be
\gamma(\vq_i,\vq_j)={1\over2}\left[\alpha(\vq_i,\vq_j)+\alpha(\vq_j,\vq_i)
\right]-\beta(\vq_i,\vq_j)=1-{\left(\vq_i.\vq_j\right)^2\over(q_i^2
q_j^2)},
\ee
which for short will be denoted $\alpha_{i,j}$ and $\gamma_{i,j}$ respectively.
Then we have,
\ba
F_2(\vq_1,\vq_2)&=&
\left({3\over4}\nu_2-{3\over 2}\right)\gamma_{1,2}+\alpha_{1,2}\\
G_2(\vq_1,\vq_2)&=&-f(\Omega_m,\Omega_\Lambda)\left[
\left({3\over4}\mu_2-{3\over 2}\right)\gamma_{1,2}+\alpha_{1,2}\right]
\ea
for the second-order solutions. Their symmetrized parts can be shown to
take the form (see Section~\ref{lagNLPT}),
\begin{equation}
         F_2^{(s)}(\vq_1,\vq_2) = \frac{1}{2}(1+\varepsilon) + \frac{1}{2} 
         \frac{\vq_1 \cdot \vq_2}{q_1 q_2} (\frac{q_1}{q_2} +
         \frac{q_2}{q_1}) + \left(\frac{2}{7} -\frac{\varepsilon }{2}
         \right) \frac{(\vq_1 \cdot \vq_2)^2}{q_1^2 q_2^2}
        \label{F2g},
\end{equation}
\begin{equation}
         G_2^{(s)}(\vq_1,\vq_2) = \varepsilon + \frac{1}{2} 
         \frac{\vq_1 \cdot \vq_2}{q_1 q_2} (\frac{q_1}{q_2} + 
         \frac{q_2}{q_1}) + \left( 1-\varepsilon \right) 
         \frac{(\vq_1 \cdot \vq_2)^2}{q_1^2 q_2^2} \label{G2g},
\end{equation}
where $\varepsilon \approx (3/7) \Omega_m^{-2/63}$ for $\Omega_m\ga
0.1$~\cite{BCHJ95}. At third order the kernel reads,
\ba
&&F_3(\vq_1,\vq_2,\vq_3)=
\mR_{1}+\nu_2\,\mR_{2}+\nu_3\,\mR_{3}+\lambda_3\,\mR_{4},
\ea
where, using the simplified notation  
$\alpha_{ij,k}=\alpha(\vq_i+\vq_j,\vq_k)$,
 $\alpha_{i,jk}=\alpha(\vq_i,\vq_j+\vq_k)$
and similar definitions for $\gamma_{ij,k}$ and $\gamma_{i,jk}$, we 
have 
\ba
\mR_{1}&=&
\left({1\over2}\alpha_{3,12}+{1\over2}\alpha_{12,3}-{1\over3}\gamma_{3,12}
\right)\alpha_{1,2}+
\left(-{3\over2}\alpha_{12,3}-{4\over3}\alpha_{3,12}
+{5\over2}\gamma_{3,12}\right)\,\gamma_{1,2}, \nonumber \\ & & \\
\mR_{2}&=&
{3\over4}\left(\alpha_{3,12}+\alpha_{12,3}-3\gamma_{3,12}\right)\gamma_{1,2},\\
\mR_{3}&=&
{3\over8}\gamma_{3,12}\,\gamma_{1,2}, \\
\mR_{4}&=&
{2\over3}\gamma_{3,12}\,\alpha_{1,2}
-\left({1\over3}\alpha_{3,12}+{1\over2}\gamma_{3,12}\right)\gamma_{1,2}.
\ea These results exhibit the explicit time and geometrical dependence
of the density field up to third order (a similar expression can be
found for $G_3$, see~\cite{Bernardeau94c}).  In Chapter~\ref{chapter5}
we examine the consequences of these results for the statistical
properties of the cosmic fields.

\subsubsection{Non-Linear Growing and Decaying Modes}
\label{tevol}

Perturbation theory describes the non-linear dynamics as a collection
of linear waves, $\de_1(\vk)$, interacting through the mode-coupling
functions $\alpha$ and $\beta$ in Eq.~(\ref{albe}). Even if the
initial conditions are set in the growing mode, after scattering due
to non-linear interactions waves do not remain purely in the growing
mode.  In the standard treatment, described above, the sub-dominant
time-dependencies that necessarily appear due to this process have
been neglected, i.e., only the fastest growing mode (proportional to
$D_1^{n}$) is taken into account at each order $n$ in PT. Here we
discuss how one can generalize the standard results to include the
full time dependence of the solutions at every order in
PT~\cite{Scoccimarro98,Scoccimarro00c}. This is necessary, for
example, to properly address the problem of transients in $N$-body
simulations in which initial conditions are set up using the
Zel'dovich Approximation (see Section~\ref{fb:LagrangianDyn}). This is
reviewed in Section~\ref{s:transients}. In addition, the approach
presented here can be useful to address evolution from non-Gaussian
initial conditions.

The equations of motion can be rewritten in a more symmetric form by
defining a two-component ``vector'' $\Psi_a(\vk,z)$, where $a=1,2$,
$z\equiv \ln a$ (we assume $\Omega_m=1$ for definiteness), and: 

\be \Psi_a(\vk,z) \equiv \Bigg( \delta(\vk,z),\ -\theta(\vk,z)/{\cal
H} \Bigg), \ee which leads to the following equations of motion (we
henceforth use the convention that repeated Fourier arguments are
integrated over)

\be \partial_z \Psi_a(\vk,z) + \Omega_{ab} \Psi_b(\vk,z) =
\gamma_{abc}(\vk,\vk_1,\vk_2) \ \Psi_b(\vk_1,z) \ \Psi_c(\vk_2,z),
\label{eom}
\ee where $\gamma_{abc}$ is a matrix whose only non-zero elements are
$\gamma_{121}(\vk,\vk_1,\vk_2)=\delta_D(\vk-\vk_1-\vk_2) \
\alpha(\vk,\vk_1)$ and
$\gamma_{222}(\vk,\vk_1,\vk_2)=\delta_D(\vk-\vk_1-\vk_2) \
\beta(\vk_1,\vk_2)$, and

\be \Omega_{ab} \equiv \Bigg[ 
\begin{array}{cc}
0 & -1 \\ -3/2 & 1/2
\end{array}        \Bigg].
\ee

The somewhat complicated expressions for the PT kernels recursion
relations in Sect.~\ref{subsec:eds} can be easily derived in this
formalism. The perturbative solutions read [see Eq.~(\ref{ptansatz})]

\be \Psi_a(\vk,z) = \sum_{n=1}^\infty {\rm e}^{n z} \ \psi_a^{(n)}(\vk)
\label{pta},
\ee which leads to

\be (n\de_{ab}+\Omega_{ab})\ \psi_b^{(n)}(\vk) =
\gamma_{abc}(\vk,\vk_1,\vk_2)  \sum_{m=1}^{n-1} \psi_b^{(n-m)}(\vk_1)
\ \psi_c^{(m)}(\vk_2).  \ee

Now, let $\sigma_{ab}^{-1}(n) \equiv n\de_{ab}+\Omega_{ab}$, then we
have:

\be \psi_a^{(n)}(\vk) = \sigma_{ab}(n) \
\gamma_{bcd}(\vk,\vk_1,\vk_2)  \sum_{m=1}^{n-1} \psi_c^{(n-m)}(\vk_1)
\ \psi_d^{(m)}(\vk_2),
\label{GGRWm}
\ee

\noindent where 

\be \sigma_{ab}(n) = \frac{1}{(2n+3)(n-1)} \Bigg[ \begin{array}{cc}
2n+1 & 2 \\ 3 & 2n \end{array} \Bigg].  \ee 

Equation~(\ref{GGRWm}) is the equivalent of the recursion relations in
Eqs.~(\ref{Fn}-\ref{Gn}), for the $n^{\rm th}$ order Fourier amplitude
solutions $\psi_a^{(n)}(\vk)$. 

To go beyond this, that is, to incorporate the transient behavior
before the asymptotics of solutions in Eq.~(\ref{pta}) are valid, it
turns out to be convenient to write down the equation of motion,
Eq.~(\ref{eom}), in integral form. Laplace transformation in the
variable $z$ leads to:

\bea \sigma_{ab}^{-1}(\omega)\  \Psi_b(\vk,\omega) = \phi_a(\vk) +
\gamma_{abc}(\vk,\vk_1,\vk_2) \oint \frac{\d\omega_1}{2\pi i} \
\Psi_b(\vk_1,\omega_1) \Psi_c(\vk_2,\omega-\omega_1), \nonumber \\
\label{eom2}
\eea

\noindent where $\phi_a(\vk)$ denote the initial conditions, that is
$\Psi_a(\vk,z=0) \equiv \phi_a(\vk)$. Multiplying by the matrix
$\sigma_{ab}$, and performing the inversion of the Laplace transform
gives~\cite{Scoccimarro00c}

\bea \Psi_a(\vk,z) &=& g_{ab}(z) \ \phi_b(\vk) + \int_0^z  \d s \
g_{ab}(z-s) \ \gamma_{bcd}(\vk,\vk_1,\vk_2)\ \Psi_c(\vk_1,s)
\Psi_d(\vk_2,s),  \nonumber \\
\label{eomi}
\eea

\noindent where the {\em linear propagator} $g_{ab}(z)$ is defined as
($c>1$ to pick out the standard retarded
propagator~\cite{Scoccimarro98}) 

\be g_{ab}(z) = \oint_{c-i\infty}^{c+i\infty} \frac{\d\omega}{2\pi i}
\sigma_{ab}(\omega) \ {\rm e}^{\omega z} = \frac{{\rm e}^z}{5} \Bigg[
\begin{array}{rr} 3 & 2 \\ 3 & 2 \end{array} \Bigg] - \frac{{\rm
e}^{-3z/2}}{5} \Bigg[ \begin{array}{rr} -2 & 2 \\ 3 & -3 \end{array}
\Bigg],
\label{prop}
\ee 

\noindent for $z\geq 0$, whereas $g_{ab}(z) =0$ for $z<0$ due to
causality, $g_{ab}(z) \rightarrow \de_{ab}$ as $z\rightarrow 0^{+}$.
The first term in Eq.~(\ref{prop}) represents the propagation of
linear growing mode solutions, where the second corresponds to the
decaying modes propagation. Equation~(\ref{eomi}) can be thought as an
equation for $\Psi_a(\vk,z)$ in the presence of an ``external source''
$\phi_b(\vk)$ with prescribed statistics given by the initial
conditions\footnote{This is essentially a field-theoretic description
of gravitational instability, non-linear corrections can be thought as
loop corrections to the propagator and the vertex given by the
$\gamma_{abc}$ matrix, see~\cite{Scoccimarro00c} for details.}. It
contains the full time dependence of non-linear solutions, as will be
discussed in detail in Sect.~\ref{s:transients}. To recover the
standard (asymptotic) time dependence one must take the initial
conditions to be set in the growing mode, $\phi_b \propto (1,1)$,
which vanishes upon contraction with the second term in
Eq.~(\ref{prop}), and reduces to the familiar linear scaling
$\phi_a(z) = {\rm e}^z \phi_a(0) = a(\tau)\ \phi_a(0)$; and, in
addition, set the lower limit of integration in Eq.~(\ref{eomi}) to
$s=-\infty$, to place initial conditions ``infinitely far away'' in
the past.

\subsection{Lagrangian Dynamics}
\label{fb:LagrangianDyn}

So far we have dealt with density and velocity fields and their
equations of motion.  However, it is possible to develop non-linear PT
in a different framework, the so-called Lagrangian scheme, by
following the trajectories of particles or fluid
elements~\cite{Zeldovich70,Buchert89,MABPR91}, rather than studying
the dynamics of density and velocity fields\footnote{It is also
possible to study Lagrangian dynamics of density and velocity fields
following the fluid elements, by using the convective derivative $D/Dt
\equiv \partial/\partial t + \vu\cdot \nabla$ in the equations of
motion, Eqs.~(\ref{continuity}-\ref{euler}).  We will not discuss this
possibility here, but e.g. see~\cite{BeJa94,HuBe96}}.  In Lagrangian
PT\footnote{For reviews of Lagrangian PT, see
e.g.~\cite{Buchert96,Bouchet96}.}, the object of interest is the
displacement field ${\bf \Psi}(\vq)$ which maps the initial particle
positions $\vq$ into the final Eulerian particle positions $\vx$,

\begin{equation}
\vx(\tau) = \vq + {\bf \Psi}(\vq,\tau).
\label{Lmap}
\end{equation}
The equation of motion for particle trajectories $\vx(\tau)$ is then 

\begin{equation}
\frac{\d^2 \vx}{\d \tau^2} + {\cal H}(\tau) \ \frac{\d \vx}{\d \tau}= -
\del \Phi, 
\end{equation}
where $\Phi$ denotes the gravitational potential, and $\del$ the
gradient operator in Eulerian coordinates $\vx$. Taking the
divergence of this equation we obtain 

\begin{equation}
J(\vq,\tau)\ \del \cdot \Big[ \frac{\d^2 {\bf \Psi}}{\d \tau^2} + {\cal
H}(\tau) \ \frac{\d {\bf \Psi}}{\d \tau} \Big] = \frac{3}{2} \Omega_m {\cal H}^2
(J-1),
\label{leom}
\end{equation}

where we have used Poisson equation together with the fact that the 
density field obeys $\bar{\rho}\ (1+\de(\vx)) \d^{3}x = \bar{\rho}\ \d^{3}q$, 
thus

\begin{equation}
1+\de(\vx) =\frac{1}{{\rm Det}\Big( \de_{ij}+ \Psi_{i,j} \Big)} \equiv 
\frac{1}{J(\vq,\tau)},
\label{dlag}
\end{equation}

where $\Psi_{i,j} \equiv \partial \Psi_i /\partial \vq_j$, and
$J(\vq,\tau)$ is the Jacobian of the transformation between Eulerian
and Lagrangian space. Note that when there is shell crossing, i.e.
fluid elements with different initial positions $\vq$ end up at the
same Eulerian position $\vx$ through the mapping in Eq.~(\ref{Lmap}),
the Jacobian vanishes and the density field becomes singular. At these
points the description of dynamics in terms of a mapping does not hold
anymore. 

Equation~(\ref{leom}) can be fully rewritten in terms of Lagrangian 
coordinates by using that $\del_i = ( \de_{ij}+ \Psi_{i,j})^{-1} 
\del_{q_j}$, where $\del_q \equiv \partial /\partial \vq$ denotes the 
gradient operator in Lagrangian coordinates. The resulting non-linear 
equation for ${\bf \Psi}(\vq)$ is then solved perturbatively, expanding 
about its linear solution.

\subsection{Linear Solutions and the Zel'dovich Approximation}

The linear solution of Eq.~(\ref{leom})
\begin{equation}
\del_q \cdot {\bf \Psi}^{(1)}= -D_1(\tau) \ \de(\vq),
\label{Psi1}
\end{equation}
where $\de(\vq)$ denotes the density field imposed by the initial 
conditions and $D_1(\tau)$ is the linear growth factor, which obeys 
Eq.~(\ref{D1}). We implicitly assume that vorticity vanishes, then 
Eq.~(\ref{Psi1}) completely determines the displacement field to 
linear order. Linear Lagrangian solutions have the property that they 
become exact for local one-dimensional motion, i.e. when the two 
eigenvalues of the velocity gradient along the trajectory 
vanish~\cite{Buchert89}. Note that the evolution of fluid elements at 
this order is {\em local}, i.e. it does not depend on the behavior of 
the rest of fluid elements. 

The {\em Zel'dovich Approximation} (hereafter ZA)~\cite{Zeldovich70}
consists in using the linear displacement field as an approximate
solution for the dynamical equations\footnote{Rigorously, the ZA results
from using the linear displacement field with the constraint that at
large scales one recovers linear Eulerian PT~\cite{Buchert92}.}. It
follows from Eq.~(\ref{dlag}) that the local density field reads,
\begin{equation}
        1+\de(\vx,\tau) = \frac{1}{[1- \lambda_{1} D_1(\tau)][1- 
        \lambda_{2}D_1(\tau)][1- \lambda_{3}D_1(\tau)]} 
        \label{dlag2},
\end{equation}
where $\lambda_{i}$ are the local eigenvalues of the tidal tensor
$\Psi_{i,j}$. {}From this expression we can see that depending on the
relative magnitude of these eigenvalues, the ZA leads to planar
collapse (one positive eigenvalue larger than the rest), filamentary
collapse (two positive eigenvalues larger than the third), or
spherical collapse (all eigenvalues positive and equal). If all
eigenvalues are negative, then the evolution corresponds to an
underdense region, eventually reaching $\de=-1$. For Gaussian initial
conditions, it is possible to work out the probability distribution
for the eigenvalues~\cite{Doroshkevich70}, which leads through the
non-linear transformation in Eq.~(\ref{dlag2}) to a characterization
of the one-point statistical properties of the density field. These
results will be discussed in Section~\ref{pdfza}.

\subsection{Lagrangian Perturbation Theory}
\label{lagNLPT}

Unlike in Eulerian PT, there is no known recursive solution for the
expression of the order by order cosmic fields in Lagrangian PT, even
for the Einstein-de Sitter case. One reason for that is that beyond
second order, even though one can assume an irrotational flow in
Eulerian space, this does not imply that the displacement field is
irrotational~\cite{Buchert94}. It has been stressed that already
second-order Lagrangian PT for the displacement field (hereafter
2LPT), does provide a remarkable improvement over the ZA in
describing the global properties of density and velocity
fields~\cite{BMW94,MBW95,BCHJ95} and in most practical cases the
improvement brought by third-order Lagrangian PT is
marginal~\cite{BMW94,MBW95}.

One way to understand this situation is to recall that the Lagrangian 
picture is intrinsically non-linear in the density field (e.g. see 
Eq.~(\ref{dlag})), and a small perturbation in Lagrangian fluid 
element paths carries a considerable amount of non-linear information 
about the corresponding Eulerian density and velocity fields. In 
particular, as we shall see below, a truncation of Lagrangian PT at a 
fixed order, yields non-zero Eulerian PT kernels at every order. 
However, as we shall review in the next few chapters, this is not 
always an advantage, particularly when dealing with initial conditions 
with enough small-scale power where shell crossing is significant. In 
these cases, Lagrangian PT generally breaks down at scales larger than 
Eulerian PT.

The reason for the remarkable improvement of 2LPT over ZA is in fact 
not surprising. The solution of Eq.~(\ref{leom}) to second order 
describes the correction to the ZA displacement due to gravitational 
tidal effects, that is, it takes into account the fact that 
gravitational instability is non-local. It reads

\begin{equation}
\del_q \cdot {\bf \Psi}^{(2)}= \frac{1}{2} D_2(\tau) \sum_{i \neq j}
(\Psi_{i,i}^{(1)} \Psi_{j,j}^{(1)} - \Psi_{i,j}^{(1)} \Psi_{j,i}^{(1)}),
\label{Psi2}
\end{equation}

where $D_2(\tau)$ denotes the second-order growth factor, which for 
$0.1 \leq \Omega_m \leq 3$ ($\Omega_\Lambda=0$) obeys

\begin{equation}
D_2(\tau) \approx -\frac{3}{7} D_1^2(\tau),
\end{equation}

or more precisely

\begin{equation}
D_2(\tau) \approx -\frac{3}{7} D_1^2(\tau) \ \Omega_m^{-2/63}, 
\end{equation}

to better than 7\% and 0.5\% respectively~\cite{BJCP92}, whereas for 
flat models with non-zero cosmological constant $\Omega_\Lambda$ we have for 
$0.01 \leq \Omega_m \leq 1$

\begin{equation}
D_2(\tau) \approx -\frac{3}{7} D_1^2(\tau) \ \Omega_m^{-1/143}, 
\end{equation}

to better than 0.6\%~\cite{BCHJ95}.  Since Lagrangian solutions up to
second-order are curl-free\footnote{This is assuming that initial
conditions are in the growing mode, for a more general treatment
see~\cite{BuEh93}.}, it is convenient to define Lagrangian potentials
$\phi^{(1)}$ and $\phi^{(2)}$ so that in 2LPT

\begin{equation}
\vx(\vq) = \vq -D_1\ \del_q \phi^{(1)} + D_2\ \del_q \phi^{(2)},
\label{dis2}
\end{equation}

and the velocity field then reads 

\begin{equation}
{\vu} = -D_1\ f_1\ {\cal H}\ \del_q \phi^{(1)} + D_2\ f_2\ {\cal H}\ 
\del_q \phi^{(2)},
\label{vel2}
\end{equation}
 
where the logarithmic derivatives of the growth factors $f_i \equiv 
(\d\ln D_i)/(\d \ln a)$ can be approximated for open models with $0.1 
\leq \Omega_m \leq 1$ by

\begin{equation}
f_1 \approx \Omega_m^{3/5}, \ \ \ \ \ f_2 \approx 2 \ \Omega_m^{4/7},
\end{equation} 

to better than 2\%~\cite{Peebles76} and 5\%~\cite{BCHJ95}, respectively. 
For flat models with non-zero cosmological constant $\Omega_\Lambda$ we have 
for $0.01 \leq \Omega_m \leq 1$

\begin{equation}
f_1 \approx \Omega_m^{5/9}, \ \ \ \ \ f_2 \approx 2 \ \Omega_m^{6/11},
\end{equation} 

to better than 10\% and 12\%, respectively~\cite{BCHJ95}. The accuracy
of these two fits improves significantly for $\Omega_m \geq 0.1$, in
the relevant range according to present observations.  Summarizing,
the time-independent potentials in Eqs.~(\ref{dis2}) and (\ref{vel2})
obey the following Poisson equations~\cite{BMW94}

\label{poisson2lpt}
\begin{eqnarray}
\del_q^2 \phi^{(1)}(\vq) &=& \de(\vq), 
\label{phi1} \\ 
\del_q^2 \phi^{(2)}(\vq)&=& \sum_{i>j}
[\phi_{,ii}^{(1)}(\vq)\ \phi_{,jj}^{(1)}(\vq) - (\phi_{,ij}^{(1)}(\vq))^2].
\label{phi2}
\end{eqnarray}

It is possible to improve on 2LPT by going to third-order in the
displacement field (3LPT), however it becomes more costly due to the
need of solving three additional Poisson
equations~\cite{Buchert94,Catelan95}.  Third-order results give a
better behavior in underdense regions~\cite{BCHJ95} and lead to
additional substructure in high-density regions~\cite{BKKS97}. 
Detailed comparison of Lagrangian PT at different orders against
numerical simulations is given in~\cite{BCHJ95,KBM97}.

\subsection{Non-Linear Approximations}
\label{sec:nonlinearapp}

When density fluctuations become strongly non-linear, PT breaks down
and one has to resort to numerical simulations to study their
evolution. However, numerical simulations provide limited physical
insight into the physics of gravitational clustering. On the other
hand, many non-linear approximations to the equations of motion have
been suggested in the literature which allow calculations to be
extrapolated to the non-linear regime. However, as we shall see, it
seems fair to say that these approximations have mostly been useful to
gain understanding about different aspects of gravitational clustering
while quantitatively none of them seem to be accurate enough for
practical use. Rigorous PT has provided a very useful way to benchmark
these different approximations in the weakly non-linear regime.

In general, most non-linear approximations can be considered as
different assumptions (valid in linear PT) that replace Poisson's
equation~\cite{MuSt94}. These modified dynamics, are often
local, in the sense described above for the ZA, in order to 
provide a simpler way of calculating the evolution of perturbations
than the full non-local dynamics. 

Probably the best known of non-linear approximations is the ZA, which
in Eulerian space is equivalent to replacing the Poisson equation by
the following ansatz~\cite{MuSt94,HuBe96}

\begin{equation}
\vu(\vx,\tau)=-\frac{2f}{3\Omega_m{\cal H}(\tau)} \nabla\Phi(\vx,\tau),
\label{eZA}
\end{equation}

which is the relation between velocity and gravitational potential
valid in linear PT. Conservation of momentum (assuming for
definiteness $\Omega_m=1$) then becomes [see Eq.~(\ref{euler})]

\begin{equation}
{\partial \vu(\vx,\tau) \over{\partial \tau}} - \frac{{\cal H}(\tau)}{2}\
\vu(\vx,\tau) + \vu(\vx,\tau) \cdot \nabla \vu(\vx,\tau) = 0
\label{eulerZA}.
\end{equation}  

It is straightforward to find the PT recursion relations using these
equations of motion~\cite{ScFr96a}, the result for the
density field kernel is particularly simple~\cite{GrWi87}

\begin{equation}
F_n^{(s)} (\vq_1,...,\vq_n)= \frac{1}{n!} \frac{\vk\cdot\vq_1}{q_1^2}
\ldots \frac{\vk\cdot\vq_n}{q_n^2},
\label{FnZA}
\end{equation}

where $\vk \equiv \vq_1 + \ldots + \vq_n$. As we mentioned before, the ZA
is a local approximation and becomes the exact dynamics in
one-dimensional collapse. It is also possible to formulate local
approximations that besides being exact for planar collapse like the
ZA, are also exact for spherical~\cite{BeJa94} and even
cylindrical collapse~\cite{HuBe96}. However, their
implementation for the calculation of statistical properties of
density and velocity fields is not straightforward. 

A significant shortcoming of the ZA is the fact that after shell
crossing (``pancake formation''), matter continues to flow throughout
the pancake without ever turning around, washing out structures at
small scales. This can be fixed phenomenologically by adding some
small effective viscosity to Eq.~(\ref{eulerZA}), which then becomes
the Burgers' equation\footnote{An attempt to see how this equation
might arise from the physics of multi streaming has been given
in~\cite{BuDo98}.}

\begin{equation}
{\partial \vu(\vx,\tau) \over{\partial \tau}} - \frac{{\cal H}(\tau)}{2}\
\vu(\vx,\tau) + \vu(\vx,\tau) \cdot \nabla \vu(\vx,\tau) = \nu \nabla^2
\vu(\vx,\tau) 
\label{adhesion}.
\end{equation}  

This is the so-called {\em adhesion approximation}~\cite{GSS89}. This
equation has the nice property that for a potential flow it can be
reduced to a linear diffusion equation, and therefore solved
exactly. Given the initial conditions, this can be used to predict the
location of pancakes and clusters, giving good agreement when compared
to numerical simulations~\cite{KPS90}. More detailed comparisons with
numerical simulations for density field statistics show an improvement
over the ZA at small scales~\cite{WeGu90}, however, at weakly
non-linear scales the adhesion approximation is essentially equal to
the ZA.

The {\em linear potential
approximation}~\cite{BSV93,BaPa94} assumes that the
gravitational potential remains the same as in the linear regime,
therefore

\begin{equation}
\nabla^2\Phi(\vx,\tau) = \frac{3}{2}\Omega_m {\cal H}^2(\tau)
\de_1(\vx,\tau), \label{lpa}
\end{equation}  

where $\de_1(\vx,\tau) = D_1^{(+)}(\tau) \de_1(\vx)$ is the linearly
extrapolated density field. The idea behind this approximation is that
since $\Phi \propto \de/k^2$, the gravitational potential is dominated
by long-wavelength modes more than the density field, and therefore it
ought to obey linear PT to a better approximation. 

In the {\em frozen flow approximation}~\cite{MLMS92}, the velocity
field is instead assumed to remain linear, 

\begin{equation}
\theta(\vx,\tau) = -{\cal H}(\tau) f(\Omega_m,\Omega_\Lambda) 
\de_1(\vx,\tau), \label{ff}
\end{equation}

i.e. the velocity field kernels $G_n^{(s)}\equiv 0$ ($n>1$). In the
next chapters we will briefly review how these different
approximations compare in the weakly non-linear
regime~\cite{MuSt94,MSS94,BSBC94,ScFr96a}, see e.g. Table~4 in
Chapter~5.

\subsection{Numerical Simulations}
\label{sec:numsim}
\subsubsection{Introduction}

Cosmological dark matter simulations have become a central tool in
predicting the evolution of structure in the universe well into the
non-linear regime.  Current state of the art numerical simulations can
follow the dynamics of about $10^9$ particles (see
e.g.~\cite{Couchman97}), which although impressive, is still tens of
orders of magnitude smaller than the number of dark matter particles
expected in a cosmological volume, as mentioned in the introduction.

However, this is not an insurmountable limitation.  As we discussed in
section~\ref{sec:vlasov}, in the limit that the number of particles
$N\gg 1$, collisionless dark matter obeys the Vlasov equation for the
distribution function in phase space, Eq.~(\ref{vlasov}).  The task of
numerical simulations is to sample this distribution by partitioning
phase space into $N$ elementary volumes, ``particles'' with positions,
velocities and (possibly different) masses $m_i$, $i=1,\cdots,N$, and
following the evolution of these test particles due to the action of
gravity and the expansion of the universe (technically, these
particles obey the equations of the characteristics of the Vlasov
equation).  The number of particles $N$ fixes the {\em mass
resolution} of the numerical simulation.

Each particle $i$ can be thought of as carrying a ``smooth'' density
profile, which can be viewed as a ``cloud'' of typical size
$\epsilon_i$.  The parameter $\epsilon_i$ is called the softening
length (associated to particle $i$).  In general, $\epsilon_i \propto
m_i^{1/3}$.  This softening is introduced to suppress interactions
between nearby particles in order to reduce $N$-body relaxation, which
is an artifact of the discrete description of the distribution
function.  It fixes the {\em spatial resolution} of the simulation. 
In general it is chosen to be a small fraction of the (local or
global) mean inter-particle separation, but this can vary
significantly depending on the type of code used.

In this section, we briefly discuss methods used to solve numerically
the Vlasov equation. A complete discussion of N-body methods is beyond
the scope of this work, we shall only describe the most common methods
closely following~\cite{Colombi01}; for a comprehensive review see
e.g.~\cite{Bertschinger98}.

The basic steps in an $N$-body simulation can be summarized as follows: 
\begin{enumerate}
\item[(i)] implementation of initial conditions (\cite{KlSh83,EDWF85},
see e.g.~\cite{Bertschinger01} and references therein for recent
developments); \item[(ii)] calculation of the force by solving the
Poisson equation; \item[(iii)] update of positions and velocities of
particles; \item[(iv)] diagnostics, e.g. tests of energy conservation;
\item[(v)] go back to (ii) until simulation is completed.
\end{enumerate}

In general, step (iii) is performed with time integrators accurate to
second order, preferably symplectic (i.e. that preserve phase-space
volume). The Leapfrog integrator (e.g.,~\cite{HoEa81}), where
velocities and positions are shifted from each other by half a
time-step, is probably the most common one. The Predictor-Corrector
scheme is also popular since it allows easy implementation of
individual, varying time-step per particle
(e.g.,~\cite{SYW00}). Low-order integrators are used mostly to
minimize the storage of variables for a large number of particles
whose orbits must be integrated and to reduce the cost of the force
calculation. Because of the chaotic nature of gravitational dynamics
it is not feasible to follow very accurately individual particle
orbits but only to properly recover the properties of bound objects in
a statistical sense.

All the methods that we describe in what follows mainly differ in the
calculation of the force applied to each particle or, in other words,
in how the Poisson equation is solved.

\subsubsection{Direct Summation}
Also known as Particle-Particle (PP) method (e.g.,~\cite{ATG79}), it
consists in evaluating the force on each particle by summing directly
the influence exerted on it by all neighbors. This method is robust
but very CPU consuming: scaling as ${\cal O}(N^2)$, it allows a small
number of particles, typically $N \sim 10^3 - 10^5$. It was revived
recently by the development of special hardware dedicated to the
computation of the Newtonian force (e.g.~\cite{MTES97}), mostly used
for stellar dynamics calculations (but see e.g.~\cite{FuMa97} for a
cosmological application).

\subsubsection{The Tree Algorithm}
The tree code is the most natural improvement of the PP method. It
uses the fact that the influence of remote structures on each particle
can be computed by performing a multipole expansion on clusters
containing many particles. With appropriate selection of the clusters,
the expansion can be truncated at low order. Therefore, the list of
interactions on each particle is much shorter than in the PP method,
of order $\sim \log N$, resulting in a ${\cal O}(N \log N)$ code. The
practical implementation of the tree-code consists in decomposing
hierarchically the system on a tree structure, which can be for
example a mutually nearest neighbor binary tree
(e.g.,~\cite{Appel85}), or a space balanced Oct tree in which each
branch is a cubical portion of space (e.g.,
\cite{BaHu86,Hernquist87,BoHe88}). Then a criterion is applied to see
whether or not a given cluster of particles has to be broken into
smaller pieces (or equivalently, if it is necessary to walk down the
tree).

Various schemes exist (e.g.,~\cite{SaWa94a}), the simplest one for the
Oct tree~\cite{BaHu86} consisting in subdividing the cells until the
condition $s/r \leq \theta$ is fulfilled, where $s$ is the size of the
cell, $r$ is the distance of the cell center of mass to the particle
and $\theta$ is a tunable parameter of order unity.

The tree data structure has many advantages: (i) the CPU spent per
time-step does not depend significantly on the degree of clustering of
the system; (ii) implementation of individual time-steps per particle
is fairly easy and this can speed up the simulation significantly;
(iii) the use of individual masses per particle allows ``zooming'' in
a particular region, for example a cluster, a galaxy halo or a void:
the location of interest is sampled accurately with high resolution
particles (with small mass), while tidal effects are modeled by low
resolution particles of mass increasing with distance from the high
resolution region; (iv) implementation on parallel architectures with
distributed memory is relatively straightforward (e.g.,
\cite{SaWa94b,Dubinski96,SYW00}). However, tree-codes are rather
demanding in memory (25-35 words per particles, e.g.,
\cite{Couchman97}) and accurate handling of periodic boundaries (e.g.,
\cite{HBS91}) is costly.

Typically, simulations using the tree-code can involve up to $\sim
10^7-10^8$ particles if done on parallel supercomputers. They have
high spatial resolution, of order $\epsilon \sim \lambda/(10-20)$,
where $\lambda$ is the mean inter-particle distance.

\subsubsection{The PM Algorithm}
In the Particle-Mesh (PM) method (e.g.
see~\cite{HoEa81,DKNPSS80,Melott83,BAP85}), the mass of each particle
is interpolated on a fixed grid of size $N_{\rm g}$ (with $N_{\rm
g}^3$ sites) to compute the density.  The Poisson equation is solved
on the grid, generally by using a Fast Fourier transform, then forces
are interpolated back on the particles.  Implementing a PM code is
thus rather simple, even on parallel architectures.  Scaling as ${\cal
O}(N, N_{\rm g}^3 \log N_{\rm g})$, PM simulations have generally the
advantage being low CPU consumers and require reasonable amount of
memory.  Thus, a large number of particles can be used, $N \sim
10^7-10^9$, and typically $N_{\rm g}=N^{1/3}$ or $2 N^{1/3}$.  The
main advantage and weakness of the PM approach is its low spatial
resolution.  Indeed, the softening parameter is fixed by the size of
the grid, $\epsilon \sim L/N_{\rm g}$, where $L$ is the size of the
box: large softening length reduces the effects of $N$-body relaxation
and allows good phase-space sampling, but considerably narrows the
available dynamic scale range.  To achieve a spatial resolution
comparable to that of a tree-code while keeping the advantage of the
PM code, very large values of $N_{\rm g}$ and $N$ would be needed,
implying a tremendous cost both in memory and in CPU.

\subsubsection{Hybrid Methods}
To increase spatial resolution of the PM approach, several improvements
have been suggested. 

The most popular one is the P$^3$M code (PP$+$PM) where the PM force
is supplemented with a short-range contribution obtained by direct
summation of individual interactions between nearby particles (e.g.,
\cite{HoEa81,EDWF85}). Implementation of this code on a parallel
supercomputer (T3E) produced a very large cosmological simulation with
$10^9$ particles in a ``Hubble'' volume of size $L=2000 h^{-1}$
Mpc~\cite{MCPP98}. The main caveat of the P$^3$M approach is that as
the system evolves to a more clustered state, the time spent in
calculation of PP interactions becomes increasingly significant. To
reduce the slowing-down due to PP interactions, it was proposed to use
a hierarchy of adaptive meshes in regions of high particle
density~\cite{Couchman91}, giving birth to a very efficient $N$-body
code, the Adaptive P$^3$M (AP$^3$M).

Instead of direct PP summations to correct the PM force for short
range interactions, it is possible to use a tree algorithm in high
density regions~\cite{Xu95} or in all PM cells~\cite{Bagla99}
similarly as in the P$^3$M code. Both these methods are potentially
faster than their P$^3$M competitor.

In the same spirit as in AP$^3$M, but without the PP part, another
alternative is to use Adaptive Mesh Refinement (AMR): the PM mesh is
increased locally when required with a hierarchy of nested rectangular
sub-grids (e.g.,~\cite{Villumsen89,ANC94,JDC94,GCW97}). The forces can
be computed at each level of the hierarchy by a Fourier transform with
appropriate boundary conditions. In fact, the sub-grids need not be
rectangular if one uses Oct tree structures, which is theoretically
even more efficient. In this adaptive refinement tree (ART)
method~\cite{KKK97}, the Poisson equation is solved by relaxation
methods (e.g.,~\cite{HoEa81,PTVF92}).

Finally, it is worth mentioning a Lagrangian approach, which consists
in using a mesh with fixed size like in the PM code, but moving with
the flow so that resolution increases in high density regions and
decreases elsewhere~\cite{Gnedin95,Pen95}. However, this potentially
powerful method presents some difficulties, e.g. mesh distortions may
induce severe force anisotropies.

\clearpage

\section{\bf Random Cosmic Fields and their Statistical Description}
\label{fb:Stoch}

In this chapter we succinctly recall current ideas about the physical
origin of stochasticity in cosmic fields in different cosmological
scenarios. We then present the statistical tools that are commonly
used to describe random cosmic fields such as power spectra,
probability distribution functions, moments and cumulants, and give
some mathematical properties of interest.

\subsection{The Need for a Statistical Approach}

As we shall review in detail in the following chapters, the current
explanation of the large-scale structure of the universe is that the
present distribution of matter on cosmological scales results from the
growth of primordial, small, seed fluctuations on an otherwise
homogeneous universe amplified by gravitational instability.  Tests of
cosmological theories which characterize these primordial seeds are
not deterministic in nature but rather statistical, for the following
reasons. First, we do not have direct observational access to
primordial fluctuations (which would provide definite initial
conditions for the deterministic evolution equations). In addition,
the time-scale for cosmological evolution is so much longer than that
over which we can make observations, that is not possible to follow
the evolution of single systems. In other words, what we observe
through our the past light cone is different objects at different
times of their evolution, therefore testing the evolution of structure
must be done statistically.

The observable universe is thus modeled as a stochastic realization of
a statistical ensemble of possibilities.  The goal is to make
statistical predictions, which in turn depend on the statistical
properties of the primordial perturbations leading to the formation of
large-scale structures.  Among the two classes of models that have
emerged to explain the large-scale structure of the universe, the
physical origin of stochasticity can be quite different and thus give
rise to very different predictions.

The most widely considered models, based on the inflationary
paradigm~\cite{Guth81}, generically give birth to
adiabatic\footnote{As opposed to {\em isocurvature} fluctuations which
is a set of individual perturbations such that the total fluctuation
amplitude vanishes.  In the adiabatic case, the total amplitude does
not vanish and this leads to perturbations in the spatial curvature.}
Gaussian initial fluctuations, at least in the simplest single-field
models~\cite{Starobinsky82,Hawking82,GuPi82,BST83}.  In this case the
origin of stochasticity lies on quantum fluctuations generated in the
early universe; we will consider this case in more detail below. 
However, one should keep in mind that inflation is not necessarily the
only mechanism that leads to Gaussian, or almost Gaussian, initial
conditions.  For instance, topological defects based on the non-linear
$\sigma$-model in the large $N$-limit would also give Gaussian initial
conditions~\cite{TuSp91,Jaffe94}.  And in general the central limit
theorem ensures that such initial conditions are likely to happen in
very broad classes of models.

The second class of models that have been developed for structure
formation are based on topological defects, of which cosmic strings
have been studied in most detail.  In this case the origin of
stochasticity lies on thermal fluctuations of a field that undergoes a
phase transition as the universe cools, and is likely to obey
non-Gaussian properties.  Note however that these two classes of
models do not necessarily exclude each other.  For instance, formation
of cosmic strings are encountered in specific models of
inflation~\cite{BDDR98,BDP98,Jeannerot99}.  There are also models
inspired by duality properties of superstring theories, in which an
inflationary phase can be encountered but structure formation is
caused by the quantum fluctuations of the axion
field\footnote{However, this generally leads to isocurvature
fluctuations rather than adiabatic.}~\cite{Veneziano97,CEW97,BMUV98}
rather than the inflaton field.  With such a mechanism the initial
metric fluctuations will not obey Gaussian statistics.

\subsubsection{Physical Origin of Fluctuations from Inflation}

In models of inflation the stochastic properties of the fields
originate from quantum fluctuations of a scalar field, the inflaton. It
is beyond the scope of this review to describe inflationary models in
any detail. We instead refer the reader to recent reviews for a
complete discussion~\cite{LiLy93,LiLy00,LyRi99}.  It is worth however
recalling that in such models (at least for the simplest single-field
models within the slow-roll approximation) all fluctuations originate
from scalar adiabatic perturbations.  During the inflationary phase
the energy density of the universe is dominated by the density stored
in the inflaton field. This field has quantum fluctuations that can be
decomposed in Fourier modes using the creation and annihilation
operators $a^{\dag}_{\vk}$ and $a_{\vk}$ for a wave mode $\vk$,
\be
\de\varphi=
\int\d^3 \vk\left[a_{\vk}\,\psi_k(t)\,\exp(\ii\vk.\vx)+
a^{\dag}_{\vk}\,\psi_k^*(t)\,\exp(-\ii\vk.\vx)\right].
\ee
The operators obey the standard commutation relation,
\be
[a_{\vk},a^{\dag}_{-\vk'}]=\de_D(\vk+\vk'),
\label{fb:comm}
\ee

and the mode functions $\psi_k(t)$ are obtained from the Klein-Gordon
equation for $\varphi$ in an expanding Universe. We give here its
expression for a de-Sitter metric (i.e. when the spatial sections are
flat and $H$ is constant),

\be \psi_k(t)={H\over (2\,k)^{1/2}\,k}\, \left(\ii+{k\over
\,a\,H}\right)\,\exp\left[{\ii\,k\over a\,H}\right],
\label{fb:psik}
\ee where $a$ and $H$ are respectively the expansion factor and the
Hubble constant that are determined by the overall content of the
Universe through the Friedmann equations,
Eqs.~(\ref{friedmann1}-\ref{friedmann2}).

When the modes exit the Hubble radius, $k/(aH)\ll1$, one can see from
Eq.~(\ref{fb:psik}) that the dominant mode reads, \be
\varphi_{\vk}\approx {iH\over \sqrt{2}
k^{3/2}}\,\left(a_{\vk}+a_{-\vk}^{\dag}\right),\ \
\de\varphi=\int\d^3\vk\,\varphi_{\vk}\,e^{\ii\,\vk.\vx}.
\label{fb:phikinfl}
\ee Thus these modes are all proportional to
$a_{\vk}+a_{-\vk}^{\dag}$.  One important consequence of this is that
the quantum nature of the fluctuations has
disappeared~\cite{GuPi85,KPS98,KLPS98}: any combinations of
$\varphi_{\vk}$ commute with each other.  The field $\varphi$ can then
be seen as a classic stochastic field where  {\em ensemble averages
identify with vacuum expectation values}, \ba \langle
... \rangle\equiv\langle 0\vert...\vert0\rangle.  \ea

After the inflationary phase the modes re-enter the Hubble
radius. They leave imprints of their energy fluctuations in the
gravitational potential, the statistical properties of which can
therefore be deduced from Eqs. (\ref{fb:comm}, \ref{fb:phikinfl}).
All subsequent stochasticity that appears in the cosmic fields can
thus be expressed in terms of the random variable $\varphi_{\vk}$. 

\subsubsection{Physical Origin of Fluctuations from Topological Defects}

In models of structure formation with topological defects,
stochasticity originates from thermal fluctuations.  One important
difficulty in this case is that topological defects generally behave
as active seeds, and except in some special cases (see for
instance~\cite{DKM99}), the dynamical evolution of these seeds is
nonlinear and nonlocal, hence requiring heavy numerical calculation
for their description. This is in particular true for cosmic strings
that form a network whose evolution is extremely complex (see for
instance~\cite{BoBe90}). Therefore in this case it is not possible to
write down in general how the stochasticity in cosmic fields relates
to more fundamental processes. See~\cite{ViSh00} for a review of the
physics of topological defects. Current observations of multiple
acoustic peaks in the power spectrum of microwave background
anisotropies severely constrain significant contributions to
perturbations from active seeds~\cite{BOOMERANG,DASI,MAXIMA}.

\subsection{Correlation Functions and Power Spectra}

{}From now on, we consider a cosmic scalar field whose statistical
properties we want to describe.  This field can either be the cosmic
density field, $\de(\vx)$, the cosmic gravitational potential, the
velocity divergence field, or any other field of interest.

\subsubsection{Statistical Homogeneity and Isotropy}
\label{sec:stathomiso}

A random field is called {\em statistically homogeneous}\footnote{This
is in contrast with a {\em homogeneous} field, which takes the same
value everywhere in space.} if all the joint multipoint {\em probability
distribution functions} $p(\de_1, \de_2, \dots)$ or its {\em moments},
ensemble averages of local density products,  remain the same
under translation of the coordinates $\vx_1, \vx_2, \dots$ in space
(here $\de_i \equiv \de(\vx_i)$).  Thus the probabilities depend only
on the relative positions.  A stochastic field is called {\em
statistically isotropic} if $p(\de_1, \de_2, \dots)$ is invariant
under spatial rotations.  We will assume that cosmic fields are
statistically homogeneous and isotropic, as predicted by most
cosmological theories. The validity of this assumption can and should
be tested against the observational data. Examples of primordial
fields which do not obey statistical homogeneity and isotropy are
fluctuations in compact hyperbolic spaces (see
e.g.~\cite{BPS00}). Furthermore, redshift distortions in galaxy
redshift surveys introduce significant deviations from statistical
isotropy and homogeneity in the redshift-space density field, as will
be reviewed in Chapter~7.

\subsubsection{Two-Point Correlation Function and Power Spectrum}
\label{sec:sec3.2.2}

The two-point correlation function is defined as the joint ensemble
average of the density at two different locations,
\be
\xi(r)=\mg\de(\vx)\de(\vx+\vr)\md, 
\label{eq:xidef}
\ee
which depends only on the norm of $\vr$ due to statistical homogeneity
and isotropy. The density contrast $\de(\vx)$ is usually written in
terms of its Fourier components,
\be
\de(\vx)=\int{\d^3\vk}\,\de(\vk)\,\exp(\ii\vk\cdot \vx).
\label{dek}
\ee
The quantities  $\de(\vk)$ are then complex random variables.
As  $\de(\vx)$ is real, it follows that 
\beq{
\de(\vk)=\de^*(-\vk).
} The density field is therefore determined entirely by the
statistical properties of the random variable $\de(\vk)$. We can
compute the correlators in Fourier space,
\be
\mg\de(\vk)\de(\vk')\md=\int\frac{\d^3\vx}{(2\pi)^3} 
\frac{\d^3\vr}{(2\pi)^3}\,\mg\de(\vx)\de(\vx+\vr)\md\,
\exp[-\ii(\vk+\vk')\cdot\vx-\ii\vk'\cdot\vr]
\ee
which gives,
\ba
\mg\de(\vk)\de(\vk')\md&=&\int \frac{\d^3\vx}{(2\pi)^3} 
\frac{\d^3\vr}{(2\pi)^3}\,\xi(r)\, 
\exp[-\ii(\vk+\vk')\cdot \vx-\ii\vk'\cdot\vr]\nonumber\\
&=&\de_D(\vk+\vk')\int \frac{\d^3\vr}{(2\pi)^3}\,
\xi(r)\, \exp(\ii \vk\cdot \vr)
\nonumber\\ 
&\equiv&\de_D(\vk+\vk')\,P(k),
\label{pk}
\ea
where $P(k)$ is by definition the density {\em power spectrum}. The
inverse relation between two-point correlation function and power
spectrum thus reads

\be
\xi(r) = \int \d^3\vk\,P(k) \exp(\ii \vk \cdot \vr).
\label{xipk}
\ee

There are basically two conventions in the literature regarding the
definition of the power spectrum, which differ by a factor of
$(2\pi)^3$. In this review we use the convention in
Eqs.~(\ref{fourier}), (\ref{dek}) and (\ref{pk}) which lead to
Eq.~(\ref{xipk}). Another popular choice is to reverse the role of
$(2\pi)^3$ factors in the Fourier transforms, i.e. $\de(\vk) \equiv
\int \d^3\vr \exp(-\ii \vk \cdot \vr) \de(\vr)$, and then modify 
Eq.~(\ref{pk}) to read $\mg \de(\vk) \de(\vk')\md \equiv (2\pi)^3
\de_D(\vk+\vk')\,P(k)$, which leads to $k^3 P(k)/(2\pi^2)$ 
being the contribution per logarithmic wavenumber to the variance, rather 
than $4\pi k^3 P(k)$ as in our case.

\subsubsection{The Wick Theorem for Gaussian Fields}

The power spectrum is a well defined quantity for almost all
homogeneous random fields. This concept becomes however extremely
fruitful when one considers a {\em Gaussian} field.  It means that any
joint distribution of local densities is Gaussian distributed. Any
ensemble average of product of variables can then be obtained by
product of ensemble averages of pairs.  We write explicitly this
property for the Fourier modes as it will be used extensively in this
work,
\ba
\mg\de(\vk_1)\dots\de(\vk_{2p+1})\md&=&0\label{fb:Wick1}\\
\mg\de(\vk_1)\dots\de(\vk_{2p})\md&=&
\sum_{\rm all\ pair\ associations}\ \ \prod_{p\ \rm pairs\ (i,j)}
\mg\de(\vk_i)\de(\vk_j)\md \label{fb:Wick2}
\ea
This is the {\em Wick theorem}, a fundamental theorem for classic and
quantum field theories.

The statistical properties of the random variables $\de(\vk)$ are then
entirely determined by the shape and normalization of $P(k)$.  A
specific cosmological model will eventually be determined e.g. by the
power spectrum in the linear regime, by $\Omega_m$ and
$\Omega_\Lambda$ only as long as one is only interested in the dark
matter behavior\footnote{Note that there are now emerging models with
a non-standard vacuum equation of state, the so-called quintessence
models~\cite{RaPe88,ZWS99}, in which the vacuum energy is that of a
non-static scalar field. In this case the detailed behavior of the
large-scale structure growth will depend on the dynamical evolution of
the quintessence field.}.

As mentioned in  the previous section, in the  case of an inflationary
scenario  the   initial  energy   fluctuations  are  expected   to  be
distributed           as          a           Gaussian          random
field~\cite{Starobinsky82,Hawking82,GuPi82,BST83}.     This    is    a
consequence of the commutation  rules given by Eq.~(\ref{fb:comm}) for
the creation and annihilation operators for a free quantum field. They
imply that
\begin{equation}
\left[\left(a_{\vk}+a_{-\vk}^{\dag}\right),
\left(a_{\vk'}+a_{-\vk'}^{\dag}\right)\right]=\de_D(\vk+\vk').  
\end{equation} 

As a consequence of this, the relations in
Eqs.~(\ref{fb:Wick1}-\ref{fb:Wick2}) are verified for $\varphi_{\vk}$
for all modes that exit the Hubble radius, which long afterwards come
back in as classical stochastic perturbations. These properties
obviously apply also to any quantities linearly related to
$\varphi_{\vk}$.

\subsubsection{Higher-Order Correlators: Diagrammatics}

In general it is possible to define higher-order correlation
functions.  They are defined as the {\em connected} part (denoted with
subscript $c$) of the joint ensemble average of the density in an
arbitrarily number of locations. They can be formally written,
\ba
\xi_N(\vx_1,\dots,\vx_N)&=&\langle
\de(\vx_1),\dots,\de(\vx_N)\rangle_c\\
&\equiv&\langle\de(\vx_1),\dots,\de(\vx_N)\rangle-\nonumber\\
&&\ \ \ -
\sum_{\mS\in {\cal \mP}\left(\{\vx_1,\dots,\vx_n\}\right)}
\prod_{s_i\in\mS}\xi_{\#s_i}(\vx_{s_i(1)},\dots,
\vx_{s_i(\#s_i)})
\label{fb:xindef}
\ea
where the sum is made over the proper partitions (any partition except
the set itself) of $\{\vx_1,\dots,\vx_N\}$ and $s_i$ is thus a subset
of $\{\vx_1,\dots,\vx_N\}$ contained in partition $\mS$.  When the
average of $\de(\vx)$ is defined as zero, only partitions that contain
no singlets contribute.

\begin{figure}
\vspace{4 cm}
\special{hscale=80 vscale=80 voffset=0 hoffset=30 psfile=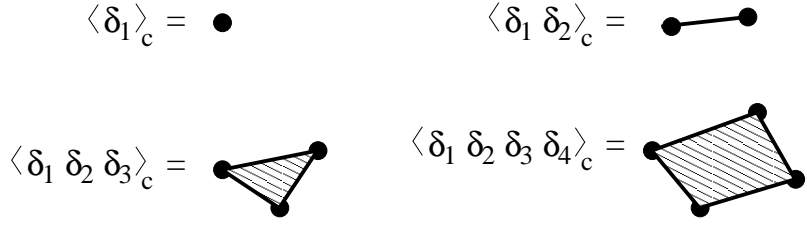}
\caption{Representation of the connected part of the moments.}
\label{fb:conn}
\end{figure}

\begin{figure}
\vspace{2 cm}
\special{hscale=70 vscale=70 voffset=0 hoffset=30 psfile=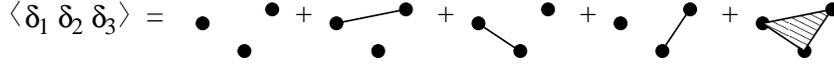}
\caption{Writing of the three-point moment in terms of connected parts.}
\label{fb:3pts}
\end{figure}

\begin{figure}
\vspace{6 cm}
\special{hscale=70 vscale=70 voffset=0 hoffset=5 psfile=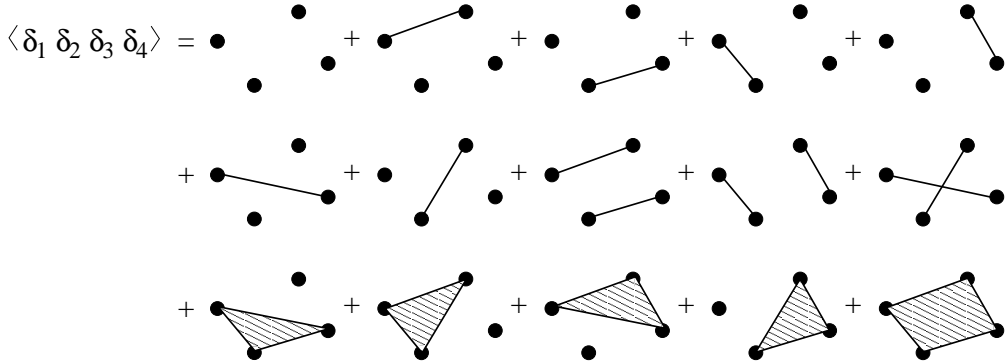}
\caption{Same as Fig. \ref{fb:3pts} for the four-point moment.}
\label{fb:4pts}
\end{figure}

The decomposition in connected and non-connected parts can be easily
visualized. It means that any ensemble average can be decomposed in
a product of connected parts. They are defined for instance in
Fig.~\ref{fb:conn}. The tree-point moment is ``written'' in
Fig.~\ref{fb:3pts} and the four-point moment in Fig.~\ref{fb:4pts}.

\begin{figure}
\vspace{2 cm}
\special{hscale=80 vscale=80 voffset=0 hoffset=10 psfile=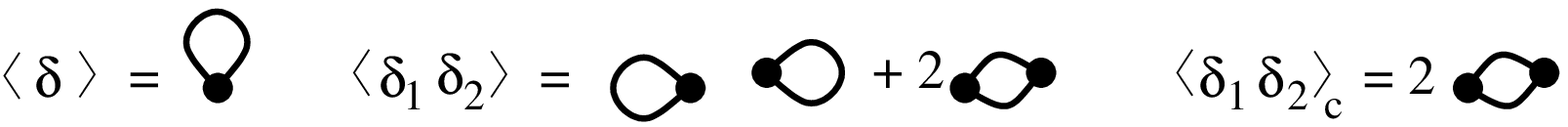}
\caption{Disconnected and connected part of the two-point function of
the field $\de$ assuming it is given by $\de=\phi^2$ with $\phi$ Gaussian.}
\label{fb:del2stat}
\end{figure}

In case of a Gaussian field all connected correlation functions are
zero except $\xi_2$. This is a consequence of Wick's theorem. As a
result the only non-zero connected part is the two-point correlation
function. An important consequence is that the statistical properties
of any field, not necessarily linear, built from a Gaussian field
$\de$ can be written in terms of combinations of two-point functions
of $\de$. Note that in a diagrammatic representation the connected
moments of any of such field is represented by a {\em connected}
graph. This is illustrated in Fig.~\ref{fb:del2stat} for the field
$\de=\phi^2$: the connected part of the 2-point function of this field
is obtained by all the diagrams that explicitly join the two
points. The other ones contribute to the moments, but not to its
connected part.

The connected part has the important property that it vanishes when
one or more points are separated by infinite separation. In addition,
it provides a useful way of characterizing the statistical properties,
since unlike unconnected correlation functions, each connected
correlation provides independent information.

These definitions can be extended to Fourier space. Because of
homogeneity of space $\langle\de(\vk_1)\dots\de(\vk_N)\rangle_c$ is
always proportional to $\de_D(\vk_1+\dots+\vk_N)$. Then we
can define $P_N(\vk_1,\dots,\vk_N)$ with
\be
\langle\de(\vk_1)\dots\de(\vk_N)\rangle_c=
\de_D(\vk_1+\dots+\vk_N)\,P_N(\vk_1,\dots,\vk_N).
\label{fb:Pndef}
\ee
One particular case that will be discussed in the following is for
$n=3$, the bispectrum, which is usually denoted by
$B(\vk_1,\vk_2,\vk_3)$.

\subsubsection{Probabilities and Correlation Functions}
\label{sec:probcorr}

Correlation functions are directly related to the multi-point
probability function, in fact they can be defined from them. Here we
illustrate this for the case of the density field, as these results
are frequently used in the literature. The physical interpretation of
the two-point correlation function is that it measures the excess over
random probability that two particles at volume elements $dV_1$ and
$dV_2$ are separated by distance $x_{12} \equiv |\vx_1-\vx_2|$,

\be
\label{dp12}
\d P_{12} = n^2 [1+\xi(x_{12})] \d V_1 \d V_2,
\ee

where $n$ is the mean density. If there is no clustering (random
distribution), $\xi=0$ and the probability of having a pair of
particles is just given by the mean density squared, independently of
distance. Since the probability of having a particle in $dV_1$ is $n
dV_1$, the conditional probability that there is a particle at $dV_2$
{\em given} that there is one at $dV_1$ is

\be
\label{dpc12}
\d P(2|1)= n  [1+\xi(x_{12})] \d V_2.
\ee

The nature of clustering is clear from this expression; if objects are
clustered ($\xi(x_{12})>0$), then the conditional probability is
enhanced, whereas if objects are anticorrelated ($\xi(x_{12})<0$) the
conditional probability is suppressed over the random distribution
case, as expected. Similarly to Eq.~(\ref{dp12}), for the three-point
case the probability of having three objects is given by

\be
\label{dp123}
\d P_{123} = n^3
[1+\xi(x_{12})+\xi(x_{23})+\xi(x_{31})+\xi_3(x_{12},x_{23},x_{31})]
\d V_1 \d V_2 \d V_3,
\ee

where $\xi_3$ denotes the three-point (connected) correlation
function. If the density field were Gaussian, $\xi_3=0$, and all
probabilities are determined by $\xi(r)$ alone. Analogous results hold
for higher-order correlations (e.g. see~\cite{Peebles80}).

\subsection{Moments, Cumulants and their Generating Functions}
\label{sec:sec3.3}

\subsubsection{Moments and Cumulants}
\label{sec:sec3.3.1}

One particular case for Eq.  (\ref{fb:xindef}) is when all points are
at the same location.  Because of statistical homogeneity
$\xi_p(\vx,\dots,\vx)$ is independent on the position $\vx$ and it
reduces to the cumulants of the one-point density probability
distribution functions, $\langle\de^p\rangle_c$.  The relation
(\ref{fb:xindef}) tells us also how the cumulants are related to the
moments $\langle\de^p\rangle$.  For convenience we write here the
first few terms, 
\ba \cum{}&=&\mom{}\nonumber\\
\cum2 &=& \sigma^2 = \mom2 - \cum{}^2  \nonumber \\*
\cum3 &=& \mom3 -3 \cum2 \cum{} - \cum{}^3  \label{nonc_c} \\*
\cum4 &=& \mom4-4 \cum3 \cum{}-3 \cum2^2 -6 \cum2 \cum{}^2-\cum{}^4\nonumber\\*
\cum5 &=& \mom5-5 \cum4 \cum{}-10 \cum3 \cum2-10 \cum3 \cum{}^2
-15\cum2^2\cum{}\nonumber\\*
&&-10 \cum2 \cum{}^3-\cum{}^5  \nonumber 
\ea In most cases $\mom{}=0$ and the above equations simplify
considerably.  In the
following we usually denote $\sigma^2$ the local second order
cumulant.  The Wick theorem then implies that in case of a Gaussian
field $\sigma^2$ is the only non-vanishing cumulant.

It is important to note that the local PDF is essentially
characterized by its {\em cumulants} which constitute a set of
independent quantities. This is important since in most of
applications that follow the higher-order cumulants are small compared
to their associated moments. Finally, let's note that a useful
mathematical property of cumulants is that $\langle (b \de)^n
\rangle_c = b^n \langle \de^n
\rangle_c$, and $\langle (b+\de)^n \rangle_c = \langle \de^n
\rangle_c$ where $b$ is an ordinary number.

\subsubsection{Smoothing}
The density distribution is usually smoothed with a filter $W_R$ of a
given size, $R$, commonly a top-hat or a Gaussian window.  Indeed,
this is required by the discrete nature of galaxy catalogs and
$N$-body experiments used to simulate them.  Moreover, we shall see
later that the scale-free nature of gravitational clustering implies
some remarkable properties about the scaling behavior of the smoothed
density distribution.  The quantities of interest are then the moments
$\langle \delta_R^p
\rangle$ and
the cumulants $\langle \delta_R^p \rangle_c$ of the smoothed density
field
\be
\de_R(\vx)=\int W_R(\vx'-\vx)\de(\vx')\d^3\vx'.
\ee
Note that for the top hat window, 
\be
\langle\de_R^p\rangle_c=\int_{v_R}\xi_p(\vx_1,\dots,\vx_p){\d^{\cal D}\vx_1\dots
\d^{\cal D}\vx_p\over v_R^p}
\ee
(where ${\cal D}=2$ or 3 is the dimension of the field) is nothing
but the average of the $N$-point correlation function over the
corresponding cell of volume $v_R$.

For a smooth field, equations in Sect.~\ref{sec:sec3.3.1} are valid
for $\delta$ as well as $\delta_R$.  Some corrections are required if
$\delta$ is a sum of Dirac delta functions as in real galaxy catalogs.
We shall come back to this in Chapter~\ref{sec:chapter7}.

In the remaining of this chapter, we shall omit the subscript $R$
which stands for smoothing, but it will be implicitly assumed.

\subsubsection{Generating Functions}
\label{sec:sec3.3.2}

It is convenient to define a function from which all moments can be
generated, namely the {\em moment generating function} defined by
\be
\mM(t)\equiv\sum_{p=0}^{\infty}{\mom{p}\over p!}t^p=\int_{-\infty}^{+\infty}
p(\de)e^{t\de}\d\de=\langle \exp(t\de)\rangle.
\label{fb:Lap}
\ee
The moments can obviously obtained by subsequent derivatives of this
function at the origin $t=0$. A cumulant generating function can
similarly be defined by
\be
\mC(t)\equiv\sum_{p=2}^{\infty}{\cum{p}\over p!}t^p.
\ee
A fundamental result is that the cumulant generating function is given
by the logarithm of the moment generation function (see e.g. appendix
D in~\cite{BDFN92} for a proof)

\be
\mM(t)=\exp[\mC(t)].
\ee
In case of a Gaussian PDF, this is straightforward to check since
$\langle \exp(t\de)\rangle=\exp(\sigma^2t^2/2)$.

\subsection{Probability Distribution Functions}

The probability distribution function (PDF) of the local density can
be obtained from the cumulant generating function by inverting Eq. 
(\ref{fb:Lap})\footnote{However, it may happen that the moment or
cumulant generating function is not defined because of the lack of
convergence of the series in Eq.~(\ref{fb:Lap}).  In this case the PDF
is not uniquely defined by its moments.  In particular, this is the
case for the log-normal distribution.  There are indeed other PDF's
that have the same moments~\cite{Heyde63}.}.  This inverse relation
involves the inverse Laplace transform, and can formally be written in
terms of an integral in the complex plane (see \cite{BaSc89a} and
Appendix~\ref{app:pdedelta} for a detailed account of this relation),
\be P(\de)=\int_{-\ii\infty}^{\ii\infty}{\d t\over 2\pi \ii}
\exp[t\de+\mC(t)].
\label{fb:PDFmC}
\ee
For a Gaussian distribution the change of variable $t\to\ii t$
gives the familiar Gaussian integral.

This can be easily generalized to multidimensional PDF's. We then
have
\be
P(\de_1,\dots,\de_p)=
\int_{-\ii\infty}^{\ii\infty}{\d t_1\over 2\pi \ii}\ldots
\int_{-\ii\infty}^{\ii\infty}{\d t_q\over 2\pi \ii}
\exp\left[\left(\sum_{q=1}^p t_q\de_q\right)+\mC(t_1,\dots,t_p)\right],
\ee
with
\be
\mC(t_1,\dots,t_p)=\sum_{q_1,\dots,q_p}
\langle\de_1^{q_1}\dots\de_p^{q_p}\rangle_c\ 
{t_1^{q_1}\dots t_p^{q_p}\over q_1!\dots q_p!}.
\label{eq:cumulmulti}
\ee

\subsection{Weakly Non-Gaussian Distributions: Edgeworth Expansion}
\label{sec:edgeworth}

Throughout this review we will be often dealing with  
fields that depart only weakly from a Gaussian distribution.
To be more specific, they depart in such a way that
\be
\cum{p}\sim\sigma^{2p-2}
\ee
when $\sigma$ is small\footnote{This is a consequence of Gaussian
initial conditions and the fact that non-linearities in the equations
of motion are quadratic, see Chapter~\ref{chapter4}.}. It is then
natural to define the coefficient $S_p$ as
\be
S_p={\cum{p}\over\sigma^{2p-2}}.
\ee
(similar definitions will be introduced subsequently for the other
fields).  Introducing the $S_p$ generating function (sometimes also
called the cumulant generating function) with,
\be
\varphi(y)=\sum_{p=2}^{\infty}S_p{(-1)^{p-1}\over p!}y^p=
-\sigma^2\mC(-y/\sigma^2)
\label{eq:varphieq}
\ee
we get from Eq. (\ref{fb:PDFmC}),
\be
P(\de)\d\de={\d\de \over 2\pi\ii\sigma^2}
\int_{-\ii\infty}^{+\ii\infty} \d y
\exp\left[-{\varphi(y)\over\sigma^2}+
{\de y\over \sigma^2}\right].
\label{fb:PDFvarphi}
\ee
Then a number of approximations and truncations can be applied to this
expression to decompose the local PDF.  This leads to the Edgeworth
form of the Gram-Charlier series~\cite{KeSt94} applied to statistics of
weakly nonlinear fields. This expansion was derived initially 
in~\cite{LonguetHiggins63,LonguetHiggins64} and later proposed in
cosmological contexts~\cite{ScBe91,BeKo95,JWACB95}

The Edgeworth expansion can be derived from Eq.~(\ref{fb:PDFvarphi})
of the density PDF assuming that the density contrast $\de$ is of the
order of $\sigma$ and small. The relevant values of $y$ are then also
of the order of $\sigma$ and are thus expected to be small. It is then
legitimate to expand the function $\varphi(y)$
\ba
\varphi(y)\approx
-{1\over 2} y^2+{S_3\over 3!}\ y^3-{S_4\over 4!}\ y^4+{S_5\over 5!}\ y^5 \pm
\dots.
\label{fb:varphi}
\ea
To calculate the density PDF, we substitute the expansion
(\ref{fb:varphi}) into the integral in Eq.~(\ref{fb:PDFvarphi}).  Then
we make a further expansion of the {\sl non-Gaussian} part of the
factor $\exp\left[-{\varphi(y)/\sigma^2}\right]$ with respect to both
$y$ and $\sigma$ assuming they are of the same order.

Finally, collecting the terms of the same order in $\sigma$ we obtain
the so-called Edgeworth form of the Gram-Charlier series for density
PDF,
\ba
P(\de)\d\de&=&{1 \over (2\pi\sigma^2)^{1/2}}
\exp\left( -\nu^2/2\right)\times\nonumber\\
&&\times\biggl[1 + \sigma {S_3 \over 6} H_3\left(\nu\right)
+\sigma^2\biggl( {S_4 \over 24} H_4\left(\nu\right)
+{S_3^2\over72}H_6\left(\nu\right)
 \biggr)\biggr.\nonumber\\
&&\biggl.+\sigma^3\biggl({S_5\over 120} H_5(\nu)+
{S_4 S_3\over 144} H_7(\nu)+ {S_3^3\over 1296} H_9(\nu)\biggr)
+ ... \biggr] \d\dta,
\label{fb:edge}
\ea
where $\nu=\de/\sigma$ and $H_n(\nu)$ are the Hermite polynomials 
\ba
H_n(\nu)&\equiv&
(-1)^n\exp(\nu^2/2) {\d^n\over\d\nu^n}\exp(-\nu^2/2)\nonumber\\
&=&\nu^n-{n(n-1)\over 1!}\ {\nu^{n-2}\over 2}
+{n(n-1)(n-2)(n-3)\over 2!}\ {\nu^{n-4}\over 2^2}-\dots,
\label{fb:Hermit}
\ea
thus 
\ba
H_3(\nu)&=&\nu^3-3\nu,\\
H_4(\nu)&=&\nu^4-6\,\nu^2+3,\\
H_5(\nu)&=&\nu^5-10\,\nu^3+15\,\nu,\\
&&\dots\nonumber
\ea
This is a universal form for any slightly non-Gaussian field,
i.e. when $\sigma$ is small and $S_p$ are finite. Note that the
parameters $S_p$ might vary weakly with $\sigma$ affecting the
expansion (\ref{fb:edge}) beyond the third-order term (see
\cite{BeKo95}).

With such an approach, it is possible to get an approximate form of
the density PDF from a few known low-order cumulants.  This method is
irreplaceable when only a few cumulants have been derived from first
principles. However, it is important to note that this expansion is
valid only in the {\sl slightly } non-Gaussian regime. The validity
domain of the form (\ref{fb:edge}) is limited to finite values of
$\de/\sigma$, typically $\de/\sigma \la 0.5$. 

A well-known problem with the Edgeworth expansion is that it does not
give a positive definite PDF, in particular this manifests itself in
the tails of the distribution. To improve this behavior, an
Edgeworth-like expansion about the Gamma PDF (which has exponential
tails) has been explored in~\cite{GFE00}. To bypass the positivity
problem, it was proposed to apply the Edgeworth expansion to the
logarithm of the density instead of the density
itself~\cite{Colombi94}. With this change of variable, motivated by
dynamics~\cite{CoJo91}, the approximation works well even into the
nonlinear regime for $\sigma^2 \la 10$~\cite{Colombi94,UeYo96}.

Extensions of Eq.~(\ref{fb:edge}) have been written for joint PDF's
\cite{LonguetHiggins64,Lokas98}. Note that it can be done only when 
the cross-correlation matrix between the variables is regular
(see \cite{BCLSK99} for details).

\clearpage 
                                                                     
\section{\bf From Dynamics to Statistics: $N$-Point Results}
\label{chapter4}

A general approach to go from dynamics to statistics would be to solve
the Vlasov equation from initial conditions for the phase-space
density function $f(\vx,\vp)$ given by a stochastic process such as
inflation. Correlation functions in configuration space reviewed in
Chapter~3 can be trivially extended to phase-space, and the Vlasov
equation yields equations of motion for these phase-space correlation
functions. The result is a set of coupled non-linear
integro-differential equations, the so-called BBGKY
hierarchy\footnote{after N. N. Bogoliubov, M. Born, H. S. Green,
J. G. Kirkwood and J. Yvon, who independently obtained the set of
equation between 1935 and 1962. Rigorously, this route from the Vlasov
equation to the BBGKY equations is restricted to the so-called ``fluid
limit'' in which the number of particles is effectively infinite and
there are no relaxation effects.}, in which the one-point density is
related to the two-point phase-space correlation function, the
two-point depends on the three-point, and so forth. However, as
mentioned in Chapter~2, if we restrict ourselves to the single stream
regime study of the Vlasov equation reduces to studying the evolution
of the density and velocity fields given by the continuity, Euler and
Poisson equations. Therefore, all we have to consider in this case is
the correlation functions of density and velocity fields.

In this chapter, we review how the results discussed in Chapter~2
about the time evolution of density and velocity fields can be used to
understand the evolution of their statistical properties,
characterized by correlation functions as summarized in the previous
chapter. Most of the calculations will be done assuming Gaussian
initial conditions; in this case the main focus is in quantitative
understanding of the emergence of non-Gaussianity due to non-linear
evolution. In Sect.~\ref{ngic} we discuss results derived from
non-Gaussian initial conditions. In Chapter~5 we present, with similar
structure, analogous results for one-point statistics, with emphasis
on the evolution of local moments and PDF's.

\subsection{The Weakly Non-Linear Regime: ``Tree-Level'' PT}
\label{sec:TLPT}

\subsubsection{Emergence of Non-Gaussianity}
\label{emng}

If the cosmic fields are Gaussian, their power spectrum $P(k,\tau)$,
 
\begin{equation}
        \Big< \tilde{\delta}({\bf k},\tau) \tilde{\delta}
({\bf k}',\tau)\Big>_c = \delta_D({\bf k}+{\bf k}') P(k,\tau)
        \label{powerspectrum}.
\end{equation}

(or, equivalently, their two-point correlation function) completely
describes the statistical properties. However, as we saw in Chapter 2,
the dynamics of gravitational instability is non-linear, and therefore
non-linear evolution inevitably leads to the development of
non-Gaussian features.

The statistical characterization of non-Gaussian fields is, in
general, a non-trivial subject. As we discussed in the previous
chapter, the problem is that in principle all $N-$point correlation
functions are needed to specify the statistical properties of cosmic
fields. In fact, for general non-Gaussian fields, it is not clear that
correlation functions (either in real or Fourier space) are the best
set of quantities that describes the statistics in the most useful
way.

The situation is somewhat different for gravitational clustering from
Gaussian initial conditions. Here it is possible to calculate in a
model-independent way precisely how the non-Gaussian features arise,
and what is the most natural statistical description. In particular,
since the non-linearities in the equations of motion are quadratic,
gravitational instability generates connected higher order correlation
functions that scale as $\xi_{N} \propto \xi_{2}^{N-1}$ at large
scales, where $\xi_{2}\ll 1$ and PT applies~\cite{Fry84b}. This
scaling can be naturally represented by connected tree diagrams, where
each link represents the two-point function (or power spectrum in
Fourier space), since for $N$ points $(N-1)$ links are necessary to
connect them in a tree-like fashion.

As a consequence of this scaling, the so-called {\em hierarchical
amplitudes} $Q_{N}$ defined by

\begin{equation}
        Q_{N} \equiv \frac{\xi_{N}}{\sum_{\rm labelings} \ 
        \prod_{{\rm edges}~ij}^{N-1}\ \xi_{2}(r_{ij})},
        \label{Qn}
\end{equation}

where the denominator is given by all the topological distinct tree
diagrams (the different $N^{N-2}$ ways of drawing $N-1$ links that
connect $N$ points), are a very useful set of statistical quantities
to describe the properties of cosmic fields. In particular, they are
independent of the amplitude of the two-point function, and for
scale-free initial conditions they are independent of overall scale.
As we shall see, the usefulness of these statistics is not just
restricted to the weakly non-linear regime (large scales); in fact,
there are reasons to expect that in the opposite regime, at small
scales where $\xi_{2}\gg 1$, the scaling $\xi_{N} \propto
\xi_{2}^{N-1}$ is recovered. In this sense, the hierarchical
amplitudes $Q_{N}$ (and their one-point cousins, the $S_{p}$
parameters) are the most natural set of statistics to describe the
non-Gaussianity that results from gravitational clustering.

 \begin{figure}
 \centering
 \centerline{\epsfxsize=4. truecm \epsfysize=1.5 truecm 
 \epsfbox{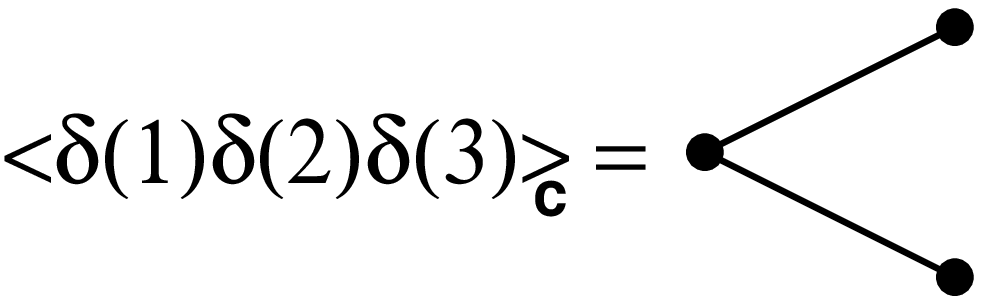}}
 \caption{ Tree diagrams for the three-point function or bispectrum. }
 \label{fig4_1}
 \end{figure}
 \begin{figure}
 \centering
 \centerline{\epsfxsize=8. truecm \epsfysize=1.5 truecm 
 \epsfbox{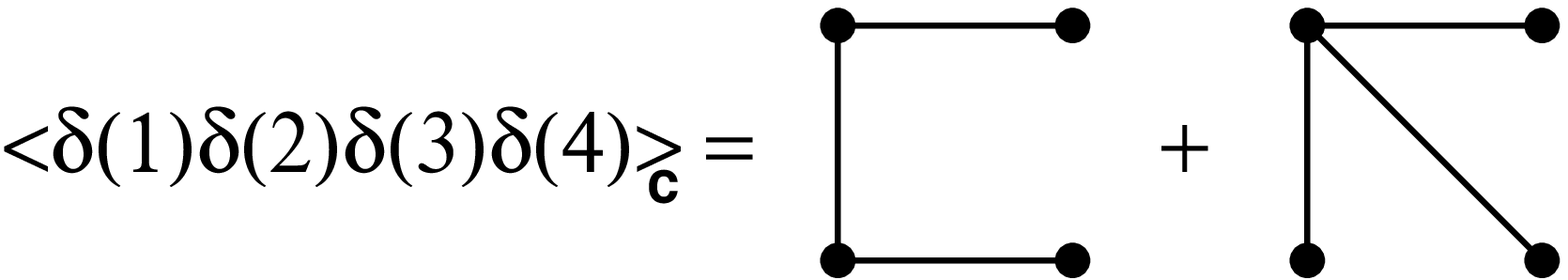}}
 \caption{ Tree diagrams for the four-point function or trispectrum. }
 \label{fig4_2}
 \end{figure}

Figures~\ref{fig4_1} and~\ref{fig4_2} show the tree diagrams that
describe the three- and four-point function induced by gravity. As we
already said, $N-1$ links (representing $\xi_2$) are needed to
describe the connected $N$-point function, and furthermore, the number
of lines coming out of a given vertex is the order in PT that gives
rise to such a diagram. For example, the diagram in Fig.~\ref{fig4_1}
requires linear and second order PT, representing
$\langle\de_2(1)\de_1(2)\de_1(3)\rangle_c$ 
(as in Chapter 2, subscripts describe the
order in PT). On the other hand, the diagrams in Fig.~\ref{fig4_2}
require up to third-order in PT. The first term represents
$\langle\de_1(1)\de_2(2)\de_2(3)\de_1(4)\rangle_c$ whereas the second describes
$\langle\de_1(1)\de_3(2)\de_1(3)\de_1(4)\rangle_c$. 

In general, a consistent calculation of the connected $p-$point
function induced by gravity to leading order (``tree-level'') requires
from first to $(p-1)^{\rm th}$ order in PT~\cite{Fry84b}. At large
scales, where $\xi_2 \ll 1$, tree-level PT leads to hierarchical
amplitudes $Q_N$ which are independent of $\xi_2$. As $\xi_2
\rightarrow 1$, there are corrections to tree-level PT which describe
the $\xi_2$ dependence of the $Q_N$ amplitudes. These are naturally
described in terms of diagrams as well, in particular, the next to
leading order contributions (``one-loop'' corrections) require from
first to $(p+1)^{\rm th}$ order in
PT~\cite{ScFr96a}. These are represented by one-loop
diagrams, i.e. connected diagrams where there is one closed loop. The
additional link over a tree diagram required to form a closed loop
leads to $Q_N \propto \xi_2$.

\begin{figure} \centering \centerline{\epsfxsize=10. truecm
 \epsfysize=1. truecm \epsfbox{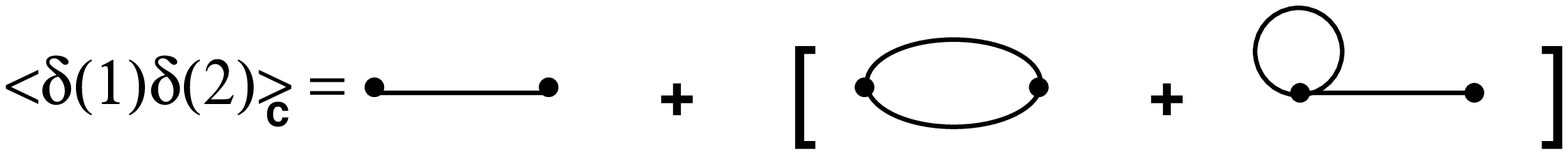}} 
\caption{ Diagrams for the two-point function or power spectrum up to
 one-loop.  See Eqs.~(\protect \ref{P22}) and (\protect \ref{P13}) for
 one-loop diagram amplitudes.}  
\label{fig4_3} 
\end{figure}

\begin{figure}
 \centering
 \centerline{\epsfxsize=12. truecm \epsfysize=2. truecm 
 \epsfbox{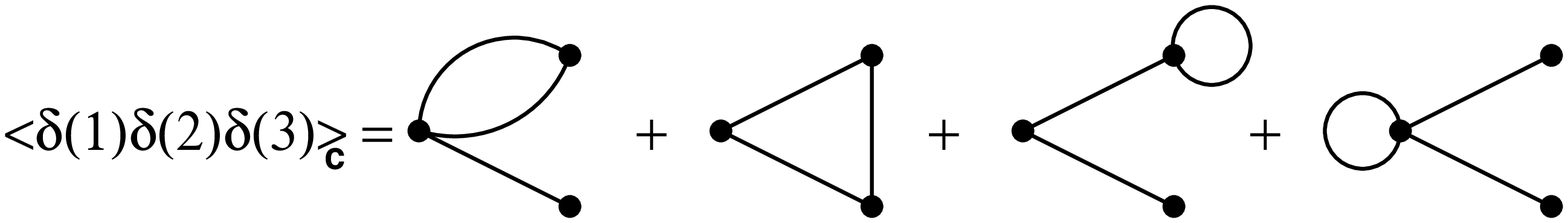}}
 \caption{ Diagrams for the three-point function or 
bispectrum up to one-loop. }
 \label{fig4_4}
 \end{figure}

Figures~\ref{fig4_3} and~\ref{fig4_4} show the one-loop diagrams for
the power spectrum and bispectrum. The one-loop corrections to the
power spectrum (the two terms in square brackets in Fig.~\ref{fig4_3})
describe the non-linear corrections to the linear evolution, that is,
the effects of mode-coupling and the onset of non-linear structure
growth. Recall that each line in a diagram represents the power
spectrum $P^{(0)}(k)$ (or two-point function) of the {\em linear}
density field. As a result, the one-loop power spectrum scales
$P^{(1)}(k) \propto P^{(0)}(k)^2$.

Are all these diagrams really necessary? In essence, what the
diagrammatic representation does is to order the contributions of the
same order irrespective of the statistical quantity being
considered. For example, it is not consistent to consider the
evolution of the power spectrum in second-order PT (second term in
Fig~\ref{fig4_3}) since there is a contribution of the {\em same}
order coming from third-order PT (third term in
Fig~\ref{fig4_3}). Instead, one should consider the evolution of the
power spectrum to ``one-loop'' PT (which includes the two
contributions of the same order, the terms in square brackets in
Fig~\ref{fig4_3}).  A similar situation happens with the connected
four-point function induced by gravity; it is inconsistent to
calculate it in second-order PT (first term in Fig~\ref{fig4_2}),
rather a consistent calculation of the four-point function to leading
order requires ``tree-level'' PT (which also involves third-order PT,
i.e. the second term in Fig~\ref{fig4_2}).

We will now review results on the evolution of different statistical 
quantities in tree-level PT.


\subsubsection{Power Spectrum Evolution in Linear PT}

The simplest (trivial) application of PT is the leading order
contribution to the evolution of the power spectrum. Since we are
dealing with the two-point function in Fourier space ($N=2$), only
linear theory is required, that is, the connected part is just given
by a single line joining the two points.

In this review we are concerned about time evolution of the cosmic
fields during the matter domination epoch. In this case, as we
discussed in Chapter 2, diffusion effects are negligible and the
evolution can be cast in terms of perfect fluid equations that
describe conservation of mass and momentum. In this case, the
evolution of the density field is given by a simple time-dependent
scaling of the ``linear'' power spectrum

\begin{equation}
        P(k,\tau) = [D_{1}^{(+)}(\tau)]^{2}\ P_L(k)
        \label{pklinear}
\end{equation}

where $D_{1}^{(+)}(\tau)$ is the growing part of the linear growth
factor.  One must note, however, that the ``linear'' power spectrum
specified by $P_L(k)$\footnote{We denote the linear power spectrum
interchangeably by $P_L(k)$ or by $P^{(0)}(k)$.} derives from the
linear evolution of density fluctuations through the radiation
domination era and the resulting decoupling of matter from radiation.
This evolution must be followed by using general relativistic
Boltzmann numerical codes~\cite{PeYu70,BoEf84,MaBe95,SeZa96}, although
analytic techniques can be used to understand quantitatively the
results~\cite{HuSu95,HuSu96}.  The end result is that

\begin{equation}
        P_L(k) = k^{n_{p}}\ T^{2}(k)
        \label{transfer}
\end{equation}

where $n_{p}$ is the primordial spectral index ($n_{p}=1$ denotes the
canonical scale-invariant
spectrum~\cite{Harrison70,Zeldovich72,PeYu70}\footnote{This
corresponds to fluctuations in the gravitational potential at the
Hubble radius scale that have the same amplitude for all modes,
i.e. the gravitational potential has a power spectrum $P_\varphi \sim
k^{-3}$, as predicted by inflationary models, see
Eq.~(\ref{fb:phikinfl}).}), $T(k)$ is the transfer function that
describes the evolution of the density field perturbations through
decoupling ($T(0) \equiv 1$).  It depends on cosmological parameters
in a complicated way, although in simple cases (where the baryonic
content is negligible) it can be approximated by a fitting function
that depends on the shape parameter $\Gamma \equiv
\Omega_m h$~\cite{BoEf84,BBKS86}. For the adiabatic cold dark matter (CDM)
scenario, $T^{2}(k) \rightarrow \ln^{2}(k)/k^4$ as $k
\rightarrow \infty$, due to the suppression of fluctuations growth 
during the radiation dominated era, see e.g.~\cite{Efstathiou90} for a
review.

\subsubsection{The Bispectrum induced by Gravity}
\label{bispgrav}

We now focus on the non-linear evolution of the three-point cumulant
of the density field, the bispectrum $ B({\bf k}_1,{\bf k}_2,\tau)$,
defined by (compare with Eq.~\ref{powerspectrum})

\begin{equation}
        \Big< \tilde{\delta}({\bf k}_1,\tau) \tilde{\delta}
({\bf k}_2,\tau) \tilde{\delta}({\bf k}_3,\tau) \Big>_c = 
\delta_D({\bf k}_1+{\bf k}_2+{\bf k}_3) \ B({\bf k}_1,{\bf k}_2,\tau)
        \label{bispectrum},
\end{equation} 

\noindent As we discussed already, it is convenient to define the
reduced bispectrum $Q$ as follows~\cite{FrSe82,Fry84b}

\begin{equation}
    \tilde{Q} \equiv \frac{ B({\bf k}_1,{\bf k}_2,\tau)}{ P(k_1,\tau)
     P(k_2,\tau) + P(k_2,\tau) P(k_3,\tau) + P(k_3,\tau) P(k_1,\tau) }
     \label{q},
\end{equation} 

\noindent which has the desirable property that it is scale and time
independent to lowest order (tree-level) in non-linear PT,

\begin{equation}
\tilde{Q}^{(0)} = \frac{ 2 F_2(\vk_1,\vk_2) P(k_1,\tau)P(k_2,\tau) + {\rm
cyc.}}{P(k_1,\tau) P(k_2,\tau) + P(k_2,\tau) P(k_3,\tau) + P(k_3,\tau)
P(k_1,\tau) } \label{qtree},
\end{equation} 
 
where $F_2(\vk_1,\vk_2)$ denotes the second-order kernel obtained from
the equations of motion, as in Section~\ref{subsec:eds}. Recall that
this kernel is very insensitive to cosmological parameters [see
Eq.(\ref{F2g})], as a consequence of this, the tree-level reduced
bispectrum $\tilde{Q}^{(0)}$ is almost independent of
cosmology~\cite{Fry94a,HBCJ95}. In addition, from Eq.~(\ref{qtree}) it
follows that $\tilde{Q}^{(0)}$ is independent of time and
normalization~\cite{Fry84b}.  Furthermore, for scale-free initial
conditions, $P_L(k) \propto k^{n}$, $\tilde{Q}^{(0)}$ is also independent of
overall scale.  For the particular case of equilateral configurations
($k_{1}=k_{2}=k_{3}$ and $\hat{k}_{i} \cdot
\hat{k}_{j}=-0.5$ for all pairs), $\tilde{Q}^{(0)}$ is independent of spectral
index as well, $\tilde{Q}^{(0)}_{EQ}=4/7$. In general, for scale-free initial
power spectra, $\tilde{Q}^{(0)}$ depends on configuration shape through, e.g.,
the ratio $k_{1}/k_{2}$ and the angle $\theta$ defined by $\hat{k}_{1}
\cdot \hat{k}_{2} =\cos{\theta}$. In fact, since
bias between the galaxies and the underlying density field is known to
change this shape dependence~\cite{FrGa93}, measurements of the
reduced bispectrum $Q$ in galaxy surveys could provide a measure of
bias which is insensitive to other cosmological
parameters~\cite{Fry94a}, unlike the usual determination from peculiar
velocities which has a degeneracy with the density parameter
$\Omega_m$. We will review these applications in
Chapter~\ref{chapter9}.

\begin{figure}[t!]
\centering
\centerline{\epsfxsize=8. truecm \epsfysize=8. truecm 
\epsfbox{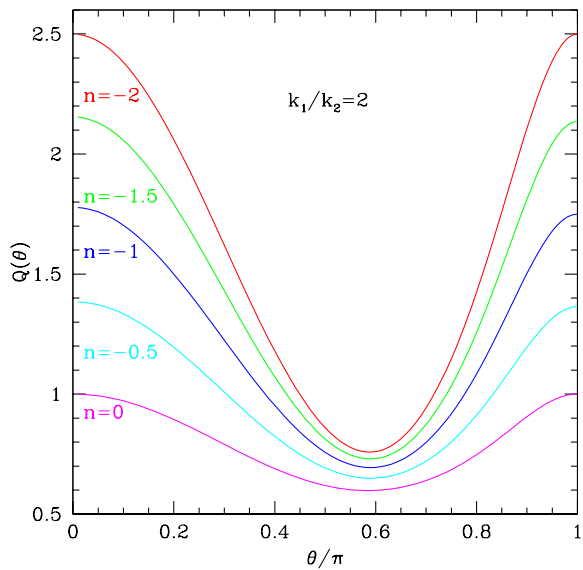}}
\caption{The tree-level reduced bispectrum $\tilde{Q}^{(0)}$ for triangle
configurations given by $k_{1}/k_{2}=2$ as a function of the angle
$\theta$ ($\hat{k}_{1} \cdot \hat{k}_{2} =\cos{\theta}$). The
different curves correspond to spectral indices $n=-2,-1.5,-1,-0.5,0$
(from top to bottom)}
\label{figbitree}
\end{figure}

Figure~\ref{figbitree} shows $\tilde{Q}^{(0)}$ for the triangle configuration
given by $k_{1}/k_{2}=2$ as a function of the angle $\theta$ between
these wavevectors ($\cos \theta \equiv \hat{\vk}_1 \cdot \hat{\vk}_2$) for
different spectral indices.  The shape or configuration dependence of
$\tilde{Q}^{(0)}$ comes from the second order perturbation theory kernel
$F_2^{(s)}$ (see Eqs.~(\ref{qtree}) and (\ref{Btree})) and can be
understood in physical terms as follows. {}From the recursion relations
given in Chapter 2, we can write

\begin{equation}
 F_2^{(s)}({\bf k}_1,{\bf k}_2)= \frac{5}{14} \ 
    \Big[  \alpha({\bf k}_2, {\bf k}_1) + \alpha({\bf k}_1,{\bf
    k}_2) \Big] + \frac{2}{7} \ \beta({\bf k}_1,{\bf k}_2)
     \label{F2phys},
\end{equation} 

\noindent with $\alpha$ and $\beta$ defined in Eq.~(\ref{albe}). 
The terms in square brackets contribute a constant term, independent
of configuration, coming from the $\theta \times \delta$ term in the
equations of motion, plus terms which depend on configuration and
describe gradients of the density field in the direction of the flow
(i.e., the term ${\bf u} \cdot
\nabla \delta$ in the continuity equation). Similarly, the last term
in Eq.~(\ref{F2phys}) contributes configuration dependent terms which
come from gradients of the velocity divergence in the direction of the
flow (due to the term $({\bf u} \cdot \nabla) {\bf u}$ in Euler's
equation). Therefore, the configuration dependence of the bispectrum
reflects the anisotropy of structures and flows generated by
gravitational instability.  The enhancement of correlations for
collinear wavevectors ($\theta =0, \pi$) in Figure~\ref{figbitree},
reflects the fact that gravitational instability generates density and
velocity divergence gradients which are mostly parallel to the
flow~\cite{Scoccimarro97}. The dependence on the spectrum is also easy
to understand: models with more large-scale power (smaller spectral
indices $n$) give rise to anisotropic structures and flows with larger
coherence length, which upon ensemble averaging leads to a more
anisotropic bispectrum.

\subsubsection{The Three-Point Correlation Function}
\label{sec:threptcor}

\begin{figure}[t!]  \centering \centerline{\epsfxsize=8. truecm
\epsfysize=8. truecm \epsfbox{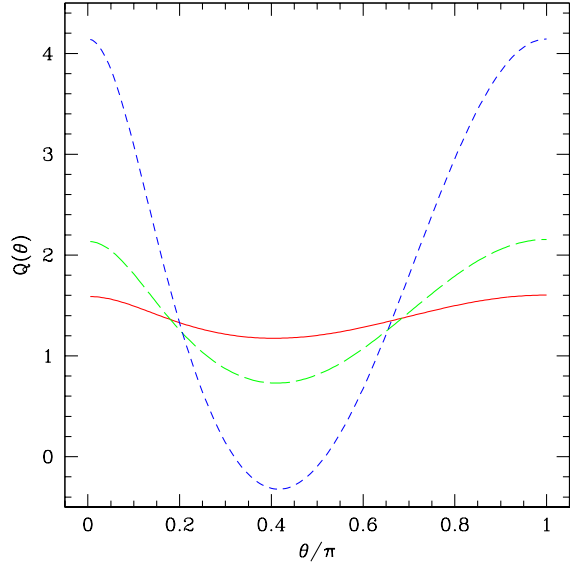}} \caption{The tree-level
three-point amplitude in real space $Q^{(0)}$ for triangle
configurations given by $r_{12}/r_{23}=2$ as a function of the angle
$\theta$ ($\hat{r}_{12} \cdot \hat{r}_{23} =\cos{\theta}$). The
different curves correspond to spectral indices $n=-2,-1.5,-1$ (from
top to bottom at $\theta=0.4\pi$)} \label{figQreal} \end{figure}

The three-point function $\xi_3$ can be found straightforwardly by
Fourier transformation of the bispectrum, leading to 

\begin{eqnarray}
\xi_3(\vx_1,\vx_2,\vx_3) 
&=& \Bigg[ \frac{10}{7} \xi(x_{13}) \xi(x_{23})+\nabla \xi(x_{13}) \cdot
\nabla^{-1} \xi(x_{23}) \nonumber \\
& & \hspace{-2cm}
+ \nabla \xi(x_{23}) \cdot
\nabla^{-1} \xi(x_{13}) + \frac{4}{7} \Big(\nabla_a \nabla_b^{-1}
\xi(x_{13}) \Big) \Big(\nabla_a \nabla_b^{-1}
\xi(x_{23}) \Big) \Bigg] + {\rm cyc.},
\end{eqnarray}

where the inverse gradient is defined by the Fourier representation

\begin{equation}
\nabla^{-1} \xi(x) \equiv -i \int \d^3\vk \exp(i \vk \cdot \vx)\ 
\frac{\vk}{k^2} P(k).
\end{equation}

For scale-free initial conditions, $P(k) \propto k^n$, $\xi(x) \propto
x^{-(n+3)}$ (with $n<0$ for convergence), and thus

\begin{eqnarray}
\xi_3(\vx_1,\vx_2,\vx_3)&= &
\Bigg[\frac{10}{7}+ \frac{n+3}{n} (\hat{x}_{13} \cdot
\hat{x}_{23}) \Big( \frac{x_{23}}{x_{13}}+\frac{x_{13}}{x_{23}}
\Big)\nonumber \\ 
& & \hspace{-2cm} + \frac{4}{7} \Big[\frac{3-2(n+3)+(n+3)^2
(\hat{x}_{13} \cdot \hat{x}_{23})^2}{n^2} \Big] \Bigg] \xi(x_{13})
\xi(x_{23}) + {\rm cyc.}.
\end{eqnarray}

Similarly to Fourier space, we can define the three-point amplitude in
real space $Q$\footnote{In this case, however, one must be careful 
not to use such a statistic for scales near the zero-crossing of 
$\xi(r)$~\cite{BuKa99}.},

\begin{equation}
Q=\frac{\xi_3(\vx_1,\vx_2\vx_3)}{\xi(x_{12})\xi(x_{23})+\xi(x_{23})\xi(x_{31})+
\xi(x_{31})\xi(x_{12})},
\label{Qreal}
\end{equation}

which is shown in Fig.~\ref{figQreal} for spectral indices
$n=-2,-1.5,-1$ (solid, dashed and short-dashed, respectively).  Note
that in real space the three-point amplitude $Q$ has a stronger shape
dependence for spectra with more power on small scales (larger
spectral index $n$), unlike the case of Fourier space.  This is
because scales are weighted differently.  Since $\xi(x)$ is actually
equivalent to $k^3 P(k)$ rather than $P(k)$, using $\xi(x)/x^3$ to
define $Q$ in real space rather than $\xi(x)$ leads to a similar
behavior with spectral index than in Fourier space.

Note that for scale-free initial conditions, the three-point amplitude
for equilateral triangles reduces to the following simple expression
as a function of spectral index $n$,
\be
Q_{\rm EQ} = \frac{18n^2+19n-3}{7n^2}.
\ee

\begin{figure}[t!]
\centerline{ {\epsfxsize=14.cm \epsfbox{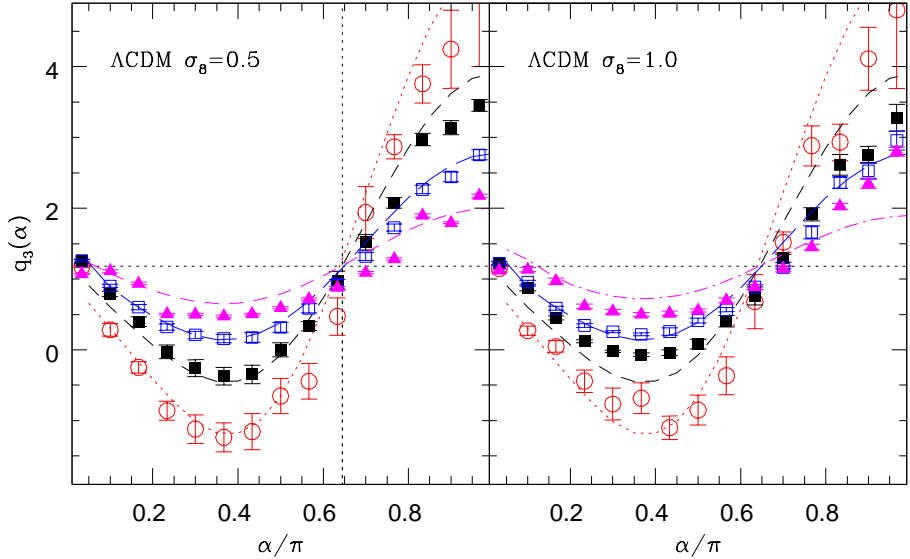}}} \vskip -3.cm
\caption{Tree-level PT predictions of the three-point amplitude
$Q^{(0)}$ in the $\Lambda CDM$ model for triangle configurations given
by $r_{12}/r_{23}=1$ as a function of the angle $\alpha$
($\hat{r}_{12} \cdot \hat{r}_{23} =\cos{\alpha}$).  The different
curves correspond to different triangle sides $r_{12}= 6, 12, 18, 24$
Mpc/h (from top to bottom at $\theta=0.4\pi$).  Symbols with error
bars correspond to measurements in numerical simulations at
$\sigma_8=0.5$ (left panel) and $\sigma_8=1.0$ (right panel).  From
\cite{BaGa01}.}
\label{figQcdm} \end{figure}

Figure~\ref{figQcdm} shows a comparison of the tree-level PT
prediction for $Q_3$ in $\Lambda CDM$ models (lines) with the fully
non-linear values of $Q_3$ measured in N-body simulations (symbols
with error bars).  Even on the earlier outputs ($\sigma_8 =0.5$, left
panel) corrections to the tree-level results become important at
scales $r_{12} < 12$ Mpc/h.  At larger scales there is an excellent
agreement with tree-level PT. This seems in contradiction with claims
in~\cite{JiBo97}, but note that for the later outputs ($\sigma_8
=1.0$, right panel) non-linear corrections can be significant at very
large scales $r_{12} < 18$ Mpc/h, so that for precision measurements
one needs to take into account the loop corrections (see~\cite{BaGa01}
for more details).

\subsection{The Transition to the Non-Linear Regime: 
``Loop Corrections''}

\subsubsection{One-Loop PT and Previrialization}
\label{previr}

In the previous section we discussed the leading order contribution to
correlations functions, and found that these are given by tree-level
PT, resulting in the linear evolution of the power spectrum and in
hierarchical amplitudes $Q_N$ independent of the amplitude of
fluctuations. Higher-order corrections to tree-level PT (organized in
terms of ``loop'' diagrams) can in principle be calculated, but what
new physics do they describe? Essentially one-loop PT describes the
first effects of mode-mode coupling in the evolution of the power
spectrum, and the dependence of the hierarchical amplitudes $Q_N$ on
$\xi_2$ . It also gives a quantitative estimate of where tree-level PT
breaks down, and leads to a physical understanding of the transition
to the non-linear regime.

One the main lessons learned from one-loop PT is the fact that
non-linear growth of density and velocity fields can be {\em slower}
than in linear PT, in contrast with e.g. the spherical collapse model
where non-linear growth is always faster than linear. This effect, is
due to tidal effects which lead to non-radial motions and thus less
effective collapse of perturbations. This was conjectured as a
possibility and termed ``previrialization''~\cite{DaPe77}; numerical
simulations however showed evidence in favor~\cite{ViDa86,Peebles90}
and against~\cite{EvCr92} this idea. The first quantitative
calculation of the evolution of power spectra beyond linear theory for
a wide class of initial conditions and comparison with numerical
simulations was done in~\cite{SuSa91}, where it was shown that one-loop
corrections to the linear power spectrum can be either negative or
positive depending on whether the initial spectral index was larger or
smaller than $n\approx -1$. Subsequent work confirmed these
predictions in greater detail~\cite{MSS92,LJBH96,ScFr96b}; in
particular, the connection between one-loop corrections to the power
spectrum and previous work on previrialization was first emphasized
in~\cite{LJBH96}. In fact, a detailed investigation shows that
one-loop PT predicts the change of behavior to occur at $n\approx
-1.4$~\cite{ScFr96b}, and divergences appear for $n\ga-1$ which must
be cutoff at some small-scale in order to produce finite results. We
shall come back to this problem below.

In addition, one-loop corrections to the bispectrum show a very
similar behavior with initial spectral
index~\cite{Scoccimarro97,SCFFHM98}. For $n \la -1.4$ one-loop
corrections increase the configuration dependence of $Q$, whereas in
the opposite case they tend to flatten it out. These results for
scale-free initial conditions are relevant for understanding other
spectra. Indeed, calculations for CDM
spectra~\cite{BaEf94,JaBe94,SCFFHM98} showed that the non-linear power
spectrum is smaller than the linear one close to the non-linear scale, where
the effective spectral index is $n \ga -1$. Furthermore, these results
give insight into the evolution of CDM-type of initial spectra:
transfer of power happens from large to small scales because more
positive spectral indices evolve slower than negative ones. In
fact, as a result, non-linear evolution drives the non-linear
power spectrum closer to the critical index $n\approx
-1$~\cite{ScFr96b,BaPa97}.

\subsubsection{The One-Loop Power Spectrum}
\label{sec:1Lpk}

As mentioned above, one-loop corrections to power spectrum (or
equivalently to the two-point correlation function) have been
extensively studied in the literature
\cite{Juszkiewicz81,Vishniac83,JSB84,Coles90,SuSa91,MSS92,JaBe94,BaEf94,LJBH96,ScFr96b}\footnote{
Multi-loop corrections to the power spectrum were considered
in~\protect\cite{Fry94b}, including the full contributions up to 2
loops and the most important terms at large $k$ in 3- and 4-loop
order.}. We now briefly review these results.

We can write the power spectrum up to one-loop corrections as

\begin{equation}
        P(k,\tau) = P^{(0)}(k,\tau) + P^{(1)}(k,\tau) + \ldots 
        \label{Ploopexp},
\end{equation}

\noindent where the superscript $(n)$ denotes an $n$-loop
contribution, the tree-level ($0$-loop) contribution is just the
linear spectrum,

\begin{equation}
        P^{(0)}(k,\tau) = [D^{(+)}_1]^2\ P_L(k)  
        \label{P^(0)},
\end{equation}

\noindent and the one-loop contribution consists of two terms (see
Fig.~\ref{fig4_3}),

\begin{equation}
        P^{(1)}(k,\tau) = P_{22}(k,\tau) + P_{13}(k,\tau) 
        \label{P^(1)},
\end{equation}
with 
\begin{eqnarray}
P_{22}(k,\tau) & \equiv & 2 \int [F_2^{(s)}({\bf k}-{\bf q},{\bf
q})]^2 P_L(|{\bf k}-{\bf q}|,\tau) P_L(q,\tau) \d^3\vq
\label{P22}, \\ 
P_{13}(k,\tau) & \equiv & 6 \int F_3^{(s)}({\bf k},{\bf q},-{\bf q})
P_L(k,\tau) P_L(q,\tau) \d^3\vq \label{P13}.
\end{eqnarray}

\noindent Here $P_{ij}$ denotes the amplitude given by a connected
diagram representing the contribution from $\langle \delta_i \delta_j
\rangle_c$ to the power spectrum. We have assumed Gaussian initial
conditions, for which $P_{ij}$ vanishes if $i+j$ is odd. Note the
different structure in the two contributions, Eq.~(\ref{P22}) is
positive definite and describes the effects of mode-coupling between
waves with wave-vectors $\vk-\vq$ and $\vq$, i.e. if $P_L(k)=0$ for
$k>k_c$, then $P_{22}(k)=0$ only when $k>2k_c$. On the other hand,
Eq.~(\ref{P13}) is in general negative (leading to the effects of
previrialization mentioned above) and does not describe mode-coupling,
i.e. $P_{13}(k)$ is proportional to $P_L(k)$. This term can be
interpreted as the one-loop correction to the propagator in
Eq.~(\ref{prop})~\cite{Scoccimarro00c}, i.e. the nonlinear correction to
the standard $a(\tau)$ linear growth.

\begin{table}
\begin{center}
\begin{tabular}{ccc}
\hline
$n$ & $P_{13}/(\pi A^2 a^4)$ & $P_{22}/(\pi A^2 a^4)$ \\ 
\hline 
$ 1$  & $-\frac{122}{315} k^3 k_c^2$  &  $\frac{18}{49} k^4 k_c$ \\ 
$0$  & $-\frac{244}{315} k^2 k_c$  &  $\frac{29\pi^2}{196} k^3$ \\ 
$-1$  & $\frac{128}{225} k -\frac{4}{3}k
\ln\frac{k_c}{\epsilon}-\frac{176}{315}k \ln\frac{k}{k_c}$  
&  $\frac{80}{147} k + \frac{4}{3}k \ln\frac{k}{\epsilon}$ \\ 
$-2$  & $\frac{5\pi^2}{28 k}-\frac{4}{3\epsilon} $  &  
$\frac{75\pi^2}{196k} +\frac{4}{3\epsilon}$ \\ 
\hline 
\end{tabular}
\caption{Contributions to the one-loop power spectrum as a function 
of spectral index $n$.}
\label{1LP}
\end{center}
\end{table}

The structure of these contributions can be illustrated by their
calculation for scale-free initial conditions, where the linearly
extrapolated power spectrum is $P_L(k) = A a^2 k^n$, shown in
Table~\ref{1LP}. The linear power spectrum is cutoff at low
wavenumbers (infrared) and high wavenumbers (ultraviolet) to control
divergences that appear in the calculation; that is, $P_L(k)=0$ for
$k<\epsilon$ and $k>k_c$. These results assume $k \gg \epsilon$ and $k
\ll k_c$, otherwise there are additional terms~\cite{MSS92,ScFr96b}.

The general structure of divergences is that for $n \le -1$ there are
infrared divergences that are caused by terms of the kind $\int
P(q)/q^2 \d^3q$; these are cancelled when the partial contributions are
added. In fact, it is possible to show that this cancellation still
holds for leading infrared divergences to arbitrary number of
loops~\cite{JaBe96}.  It was shown in~\cite{ScFr96a} that this
cancellation is general, infrared divergences arise due to the rms
velocity field (whose large-scale limit variance is $\int P(q)/q^2
\d^3q$), but since a homogeneous flow cannot affect equal-time
correlation functions because of Galilean invariance of the equations
of motion, these terms must cancel at the end.

Ultraviolet divergences are more harmful. We see from Table~\ref{1LP}
that as $n\ge -1$ the $P_{13}$ contribution becomes ultraviolet
divergent (and when $n\ge 1$ for $P_{22}$ as well), but in this case
there is no cancellation. Thus, one-loop corrections to the power
spectrum are meaningless at face value for scale-free initial
conditions with $n\ge -1$. Furthermore, one-loop corrections to the
bispectrum are also divergent for scale-free initial spectra as $n \to
-1$.  Of course, it is possible that these divergences are cancelled
by higher-order terms, but to date this has not been
investigated. This seems a rather academic problem, since no linear
power spectrum relevant in cosmology is scale-free, and for CDM-type
spectra there are no divergences. On the other hand, understanding
this problem may shed light on aspects of gravitational clustering in
the transition to the non-linear regime.

\begin{figure}[t!]
\centering
\centerline{\epsfxsize=8.  truecm \epsfysize=8.  truecm 
\epsfbox{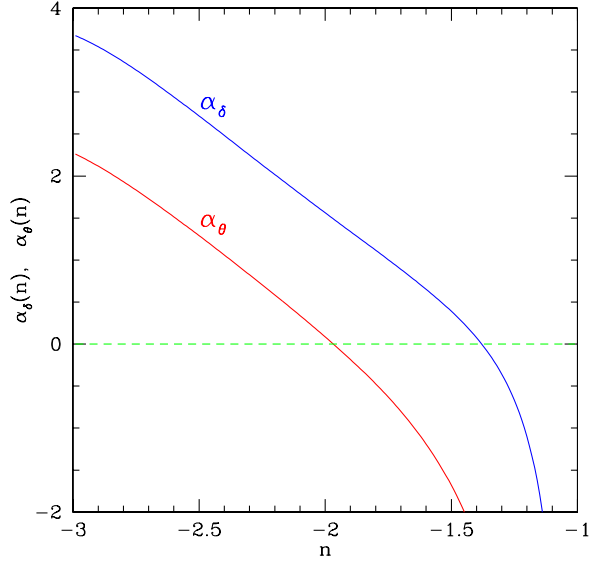}} 
\caption{One-loop corrections to the power spectrum of the density
field as a function of spectral index [see
Eq.~(\protect{\ref{1LDk}})]. Also shown is the one-loop corrections to
the velocity divergence power spectrum, $\alpha_\theta(n)$. Note that
non-linear effects can slow down the growth of the velocity power
spectrum for a broader class of initial conditions than in case of the
density field.}
\label{alphafn}
\end{figure}

To characterize the degree of non-linear evolution when including
one-loop corrections, it is convenient to define a physical scale from
the linear power spectrum, the {\em non-linear scale} $R_0$, as the
scale where the smoothed linear variance is unity,

\begin{equation}
         \sigma^2_\ell(R_0) = \int \d^{3}\vk \ P_L(k,\tau) \ W^{2}(k
         R_{0}) \equiv 1 \label{R0}.
\end{equation}

For scale-free initial conditions and a Gaussian filter,
$W(x)=\exp(-x^2/2)$, Eq.~(\ref{R0}) gives $R_0^{n+3}= 2\pi Aa^2
\Gamma[(n+3)/2]$.  This is related to the non-linear scale 
defined from the power spectrum, $\Delta(k_{nl}) =4\pi k_{nl}^3
P(k_{nl})=1$ by 

\be
k_{nl}R_0=\Gamma[(n+5)/2].
\label{nlscale}
\ee

Figure~\ref{alphafn} displays the one-loop correction to the power
spectrum in terms of the function $\alpha_\de (n)$ defined by

\begin{equation}
        \Delta(k) \equiv \frac{2 (k
        R_{0})^{n+3}}{\Gamma[(n+3)/2]} \Big[ 1 + \alpha_\de(n) \ (k
        R_{0})^{n+3} \Big] \label{1LDk},
\end{equation}

which measures the strength of one-loop corrections (and similarly for
the velocity divergence spectrum replacing $\alpha_\de$ by
$\alpha_\te$).  This function has been calculated using the technique
of dimensional regularization in~\cite{ScFr96b} (see
Appendix~\ref{DimReg} for a brief discussion of this).  {}From
Fig.~\ref{alphafn} we see that loop corrections are significant with
$\alpha_\de$ close to unity or larger for spectral indices $n \la
-1.7$.  For $n_c \approx -1.4$ one-loop corrections to the power
spectrum vanish (and for the bispectrum as well~\cite{Scoccimarro97}). 
For this ``critical'' index, tree-level PT should be an excellent
approximation.  One should keep in mind, however, that the value of
the critical index can change when higher-order corrections are taken
into account; particularly given the proximity of $n_c$ to $n=-1$
where ultraviolet divergences drive $\alpha \rightarrow -\infty$.  On
the other hand, recent numerical results agree very well with $n_c
\approx -1.4$, at least for redshifts $z \sim 3$ evolved from CDM-like
initial spectra~\cite{ZSH01}.

Figure~\ref{alphafn} also shows the one-loop correction coefficient
$\alpha_\te$ for the velocity divergence spectrum. We see that
generally velocities grow much slower than the density field when
non-linear contributions are taken into account. For $n\ga-1.9$
one-loop PT predicts that velocities grow slower than in linear
PT. Although this has not been investigated in detail against
numerical simulations, the general trend makes sense: tidal effects
lead to increasingly non-radial motions as $n$ increases, thus the
velocity divergence should grow increasingly slower than in the linear
case.

\begin{figure}
\centering
\centerline{\epsfxsize=8. truecm \epsfysize=8. truecm 
\epsfbox{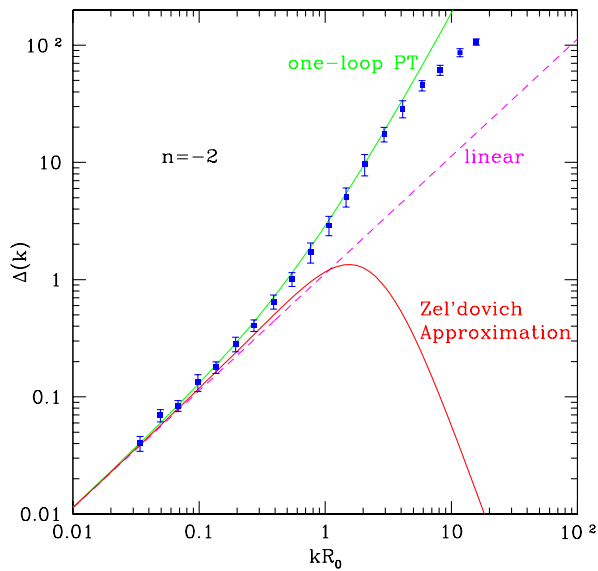}}
\caption{The power spectrum for $n=-2$ scale-free initial conditions. 
Symbols denote measurements in numerical simulations
from~\cite{SCFFHM98}. Lines denote linear PT, one-loop PT
[Eq.~(\ref{1LDk})] and the Zel'dovich Approximation results
[Eq.~(\ref{pkza})], as labeled.}
\label{fig_pow_ptza}
\end{figure}

\begin{figure}[t!]
\centering
\centerline{\epsfxsize=10.  truecm \epsfysize=10.  truecm 
\epsfbox{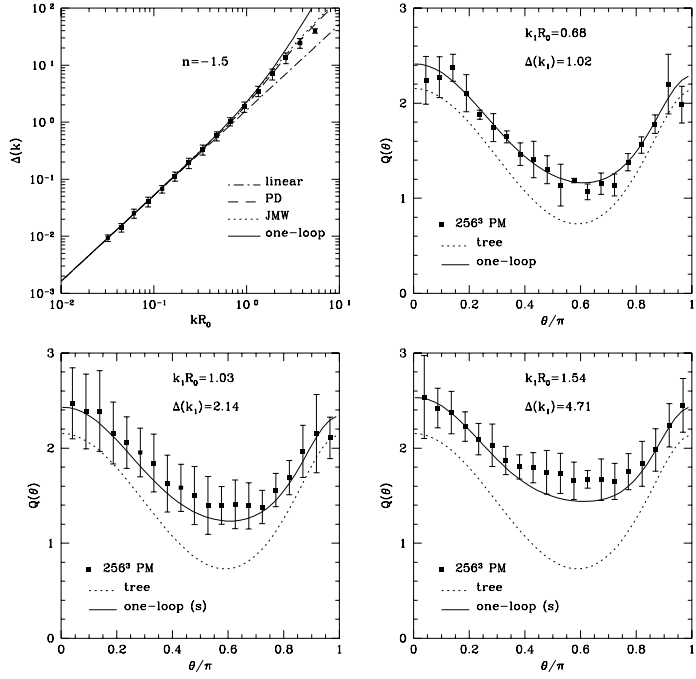}} 
\caption{ The left top panel shows the non-linear power spectrum as a
function of scale for $n=-1.5$ scale-free initial conditions.  Symbols
denote measurements in numerical simulations, whereas lines show the
linear, and the fitting formulas of~\cite{JMW95,PeDo96} and one-loop
perturbative results, as indicated.  The other three panels show the
reduced bispectrum $Q$ for triangle configurations with
$k_{1}/k_{2}=2$, as a function of the angle $\theta$ between $\vk_1$
and $\vk_2$, in numerical simulations and for tree-level and one-loop
PT.  The panels correspond to stages of non-linear evolution
characterized by $\Delta(k_1)$. Taken from~\cite{SCFFHM98}.}
\label{fig_1LBnm1p5}
\end{figure}

Figure~\ref{fig_pow_ptza} compares the results of one-loop corrections
for $n=-2$ against numerical simulations, whereas the top left panel
in Fig.~\ref{fig_1LBnm1p5} shows results for $n=-1.5$. In both cases
we see very good agreement even into considerably non-linear scales
where $\Delta(k)\sim 10-100$, providing a substantial improvement over
linear PT. Also note the general trend, in agreement with numerical
simulations, that non-linear corrections are significantly larger for
$n=-2$ than for $n=-1.5$.

\subsubsection{The One-Loop Bispectrum}
\label{sec:bi}

The loop expansion for the bispectrum, $B = B^{(0)} + B^{(1)} +
\ldots$, is given by the tree-level part $B^{(0)}$ in terms a single
diagram from second-order PT (see Fig.~\ref{fig4_1}) plus its
permutations over external momenta (recall that $\vk_1+\vk_2+
\vk_3\equiv 0$)

\begin{eqnarray}
B^{(0)} &\equiv& 2 P_L(k_1) P_L(k_2) F_2^{(s)}({\bf k}_1,{\bf
k}_2) + 2 P_L(k_2) P_L(k_3) F_2^{(s)}({\bf k}_2,{\bf k}_3) \nonumber
\\ & & + 2 P_L(k_3) P_L(k_1) F_2^{(s)}({\bf k}_3,{\bf k}_1)
\label{Btree}.
\end{eqnarray}

\noindent The one-loop contribution consists of four distinct diagrams
involving up to fourth-order solutions~\cite{Scoccimarro97,SCFFHM98}

\begin{equation}
B^{(1)} \equiv B_{222} + B_{321}^{I} + B_{321}^{II} + B_{411}
        \label{B1loop},
\end{equation}

\noindent where:

\label{1loopB}
\begin{eqnarray}
         B_{222} & \equiv & 8 \int \d^{3}\vq P_L(q,\tau) F_2^{(s)}(-{\bf
         q},{\bf q}+{\bf k}_1) P_L(|{\bf q}+{\bf k}_1|,\tau) \nonumber
         \\ & & \ \ \ \times F_2^{(s)}(-{\bf q}- {\bf k}_1,{\bf
         q}-{\bf k}_2) P_L(|{\bf q}-{\bf k}_2|,\tau) F_2^{(s)}({\bf
         k}_2-{\bf q},{\bf q}) \label{B222}, \\ & & 
        \nonumber \\ B_{321}^I & \equiv & 6
         P_L(k_3,\tau) \int \d^{3}\vq P_L(q,\tau) F_3^{(s)}(-{\bf q},{\bf
         q}-{\bf k}_2,-{\bf k}_3 ) P_L(|{\bf q}-{\bf k}_2|,\tau)
         \nonumber \\ & & \ \times F_2^{(s)}({\bf q},{\bf k}_2-{\bf
         q}) + {\rm permutations} \label{B321I}, \\ & & 
        \nonumber \\ B_{321}^{II} &
         \equiv & 6 P_L(k_2,\tau) P_L(k_3,\tau) F_2^{(s)}({\bf
         k}_2,{\bf k}_3) \int \d^{3}\vq P_L(q,\tau) F_3^{(s)}({\bf
         k}_3,{\bf q},-{\bf q}) \nonumber \\ & & + {\rm permutations}
         \label{B321II}, \\ & & 
        \nonumber \\ B_{411} & \equiv & 12 P_L(k_2,\tau)
         P_L(k_3,\tau) \int \d^{3}\vq P_L(q,\tau) F_4^{(s)}({\bf q},-{\bf
         q},-{\bf k}_2,-{\bf k}_3) \nonumber \\ & & + {\rm
         permutations} \label{B411}.
\end{eqnarray}

\noindent  For the reduced bispectrum $\tilde{Q}$ [see Eq.~(\ref{q})], 
the loop expansion  yields:
 
\begin{equation}
    \tilde{Q}\equiv \frac{ B^{(0)} +
        B^{(1)} + \ldots}
    { \Sigma^{(0)} + 
        \Sigma^{(1)} + \ldots }
        \label{Qexpand},
\end{equation} 

\noindent where 
$\Sigma^{(0)}\equiv
P_L(k_1)P_L(k_2)+P_L(k_2)P_L(k_3)+P_L(k_3)P_L(k_1)$, and its one-loop
correction $\Sigma^{(1)}
\equiv P^{(0)}(k_1) P^{(1)}(k_2)+{\rm permutations}$ (recall $P^{(0)}
\equiv P_L$). For large scales, it is possible to expand 
$\tilde{Q} \equiv  \tilde{Q}^{(0)} + \tilde{Q}^{(1)} +~\ldots $, which gives:

\begin{equation}
\tilde{Q}^{(0)}\equiv \frac{ B^{(0)}}{ \Sigma^{(0)} }\ \ \ \ \ 
\tilde{Q}^{(1)}\equiv \frac{ B^{(1)}-\tilde{Q}^{(0)}\Sigma^{(1)}}{\Sigma^{(0)}}
\label{q1l}.
\end{equation}

\noindent Note that $\tilde{Q}^{(1)}$ depends on the normalization of the
linear power spectrum, and its amplitude increases with time
evolution.  For initial power-law spectra $P_L(k)=Aa^2 k^{n}$ with
$n=-2$ the calculation using dimensional regularization (see
Appendix~\ref{DimReg}) yields a closed form; otherwise the result can
be expressed in terms of hypergeometric functions of two
variables~\cite{Scoccimarro97} or computed by direct numerical
integration~\cite{SCFFHM98}.

Figure~\ref{fig_1LBnm1p5} shows the predictions of one-loop PT
compared to N-body simulations for scale-free initial conditions with
$n=-1.5$. In the top right panel, we see that the predictions of
Eq.~(\ref{q1l}) agree very well with simulations at the nonlinear
scale. In the bottom panels, where $\Delta>1$, we have used
Eq.~(\ref{Qexpand}) instead of Eq.~(\ref{q1l}). At these scales
Eq.~(\ref{Qexpand}) {\em saturates}, that is, the one-loop quantities
$B^{(1)}$ and $\Sigma^{(1)}$ dominate over the corresponding
tree-level values and further time evolution does not change much the
amplitude $Q$, because $B^{(1)}$ and $\Sigma^{(1)}$ have the same
scale and, by self-similarity, time-dependence. At even more
non-linear scales, simulations show that the configuration dependence
of the bispectrum is completely washed out~\cite{SCFFHM98}.

\begin{figure}[t]
\begin{tabular}{cc}
{\epsfysize=6.5truecm \epsfbox{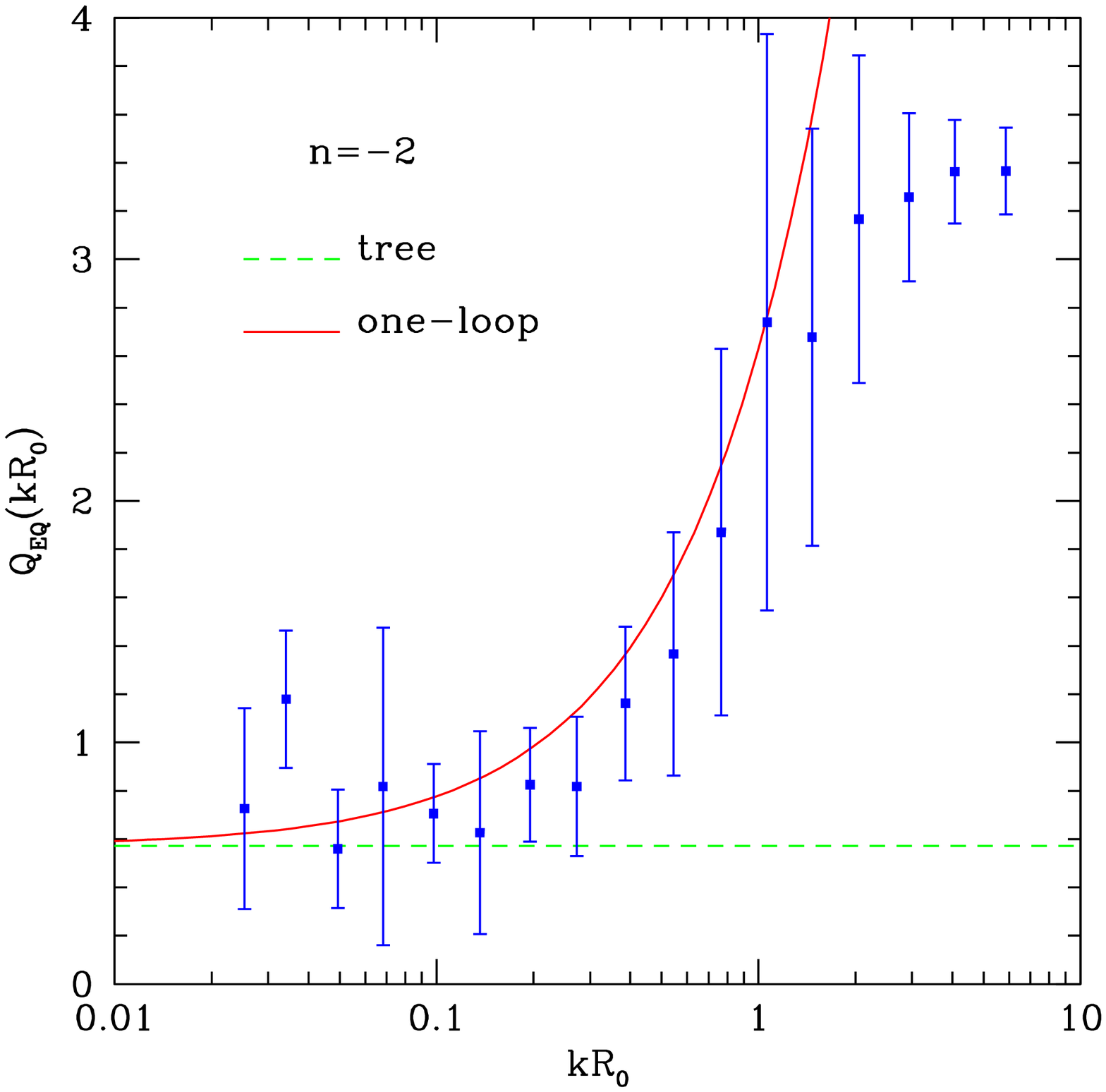}}&
{\epsfysize=6.5truecm \epsfbox{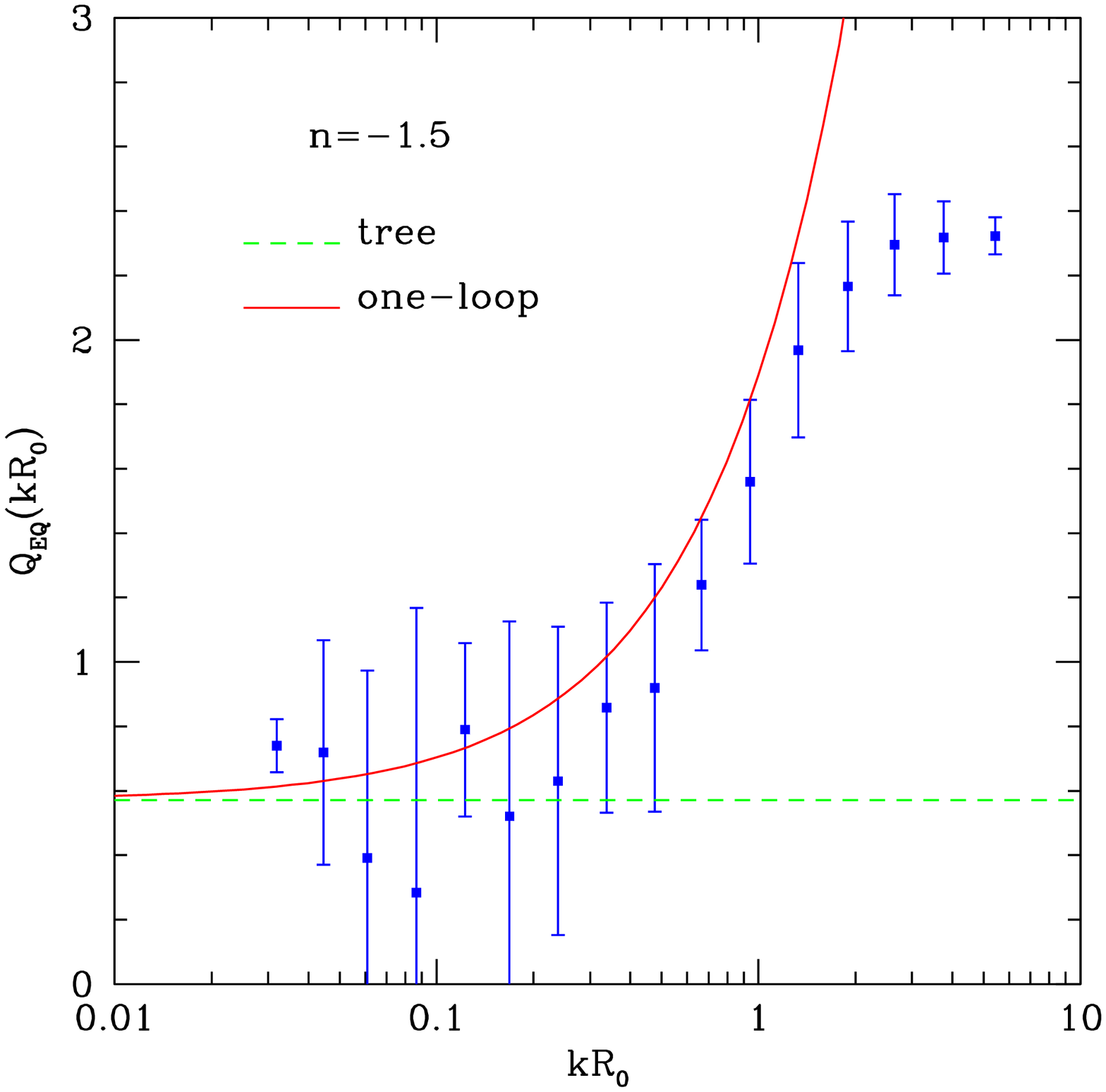}}
\end{tabular}
\caption{One-loop bispectrum predictions for equilateral
configurations for scale-free spectra with $n=-2$, Eq.~(\ref{Qeqnm2}),
and $n=-1.5$, Eq.~(\ref{Qeqnm1p5}), against N-body simulations
measurements from~\cite{SCFFHM98}. Error bars come from different
output times, assuming self-similarity, see Sect.~\ref{sss}. This
might not be well obeyed for $n=-2$, due to the importance of
finite-volume effects for such a steep spectrum, particularly at late
times, see~\cite{MaFr00a} and discussion in
Sect.~\ref{sec:cosmicinsim}.}
\label{fig_1LBeq_sf} 
\end{figure}

Using the one-loop power spectrum for $n=-2$ given in Table~\ref{1LP},
$P^{(1)}(k) =A^2a^4\ 55\pi^3/(98k)$, $\tilde{Q}^{(1)}$ follows from
Eq.~(\ref{q1l}). The calculation can be done
analytically~\cite{Scoccimarro97}; for conciseness we reproduce here
only the result for equilateral configurations,

\be
\tilde{Q}_{EQ}=\frac{4}{7}+\frac{1426697}{3863552}\pi^{3/2} kR_0= 
0.57 [1+3.6\ kR_0],\ \ \ \ \ (n=-2)
\label{Qeqnm2}
\ee
 
and for $n=-1.5$ we have from numerical integration~\cite{SCFFHM98}

\be
\tilde{Q}_{EQ}=\frac{4}{7}+1.32 (kR_0)^{3/2}= 
0.57 [1+2.316\ (kR_0)^{3/2}],\ \ \ \ \ (n=-1.5)
\label{Qeqnm1p5}
\ee

\begin{figure}[t!]
\begin{tabular}{cc}
{\epsfysize=6.truecm \epsfbox{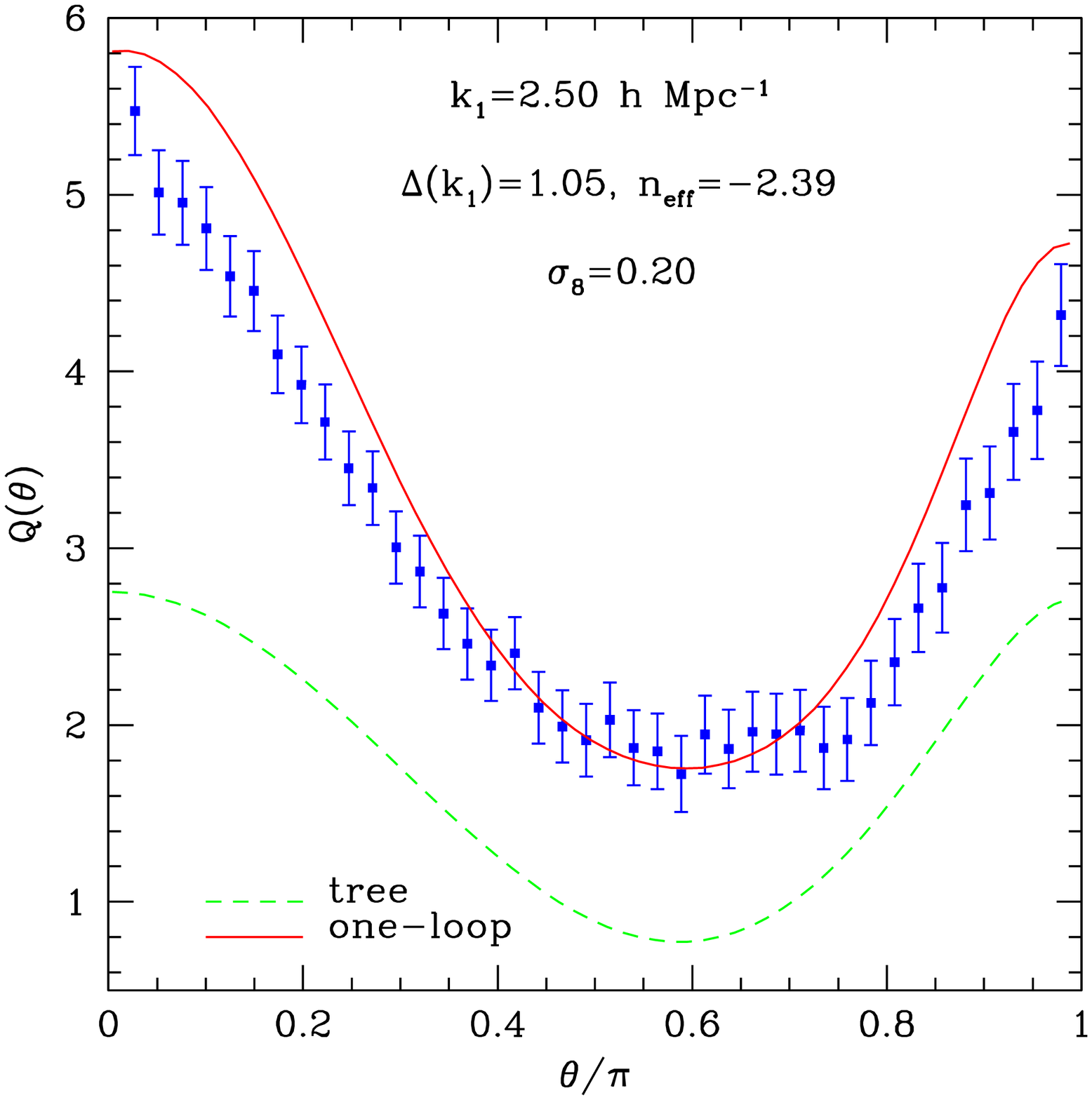}}&
{\epsfysize=6.truecm \epsfbox{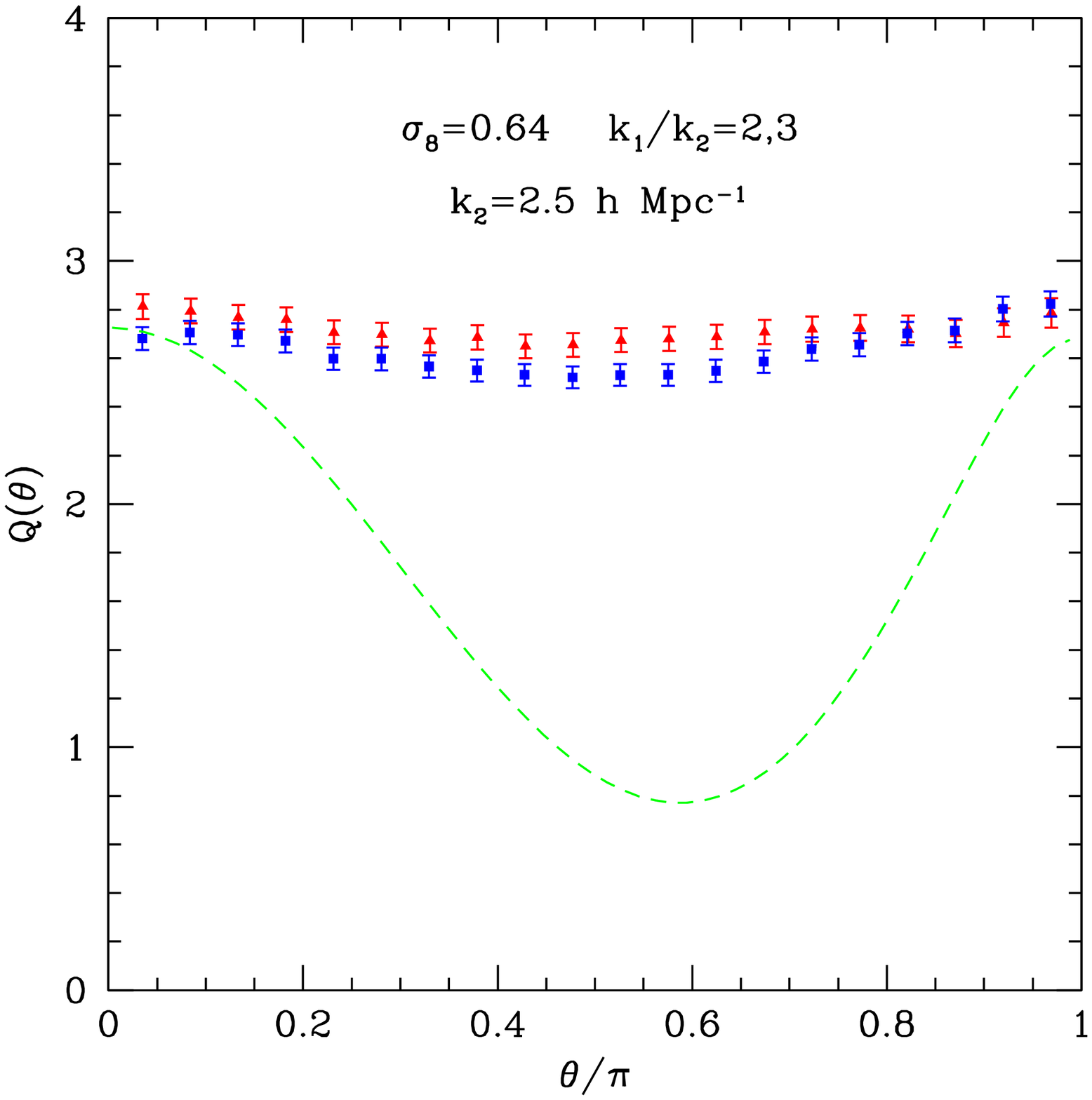}}
\end{tabular}
\caption{The left panel shows the one-loop bispectrum predictions for 
CDM model at scales approaching the non-linear regime, for $k_1/k_2=2$
and $\Delta \approx 1$ (left) against numerical
simulations~\cite{SCFFHM98}. The right panel shows the saturation of
$\tilde{Q}$ at small scales in the highly non-linear regime, for two
different ratios for $k_1/k_2=2,3$ and $\Delta \protect\ga
100$~\cite{ScFr99}. Dashed lines in both panels correspond to
tree-level PT results.}
\label{fig_Qcdm} 
\end{figure}

Figure~\ref{fig_1LBeq_sf} compares these results against N-body
simulations. We see that despite the strong corrections, with one-loop
coefficients larger than unity, one-loop predictions are accurate even
at $kR_0 =1$. As we pointed out before, many of the scale-free results
carry over to the CDM case taking into account the effective spectral
index. Figure~\ref{fig_Qcdm} illustrates the fact that one-loop
corrections can increase quite significantly the configuration
dependence of the bispectrum at weakly non-linear scales (left panel)
when the spectral index is $n<-2$, in agreement with numerical
simulations. On the other extreme, in the highly non-linear regime
(right panel), the bispectrum becomes effectively independent of
triangle shape, with amplitude that approximately matches that of
colinear amplitudes in tree-level PT.

Based on results from N-body simulations, it has been pointed out
in~\cite{FMS93} (see also~\cite{FMS95}) that for $n=-1$ nonlinear
evolution tends to ``wash out'' the configuration dependence of the
bispectrum present at the largest scales (and given by tree-level
perturbation theory), giving rise to the so-called hierarchical form
$Q \approx const$ in the strongly non-linear regime (see
Sect.~\ref{sec:HM}).  One-loop perturbation theory must predict
this feature in order to be a good description of the transition to
the nonlinear regime.  In fact, numerical
integration~\cite{Scoccimarro97} of the one loop bispectrum for
different spectral indices from $n=-2$ to $n=-1$ shows that there is a
change in behavior of the nonlinear evolution: for $n \la -1.4$ the
one-loop corrections {\it enhance} the configuration dependence of the
bispectrum, whereas for $n \ga -1.4$, they tend to cancel it, in
qualitative agreement with numerical simulations.  Note that this
``critical index'' $n_{c} \approx -1.4$ is the same spectral index at
which one-loop corrections to the power spectrum vanish, marking the
transition between faster and slower than linear growth of the
variance of density fluctuations.

\subsection{The Power Spectrum in the Zel'dovich Approximation}

The Zel'dovich approximation (ZA,~\cite{Zeldovich70}) is one of the
rare cases in which exact (non-perturbative) results can be
obtained. However, given the drastic approximation to the dynamics,
these exact results for the evolution of clustering statistics are of
limited interest due to their restricted regime of validity. The
reason behind this is that in the ZA when different streams cross they
pass each other without interacting, because the evolution of fluid
elements is local. As a result, high-density regions become washed
out. Nonetheless, the ZA often provides useful insights into
non-linear behavior. 

For Gaussian initial conditions, the full
non-linear power spectrum in the ZA can be obtained as
follows~\cite{BoCo88,MHP93,ScBa95,FiNu96,TaHa96}. Changing from
Eulerian to Lagrangian coordinates, the Fourier transform of the
density field is $\de(\vk)=\int \d^3 \vq \exp[i \vk \cdot (\vq+\Psi)]$,
where $\Psi(\vq)$ is the displacement field. The power spectrum is thus

\be
P(k)= \int \d^3 q \exp(i\vk\cdot\vq) \lexp \exp(i\vk\cdot\Delta\Psi)
\rexp, 
\label{PkZA}
\ee

where $\Delta\Psi \equiv \Psi(\vq_1)-\Psi(\vq_2)$ and
$\vq=\vq_1-\vq_2$. For Gaussian initial conditions the ZA displacement
is a Gaussian random field, so Eq.~(\ref{PkZA}) can be evaluated in
terms of the two-point correlator of $\Psi(\vq)$. An analytic result for
the power spectrum in the ZA has been obtained in~\cite{TaHa96} for
scale-free initial conditions with $-3\leq n \leq -1$. For $n=-2$ it
is

\be
\Delta(k)= \frac{(k/k_{nl})}{[1+{\pi^2 \over 64} (\frac{k}{k_{nl}})^2
]^2} \times \Bigg( 1+ {3\pi^2 \over 64}
\frac{(k/k_{nl})}{\sqrt{1+{\pi^2\over 64} (\frac{k}{k_{nl}})^2}}
\Bigg), \label{pkza}
\ee

where the non-linear wavenumber obeys $\Delta_L(k_{nl})=1$. This
result is shown in Fig.~\ref{fig_pow_ptza} (note that in the figure we
use $R_0$ to characterize the non-linear scale,
$k_{nl}R_0=\Gamma[(n+5)/2$]), together with the prediction of one-loop
PT, linear theory and measurements in N-body simulations (symbols with
error bars). Clearly the lack of power at small scales due to
shell-crossing makes the ZA prediction a poor description of the
non-linear power spectrum. Attempts have been made in the literature
to truncate the small-scale power in the initial conditions and then
use ZA~\cite{CMS93}, this improves the cross-correlation coefficient
between ZA and $N$-body simulation density
fields~\cite{CMS93,BMW94,MBW95} but it does not bring the power
spectrum into agreement~\cite{BMW94,MBW95}. Similar results for the
effect of shell crossing on the power spectrum hold for 2LPT and 3LPT,
see e.g.~\cite{BMW94,MBW95,KBM97}.

\subsection{Non-Gaussian Initial Conditions}
\label{ngic}

\subsubsection{General Results}

So far we have discussed results for Gaussian initial conditions. When
the initial conditions are not Gaussian, higher-order correlation
functions are non-zero from the beginning and their evolution beyond
linear PT is non-trivial~\cite{FrSc94}. Here we present a brief
summary of the general results for the power spectrum and bispectrum,
in the next section we discuss the application to the $\chi^2$ model,
for which correlation functions beyond linear perturbation theory have
been derived~\cite{Scoccimarro00}. This belongs to the class of
dimensional scaling models, in which the hierarchy of initial
correlation functions obey $\xi_N \sim \xi_2^{N/2}$. Another
dimensional scaling model that has been studied is the non-linear
$\sigma$-model~\cite{Jaffe94}. In addition, hierarchical scaling
models, where $\xi_N \sim \xi_2^{N-1}$ as generated by gravity from
Gaussian initial conditions, have been studied
in~\cite{LuSc93,VWHK00}. Most quantitative studies of non-Gaussian
initial conditions, however, have been done using one-point statistics
rather than correlation functions, we review them in
Sect.~\ref{ngic2}.

It is worth emphasizing that the arguments developed in this section
(and in Sect.~\ref{ngic2}) are valid only if the history of density
fluctuations can be well separated into two periods, (i) imprint of
non-Gaussian initial fluctuations at very early times, where $\sigma_I
\ll 1$, and then (ii) growth of these fluctuations due to 
gravitational instability. This is a good approximation for most
physically motivated non-Gaussian models.

Let us consider the evolution of the power spectrum and bispectrum
from arbitrary non-Gaussian initial
conditions\footnote{See~\cite{VeHe01} for a recent study of the
trispectrum for non-Gaussian initial conditions.}.  The first
non-trivial correction to the linear evolution of the power spectrum
involves second-order PT, since $\langle \de^2 \rangle = \langle
(\de_1+\de_2+\ldots)^2 \rangle \approx \langle \de_1^2 \rangle + 2
\langle \de_1 \de_2 \rangle +\ldots$; the second term which vanishes
for the Gaussian case (since $\langle \de_1 \de_2 \rangle \sim \langle
\de_1^3 \rangle$) leads instead to\footnote{See Sect.~\ref{ngic2} for
additional explanation of the new contributions that appear due to
primordial non-Gaussianity.}

\begin{equation}
P(k) = P^I(k) + 2 \int \d^3 \vq\ F_2(\vk+\vq,-\vq)\ B^I(\vk,\vq),
\label{powerNG} 
\end{equation}

which depends on the initial bispectrum $B^I$, and similarly for the
non-linear evolution of the bispectrum

\begin{eqnarray}
B_{123}^{(0)} &=& B_{123}^I + B_{123}^G + \int \d^3\vq\
F_2(\vk_1+\vk_2-\vq,\vq)\ P_4^I(\vk_1,\vk_2,\vk_1+\vk_2-\vq,\vq),
\nonumber \\ & & 
\label{BispNG}
\end{eqnarray}

where $B_{123}^I$ denotes the contribution of the initial bispectrum,
scaled to the present time using linear PT, $B_{123}^I(\tau) \propto 
[D_{1}^{(+)}(\tau)]^{3}$, $B_{123}^G$ represents the usual
gravitationally induced bispectrum, Eq.~(\ref{qtree}), and the last
term represents the contribution coming from the initial trispectrum
linearly evolved to the present, $P_4^I$ given by

\bea
\langle
\de^I(\vk_1)\de^I(\vk_2)\de^I(\vk_3)\de^I(\vk_4)\rangle_c &\equiv &
\de_D(\vk_1+\vk_2+\vk_3+\vk_4) \  P_4^I(\vk_1,\vk_2,\vk_3,\vk_4). 
\nonumber \\ \label{trisp}
\eea
 
Clearly, the complicated term in Eq.~(\ref{BispNG}) is the last one,
which involves a convolution of the initial trispectrum with the
second-order PT kernel $F_2(\vk_1,\vk_2)$. Note that only the first
term scales as $[D_{1}^{(+)}(\tau)]^{3}$, the last two terms have the
same scaling with time, $[D_{1}^{(+)}(\tau)]^{4}$, and therefore
dominate at late times. The structure of these contributions is best
illustrated by considering a specific model, as we now do.

\subsubsection{$\chi^2$ Initial Conditions}
\label{ssec:chisq}

An example that shows how different the bispectrum can be in models
with non-Gaussian initial conditions, is the chi-squared
model~\cite{Peebles99a,Peebles99b}. There are in fact a number of
inflationary models in the literature that motivate $\chi^2$ initial
conditions~\cite{KBHP89,AMM97,LiMu97,Peebles97}. It is also possible
that this particular model may be a good representation of the general
behavior of dimensional scaling models, and thus provide valuable
insight. In this case, the density field after inflation is
proportional to the square of a Gaussian scalar field $\phi(\vx)$,
$\rho(\vx) \propto \phi(\vx)^2$. The initial correlations are easiest
calculated in real space~\cite{Peebles99b}

\bea
\xi^I_2 &=& 2 \frac{\xi_\phi^2(r)}{\sigma_\phi^4},
\label{xiI}\\
\xi_3^I &=& 2^{3/2} \sqrt{ \xi_2^I(r_{12}) \xi_2^I(r_{23}) \xi_2^I(r_{31})},
\label{zetaI}\\
\xi_4^I &=& 4 \Bigg[ 
\sqrt{ \xi_2^I(r_{12}) \xi_2^I(r_{23}) \xi_2^I(r_{34}) \xi_2^I(r_{41}) } + 
\sqrt{ \xi_2^I(r_{12}) \xi_2^I(r_{24}) \xi_2^I(r_{43}) \xi_2^I(r_{31}) } + 
\nonumber \\ &+& 
\sqrt{ \xi_2^I(r_{13}) \xi_2^I(r_{32}) \xi_2^I(r_{24}) \xi_2^I(r_{41}) } \Bigg],
\label{etaI}
\eea

where $r_{ij}\equiv |\vr_i-\vr_j|$. However, non-linear corrections
are more difficult to calculate in real space~\cite{FrSc94}, so we
turn to Fourier space.  The initial density power spectrum and
bispectrum read (a similar expression holds for the trispectrum,
see~\cite{Scoccimarro00}),

\begin{equation}
P^I(k) = 2 \int \d^3 \vq\ P_{\phi}(q) P_{\phi}(|\vk-\vq|),
\label{powerCHI} 
\end{equation}

\begin{equation}
B^I(k_1,k_2,k_3) = 12 \int \d^3 \vq\ P_{\phi}(q)P_{\phi}(|\vk_1-\vq|)
P_{\phi}(|\vk_2+\vq|),  
\label{bispCHI} 
\end{equation}

where $P_{\phi}(k)$ denotes the power spectrum of the $\phi$
field. For scale-free spectra, $P_{\phi}(k) \propto k^{n_\phi}$,
$P^I(k) \propto k^{2n_\phi+3}$, with amplitude calculable in terms of
Gamma functions; similarly the bispectrum can be expressed in terms of
hypergeometric functions~\cite{Scoccimarro00}. To calculate the
hierarchical amplitude to tree-level we also need the next-to-leading
order evolution of the power spectrum, that is Eq.~(\ref{powerNG}),
which depends on the initial bispectrum, Eq.~(\ref{bispCHI}). A simple
analytic result is obtained for the particular case, $P_{\phi}(k)=A
k^{-2}$, not too far from the ``canonical'' $n_{\phi} = -2.4$
(e.g. giving $n=-1.8$,~\cite{Peebles99a,Peebles99b}),
then~\cite{Scoccimarro00}

\begin{equation}
P^I(k) = \frac{2 \pi^3 A^2}{k}+ \frac{96 \pi^4 A^3}{7},
\ \ \ \ \ \ \ \ \ \ B^I(k_1,k_2,k_3) =
\frac{12 \pi^3 A^3}{k_1 k_2 k_3}, 
\label{powbispCHI} 
\end{equation}

\noindent Defining the non-linear scale $k_{nl}$ from the linear power
spectrum as usual, $4\pi k_{nl}^3 P_L(k_{nl})=\Delta_L(k_{nl})=1$, it
follows that

\begin{equation}
\Delta(k) = \left( \frac{k}{k_{nl}} \right)^2\ \left( 1 +
\frac{24}{7\sqrt{2} \pi} \frac{k}{k_{nl}} \right),
\label{delCHI} 
\end{equation}

Then the tree-level hierarchical amplitude reads~\cite{Scoccimarro00},

\bea
\tilde{Q}_{123} &=& \frac{4\sqrt{2}}{\pi}\ \frac{k_{nl}}{k_1+ k_2+ k_3}-
\frac{192}{7\pi^2} \frac{k_1 k_2+k_2 k_3+k_3 k_1}{(k_1+ k_2+ k_3)^2} +
\tilde{Q}_{123}^G + \tilde{Q}_{123}(P_4),  \nonumber \\ & & 
\label{QCHI} 
\eea

where $\tilde{Q}_{123}^G$ denotes the hierarchical amplitude obtained
from Gaussian initial conditions, and $\tilde{Q}_{123}(P_4)$ denotes
the contribution from the last term in Eq.~(\ref{BispNG}) which is
difficult to calculate analytically. In particular, for equilateral
configurations $\tilde{Q}_{eq}^I = (4\sqrt{2}/3\pi) (k_{nl}/k)$. On
the other hand, for Gaussian initial conditions,
$\tilde{Q}_{eq}^G=4/7$ independent of spectral index; similarly there
is a contribution from non-Gaussian initial conditions that is scale
independent, $\delta \tilde{Q}_{eq} = -64/7\pi^2$. Since
$\tilde{Q}_{123}(P_4)$ is also independent of scale, it turns out that
the signature of this type of non-Gaussian initial conditions is that
$\tilde{Q}_{123}$ shows a strong scale dependence at large scales as
$k/k_{nl} \rightarrow 0$. This is not just a peculiar property of this
particular model, but rather of any non-Gaussian initial conditions
with dimensional scaling\footnote{See Sect.~\ref{ngic2} for a more
detailed discussion of this point and its generalizations.}. Note also
that $\tilde{Q}^I$ shows, in some sense, the opposite configuration
dependence from $\tilde{Q}^G$, for triangles where $k_1/k_2=2$ as in
Fig~\ref{figbitree}, $\tilde{Q}^I(\theta)$ is an increasing function
of $\theta$, as expected from the scale dependence, in particular
$\tilde{Q}^I(\pi)/\tilde{Q}^I(0) =3/2$.

\begin{figure}[t]
\begin{tabular}{cc}
{\epsfysize=6.5truecm \epsfbox{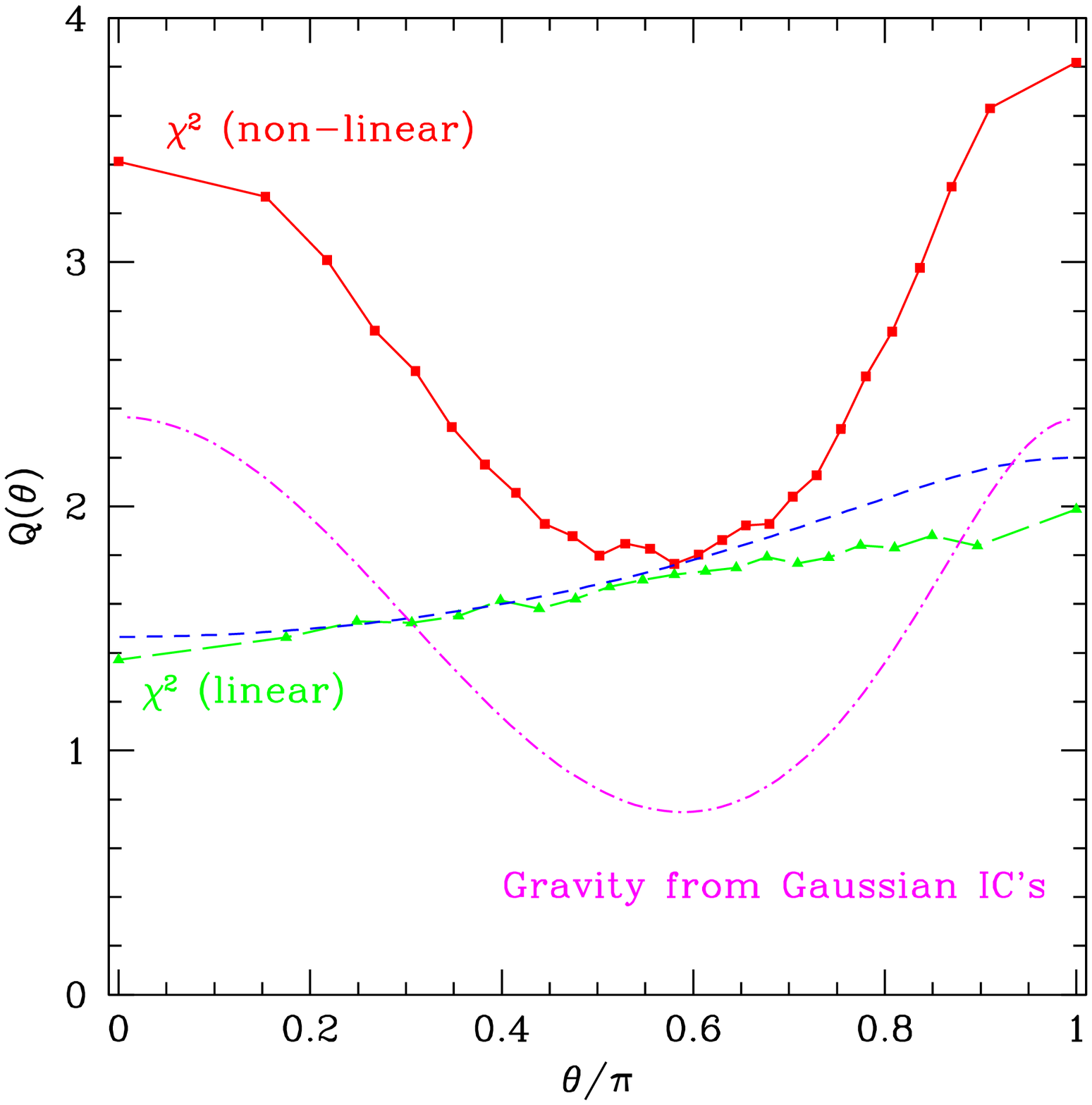}}&
{\epsfysize=6.5truecm \epsfbox{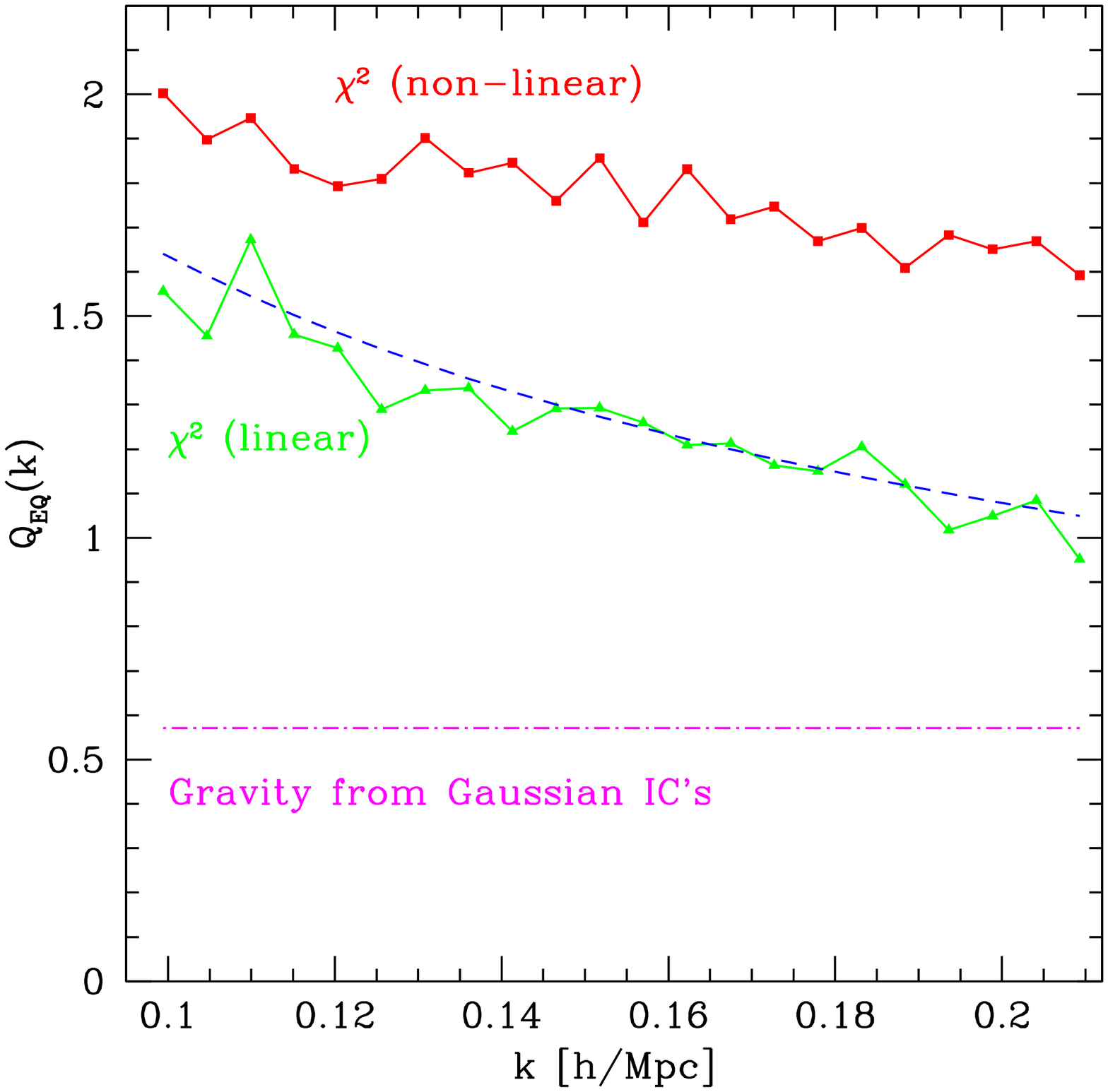}}
\end{tabular}
\caption{The reduced bispectrum $\tilde{Q}$ for triangles with sides
$k_1=0.068$ h/Mpc and $k_2=2 k_1$ as a function of the angle $\theta$
between $\vk_1$ and $\vk_2$ (left panel). Right panel shows $\tilde{Q}$ for
equilateral triangles as a function of scale $k$.  Triangles denote
linear extrapolation from $\chi^2$ initial conditions, whereas square
symbols show the result of non-linear evolution. Dot-dashed lines show
the predictions of non-linear PT from Gaussian initial conditions with
the same initial power spectrum as the $\chi^2$ model.}
\label{figiso} 
\end{figure}

Figure~\ref{figiso} shows the results of using 2LPT (see
Sect.~\ref{lagNLPT}) evolved from $\chi^2$ initial
conditions~\cite{Scoccimarro00}. The auxiliary Gaussian field $\phi$
was chosen to have a spectral index $n_{\phi}=-2.4$, leading to
$n=-1.8$ as proposed in~\cite{Peebles99a}. The amplitude of the power
spectrum has been chosen to give $k_{nl} \equiv 0.33$ h/Mpc. The
dashed lines in Fig.~\ref{figiso} (left panel) show the predictions of
the first term in Eq.~(\ref{QCHI}) for the reduced bispectrum at
$k_1=0.068$ h/Mpc, $k_2=2 k_1$, as a function of angle $\theta$
between $\vk_1$ and $\vk_2$.  This corresponds to $n=-1$, however, it
approximately matches the numerical results (triangles, $n=-1.8$). The
latter show less dependence on angle, as expected because the scale
dependence in the $n=-1.8$ case ($\tilde{Q}^I \propto k^{-0.6}$) is
weaker than for $n=-1$ ($\tilde{Q}^I\propto k^{-1}$). The right panel
in Fig.~\ref{figiso} shows equilateral configurations as a function of
scale for $\chi^2$ initial conditions (triangles) and
$\tilde{Q}^I_{eq}(k) = 0.8 (k/k_{nl})^{-0.6}$ (dashed lines), where
the proportionality constant was chosen to fit the numerical result,
this is slightly larger than the prediction in the first term of
Eq.~(\ref{QCHI}) for $n=-1$ equilateral configurations, and closer to
the real-space result $Q_{eq}(x) = 0.94 (x/x_{nl})^{0.6}$.

The behavior of the $\chi^2$ bispectrum is notoriously different from
that generated by gravity from Gaussian initial conditions for
identical power spectrum (dot-dashed lines in
Fig.~\ref{figiso})~\cite{FrGa99}. The structures generated by squaring
a Gaussian field roughly correspond to the underlying Gaussian
high-peaks which are mostly spherical, thus the reduced bispectrum is
approximately flat. In fact, the increase of $\tilde{Q}^I$ as $\theta
\rightarrow \pi$ seen in Fig.\ref{figiso} is basically due to the
scale dependence of $\tilde{Q}^I$, i.e. as $\theta \rightarrow \pi$, the side
$k_3$ decreases and thus $\tilde{Q}^I$ increases.

As shown in Eq.~(\ref{QCHI}), non-linear corrections to the bispectrum
are significant at the scales of interest, so linear extrapolation of
the initial bispectrum is insufficient to make comparison with current
observations. The square symbols in left panel of Fig.~\ref{figiso}
show the reduced bispectrum after non-linear corrections are
included. As a result, the familiar dependence of $\tilde{Q}_{123}$ on
the triangle shape due to the dynamics of large-scale structures is
recovered , and the scale dependence shown by $\tilde{Q}^I$ is now
reduced (right panel in Fig.~\ref{figiso}). However, the differences
between the Gaussian and $\chi^2$ case are very obvious: the $\chi^2$
evolved bispectrum has an amplitude about 2-4 times larger than that
of an initially Gaussian field with the same power
spectrum. Furthermore, the $\chi^2$ case shows residual scale
dependence that reflects the dimensional scaling of the initial
conditions. These signatures can be used to test this model against
observations~\cite{FrGa99,SFFF01,FFFS01}, as we shall discuss in
Sect.~8.

\subsection{The Strongly Non-Linear Regime}
\label{snlr}

In this section we consider the behavior of the density and velocity
fields in the strongly non-linear regime, with emphasis on the
connections with PT. Only a limited number of relevant results are
known in this regime, due to the complexity of solving the Vlasov
equation for the phase-space density distribution. These results,
based on simple arguments of symmetry and stability, lead however to
valuable insight into the behavior of correlations at small scales.

\subsubsection{The Self-Similar Solution}
\label{sss}

The existence of self-similar solutions relies on two assumptions
within the framework of collisionless dark matter clustering, 

\ben
\item There are no characteristic time-scales, this requires 
$\Omega_m=1$ where the expansion factor scales as a power-law, $a \sim t^{2/3}$. 
\item There are no characteristic length-scales. This implies 
scale-free initial conditions, e.g. Gaussian with initial spectrum
$P_I(k) \sim k^n$.
\een

Since gravity is scale-free, there are no scales involved in the
solution of the coupled Vlasov and Poisson equations. As a result of
this, the Vlasov equation admits self-similar solutions
with~\cite{DaPe77}

\begin{equation}
f(\vx,\vp,t)=t^{-3-3\alpha}\hat{f}\left(\vx/t^{\alpha},
\vp/t^{\beta+1/3}\right), 
\end{equation}

where $\beta=\alpha+1/3$ and $t$ is cosmic time. Integration over
momentum leads to correlation functions that are only functions of the
self-similarity variables $\vs_i \equiv \vx_i/t^{\alpha}$, in
particular the two-point correlation function reads,
\be
\xi(\vx,t)=f_2\left({x\over t^{\alpha}}\right),
\label{fb:xiselfsim}
\ee 
and similarly for higher-order correlation functions,
e.g. $\xi_3(\vx_1,\vx_2,\vx_3,t)=f_3(\vs_1,\vs_2,\vs_3)$. Note that
this solution holds in all regimes, from large to small scales. Using
the large-scale behavior expected from linear PT, it is then possible
to compute the index $\alpha$, requiring that $\xi_L(\vx,a) \sim a^2
x^{-(n+3)}$ be a function only of the self-similarity variable
$xt^{-\alpha}$ leads to

\be
\alpha={4\over 3(n+3)}. 
\label{fb:alval}
\ee

Note that the self-similar scaling of correlation functions can also
be obtained from the fluid equations of motion~\cite{ScFr96b}, as
expected since only symmetry arguments (which have nothing to do with
shell crossing) are involved\footnote{For $n=-2$, where finite volume
effects become very important, self-similarity has been difficult to
obtain in numerical simulations.  However, even in this case current
results show that self-similarity is obeyed~\cite{JaBe98}.}. 
Self-similarity reduces the dimensionality of the equations of motion;
it is possible to achieve further reduction by considering symmetric
initial conditions, e.g. planar, cylindrical or spherical.  In these
cases, exact self-similar solutions can be found by direct numerical
integration, see e.g.~\cite{FiGo84,Bertschinger85}.  Although this
provides useful insight about the non-linear behavior of isolated
perturbations, it does not address the evolution of correlation
functions.  Detailed results for correlation functions in the
non-linear regime can however be obtained by combining the
self-similar solution with stable clustering arguments, as we now
discuss.

\begin{figure}[t!]
\centering
\centerline{\epsfxsize=8truecm\epsfysize=8truecm\epsfbox{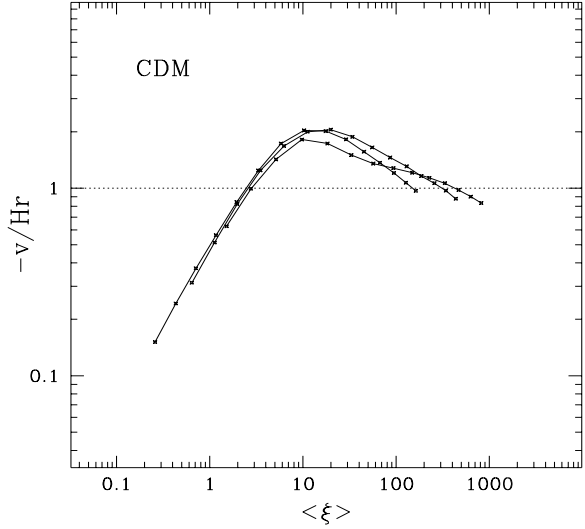}}
\caption{The ratio of the mean pair (peculiar) velocity to the Hubble
velocity, $-u/Hx$, as a function of the mean correlation function
$\xi_{\rm av}$ for a CDM model.  The pair conservation equation is
used to solve for $-u/Hx$ using the evolution of $\xi_{\rm
av}(a,x)$. The three curves are for $a=0.3, 0.6, 0.8$. They would
coincide for a scale-free spectrum. They seem to
approach the stable clustering value $-u/Hx=1$ for $\xi_{\rm
av}>200$. Taken from~\protect\cite{Jain97}.}
\label{figstcl1}
\end{figure}

\subsubsection{Stable Clustering}
\label{stcl}

Stable clustering asserts that at small scales, high-density regions
decouple from the Hubble expansion and their physical size is stable,
i.e. it does not change with time~\cite{DaPe77}. This implies that the
relative motion of particles within gravitationally bound structures
should compensate on average the Hubble expansion. Following this idea
general relations can be obtained for the behavior of the two-point
correlation function from the continuity equation alone.  Indeed, from
Eq.~(\ref{continuity}) it follows that

\begin{eqnarray}
\frac{\partial \xi_{12}}{\partial \tau} &=&
\frac{\partial}{\partial \tau} \lexp (1+\de(\vx_1))(1+\de(\vx_2)) 
\rexp\nonumber\\
 &=& \lexp -\nabla_1 [(1+\de(\vx_1))\vu(\vx_1)](1+\de(\vx_2)) \rexp
\nonumber\\
&&\hspace{+1cm}-\lexp 
(1+\de(\vx_1)) \nabla_2 [(1+\de(\vx_2)) \vu(\vx_2) ] \rexp.
\end{eqnarray}

Pulling out the derivatives using statistical homogeneity, we arrive
to the pair conservation equation~\cite{DaPe77}

\be
\frac{\partial \xi_{12}}{\partial \tau} + \nabla_{12}\cdot 
[ \vu_{12} (1+\xi_{12})]=0,
\label{paircons}
\ee
where the pairwise velocity is defined as
\be
\vu_{12} \equiv \frac{\lexp (1+\de(\vx_1))(1+\de(\vx_2))(\vu(\vx_1)-\vu(\vx_2))
\rexp}{\lexp (1+\de(\vx_1))(1+\de(\vx_2)) \rexp}.
\label{v12}
\ee

In the non-linear regime, $\xi \gg 1$, stable clustering implies that
the pairwise velocity exactly cancels the Hubble flow, $\vu_{12}=-
{\cal H} \vx_{12}$. Under this assumption, Eq.~(\ref{paircons}) can be
readily solved to yield

\be
\xi(x,\tau) \approx 1+\xi(x,\tau) = a^3(\tau) f_2(a\,x),
\label{scxi}
\ee

which means that the probability of having a neighbor at a fixed
physical separation, Eq.~(\ref{dpc12}), becomes independent of
time. Equation~(\ref{paircons}) can be rewritten as,
\be
-\frac{u_{12}(x)}{{\cal H} x} = {1\over 3(1+\xi(x))} \frac{\partial
\xi_{\rm av}(x)}{\partial \ln a},
\label{xibarpc}
\ee
which shows that the pairwise velocity is intimately related to the
behavior of the two-point function. Here we defined the average
two-point function as
\be
\xi_{\rm av}(x)=\frac{3}{x^3} \int_0^x x'^2 \d x' \xi(x')
\ee
and $u_{12}$ is the norm of $\vu_{12}$ that can only be along the 
$\vx_2-\vx_1$ direction.

{}From Eq.~(\ref{xibarpc}) it follows that if the time evolution is
modeled as following linear PT, then the rhs becomes $2f \xi_{\rm
av}/3$.  As $\xi_{\rm av}\ga1$, $\xi_{\rm av}$ grows faster than
linear theory and thus pairwise velocities overcompensate the Hubble
flow; this leads to the well-known ``shoulder'' (a sudden increase of
slope) in the two-point correlation function~\cite{GoRe75}.  These
regimes are illustrated in
Fig.~\ref{figstcl1}\footnote{See~\cite{FuSu01} for a recent study of
the time dependence of the pairwise velocity in the non-linear regime
due to merging.}.  {}From Eq.~(\ref{xibarpc}) it is also clear that a
way to model the evolution of the two-point correlation function is by
modeling the dependence of pairwise velocities on $\xi_{\rm
av}$~\cite{HKLM91,NiPa94,JSD99,FJFDJ99,CJSB01}.  The analysis of high
resolution N-body simulations ~\cite{JSD99} run by the Virgo
Consortium ~\cite{JFPTCWCPEN98} show that the slope of $\xi_2(r)$
indeed exhibits a ``shoulder'' in the form of an inflection point
$d^2\xi_2(r)/dr^2 = 0$ at separation $r_*$ close to the correlation
length $r_0$ where $\xi_2(r_0)=1$.  This property has been recently
corroborated for different initial power-spectrum
shapes~\cite{GaJu01}.  The equality between $r_*$ and $r_0$ is related
to the fact that loop corrections become important close to the
non-linear scale in CDM models at $z=0$, giving rise to a change in
slope.  For models where the spectral index at the non-linear scale is
very negative (such as CDM models at high redshift, $z\sim 3$, see
e.g.~\cite{ZSH01}), loop corrections can be very large (see
Fig.~\ref{alphafn}), and the non-linear scale $r_{0}$ can be much
smaller than that where loop corrections become important (related to
$r_{*}$).

A similar approach can be used to obtain the behavior of higher-order
correlation functions under additional stable clustering
conditions~\cite{Peebles80,Jain97}. The starting point is again the
continuity equation, Eq.~(\ref{continuity}), and for the three-point
case we have

\be
\frac{\partial h_{123}}{\partial \tau} = - \lexp \nabla_1 \cdot (A_{123} 
\vu_1) + \nabla_2 \cdot (A_{123} \vu_2) + \nabla_3 \cdot (A_{123} \vu_3) 
\rexp, 
\ee

where $A_{123} \equiv (1+\de(\vx_1))(1+\de(\vx_2))(1+\de(\vx_3))$ 
and $h_{123} \equiv
\lexp A_{123} \rexp = 1+\xi_{12}+\xi_{23}+\xi_{31}+\xi_{123}$. 
Analogous calculations to the two-point case show that 

\be
\frac{\partial h_{123}}{\partial \tau} + 
\nabla_{12} \cdot (\vw_{12,3}\  h_{123}) + 
\nabla_{23} \cdot (\vw_{23,1}\  h_{123}) =0,
\label{triplet}
\ee

where 

\be
\vw_{12,3} \equiv \frac{ \lexp A_{123}\ (\vu_1-\vu_2) \rexp}{h_{123}},
\ee

and similarly for $\vw_{23,1}$. Note that these three-body weighted
pairwise velocities are actually three-point quantities~\cite{Jain97},
since a third object is involved, so they are different from
Eq.~(\ref{v12}). However, in the same spirit as in the two-point case,
if we assume that stable clustering leads to $\vw_{ij,k}=-{\cal H}
\vx_{ij}$ {\em independently of the position of object $k$}, it
follows that the solution of Eq.~(\ref{triplet}) is

\be
\xi_3(\vx_1,\vx_2,\vx_3) \approx h_{123} = a^6(\tau)\,
f_3(a\,\vx_{1},a\,\vx_{2},a\,\vx_{3}), 
\label{sczeta}
\ee

and thus the probability of having two neighbors at a fixed physical
separation $a \vx_{12}$ and $a \vx_{23}$ from a given object at
$\vx_2$, becomes independent of time [e.g. see
Eqs.~(\ref{dpc12}-\ref{dp123})]. Similar results hold for higher-order
$N$-point correlation functions $\xi_N$~\cite{Peebles80}, and imply
that $\xi_N/\xi_2^{N-1}$ as a function of physical separation become
independent of time in the highly non-linear regime ($1 \ll \xi_2 \ll
\ldots \ll \xi_N$). Note however that the additional stability
conditions such as $\vw_{12,3} \approx -{\cal H} \vx_{12}$ have not
been so far tested against numerical simulations.

\subsubsection{Scale Invariance}

The joint use of stable clustering arguments and the self-similar
solution leads to scale-invariant correlation functions in the
non-linear regime, with precise predictions for the power-law
indices. Equations~(\ref{fb:xiselfsim}) and~(\ref{scxi}) impose
that $f_2(x)$ follows a power law in $x$,
\be
\xi(x)\sim x^{-\gamma}
\label{fb:xiplaw}
\ee
and matching the time dependences it follows that 
\be
\gamma={6\over 3\alpha+2}={3(n+3)\over(n+5)}.
\label{gamma_n}
\ee

Thus, self-similarity plus stable clustering fixes the full time and
spatial dependence of the two-point correlation function in the
non-linear regime in terms of the initial conditions~\cite{DaPe77}. 

A simple generalization of this argument is to assume that in the
non-linear regime $\vu_{12}=-h {\cal H} \vx_{12}$, where $h$ is some
constant, not necessarily unity. In this case, Eq.~(\ref{scxi})
becomes $\xi(x,\tau) = a^{3h}(\tau) f(a^h\,x)$, and this leads to
$\gamma=3h(n+3)/[2+h(n+3)]$~\cite{Padmanabhan96,YaGo97}. Interestingly,
if $h(n+3)$ is a constant independent of spectral index $n$, then the
slope of the two-point correlation function becomes independent of
initial conditions\footnote{A more detailed analysis of the BBGKY
hierarchy shows that, in the absence of self-similarity, power-law
solutions for the two-point function in the non-linear regime exist,
but their relation to the initial spectral index depends on $h$, the
scaling of $\xi_3$ in terms of $\xi_2$ and the skewness of the
velocity distribution. Furthermore, perturbations away from
self-similarity may not be
stable~\cite{RuFr92,YaGo97,YaGo98}.}. Current scale-free simulations
do not see evidence for a spectral index dependence of the asymptotic
value of pairwise velocities and are in reasonable agreement with
stable clustering~\cite{CBH96,Jain97,CoPe98}, although the dynamic
range in the highly non-linear regime is still somewhat limited. For a
different point of view see~\cite{PCOS96}.

The behavior of the higher-order correlation functions can similarly
be constrained. Since stable clustering implies that $Q_N \sim
\xi_N/\xi_2^{N-1}$ is independent of time, adding self-similarity
leads to $Q_N$ being independent of overall scale as well; this leads
to a scaling relation for higher-order correlations that can be
formulated in general as,
\be
\label{SIM}
\xi_N(\lambda \vx_1,...,\lambda \vx_N) = \lambda^{-(N-1)\gamma}
\ \xi_N(\vx_1,...,\vx_N), 
\label{fb:xiNplaw}
\ee 
where $\gamma$ is the index of the two-point function,
Eq.~(\ref{gamma_n}).  As a result, self-similarity plus stable
clustering does not fix completely the behavior of the three-point and
higher-order correlation functions. Although $Q_N$ does not depend on
the overall scale, it does in principle depend on the configuration of
the $N$ points, i.e. it can depend on ratios such as
$x_{12}/x_{23}$. This is the same as in tree-level PT, where $Q_3$
depends on the triangle shape (Figs.~\ref{figbitree}
and~\ref{figQreal}).

We should at this point reconsider the results in this section from
the point of view of the dynamics of gravitational instability. The
equations of motion for the two and three-point correlation functions,
Eqs.~(\ref{paircons}) and~(\ref{triplet}), which express conservation
of pairs and triplets, were obtained from the equation of continuity
alone. These are rigorous results. The validity of self-similarity is
also rigorous for scale-free initial conditions in a $\Omega_m=1$
universe. On the other hand, the conditions of stable clustering are
only a (physically motivated) ansatz, and they replace what might be
obtained by solving the remaining piece of the dynamics, i.e. momentum
conservation, in the highly non-linear regime. Note however that the
conditions of stable clustering can only be part of the story for
higher-order correlation functions since these do not explain why e.g.
$Q_3$ tends to become constant independent of triangle configuration
in the non-linear regime.

\subsubsection{The Non-Linear Evolution of Two-Point Statistics}
\label{nlevtp}

Self-similarity gives a powerful constraint on the space and time
evolution of correlation functions, by requiring that these depend
only of the self-similarity variables. However, different initial
spectra can lead to very different functions of the self-similarity
variables. Hamilton et al.~\cite{HKLM91} suggested a useful way of
thinking about the non-linear evolution of the two-point correlation
function, by which the evolution from different initial spectra can
all be described by the same (approximately) universal formula,
obtained empirically by fitting to numerical simulations.

The starting point is conservation of pairs, Eq.~(\ref{paircons}),
which implies

\be
\frac{\partial [x^3 (1+\xi_{\rm av})]}{\partial \tau} + u_{12} \frac{\partial  
[x^3 (1+\xi_{\rm av})]}{\partial x}=0.
\ee 
Thus, a sphere of radius $x$ such that $x^3 (1+\xi_{\rm av}) \equiv
x_L^3$ is independent of time will contain the same number of
neighbors throughout non-linear evolution.  At early times, when
fluctuations are small, $x_L \approx x$; as clustering develops and
becomes non-linear, $x$ becomes smaller than $x_L$. This motivated the
ansatz that the non-linear average two-point correlation function at
scale $x$ should be a function of the linear one at scale
$x_L$~\cite{HKLM91}

\be
\xi_{\rm av}(x,\tau)= {\cal F}_{map} [\xi_{\rm av\,L}(x_L,\tau)],
\label{hklm}
\ee

where the mapping ${\cal F}_{map}$ was assumed to be universal,
i.e. independent of initial conditions. Using more recent numerical
simulations~\cite{JMW95} showed that there is a dependence of ${\cal
F}_{map}$ on spectral index (particularly as $n<-1$); in
addition~\cite{PeDo94} extended the mapping above to the power
spectrum and arbitrary $\Omega_m$ and $\Omega_\Lambda$. In this case,
the non-linear power spectrum at scale $k$ is assumed to be a function
of the linear power spectrum at scale $k_L$, such that
$k=[1+\Delta(k)]^{1/3} k_L$, where $\Delta(k) \equiv 4\pi k^3 P(k)$,

\be
\Delta(k,\tau)= {\cal F}_{n,\Omega_m,\Omega_\Lambda} [\Delta(k_L,\tau)],
\label{hklm2}
\ee

where it is emphasized that the mapping depends on spectral index and
cosmological parameters. Several groups have reported improved fitting
formulae that take into account these extra
dependences~\cite{JMW95,BaGa96,PeDo96}. In the most often used
version, the fitting function ${\cal F}_{map}$ contains 5 free
functions of the spectral index $n$ which interpolate between ${\cal
F}_{map}(x) \approx x$ in the linear regime and ${\cal F}_{map}
\approx x^{3/2}$ in the non-linear regime where stable clustering is
assumed to hold~\cite{PeDo96}

\be
{\cal F}_{map}(x) = x \Big[ \frac{1+B \beta x + [Ax]^{\alpha
\beta}}{1+ [(Ax)^{\alpha} g^3(\Omega)/(Vx^{1/2})]^\beta}
\Big]^{1/\beta},
\ee

where $A=0.482 (1+n/3)^{-0.947}$, $B=0.226 (1+n/3)^{-1.778}$,
$\alpha=3.310 (1+n/3)^{-0.244}$, $\beta=0.862 (1+n/3)^{-0.287}$,
$V=11.55 (1+n/3)^{-0.423}$, and the linear growth factor has been
written as $D_1=a g(\Omega)$ with $g(\Omega)={5\over 2}
\Omega_m/[\Omega_m^{4/7}-\Omega_\Lambda+(1+\Omega_m/2)(1+ 
\Omega_\Lambda/70)]$~\cite{CPT92}. For models which are not scale free, 
such as CDM models, the spectral index is taken as $n(k_L) \equiv
[\d\ln P/\d\ln k](k=k_L/2)$~\cite{PeDo96}. Extensions of this approach
to models with massive neutrinos are considered in~\cite{Ma98}; for a
description of the non-linear evolution of the bispectrum along these
lines see~\cite{ScCo00}.

The ansatz that the non-linear power spectrum at a given scale is a
function of the linear power at larger scales is a reasonable first
guess, but this cannot be expected to hold in detail. First, as we
described in Section~\ref{sec:1Lpk}, mode-coupling leads to a transfer
of power from large to small scales (in CDM spectra with decreasing
spectral index as a function of scale) and the resulting small-scale
power has a contribution from a range of scales in the linear power
spectrum. In addition, the mapping above is only based on the pairs
conservation equation, and thus only takes into account mass
conservation. The conditions of validity of the HKLM mapping have been
explored in~\cite{NiPa94}, where it is shown that {\em if} the scaled
pairwise velocity $u_{12}/({\cal H} x_{12})$ is {\em only a function}
of the average correlation function, $u_{12}/({\cal H}
x_{12})=H(\xi_{\rm av})$, then conservation of pairs implies

\be
\xi_{\rm av\,L}(x_L)= \exp \Big[ {2\over 3} \int^{\xi_{\rm av}(x)} \frac{\d s}{H(s) 
(1+s)} \Big],
\ee

where $x_L$ and $x$ are related as in the HKLM mapping. In linear PT,
$H=2\xi_{\rm av}/3$, and if stable clustering holds $H=1$. In general
however $H$ cannot be strictly a function of $\xi_{\rm av}$ alone
(e.g. due to mode-coupling in the weakly non-linear regime). A recent 
numerical model for the evolution of the pairwise velocity is given 
in~\cite{CJSB01}, which is used to model the non-linear evolution of 
the average correlation function.

\subsubsection{The Hierarchical Models}
\label{sec:HM}

The absence of solutions of the equations of motion in the non-linear
regime has motivated the search for consistent relations between
correlation functions inspired by observations of galaxy clustering
and the symmetries of dynamics, i.e. the self-similar solution. The
most common example is the so-called {\em hierarchical model} for the
connected $p$-point correlation function~\cite{GrPe77,Fry84a} which
naturally obeys the scaling law (\ref{fb:xiNplaw}),

\ba
\xi_N(\vx_1,\dots,\vx_N)
&=& \sum_{a=1}^{t_N} Q_{N,a} \sum_{\rm labelings}\
 \prod_{\rm edges}^{N-1} \xi_{AB}\ . 
\label{HM} 
\ea 

The product is over $N-1$ edges that link $N$ objects (vertices) $A,
B, ...$, with a two-point correlation function $\xi_{XY}$ assigned to
each edge. These configurations can be associated with `tree' graphs,
called $N$-trees. Topologically distinct $N$-trees, denoted by $a$, in
general have different amplitudes, denoted by $Q_{N,a}$, but those
configurations which differ only by permutations of the labels
1,...,$N$ (and therefore correspond to the same topology) have the
same amplitude.  There are $t_N$ distinct $N$-trees ($t_3=1$, $t_4$=2,
etc., see~\cite{Fry84b,BSS94}) and a total of $N^{N-2}$ labeled trees.

In summary, the {\it hierarchical model} represents the connected
$N$-point functions as sums of products of $(N-1)$ two-point
functions, introducing at each level only as many extra parameters
$Q_{N,a}$ as there are distinct topologies.  In a  {\em degenerate
hierarchical model}, the amplitudes $Q_{N,a}$ are furthermore
independent of scale and configuration.  In this case, $Q_{N,a}=Q_N$,
and the hierarchical amplitudes $S_N \simeq N^{N-2}\ Q_N$.  In the
general case, it can be expected that  the amplitudes $Q_N$ depend on
overall scale and configuration. For example, for Gaussian initial
conditions, in the {\it weakly} non-linear regime, $\sigma^2 \ll 1$,
perturbation theory predicts a clustering pattern that is hierarchical
but not degenerate.

It is important to note that if the degenerate hierarchical holds in
the nonlinear regime, the $Q_N$'s should obey positivity
constraints. By requiring that the fluctuations of the number density
of neighbors should be positive, it follows that~\cite{Peebles80}
\be
Q_3\ge {1\over 3}.
\ee
This constraint was latter generalized through Schwarz inequalities 
in \cite{Fry84a} to get,
\be
(2\,M)^{2M-2}Q_{2M}\,(2\,N)^{2N-2}Q_{2N}\ge\left[(M+N)^{M+N-2}Q_{M+N}\right]^2
\label{fb:QNMineq1}
\ee
where $M$ and $N$ are integers or odd half-integers.
Similar constraints\footnote{A more physically motivated constraint
can be derived by imposing that cluster points be more correlated
than field points~\cite{HaGo88,Hamilton88b}. It leads to 
\begin{equation}
\label{CC>GG}
Q_p \geq \frac{1}{2} \Big(\frac{p-1}{p}\Big)^{p-3}\ Q_{p-1} \geq
\ldots \geq \frac{p!}{2^{p-1} p^{p-2}}  
\end{equation}
which appear more stringent than the constraints above. These
constraints are saturated in the model of Eq.(\ref{Qham}) with
$Q=1/2$.}  have been derived in~\cite{BeSc99},
\begin{equation}  
(N+2)^N Q_{N+2}\,N^{N-2}Q_N\ge\left[(N+1)^{N+1}Q_{N+1}\right]^2.
\label{fb:QNMineq2}
\end{equation}
There is no proof, not even indications, that any model fulfilling
these constraints is mathematically valid. This is a serious
limitation for building such models.

Using the BBGKY hierarchy obtained from the Vlasov equation and
assuming a hierarchical form similar to Eq.~(\ref{HM}) for the {\em
phase-space $N$-point distribution function} in the stable clustering
limit Fry~\cite{Fry82,Fry84a} obtained ($N\geq 3$)

\begin{equation}
\label{Qfry}
Q_N = Q_{N,a} = \frac{1}{2}\ \Big(\frac{N}{N-1}\Big)\
\Big(\frac{4Q_3}{N}\Big)^{N-2}; 
\end{equation}

in this case, different tree diagrams all have the same amplitude,
i.e., the clustering pattern is degenerate. On the other hand,
Hamilton~\cite{Hamilton88}, correcting an unjustified symmetry
assumption in~\cite{Fry82,Fry84a}, instead found

\begin{equation}
\label{Qham}
Q_{N,{\rm snake}} = Q_3^{N-2}, \ \ \ \ \ Q_{N,{\rm star}}=0 
\end{equation}
where ``star'' graphs correspond to those tree graphs in which one
vertex is connected to the other $(N-1)$ vertices, the rest being
``snake'' graphs (if $Q_3=1/2$ this corresponds to the Rayleigh-L\'evy
random walk fractal described in~\cite{Peebles80}). 
Summed over the snake graphs, (\ref{Qham}) yields

\begin{equation}
Q_N= \frac{N!}{2}\ \Big(\frac{Q_3}{N}\Big)^{N-2}.
\end{equation}

Unfortunately, as emphasized in~\cite{Hamilton88}, these results are
{\em not} physically meaningful solutions to the BBGKY hierarchy, but
rather a direct consequence of the assumed factorization in {\em
phase-space}.  As a result, this approach leads to unphysical
predictions such as that cluster-cluster correlations are equal to
galaxy-galaxy correlations to all orders.  It remains to be seen
whether physically relevant solutions to the BBGKY hierarchy which
satisfy Eq.~(\ref{HM}) really do exist.  Despite these shortcomings,
the results in Eq.~(\ref{Qfry}) and Eq.~(\ref{Qham}) are often quoted
in the literature as physically relevant solutions to the BBGKY
hierarchy!

Another phenomenological assumption on the parameters $Q_{N,a}$, which
has the virtue of being closer to the mathematical structure found in
PT, is provided by the {\em tree hierarchical
model}~\cite{BeSc92,MMC00,BeSc99}. In this case the parameters
$Q_{N,a}$ are obtained by the product of weights $\nu_i$ associated to
each of the vertex appearing in the tree structure,
\be
Q_{N,a}=\Pi_i\nu_i^{d_i(a)}.
\label{fb:TreeHM}
\ee   In this expression the product is made over all vertices
appearing in configuration $a$, $\nu_i$ is weight of the vertex
connected to $i$ lines and $d_i(a)$ is the number of such
vertices. The parameter $Q_{N,a}$ is therefore completely specified by
the star diagram amplitudes. This pattern is analogous to what emerges
from PT at large scales, although the parameters $Q_{N,a}$ are here
usually taken to be constant, independent of scale and configuration.
But even in the absence of this latter hypothesis the genuine tree
structure\footnote{In the sense that any part of the diagram can be
computed irrespectively of the global configuration.} of the tree
hierarchical model turned out to be very useful for phenomenological
investigations (see~\cite{BeSc99} and Sect.~\ref{sec:bias}).

\subsubsection{Hyperextended Perturbation Theory}
\label{sec:HEPT}

More direct connections with PT results have been proposed to build
models of non-linear clustering. One is known as the ``hyperextended
perturbation theory'' (HEPT,~\cite{ScFr99})\footnote{A more
phenomenological model, EPT (Extended Perturbation Theory), is
presented in Sect.~\ref{sec:EPT}.}. Its construction is based on the
observation that colinear configurations play a special role in
gravitational clustering, which become apparent in the discussion on
the bispectrum loop corrections (see Sect.~\ref{sec:bi}).  They
correspond to matter flowing parallel to density gradients, thus
enhancing clustering at small scales until eventually giving rise to
bound objects that support themselves by velocity dispersion
(virialization).  {\em HEPT conjectures that the ``effective'' $Q_N$
clustering amplitudes in the strongly non-linear regime are the same
as the weakly non-linear (tree-level PT) colinear amplitudes}, as
shown in Fig.~\ref{fig_Qcdm} to hold well for three-point correlations.

\begin{figure}[t]
\begin{center}
\begin{tabular}{cc}
{\epsfysize=6.5truecm \epsfbox{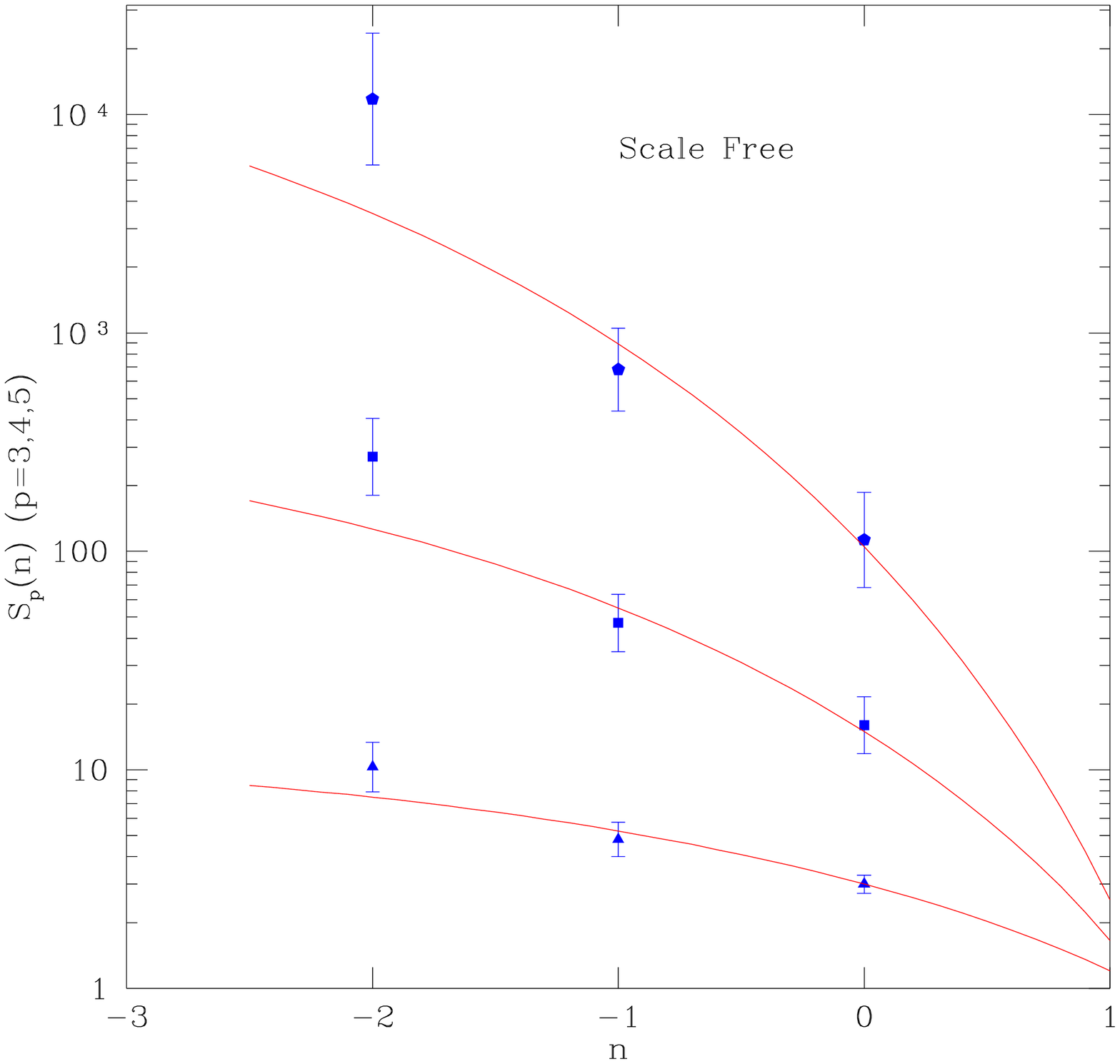}}&
{\epsfysize=6.5truecm \epsfbox{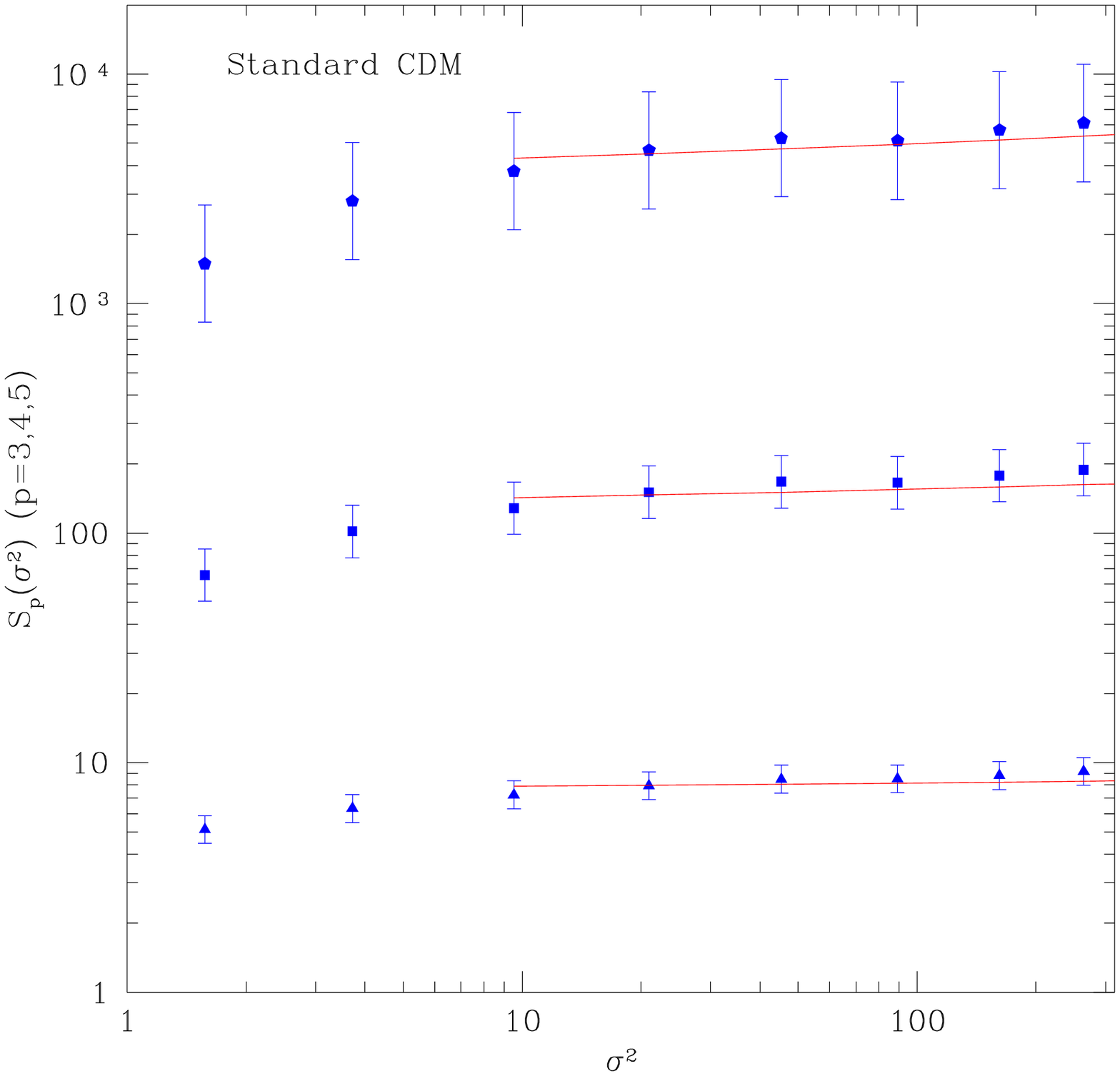}}
\end{tabular}
\end{center}
\caption{HEPT compared
to N-body simulations for scale-free initial conditions (left) and CDM
(right). }
\label{fig_hept} 
\end{figure}

Note that by effective amplitudes $Q_N^{\rm eff}$ the overall
magnitude of $Q_N$ is understood: it is possible that $Q_N$, for
$N>3$, although independent of overall scale, is a function of
configuration.  To calculate the resulting $S_N$ parameters, it is
further assumed that $S_N \simeq N^{N-2}\ Q_N^{\rm eff}$, that is, the
$S_N$ are given by the typical configuration amplitude $Q_N^{\rm eff}$
times the total number of labeled trees, $N^{N-2}$, neglecting a small
correction due to smoothing~\cite{BSS94}. The resulting non-linear
$S_N$ amplitudes follow from tree-level PT~\cite{ScFr99}

\be
\label{s3}
S_3^{\rm sat}(n) = 3\ Q_3^{\rm sat}(n) = 3\ \frac{4-2^n}{1+2^{n+1}},
\ee

\be
\label{s4}
S_4^{\rm sat}(n) = 16\ Q_4^{\rm sat}(n) = 8\ \frac{54-27\ 2^n+2\ 3^n +
6^n}{(1+6\ 2^n + 3\ 3^n + 6\ 6^n)}.
\ee

\be
\label{s5}
S_5^{\rm sat}(n) = 125\ Q_5^{\rm sat}(n) = \frac{125}{6}\ \frac{N(n)}{D(n)}
\ee

\noindent where $n$ is the spectral index, obtained from $(n+3) \equiv
- \d\ln \sigma^2_L(R) / \d\ln R$, $N=1536-1152 2^n+128 3^n+66 4^n+64
6^n-9 8^n -2 12^n -24^n$, $D=1+12 2^n+12 3^n+16 4^n+24 6^n+24 8^n+12
12^n+24 24^n$.  One can check that these $Q_N$ amplitudes satisfy the
above positivity constraints,
Eqs.~(\ref{fb:QNMineq1},\ref{fb:QNMineq2}) and even the constraint in
Eq.~(\ref{CC>GG}) as long as $n \la 0.75$, which is well within the
physically interesting range.

The left panel of Fig.~\ref{fig_hept} shows a comparison of these
predictions with the numerical simulation measurements in~\cite{CBH96}
for scale-free initial conditions with $\Omega_m=1$. The plotted
values correspond to the measured value of $S_p$ when the non-linear
variance $\sigma^2=100$. We see that the N-body results are generally
in good agreement with the predictions of HEPT, Eqs.~(\ref{s3}),
(\ref{s4}) and (\ref{s5}), keeping in mind that for $n=-2$
finite-volume corrections to the $S_p$ measured in the simulations are
quite large and thus uncertain (see Sect.~\ref{sec:cosmicinsim}).
The right panel shows a similar comparison of HEPT with numerical
simulations in the non-linear regime for the SCDM model ($\Gamma
=0.5$, $\sigma_8=0.34$,~\cite{CBS94}). The agreement between the
N-body results and the HEPT predictions is excellent in this case.
The small change in predicted value of $S_p$ with scale is due to the
scale-dependence of the linear CDM spectral index.

Is interesting to note that for $n=0$, HEPT predicts $S_p=(2p-3)!!$,
which agrees exactly with the excursion set model developed
in~\cite{Sheth98b} for white-noise Gaussian initial fluctuations. In
this case, the one-point PDF yields an inverse Gaussian distribution,
which has been shown to agree well in the non-linear regime when
compared to numerical simulations~\cite{Sheth98b}. This remarkable
agreement between HEPT and the excursion set model deserves further
study.

\clearpage 
                                                                      
\section{\bf From Dynamics to Statistics: The Local Cosmic Fields}
\label{chapter5}

We have seen in Section~4 that the non-linear nature of gravitational
dynamics leads, through mode coupling effects, to the emergence of
non-Gaussianity. In the previous section we have explored the behavior
of multi-point correlation functions. Here we present statistical
properties related to the local density contrast in real space.  We
first describe the results that have been obtained for the moments of
the local density field. In particular we show how to compute the full
cumulant generating function of the one-point density contrast at tree
level. Results including loop corrections are given when
known. Finally, we present techniques for the computation of the
density PDF and various applications of these results.  When dealing
with smoothed fields, we shall assume that filtering is done with a
top-hat window unless specified otherwise.

\subsection{The Density Field Third Moment: Skewness}
\subsubsection{The Unsmoothed Case}

The first non-trivial moment that emerges due to mode coupling is the
third moment of the local density probability distribution function,
characterized by the skewness parameter. The computation of the
leading order term of $\mg\delta^3\md$ is obtained through the
expansion $\mg\delta^3\md=\mg (\delta^{(1)}+\delta^{(2)}+\dots
)^3\md.$ When the terms that appear in this formula are organized in
increasing powers of the local linear density, we have
$\mg\delta^3\md=\mg (\delta^{(1)})^3\md+3\,
\mg (\delta^{(1)})^2\,\delta^{(2)}\md+\dots$, 
where the neglected terms are of higher-order in PT. The first term of
this expansion is identically zero for Gaussian initial
conditions. The second term is therefore the leading order,
``tree-level'' in diagrammatic language (see
Section~\ref{sec:TLPT}). We then have\footnote{For simplicity,
calculations in this section are done for the Einstein-de Sitter case,
$\Omega_m=1$.}

\ba \mg\delta^3\md&\approx&
3\,\mg\left(\delta^{(1)}\right)^2\,\delta^{(2)}\md\\ &=&
3\,a^4\,\int{\d^3\vk_1}\dots\int{\d^3\vk_4}\,
\mg\delta_1(\vk_1)\,\delta_1(\vk_2)\,\delta_1(\vk_3)\,\delta_1(\vk_4)\md
\times\nonumber\\ && \,F_2(\vk_2,\vk_3)\,
\exp[\ii(\vk_1+\vk_2+\vk_3+\vk_4)\cdot\vx]. 
\ea 

For Gaussian initial conditions, linear Fourier modes $\de_1(\vk)$ can
only correlate in pairs [Eq.~(\ref{fb:Wick2})].  If $\vk_2$ and
$\vk_3$ are paired, the integral vanishes [because $\mg \de \md=0$,
see the structure of the kernel $F_2$ in Eq.~(\ref{F2})]. The other
two pairings give identical contributions, and thus

\be \mg\delta^3\md=
6\,a^4\,\int{\d^3\vk_1}\,\int{\d^3\vk_2}\,P(k_1)\,P(k_2)\,F_2(\vk_1,\vk_2).
\ee 

Integrating over the angle between $\vk_1$ and $\vk_2$ leads to
$\mg\delta^3\md={(34/7)}\mg\delta^2\md^2$~\cite{Peebles80}.  For the
reasons discussed in Sect.~\ref{emng}, it is convenient to rescale the
third moment and define the skewness parameter $S_3$ (see Sect.~2),

\be S_3\equiv{\mg\delta^3\md\over \mg\delta^2\md^2}={34\over 7}+{\cal
O}(\sigma^2).  \label{s3u}
\ee

The skewness measures the tendency of gravitational clustering to
create an asymmetry between underdense and overdense regions (see
Fig.~\ref{SmoothEff}). Indeed, as clustering proceeds there is an
increased probability of having large values of $\de$ (compared to a
Gaussian distribution), leading to an enhancement of the high-density
tail of the PDF. In addition, as underdense regions expand and most of
the volume becomes underdense, the maximum of the PDF shifts to
negative values of $\de$. {}From Eq.~(\ref{fb:edge}) we see that the
maximum of the PDF is in fact reached at

\be
\label{s3demax}
\de_{{\rm max}} \approx -\frac{S_{3}}{2}\ \sigma^{2},
\ee
to first order in $\sigma$. We thus see that the skewness factor
$S_{3}$ contains very useful information on the shape of the PDF.

\subsubsection{The Smoothed Case}

At this stage however the calculation in Eq.~(\ref{s3u}) is somewhat
academic because it applies to the statistical properties of the
local, unfiltered, density field. In practice the fields are always
observed at a finite spatial resolution (whether it is in an
observational context or in numerical simulations). The effect of
filtering, which amounts to convolving the density field with some
window function, should be taken into account in the computation of
$S_3$. The main difficulty lies in the complexity this brings into the
computation of the angular integral. To obtain the skewness of the
local filtered density, $\delta_R$, one indeed needs to calculate,
\begin{equation}
\mg\delta_R^3\md=
3\,\mg\left(\delta_R^{(1)}\right)^2\,\delta_R^{(2)}\md 
\end{equation} 
with 
\begin{eqnarray}
\delta_R^{(1)}&=&a\,\disp{\int{\d^3\vk}\,\delta(\vk)\,
\exp[\ii\vk\cdot\vx]\,W_3(k_1\,R)},\\
\delta_R^{(2)}&=&a^2\,\disp{\int{\d^3\vk_1}\,\int{\d^3\vk_2}\,
\delta(\vk_1)\,\delta(\vk_2)\,
\exp[\ii(\vk_1+\vk_2)\cdot\vx]}\times\nonumber\\ 
&&\disp{F_2\left(\vk_1,\vk_2\right)\,W_3(\vert\vk_1+\vk_2\vert\,R)},
\label{fb:delta2R}
\end{eqnarray}
where $W_3(k)$ is the 3D filtering function in Fourier space.  It
leads to the expression for the third moment,    
\ba
\mg\delta_R^3\md&=&
6\,a^4\,\int{\d^3\vk_1}\,\int{\d^3\vk_2}\,P(k_1)\,P(k_2)\,
W_3(k_1\,R)\,W_3(k_2\,R) \times\nonumber\\
&&F_2(\vk_1,\vk_2)\,W_3(\vert\vk_1+\vk_2\vert\,R),     
\ea   

so that the relative angle between $\vk_1$ and $\vk_2$ appears in both
$F_2$ and $W_3$. The result depends obviously on the filtering
procedure. It turns out that the final result is simple for a top-hat
filter in real space. In this case,

\be W_3(k)=\sqrt{3\pi\over 2}{J_{3/2}(k)\over k^{3/2}}= {3\over
k^3}\left[\sin(k)-k\,\cos(k)\right].   \ee

Following the investigations initiated in \cite{JBC93} for the
properties of the top-hat window function\footnote{These properties
have been obtained from the summation theorem of Bessel functions, see
e.g.~\protect\cite{Watson44}. Such relations hold in any space
dimension for top-hat filters.} it can be shown
(see~\cite{Bernardeau94c} and Appendix~\ref{tophatgeom}),

\ba 
\int{\d\Omega_{12}\over
4\pi}\,W_3(\vert\vk_1+\vk_2\vert)\left[1- {(\vk_1\cdot\vk_2)^2\over
k_1^2\,k_2^2}\right]&=& {2\over 3}\,W_3(k_1)\,W_3(k_2)\\
\int{\d\Omega_{12}\over 4\pi}\,W_3(\vert\vk_1+\vk_2\vert)\left[1+
{\vk_1\cdot\vk_2\over k_1^2}\right]&=& \nonumber \\
  & & \hskip -2cm W_3(k_1)\left[W_3(k_2)+{1\over
3} k_2\,W_3'(k_2)\right].  \ea 

It is easy to see that $F_2$ can be expressed with the help of the two
polynomials involved in the preceding relations. One finally
obtains~\cite{Bernardeau94c}, 

\be 
S_3={34\over7}+{\d\log\sigma^2(R)\over\d\log R}.
\label{s3ex}  
\ee

The skewness thus depends on the power spectrum shape (mainly at the
filtering scale). For a power-law spectrum, $P(k)\propto k^n$, it
follows that $S_3={34/7}-(n+3)$~\cite{JBC93}. Galaxy surveys indicate
that the spectral index $n$ is of the order of $n\approx -1.5$ close
to the non-linear scale.  Comparisons with numerical simulations have
shown that the prediction of Eq.~(\ref{s3ex}) is very accurate, as can
be seen in Fig.~\ref{fb:figSpGazta}.

\subsubsection{Physical Interpretation of Smoothing}

\begin{figure}[t!]
\centering
\centerline{\epsfysize=14truecm\epsfbox{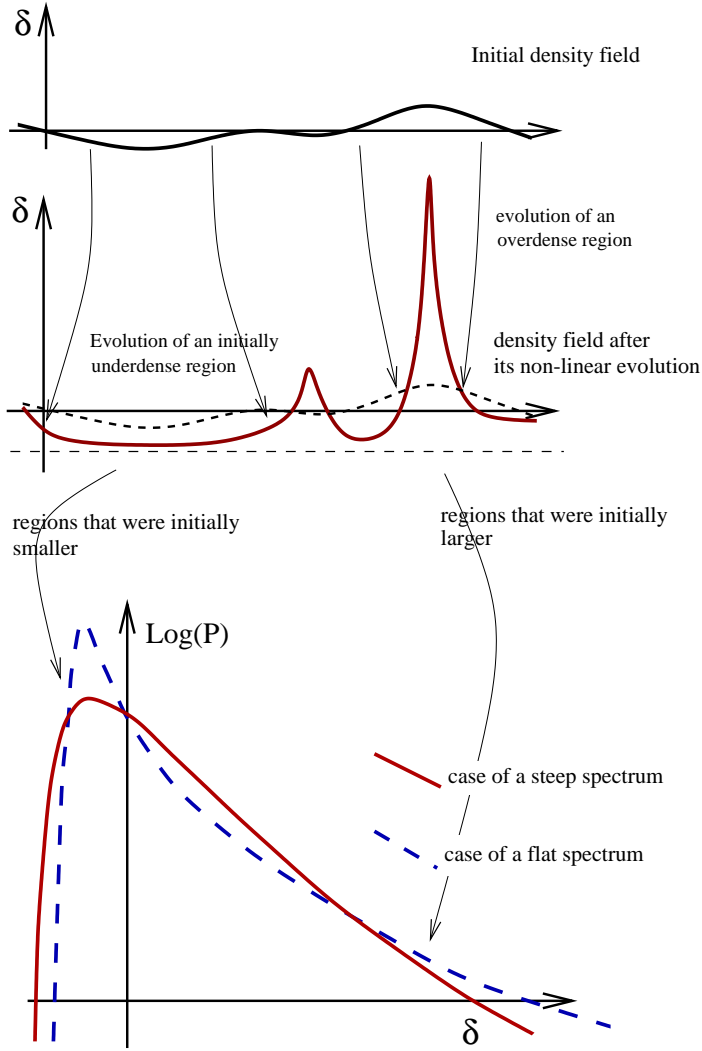}}
\caption{Skewness is a measure of the asymmetry of the local density
distribution function. It appears because underdense regions evolve
less rapidly than  overdense regions as soon as nonlinearities start
to play a role. The dependence of skewness with the shape of the power
spectrum comes from a mapping between Lagrangian space, in which the
initial size of the perturbation is determined, and Eulerian
space. For a given filtering scale $R$, overdense regions come from
the collapse of regions that had initially a larger size, whereas underdense
regions come from initially smaller regions. As a result, the skewness
is expected to be smaller  for power spectra with more small scale
fluctuations (steep spectra case, that is when $k^3\,P(k)$ is rapidly
increasing with $k$).}
\label{SmoothEff}
\end{figure}

To understand the dependence of the skewness parameter with power
spectrum shape it is very instructive to examine in detail the nature
of the contributions that appear when the filtering effects are taken
into account.

For this purpose let us consider the same problem in Lagrangian space.
If one calculates $J^{(2)}$, the second-order expansion of the
Jacobian, one obtains [from Eqs.~(\ref{leom},\ref{Psi2}) and assuming
$\Omega_m=1$],
\ba
&J^{(2)}=a^2\,\disp{{2\over7}\int{\d^3\vk_1}\,\int{\d^3\vk_2}\,
\delta(\vk_1)\,\delta(\vk_2)\,
\exp[\ii(\vk_1+\vk_2)\cdot\vq]} 
\disp{\left[1-{(\vk_1\cdot\vk_2)^2\over k_1^2 k_2^2}\right]}.\nonumber\\  
\ea  

This gives for the density [e.g. Eq.~(\ref{dlag})], once the Jacobian
(which is a direct estimation of the volume) has been filtered at a
given {\it Lagrangian} scale $R$,

\ba
&\delta_R^{(2)}=\disp{\int{\d^3\vk_1}\,\int{\d^3\vk_1}\,
a^2\,\delta(\vk_1)\,\delta(\vk_2)\,
\exp[\ii(\vk_1+\vk_2)\cdot\vq]}\times\nonumber\\ &\disp{
\left[W(k_1\,R)\,W(k_2\,R)-{2\over
7}W(\vert\vk_1+\vk_2\vert\,R)\,\left(1- {(\vk_1\cdot\vk_2)^2\over k_1^2
k_2^2}\right)\right]}.   
\ea  

Because smoothing effects are calculated in Lagrangian space (denoted
by $\vq$), this expression is different from the Eulerian space
filtering result, Eq.~(\ref{fb:delta2R}). In fact, it follows that
$S_3^{\rm Lag}={34/ 7}$ even when filtering effects are taken into
account.  The mere fact that one does not obtain the same result
should not be surprising. In this latter case the filtering has been
made at a given {\it mass} scale. The difference between the two
calculations comes from the fact that the larger the mass of a region
initially is, the smaller the volume it occupies will be.  Filtering
at a fixed Eulerian scale therefore mixes different initial mass
scales. The asymmetry will then be less than one could have expected
because, for a standard hierarchical spectrum, larger mass scales
correspond to smaller fluctuations.

\subsubsection{Dependence of the Skewness on Cosmological Parameters}

\begin{figure}[t!]
\centering
\centerline{\epsfysize=6truecm\epsfbox{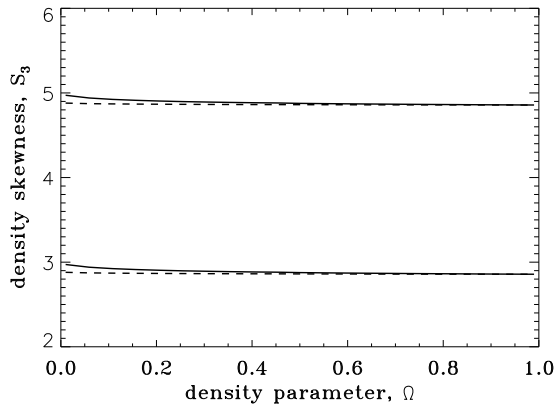}}
\caption{The skewness $S_3$ as a function of $\Omega_m$ for
zero-$\Omega_{\Lambda}$ Universes (solid lines) and flat universes
with $\Omega_m+\Omega_{\Lambda}=1$ (dashed lines). The upper and lower
curves correspond to a power law spectrum with $n=-3$ and $n=-1$,
respectively.}
\label{fb:S3Omega}
\end{figure}

As the skewness is induced by gravitational dynamics, it is important
to know how much it can depend on cosmological parameters.  In general
the parameter $S_3$ depends on the growth rate of the second-order
PT solution, see Sect.~\ref{sec:codeno}, through
\be
S_3=3\,\nu_2+{\d\log\sigma^2(R)\over \d\log R}.
\ee
Explicit calculations~\cite{BJCP92} have shown that $\nu_2$ can be
well approximated by
\be \nu_2\approx{4\over 3}+{2\over7}\Omega_m^{-2/63},
\ee
obtained by expansion about $\Omega_m=1$ for
$\Omega_{\Lambda}=0$\footnote{But it is valid for all values of
$\Omega_m$ of cosmological interest.}.  We then have the following result,
\be
S_3={34\over 7}+{6\over 7}\left(\Omega_m^{-0.03}-1\right)-(n+3).
\ee
A similar result follows when $\Omega_\Lambda\neq 0$,
see~\cite{Bernardeau94c,HBCJ95} and also ~\cite{FoGa98b}.  
In practice, for current
applications to data, such a small dependence on cosmological
parameters can simply be ignored, as illustrated in
Fig.~\ref{fb:S3Omega}.  This turns out to be true even when
cosmologies with non-standard vacuum equation of state are considered
(e.g. quintessence models) ~\cite{KaBu99,GaLo01,BeBe01}.

\subsubsection{The Skewness of the Local Velocity Divergence}

The skewness of the velocity divergence can obviously be calculated in
a similar fashion. However, because of the overall
$f(\Omega_m,\Omega_{\Lambda})$ factor for the linear growth of
velocities, it is natural to expect that the velocity divergence
skewness parameter, $T_3$, has a significant $\Omega_m$
dependence~\cite{BJDB95}. In general,
\be
T_3\equiv{\langle\theta^3\rangle\over\langle\theta^2\rangle^2}=
-{1\over f(\Omega_m,\Omega_{\Lambda})}\left[3\mu_2+
{\d\log\sigma^2(R)\over \d\log R}\right].
\ee
Taking into account the specific time dependence of $\mu_2$ we get,
\be
T_3=
-{1\over f(\Omega_m,\Omega_{\Lambda})}\left[2+{12\over 7}\Omega_m^{-1/21}
+{\d\log\sigma^2(R)\over \d\log R}\right],
\label{eq:t3accu}
\ee
which within a very good accuracy implies that $T_3\approx
-[26/7-(n+3)]/\Omega_m^{0.6}$ for a power-law spectrum.  This makes
the dimensionless quantity $T_3$ a very good candidate for a
determination of $\Omega_m$ independent of galaxy biasing.  Attempts
to carry out such measurements, however, faced very large systematics
in the data~\cite{BJDB95}. So far no reliable constraints have been
drawn from this technique.

\subsection{The Fourth-Order Density Cumulant: Kurtosis}

The previous results can be applied to any low-order cumulants of the
cosmic field.  Fry~\cite{Fry84b} computed the fourth cumulant of the
cosmic density field, but without taking into account the filtering
effects. These were included later for top-hat~\cite{Bernardeau94c}
and Gaussian filters~\cite{LJWB95}.

Formally the fourth-order cumulant of the local density is given by,
\ba 
\mg\delta^4\md_c&\equiv&\mg\delta^4\md-3\,\mg\delta^2\md^2\\
&=&12\,\mg(\delta^{(1)})^2\,(\delta^{(2)})^2\md_c+
4\,\mg(\delta^{(1)})^3\,\delta^{(3)}\md_c.\nonumber 
\ea 
In these equations it is essential to take the connected part
only. There are terms that involve loop corrections to the variance
that are of the same order in $\sigma$ but they naturally cancel when
the non-connected part of the fourth moment is subtracted out. The
consequence is that,
\be
\mg\delta^4\md_c\sim\mg\delta^2\md^3,
\ee
and one can define the kurtosis parameter $S_4$,
\be
S_4\equiv {\mg\delta^4\md_c/\mg\delta^2\md^3}.
\ee
This equation allows one to compute the leading part of $S_4$ in the
weakly non-linear regime. In general $S_4$ can be expressed in terms
of the functions $D_1$, $\nu_2$ and $\nu_3$. This can be obtained by
successive applications of the geometrical properties of the top-hat
window function (see
\cite{Bernardeau94c} and appendix~\ref{tophatgeom} for details).  We
have,

\ba S_4&=&4\nu_3+12\nu_2^2+
\left(14\nu_2-2\right){\d\log[\sigma^2(R_0)]\over \d\log R_0}+
\nonumber\\ &&+{7\over3}\left({\d\log[\sigma^2(R_0)]\over \d\log
R_0}\right)^2 +{2\over3}{\d^2\log[\sigma^2(R_0)]\over \d\log^2 R_0}.
\ea 

For a power law spectrum of index $n$ this leads to 

\ba
S_4&=&{60712\over 1323}-{62\over3}(n+3)+ {7\over3}(n+3)^2.  
\ea 

This result is exact for an Einstein-de Sitter universe. It is
extremely accurate, within a few per cent for all models of
cosmological interest. Similar results can be obtained for the
velocity divergence.

\subsection{Results for Gaussian Smoothing Filters}

So far we have been giving results for a top-hat filter only. The
reason is that they can be given in a closed form for any shape of the
power spectrum.  Another quite natural filter to choose is the
Gaussian filter.  In this case however there are no simple closed forms
that are valid for any power spectrum shape. Results are known for
power-law spectra only~\cite{JBC93,Matsubara94b,LJWB95}. 

The principle of the calculation in this case is to decompose the
angular part that enters in the window function as a sum of Legendre
functions,
\be
e^{-\vp\cdot\vq}=e^{-pq\cos\varphi}=
\sum_{m=0}^{\infty}(-1)^m(2 m+1)\sqrt{\pi\over2pq}I_{m+{1\over2}}(pq)P_m(\cos\varphi),
\ee
where $I_{m+{1\over2}}(pq)$ are Bessel functions.  The integration
over $\varphi$ is made simple by the orthogonality relation between
the Legendre polynomials. Finally each term appearing in the
decomposition of the Bessel function,
\be 
I_{\nu}(z)=\sum_{m=0}^{\infty}{1\over m!\Gamma(\nu+m+1)}\left(z\over2\right)^{\nu+2m},
\ee
can be integrated out for power-law spectra since,
\be
\int_0^{\infty}q^{\alpha}e^{-q^2}\d q={1\over 2}\Gamma\left({\alpha+1\over 2}\right),
\ee
which after resummation leads to hypergeometric functions of the kind
$\,_2F_1$. Eventually the result for $S_3$ is

\ba
S_3 &=&3\,_2F_1\left({n+3\over2},{n+3\over 2},{3\over 2},{1\over
4}\right) - \left(n+{8\over 7}\right)\,_2F_1\left({n+3\over2},{n+3\over 2},{5\over 2},{1\over 4}\right). \nonumber \\
\ea
and similarly the velocity skewness is
\ba
T_3 &=& -3\,_2F_1\left({n+3\over2},{n+3\over 2},{3\over 2},{1\over
4}\right) +
\left(n+{16\over 7}\right)\,_2F_1\left({n+3\over2},{n+3\over 2},{5\over 2},{1\over 4}
\right).\nonumber \\
\ea
This result is exact for an Einstein-de Sitter Universe but obviously,
as for the top-hat filter, $S_3$ is expected to depend only weakly on
cosmological parameters and the dominant dependence of $T_3$ is that
proportional to $1/f(\Omega_m)$.  The result for $S_3$ is shown as a
dashed line in Fig.~\ref{fb:AllSp}.

The kurtosis cannot be calculated in closed form even for power-law
spectra (although a semi-analytic formula can be
given~\cite{LJWB95}). However there exists a simple prescription that
allows one to get an approximate expression for the kurtosis.  It
consists in using the formal expression of the kurtosis obtained for a
top-hat filter but calculated for $n=n_{\rm eff}$ such that it gives
the correct value for the skewness. Such a prescription has been found
to give accurate results, about 1\% accuracy for $n=-1$~\cite{LJWB95}.

\subsection{The Density Cumulants Hierarchy}
\label{fb:CorHierarchies}

In general the nonlinear couplings are going to induce non-zero cumulants at
any order. We can define~\cite{GGRW86}
\begin{equation}
S_p\equiv{\mg\delta^p\md_c/\mg\delta^2\md^{p-1}},
\label{fb:Sp}
\end{equation}
that generalizes the $S_3$ and $S_4$ parameters considered in the
previous section.  All these quantities are finite (and non-zero) at
large scales for Gaussian initial conditions and can in principle be
computed from PT expansions.  However, the direct calculation of $S_p$
becomes extremely difficult with increasing order $p$ due to the
complexity of the kernels $F_p$ and $G_p$. Fortunately, it turns out
to be possible to take great advantage of the close relationship
between the $S_p$ parameters and the vertices $\nu_p$ describing the
spherical collapse dynamics, as described in Sect. \ref{subsec:eds},
to compute the $S_p$ parameters for any $p$.

In the derivation presented here we adopt a pedestrian approach for
building, step by step, the functional shape of the cumulant
generating function.  A more direct approach has recently been
developed in \cite{Valageas01a,Valageas01b} in which the generating
function of the cumulant is obtained directly, via a saddle-point
approximation in the computation of the cumulant generating function
which corresponds to its tree-order calculation.  This approach avoids
technical difficulties encountered in the computation of the
Lagrangian space filtering properties and in the Lagrangian-Eulerian
mapping and is certainly an interesting complementary view to what we
present here.

\subsubsection{The Unsmoothed Density Cumulant Generating Function}
\label{sec:snnun}

The computation of $S_p$ coefficients is based on the property that
each of them can be decomposed into a sum of product of ``vertices'',
at least when filtering effects are not taken into account.  As seen
before, $S_4=12\,\nu_2^2+4\,\nu_3$.  This property extends to all
orders, so that the $S_p$ parameters can be expressed as functions of
$\nu_q$'s only ($q=2,\ldots,p-1$).  Note that the vertices $\nu_p$
defined in Eq.~(\ref{nu_n}) as angular averages of PT kernels
correspond to
\begin{equation}
\nu_p=\mg\delta^{(p)}[\delta^{(1)}]^p\md_c/ \mg[\delta^{(1)}]^2\md^p.
\label{eq:defver}
\end{equation}

\begin{figure}
\centerline{\epsfysize=1.8truecm\epsfbox{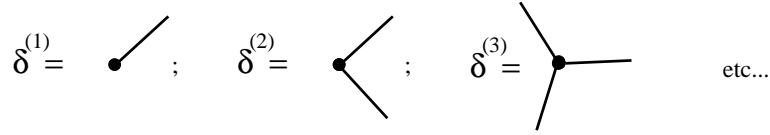}}
\caption{Diagrammatic representation of $\delta^{(p)}$. Each line stands
for a factor $\delta(\vk)$.}
\label{deltap}
\end{figure}

This decomposition of $S_p$ into a sum of product of vertices can be
observed easily in a graphical representation. Indeed
\begin{equation}
\mg \delta^p\md_c=\sum_{q_i}
\mg \delta^{(q_1)}\dots\delta^{(q_p)}\md_c,
\end{equation}
where each $\delta$ has been expanded in PT.  Each $\delta^{(q)}$
contains a product of $q$ random Gaussian variables $\delta(\vk)$.
Each of these points can be represented by one dot, so that when the 
ensemble average is computed, because of the Wick theorem, dots are
connected pairwise.  The $\delta^{(q)}$ therefore can be represented
as in Fig.~\ref{deltap} with $q$ outgoing lines.

\begin{figure}
\centerline{\epsfysize=2.2truecm\epsfbox{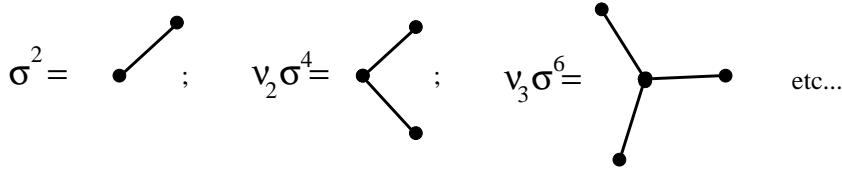}}
\caption{Computation of the simplest  graphs. Each line represents a factor
$\sigma^2$.  Vertices are  obtained from  the  angular average  of the  wave
vectors leaving $\nu_p$.}
\label{nup}
\end{figure}

\begin{figure}
\centerline{\epsfysize=2.2truecm\epsfbox{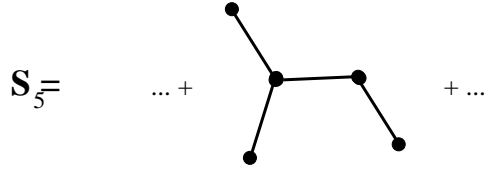}}
\caption{A graph contributing to $S_5$.}
\label{exS5}
\end{figure}

Diagrams  that contribute to the  leading order  of $S_p$  are  those which
contain enough dots so that  a connected diagram that minimizes the number
of links can be built.  The number of links for connecting $p$ points is 
$p-1$, we should then have $\sum_i q_i=2(p-1)$ so that
\begin{equation}
S_p=\disp{\sum_{{\rm graphs},\ \sum_i q_i=2(p-1)}\mg
\delta^{(q_1)}\dots\delta^{(q_p)}\md_c/
\mg\left[\delta^{(1)}\right]^2\md^{p-1}}.
\end{equation}
An example of such a graph for $S_5$ is shown in Fig. \ref{exS5}.

It is worth noting that all these diagrams are trees, so that the
integration over the wave vectors can be made step by
step\footnote{This is possible however only when smoothing effects are
neglected.}. Then the value of each diagram is obtained by assigning
each line to the value of $\sigma^2$ and each vertex to $\nu_p$
depending on the number $p$ of lines it is connected to, see
e.g. Fig.~\ref{nup}.

This order by order decomposition can actually be replaced by a
functional relation at the level of the generating functions.  If we
define the generating function of $S_p$ as
\begin{equation}
\varphi(y)=\sum_{p=1}^{\infty}-S_p\,{(-y)^p\over p!},\ \ \
(S_1=S_2\equiv 1),
\label{fb:phiy}
\end{equation}
and the vertex generating function as
\begin{equation}
\mGd(\tau)=\sum_{p=1}^{\infty}{\nu_p}\,{(-\tau)^p\over p!},
\label{fb:mGd}
\end{equation}
it is possible to show that $\varphi$ and $\mGd$ are related to each other
through the system of equations 
\begin{eqnarray}
\varphi(y)&=&y\,\mGd[\tau(y)]+{1\over2}\tau^2(y),\label{fb:phieq}\\
\tau(y)&=&-y\,\mGd'[\tau(y)]. \label{taueq}
\end{eqnarray}

\begin{figure}
\centerline{\epsfysize=1.8truecm\epsfbox{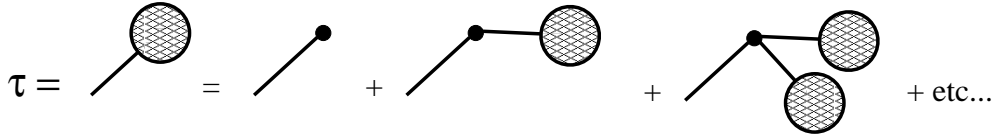}}
\caption{Graphical representation of Eq.~(\protect\ref{taueq}), $\tau$
is the generating function of graphs with one external line.}
\label{taugraph}
\end{figure}

The demonstration of these equations is not straightforward and is
given in Appendix~\ref{LegTrans}.  To get some insight about these two
equations, one can note that $\tau$ is the conjugate variable to the
one-line vertex (that is $\nu_1$, set to unity at the end of the
calculation).  As such, it corresponds to the generating function of
all graphs with {\em one} external line.  It is then solution of an
implicit equation, illustrated in Fig.~\ref{taugraph}, which
corresponds to Eq.~(\ref{taueq}).  Naturally, it involves the vertex
generating function.  It is to be noted however that in this
perspective the equations (\ref{fb:phieq},\ref{taueq}) and the
parameter $y$ have no intrinsic physical interpretation.  It has been
pointed out recently in~\cite{Valageas01a,Valageas01b} that this
system can actually be obtained directly from a saddle-point
approximation in the computation of the local density contrast PDF. It
gives insights into the physical meaning of the solutions of
Eq.~(\ref{taueq}).  We will come back to this point in
Sect.~\ref{fb:TheDensityPDF}.

Recall that  vertices describe the spherical collapse dynamics (see
Sect.~\ref{subsec:eds}), thus $\mGd(\tau)$ corresponds to the density
contrast of collapsing structures with spherical symmetry when
$(-\tau)$ is its linear density contrast. The first few values of
$\nu_p$ can then be easily computed,
\begin{equation}
\nu_2={34\over 21},\ \  \nu_3={682\over189},\ \  \nu_4={446440\over
43659},
\end{equation}
which implies,
\begin{eqnarray}
S_3&=&3\nu_2={34\over7};\\
S_4&=&4\nu_3+12\nu_2^2={60\,712\over1\,323}\approx 45.89;\\
S_5&=&5\nu_4+60\nu_3\nu_2+60\nu_2^3={200\,575\,880\over
305\,613}\approx 656.3;\\
S_6&=&6\nu_5+120\nu_4\nu_2+90\nu_3^2+720\nu_3\nu_2^2+360\nu_2^4\approx
12,700\\ &&\dots\nonumber
\end{eqnarray}
At this stage however, the effects of filtering have not been taken
into account.

\subsubsection{Geometrical Properties of Smoothing in Lagrangian Space}

As the examination of the particular case of $S_3$ has shown, the
smoothing effects for a top-hat filter are entirely due to the mapping
between Lagrangian space and Eulerian space.  This can be generalized
to any order~\cite{Bernardeau94a}.

The Lagrangian space dynamics is jointly described by the displacement
field (that plays a role similar to the velocity field) and the
Jacobian, whose inverse gives the density.  The latter can be expanded
with respect to the initial density contrast,
\begin{equation}
J(\vq)=1+J^{(1)}(\vq)+J^{(2)}(\vq)+\dots
\end{equation}
At a given order we will have\footnote{We assume $\Omega_m=1$, but the
calculations trivially extend to all cosmologies.}, \ba J^{(p)}(\vq)& =
&a^p \int{\d^3\vk_1\over (2\pi)^{3/2}}\dots {\d^3\vk_p\over
(2\pi)^{3/2}}\,J_p(\vk_1,\dots,\vk_p)
\exp[\ii\vq\cdot(\vk_1+\dots+\vk_p)].\nonumber \\ \ea The Jacobian is
actually given by the determinant of the deformation tensor, obtained
from the first derivative of the displacement field, $\vPsi$, see
Eq.~(\ref{dlag}).  The precise relation is
\begin{eqnarray}
J(\vq)&\equiv&\left\vert{\partial\vx\over\partial \vq}\right\vert
=1+\gradq\cdot\vPsi+{1\over 2}\left[\left(\gradq\cdot\vPsi\right)^2-
\sum_{ij}\Psi_{i,j}\Psi_{j,i}\right]\nonumber\\ &&+{1\over 6}
\left[\left(\gradq\cdot\vPsi\right)^3-3\gradq\cdot\vPsi\sum_{ij}\Psi_{i,j}
\Psi_{j,i}+2 \sum_{ijk}\Psi_{i,j}\Psi_{j,k}\Psi_{k,i}\right].
\end{eqnarray}

The equations of motion are closed by the Euler equation,
Eq.~(\ref{leom}).  This shows that the kernels of the Jacobian
expansion are built recursively from the function $\beta(\vk_1,\vk_2)=
1-(\vk_1\cdot\vk_2)^2/(k_1 k_2)^2$ and
\begin{eqnarray}
\eta(\vk_1,\vk_2,\vk_3)&=    1-\left({\vk_1\cdot\vk_2\over   k_1
k_2}\right)^2 -&\left({\vk_2\cdot\vk_3\over  k_2  k_3}\right)^2
-\left({\vk_3\cdot\vk_1\over  k_3 k_1}\right)^2\nonumber\\
&&+2\,{\vk_1\cdot\vk_2\  \vk_2\cdot\vk_3\ \vk_3\cdot\vk_1\over k_1^2 k_2^2 k_3^2}.
\end{eqnarray}
We  have  seen  previously  that  a top-hat  filter    commutes  with
$\beta$. It can also be shown that,
\begin{eqnarray}
&\int{\d \Omega_1\over 4\pi}{\d\Omega_2\over 4\pi}{\d\Omega_3\over
4\pi}\ \WTHt\eta(\vk_1,\vk_2,\vk_3)= \nonumber\\ &={2\over
9}\,W(k_1\,R)\ W(k_2\,R)\,W(k_3\,R).
\end{eqnarray}
Here again, an exact ``commutation property'' is observed.  Successive
applications of these geometrical properties\footnote{This
demonstration is incomplete here because the displacement in
Lagrangian space is not in general potential (see~\cite{Bernardeau94b}
for a more complete demonstration).} then imply
that~\cite{Bernardeau94b},
\begin{eqnarray}
j_p&\equiv&\overline{J_p(\vk_1,\dots,\vk_p)\,
W(\vert\vk_1+\dots+\vk_p\vert R)}\\
\label{fb:jpdef}
&=& \overline{J_p(\vk_1,\dots,\vk_p)\,}\, W(k_1\,R)\dots W(k_p\,R),
\label{jpcom}
\end{eqnarray}
where a bar denotes angular averaged quantities.  This is a
generalization of the results obtained for parameter $S_3$, which has
been found to be insensitive to filtering effects in Lagrangian space
(for a top-hat filter only).

\subsubsection{Lagrangian to Eulerian Space Mapping: Smoothed Case}
\label{sec:lagtoeul}

As for the skewness $S_3$, a mapping between Lagrangian and Eulerian
space should permit one to calculate the $S_p$'s at any order $p$.

The hierarchy in Eq.~(\ref{fb:jpdef}) gives implicitly the cumulant
generating function of the {\em volume} distribution function for a
{\em fixed mass scale}.  One can then make the following remark: the
probability that a mass $M$ occupies a volume larger than $V$ is also
the probability that a volume $V$ contains a mass lower than $M$.  It
suffices for that to consider concentric spheres around a given point
$\vx_0$\footnote{This statement is however rigorous for centered
probabilities only.}. It is therefore possible to relate the real
space density PDF to the Lagrangian space one. At this stage however
we are only interested in the leading order behavior of the
cumulants. We can then notice that, in the small variance limit, the
one-point density PDF formally given by Eq.~(\ref{fb:PDFvarphi}), can
be calculated by the steepest descent method. The saddle point
position is given by the equation, ${\d\varphi(y)/ \d y}=\delta$, and
in addition we have ${\d\varphi(y)/ \d y}=\mGd(\tau)$, when $\tau$ is
given implicitly by Eq.~(\ref{taueq}). The saddle-point position is
therefore obtained by a simple change of variable from the linear
density $\tau$ to the nonlinear density contrast $\delta$.  It implies
that the one-point PDF is roughly given by
\begin{equation}
p(\delta)\d\delta\sim\exp\left(-{\tau^2\over 2\,\sigma^2}\right)\d\delta
\end{equation}
with a weakly $\delta$-dependent prefactor. It is important to note
that the {\em leading order} cumulants of this PDF do {\em not} depend
on these prefactors. They are entirely encoded in the $\tau$-$\delta$
relation.

As suggested in the previous paragraph, if we now identify
$p_E(\delta>\delta_0)$ and $p_L(\delta<\delta_0)$ (one being computed
at a fixed real space radius, the other at a fixed mass scale) we
obtain a consistency relation 
\begin{equation}
-{\tau_E^2\over 2\sigma^2(R)}= -{\tau_L^2\over
2\sigma^2\left[(1+\delta)^{1/3}\,R\right]},
\end{equation}
so that the two have the same leading-order cumulants.  Here and in
the following we use indices $L$ or $E$ for variables that live
respectively in Lagrangian space or Eulerian space. More precisely we
denote by $\varphi^L$ the cumulant generating function in Lagrangian
space and $\mGd^L$ the corresponding vertex generating function.  In
Eulerian space we use the $E$ superscript\footnote{It is always
possible to assume that there exists a function $\mGd^E$ associated to
$\varphi^E$, even if there is no associated diagrammatic
representation, assuming the same formal functional relation between
them.}. In the previous equation, the density contrast is a parameter
given a priori. The variables $\tau_E$ and $\tau_L$ depend formally on
$\delta$ through the saddle-point equations,
\begin{equation}
\delta=\mGd^L(\tau_L)=\mGd^E(\tau_E),
\end{equation}
and in Lagrangian space $\sigma$ is taken at the mass scale
corresponding to the density contrast $\delta$ ($\sigma$ is computed a
priori in Eulerian space).

{}From these equations we can eliminate $\tau_L$ to get an implicit
equation between $\mGd^E$ and $\tau_E$,
\begin{equation}
\mGd^E(\tau_E)=\mGd^L\left({\sigma\left[(1+\mGd^E(\tau_E))^{1/3}\,R\right]\over
\sigma(R)}\tau_E\right),
\label{fb:mGdE}
\end{equation}
where $\mGd^L(\tau_L)$ is known and is obtained from spherical
collapse dynamics.  The cumulant generating function, $\varphi^E(y)$,
is then built from $\mGd^E(\tau_E)$ the same way as $\varphi^L(y)$ was
from $\mGd^L(\tau_L)$ [Eqs.~(\ref{fb:phieq}) and (\ref{taueq})].

\begin{figure}
\centering{ \epsfysize=6.5truecm\epsfbox{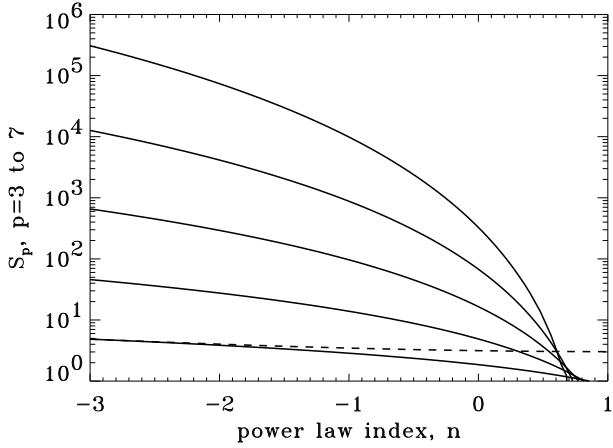} }
\caption{The predicted $S_p$ parameters for  power law spectra as
functions of the spectral index. The results are shown for top-hat
filter except for the dashed line which corresponds to the skewness
for a Gaussian filter.}
\label{fb:AllSp}
\end{figure}

\begin{figure}
\vspace{10cm} \special{hscale=60 vscale=60 voffset=0 hoffset=0
psfile=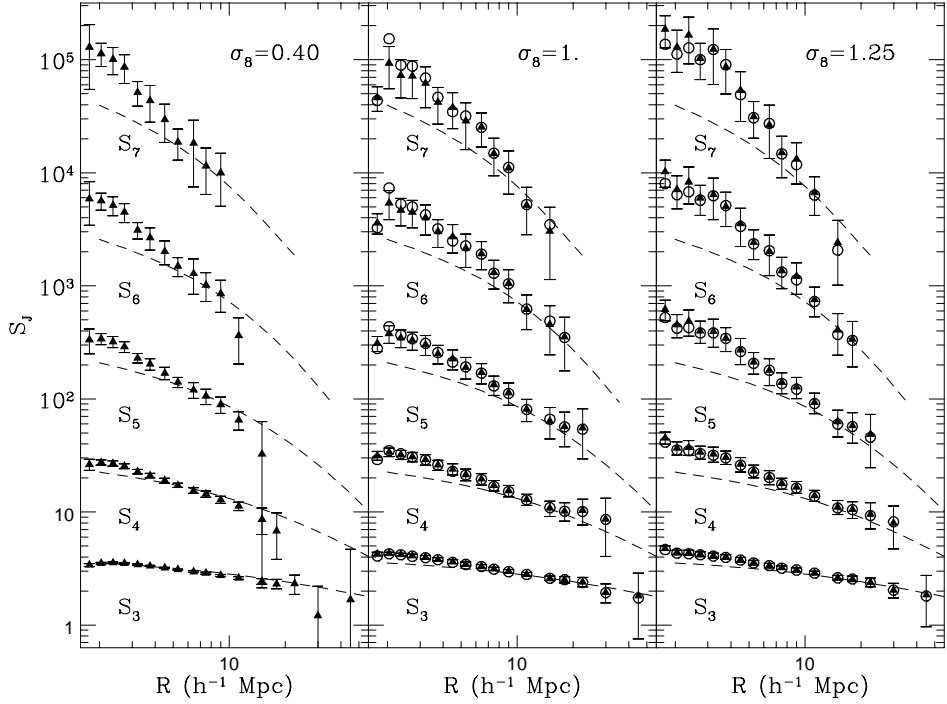}
\caption{The  $S_p$   parameters  for  $3\le   p\le7$.  Comparisons
between theoretical    predictions   and    results from   numerical
simulations (from~\cite{BGE95}) ($\sigma_8$ is the linear variance in
a sphere of radius $8\,h^{-1}$ Mpc).}
\label{fb:figSpGazta}
\end{figure}

Expanding this function around $y=0$ leads to explicit expressions for
the first few values of $S_p$.  They can be written as functions of
successive logarithmic derivatives of the variance,
\begin{equation}
\gamma_p\equiv{\d^p\log\sigma^2(R)\over\d\log^p R},
\label{fb:gammap}
\end{equation}
and read
\begin{eqnarray}
S_3&=&{{34}\over 7} + {\gm_1}, \label{eq:s3ptpt}\\ S_4&=&
{{60712}\over {1323}} + {{62\,{\gm_1}}\over 3} +
{{7\,{{{\gm_1}}^2}}\over 3} + {{2\,{\gm_2}}\over 3},\\ S_5&=&
{{200575880}\over {305613}} + {{1847200\,{\gm_1}}\over {3969}} +
{{6940\,{{{\gm_1}}^2}}\over {63}} + {{235\,{{{\gm_1}}^3}}\over
{27}}\nonumber\\ &&\ \ + {{1490\,{\gm_2}}\over {63}} +
{{50\,{\gm_1}\,{\gm_2}}\over 9} +  {{10\,{\gm_3}}\over {27}},\\ S_6&=&
{\frac{351903409720}{27810783}} +  {\frac{3769596070\,\gm_1}{305613}}
+  {\frac{17907475\,{{\gm_1}^2}}{3969}} + \nonumber\\ &&
{\frac{138730\,{{\gm_1}^3}}{189}} +  {\frac{1210\,{{\gm_1}^4}}{27}} +
{\frac{3078965\,\gm_2}{3969}} +  {\frac{23680\,\gm_1\,\gm_2}{63}} +
\nonumber\\ &&  {\frac{410\,{{\gm_1}^2}\,\gm_2}{9}} +
{\frac{35\,{{\gm_2}^2}}{9}} +  {\frac{3790\,\gm_3}{189}} +
{\frac{130\,\gm_1\,\gm_3}{27}} +  {\frac{5\,\gm_4}{27}},
\label{eq:s6ptpt}\\
&&\dots\nonumber
\end{eqnarray}
For a power-law spectrum, these coefficients depend only on spectral
index $n$, through $\gamma_1=-(n+3)$ and $\gamma_i=0$ for $i\ge 2$.
They are plotted as functions of $n$ in Fig.~\ref{fb:AllSp}. They all
appear to be decreasing functions of $n$. The above predictions were
compared against numerical experiments, as illustrated in
Fig.~\ref{fb:figSpGazta} for CDM.  The agreement between theory and
measurements is close to perfect as long as the variance is below
unity.  It is quite remarkable to see that the validity domain of PT
results does not deteriorate significantly when the cumulant order
increases.

\subsection{One-Loop Corrections to One-Point Moments}
\label{Sect:OneLoopMoments}

We now consider results that include the dependence of $S_p$
parameters on the variance. Due to the complexity of these
calculations, only few exact results are known, but there are useful
approximate results from the spherical collapse model.

\begin{table}
\caption{Tree-level and one-loop corrections predicted by various
non-linear approximations.}
\label{comp}
\vspace{3mm}
\begin{center}
\begin{tabular}{cccccc}
\hline Moment expansions & $s_{2,4}$ & $S_{3,0}$ & $S_{3,2}$ &
$S_{4,0}$ & $S_{4,2}$ \\ \hline  FFA, Unsmoothed & 0.43 & 3    & 1
& 16    & 15.0 \\  LPA, Unsmoothed & 0.72 & 3.40 & 2.12 & 21.22 &
37.12 \\ ZA, Unsmoothed & 1.27 & 4    & 4.69 & 30.22 & 98.51 \\
Exact\ PT, Unsmoothed                  & 1.82 & 4.86 & 9.80 & 45.89 &
$-$ \\  Exact\ PT, Top-Hat Smoothing, $n=-2$   & 0.88 & 3.86 & 3.18 &
27.56 & $-$ \\  Exact\ PT, Gaussian Smoothing, $n=-2$  & 0.88 & 4.02 &
3.83 & 30.4  & $-$ \\  \hline 
\end{tabular}
\end{center}
\vspace{1mm}
\end{table}

\subsubsection{Exact Results}

To get loop corrections for the one-point density moments, it is
necessary to expand both the second moment and the higher-order
moments with respect to the linear variance $\sigma_L$, \be \sigma^2=
\sigma_L^2 + \sum_{n=3}^\infty s_{2,n}\ \sigma_L^n, \ee and \be
S_p(\sigma_L) = S_{p,0}+\sum_{n=1}^\infty S_{p,n}^L\ \sigma_L^n. 
\label{eq:s3sc}
\ee Note that for Gaussian initial conditions, the contributions with
$n$ odd vanish. The $S_p$ parameters can also be expanded with respect
to the non-linear variance,
\begin{equation}
S_p(\sigma) = S_{p,0}+\sum_{n=1}^\infty S_{p,n}\ \sigma^n,
\end{equation}
and it is easy to see that, $S_{p,2}=S_{p,2}^L$,
$S_{p,4}=S_{p,4}^L-S_{p,2}^L\,s_{2,4}$, etc... for Gaussian initial
conditions. Table~\ref{comp} shows the results of one-loop corrections
in various approximations to the dynamics described in
Sect.~\ref{sec:nonlinearapp} (frozen flow approximation, FFA; linear
potential approximation, LPA; and ZA), and exact
PT~\cite{ScFr96a}. These results, however, ignore the effects of
smoothing which, as is known from tree-level results, are significant.

Taking into account smoothing effects in the exact PT framework has
only been done numerically for the case $n=-2$, where the one-loop
bispectrum yields a closed form~\cite{Scoccimarro97}. The resulting
one-loop coefficients are shown in Table~\ref{comp} as well, for
top-hat and Gaussian smoothing. When $n\ge -1$, one-loop corrections
to $S_3$ diverge, as for the power spectrum and bispectrum.

\subsubsection{The Spherical Collapse Model Approximation}
\label{sec:SCmodel}

\begin{table}
\caption{Values for the higher-order perturbative contributions in the
SC model for the unsmoothed ($n=-3$) and smoothed ($n=-2,-1,0$)
density fields, for a top-hat filter an a power-law power spectrum.
When known exact one-loop results are quoted in brackets.  More
details can be found in~\cite{FoGa98a}.}
\label{nlsc}
\vspace{.3cm}
\begin{center}
\begin{tabular}{ccccc}
\hline { SC} & {Unsmoothed} & \multicolumn{3}{c}{ Smoothed} \\  \hline
& $n=-3$ & $n=-2$ & $n=-1$ & $n=0$ \\  \hline $s_{2,4}$   &
1.44\,[1.82]& 0.61\,[0.88]& 0.40\,[$\infty$] &  0.79\,[$\infty$] \\
$s_{2,6}$   & 3.21 & 0.34 & 0.05 &  0.68 \\  \hline $S_{3,0}$ & 4.86 &
3.86 & 2.86 & 1.86 \\  $S_{3,2}^L$ & 10.08\,[9.80] & 3.21\,[3.18] &
0.59\,[$\infty$] & -0.02\,[$\infty$] \\  $S_{3,4}^L$ & 47.94 & 3.80 &
0.07 & 0.06  \\  $S_{4,0}$ & 45.89 & 27.56 & 13.89 & 4.89  \\
$S_{4,2}^L$ & 267.72 & 63.56 & 7.39 & -0.16 \\  $S_{4,4}^L$ & 2037.2 &
138.43 & 1.99 & 0.31 \\  \hline 
\end{tabular}
\end{center}
\end{table}

Given the complexity of loop calculations, approximate expressions
have been looked for. The so-called spherical collapse (SC) model
prescription~\cite{FoGa98a} provides a nice and elegant way for
getting {\em approximate} loop corrections for the local
cumulants\footnote{Another prescription, which turns out to be not as
accurate, is given in~\cite{PrSc97}.}.

This model consists in assuming that shear contributions in the
equations of motion in Lagrangian space can be neglected, which
implies that density fluctuations grow locally according to spherical
collapse dynamics.  In this case, the cumulants can be obtained by a
simple nonlinear transformation of the local Lagrangian density
contrast $\delta$, \be \delta=(1+\mGd(-\delta_{\rm lin}))\
\mg\left[1+\mGd(-\delta_{\rm lin})\right]^{-1} \md_L -1,
\label{fb:SCmodel}
\ee expressed in terms of the linear density contrast $\delta_{\rm
lin}$ assumed to obey Gaussian statistics.  Note that the ensemble
average in Eq. (\ref{fb:SCmodel}) is computed in Lagrangian
space\footnote{Which means that all matter elements are equally
weighted, instead of volume elements.}. Given the fact that the usual
ensemble average in Eulerian space is related to the Lagrangian one
through $\mg X \md_L\equiv \mg (1+\delta)\,X\md$, the normalization
factor $\mg\left[1+\mGd(\delta_{\rm lin})\right]^{-1}\md_L$ is
required to obey the constraint that $\mg (1+\delta)^{-1} \md_L= \mg 1
\md_E=1$.

For Gaussian initial conditions, the SC model reproduces the
tree-level results. Its interest comes from the fact that estimates of
loop corrections can be obtained by pursuing relatively simple
calculations to the required order. In addition, as we shall see in
the next section, it allows a straightforward extension to
non-Gaussian initial conditions.  The smoothing effects, as shown from
calculations exact up to tree level, introduce further complications
but can be taken into account by simply changing the vertex generating
function $\mGd$ in Eq.~(\ref{fb:SCmodel}) to the one found in
Eq.~(\ref{fb:mGdE}). Rigorously, this equation is valid only at tree
level: its extension to loop corrections in the SC model can hardly be
justified\footnote{In the SC model, the kernels in the Jacobian of the
mapping from Lagrangian to Eulerian space present no angular
dependence, and this is actually incompatible with the commutation
property in Eq.~(\ref{jpcom}).},  but turns out to be a good
approximation.
 
\begin{figure} 
\begin{tabular}{cc}
{\epsfysize=6.5truecm \epsfbox{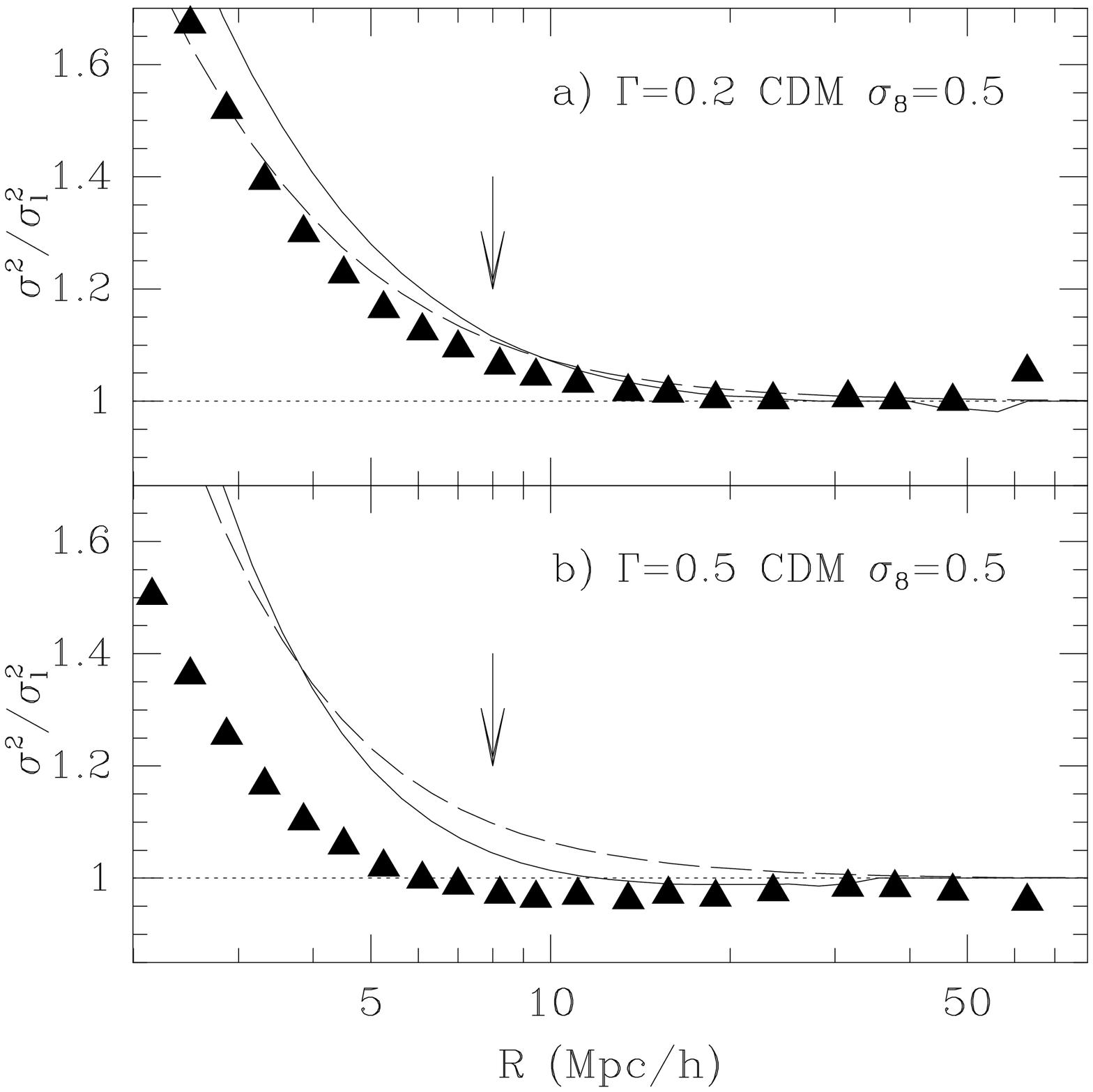}}& {\epsfysize=7. truecm
\epsfbox{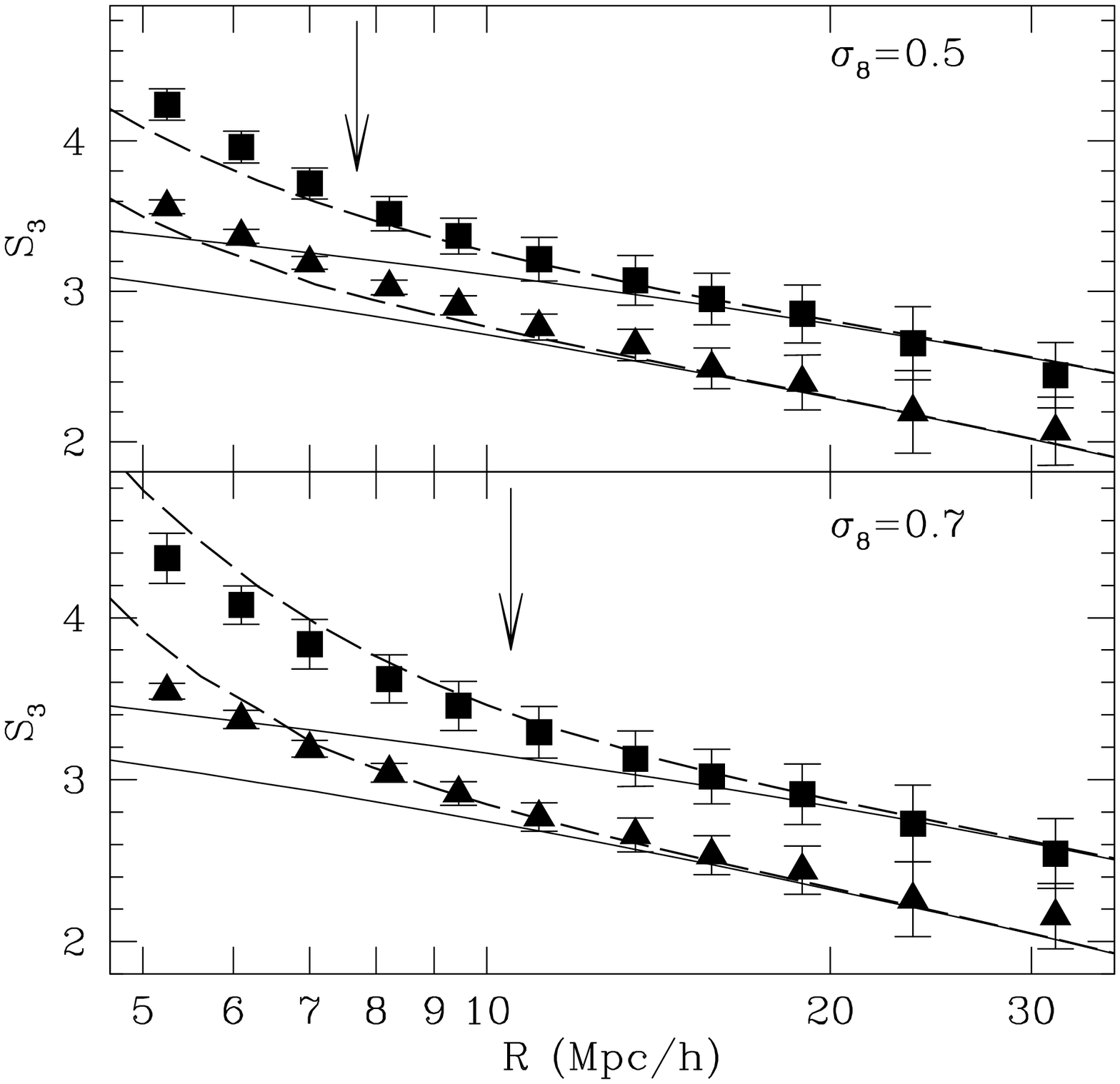}}
\end{tabular}
\caption[junk]{Non-linear evolution of the variance (left panels)  and
of the skewness parameter $S_3$ (right panels) from 10 realizations of
flat CDM $N$-body simulations. Two models are considered, $\Lambda$CDM
with $\Omega_m+\Omega_\Lambda=1$ and $\Gamma=0.2$, and SCDM with
$\Omega_m=1$ and $\Gamma=0.5$, where $\Gamma$ is the shape parameter
of the power-spectrum~\cite{EBW92}. In the left panels, symbols show
the ratio of the non-linear to the linear variance as a function of
smoothing radius. The value of $\Gamma$ is indicated on the panels,
while $\sigma_8^2$ stands for the linear variance in a sphere of
radius $8\,h^{-1}$ Mpc.  The SC model predictions are shown as a
short-dashed line while one-loop PT predictions are shown as a solid
line. The arrows indicate where $\sigma_l=0.5$. In the right panels,
the output times correspond to $\sigma_8=0.5$ (top) and $\sigma_8=0.7$
(bottom). Squares and triangles correspond to measurements in
$\Gamma=0.2$ and $\Gamma=0.5$ simulations, respectively.  Each case is
compared to the corresponding PT tree-level predictions (solid lines)
and SC model (long-dashed). {}From~\cite{FoGa98a}.}
\label{numSC}
\end{figure}

When comparisons are possible, the SC model is seen to provide
predictions that are in good agreement with exact PT results (see
Table~\ref{nlsc}), in particular for the $S_p$ parameters. Indeed, for
the variance (or cumulants), the SC prescription does not work as well
(see e.g. Fig.~\ref{numSC}). The reason for this are tidal
contributions, which are neglected in the SC approximation and lead to
the previrialization effects discussed for the exact PT case in
Sect.~\ref{previr}. Tidal effects tend to cancel for the $S_p$ because
of the ratios of cumulants involved. In the SC prescription no
divergences are found for $n\ge -1$, thus the interpretation of those
remains unresolved.

When tested against numerical simulations, the SC model provides a
good account of the departure from tree-level results as illustrated
by Fig.~\ref{numSC} for CDM models (see also Fig.~\ref{sc_ept} in
Sect.~\ref{sec:EPT}).

\subsection{Evolution from Non-Gaussian Initial Conditions}
\label{ngic2}

We now discuss the effects of non-Gaussian initial conditions on the
evolution of smoothed moments of the density field. As pointed out in
Sect.~\ref{ngic}, this is a complicated subject due to the infinite
number of possible non-Gaussian initial conditions. For this reason,
there are few general results, and only some particular models have
been worked out in detail. Early work concentrated on numerical
simulation studies~\cite{MMLM91,WeCo92,CMLMM93} of models with
positive and negative primordial skewness and comparison with
observations. In addition, a number of studies considered the
evolution of higher-order moments from non-Gaussian initial conditions
given by cosmic strings~\cite{Colombi93,ASWA98} and texture
models~\cite{GaMa96} using numerical simulations. Recently,
measurements of higher-order moments in numerical simulations with
$\chi^2_N$ initial conditions with $N$ degrees of freedom were given
in~\cite{White99}.

General properties of one-point moments evolved from non-Gaussian
initial conditions were considered using PT
in~\cite{FrSc94,Jaffe94,ChBo96,GaFo98,DJKU00}. To illustrate the main
ideas, let us write the PT expression for the first one-point moments:
\begin{eqnarray}
\lexp \de^2 \rexp &=& \lexp \de_1^2 \rexp +  \Big[ 2 \lexp \de_1 \de_2
\rexp \Big] +  \lexp \de_2^2 \rexp +2  \lexp \de_1 \de_3 \rexp + {\cal
O}(\sigma^5),
\label{varng}\\
\lexp \de^3 \rexp &=& \Big[\lexp \de_1^3 \rexp \Big]  + 3 \lexp
\de_1^2 \de_2 \rexp +  \Big[ 3 \lexp \de_2^2 \de_1 \rexp +3  \lexp
\de_1^2 \de_3 \rexp \Big] + {\cal O}(\sigma^6),
\label{m3ng}\\
\lexp \de^4 \rexp &=& \lexp \de_1^4 \rexp +  \Big[ 4 \lexp \de_1^3
\de_2 \rexp \Big] +  6 \lexp \de_1^2 \de_2^2 \rexp +4 \lexp \de_1^3
\de_3 \rexp +   {\cal O}(\sigma^7),
\label{m4ng}
\end{eqnarray}

where we simply use the PT expansion $\de =\de_1+\de_2+\ldots$. 
Square brackets denote terms which scale as odd-powers of $\de_1$, and
thus vanish for Gaussian initial conditions.  A first general remark
one can make is that these additional terms give to non-Gaussian
initial conditions a different {\em scaling} than for the Gaussian
case~\cite{FrSc94,ChBo96}.  In addition, the other terms in the
skewness have contribution from non-Gaussian initial conditions as
well; this does not modify the scaling of these terms but it can
significantly change the amplitude.  When dealing with non-Gaussian
initial conditions, the time-dependence and scale dependence must be
considered separately.  To illustrate this, consider the evolution of
the $S_p$ parameters as a function of smoothing scale $R$ and redshift
$z$, assuming for simplicity $\Omega_m=1$ [so that the growth factor
is $a(z)=(1+z)^{-1}$], at largest scales where linear PT applies we
have

\be S_p(R,z) \sim (1+z)^{p-2} S_p^I(R).
\label{Spscaling}
\ee

For dimensional scaling models, where the initial conditions satisfy
$S_p^I(R)\sim [\sigma_I(R)]^{2-p}$, this implies $S_p(R,z) \sim
[\sigma_I(R,z)]^{2-p}$; that is, the $S_p$ parameters scale as inverse
powers of the variance {\em at all times}. Note, however, that
Eq.~(\ref{Spscaling}) is more general, it implies that irrespective of
scaling considerations, {\em in non-Gaussian models the $S_p$
parameters should be an increasing function of redshift}; this can be
used to constrain primordial non-Gaussianity from
observations\footnote{Such a method is potentially extremely powerful,
as galaxy biasing would tend if anything to actually {\em decrease}
the $S_p$ parameters with $z$, as bias tends to become larger in the
past, see e.g.~\cite{SPLO01} and discussion in Chapter~8.}. However,
we caution that, as mentioned in Sect.~\ref{ngic}, all these arguments
are valid if the non-Gaussian fluctuations were generated at early
times, and their sources are not active during structure formation.

At what scale does the approximation of linear perturbation theory,
Eq.~(\ref{Spscaling}), break down? The answer to this question is of
course significantly model dependent, but it is very important in
order to constraint primordial non-Gaussianity. Indeed, we can write
the second and third moments from Eqs.~(\ref{powerNG})
and~(\ref{BispNG})

\begin{eqnarray}
\sigma^2(R) &=& \sigma^2_I(R) + 2 \int \d^3 \vk\, W^2(kR)\,\int \d^3
\vq\, F_2(\vk+\vq,-\vq)\ B^I(\vk,\vq) ,
\label{sigNG} \\
\lexp \de^3(R) \rexp &=& \lexp \de^3_I(R) \rexp + \lexp \de^3_G(R)
\rexp + \int \d^3\vk_1 \int \d^3 \vk_2\, W(k_1R) W(k_2R) W(k_{12}R)
\nonumber \\ & & \times \int \d^3\vq\,F_2(\vk_1+\vk_2-\vq,\vq)\
P_4^I(\vk_1,\vk_2,\vk_1+\vk_2-\vq,\vq) ,
\label{d3NG}
\end{eqnarray}

where $k_{12} \equiv |\vk_1+\vk_2|$, $B^I$ and $P_4^I$ denote the
initial bispectrum and trispectrum, respectively, and the subscript
``$G$'' denotes the usual contribution to the third moment due to
gravity from Gaussian initial conditions. Therefore, as discussed in
Sect.~\ref{ngic} for the bispectrum, corrections to the linear
evolution of $S_3$ depend on the relative magnitude of the initial
bispectrum and trispectrum compared to the usual gravitationally
induced skewness.

\begin{table}
\caption{Values of the higher-order perturbative contributions in the
SC model from non-Gaussian initial conditions with $B_J=1$ for the
unsmoothed ($n=-3$) and smoothed ($n=-2,-1,0$) density fields for a
top-hat window and a power-law spectrum.}
\label{ngicsc}
\vspace{.3cm}
\begin{center}
\begin{tabular}{ccccc}
\hline SC & Unsmoothed & \multicolumn{3}{c}{Smoothed} \\  \hline
$B_J=1$& $n=-3$ & $n=-2$ & $n=-1$ & $n=0$  \\ \hline  $s_{2,3}$   &
0.62 & 0.29 & -0.05 &  -0.38 \\  $s_{2,4}$   & 1.87 & 0.74 & 0.44 &
0.98 \\ $s_{2,5}$   & 3.36 & 0.60 & -0.05 &  -1.05 \\  \hline
$S_{3,0}$ & 5.05 & 4.21 & 3.38 & 2.55 \\  $S_{3,1}^L$ & 7.26 & 3.91 &
1.55 & 0.19 \\  $S_{3,2}^L$ & 23.53 & 7.37 & 1.18 & 0.20  \\  \hline
$S_{4,-1}^L$ & 19.81 & 16.14 & 12.48 & 8.81  \\  $S_{4,0}$ & 85.88 &
52.84 & 28.31 & 12.27 \\  $S_{4,1}^L$ & 332.51 & 128.51 & 32.83 & 2.70
\\  \hline
\end{tabular}
\end{center}
\vspace{.1cm}
\end{table}

This model dependence can be parametrized in a very useful way under
the additional assumption of spherical symmetry. In the spherical
collapse model, it is possible to work out entirely the perturbation
expansion for one-point moments from non-Gaussian initial conditions,
but the solutions are not exact as discussed further 
below\footnote{Some additional results have been recently obtained 
for the PDF from specific type of non-Gaussian initial conditions, 
see~\cite{Valageas01c}.}. Consider
non-Gaussian initial conditions with dimensional scaling. To take into
account non-Gaussian terms, one has to rewrite Eq.~(\ref{eq:s3sc}) as
\be S_p(\sigma_L) = \sum_{n=-p+2}^{-1} S_{p,n}^L\,\sigma_L^n
+S_{p,0}+\sum_{n=1}^{\infty} S_{p,n}^L\,\sigma_L^n,
\label{eq:s3scng}
\ee where $\sigma_L=\sigma_I$ is given by linear theory as in
Eq.~(\ref{sigNG}). The first non-vanishing perturbative contributions
to the variance, skewness and kurtosis read~\cite{GaFo98} \bea s_{2,3}
&=& \left[\frac{1}{3} S_3^G-1\right]\,B_3,  \nn \\  s_{2,4} &=& 3 -
{4\over 3}\,S_3^G - {5 \over 18}\,\left(S_3^G\right)^2 +
\frac{1}{4}{S_4^G} + \left[1 - \frac{1}{2}{S_3^G} -
\frac{1}{12}{\left(S_3^G\right)^2} + \frac{1}{12}{S_4^G}\right]\,B_4,
\nn \\ S_{3,-1}^L &\equiv& S_3^{(0)} = {B_3},  \nn \\  S_{3,0} &=&
S_3^G -2\,\left[\frac{1}{3}{S_3^G}-1\right]\,B_3^2 +
\left[\frac{1}{2}{S_3^G}\,-1\right]\,B_4,  \nn \\  S_{3,1}^L &=&
\left[\frac{1}{6}{S_3^G} -{17 \over 18}\,\left(S_3^G\right)^2 +
{5\over 8}\,S_4^G \right]\,B_3  + \left[3\,- 2\,S_3^G +
\frac{1}{3}{\left(S_3^G\right)^2} \right]\,B_3^3  \nn \\  &+& \left[-
4 + {8\over 3}\,S_3^G - \frac{1}{6}{\left(S_3^G\right)^2} -
\frac{1}{6}{S_4^G} \right]\,B_3\,B_4  \nn \\ &+&\left[1 - {2\over
3}\,S_3^G - \frac{1}{12}{\left(S_3^G\right)^2} + \frac{1}{8}{S_4^G}
\right]\,B_5,  \nn \\  S_{4,-2} &\equiv& S_4^{(0)} = {B_4}, \nn \\
S_{4,-1}^L &=& 4\,S_3^G\,B_3 + \left[3 - S_3^G\right]\,B_3\,B_4 +
\left[{2\over 3}\,S_3^G - 1\right]\,B_5,  \nn \\  S_{4,0} &=& S_4^G +
\left[3 + 7\,S_3^G - {14\over 3}\,\left(S_3^G\right)^2 + {3\over
2}\,S_4^G \right]\,B_3^2  \nonumber \\ &+& \left[-1 - {10\over
3}\,S_3^G + \frac{1}{6}{\left(S_3^G\right)^2 } + {5\over 4}\,S_4^G
\right]\,B_4  \nn \\ &+& \left[6 - 4\,S_3^G + {2\over
3}\,\left(S_3^G\right)^2 \right]\,B_3^2\,B_4 + \left[- 3\, + {3\over
2}\,S_3^G + \frac{1}{4}{\left(S_3^G\right)^2} - \frac{1}{4}{S_4^G}
\right]\,B_4^2  \nn \\ &+& \left[- 3\, + 3\,S_3^G\, - {2\over
3}\,\left(S_3^G\right)^2 \right]\,B_3\,B_5 + \left[1\, - {5\over
6}\,S_3^G - \frac{1}{18}{\left(S_3^G\right)^2} + \frac{1}{6}{S_4^G}
\right]\,B_6.  \nn \\ 
\label{sj_dim}
\eea  Here the non-Gaussianity in the initial conditions is
characterized via the dimensionless scaling amplitudes \be B_p \equiv
\frac{\lexp \de^p_I \rexp_c}{\sigma^p_I}.
\label{eq:Bpdef}
\ee For non-Gaussian initial conditions seeded by topological defects
such as textures~\cite{TuSp91,GaMa96} or cosmic
strings~\cite{Colombi93,ASWA98}, $B_p$ is expected to be of order
unity\footnote{For cosmic strings, this statement is valid if the
scale considered is sufficiently large, $R \ga 1.5 (\Omega_m
h^2)^{-1}$ Mpc, see~\cite{ASWA98} for details.}.  For reference,
Table~\ref{ngicsc} lists these results for $B_p=1$ and power-law
initial spectra as a function of spectral index $n$, in this case it
is clear that non-linear corrections to the linear result,
Eq.~(\ref{Spscaling}), can be very important even at large
scales. Even more so, $\chi^2$ initial conditions (with spectral index
such that it reproduces observations) have $B_3 \approx 2.5$ and $B_4
\approx 10$~\cite{Peebles99b,White99}; therefore non-linear
corrections are particularly strong~\cite{GaFo98,Scoccimarro00}.

\begin{figure}[t!]
\begin{tabular}{cc}
{\epsfxsize=6.8truecm\epsfysize=6.8truecm\epsfbox{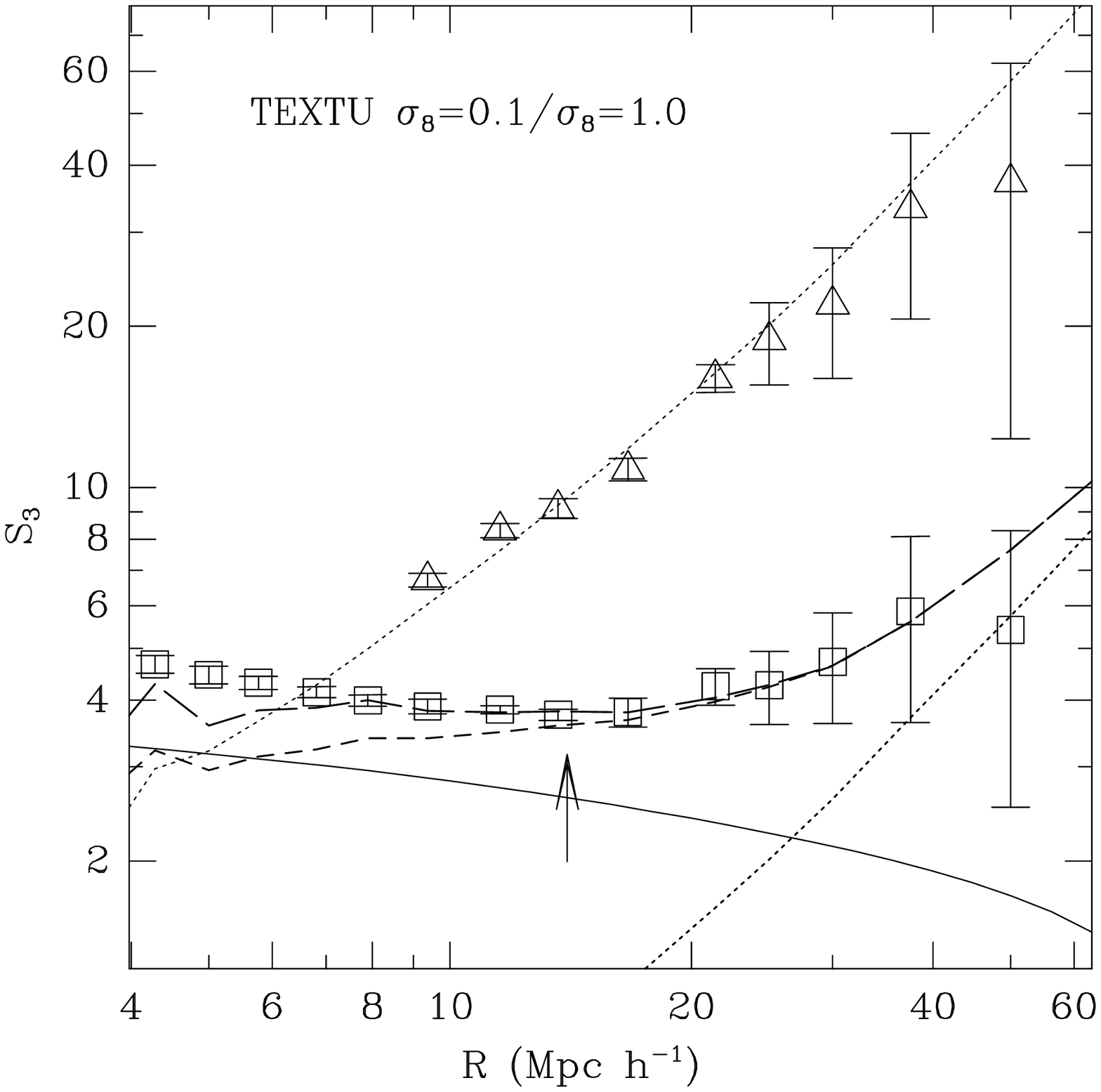}}&
{\epsfxsize=6.8truecm\epsfysize=6.8truecm\epsfbox{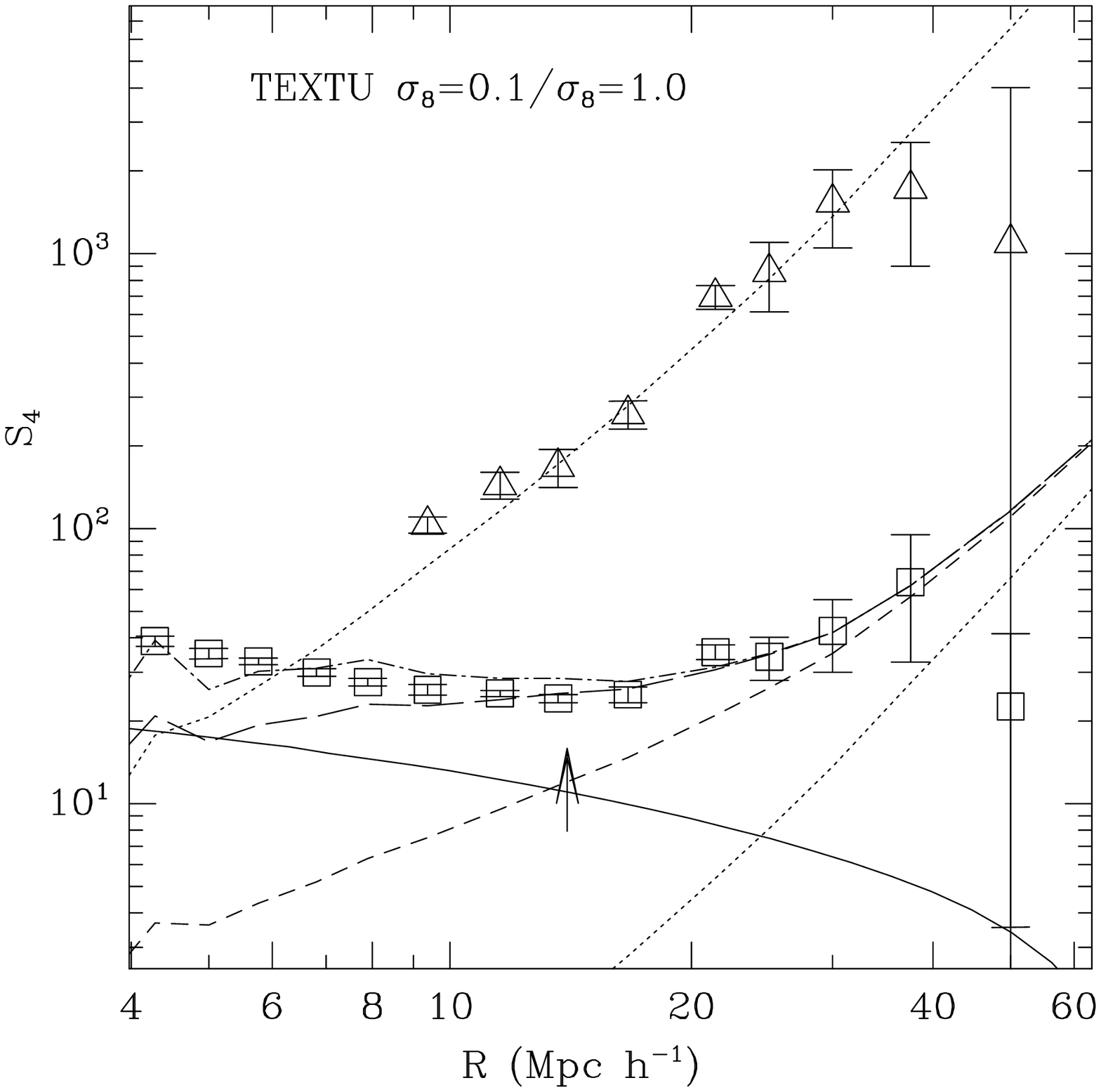}}
\end{tabular}
\caption{The  skewness and kurtosis, $S_3$ and $S_4$, for texture-like
non-Gaussian models.  The triangles show the initial conditions
($\sigma_8=0.1$), which are fitted well by the dimensional scaling,
$S_3= B_3/\sigma$ and $S_4=B_4/\sigma^2$ with $B_3=B_4\simeq 0.5$,
shown as the upper dotted line. Squares show $S_3$ and $S_4$ for a
later output corresponding to $\sigma_8=1.0$.  The SC predictions for
the $\sigma_8=1$ output are shown as short-dashed (including the
second order contribution) and long-dashed line (including the third
order).  The continuous line shows the corresponding tree-level PT
prediction for Gaussian initial conditions. The lower dotted lines
correspond to the linear theory prediction. In right panel the dot
long-dashed line displays the SC prediction including the 4th
perturbative contribution. {}From~\cite{GaFo98}.}
\label{S3text}
\end{figure}

When compared to exact PT calculations or to measurements in numerical
simulations, the SC model is seen to provide quite accurate
predictions.  This is illustrated by Fig.~\ref{S3text} for the skewness
and kurtosis in texture models~\cite{GaFo98}.  These parameters evolve
slowly from non-Gaussian initial conditions towards the (Gaussian)
gravitational predictions.  However, even at present time, a
systematic shift can be observed in Fig.~\ref{S3text} between the
Gaussian and the non Gaussian case, well described by the SC
predictions taken at appropriate order. The main signature of
non-Gaussianity remains at the largest scales, where the $S_p$
parameters show a sharp increase: this is the scaling regime of
Eq.~(\ref{Spscaling}) where observations can best constrain
non-Gaussianity~\cite{SiJu91,DJKU00}. This is explicitly illustrated
in Sect.~8.

\subsection{Transients from Initial Conditions}
\label{s:transients}

The standard procedure in numerical simulations is to set up the
initial perturbations, assumed to be Gaussian, by using the Zel'dovich
approximation (ZA,~\cite{Zeldovich70}). This gives a useful
prescription to perturb the positions of particles from some initial
homogeneous pattern (commonly a grid or a ``glass''~\cite{White96})
and assign them velocities according to the growing mode in linear
perturbation theory. In this way, one can generate fluctuations with
any desired power spectrum and then numerically evolve them forward in
time to the present epoch.

Although the ZA correctly reproduces the linear growing modes of
density and velocity perturbations, non-linear correlations are known
to be inaccurate when compared to the exact
dynamics~\cite{GrWi87,JBC93,Bernardeau94c,CaMo94,JWACB95}, see also
Table~\ref{nlsc}.  This implies that it may take a non-negligible
amount of time for the exact dynamics to establish the correct
statistical properties of density and velocity fields. This transient
behavior affects in greater extent statistical quantities which are
sensitive to phase correlations of density and velocity fields; by
contrast, the two-point function, variance, and power spectrum of
density fluctuations at large scales can be described by linear
perturbation theory, and are thus unaffected by the incorrect
higher-order correlations imposed by the initial conditions.

In Sect.~\ref{tevol} we presented the solutions involving the full
time dependence from arbitrary initial
conditions~\cite{Scoccimarro98}. Again, we assume $\Omega_m=1$ for
simplicity. The recursion relations for PT kernels including
transients results from using the following ansatz in Eq.~(\ref{eomi}),

\bea
\label{pt_exp3}
        \Psi_a^{(n)}(\vk,z) &= &\int \d^3\vk_1 \ldots \int \d^3\vk_n \
\dD_n \ {\cal F}_a^{(n)} (\vk_1, \ldots, \vk_n;z)  \delta_1(\vk_1)
\cdots \delta_1(\vk_n), \nonumber\\ \label{Phi_n}  \eea

where $a=1,2$, $z\equiv \ln a(\tau)$ with $a(\tau)$ the scale factor,
and the $n^{\rm th}$ order solutions for density and velocity fields
are components of the vector $\Psi_b$, i.e.  $\Psi_1^{(n)} \equiv
\de_n$, $\Psi_2^{(n)} \equiv \te_n$. In Eq.~(\ref{Phi_n}), $\dD_n
\equiv \de_D(\vk-\vk_1-\ldots-\vk_n)$. 

The kernels ${\cal F}_a^{(n)}$ now depend on time and reduce to the
standard ones when transients die out, that is ${\cal
F}_1^{(n)}\rightarrow F_n$, ${\cal F}_2^{(n)}\rightarrow G_n$ when $z
\rightarrow \infty$. Also, Eq.~(\ref{pt_exp3}) incorporates in a
convenient way initial conditions, i.e. at $z=0$, ${\cal F}_a^{(n)} =
{\cal I}_a^{(n)}$, where the kernels ${\cal I}_a^{(n)}$ describe the
initial correlations imposed at the start of the simulation. For the
ZA we have

\be {\cal I}_1^{(n)}=F_n^{\rm ZA}, \ \ \ \ \ {\cal I}_2^{(n)}=G_n^{\rm
ZA}.  \ee 

Although most existing initial conditions codes use the ZA
prescription to set up their initial conditions, there is another
prescription to set initial velocities suggested in~\cite{EDWF85},
which avoids the high initial velocities that result from the use of
ZA because of small-scale density fluctuations approaching unity when
starting a simulation at low redshifts. This procedure corresponds to
recalculate the velocities from the gravitational potential due to the
perturbed particle positions, obtained by solving again Poisson
equation after particles have been displaced according to the ZA.
Linear PT is then applied to the density field to obtain the
velocities, which implies instead that the initial velocity field is
such that the divergence field $\Theta(x) \equiv \theta(\vx)/(-f \
{\cal H})$ has the same higher-order correlations as the ZA density
perturbations. In this case, 

\be {\cal I}_1^{(n)}=F_n^{\rm ZA}, \ \ \ \ \ {\cal I}_2^{(n)}=F_n^{\rm
ZA}.  \ee The recursion relations for ${\cal F}_a^{(n)}$, which solve
the non-linear dynamics at arbitrary order in PT, can be obtained by
replacing Eq.~(\ref{Phi_n}) into Eq.~(\ref{eomi}), which
yields~\cite{Scoccimarro98}

\bea
\label{pt_recrel_sd}
{\cal F}_a^{(n)}(\vk_1,  \ldots , \vk_n;z) &=&  {\rm e}^{-n z} \
g_{ab}(z) \ {\cal I}_b^{(n)}(\vk_1,  \ldots , \vk_n) \nonumber\\
&&\hspace{-2cm}+ \sum_{m=1}^{n-1} \int_0^z \d s  \ {\rm e}^{n(s-z)} \
g_{ab}(z-s)  \gamma_{bcd}(\vk^{(m)},\vk^{(n-m)}) \nonumber\\
&&\hspace{-1cm}\times{\cal F}_c^{(m)}(\vk_1,  \ldots ,  \vk_m;s) \
{\cal F}_d^{(n-m)}(\vk_{m+1}, \ldots , \vk_n;s), \eea where we have
assumed the summation convention over repeated indices, which run
between 1 and 2. Equation~(\ref{pt_recrel_sd}) reduces to the standard
recursion relations for Gaussian initial conditions (${\cal
I}_a^{(n)}=0$ for $n>1$) when transients are neglected, i.e. the time
dependence of ${\cal F}_a^{(n)}$ is neglected and the lower limit of
integration is replaced by $s=-\infty$.  Also, it is easy to check
from Eq.~(\ref{pt_recrel_sd}) that if ${\cal I}_a^{(n)} =
(F_{n},G_{n}) $, then ${\cal F}_a^{(n)} = (F_{n},G_{n})$, as it should
be.  Note that PT kernels in Eq.~(\ref{pt_recrel_sd}) are no longer a
separable function of wave-vectors and time.

{}From the recursion relations given by Eq.~(\ref{pt_recrel_sd}), it
is possible to find the recursion relations for the smoothed vertices
$\nu_{n}$ and $\mu_{n}$ as functions of scale factor $a$ and smoothing
scale $R$, and therefore infer the values of the cumulants as
functions of the $\gamma_p$'s [Eq.~(\ref{fb:gammap})] similarly as in
Sect.~\ref{fb:CorHierarchies}, but with additional dependence with the
scale factor. For the skewness parameters, one finds in the
Einstein-de Sitter case

\label{p=3za}
\bea S_3(a) &=& \frac{[4+ \gamma_1]}{a}  + \Big\{ \frac{34}{7} +
\gamma_1 \Big\} - \frac{\gamma_1 +\frac{26}{5}}{a}  +
\frac{12}{35a^{7/2}},\label{S3g}\\ &=& \frac{34}{7} + \gamma_1 -
\frac{6}{5 a}+ \frac{12}{35 a^{7/2}},
\label{S3g2} \\ 
T_3(a) &=& -\frac{[2+ \gamma_1]}{a}  - \Big\{ \frac{26}{7} + \gamma_1
\Big\} + \frac{\gamma_1 +\frac{16}{5}}{a} + \frac{18}{35
a^{7/2}},\label{T3g}\\ &=& -\frac{26}{7} - \gamma_1 + \frac{6}{5 a}+
\frac{18}{35 a^{7/2}},
\label{T3g2}
\eea where we have assumed ZA initial velocities. On the other hand,
for initial velocities set from perturbed particle positions, we have:
\label{p=3}
\bea S_3(a) &=& \frac{[4+ \gamma_1]}{a}  + \Big\{ \frac{34}{7} +
\gamma_1 \Big\} - \frac{\gamma_1+\frac{22}{5}}{a}- \frac{16}{35
a^{7/2}},\\ &=& \frac{34}{7} + \gamma_1 - \frac{2}{5 a}- \frac{16}{35
a^{7/2}} ,
\label{S3} \\ 
T_3(a) &=& -\frac{[4+ \gamma_1]}{a}  - \Big\{ \frac{26}{7} + \gamma_1
\Big\}+ \frac{\gamma_1+\frac{22}{5}}{a} -  \frac{24}{35 a^{7/2}},\\
&=& - \frac{26}{7} - \gamma_1 + \frac{2}{5 a}- \frac{24}{35 a^{7/2}}. 
\label{T3}
\eea  For $\Omega_m \neq 1$, these expressions are approximately valid
upon replacing the scale factor $a$ by the linear growth factor
$D_1(\tau)$. The first term in square brackets in Eqs.~(\ref{S3g})
and~(\ref{T3g}) represents the initial skewness given by the
ZA~(e.g.~\cite{Bernardeau94c}), which decays with the expansion as
$a^{-1}$, as expected from the discussion on non-Gaussian initial
conditions in the previous section. The second and remaining terms in
Eqs.~(\ref{S3g}) and~(\ref{T3g}) represent the asymptotic exact values
(in between braces) and the transient induced by the exact dynamics
respectively; their sum vanishes at $a=1$ where the only correlations
are those imposed by the initial conditions. Similar results to these
are obtained for higher-order moments, we refer the reader
to~\cite{Scoccimarro98} for explicit expressions. Note that for
scale-free initial conditions, the transient contributions to $S_p$
and $T_p$ break self-similarity. Transients turn out to be somewhat
less important for velocities set from perturbed particle positions,
than in the ZA prescription, as in this case higher-order correlations
are closer to those in the exact dynamics.

\begin{figure}[t]
\begin{tabular}{cc}
{\epsfysize=6.5truecm \epsfbox{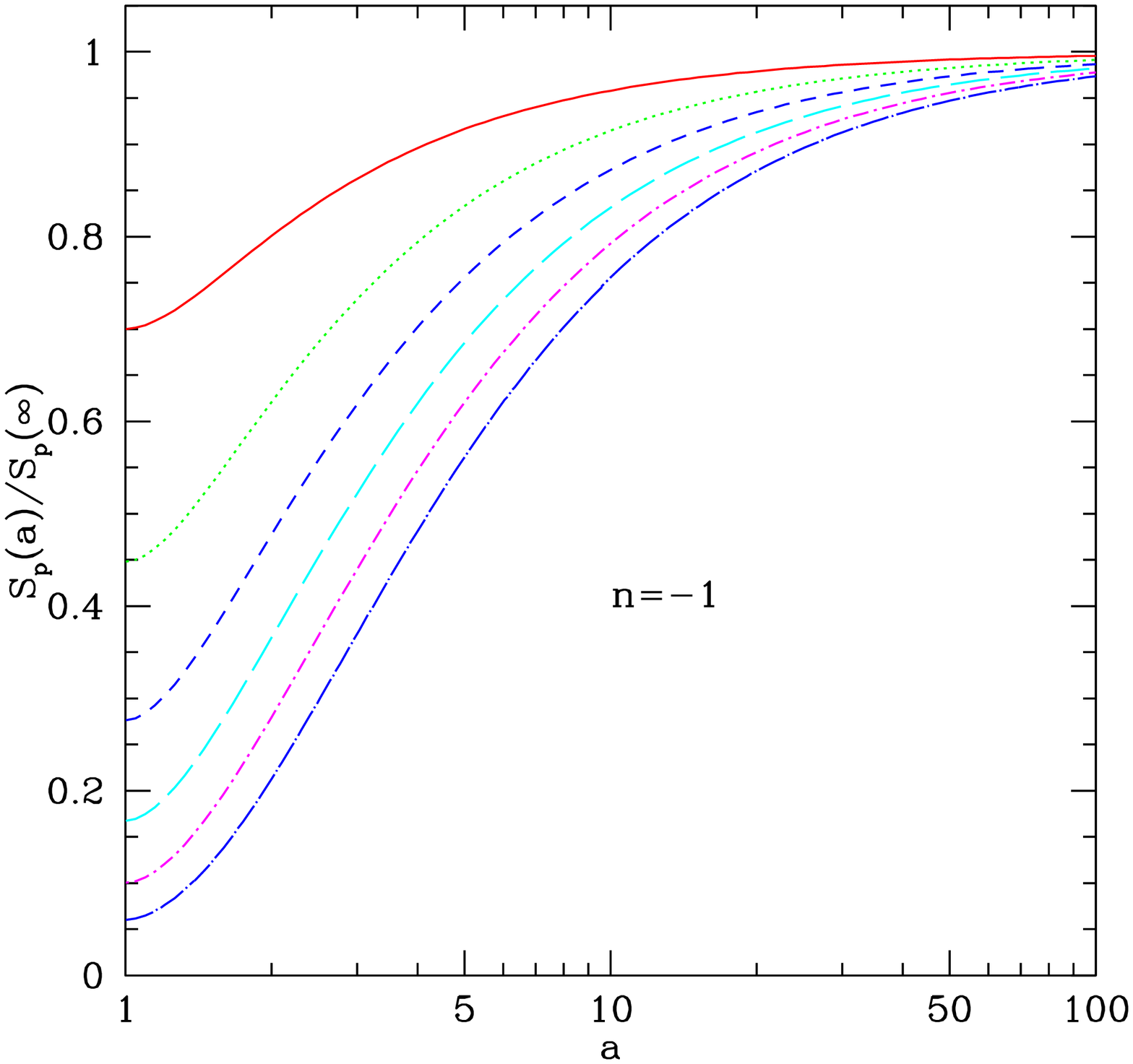}}&
{\epsfysize=6.5truecm \epsfbox{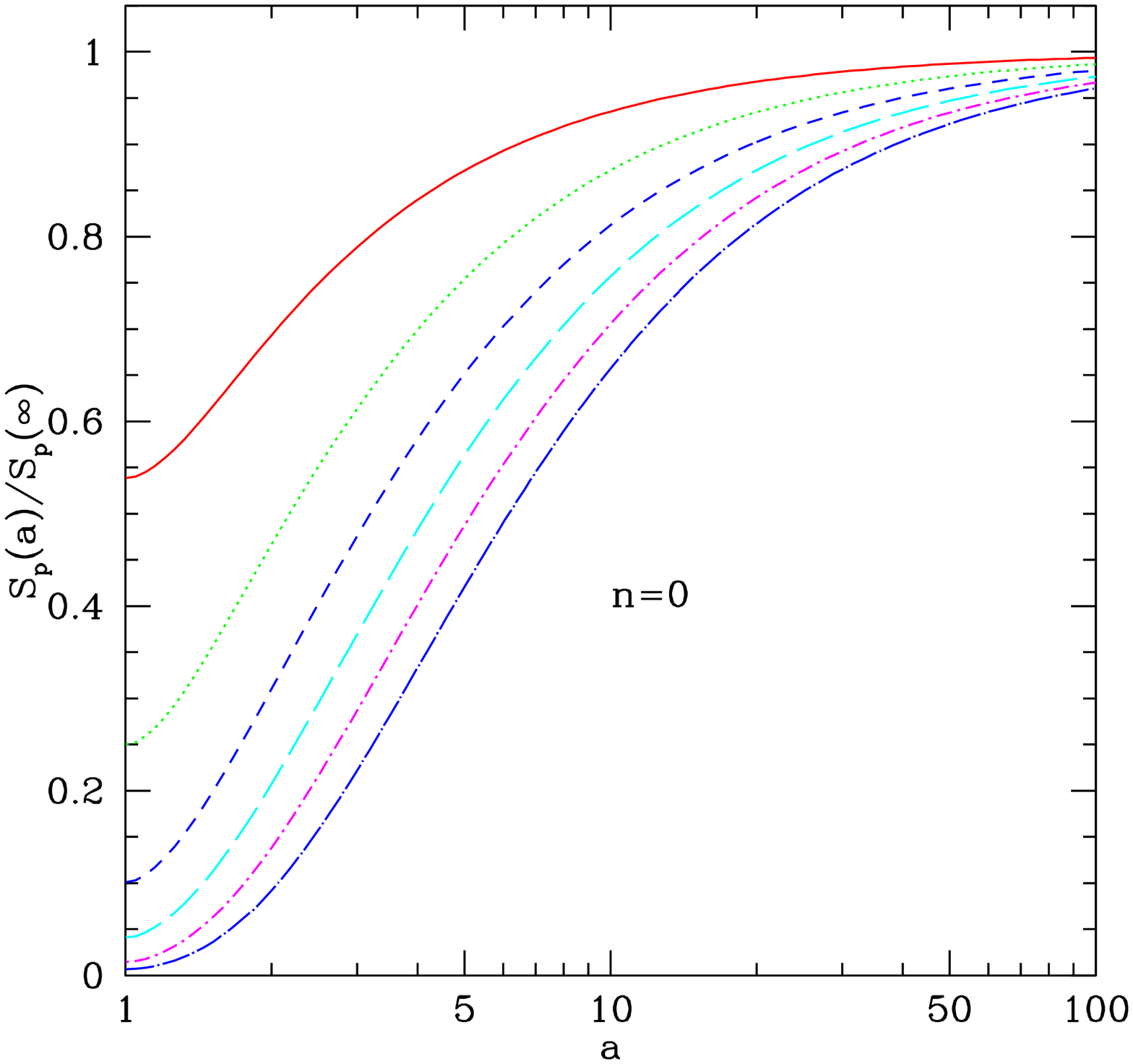}}
\end{tabular}
\caption{The ratio of the tree-level $S_p$ parameters at scale factor
$a$ to their asymptotic exact dynamics value for scale-free initial
spectra with spectral indices $n=-1,0$. {}From top to bottom
$p=3,\ldots,8$.  The values at $a=1$ represent those set by the ZA
initial conditions.}
\label{trans1} 
\end{figure}

Figure~\ref{trans1} illustrates these results for the skewness and
higher-order $S_p$ parameters as functions of scale factor $a$ for
different spectral indices, assuming that velocities are set as in the
ZA. The plots show the ratio of $S_p(a)$ to its ``true'' asymptotic
value predicted by PT, $S_p(\infty)$, for $3 \leq p \leq 8$. The
values at $a=1$ correspond to the ratio of ZA to exact dynamics
$S_p$'s, which becomes smaller as either $p$ or $n$ increases. For the
skewness, it takes as much as $a=6$ for $n=0$ to achieve 10\% of the
asymptotic exact PT value, whereas spectra with more large-scale
power, where the ZA works better, require less expansion factors to
yield the same accuracy. As $p$ increases, however, the transients
become worse and at $p=8$ an expansion by a factor $a=40$ is required
for $n=0$ to achieve $10\% $ accuracy in $S_{8}$. This suggests that
the tails of the PDF could be quite affected by transients from
initial conditions.

\begin{figure}[t!]
\centering
\vspace{1cm}
\centerline{\epsfxsize=13.5truecm\epsfbox{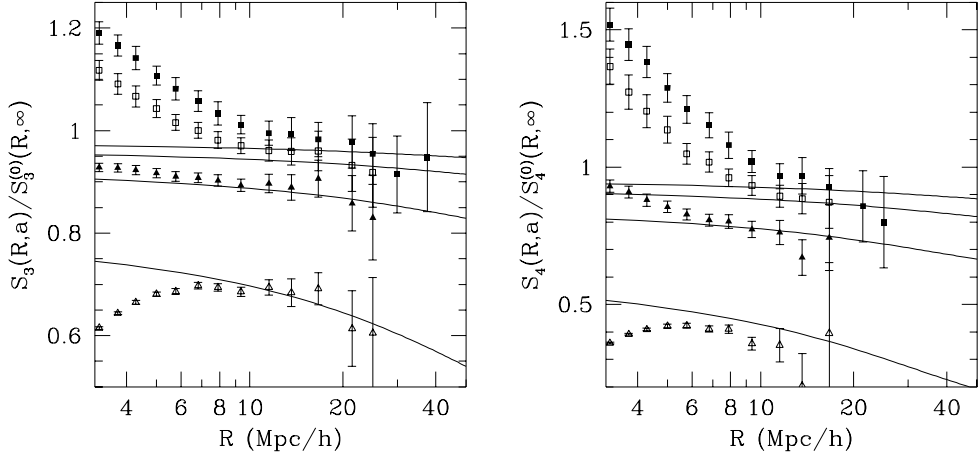}}
\caption{Symbols show the ratio of the $S_p$ parameters for different
scale factor $a$ (simulation began at $a=1$) measured in SCDM
numerical simulations~\protect\cite{BGE95} to their asymptotic
tree-level exact dynamics value as a function of smoothing scale
$R$. Symbols represent $a=1$ (open triangles), $a=1.66$ (filled
triangles), $a=2.75$ (open squares) and $a=4.2$ (filled
squares). Error bars denote the variance of measurements in 10
realizations. Solid lines correspond to the predictions of transients
in tree-level PT, expected to be valid at large scales.}
\label{trans2}
\end{figure}

Figure~\ref{trans2} presents a comparison of the perturbative
predictions for transients in $S_p$ parameters with the standard CDM
numerical simulations measurements of~\cite{BGE95}. In this case,
initial velocities are set as in~\cite{EDWF85} rather than using the
ZA. The error bars in the measurements correspond to the variance over
10 realizations.  If there were no transients and no other sources of
systematic uncertainties, all the curves would approach unity at large
scales, where tree-level PT applies. Unfortunately, there are other
sources of systematic uncertainties which prevents a clean test of the
transients predictions from PT, as we now briefly discuss, but more
details will be given in Sect.~\ref{sec:9}.

The different symbols correspond to different outputs of the
simulation: open triangles denote initial conditions ($a=1$, $\sigma_8
= 0.24$), solid triangles ($a=1.66$, $\sigma_8=0.40$), open squares
($a=2.75$, $\sigma_8=0.66$), and solid squares ($a=4.2$,
$\sigma_8=1.0$). For the initial conditions measurements (open
triangles) there is some disagreement with the ZA predictions,
especially at small scales, due to discreteness effects, which have
not been corrected for. The initial particle arrangement is a grid,
therefore the Poisson model commonly used to correct for discreteness
is not necessarily a good approximation (see~\cite{BGE95} for further
discussion of this point and Sect.~\ref{sec:relsof} below).  The
second output time (solid triangles) is perhaps the best for testing
the predictions of transients: discreteness corrections become much
smaller due to evolution away from the initial conditions, and the
system has not yet evolved long enough so that finite volume
corrections are important (see also Sect.~\ref{sec:cosmicinsim}). For
$S_3$ we see excellent agreement with the predictions of
Eq.~(\ref{S3}), with a small excess at small scales due to non-linear
evolution away from the tree-level prediction. For $p>3$ the numerical
results show a similar behavior with increased deviation at small
scales due to non-linear evolution, as expected. For the last two
outputs we see a further increase of non-linear effects at small
scales, then a reasonable agreement with the transients predictions,
and lastly a decrease of the numerical results compared to the PT
predictions at large scales due to finite volume effects, which
increase with $\sigma_8$, $R$, and $p$~\cite{CBS94,BGE95,CBH96,MBMS99}.

\subsection{The Density PDF}
\label{fb:TheDensityPDF}

Up to now, we have given exhaustive results on the local density
moments. In the following we show how these results can be used to
reconstruct the one-point density PDF's~\cite{Bernardeau94a}.

\subsubsection{Reconstruction of the PDF from the Generating Function}

We use here the relation between the probability distribution function
and the generating function $\varphi(y)$, Eq.~(\ref{fb:PDFvarphi}).
To be able to use such a relation one needs a supplementary
non-trivial hypothesis. Indeed $\varphi(y)$ is a priori $\sigma$
dependent through every $S_p$ parameter. We assume here that we have
\begin{equation}
\varphi(y,\sigma)\to\varphi(y)\ \ {\rm when}\ \ \sigma\to 0,
\label{fb:PhiUni}
\end{equation}
in an {\em uniform} way as suggested by numerical simulation results
on $S_p$.  No proof has however been given of such a property.  It has
even been challenged recently by calculations presented
in~\cite{Valageas01b,Valageas01d}, which suggest that
$\varphi(y,\sigma)$ is not analytic at $y\to0^-$ for finite values of
$\sigma$.  That would affect results presented below (in particular
the shape of the large density tails).  In the following we will
ignore these subtleties and assume that, when the variance is small
enough, it is legitimate to compute the density PDF from,
\begin{equation}
p(\delta)\d\delta=\int_{-\ii\infty}^{+\ii\infty} {\d y\over
2\pi\ii\sigma^2}\exp\left[-{\varphi(y)\over \sigma^2} +{y\dta\over
\sigma^2}\right] \d\delta
\label{fb:pdeltanum}
\end{equation}
where $\varphi(y)$ is given by the system (\ref{fb:phieq},\ref{taueq})
by analytic continuation from the point $\varphi(0)=0$.

{}From this equation numerous results  can be obtained. The different
forms of $p(\delta)$ have been described in detail in
\cite{BaSc89a,BaSc89b}. Taking advantage of the approximation
$\sigma\ll1$ one can apply the saddle-point approximation to get
\begin{equation}
p(\delta)\d\delta={\d\delta\over -\mG_{\delta}'(\tau)}
\left[{1-\tau\mG_{\delta}''(\tau)/\mG_{\delta}'(\tau)  \over 2\pi
\sigma^2 }\right]^{1/2}\  \exp\left(-{\tau^2\over 2\sigma^2}\right),\
\  \mG_{\delta}(\tau)=\dta.
\label{fb:pdeltasaddle}
\end{equation}

\begin{table}
\caption{Parameters of the singularity (\ref{phising}) for different
values of the spectral index $n$ (there is no singularity for
$n\ge0$).}  
\vspace{.3cm} \centerline{
\begin{tabular}{cccccc}
\hline $n$&$\dta_c$&$y_s$&$r_s$&$a_s$&$\varphi_s$\\ \hline -3
&0.656&-0.184& 1.66&1.84 &-0.030\\ -2.5&0.804&-0.213& 1.80&2.21
&-0.041\\ -2  &1.034&-0.253& 2.03&2.81 &-0.058\\ -1.5&1.44 &-0.310&
2.44&3.93 &-0.093\\ -1  &2.344&-0.401& 3.34&6.68 &-0.172\\
-0.5&5.632&-0.574& 6.63&18.94&-0.434\\ \hline
\end{tabular}}
\label{fb:tabSing}
\end{table}

This solution is valid when $\dta\le\dta_c$ where $\dta_c$ is the
value of the density contrast for which
$1=\tau\mG_{\dta}''(\tau)/\mG_{\dta}'(\tau)$.  Here function
$\mG_{\delta}(\tau)$ is equal to $\mG_{\delta}^L(\tau)$ or
$\mG_{\delta}^E(\tau)$ whether one works in Lagrangian space or
Eulerian space while taking smoothing into account
(Sect.~\ref{sec:lagtoeul}).

When $\dta$ is larger than $\dta_c$ the saddle point approximation is
no longer valid.  The shape of $p(\delta)$ is then determined by the
behavior of $\varphi(y)$ near its singularity on the real axis,
\begin{equation}
\varphi(y)\simeq \varphi_s+r_s(y-y_s)-a_s(y-y_s)^{3/2},\label{phising}
\end{equation}
and we have
\begin{equation}
p(\dta)\d\dta={3\,a_s\sigma \over 4\sqrt{\pi}}
(1+\dta-r_s)^{-5/2}\, \exp\left[{-\vert y_s \vert
\dta/\sigma^2+\vert\varphi_s \vert/\sigma^2}\right]\d\dta.
\label{fb:pdeltatail}
\end{equation}
Table~\ref{fb:tabSing} gives the parameters describing the singularity
corresponding to different values of the spectral index, for the PDF
of the smoothed density field in Eulerian space\footnote{The case
$n=-3$ corresponds as well to the PDF in Lagrangian space or to the
unsmoothed case.}.  One sees that the shape of the cut-off is very
different from that of a Gaussian distribution.  This shape is due to
the analytic properties of the generating function $\varphi(y)$ on the
real axis.  We explicitly assume here that the Eq.~(\ref{fb:PhiUni})
is valid, in particular that the position of the first singularity is
at finite distance from the origin when $\sigma$ is finite.  It has
been pointed out in~\cite{Valageas01d} that the equation (\ref{taueq})
admits a second branch for $y_s<y<0$ which cannot ignored in the
computation of the density PDF for finite values of $\sigma$.  In
practice its effect is modest.  It however affects the analytical
properties of $\varphi(y)$ and therefore the shape of the large
density tail, Eq.~(\ref{fb:pdeltatail}).

\begin{figure}
\vspace{8 cm} \special{hscale=80 vscale=80 voffset=0 hoffset=-20
psfile=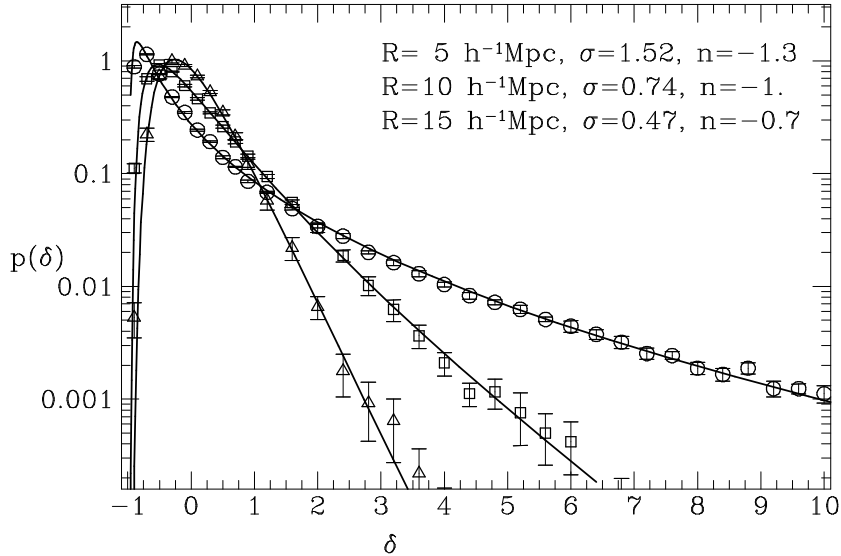}
\caption{ Comparison between predictions  of tree-level  PT with
results of $N$-body simulations in the standard CDM model [predictions
were calculated assuming
Eq.~(\ref{fb:Gdsimp})]. {}From~\cite{Bernardeau94a}.}
\label{fb:PDFnum}
\end{figure}

Numerically it is always possible to integrate
Eq.~(\ref{fb:pdeltanum}) without using the saddle-point
approximation. It is then useful to take advantage of the weak
$\Omega_m$ and $\Omega_{\Lambda}$ dependence of the vertex generating
function. In particular one can use
\begin{equation}
\mGd^L(\tau)=\left(1+{2\,\tau\over 3}\right)^{-3/2}-1,
\label{fb:Gdsimp}
\end{equation}
which is the exact result for the spherical collapse dynamics when
$\Omega_m\to 0$, $\Omega_{\Lambda}=0$. This leads to slight
over-estimation of the low-order vertex [in this case $S_3=5-(n+3)$
for instance] but the power-law behavior at large $\tau$ is correctly
reproduced. For this $\mGd^L$ and for a power-law spectrum $\tau$ can
be explicitly written in term of $\mGd^E$. It is interesting to note
that for $n=0$ there is no singularity, the saddle point approximation
reduces to Eq. (\ref{fb:pdeltasaddle}) and the Eulerian PDF of the
smoothed density field reads,
\begin{eqnarray}
p_{n=0}(\delta)\d\delta&=&
\sqrt{(1+\delta)^{-5/3}+(1+\delta)^{-7/3}}\nonumber\\
&&\times\exp\left[-{9 \left((1+\delta)^{2/3}-1\right)^2\over
8(1+\delta)^{1/3}\sigma^2} \right] {\d\delta\over2\sqrt{\pi}\sigma}.
\end{eqnarray}

One can also obtain the PDF from the SC model using the local
lagrangian mapping \cite{GaCr99,ScGa01}.  The PDF's that are obtained
are in good agreement with results of numerical simulations. In
Fig. \ref{fb:PDFnum}, PT predictions for different smoothing scales
are compared to measurements in a P$^3$M simulation for the standard
CDM model. The predicted shape for the PDF (computed from the measured
variance and known linear spectral index) is in remarkable agreement
with the $N$-body results.

\subsubsection{Dependence on Cosmological Parameters}

\begin{figure}
\centering{\epsfysize=5.5truecm\epsfbox{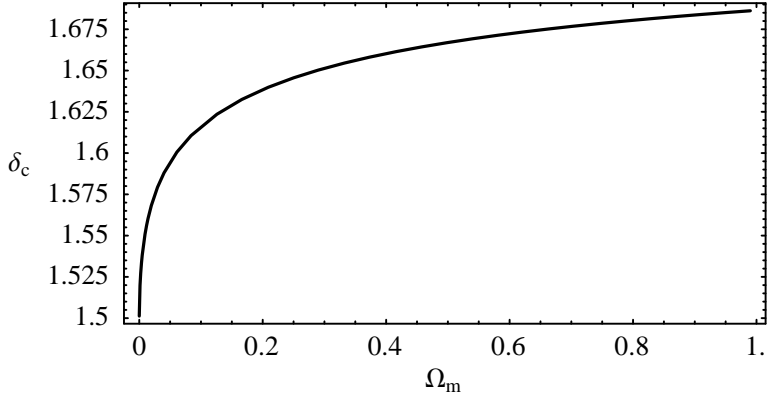}}
\caption{Variation of the position of the critical (linear) value for
the density contrast as a function of $\Omega_m$ for open cosmologies.}
\label{fb:SphCollSing}
\end{figure}

The dependence of the shape of the PDF on cosmological parameters is
entirely contained in the spherical collapse dynamics when the density
field is expressed in terms of the linear density contrast. It can be
examined for instance in terms of the position of the critical density
contrast, $\de_c$.  The variation of $\de_c$ with cosmology is rather
modest as shown in Fig.~\ref{fb:SphCollSing} for $\Omega_\Lambda=0$.
This results applies also to the overall shape of $\mGd$ (see
\cite{Bernardeau94a,Bernardeau94b}), for which the dependence on
cosmological parameters remains extremely weak, at percent level. This
extends what has be found explicitly for the $S_3$ and $S_4$
parameters.

\subsubsection{The PDF in the Zel'dovich Approximation}
\label{pdfza}

For approximate dynamics such as ZA the previous construction can also
be done. It follows exactly the same scheme and the tree-order
cumulant generating function can be obtained through the ZA spherical
collapse dynamics~\cite{MSS94,BeKo95}\footnote{Extension to other
non-linear approximations discussed in Sect.~\ref{sec:nonlinearapp} is
considered as well in~\cite{MSS94}. In addition, recent works have
focussed on the PDF generated by second-order PT~\cite{TaWa00,WaTa01};
however, these neglect the effects of smoothing.}. It is given by
\begin{equation}
\mG_\delta^{\rm ZA}=\left(1-{\tau\over 3}\right)^{-3}.
\end{equation}
One could then compute the Laplace inverse transform of the cumulant
generating function to get the one-point density PDF. As in the
previous case, this result is not exact in the sense that it is based
on the leading order result for the cumulants.

In case of the ZA it is actually possible to do an a priori much more
accurate calculation with a direct approach. Indeed, the local density
contrast neglecting filtering effects is given by the inverse Jacobian
of the deformation tensor, Eq.~(\ref{dlag2}), and the joint PDF of the
eigenvalues can then be explicitly calculated~\cite{Doroshkevich70}
\bea p(\lambda_1,\lambda_2,\lambda_3)&=&\frac{5^{5/2} 27}{8\pi
\sigma^6} (\lambda_3-\lambda_1)
(\lambda_3-\lambda_2)(\lambda_2-\lambda_1) \nonumber \\
&&\hspace{-2cm} \times \exp
\left\{\left[-3(\lambda_1+\lambda_2+\lambda_3)^2 -\frac{15}{2}
(\lambda_1 \lambda_2 + \lambda_1 \lambda_3 + \lambda_2
\lambda_3)\right]/\sigma^2\right\}  \eea
where we have assumed that $\lambda_1<\lambda_2<\lambda_3$. {}From
this it is possible to compute the shape of the one-point density
PDF~\cite{KBGND94,BeKo95},
\begin{eqnarray}
p(\delta) &=& {{9\,5^{3/2}} \over {4\pi\,N_s(1+\delta)^3\,\sigma^4}}
 \int_{3/(1+\delta)^3}^{\infty}\d s \,e^{-{(s-3)^2 / 2
 \sigma^2}}\nonumber\\  &&\times \left( 1+ e^{-{6s/ \sigma^2}}\right)
 \,\left( e^{-{\beta_1^2 / 2\sigma^2}} +e^{-{\beta_2^2 / 2\sigma^2}}
 -e^{-{\beta_3^2 / 2\sigma^2}}  \right)\\  \beta_n (s) &\equiv& s
 \,5^{1/2} \left( {1\over2} +\cos\left[{2\over3}(n-1)\pi +{1\over3}
 \arccos \left({54\over s^3(1+\delta)^3} -1 \right)\right]\right) ,
\end{eqnarray}
where  $N_s$ is the mean number of streams; $N_s=1$ in the single
stream regime. 

The above prediction for the PDF is however of limited value because,
in the absence of smoothing, there is an accumulation of density
values at infinity. This is due to the fact that there is always a
finite probability of forming caustics (where the Jacobian
vanishes). An unfortunate consequence of this is that the moments of
this distribution are always infinite! This does not, however,
contradict the results given in Sect.~\ref{Sect:OneLoopMoments} as
shown in~\cite{BeKo95}: when a cut-off is applied to the large density
tail, the moments remain finite, and behave as expected from the PT
calculations. This has been explicitly verified up to one-loop
order~\cite{ScFr96a}.

\subsection{Two-Dimensional Dynamics}
\label{sec:sec2Ddyn}

The case of gravitational instability in two spatial dimensions (2D)
might be viewed as quite academic. It is however worth investigating
for different reasons: (i) it is a good illustration of the general
method; (ii) numerical simulations in 2D dynamics can be done with a
much larger dynamical range than in 3D; and, perhaps most importantly,
(iii) the 2D results turn out to be of direct use to study statistical
properties of the projected density (Sect.~\ref{sec:projeff}),
relevant for observations of angular clustering and weak gravitational
lensing.

The dynamics we are interested in corresponds actually to density
fluctuations embedded in a 3D space but which are uniform along one
direction. The general equations of motion are left unchanged; here,
we consider again only the Einstein-de Sitter case.

Let us review the different stages of the
calculation~\cite{Bernardeau95}. For the naked vertices, without
smoothing effects, the only change introduced is due to the
$\cos^2(\vk_1,\vk_2)$ factor that in 2D averages to $1/2$ instead of
$1/3$.  The resulting recursion relations between the vertices $\nu_n$
and $\mu_n$ then read,
\begin{equation}
    \nu_{n} = \sum_{m=1}^{n-1} {n \choose m}
    \frac{\mu_{m}}{(2n+3)(n-1)} \left[ (2n+1) \nu_{n-m} + \mu_{n-m}
    \right],
\label{fb:nunrec2D}
\end{equation}
\begin{equation}
    \mu_{n} = \sum_{m=1}^{n-1} {n \choose m}
    \frac{\mu_{m}}{(2n+3)(n-1)} \left[ 3 \nu_{n-m} + n \mu_{n-m}
    \right],
\label{fb:munrec2D}
\end{equation}
instead of Eqs.~(\ref{fb:nunrec}) and (\ref{fb:munrec}). No simple
solution for the generating function of $\nu_n$, $\mGdd(\tau)$, is
known although it again corresponds to the equation describing the
``spherical'' collapse in 2D\footnote{To our knowledge there is no
closed analytical solution for the 2D spherical collapse.}. It can
however be shown that $\mGdd(\tau)-1\sim \tau^{-(\sqrt{13}-1)/2}$ when
$\tau\to\infty$, and the expression
\begin{equation}
\mGdd(\tau)=\left(1+{\tau\over\nu}\right)^{-\nu}-1\ \ \ \hbox{with}\ \
\nu= {\sqrt{13}-1\over 2}
\end{equation}
provides a good fit.  More precisely one can rigorously calculate the
expansion of $\mGd(\tau)$ near $\tau=0$ and it reads
\begin{equation}
\mGdd(\tau)=-\tau+{12\over14}\tau^2-{29\over
42}\tau^3+{79\over147}\tau^4 -{2085\over5096}\tau^5+\dots
\end{equation}
The resulting values for the $S_p^{\rm 2D}$ parameters when smoothing
is neglected are $S^{\rm 2D}_3={36/7}$, $S^{\rm 2D}_4={2540/49}$,
$S^{\rm 2D}_5=793$, $S^{\rm 2D}_6=13370$. When filtering is taken into
account the vertex generating function becomes\footnote{In 2D dynamics
if $P(k)\sim k^n$ then $\sigma(R)\propto R^{-(n+2)}$.},
\begin{equation}
\mGd^E(\tau)=\mGdd\left(\tau \left[1+\mGd^E(\tau)\right]^{-2-n}\right),
\end{equation}
for a power-law spectrum of index $n$. This leads
to~\cite{Bernardeau95}
\begin{eqnarray}
S^{\rm 2D}_3&=&{{36}\over 7} - {{3\,(n+2)}\over 2};\\ S^{\rm
2D}_4&=&{{2540}\over {49}}-33\,(n+2)+{{21\,{(n+2)^2}}\over4};\\ S^{\rm
2D}_5&=&{{271960}\over {343}} - {{38900\,\left( n+2 \right) }\over
{49}} +  {{3705\,{{\left( n+2 \right) }^2}}\over {14}} -
{{235\,{{\left( n+2 \right) }^3}}\over 8};\\ S^{\rm
2D}_6&=&{{510882660}\over {31213}} - {{7721415\,\left( n+2 \right)
}\over {343}} +  {{2272395\,{{\left( n+2 \right) }^2}}\over {196}}
\nonumber\\ && -{{74205\,{{\left( n+2 \right) }^3}}\over {28}} +
{{1815\,{{\left( n+2 \right) }^4}}\over 8}.
\end{eqnarray}
Obviously, these results can also be obtained from a direct
perturbative calculation using the geometrical properties of the 2D
top-hat window function given in Appendix~\ref{tophatgeom}. The
position and shape of the singularity is also changed in 2D dynamics.
In Table~\ref{fb:2dphising} we give the parameters of the singularity
in $\varphi(y)$.

\begin{table}
\caption{Parameters of the singularity, Eq.~(\ref{phising}), for the
2D case. There is no singularity for $n \geq 0$.}
\label{tab2}
\vspace{.3cm} \centerline{
\begin{tabular}{@{}ccccc}
\hline $n$ &$y_s$   &$\varphi_s$&$r_s$& $a_s$\\ \hline -2 & -0.172&
-0.197& 1.60& -1.72\\ -1.5& -0.212& -0.252& 1.81& -2.25\\ -1 & -0.277&
-0.350& 2.23& -3.41\\ -0.5& -0.403& -0.581& 3.55& -7.73\\ \hline
\end{tabular}}
\label{fb:2dphising}
\end{table}

\subsection{The Velocity Divergence PDF}
\label{fb:TPDF}

So far our description has been focussed on the density field. The
structure of the equations for the velocity divergence is the same as
for the local density. We briefly account here for the results that
have been obtained at tree level for the velocity
divergence~\cite{Bernardeau94a}. Loop corrections with exact PT are
discussed in e.g.~\cite{ScFr96a}. Note that the SC model approximation
described in Sect.~\ref{sec:SCmodel} does not do as well as for the
density contrast, due to tidal contributions\footnote{Velocities are
more affected by previrialization effects, as shown in
Fig.~\protect\ref{alphafn}.}, but can provide again approximate loop
corrections for the cumulants while still giving exact tree-level
results~\cite{FoGa98b}.

\subsubsection{The Velocity Divergence Cumulants Hierarchy}

In what follows, we assume that the velocity divergence is expressed
in units of the conformal expansion rate, ${\cal H}=a H$.  For
convenience, we define the vertex generating function for the velocity
divergence as
\begin{equation}
   \mGv(\tau)\equiv -f(\Omega_m,\Omega_\Lambda) \sum_{p\geq 1} \mu_p
   \frac{(-\tau)^p}{p!} \equiv \sum_{p\geq 1} \breve{\mu}_p
   \frac{(-\tau)^p}{p!}.
\end{equation}
This definition corresponds to slightly different vertices from those
given by Eq.~(\ref{mu_n}), 
\begin{equation}
    \breve{\mu}_p \equiv \langle \theta^{(n)} [ \delta^{(1)} ]^p
\rangle_c / \langle [ \delta^{(1)} ]^2 \rangle^p.
\end{equation}
When the filtering effect is not taken into account the vertex
generating function can be obtained from the one of the density
field. {}From the continuity equation we have~\cite{Bernardeau92b},
\begin{eqnarray}
&&\mG_{\theta}(a,\tau)= \nonumber\\
&&\ \ \ -{\left[a{\d \over \d a}\mG_{\dta}(a,\tau)+
f(\Omega_m,\Omega_{\Lambda})
\tau{\d\over \d \tau}\mG_{\dta}(a,\tau)\right]}
\left[ 1+\mG_{\dta}(a,\tau)\right]^{-1}.
\label{fb:Gv}
\end{eqnarray}
One can use the fact that function $\mG_{\delta}(a,\tau)$ is nearly
insensitive to the values of $\Omega_m$ and $\Omega_\Lambda$ to obtain
a simplified form for the function $\mGv(a, \tau)$, 
\begin{equation}
\mGv(\tau)\approx-f(\Omega_m,\Omega_{\Lambda})
\tau{\d\over \d \tau}\mG_{\dta}^L(\tau)/
\left[1+\mG_{\dta}^L(\tau)\right], 
\label{fb:GvOdep}
\end{equation}
so that $\mGv(\tau)\approx{f(\Omega_m,\Omega_{\Lambda})
\tau\left(1+{2\tau/3}\right)^{-1}}$ if approximation in Eq.~(\ref{fb:Gdsimp})
is used.  This in fact fully justifies the definition of the vertices
$\mu_p$ which are seen to be almost independent of the cosmological
parameters, as already discussed in Sect.~\ref{sec:codeno}.

{}From now on, we use again for clarity the Lagrangian and Eulerian
superscripts, in particular $\mGv^L \equiv \mGv$, $\mG_{\dta}^L
\equiv \mG_{\dta}$.  Including filtering effects 
requires taking into account the mapping from Lagrangian to Eulerian
space, as explained in Sect.~\ref{sec:lagtoeul}. As a consequence of
this we have
\begin{equation}
\mGv^E(\tau)=
\mGv^L\left[\tau\ 
{\sigma\left([1+\mG_{\dta}^{E}(\tau)]^{1/3}R\right)
\over\sigma(R)}\right],
\label{eq:mapppp}
\end{equation}
which amounts to say that the velocity divergence should be calculated
at the correct mass scale. This remapping does not further complicate
the dependence on cosmological parameters:
$\mGv^E(\tau)/f(\Omega_m,\Omega_{\Lambda})$ remains independent of
$(\Omega_m,\Omega_\Lambda)$ to a very good accuracy.

It is possible to derive the cumulants $T_p$ from the implicit
Eq.~(\ref{eq:mapppp}), relying on the usual relations given in
Sect.~\ref{sec:snnun} between the the cumulants and what would be the
genuine intrinsic velocity divergence vertices, $ \mu_p^{{\rm intr}}
\equiv \langle \theta^{(n)} [ \theta^{(1)} ]^p
\rangle_{c,E} / \langle [ \theta^{(1)} ]^2 \rangle^p_E$
that are straightforwardly related to $\breve{\mu}_p^{E}$  through $
\mu_p^{{\rm intr}}= \breve{\mu}_p^{E}
[-f(\Omega_m,\Omega_\Lambda) ]^{-p}$. The corresponding vertex
generating function, $\mGv^{{\rm intr}}(\tau)$, is given by
$\mGv^{{\rm intr}}(\tau)= \mGv^E[-f(\Omega_m,\Omega_\Lambda)\tau]$
together with Eqs.~(\ref{fb:phiy}), (\ref{fb:phieq}) and
(\ref{taueq}), and replacing $S_p$ with $T_p$ and $\mG_\delta$ with
$\mGv^{{\rm intr}}$, can be used to compute the velocity divergence
cumulant parameters. For an Einstein-de Sitter universe, the first two
read 
\begin{eqnarray}
T_3(\Omega_m=1,\Omega_\Lambda=0)&=&-\left({{26}\over 7} + {\gm_1}\right),\\
T_4(\Omega_m=1,\Omega_\Lambda=0)&=&{{12088}\over {441}} + {{338\,{\gm_1}}\over {21}} + 
  {{7\,{{{\gm_1}}^2}}\over 3} + {{2\,{\gm_2}}\over 3},\\
\dots, \nonumber
\end{eqnarray}
where the parameters $\gamma_p$ are given by Eq.~(\ref{fb:gammap}).
Furthermore, the dependence on cosmological parameters is
straightforwardly given by\footnote{To be compared, for example, to the
more accurate result given for $T_3$ in Eq.~(\ref{eq:t3accu}).}
\begin{equation}
T_p(\Omega_m,\Omega_{\Lambda})
\approx{1\over f(\Omega_m,\Omega_{\Lambda})^{(p-2)}}
T_p(\Omega_m=1,\Omega_{\Lambda}=0),
\label{fb:TnOdep}
\end{equation}
which implies a relatively strong $\Omega_m$ dependence for the shape
of $p(\theta)$ as we now discuss. 

\subsubsection{The Shape of the PDF}

The above line of arguments provides a general rule for the dependence
of the PDF on cosmological parameters:
\begin{equation}
p\left[ f(\Omega_m,\Omega_{\Lambda}),\theta,\sigma_{\theta} \right]
\d\theta\approx  
\ \ p\left[1,{\theta\over f(\Omega_m,\Omega_{\Lambda}) },
{\sigma_{\theta}\over
f(\Omega_m,\Omega_{\Lambda})} \right]{\d\theta\over 
f(\Omega_m,\Omega_{\Lambda}) }.
\end{equation}
Otherwise, the PDF can be calculated exactly the same way as for the
density contrast.

The case $n=-1$ is worth further investigations since it is then
possible to derive a closed form that fits extremely well the exact
numerical integration, similarly as for the PDF of $\delta$ for
$n=0$. This approximation is based on the approximate form in
Eq.~(\ref{fb:Gdsimp}) for the function $\mG^L_{\dta}$. With $n=-1$ it
leads to
\begin{equation}
\mG^{E,n=-1}_{\delta}(\tau)=\left[-{\tau\over3}+\left(1+{\tau^2\over9}
\right)\right]^3-1.
\end{equation}
One can then show that
\begin{equation}
\mG^{E,n=-1}_{\theta}(\tau)=f(\Omega_m,\Omega_{\Lambda})\left[ \tau
          \left(1+{\tau^2 \over 9}\right)^{1/2}-{\tau^2 \over 3} \right].
\end{equation}
The calculation of the PDF of the velocity divergence from the
saddle-point approximation [e.g. Eq.~(\ref{fb:pdeltasaddle})] then
leads to the expression,
\begin{equation}
p(\theta)\d\theta={([2\kappa-1]/\kappa^{1/2}+[\lambda-1]/\lambda^{1/2})^{-3/2} 
\over \kappa^{3/4} (2\pi)^{1/2} \sigma_{\theta}}
\exp\left[-{\theta^2\over 2\lambda\sigma_{\theta}^2}\right]\d\theta,
\label{fb:thetaPDF}
\end{equation}
with 
\begin{equation}
\kappa=1+{\theta^2\over 9\lambda f(\Omega_m,\Omega_\Lambda)^2},
\ \ \ \lambda=1-{2\theta\over 3 f(\Omega_m,\Omega_\Lambda)},
\end{equation}
where $\theta$ is expressed in units of the conformal expansion rate,
${\cal H}$.

\begin{figure}[t]
\begin{tabular}{cc}
{\epsfxsize=6.5truecm\epsfysize=6.5truecm\epsfbox{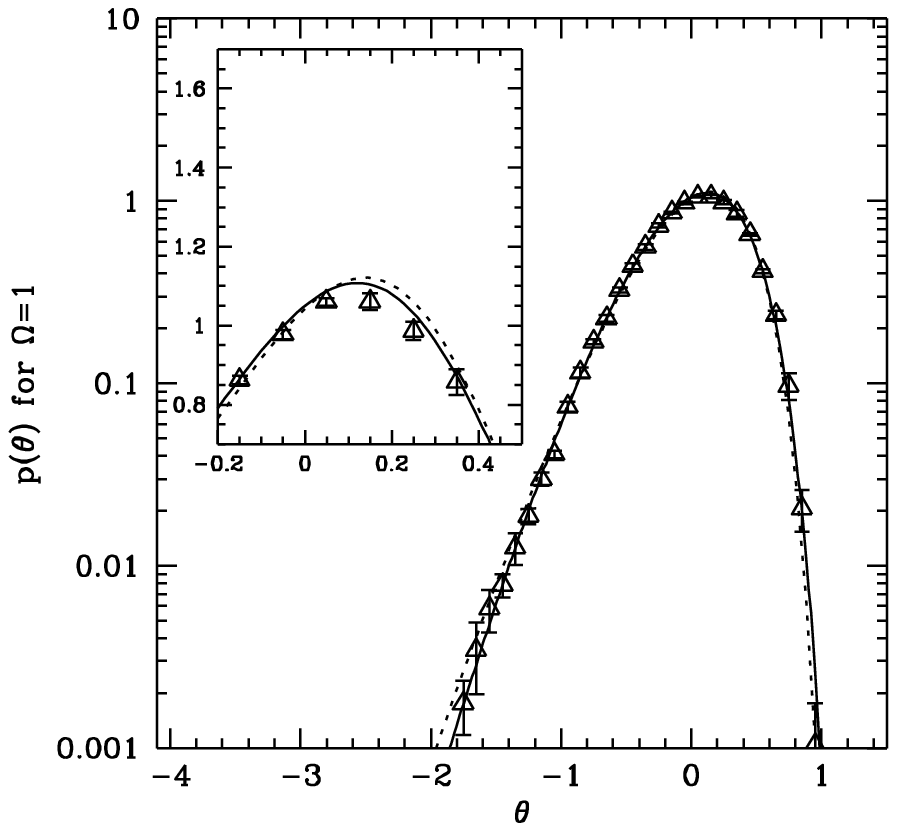}}&
{\epsfxsize=6.5truecm\epsfysize=6.5truecm\epsfbox{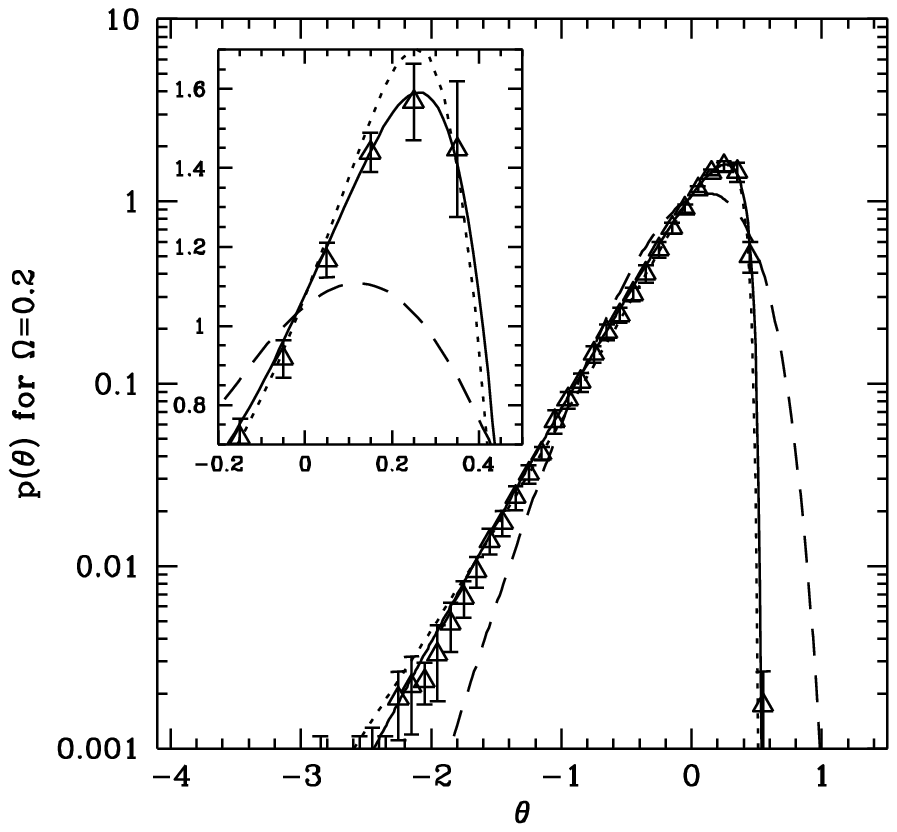}}
\end{tabular}
\caption{The PDF  of the velocity divergence for two different
values of $\Omega_m$ ($\Omega_m=1$, left panel and $\Omega_m=0.2$,
right panel).  The dotted lines correspond to the approximate analytic
fit [Eq.~(\ref{fb:thetaPDF})] and the solid lines to the theoretical
predictions obtained from a direct numerical integration of the
inverse Laplace transform with $n=-0.7$. In right panel the dashed
line is the prediction for $\Omega_m=1$ and the same $\sigma_\theta \approx
0.4$.  {}From~\cite{BVDWHB97}.}
\label{fb:div_pdf}
\end{figure}

\subsubsection{Comparison with $N$-Body Simulations}

Measurements in numerical simulations turn out to be much more
non-trivial for the velocity field than for the density field. The
reason is that in $N$-body simulations, the density field is traced by
a Poisson realization. Although it suffices to count points, in grid
cells for instance, to get the filtered density\footnote{Corrected
for discreteness effects using factorial moments as discussed in
Sect.~\ref{sec:sec5}.}, the velocity field is only known in a
non-uniform way where particles happen to be. Therefore, simple
averages of velocities do not lead to good estimations of the
statistical properties one is interested in, especially when the
number density of particles is small.

For this purpose specific methods have been developed to deal with
velocity field statistics~\cite{BeVDW96}. The idea is to use
tessellations to obtain a continuous description of the velocity field;
two alternative prescriptions have been proposed. One makes use of the
Voronoi tessellation; in this case the velocity is assumed to be
uniform within each Voronoi cell, in other words, the local velocity
at any space point is the one of the closest particle.  The second
method makes use of the Delaunay tessellation. In this case the local
velocity is assumed to vary linearly within each Delaunay tetrahedron
(such ensemble of tetrahedra forms a unique partition of space); the
local velocity is then defined by a linear combination of its closest
neighbors, see~\cite{BeVDW96,BVDWHB97} for details.

These methods have been applied to results of numerical
simulations~\cite{BVDWHB97,KCPR00}. Comparisons between theoretical
predictions, in particular the form (\ref{fb:thetaPDF}), and the
measurements are shown in Fig.~\ref{fb:div_pdf}.  The simulation used
here is a PM simulation with a scale-free spectrum with $n=-1$. The
prediction, Eq.~(\ref{fb:thetaPDF}), gives a good account of the shape
of the divergence PDF, especially in the tails. The detailed behavior
of the PDF near its maximum requires a more exact computation. We
obtained it here by an exact inverse Laplace computation using
Eq.~(\ref{fb:Gdsimp}) for the density vertex generating function [and
Eq.~(\ref{fb:GvOdep})] to get the velocity vertices. Because this
expression does not accurately predict the low-order
cumulants\footnote{For example, $T_3=4-(n+3)$ instead of
$T_3=26/7-(n+3)$.} the integration has been made with $n=-0.7$,instead
of $n=-1$, to compensate for this problem. The agreement with
simulations is quite remarkable.

\subsection{The Velocity-Density Relation}
\label{fb:VDrel}

\begin{figure}
\centering{\epsfxsize=9.5truecm\epsfbox{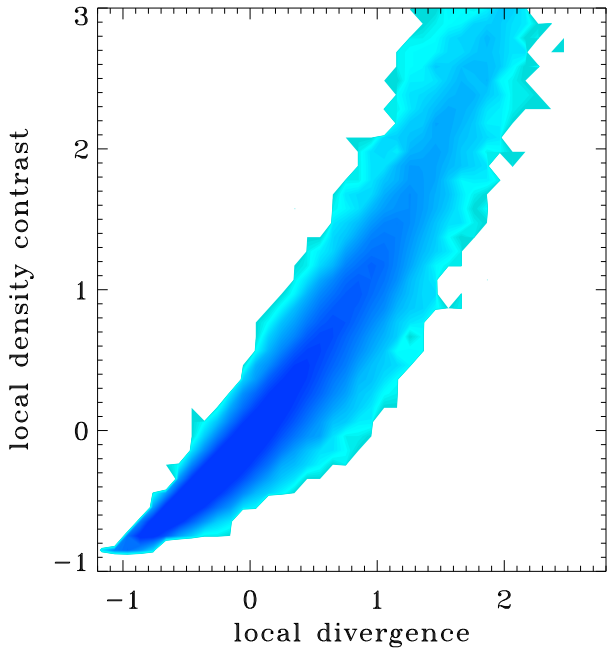}}
\caption{Example of a joint PDF of the density and the velocity divergence. 
The color is in logarithmic scale, the smoothing scale is 15 Mpc/h,
the spectrum is scale-free with $n=-1.5$, and $\sigma_8\equiv 1$,
see~\cite{BCLSK99} for details.}
\label{DDPDF}
\end{figure}

PT also allows one to consider multivariate PDF's such as the joint
distribution of the local density contrast and the local divergence
$\theta$.  An example of such PDF is shown in Fig.~\ref{DDPDF}. It
illustrates in particular the fact that the local density and local
divergence do not follow in general a one to one correspondence, as it
would be the case in linear perturbation theory. Deviations from this
regime induce not only a nonlinear relation between $\delta$ and
$\theta$, i.e.~a bending in the $\delta$-$\theta$ relation, but also a
significant scatter.

In general the statistical properties of these two fields can be studied
through their joint cumulants, $\langle
\delta^p\theta^q\rangle_c$. Similarly to cases involving only one variable
it is possible to compute such quantities at leading order, or at next
to leading order (involving loop corrections) in PT. One can define
the parameters $U_{p\,q}$ as,
\begin{equation}
\langle \delta^p\theta^q\rangle_c=U_{p\,q}
\langle \delta^2\rangle^{p+q-1},
\label{fb:Cpq}
\end{equation}
where $\theta$ is expressed in units of the conformal expansion rate,
${\cal H}$.  The $U_{p\,q}$'s are finite (and non-zero) at large
scales for Gaussian initial conditions and can be easily computed at
tree order.  Their calculation follows a tree construction from the  
vertices $\nu_p$ and $\mu_q$. For instance, one obtains
\begin{eqnarray}
U_{1\,1}&=&\nu_1\breve{\mu}_1=\breve{\mu}_1=-f(\Omega_m,\Omega_{\Lambda}), \nonumber\\
U_{2\,1}&=&2\nu_2\breve{\mu}_1+\breve{\mu}_2, \nonumber\\
U_{3\,1}&=&3\nu_3\breve{\mu}_1+\breve{\mu}_3+6\nu_2^2\breve{\mu}_1+6\nu_2\breve{\mu}_2,\nonumber\\
U_{2\,2}&=&2\nu_3\breve{\mu}_1^2+2\breve{\mu}_3\breve{\mu}_1
       +8\nu_2\breve{\mu}_2\breve{\mu}_1+2\nu_2^2\breve{\mu}_1^2+2\breve{\mu}_2^2. \nonumber
\end{eqnarray}
with $\breve{\mu}_p \equiv -f(\Omega_m,\Omega_{\Lambda})\mu_p$.

These expressions are straightforward when the smoothing effects are
not taken into account. They are still true otherwise, but they rely
on the fact that the same mapping applies to the density and the
velocity divergence.  More generally it is possible to derive
explicitly the generating function of the joint cumulants. The
demonstration is presented in Appendix.~\ref{LegTrans2}.

An interesting application of these results is the computation of the
joint density-velocity PDF. Assuming that the leading order
contributions to cumulants provide a reliable description, we have
\begin{eqnarray}
p(\delta,\theta)\d\delta\d\theta&=&     \int_{-\ii\infty}^{+\ii\infty}{\d
y_1\over  2\pi\ii}  \int_{-\ii\infty}^{+\ii\infty}{\d y_2\over  2\pi\ii}
\exp\left[{\delta\,y_1\over   \sigma^2}+{\theta\,y_2\over   \sigma^2}-
{\varphi(y_1,y_2)\over \sigma^2}\right],\\
\varphi(y_1,y_2)&=&y_1\mGd(\tau)+y_2{\mG}_{\theta}(\tau)-{1\over
2}y_1\tau{\d\over\d\tau}\mGd(\tau)-{1\over
2}y_2\tau{\d\over\d\tau}{\mG}_{\theta}(\tau),\nonumber\\
\tau&=&-y_1{\d\over\d\tau}\mGd(\tau)-y_2{\d\over\d\tau} {\mG}_\theta(\tau),\nonumber
\end{eqnarray}
where $\sigma^2$ is the variance of the {\it density} field.

As a consequence of this relation one can compute {\em constrained}
averages such as the expectation value of $\theta$ under the
constraint that the local density is known,
$\langle\theta\rangle_{\delta}$.  For a vanishing variance (that is,
at tree level) the result turns out to be extremely simple and
reads~\cite{Bernardeau92a},
\begin{equation}
\langle\theta\rangle_{\delta}=
\mGv(\tau),\ \ \ {\rm with}\ \ \mGd(\tau)=\delta.
\end{equation}
This relation can obviously be inverted to get
$\langle\delta\rangle_{\theta}$.  It is interesting to note that this
result is not quantitatively changed when top-hat smoothing effects
are taken into account (nor it depends on the shape of the power
spectrum), which is not true anymore with Gaussian
smoothing~\cite{ChLo97}.

A more pedestrian approach should be used when the variance is not
negligible: 
\begin{eqnarray}
\langle\delta\rangle_{\theta}&=&
a_0+a_1\,\theta+a_2\,\theta^2+a_3\,\theta^3+\dots\\
\langle\theta\rangle_{\delta}&=&
r_0+r_1\,\delta+r_2\,\delta^2+r_3\,\delta^3+\dots
\end{eqnarray}
Computations should be made order by order and it becomes inevitable
to introduce next-to-leading order corrections, i.e.~loop corrections.

\begin{table}
\caption{
The coefficients $a_1,\dots, a_3$ and $r_1,\dots, r_3$ as functions of
the spectral index $n$ for scale-free power spectra and Gaussian
smoothing. Results are given at leading order, except for $a_1$ and
$r_1$ for which one-loop corrections are included when available
(correction is infinite for $n\ge -1$).}
\label{fb:aprptable}
\vspace{.3cm}
\centering{
\begin{tabular}{ccllcll} \hline 
index $n$ &$a_1$&\ \ \ $a_{2}$ &\ \ \ $a_{3}$ &$r_1$&$r_{2}$&$r_{3}$\\
\hline
-3.0 & --        & 0.190&-0.0101 &1+0.3$\,\sigma^2 $&-0.190&0.0826\\
-2.5 & --        &0.192 &-0.00935 &1+0.202$\,\sigma^2$&-0.192&0.0822\\
-2.0 & 1-0.172$\,\sigma_\theta^2$&0.196&-0.00548 &1+0.077$\,\sigma^2$&-0.196&0.0821\\
-1.5 & 1+0.187$\,\sigma_\theta^2$&0.203&-0.000127&1-0.296$\,\sigma^2$&-0.203&0.0822\\
-1.0 & $1+[\infty]$      &0.213 & 0.00713  &$1+[\infty]$     &-0.213&0.0835\\
-0.5 & $1+[\infty]$      &0.227 & 0.0165  &$1+[\infty]$     &-0.227&0.0865\\
 0  & $1+[\infty]$      &0.246 & 0.0279  &$1+[\infty]$     &-0.246&0.0928\\
 0.5 & $1+[\infty]$      &0.270 & 0.0408  &$1+[\infty]$     &-0.270&0.1051\\
 1.0 & $1+[\infty]$      &0.301 & 0.0532  &$1+[\infty]$     &-0.301&0.1283\\
\hline
\end{tabular}}
\end{table}

The coefficients $a_0,\dots, a_3$ and $r_0,\dots, r_3$ have been
computed explicitly up to third-order in
PT~\cite{ChLo97,CLPN98,BCLSK99}.  It is to be noted that at leading
order one has $a_0=-a_2\,\sigma_\theta^2$ and $r_0=-r_2\,\sigma^2$ to
ensure that the global ensemble average of $\theta$ and $\delta$
vanish.  Note also that the third-order PT results for $a_1$ and $r_1$
involve a loop correction that diverges for $n\ge -1$.  The known
results are given in Table~\ref{fb:aprptable} for the Einstein-de
Sitter case and Gaussian smoothing.

The $\Omega_m$ dependence of these coefficients can be explicitly
derived.  For instance, the coefficient $r_2$ can be expressed in
terms of the skewness of the two fields (at leading order only), which
leads to
$r_2={f(\Omega_m,\Omega_{\Lambda})}(S_3+f(\Omega_m,\Omega_{\Lambda})T_3)/6$.
For a top-hat filter, $r_2$ is always given by
$f(\Omega_m,\Omega_{\Lambda})\,4/21$ and, for a Gaussian window it
varies slightly with the power spectrum index but shows a similarly
strong $f(\Omega_m,\Omega_{\Lambda})$ (and therefore $\Omega_m$)
dependence. Comparisons with numerical simulations have demonstrated
the accuracy and robustness of these predictions (except for the loop
terms)~\cite{BCLSK99,KCPR00}.
 
Such results are of obvious observational interest, since one can in
principle measure the value of $\Omega_m$ from velocity-density
comparisons, see~\cite{Dekel94}.  In particular a detailed analysis of
the curvature in the $\delta-\theta$ relation (through $a_2$ or $r_2$)
would provide a way to break the degeneracy between biasing parameters
(Sect.~\ref{sec:bias}) and
$\Omega_m$~\cite{Chodorowski99,BCLSK99}\footnote{The scatter in this
relation seen in Fig.~\ref{DDPDF} can be reduced by including also
off-diagonal components of the velocity deformation
tensor~\cite{Gramann93,MaYa95,Chodorowski97}.}.  Moreover, these
results can be extended to take into account redshift distortion
effects (Sect.~\ref{sec:reddis}) as described in~\cite{Chodorowski00}. 
The main practical issue is that current velocity surveys are not
sufficiently large to provide accurate density-velocity comparisons
going beyond linear PT.

It is finally worth noting that these investigations are also useful for
detailed analysis of the Lyman-$\alpha$ forest~\cite{NuHa99}.

\subsection{The Two-Point Density PDF}
\label{sec:jpdf}

Perturbation theory can obviously be applied to any combination of the
density taken at different locations. In particular, for sound cosmic
error computations (see Chapter~\ref{sec:chapter7}) the bivariate
density distribution is an important quantity that has been
investigated in some detail.

The object of this sub-section is to present the exact results that
have been obtained at tree-level for the two-point density
cumulants~\cite{Bernardeau96}. We consider the joint densities at
positions $\vx_1$ and $\vx_2$ and we are interested in computing the
cumulants $\mg\delta^p(\vx_1)\delta^q(\vx_2)\md_c$ where the field is
supposed to be filtered at a given scale $R$. In general such
cumulants are expected to have quite complicated expressions,
depending on both the smoothing length $R$ and the distance
$\vert\vx_1-\vx_2\vert$. We make here the approximation that the
distance between the two points is large compared to the smoothing
scale. In other words, we neglect  short-distance effects.

Let us define the parameters $C_{p\,q}$ with,
\be
C_{p\,q}={\mg\delta^p(\vx_1)\delta^q(\vx_2)\md_c\over
\mg\delta(\vx_1)\delta(\vx_2)\md\ \mg\delta^2\md^{p+q-2}}.
\label{fb:Spqdef}
\ee
Because of the tree structure of the correlation hierarchy, we expect
the coefficients $C_{p\,q}$ to be finite in both the large distance
limit and at leading order in the variance.  This expresses the fact
that among all the diagrams that connect the two cells, the ones that
involve only one line between the cells are expected to be dominant in
cases when $\mg\delta(\vx_1)
\delta(\vx_2)\md\ll\mg\delta^2\md$. 

\begin{figure}
\centering{
\epsfysize=2.6truecm\epsfbox{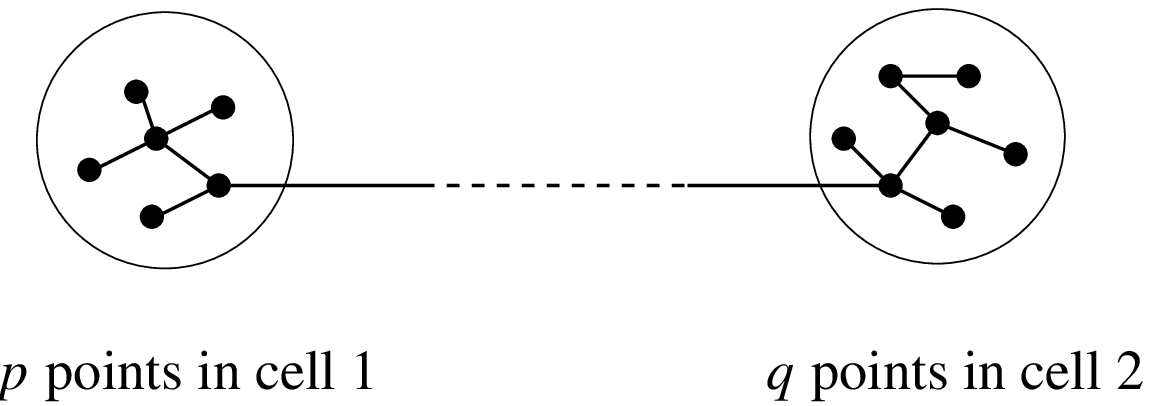}}
\caption{Structure  of the coefficient  $C_{p\,q}$ in  large separation
limit: $C_{p\,q}$ is given by the sum of all possible trees
joining $p$ points in first cell to $q$ points in the second with
only one crossing line. The  sums can  be done separately  on
each side  leading to $C_{p\,q}=C_{p\,1}\,C_{q\,1}$.}
\label{SpqStruct}
\end{figure}

The next remarkable property is directly due to the tree structure of
the high-order correlation functions. The coefficients $C_{p\,q}$ are
dimensionless quantities, that correspond to some geometrical averages
of trees. It is quite easy to realize (see Fig. \ref{SpqStruct}) that
such averages can be factorized into two parts, corresponding to the
end points of the line joining the two cells. In other words one
should have,
\begin{equation}
C_{p\,q}=C_{p\,1}C_{q\,1}.
\label{eq:cpqfac}
\end{equation}

This factorization property is specific to tree structures.  It was
encountered originally in previous work in the fully non-linear
regime~\cite{BeSc91}. It has specific consequences on the behavior of
the two-point density PDF, namely we expect that,
\begin{eqnarray}
p\left[\rho(\vx_1),\rho(\vx_2)\right]&=&p\left[\rho(\vx_1)\right]\,
p\left[\rho(\vx_2)\right] 
\left(1+b[\rho(\vx_1)]\mg\delta(\vx_1)\delta(\vx_2)\md
b[\rho(\vx_2)]\right). \nonumber\\
\end{eqnarray}
The joint density PDF is thus entirely determined by the shape of the
``bias'' function, $b(\rho)$\footnote{The interpretation of this
function as a bias function is discussed in Sect.~\ref{hierbias}.}.

The general computation of the $C_{p\,1}$ series is not
straightforward, although the tree structure of the cumulants is
indicative of a solution. Indeed the generating function $\psi(y)$ of
$C_{p\,1}$,
\begin{equation}
\psi(y)=\sum_{p=1}^{\infty}C_{p\,1}{y^n\over p!},
\end{equation}
corresponds to the generating function of the diagrams with one
external line. For exact trees this would be $\tau(y)$. However, the
Lagrangian to Eulerian mapping affects the relation between
$\varphi(y)$ and $\tau(y)$ and this should be taken into account. We
give here the final expression of $\psi(y)$, derived in detail
in~\cite{Bernardeau96},
\begin{equation}
\psi(y)=\tau(y) {\sigma(R)\over \sigma(R[1+\mG^E_{\delta}]^{1/3})}
\end{equation}
where $\tau(y)$ is solution of the implicit Eq.~(\ref{taueq}).  A
formal expansion of $\psi(y)$ with respect to $y$ gives the explicit
form of the first few coefficients $C_{p\,1}$. They can be expressed
in terms of the successive logarithmic derivatives of the variance,
$\gamma_i$ [Eq.~(\ref{fb:gammap})],
\begin{eqnarray}
C_{2\,1}&=&{68\over 21}+{\gamma_1\over 3},\\
C_{3\,1}&=&{11710\over 441}+{61\over 7}\gamma_1+{2\over 3}\gamma_1^2+
{\gamma_2\over 3},\\
C_{4\,1}&=&\frac{107906224}{305613} + \frac{90452\,\gamma_1}{441} + 
 {\frac{116\,{{\gamma_1}^2}}{3}} + {\frac{7\,{{\gamma_1}^3}}{3}} + 
 {\frac{758\,\gamma_2}{63}}\nonumber\\
&&+{\frac{20\,\gamma_1\,\gamma_2}{9}} + {\frac{2\,\gamma_3}{9}}.
\end{eqnarray}
These numbers provide a set of correlators that describe the joint
density distribution in the weakly nonlinear regime.  They generalize
the result found initially in~\cite{Fry84a} for $C_{2\,1}$.  Numerical
investigations (e.g.~\cite{Bernardeau96}) have shown that the large
separation approximation is very accurate even when the cells are
quite close to each other.

For a comparison of the above results with N-body simulations and the
spherical collapse model see ~\cite{GFC01}.

\subsection{Extended Perturbation Theories}
\label{sec:EPT}

\begin{figure} 
\begin{tabular}{cc}
\epsfysize=6.6truecm \epsfxsize=6.6truecm\epsfbox{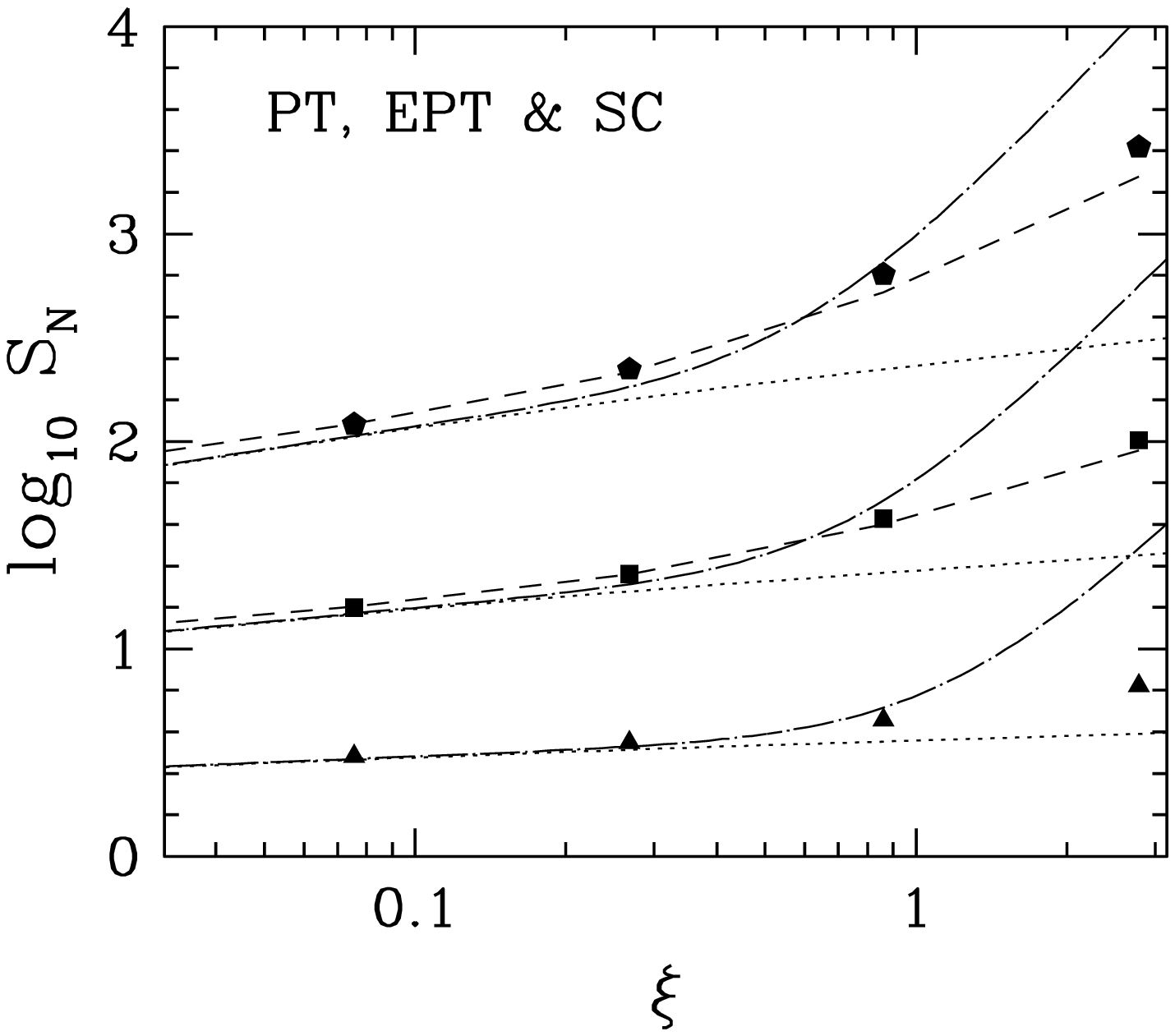}&
\epsfysize=6.6truecm \epsfxsize=6.6truecm\epsfbox{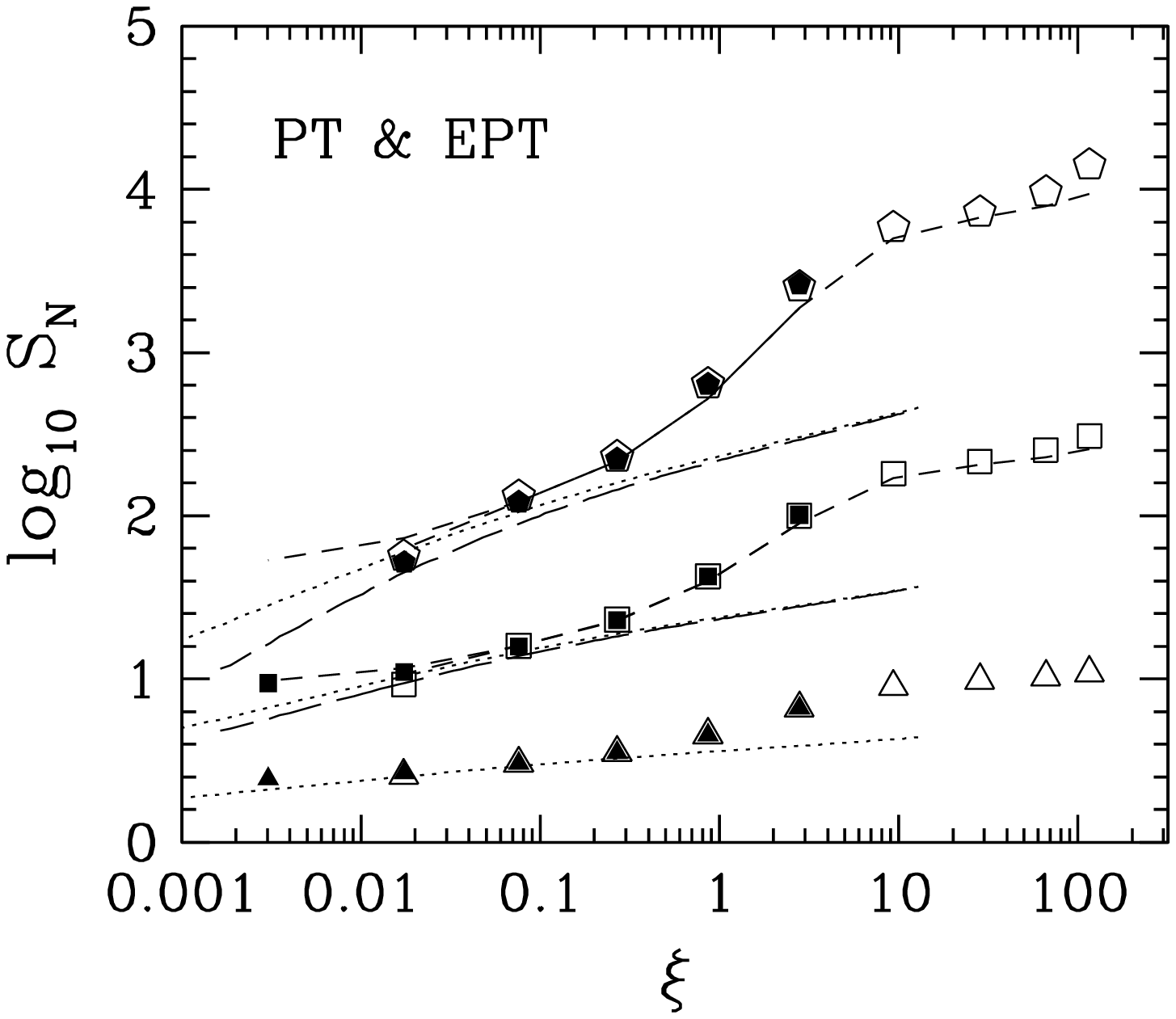}
\end{tabular}
\caption{The cumulants $S_p$ in the $\tau$CDM model as
functions of $\xib\equiv \sigma^2$, for $p=3,4$ and $5$ (with
respectively triangles, squares and pentagons) compared to tree order
PT predictions assuming a local power spectrum (dots), taking into
account spectral index variation, i.e. corrections $\gamma_p$, $p > 2$ in
Eqs.~(\ref{eq:s3ptpt}-\ref{eq:s6ptpt}) (long dashes on right panel),
EPT where $n_{\rm eff}$ is inferred from the measured $S_3$ (short
dashes) and one loop perturbation theory predictions based on the
spherical model (dots-long dashes on left panel).  {}From
\cite{CSJC99}.}
\label{sc_ept}
\end{figure}

The range of validity of perturbation theory results suggests that
they provide, on a sole phenomenological basis, a robust model for
describing the correlation hierarchy in all regimes.  In the Extended
Perturbation Theory (EPT) ansatz, the $S_p$'s are assumed to be given
by Eqs.~(\ref{eq:s3ptpt}-\ref{eq:s6ptpt}) with $\gamma_1\equiv-(n+3)$
and $\gamma_i=0,\ i\ge2$, where $n=n_p(\sigma)$ is an {\em adjustable}
parameter inferred from the {\em measured} value of $S_p$ as a
function of the measured variance $\sigma^2$:
\begin{equation}
  S_p[n=n_p(\sigma)] \equiv S_p^{\rm measured}(\sigma).
\end{equation}
As observed in~\cite{CBBH97}, for scale-free initial conditions, the 
function $n_p(\sigma)$ does not depend on cumulant order $p$ to a very
good approximation:
\begin{equation}
  n_p(\sigma) \simeq n_{\rm eff}(\sigma)
  \label{eq:neff}
\end{equation}
in any regime, from very small\footnote{Of course, in this regime
$n_{\rm eff}=n$, where $n$ is the linear spectral index.}  value of
$\sigma$ to a very large value of $\sigma$.  A simple form has been
proposed to account for these results~\cite{CBBH97},
\begin{eqnarray}
n_{\rm eff}&=&n+(n_{\rm nonlinear}-n)\,
{x^{\tau}\over x^{\tau}+x^{-\tau}}\label{fitneff}\\
x&=&\exp[\log_{10}(\sigma^2/\sigma_0^2)]\nonumber.
\end{eqnarray}
where $n_{\rm eff}$ is varying from the value of the initial power
spectrum index, $n$, to a value corresponding to the stable clustering
regime, $n_{\rm nonlinear}$. The location and the width of the
transition between these two regimes depend on the initial power
spectrum index and are described respectively by $\sigma_0$ and
$\tau$. Values of the parameters involved in Eq.~(\ref{fitneff}) are
listed in Table~\ref{tableneff} for $n$ ranging from $-2$ to
$1$. These values can be approximately obtained by the following
fitting formulae valid for $n \la -1$
\begin{eqnarray}
 n_{\rm nonlinear}(n)& \simeq &3 \frac{(n-1)}{(3+n)}, \label{eq:nnonlinear} \\
 \tau(n) & \simeq & 0.8 - 0.3\ n, \\
 \log_{10} \sigma^2_0(n) & \simeq & 0.2-0.1\ n. \label{eq:sig0ept}
\end{eqnarray}
Equation (\ref{eq:nnonlinear}) is  in good agreement with
measurements of the bispectrum~\cite{FMS93} in $N$-body simulations as
well as predictions from HEPT (Sect.~\ref{sec:HEPT}).
\begin{table}
\caption[ ]{Parameters used in fit (\ref{fitneff}).}
\label{tableneff}
\vspace{.3cm}
\begin{center}
\begin{tabular}{cccccc}
\hline
$n$ & $n_{\rm nonlinear}$ & $n_{\rm nonlinear}^-$ & $n_{\rm nonlinear}^+$ &
$\sigma_0$ & $\tau$ \\ \hline -2 & -9.5 & -12.4 & -7.22 & 1.6 & 1.4 \\ -1 &
-3 & -3.8 & -2.24 & 1.4 & 1.2 \\ 0 & -1.2 & -1.6 & -0.86 & 1.25 & 0.6 \\ +1
& -0.85 & -1.17 & -0.57 & 0.7 & 0.3 \\ \hline
\end{tabular}
\end{center}
\end{table}
For a realistic, scale dependent spectral index (such as CDM models),
the situation becomes slightly more complicated since
Eq.~(\ref{eq:neff}) is in principle not valid anymore, at least in the
weakly nonlinear regime, due to the $\gamma_p$ corrections in
Eqs.~(\ref{eq:s3ptpt}-\ref{eq:s6ptpt}), which should be taken into
account. However, these corrections are in practice quite
small~\cite{Bernardeau94a,BGE95,CSJC99} an can be neglected in a first
approximation as illustrated by the right panel of
Fig.~\ref{sc_ept}. Then, Eq.~(\ref{eq:neff}) extends as well to
non-scale-free spectra such as CDM models~\cite{CBBH97,CSJC99,SQSL99}
(see Fig.~\ref{sc_ept}). 

It is even possible to use scale-free power spectra results,
Eq.~(\ref{fitneff}), with appropriate choice of $n$ in
Eqs.~(\ref{eq:nnonlinear}-\ref{eq:sig0ept}), $n=-\gamma_1(R)-3$
obtained from the linear variance computed at smoothing scale $R$, to
obtain an approximate fit of function $n_{\rm
eff}(\sigma)$~\cite{CBBH97}.  It is worth noting as well that EPT is a
good approximation for the $S_p$'s measured in 2D galaxy catalogs,
with $n_{\rm eff}$ varying from approximately $-2$ to $-5$ depending
on the angular scale considered~\cite{SMN96}.

This description can be extended to the joint moments~\cite{SzSz97},
giving the so-called E$^2$PT framework~\cite{SCB99,CSJC99}. This
provides a reasonable description of the joint cumulants in the
nonlinear regime, but not as accurate as EPT for one-point
cumulants~\cite{CSJC99}. However, a first application suggests that
this is in disagreement with observations~\cite{SzSz97}.

Both EPT and E$^2$PT provide useful ways of describing higher-order
statistics as functions of a single parameter $n_{\rm eff}$ and can be
used for estimating cosmic errors on statistics measured in galaxy
catalogs as discussed in the next chapter. However, except in the
weakly nonlinear regime, these prescriptions lack any rigorous
theoretical background, although some elements towards their
justification can be found in HEPT (see Sect.~\ref{sec:HEPT}).

\clearpage 
\section{\bf From Theory to Observations: Estimators and Errors}
\label{sec:chapter7}

\subsection{Introduction}

This chapter focuses on issues regarding accurate estimation of
clustering statistics in large-scale galaxy surveys and their
uncertainties, in order to properly constraint theories against
observations.  We also consider applications to measurements in
$N$-body simulations, as briefly described in Sect.~\ref{sec:9}.

In many respects, the theory of estimators of large scale structure
statistics was triggered in the seventies and the early eighties by
Peebles and his collaborators.  In a series of seminal works, starting
with a fundamental paper~\cite{Peebles73}, these authors developed the
statistical theory of the two-point correlation function in real and
Fourier space, in two- and three-dimensional catalogs, including
estimates of the cosmic errors and the cosmic bias (formulated as an
integral constraint problem), followed soon by investigations on
higher-order statistics.  They used several estimators, including
count-in-cell statistics.  These results are summarized
in~\cite{Peebles80}.

Since then, and particularly in the nineties, a number of techniques
were put forward to allow a more precise testing of cosmological
theories against observations.  These include:

\ben 
\item[-] Detailed studies of two-point and higher-order correlation
functions estimators.

\item[-] Accurate estimation of errors going beyond the simple (and
often severe underestimate) Poisson error bars, to include
finite-volume effects, survey geometry and non-Gaussian contributions
due to non-linear evolution.

\item[-] The treatment of covariance between measurements at different
scales.  In order to properly test theoretical predictions, this is
equally important to an accurate treatment of errors, which are just
the diagonal elements of the covariance matrix.  Neglecting
off-diagonal elements can lead to a substantial overestimate of the
constraining power of observations (see e.g. Chapter~8).

\item[-] Implementation of techniques for data compression, error
decorrelation, and likelihood analysis for cosmological parameters
estimation.

\een

It is clear that the upcoming large-scale galaxy surveys such as
2dFGRS and SDSS will certainly have to rely heavily on these new
developments to extract all the information encoded by galaxy
clustering to constrain cosmological parameters, primordial
non-Gaussianity and galaxy formation models.  In addition to standard
second-order statistics such as the power-spectrum or the two-point
correlation function, our review focuses on higher-order statistics
for several reasons:

\ben

\item[-] As detailed in previous chapters, non-linear evolution leads
to deviations from Gaussianity, so two-point statistics are not enough
to characterize large-scale structure.  They do not contain all the
information available to constrain cosmological theories\footnote{For
example, although one could construct a matter linear power spectrum
that evolves non-linearly into the observed {\em galaxy} power
spectrum (see Fig.~\ref{pkw2apm}); it is not possible to match at the
same time the higher-order correlations at small scales (see
Fig.~\ref{s34apm}).  This implies non-trivial galaxy biasing in the
non-linear regime, as we discuss in detail in
Sects.~\ref{sec:nptang}-\ref{sec:cicang}.}.

\item[-] The additional information encoded by higher-order statistics
can be used, for example, to constrain galaxy biasing
(Sect.~\ref{sec:bias}), primordial non-Gaussianity (Sects.~\ref{ngic}
and~\ref{ngic2}) and break degeneracies present in measurements of
two-point statistics, e.g. those obtained from measurements of the
redshift-space power spectrum (Sect.~\ref{sec:reddis}).  PT provides a
framework for accomplishing this\footnote{A quantitative estimate of
how much information is added by considering higher-order statistics
is presented in~\cite{TaWa00b}.}.

\item[-] The significant improvement in accuracy for higher-order
statistics measurements expected in upcoming large scale surveys, see
e.g.~Fig~\ref{fig:cosmicSDSS} below.

\een

Needless is to say that measurements in galaxy catalogs are subject to
a number of statistical and systematic uncertainties, that must be
properly addressed before comparing to theoretical predictions,
succinctly:

\begin{enumerate}

\item[(i)] {\em Instrumental biases and obscuration:} there are
technical limitations due to the telescopes and the instruments
attached to it.  For example, in spectroscopic surveys using
multifiber devices such as the SDSS, close pairs of galaxies are not
perfectly sampled unless several passes of the same part of the sky
are done (e.g. see~\cite{BLMYZL01}).  This can affect the measurement
of clustering statistics, in particular higher-order correlations.
Also, the sky is contaminated by sources (such as stars), dust
extinction from our galaxy, etc\ldots

\item[(ii)] {\em Dynamical biases and segregation:} unfortunately it
is not always possible to measure directly quantities of dynamical
interest: in three-dimensional catalogs, the estimated object
positions are contaminated by peculiar velocities of galaxies.  In 2-D
catalogs, the effects of projection of the galaxy distribution along
the line of sight must be taken into account.  Furthermore, galaxy
catalogs sample the visible matter, whose distribution is in principle
different from that of the matter.  The resulting {\em galaxy bias}
might depend on environment, galaxy type and brightness.  Objects
selected at different distances from the observer do not necessarily
have the same properties: e.g. in magnitude-limited catalogs, the
deeper objects are intrinsically brighter.  One consequence in that
case is that the number density of galaxies decreases with distance
and thus corrections for this are required unless using volume-limited
catalogs.

\item[(iii)] {\em Statistical biases and errors:} the finite nature of
the sample induces uncertainties and systematic effects on the
measurements, denoted below as {\em cosmic bias} and {\em cosmic
error}.  These cannot be avoided (although it is possible to estimate
corrections in some cases), only reduced by increasing the size of the
catalog and optimizing its geometry.

\end{enumerate}

In this chapter, we concentrate mainly on the point (iii).  Dynamical
biases mentioned in point (ii) will be addressed in the next chapter. 
These effects can also be taken into account in the formalism, by
simply replacing the values of the statistics intervening in the
equations giving cosmic errors and cross-correlations with the
``distorted'' ones, as we shall implicitly assume in the rest of this
chapter\footnote{Of course, this step can be non trivial. 
Measurements in galaxy catalogs (Sect.~\ref{chapter9}) and in $N$-body
simulations suggest that in the nonlinear regime the hierarchical
model is generally a good approximation
(e.g.~\cite{BSD91,FMS93,CBS94,CBH96,MBMS99}), but it can fail to
describe fine statistical properties (e.g. for the power spectrum
covariance matrix~\cite{SZH99,Hamilton00}).  In the weakly nonlinear
regime, PT results including redshift distortions
(Sect.~\ref{sec:reddis}), projection along the line of sight
(Sect.~\ref{sec:projeff}) and biasing (Sect.~\ref{sec:bias}) can help
to compute the quantities determining cosmic errors, biases and
cross-correlations.  In addition to the hierarchical model, extensions
of PT to the nonlinear regime, such as EPT, E$^2$PT
(Sect.~\ref{sec:EPT}) and HEPT (Sect.~\ref{sec:HEPT}), coupled with a
realistic description of galaxy biasing can be used to estimate the
errors.}.  Segregation effects and incompleteness due to instrument
biases, obscuration or to selection in magnitude will be partly
discussed here through weighted estimators, and in Chapter~8 when
relevant.

This chapter is organized as follows.  In Sect.~\ref{basic}, we
discuss the basic concepts of {\em cosmic bias}, {\em cosmic error}
and the {\em covariance matrix}.  Before entering in technical
details, it is important to discuss the fundamental assumptions
implicit in any measurement in a galaxy catalog, namely the {\em fair
sample hypothesis}~\cite{Peebles73} and the {\em local Poisson
approximation}.  This is done in Sect.~\ref{sec:sec1}, where basic
concepts on count-in-cell statistics and discreteness effects
corrections are introduced to illustrate the ideas.  In
Sect.~\ref{sec:sec2}, we study the most widely used statistic, the
two-point correlation function, with particular attention to the {\em
Landy and Szalay} estimator~\cite{LaSz93} introduced in
Sect.~\ref{sec:estxi2}.  The corresponding cosmic errors and biases
are given and discussed in several regimes.  Section~\ref{sec:sec3} is
similar to Sect.~\ref{sec:sec2}, but treats the Fourier counterpart of
$\xi$, the power spectrum.  Generalization to higher-order statistics
is discussed in Sect.~\ref{sec:sec4a}.

Section~\ref{sec:sec5} focuses on the count-in-cell distribution
function, which probes the density field smoothed with a top-hat
window.  In that case a full analytic theory for estimators and
corresponding cosmic errors and biases is available. 
Section~\ref{sec:sec5bis} discusses multivariate counts-in-cells
statistics.  In Sect.~\ref{sec:sec6} we introduce the notion of
optimal weighting: each galaxy or fraction of space can be given a
specific statistical weight chosen to minimize the cosmic error. 
Section~\ref{sec:sec6bis} deals with cross-correlations and the shape
of the cosmic distribution function and discusses the validity of the
Gaussian approximation, useful for maximum likelihood analysis. 
Section~\ref{sec:general} reinvestigates the search for optimal
estimators in a general framework in order to give account of recent
developments.  In particular, error decorrelation and the discrete
Karhunen-Lo\`eve transforms are discussed.  Finally, Sect.~\ref{sec:9}
discusses the particular case of measurements in $N$-body simulations.

In what follows, we assume we have a $\dim$-dimensional galaxy catalog
$D$ of volume $V$ and containing $N_{\rm g}$ objects, with $N_{\rm g}
\gg 1$, corresponding to an average number density ${\bar n}_{\rm
g}=N_{\rm g}/V$.  Similarly we define a pure random catalog $R$ of
same geometry and same number of objects\footnote{Note that $R$ stands
as well for a smoothing scale, but the meaning of $R$ will be easily
determined by the context.}.  Despite the fact that we use
three-dimensional notations (${\dim}=3$) most of results below are
valid as well for angular surveys except when specified otherwise. 
Simply, $\xi(r)$ has to be replaced with $w(\theta)$, $Q_N$ with
$q_N$, etc.

\subsection{Basic Concepts}

\label{basic}

\subsubsection{Cosmic Bias and Cosmic Error}

In order to proceed we need to introduce some new notation.  If $A$ is
a statistic, its estimator will be designated by $\tA$.  The
probability $\Upsilon(\tA)$ of measuring the value $\tA$ in a galaxy
catalog (given a theory) will be called the {\em cosmic distribution
function}.  The ensemble average of $\tA$ (the average over a large
number of virtual realizations of the galaxy catalog) is

\begin{equation} \langle \tA \rangle = \int \d\tA \ \Upsilon(\tA). 
\end{equation}

Due to their nonlinear nature many estimators (such as ratios) are
biased, i.e. their ensemble average is not equal to the real value
$A$: the {\em cosmic bias} (to distinguish it from the bias between
the galaxy distribution and the matter distribution),

\begin{equation} b_A=\frac{\langle \tA \rangle - A }{A}
\label{eq:cosmicbiasdef} \end{equation}

does not vanish, except when the size of the catalog becomes infinite
(if the estimator is properly normalized).

A good estimator should have minimum cosmic bias.  It should as well
minimize the {\em cosmic error}, which is usually obtained by
calculating the variance of the function $\Upsilon$:

\begin{equation} (\Delta A)^2=\langle (\delta \tA)^2 \rangle=\int
(\delta \tA)^2\ \Upsilon(\tA)\ \d\tA, \end{equation}

with

\begin{equation} \delta \tA \equiv \tA - \langle \tA \rangle. 
\end{equation}

The cosmic error is most useful when the function $\Upsilon(\tA)$ is
Gaussian.  If this is not the case, full knowledge of the shape of the
cosmic distribution function, including its skewness, is necessary to
interpret correctly the measurements\footnote{For example, it could be
very desirable to impose in this case that a good estimator should
have minimum skewness~\cite{SOA99}.}.

\subsubsection{The Covariance Matrix}

As for correlation functions, a simple generalization of the concept
of variance is that of covariance between two different quantities;
this can be for example between two estimators $\tA$ and $\tB$

\begin{equation} {\rm Cov}(\tA,\tB)=\langle \delta \tA\ \delta \tB
\rangle=\int \delta \tA\ \delta \tB\ \Upsilon(\tA,\tB)\ \d\tA \d\tB,
\end{equation}

or simply between estimates of the same quantity at different scales;
say, for the power spectrum, the covariance matrix between estimates
of the power at $k_i$ and $k_j$ reads,

\begin{eqnarray} \label{cijp} {C}_{ij}^P&\equiv& \langle {\hat
P}(k_i){\hat P}(k_j) \rangle - \langle {\hat P}(k_i)\rangle
\langle{\hat P}(k_j) \rangle, \end{eqnarray}

where ${\hat P}(k_i)$ is the estimator of the power spectrum at a band
power centered about $k_i$.

In general, testing theoretical predictions against observations
requires knowledge of the joint covariance matrix for all the
estimators (e.g. power spectrum, bispectrum) at all scales considered.
We will consider some examples below in Sects.~\ref{covaxi},
\ref{covapk} and~\ref{sec:crossestsec}.

The cosmic error and the cosmic bias can be roughly separated in three
contributions~\cite{SzCo96} if the scale $R$ (or separation)
considered is small enough compared to the typical survey size $L$, or
equivalently, if the volume $v\equiv v_R \equiv (4/3)\pi R^3$ is small
compared to the survey volume, $V$:

\begin{enumerate}

\item[(i)] {\em Finite volume effects:} they are due to the fact that
we can have access to only a finite number of structures of a given
size in surveys (whether they are 2-D or 3-D surveys), in particular
the mean density itself is not always well determined.  These
effects are roughly proportional to the average of the two point
correlation function over the survey, $\xiav(L)$.  They are usually
designated by ``cosmic variance''.

\item[(ii)] {\em Edge effects:} they are related to the geometry of
the catalog.  In general, estimators give less weight to galaxies near
the edge than those far away from the boundaries.  As we shall see
later, edge effects can be partly corrected for, at least for
$N$-point correlation functions.  At leading order in $v/V$, they are
proportional to roughly $\xi v/V$.  Note that even 2-D surveys cannot
avoid edge effects because of the need to mask out portions of the sky
due to galaxy obscuration, bright stars, etc...  Edge effects vanish
only for $N$-body simulations with periodic boundary conditions.

\item[(iii)] {\em Discreteness effects:} one usually assumes that the
observed galaxy distribution is a discrete, local Poisson
representation of an underlying smooth field whose statistical
properties one wants to extract.  This discrete nature has to be taken
into account with appropriate corrections, not only to the mean of a
given statistic but also to the error.  Discreteness errors, which are
proportional to $1/N_{\rm g}$ at some power where $N_{\rm g}$ is the
number of objects in the catalog, become negligible for large enough
$N_{\rm g}$.

\end{enumerate}

The above separation into three contributions is convenient but
somewhat artificial, since all the effects are correlated with each
other.  For example, there are edge-discreteness effects and
edge-finite-volume effects~\cite{SzSz98}.  At next to leading order in
$R/L$, there is a supplementary edge effect contribution proportional
to the perimeter of the survey, which is most important when the
geometry of the survey is complex, and dominant when $R/L \approx
1$~\cite{Ripley88,CCDFS00}.

\subsection{Fair Sample Hypothesis and Local Poisson Approximation}

\label{sec:sec1}

\subsubsection{The Fair Sample Hypothesis}

A stochastic field is called {\em ergodic} if all information about
its multi-point probability distributions (or its moments) can be
obtained from a single realization of the field.  For example,
Gaussian fields with continuous power spectrum are
ergodic~\cite{Adler81}.

The {\em Fair Sample Hypothesis}~\cite{Peebles73} states that the
finite part of the universe accessible to observations is a fair
sample of the whole, which is represented by a statistically
homogeneous and isotropic (as defined in Sect.~\ref{sec:stathomiso})
ergodic field.  Together with the ergodic assumption, the fair sample
hypothesis states that well separated parts of the (observable)
Universe are independent realizations of the same physical process and
that there are enough of such independent samples to obtain all the
information about its probability distributions (e.g.
\cite{Peebles80,Bertschinger92}).  Under the fair sample hypothesis,
ensemble averages can be replaced with spatial averages.  In the
simplest inflationary models leading to Gaussian primordial
fluctuations, the fair sample hypothesis holds, but special cases can
be encountered in models of Universe with non-trivial global
topological properties (see e.g.~\cite{LuLa95}) where apparently well
separated parts of the Universe may be identical.

\subsubsection{Poisson Realization of a Continuous Field}
\label{sec:poirea}

In general, statistical properties of the density field are measured
in a discrete set of points, composed e.g. of galaxies or $N$-body
particles.  It is natural to assume that such point distributions
result from a Poisson realization of an underlying continuous field. 
This means that the probability of finding $N$ points in a volume $v$
at location $\vr$ is given by $P^{\rm Poisson}_N[{\bar n}_{\rm
g}v(1+\delta(\vr))]$, where $P^{\rm Poisson}_N(\barN)$ is the
probability of finding $N$ objects in a Poisson process with
expectation number $\barN={\bar n}_{\rm g} v$,

\begin{equation} P^{\rm Poisson}_N(\barN) \equiv \frac{\barN^N}{N!} e^{-\barN},
\label{eq:ppois} \end{equation}

$\delta(\vr)$ is the overall density contrast within the volume and
${\bar n}_{\rm g}$ is the average number density of the random
process.  It implies that the {\em count probability distribution
function}, hereafter CPDF, defined as the probability $P_N$ of finding
$N$ galaxies in a cell of size $R$ and volume $v$ thrown at random in
the catalog can be expressed through the convolution,

\be  P_N=\int_{-1}^{+\infty}\d\de\,p(\de)\,P^{\rm Poisson}_N
\left[\barN(1+\de) \right], \label{fb:PoiConv} \ee 

where the average number of objects per cells,  $\barN$,  reads
\begin{equation}
\barN=\sum_N N P_N.
\end{equation}

In the continuous limit, ${\bar N} \rightarrow \infty$, the CPDF of
course tends to the PDF of the underlying density field

\begin{equation}
P_N \rightarrow \frac{P[{\bar N}(1+\delta)]}{\bar N}.
\end{equation}

It is worth at this point to mention the void probability function,
$P_0$, which can be defined in discrete samples only.  From
Eqs.~(\ref{fb:PoiConv}) and (\ref{eq:ppois}), it reads

\be P_0=\int_{-1}^{+\infty}\d\de\,p(\de)\exp[-\barN(1+\de)], \ee  

which can be expressed in terms of the cumulant generating
function~\cite{White79,BaSc89a,SzSz93a} (see Sect.~\ref{sec:sec3.3}),

\be P_0=\, \exp\left[-\barN+ \mC(-\barN) \right]=\exp\left[
\sum_{n=1}^\infty {{(-\barN)^n}\over{n!}} \, \cum{n} \right]. 
\label{fb:P0}
\ee 

This property was used in practice to obtain directly the cumulant
generating function from the void probability function (e.g.,
\cite{MSC92,ElGa92,BSDFYH93}), relying on the local Poisson
approximation.

Obviously, the validity of the local Poisson approximation is
questionable.  A simple argument against it is that galaxies have an
extended size which defines zones of mutual exclusion and suggests
that at very small scales, galaxies do not follow a local Poisson
process because they must be anti-correlated.  One way to bypass this
problem is of course to choose the elementary volume such that it has
a sufficiently large size, say $\ell \ga $ a few tens of kpc.  One
might still argue that short-range physical processes depending on
environment might influence small-scale statistics in such a way that
it might be impossible to find a reasonably small scale $\ell$ for
which the Poisson process is valid.  Also, the galaxy distribution
might keep memory of initial fluctuations of the density field, even
at small, nonlinear scales, particularly in underdense regions which
do not experience shell-crossing and violent relaxation.  If for
example these initial conditions were locally fractal up to some very
small scale, obviously the local Poisson approximation would break
down.  Note on the other hand that sparse sampling
strategies~\cite{Kaiser86} which were used to build a number of galaxy
catalogs, make the samples ``closer'' to Poisson.

It is generally assumed that the observed galaxy distribution follows
the local Poisson approximation.  To our knowledge there exists no
direct rigorous check of the validity of this statement, but it is
supported indirectly, for example by the fact that the measured count
probability distribution function (CPDF, see Sect~\ref{sec:sec5}) in
galaxy catalogs compares well with models relying on the local Poisson
approximation (see, e.g.~\cite{BSDFYH93}).

In $N$-body simulations, the local Poisson assumption is in general
very good\footnote{Except when dealing with the clustering of dark
matter halos; in this case exclusion effects can lead to sub-Poisson
sampling, see e.g.~\cite{SLSDKW01}.}.  However this depends on the
statistic considered and there are some requirements on the degree of
evolution of the system into the nonlinear regime, as discussed below
in Sect.~\ref{sec:relsof}.

Under the assumption of local Poisson approximation, it is possible to
derive the correlation functions of the discrete realization in terms
of the underlying continuous one.  In particular, from
Eq.~(\ref{fb:PoiConv}) the moment generating function of the discrete
realization, ${\cal M}_{\rm disc}$, is related to that of the
continuous field, ${\cal M}$ (Sect.~\ref{sec:sec3.3.2}), by ${\cal
M}_{\rm disc}(t)={\cal M}(t)\ [\exp(t)-1]$.  This leads to the
standard expressions for moments and spectra of discrete realizations
in terms of continuous ones, e.g.
see~\cite{Layzer56,Peebles80,Fry85,SzSz93a,GaYo93,MVH97}.  Here we
give the first few low-order moments

\bea  \langle \de_n^2 \rangle &=& \frac{1}{\barN}+\xibar_2, \\ \langle
\de_n^3 \rangle &=& \frac{1}{\barN^2}+3\frac{\xibar_2}{\barN}+
\xibar_3, \eea

where $\de_n \equiv (N-\barN)/\barN$ denotes the discrete number
density contrast.  In Sect.~\ref{sec:sec5}, which discusses in more
detail count-in-cells statistics, we shall see that there exists an
elegant way of correcting for discreteness effects using factorial
moments.

Similarly, for the power spectrum and bispectrum,

\bea \langle \de_n(\vk_1) \de_n(\vk_2) \rangle &=&
\left[\frac{1}{N_{\rm g}} +P(k_1)\right]\ \de_n(\vk_{12}), \\ \langle
\de_n(\vk_1) \de_n(\vk_2) \de_n(\vk_3) \rangle &=& \left[
\frac{1}{N_{\rm g}^2}+ \frac{1}{N_{\rm g}} ( P_1+P_2+P_3) + B_{123}\right]\
\de_n(\vk_{123}), \eea
where $P_i\equiv P(k_i)$, $B_{123} \equiv B(\vk_1,\vk_2,\vk_3)$,
$\vk_{12}=\vk_1+\vk_2$ and $\vk_{123}=\vk_1+\vk_2+\vk_3$.

\subsection{The Two-Point Correlation Function}

\label{sec:sec2}

In this section, we present the traditional estimators of the
two-point correlation function based on pairs counting\footnote{For a
review on existing estimators, see, e.g.~\cite{Kerscher99,PMSSS99}.}.
We assume that the catalog under consideration is statistically
homogeneous.  Optimal weighting and correction for selection effects
will be treated in Sect.~\ref{sec:sec6}.  More elaborate estimates
taking into account cross-correlations between bins will be discussed
in Sect.~\ref{sec:sec6bis}.

\subsubsection{Estimators}

\label{sec:estxi2}

In practice, due to the discrete nature of the studied sample, the
function $\xi$ [Eq.~(\ref{eq:xidef})] is not measured at separation
exactly equal to $r$ but rather one must choose a bin, e.g.,
$[r,r+\Delta r[$.  More generally, the quantity measured is

\begin{equation} \frac{1}{G_p^\infty V_\infty^2}\int_{V_\infty}
\d^{\dim}\vr_1 \d^{\dim}\vr_2\ \Theta(\vr_1,\vr_2)\ \xi(r_{12}),
\label{eq:realxi} \end{equation}

where the function $\Theta(\vr_1,\vr_2)$ is symmetric in its arguments
(e.g.~\cite{SzSz98}).  In what follows, we assume that the function
$\Theta$ is invariant under translations and rotations,
$\Theta(\vr_1,\vr_2)=\Theta(r)$, $r=r_{12}=|\vr_1-\vr_2|$, is unity on
a domain of values of $r$, for example in the interval $[r,r+\Delta
r[$ and vanishes otherwise.  The values where $\Theta$ is non-zero
define a ``bin'' which we call $\Theta$ as well.  We assume that
$\xi(r)$ is sufficiently smooth and that the bin and the
normalization, $G_p^\infty$, are such that Eq.~(\ref{eq:realxi}) would
reduce with a good accuracy to $\xi(r)$ in a survey of very large
volume $V_\infty$.

Practical calculation of the two-point correlation function relies on
the fact that it can be defined in terms of the excess probability
over random $\delta P$ of finding two galaxies separated by a distance
(or an angle) $r$ [as discussed already in Chapter~\ref{fb:Stoch},
Eq.~(\ref{dp12})]

\begin{equation} \delta P={\bar n}_{\rm g}^2 \left[ 1+\xi(r) \right]
\delta V_1 \delta V_2, \label{eq:xidef2} \end{equation}

where $\delta V_1$ and $\delta V_2$ are volume (surface) elements and
${\bar n}_{\rm g}$ is the average number density of objects.

Let $DD$ be the number of pairs of galaxies in the galaxy catalog
belonging to the bin $\Theta$ and $RR$ defined likewise but in a
random (Poisson distributed) catalog with same geometry and same
number of objects, $N_{\rm r}=N_{\rm g}$. They read,

\begin{equation} 
DD= \int_{\vr_1 \ne \vr_2} \d^{\dim}\vr_1 \d^{\dim}\vr_2\
\Theta(\vr_1,\vr_2) \ n_{\rm g}(\vr_1)\ n_{\rm g}(\vr_2),
\label{fb:DD}
\end{equation} 

\begin{equation}  
RR=\int_{\vr_1 \ne \vr_2} \d^{\dim}\vr_1 \d^{\dim}\vr_2 \
\Theta(\vr_1,\vr_2) \ n_{\rm r}(\vr_1)\ n_{\rm r}(\vr_2),
\label{fb:RR}
\end{equation}

where $n_{\rm g}$ and $n_{\rm r}$ are local number density fields
respectively in the galaxy catalog and the random catalog:

\begin{equation}
n_{\rm g}=\sum_{j=1}^{N_{\rm g}} \delta_D(\vx-\vx_j),
\end{equation}

where $\vx_j$ are the galaxy positions and likewise for $n_{\rm r}$. 
It is easy to derive from Eq.~(\ref{eq:xidef2}) a simple estimator
commonly used in the literature~\cite{PeHa74}:

\begin{equation} \txi(r)=\frac{DD}{RR}-1.  \label{eq:simpleest}
\end{equation}

Various alternatives have been proposed to improve the estimator given
by Eq.~(\ref{eq:simpleest}), in particular to reduce the cosmic bias
induced by edge effects at large separations.  Detailed
studies~\cite{KSS99} suggest that the best of them is the Landy \&
Szalay (LS) estimator~\cite{LaSz93}\footnote{See
however~\cite{PMSSS99} for a more reserved point of view.}

\begin{equation} \txi(r)=\frac{DD-2 DR +RR}{RR},  \label{eq:LSes}
\end{equation}

where $DR$ is the number of pairs selected as previously but the first
object belongs to the galaxy sample and the second one to the random
sample

\begin{equation}
DR= \int_{\vr_1 \ne \vr_2} \d^{\dim}\vr_1 \d^{\dim}\vr_2\
\Theta(\vr_1,\vr_2) \ n_{\rm g}(\vr_1)\ n_{\rm r}(\vr_2).
\label{fb:DR}
\end{equation}

The LS estimator, which formally can be written
$(D_1-R_1)(D_2-R_2)/R_1R_2$ corresponds to the ``intuitive'' procedure
of first calculating overdensities and then expectation values; this
has the obvious generalization to higher-order correlation
functions~\cite{SzSz98}, see Sect.~\ref{sec:sec4a} for more details.

Note that the calculations of $DR$ and $RR$ can be arbitrarily
improved by arbitrarily increasing $N_{\rm r}$ and applying the
appropriate corrections to $DR$ and $RR$, i.e. multiplying $DR$ and
$RR$ by the ratio $N_{\rm g}/N_{\rm r}$ and $N_{\rm g}(N_{\rm
g}-1)/[N_{\rm r}(N_{\rm r}-1)]$ respectively, to preserve
normalization.  Actually, $DR$ and $RR$ can be computed numerically as
integrals with a different method than generating a random catalog,
the latter being equivalent to Monte-Carlo simulation.  It amounts to
replace $DR$ and $RR$ by $DF$ and $FF$ with,

\begin{equation} 
DF={\bar n}_{\rm g}\int_{\vr_1 \ne \vr_2} \d^{\dim}\vr_1
\d^{\dim}\vr_2\ \Theta(\vr_1,\vr_2)\ n_{\rm g}(\vr_1),
\label{fb:DF}
\end{equation}

\begin{equation}
FF={\bar n}^2_{\rm g}\int_{\vr_1 \ne \vr_2} \d^{\dim}\vr_1
\d^{\dim}\vr_2 \ \Theta(\vr_1,\vr_2).
\label{fb:FF}
\end{equation}
In that case, the actual measurements are performed on pixelized data.

The LS estimator is theoretically optimal with respect to both cosmic
bias and cosmic error at least in the weak correlation
limit~\cite{LaSz93}; numerical studies~\cite{KSS99} show moreover that
for practical purposes it is better than any other known estimators
based on pair counting, among those one can quote
$(DD-DR)/RR$~\cite{Hewett82}, the popular
$DD/DR-1$~\cite{DaPe83,BlAl88} and $DD RR/(DR)^2-1$~\cite{Hamilton93b}
which is actually almost as good as LS~\cite{KSS99}.  In
Sect.~\ref{sec:sec5bis} we shall mention other ways of measuring
$\xi(r)$ and higher-order correlation functions, based on multiple
counts-in-cells.

Finally, it is worth mentioning a few efficient methods used to
measure $\xi(r)$, which apply to any of the estimators discussed in
this paragraph.  The brute force approach is indeed rather slow, since
it scales typically as ${\cal O}(N_{\rm g}^2)$.  To improve the speed
of the calculation, one often interpolates the sample onto a grid and
creates a linked list where each object points to a neighbor belonging
to the same grid site.  For separations smaller than the grid step,
$\lambda$, this method scales roughly as ${\cal O}(N_{\rm g} N_{\rm
cell})$, where $N_{\rm cell}$ is the typical number of objects per
grid cell.  This approach is however limited by the the step of the
grid: measuring the correlation function at scales large compared to
$\lambda$ is rather inefficient and can become prohibitive. 
Increasing $\lambda$ makes $N_{\rm cell}$ larger and for too large
$\lambda$, the method is slow again.

Another scheme relies on a double walk in a quad-tree or a oct-tree
according to the dimension of the survey (a hierarchical decomposition
of space in cubes/squares and subcubes/subsquares,
\cite{MCGGGKNSSSW01}).  This approach is potentially powerful, since
it scales as ${\cal O}(N_{\rm g}^{3/2})$ according to its
authors~\cite{MCGGGKNSSSW01}.  It is also possible to rely on FFT's or
fast harmonic transforms at large scales~\cite{SPC01}, but it requires
appropriate treatment of the Fourier coefficients to make sure that
the quantity finally measured corresponds to the estimator of
interest, e.g. the LS estimator (see~\cite{SPC01} for a practical
implementation in harmonic space).
 
\subsubsection{Cosmic Bias and Integral Constraint of the LS Estimator}

\label{sec:secest}

The full calculation of the cosmic bias and the cosmic error of the LS
estimator was done by Landy \& Szalay~\cite{LaSz93} in the weak
correlation limit and by Bernstein~\cite{Bernstein94} for the general
case but neglecting edge effects, $r \ll L$, where $L$ is the smallest
size of the survey\footnote{It is however important to notice a subtle
difference between the two approaches: Landy \& Szalay use conditional
averages with fixed number of galaxies in the catalog $N_{\rm g}$,
while $N_{\rm g}$ is kept random in Bernstein's approach.  This
difference is analyzed in Sect.~\ref{sec:sec6bis}.}.  At leading order
in $r/L$ and assuming that the density variance at the scale of the
survey is small, the cosmic bias reads

\begin{equation} b_{\xi} \simeq \left(3-\frac{1}{\xi}\right) \xiav(L)
- 2\frac{{\breve{\xi}}_3}{\xi}-\frac{1}{2 N_{\rm g}^2}, \quad r/L,\ |\xiav(L)|,\ \left|
\frac{\xiav(L)}{\xi} \right| \ll 1,
\label{eq:cosmicbias0} \end{equation}
where

\begin{equation} \xiav(L)=\frac{1}{V^2} \int \d^{\dim}\vr_1 \d^{\dim}\vr_2
\ \xi(r) \label{eq:xiavLdef} \end{equation}

is the average of the correlation function over the survey volume (or
area).  The quantity ${\breve{\xi}}_3$ is defined as

\begin{equation} {\breve{\xi}}_3=\frac{1}{G_p V^3} \int \d^{\dim}\vr_1
\d^{\dim}\vr_2 \d^{\dim}\vr_3 \ \Theta(r_{12})\ \xi_3(r_1,r_2,r_3),
\end{equation}

where $G_p$ is the form factor defined in \cite{LaSz93} as

\begin{equation} G_p=\frac{1}{V^2} \int \d^{\dim}\vr_1 \d^{\dim}\vr_2\
\Theta(r_{12}), \label{eq:Gp} \end{equation}

i.e.~the probability of finding a pair included in the survey in bin
$\Theta$.  When $r/L$ is small enough it is simply given by  $G_p
\simeq 4\pi r^2 \Delta r/V$ (for a bin $\Theta=[r,r+\Delta r[ $).

Assuming the hierarchical model, Eq.~(\ref{HM}), we get
${\breve{\xi}}_3 \simeq 2 Q_3\,\xi\,\xiav(L)$ and the cosmic bias
simplifies to

\begin{equation} b_{\xi} \simeq \left(3-4 Q_3-\frac{1}{\xi} \right)
\xiav(L)-{1\over 2N_{\rm g}^2}, \quad r/L,\ |\xiav(L)|,\ \left|
\frac{\xiav(L)}{\xi} \right| \ll 1.  \label{eq:cosmicbias1}
\end{equation}

In the weak correlation limit, it simply reduces to~\cite{LaSz93}

\begin{equation} b_{\xi} \simeq \frac{-\xiav(L)}{\xi}, \quad |\xi|,\
|\xiav(L)| \ll 1.  \label{eq:integralconstraint} \end{equation}

The LS estimator, although designed to minimize both the cosmic error
and the cosmic bias and thus quite insensitive to edge effects and
discreteness effects, is still affected by finite-volume effects,
proportional to $\xiav(L)$ (indeed the latter cannot be reduced
without prior assumptions about clustering at scales larger than those
probed by the survey, as discussed below).  The corresponding cosmic
bias is negative, of small amplitude in the highly nonlinear regime,
but becomes significant when the separation $r$ becomes comparable to
the survey size.  In this regime, where $\xi(r)$ is expected to be
much smaller than unity, Eq.~(\ref{eq:integralconstraint}) is
generally valid: the correct value of $\xi$ is obtained by adding an
unknown constant to the measured value.  This corresponds to the so
called {\em integral constraint} problem~\cite{Peebles74c,Peebles80}. 
Physically, it arises in a finite survey because one is estimating the
mean density and fluctuations about it from the same sample, and thus
the fluctuation must vanish at the survey scale.  In other words, one
cannot estimate correlations at the survey scale since there is only
one sample available of that size.

This bias cannot be a priori corrected for unless a priori assumptions
are made on the shape of the two-point correlation function at scales
larger than those probed by the survey.  One can for instance decide
to model the two-point correlation as a power-law and do a joint
determination of all parameters~\cite{Peebles74c}.  We will come back
to this problem when discussing the case of the power spectrum, where
other corrections have been suggested, see Sect.~\ref{cbcipk}.

\subsubsection{Cosmic Error of the LS Estimator}
\label{sec:secintegra}

The general computation of the cosmic error for such estimator is
quite involved and has been derived in the literature in various
cases.  For instance, the covariance of $DD-2 DF+FF$ between two bins
$\Theta_a$ and $\Theta_b$ reads~\cite{Peebles73,Hamilton93b,Szapudi00}

\begin{eqnarray}
{\rm Cov}(DD-2 DF +FF) & = & {\bar n}_{\rm g}^4 \int \d^{\dim}\vr_1
\d^{\dim}\vr_2 \d^{\dim}\vr_3 \d^{\dim}\vr_4 \ \Theta_a(\vr_1,\vr_2)\
\Theta_b(\vr_3,\vr_4)\times  \nonumber \\ & &\hspace{-4.5cm}\left[
\xi_4(\vr_1,\vr_2,\vr_3,\vr_4) +\xi(\vr_1,\vr_3)\xi(\vr_2,\vr_4)
+\xi(\vr_1,\vr_4)\xi(\vr_2,\vr_3) \right] \nonumber \\ & & 
\hspace{-4.5cm}+4 {\bar n}_{\rm g}^3 \int \d^{\dim}\vr_1
\d^{\dim}\vr_2 \d^{\dim}r_3 \ \Theta_a(\vr_1,\vr_2) \
\Theta_b(\vr_1,\vr_3) \left[ \xi(\vr_2,\vr_3) +
\xi_3(\vr_1,\vr_2,\vr_3)\right] \nonumber \\ &  &
\hspace{-4.5cm}+2 {\bar n}_{\rm g}^2 \int \d^{\dim}\vr_1
\d^{\dim}\vr_2 \ \Theta_a(\vr_1,\vr_2)\ \Theta_b(\vr_1,\vr_2)
\left[1+\xi(\vr_1,\vr_2)\right]. 
\label{eq:cosmicerrgen}
\end{eqnarray}

This is a general expression, i.e. it applies to the two-point
correlation function as well as the power-spectrum, or any pairwise
statistics of the density field, depending on the choice of the
binning function $\Theta$.  It does not take however into account the
possible cosmic fluctuations of the denominator in the LS estimator. 
This latter effect is more cumbersome to compute because one has to
deal with moments of the inverse density.  This is possible if one
assumes that fluctuations are small.  This leads to the cosmic
covariance derived in~\cite{Bernstein94} for the LS estimator.  We
give here a simplified expression of the diagonal term, the cosmic
error:

\begin{eqnarray} \left( \frac{\Delta \xi}{\xi} \right)^2 &
\simeq & 2{\overline{\xi^2}\over \xi^2} + 4(1-2 Q_3 + Q_4) \xiav(L)
+\frac{4}{N_{\rm g}}\left[\frac{\xi_{\rm ring}(1+2Q_3 \xi)}{\xi^2}+
Q_3-1 \right] \nonumber \\ & & \hspace{-2cm}+\frac{2}{N_{\rm
g}^2}\left[\left(\frac{1}{G_p}-1\right)
\frac{1+\xi}{\xi^2}-\frac{1}{\xi}-1\right],\ \ \ r/L,\ |\xiav(L)|,\
|\xiav(L)/\xi| \ll 1, \label{eq:cosmicxi}
\end{eqnarray}

where $\overline{\xi^2}$ is the average of the square of the two-point
correlation function over the survey volume,
\begin{equation}
\overline{\xi^2}={1\over G_p^2\,V^4}
\int\d^{\dim}\vr_1\ldots\d^{\dim}\vr_4\,\Theta(r_{12})\,\Theta(r_{34})\,
\xi_2(r_{13})\,\xi_2(r_{24}),
\label{eq:overxi}
\end{equation}

and $\xi_{\rm ring}$ is the average of the two-point correlation
function for pairs inside the shell of radius $r$ and thickness
$\Delta r$

\begin{equation} \xi_{\rm ring}=\frac{1}{G_t V^3} \int \d^{\dim}\vr_1
\d^{\dim}\vr_2 \d^{\dim}\vr_3 \ \Theta(r_{12})\ \Theta(r_{13})\
\xi(r_{23}).  \end{equation}

We have introduced the new geometrical factor $G_t$ given
by~\cite{LaSz93}

\begin{equation} G_t=\frac{1}{V^3} \int \d^{\dim}\vr_1 \d^{\dim}\vr_2
\d^{\dim}\vr_3\ \Theta(r_{12}) \ \Theta(r_{13}), \label{eq:Gt}
\end{equation}

i.e. $G_t$ is the probability, given one point, of finding two others
in bin $\Theta$, for example the interval $[r,r+\Delta r[$.  As
pointed out in~\cite{Bernstein94}, $\xi_{\rm ring} \ga \xi$, but

\begin{equation} \xi_{\rm ring} \simeq \xi \end{equation}

is a good approximation.  In Eq.~(\ref{eq:cosmicxi}), a degenerate
hierarchical model (Sect.~\ref{sec:HM}) has been assumed to simplify
the results.  A more general expression can be found
in~\cite{Bernstein94} (See also~\cite{Hamilton93b,Szapudi00}.).

The finite volume errors are given by a term in $\overline{\xi^2}$ and
one proportional to $\xiav(L)$.  It is interesting to compare these
two contributions.  For a power-law spectrum of index $n$,
$\overline{\xi^2}/\xi^2$ scales like $(r/L)^{\dim}$ whereas $\xiav(L)$
scales like $(r_0/L)^{-({\dim}+n)}$ if $r_0$ is the correlation length
($\xi(r_0) \equiv 1$).  Therefore in the quasi-linear regime for which
$r\gg r_0$ and for surveys with a large number of objects, the first
term is likely to dominate (this is the case typically for wide
angular surveys), whereas for surveys which probe deeply into the
nonlinear regime the other terms are more likely to dominate.

The discreteness error is given by the term in  $1/N_{\rm g}$ which
vanishes for a randomized purely Poisson catalog. The intrinsic
Poisson error is encoded in the term  in $(1/N_{\rm g})^2$. This
estimate of the cosmic error neglects however edge effects that become
significant at scales comparable to the size of the survey.  In this
latter regime correlations are expected to be weak, and
from~\cite{LaSz93} one finds that the cosmic error is
dominated by edge-discreteness effects~\cite{SzSz98}:

\begin{equation} \left( \frac{\Delta \xi}{\xi} \right)^2 \simeq
\frac{2}{N_{\rm g}^2\xi^2} \left[ \frac{1}{G_p} - 2 \frac{G_t}{G_p^2}
+1 \right], \quad |\xi|,\ |\xiav(L)| \ll 1 . 
\label{eq:cosmicxi2} 
\end{equation}

One can note that when $r/L$ is small enough, the term in square
brackets is roughly equal to $1/G_p$ [as in Eq.  (\ref{eq:cosmicxi})],
that is the fraction of pairs available in the survey.  This is
obviously the dominant contribution of the error when the bin size
$\Delta r$ is very small.  This pure Poisson contribution can
generally be computed exactly given the geometry of the survey.

The expressions (\ref{eq:cosmicxi}) and (\ref{eq:cosmicxi2}) can be
used to estimate the full cosmic error.  This method however requires
prior assumptions about the hierarchical model parameters $Q_3$ and
$Q_4$ and for the integral of the two-point correlation function over
the survey volume, $\xiav(L)$.  For this reason, the Gaussian limit is
often used to compute errors (that is the contribution of
$\overline{\xi^2}$, e.g.~\cite{LEPM92}), but this might be a bad
approximation when $\xi \ga 1$ as we discussed
above\footnote{Figure~\ref{nb1} below, extracted from~\cite{SZH99},
illustrates that for the power-spectrum.}.

\subsubsection{The Covariance Matrix}
\label{covaxi}

As discussed above, Eq.~(\ref{eq:cosmicerrgen}) gives the cosmic
covariance matrix of the two-point correlation function assuming that
${\bar n}_{\rm g}$ is perfectly determined, while the calculation of
Bernstein~\cite{Bernstein94}, for which we gave a simplified
expression of the diagonal terms, takes into account possible
fluctuations in ${\bar n}_{\rm g}$.  We refer the reader
to~\cite{Bernstein94} for the full expression of $\vC_\xi$ which is
rather cumbersome. 

Interestingly the pure Poisson contribution vanishes for
non-overlapping bins in Eq.~(\ref{eq:cosmicerrgen}).  A simplified
formula can be obtained in the Gaussian limit where non-Gaussian and
discreteness contributions can be neglected,

\bea C_{\xi}(r_a,r_b)&&=\langle \hat\xi_2(r_a) \hat\xi_2(r_b) \rangle-
\langle\hat\xi_2(r_a)\rangle\langle\hat\xi_2(r_b)\rangle \nonumber \\
&&\hspace{-1cm} ={2\over
G_p(r_a)G_p(r_b)\,V^4}
\int\d^{\dim}\vr_1\ldots\d^{\dim}\vr_4\,\Theta_a(r_{12})\,\Theta_b(r_{34})\,
\xi_2(r_{13})\,\xi_2(r_{24}), \nonumber \\
\label{covaxir} \eea 
in particular, $C_{\xi}(r,r)=\overline{\xi^2}$
[Eq.~(\ref{eq:overxi})].  This expression can be conveniently
expressed in terms of the power spectrum.  It reads, for ${\dim}=3$,

\bea  C_{\xi}(r_a,r_b)&=& {(2\pi)^5\over V}\int k^2\d
k\,[P(k)]^2\,J_{1/2}(kr_a)\,J_{1/2}(kr_b)
\label{covaxir2} 
\eea where $J_{1/2}$ is a Bessel function.  A similar expression has
been derived for 2-D fields~\cite{EiZa01}, \bea
C_w(\theta_a,\theta_b)&=&\langle \tw_2(\theta_a) \tw_2(\theta_b) \rangle-
\langle\tw_2(\theta_a)\rangle\langle\tw_2(\theta_b)\rangle\nonumber \\
&=& \frac{2\,(2\pi)^3}{A_\Omega} \int_0^\infty k\d k\ [P(k)]^2
J_0(k\theta_a) J_0(k\theta_b), 
\label{covawtheta} 
\eea 

where $A_\Omega$ is the area of the survey, $w_2(\theta)$ represents
the angular two-point function and $\tw_2$ its estimator.

Note that as the volume/area of the survey increases, the diagonal
terms in Eq.~(\ref{covaxir}) do not, in general, become dominant
compared to the off-diagonal ones.  This is because correlation
function measurements are statistically correlated, even in the
Gaussian limit, unlike binned power spectrum measurements, e.g. see
Sect.~\ref{covapk}.

\subsubsection{Recipes for Error Calculations}

The issue of cosmic error computation is recurrent in cosmological
surveys and the previous computations clearly show that this is a
complex issue.  Various recipes have been proposed in the literature. 
A particularly popular one is the bootstrap method~\cite{BBS84}.  We
stress that bootstrap resampling is not suited for correlation
function measurements.  Indeed, as shown explicitly in
\cite{Snethlage00}, such method does not lead, in general, to a
reliable estimate of the cosmic error~\cite{PMSSS99,KSS99}.

Another popular and elementary way of estimating the errors consists
in dividing the catalog in a number of smaller subsamples of same
volume and compute the dispersion in the measurements corresponding to
each subsample~(e.g.~\cite{Gaztanaga94}).  This method is not free of
bias and generally overestimates the errors, since the obtained
dispersion is an estimator of the cosmic error on the subsamples and
not the parent catalog.  Recent studies on error
estimation~\cite{Scranton01SDSS,Zehavi01SDSS} also suggest that the
Jackknife method, which is a variant of the subsample method where the
$i^{\rm th}$ sample is obtained by {\em removing} the $i^{\rm th}$
subsample, gives a very good estimate of the cosmic error on the
two-point correlation function.  Unlike the subsample method, it does
not lead to overestimation of the cosmic error at large
scales\footnote{An alternative to these methods has been suggested by
Hamilton~\cite{Hamilton93b}, in which many realizations from a given
sample are generated by effectively varying the pair-weighting
function.}.

Of course, methods such as Jackknife and subsamples cannot lead to an
accurate estimation of finite-volume errors at the scale of the
survey, since only one realization of such a volume is available to
the observer.  This can only be achieved through a detailed
computation of the cosmic errors [Eqs.~(\ref{eq:cosmicxi}) and
(\ref{eq:cosmicxi2})] with prior assumptions about the behavior of
statistics involved at scales comparable to the survey size; or else
numerically by constructing multiple realizations of the survey, e.g.
mock catalogs relying on $N$-body simulations or simplified versions
thereof (e.g., \cite{ScSh01}).  On the other hand, methods that use
the actual data are very useful to assess systematic errors, by
comparing to other external estimates such as those just mentioned.

\subsection{The Power Spectrum}
\label{sec:sec3}

The power spectrum $P(k)$ is simply the Fourier transform of the
two-point correlation function (see Sect.~\ref{sec:sec3.2.2}), and
therefore it is formally subject to the same effects.  In fact, a
common theoretical framework can be set up for $\xi(r)$ and $P(k)$ in
order to find the best estimators (e.g.,
\cite{Hamilton97a,Hamilton97b,SzSz98}).  In practice, however, power
spectrum measurements have been undertaken mostly in linear or weakly
nonlinear scales which are subject to edge effects, difficult to
correct for.  In this section, we introduce simple (unweighted)
estimators and discuss the biases and cosmic error introduced by the
finiteness of the survey.  The techniques developed to measure $P(k)$
are numerous and sometimes very elaborate (a nice review can be found
in~\cite{THSVS98}), but most of them rely on the assumption that the
underlying statistics is Gaussian.  In this section we prefer to keep
the statistical framework general and thus restrict ourselves to
traditional estimators.  More sophisticated methods, using spatial
weighting and cross-correlations between bins, will be discussed in
Sects.~\ref{sec:sec6} and Sect.~\ref{sec:general}.

\subsubsection{Simple Estimators}

For convenience, in finite surveys the adopted normalization
convention for the Fourier transforms and the power spectra is often
different. This is the reason why in this subsection, we also adopt
following convention,

\begin{equation} \label{pkconv2} {\tilde A}(\vk)=\frac{1}{V} \int_V
\d^{\dim}\vx\ e^{-i \vk\cdot\vx} A(\vx) \end{equation}

where ${\tilde A(\vk)}$ are the Fourier modes of $A(\vx)$ and $V$ is
the survey volume (and to recover the convention used in
Eq.~(\ref{fourier}), one can simply use the formal correspondence $V
\longleftrightarrow (2\pi)^{\dim}$.)  The power spectrum is defined as
the Fourier transform of the two-point correlation function.  It
differs thus by a $V/(2\pi)^{\dim}$ normalization factor compared to
the adopted normalization in the other sections.  The higher-order
spectra are defined similarly from the higher-order correlation
functions in such a way that the functional relation between spectra
is preserved [e.g., the coefficients $\tilde{Q}$ in Eq.~(\ref{q}) are
left unchanged].

As shown in previous sections, estimating the correlation function
consists in counting pairs in bins, both in the galaxy catalog and in
random realizations with the same survey geometry.  This procedure can
be generalized to the measurement of the power-spectrum (e.g.,
\cite{FKP94}) for which the binning function $\Theta$ defined in
Sect.~\ref{sec:secest} is now different.  For one single mode the
straightforward choice would be (e.g., \cite{SzSz98})
$\Theta(\vr_1,\vr_2)= \left( e^{{\rm i}\vk\cdot(\vr_1-\vr_2)} +e^{{\rm
i}\vk\cdot(\vr_2-\vr_1)} \right)/2$.  Actual estimation of the power
is made over a $k$-bin defined for instance so that the magnitude of
wave vectors belong to a given interval $[k,k+\Delta k[$.  It means
that the function $\Theta$ to use actually reads,
\begin{equation}
\Theta(\vr_1,\vr_2)=\langle e^{{\rm
i}\vk\cdot(\vr_1-\vr_2)}\rangle_{\Theta}
\equiv\frac{1}{V_k}\int_{\vert\vk'\vert\in[k,k+\Delta k[}\d^{\dim}\vk'
e^{{\rm i}\vk'\cdot(\vr_1-\vr_2)},
\label{fb:ThetaF}
\end{equation}
where $V_k$ is the volume of the bin in $k$-space. Note that for a
rectangular shaped survey with periodic boundaries modes are
discrete and the number of modes in $V_k$ is
\begin{equation}
N_k={V\,V_k\over (2\pi)^{\dim}}.
\end{equation}
In the following we assume that $V_k$ is large enough to encompass a
sufficient number of modes to make any measurement possible.  With
this expression of $\Theta$ the quantities $DD$, $DR$, $RR$, $DF$ and
$FF$ defined in (\ref{fb:DD}-\ref{fb:FF}) where $\Theta$ is replaced
by Eq.~(\ref{fb:ThetaF}) can be used to estimate the power
spectrum~\cite{SzSz98}.

Traditionally, the estimate of the power-spectrum is done in the
following way: the density contrast is Fourier transformed directly
(e.g.~\cite{Peebles73,PeNi91,VPGH92,FDSYH93,PVGH94}):

\begin{equation}
{\hat \delta}_{\vk}=\frac{1}{V} \int \left[ \frac{{n_{\rm
g}(\vr)}}{{\bar n}_{\rm g}} -1 \right]e^{{\rm i} \vk\cdot\vx}
\d^{\dim}\vx =\frac{1}{N_{\rm g}} \sum_{j=1}^{N_{\rm g}} e^{{\rm i}
\vk\cdot \vx_j}-W_{\vk},
\end{equation}

where $W_{\vk}$ is the Fourier transform of the window function of the
survey,

\begin{equation}
W_{\vk}=\frac{1}{V} \int e^{{\rm i} \vk\cdot\vx} \d^{\dim}\vx.
\end{equation}

The power spectrum estimator is then given by,

\begin{equation}
{\hat P}(k)=\langle |\hat \delta_{\vk}|^2 \rangle_{\Theta}
-\frac{1}{N_{\rm g}}
\end{equation}

where $\langle ...\rangle_{\Theta}$ stands for summation in the
$k$-bin [e.g. Eq.(\ref{fb:ThetaF})], which can also be written

\begin{equation}
{\hat P}(k)= \frac{1}{N_{\rm g}^2} \left( DD-2DF+FF \right).
\label{eq:powest}
\end{equation}

Note that the correction for shot noise contribution is automatically
taken into account by the exclusion $\vr_1\neq \vr_2$ in the integral
$DD$.  One can see that this is analogous to the LS estimator
(\ref{eq:LSes}) in Fourier space~\cite{SzSz98}.

\subsubsection{Cosmic Bias and Integral Constraint}

\label{cbcipk}

Similarly as for the two-point correlation function, it is possible to
show that the estimator in Eq.~(\ref{eq:powest}) is
biased~\cite{Peebles73,PeNi91}, at least due to finite volume effects. 
Again this is generally described as the integral constraint problem.

The expressions for the cosmic bias can be directly inferred from
Eqs.~(\ref{eq:cosmicbias0}) and (\ref{eq:integralconstraint}).  More
specifically, at large, weakly nonlinear scales, where the Gaussian
limit is a good approximation, the cosmic bias reads~\cite{PeNi91}

\begin{equation}
b_{{\hat P}(k)} \simeq - P_*(0) \frac{\langle |W_\vk|^2
\rangle_{\Theta}}{\langle P_*(\vk) \rangle_{\Theta}}. 
\label{eq:biaspk}
\end{equation}

The quantity $P_*$ is the true power spectrum convolved with the
Fourier transform of the window function of the survey:

\begin{equation}
P_*(\vk)=P(\vk)*|W_{\vk}|^2.  \label{eq:convpk}
\end{equation}

Note that $P_*(0)$ is nothing but $\xiav(L)$ [Eq.~(\ref{eq:xiavLdef})].

At smaller scales, in the regime $k \gg 1/L$, the cosmic bias reads,

\begin{equation}
b_{{\hat P}(k)} \simeq P_*(0) \left[3- \frac{\langle |W_\vk|^2
\rangle_{\Theta} } {\langle P(\vk) \rangle_{\Theta}}\right] - \frac{ 2
\langle B_*(\vk,-\vk,0) \rangle_{\Theta} } {\langle P(\vk)
\rangle_{\Theta} }, \label{eq:biaspk1}
\end{equation}

where $B_*$ is the bispectrum (convolved with the Fourier transform of
the survey window).

In general the cosmic bias is approximated by the white noise value in
the Gaussian limit~\cite{PVGH94}

\begin{equation}
b_{{\hat P}(k)}\simeq -\langle |W_\vk|^2 \rangle_{\Theta}=-FF/N_{\rm
g}^2, \label{eq:biaspkwn}
\end{equation} 

and the corresponding correction is applied to the estimator
(\ref{eq:powest}).

An interesting approach to correct for the cosmic bias takes advantage
of the Gaussian limit expression, Eq.~(\ref{eq:biaspk}).  Since the
bias is proportional to the Fourier transform of the window of the
survey, construction of a tailored window such that $W_\vk=0$ for each
mode $\vk$ of interest makes Eq.~(\ref{eq:biaspk})
vanish~\cite{FDSYH93,THSVS98}.  However, one must keep in mind that
this procedure is approximate; even in the Gaussian limit there are
higher-order corrections to the result in Eq.~(\ref{eq:biaspk}) which
are not proportional to $W_\vk$\footnote{The cosmic bias in this
expression comes in fact from the uncertainty in the mean density
$\nbar_{\rm g}$ from the numerator in $\de=(n_{\rm g}-\nbar_{\rm
g})/\nbar_{\rm g}$; uncertainties from the denominator lead to
additional contributions, see e.g.~\cite{HuGa99}.}.

\subsubsection{The Cosmic Error}

\label{sec:secNNN}

The calculation of the cosmic error on the power-spectrum is formally
equivalent to that of the two-point correlation function.  However,
existing results assume that the average number density of galaxies in
the universe is an external parameter, i.e. the ensemble average
$\langle [\delta {\hat P}(k)]^2 \rangle$ is calculated with $N_{\rm
g}$ fixed in Eq.~(\ref{eq:powest}).  

In the limit when $k\gg 1/L$, where $L$ is the smallest size of the
survey, for the power spectrum equation~(\ref{eq:cosmicerrgen}) reads,

\begin{eqnarray}
\left[ \frac{\Delta {\hat P}(k)}{ P(k)} \right]^2 & \simeq &
\frac{2}{N_k} + \frac{{\overline T}(k,k)}{[P(k)]^2}   +
\frac{4}{N_{\rm g}}\left[ \frac{1}{N_k\,P(k)}+ \frac{{\overline
B}(k,k)}{[P(k)]^2} \right] \nonumber\\ &&+ \frac{2}{N_{\rm_g}^2}
\left[ \frac{1}{N_k\,[P(k)]^2} +  \frac{{\overline P}(k,k)}{[P(k)]^2}
\right],
\label{eq:errpow}
\end{eqnarray}

with

\begin{eqnarray}
{\overline T}(k_i,k_j) & \equiv & \langle T(\vk_1,-\vk_1,\vk_2,-\vk_2)
\rangle_{\Theta_{k_i},\Theta_{k_j}}\nonumber\\
&\equiv&
\int_{\vert\vk_1\vert\in[k_i,k_i+\Delta k_i[}\frac{d^{\dim}\vk_1}{V_{k_i}}
\int_{\vert\vk_2\vert\in[k_j,k_j+\Delta k_j[}
\frac{\d^{\dim}\vk_2}{V_{k_j}}
T(\vk_1,-\vk_1,\vk_2,-\vk_2)
\label{fb:barT}\\ 
{\overline B}(k_i,k_j) & \equiv & \langle B(\vk_1,\vk_2,-\vk_1-\vk_2)
\rangle_{\Theta_{k_i},\Theta_{k_j}}\\ {\overline P}(k_i,k_j) & \equiv
& \frac{1}{2} \langle P(\vk_1+\vk_2)+P(\vk_1-\vk_2)
\rangle_{\Theta_{k_i},\Theta_{k_j}}.
\end{eqnarray}

This result assumes that the true power spectrum is sufficiently
smooth and the bin in $k$-space thin enough that $\langle
P(k)\rangle_{\Theta_k}\simeq P(k)$, $\langle P(k)^2 \rangle_{\Theta_k}
\simeq [P(k)]^2$.  The continuous limit $N_{\rm g} \rightarrow \infty$
of Eq.~(\ref{eq:errpow}) was computed in~\cite{SZH99}, and the
Gaussian limit, $B=T=0$ in~\cite{FKP94}.

{}From the calculations of~\cite{SZH99}, one gets

\begin{equation}
{\overline T}(k,k) \simeq \frac{232}{441} [P(k)]^3 \label{eq:tkk}
\end{equation}

in the regime where PT applies, and

\begin{equation}
{\overline T}(k,k) \simeq (8 Q_{4,a} + 4 Q_{4,b}) [P(k)]^3,
\end{equation}

if the hierarchical model applies
(Sect.~\ref{sec:HM})~\cite{SZH99,Hamilton00}.  Similar calculations
can be done to evaluate ${\bar B}(k,k)$ and ${\bar P}(k,k)$.

\begin{figure*}
\begin{center} \leavevmode \epsfxsize=4.0in \epsfbox{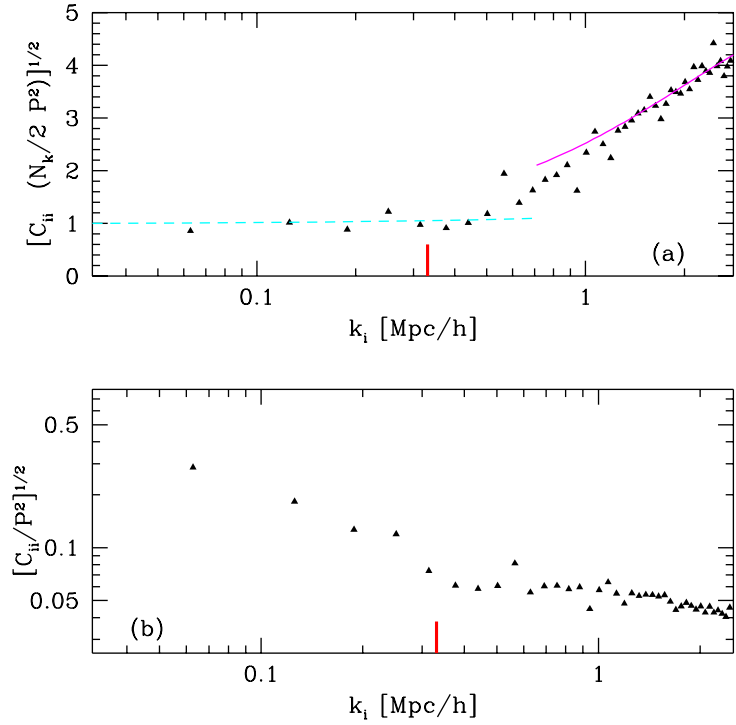}
\end{center} \caption{The top panel shows the measured cosmic error on
the power spectrum normalized by the Gaussian variance, obtained from
the dispersion over 20 PM simulations of SCDM. The dashed line shows
the predictions of PT, and the solid line the hierarchical scaling.
The bottom panel shows the fractional error in the band-power
estimates.  This fractional error scales with the size of the survey
or simulation box, the results in the figure correspond to a volume
$V_0=(100 {\, \rm h^{-1}Mpc})^3$.  Results for other volumes can be
obtained by scaling by $(V_0/V)^{1/2}$.  The vertical line on the
$x$-axis indicates the non-linear scale.  The width of shells in
$k$-space is $\Delta k=2\pi/100$ h/Mpc.} \label{nb1}
\end{figure*}

One must emphasize~\cite{MeWh99,SZH99} again the fact that the
Gaussian limit, traditionally used to compute errors and optimal
weighting (see Sect.~\ref{sec:sec6}), is invalid when $k \ga k_{\rm
nl}$, where $k_{\rm nl}$ is the transition scale to the nonlinear
regime defined from the power spectrum, $4\pi k_{\rm nl}^3 P(k_{\rm
nl}) \equiv 1$.  This is clearly illustrated by top panel of
Figure~\ref{nb1}.  It compares the measured cosmic error obtained from
the dispersion over 20 PM simulations of SCDM with the Gaussian
limit~\cite{SZH99}.  This shows that the Gaussian limit underestimates the
cosmic error, increasingly with $k/k_{\rm nl}$.  Note however that the
correction brought by Eq.~(\ref{eq:tkk}) is rather small.  As a result
the regime where the Gaussian limit is a reasonable approximation for
estimating the cosmic error extends up to values of $k/k_{\rm nl}$ of
order of a few.  This is unfortunately not true for the full cosmic
covariance matrix $C_{ij}^P \equiv
{\rm Cov}(P_{k_i},P_{k_j})$, which deviates from the
Gaussian predictions (vanishing non-diagonal terms) as soon as $k
\simeq k_{\rm nl}$~\cite{MeWh99,SZH99}, as we now discuss.

\subsubsection{The Covariance Matrix}

\label{covapk}

The covariance of the power spectrum, Eq.~(\ref{cijp}), can be easily
written beyond the Gaussian approximation neglecting shot noise and
the window of the survey~\cite{MeWh99,SZH99}\footnote{See
e.g.~\cite{Hamilton97a} for expressions including shot noise.},

\begin{eqnarray}
\label{cijp2} {C}_{ij}^P&=& 
{2 P^2(k_i)\over N_{k_i}} \de_{ij} +{\overline T}(k_i,k_j),
\end{eqnarray}
where $\de_{ij}$ is a Kronecker delta and $\overline T$ is the
bin-averaged trispectrum, (\ref{fb:barT}).

The first term in Eq.~(\ref{cijp2}) is the Gaussian contribution.  In
the Gaussian limit, each Fourier mode is an independent Gaussian
random variable.  The power estimates of different bands are therefore
uncorrelated, and the covariance is simply given by $2/N_{k_i}$ where
$N_{k_i}/2$ is the number of independent Gaussian variables.  The
second term in Eq.~(\ref{cijp2}) arises because of non-Gaussianity,
which generally introduces correlations between different Fourier
modes, and hence it is not diagonal in general.

Both terms in the covariance matrix in equation (\ref{cijp2}) are
inversely proportional to $V$ for a fixed bin size (recall that with
the adopted convention $P(k)$ scales like $1/V$ and $T$ like $1/V^3$). 
But while the Gaussian contribution decreases when $N_{k}$ increases,
the non-Gaussian term remains constant.  Therefore, when the
covariance matrix is dominated by the non-Gaussian contribution the
only way to reduce the variance of the power spectrum is to increase
the volume of the survey instead of averaging over more Fourier modes.

The importance of the non-Gaussian contribution to the
cross-correlation between band powers was studied with numerical
simulations in~\cite{MeWh99,SZH99}, in particular~\cite{MeWh99} shows
in detail that the correlations induced by non-linearities are not
negligible even at scales $k \la k_{\rm nl}$, in agreement with PT
predictions~\cite{SZH99}.  In the non-linear regime, as expected, the
cross-correlations are very strong; indeed, the cross-correlation
coefficient $r_{ij}\equiv C_{ij}/\sqrt{C_{ii}C_{jj}}$ is very close to
unity.  Predictions for $r_{ij}$ from the hierarchical ansatz using
HEPT amplitudes (see Sect.~\ref{sec:HEPT}) are in reasonable agreement
with simulations~\cite{SZH99}, although at large separations ($k_i \gg
k_j$) there are significant deviations~\cite{SZH99,Hamilton00}.

An efficient (although approximate) numerical approach to computing
the covariance matrix of the power spectrum is presented
in~\cite{ScSh01}, using a combination of 2LPT at large scales, and
knowledge about dark matter halos at small scales (see e.g.
Sect.~\ref{dhbias}-\ref{galbias}), which also allows to take into
account the effects of redshift distortions and galaxy biasing.

\subsection{Generalization to Higher-Order Correlation Functions}

\label{sec:sec4a}

Higher-order statistics such as correlation functions in real and
Fourier space were not studied in as much detail as the power spectrum
and the two-point correlation function.  In particular, there is no
accurate analytic estimate of the cosmic bias and error on such
statistics\footnote{See however the attempt in~\cite{MJB92} about
estimating the error on $\xi_3$ in various approximations.}, although
a general formalism (relying on a statistical framework set up by
Ripley~\cite{Ripley88}) which we summarize below was recently
developed by Szapudi and
collaborators~\cite{SzSz98,SzSz00,Szapudi00}.

The LS estimator presented in Sect.~\ref{sec:estxi2} for the two-point
correlation function, $\langle\delta_1 \delta_2\rangle$, can be
formally written as $(D_1-R_1)(D_2-R_2)/R_1R_2$.  As suggested
in~\cite{SzSz98}, a simple generalization for a statistic of order
$N$, for example the unconnected $N$-point correlation function, $f_N
\equiv \langle \delta_1 \ldots \delta_N \rangle$, is simply
$(D_1-R_1)(D_2-R_2)\ldots(D_N-R_N)/R_1\ldots R_N$.  More exactly,
\cite{SzSz98,Szapudi00} define symbolically an estimator $D^pR^q$ with $p+q=N$
for a function $\Theta$ symmetric in its arguments

\begin{equation}
D^pR^q = \sum \Theta(\vx_1,\ldots,\vx_p,\vy_1,\ldots,\vy_q)
\label{eq:dprq} 
\end{equation}

with $\vx_i \neq \vx_j \in D$ and $\vy_i \neq \vy_j \in R$ are objects
positions in the galaxy catalog and the random catalog respectively. 
The generalized LS estimator reads

\begin{equation}
{\hat f}_N = \frac{1}{S} \sum_i \left(\begin{array}{c} N \\ i
\end{array} \right) (-1)^{N-i} \left( \frac{D}{{\bar n}_{\rm g}}
\right)^i \left( \frac{R}{{\bar n}_{\rm r}} \right)^{N-i},
\label{eq:genLS}
\end{equation}

where the normalization number $S$ is given by

\begin{equation}
S\equiv \int \Theta(\vx_1,\ldots,\vx_N) \d^{\dim}\vx_1\ldots
\d^{\dim}\vx_N.
\end{equation}

If ${\bar n}_{\rm g}$ is determined with arbitrary accuracy the
estimator (\ref{eq:genLS}) is unbiased, optimally edge corrected in
the weak-correlation limit~\cite{SzSz98}.  For practical measurements,
however, ${\bar n}_{\rm g}$ is determined from the catalog itself, and
the integral constraint problem arises again, as described in
Sect.~\ref{sec:secintegra}.

The cosmic covariance of ${\hat f}_N$ assuming that $n_{\rm g}$ is
perfectly determined was given in~\cite{Szapudi00},
\begin{eqnarray}
{\rm Cov}(f_{N_1},f_{N_2}) & \equiv & \langle {\hat f}_{N_1,a}
{\hat f}_{N_2,b} \rangle - \langle {\hat f}_{N_1,a} \rangle
 \langle {\hat f}_{N_2,b} \rangle \nonumber \\
 & = &\frac{1}{S^2} \sum_{i,j}
\left(\begin{array}{c} N_1 \\ i \end{array} \right)
\left(\begin{array}{c} N_2 \\ j \end{array} \right) (-1)^{i+j} \left[
E(i,j,N_1,N_2) \right.  \nonumber \\ & - & \left.  {\cal S}_{0}\{
f_i(1,\ldots,i)f_j(N_1+1,\ldots,N_1+j) \} \right], \label{eq:cosmicerrorgen}
\end{eqnarray} 

with

\begin{eqnarray}
E(p_1,p_2,N_1,N_2) & \equiv & \left\langle \left( \frac{D}{{\bar
n}_{\rm g}}\right)^{p_1} \left( \frac{R}{{\bar n}_{\rm
r}}\right)^{N_1-p_1} \left( \frac{D}{{\bar n}_{\rm g}}\right)^{p_2}
\left( \frac{R}{{\bar n}_{\rm r}}\right)^{N_2-p_2} \right\rangle
\nonumber \\ &=& \sum_{i} \left(\begin{array}{c} p_1 \\ i \end{array}
\right) \left(\begin{array}{c} p_2 \\ i \end{array} \right) \ i!\
{\bar n}_{\rm g}^{-i} {\cal S}_i\{ f_{N_1+p_1+p_2-i} \}
\end{eqnarray}

where the operator $S_i$ is defined by

\begin{eqnarray}
{\cal S}_k\{g\} & \equiv & \int \d^{\dim}\vx_1\ldots
\d^{\dim}\vx_{N_1+N_2-k} \nonumber \\ & &
\Theta_a(1,\ldots,N_1)\ \Theta_b(1,\ldots,k,N_1+1,\ldots,N_1+N_2-k) \nonumber
\\ & & g(1,\ldots,p_1,N_1+1,\ldots,N_1+p_2-k), \label{eq:jenesaispas}
\end{eqnarray}
and the convention that \small ${k \choose l}$ \normalsize is nonzero
only for $k \geq 0$, $l \geq 0$ and $k \geq l$.  In these equations we
have used the short-hand notations, $1=\vx_1,\ \ldots,\ i=\vx_i$,
etc., and $g$ should be viewed as $f_i(1,\ldots,i)
f_j(N_1+1,\ldots,N_1+j)$ in Eq.~(\ref{eq:jenesaispas}) to compute the
${\cal S}_0$ term in Eq.~(\ref{eq:cosmicerrorgen}).

Equation (\ref{eq:cosmicerrorgen}) assumes that the random catalog
contains a very large number of objects, ${\bar n}_{\rm r} \rightarrow
\infty$, i.e. does not take into account errors brought by the
finiteness of $N_{\rm r}$ (see \cite{Szapudi00} for more details). 
Using a computer algebra package, one can derive from this formalism
Eq.~(\ref{eq:cosmicerrgen}).  Similar but cumbersome expression for
the three-point correlation function can be found in~\cite{Szapudi00}.

Note, as suggested in~\cite{SzSz98}, that this formalism can be
applied to Fourier space, i.e. to the power-spectrum (see~\cite{SPC01}
for a practical implementation of estimator ${\hat f}_2$ in harmonic
space) and to the bispectrum.  It can also be theoretically applied to
one-point distribution functions, such as count-in-cells, studied
below, but it was not done so far.  Therefore, we shall instead
present results relying on a more traditional approach in the next
section.

Note that for the bispectrum, some work has been done in computing its
covariance matrix and cosmic bias in particular cases. 
In~\cite{MVH97}, the bispectrum covariance matrix is estimated
including shot-noise terms and beyond the Gaussian
approximation\footnote{Estimation of the cosmic error in the Gaussian
approximation is given in~\cite{FMS93,SCFFHM98}.} by using
second-order Eulerian PT\footnote{This is however only approximate
since a consistent calculation of the connected six-point function
requires up to fifth-order Eulerian PT, a quite complicated
calculation.}.  A numerical calculation of the bispectrum covariance
matrix and the cosmic bias expected for IRAS surveys is presented
in~\cite{Scoccimarro00b} using 2LPT\footnote{This is also not a
consistent calculation of non-Gaussian terms in the covariance matrix;
however 2LPT does include significant contributions to any order in
Eulerian PT, and comparison for one-point moments suggest 2LPT is a
very good approximation~\cite{Scoccimarro98}.}.

\subsection{One-Point Distributions: Counts-in-Cells}
\label{sec:sec5}

\subsubsection{Definitions:}
The Count Probability Distribution Function (CPDF) was introduced in
\\ Sect.~\ref{sec:poirea}.  Here we give more definitions on
count-in-cells statistics, such as factorial moments and their
relation to cumulants and the CPDF in terms of generating functions. 
Some additional information can be found as well in
Appendix~\ref{app:pdedelta}.

Following the presentation in Sect.~\ref{sec:poirea}, we discuss in
more detail here an elegant way of correcting for discreteness
effects, which makes use of the {\em factorial moments}.  These are
defined as follows:
\begin{equation}
F_k\equiv \langle (N)_k \rangle = \langle N(N-1)\cdots(N-k+1) \rangle
= \sum_N (N)_k P_N. \label{eq:facmomdef}
\end{equation}
Note thus that ${\bar N} =F_1$. We have
\begin{equation}
F_k = {\bar N}^k \langle (1+\delta)^k \rangle,
\end{equation}
so $F_k/{\bar N}^k$ estimates directly 
the moment of order $k$ of the underlying (smoothed) density field.

The generating function of the counts

\begin{equation}
{\cal P}(t)\equiv \sum_N t^N P_N
\end{equation}

is related to the moment generating function through

\begin{equation}
{\cal M}(\barN t)={\cal P}(t+1).
\end{equation}

Factorial moments thus verify

\begin{equation}
F_k=\left.  \left( \frac{\partial}{\partial t} \right)^k {\cal P}(t+1)
\right|_{t=0}.
\end{equation}

It is easy to find, using Eq.~(\ref{eq:varphieq}), the following
useful recursion~\cite{SzSz93a} relating factorial moments to
quantities of physical interest, $S_p$,

\begin{equation}
S_p=\frac{\xiav_2 F_p}{N_{\rm c}^p}-\frac{1}{p}\sum_{q=1}^{p-1}
\left(\begin{array}{c} p \\ q \end{array} \right) \frac{(p-q)S_{p-q}
F_q}{N_{\rm c}^q}, \label{eq:SqfromFk}
\end{equation}

where $N_{\rm c}$ is the typical number of object in a cell in
overdense regions, $N_{\rm c} \equiv \barN\ \xiav_2$.

\subsubsection{Estimators}

\label{sec:sectesti}

In practice, the measurement of the CPDF and its factorial moments is
very simple.  It consists of throwing $C$ cells at random in the
catalog and computing

\begin{equation}
{\hat P}_N^C = \frac{1}{C} \sum_{i=1}^C \delta_{N_i,N},
\label{eq:estpn}
\end{equation}

where $\delta_{N,M}$ is the Kronecker delta function, and $N_i$
denotes the number of objects in cell ``$i$''.  Similarly, the
estimator for the factorial moment of order $k$ is

\begin{equation}
{\hat F}_k^C = \frac{1}{C} \sum_{i=1}^C (N_i)_k, \label{eq:estfk}
\end{equation}

or can be derived directly from ${\hat P}_N^C$ using
Eq.~(\ref{eq:facmomdef}).  Estimators (\ref{eq:estpn}) and
(\ref{eq:estfk}) are unbiased.  However, if one uses the relation
(\ref{eq:SqfromFk}) to compute cumulants from factorial moments, i.e.

\begin{eqnarray}
{\hat {\xiav}}&=&\frac{\tF_2}{\tF_1^2}-1, \label{eq:estxi}\\ {\hat
S_3} & = & \frac{\tF_1(\tF_3-3 \tF_1 \tF_2 + 2
\tF_1^3)}{(\tF_2-\tF_1^2)^2}, \label{eq:ests3}\\ {\hat S_4} & = &
\frac{\tF_1^2 (\tF_4 - 4\tF_3\tF_1 - 3\tF_2^2 + 12 \tF_2 \tF_1^2 -6
\tF_1^4)}{(\tF_2-\tF_1^2)^3},\label{eq:ests4}
\end{eqnarray}

the corresponding estimators are biased, because nonlinear
combinations of estimators are generally biased
(e.g.~\cite{HuGa99,SCB99}).

To reduce the bias and the errors on direct measurements of cumulants
from Eqs.~(\ref{eq:estxi}), (\ref{eq:ests3}), (\ref{eq:ests4}) it is
possible to use some prior information, for example by assuming that
the PDF of the underlying density field is given by the Edgeworth
expansion, Eq.~(\ref{fb:edge}), convolved with a Poisson distribution
to take into account discreteness, Eq.~(\ref{fb:PoiConv}).  This
procedure was actually applied to the IRAS $1.2$Jy galaxy
catalog~\cite{KiSt98}.  The advantage of such a method is that it can
be less sensitive to finite volume effects by using the shape of the
PDF near its peak (since finite volume effects mainly affect the
tails).  One disadvantage, is that the validity of the Edgeworth
expansion is quite restricted, even in the weakly non-linear regime
(see, e.g.~\cite{JWACB95}).  In particular, the PDF is not positive
definite.  Convolution with the Poisson distribution to account for
discreteness alleviates this problem for the sparse IRAS
surveys~\cite{KiSt98}; however, for applications to the next
generation of galaxy surveys this will likely not be the case. 
Another difficulty of this approach is that error estimation is not
straightforward.  On the other hand, the idea of using prior
information on the shape of the PDF to estimate moments is certainly
worth pursuing with a more detailed modeling of the density PDF.

\subsubsection{Error Propagation: Cosmic Bias vs. Cosmic Error}

\label{sec:secerrprop}

We now review the theory of error propagation in a general setting for
functions of correlated random variables, following the treatment
in~\cite{SCB99}\footnote{For a different approach, based on an
expansion in terms of the variance at the scale of the survey
see~\cite{HuGa99}.}.  This theory was actually behind the calculation
of the errors on the two-point correlation function in
Sect.~\ref{sec:sec2}.  Since the calculations are necessarily
technical, we only present computations of the cosmic bias and error
on nonlinear estimators such as those given by Eqs.~(\ref{eq:estxi}),
(\ref{eq:ests3}) and (\ref{eq:ests4}).

Let us suppose that we measure a quantity $f(\tvx)$, where $\tvx$ is a
vector of unbiased estimators, such as the factorial moments, and that
the measurement of $\tvx$ is sufficiently close to the ensemble
average $\langle \tvx \rangle=\vx$.  Then $f$ can be expanded around
the mean value

\begin{equation}
f(\tvx)=f(\vx)+\sum_k \frac{\partial f}{\partial x_k}\delta \tx_k
+\frac{1}{2} \sum_{k,l} \frac{\partial^2 f}{\partial x_k \partial
x_l}\delta \tx_k \delta \tx_l +{\cal O}(\delta x^3), \label{eq:expans}
\end{equation}

where $x_k$ is the $k$-th component of $\tvx$ and

\begin{equation}
\delta \tx_k=\tx_k-x_k.
\end{equation}

After ensemble average of Eq.~(\ref{eq:expans}) one obtains

\begin{equation}
\langle f \rangle=f(\vx)+\frac{1}{2}\sum_{k,l} \frac{\partial^2
f}{\partial x_k \partial x_l} \langle \delta \tx_k \delta \tx_l
\rangle + {\cal O}(\delta x^3).
\end{equation}

To second order the cosmic bias [Eq.~(\ref{eq:cosmicbiasdef})] thus
reads

\begin{equation}
b_f\simeq \frac{1}{2f(\vx)}\frac{\partial^2 f}{\partial x_k \partial
x_l}\langle \delta \tx_k \delta \tx_l \rangle. 
\label{eq:cosmicbiasapp}
\end{equation}

Similarly the covariance between two functions $f$ and $g$ is

\begin{equation}
{\rm Cov}(f,g)=\langle \delta {\hat f} \delta {\hat g}
\rangle=\sum_{k,l} \frac{\partial f}{\partial x_k} \frac{\partial
g}{\partial x_l} \langle \delta \tx_k \delta \tx_l \rangle + {\cal
O}(\delta x^3).  \label{eq:cosmiccross}
\end{equation}

In particular, the relative cosmic error is given by 

\begin{equation} 
\sigma_f\equiv \frac{\Delta f}{\langle f \rangle}= \sqrt{{\rm
Cov}(f,f)}/\langle f \rangle.
\end{equation}

It is important to notice the following point, from
Eqs.~(\ref{eq:cosmicbiasapp}) and (\ref{eq:cosmiccross}):

\begin{equation}
b_f \sim {\cal O}(\sigma_f^2).
\end{equation}

The range of applicability of this perturbative theory of error
propagation is $\langle \delta \tx_k \delta \tx_l \rangle/x_k x_k \ll
1$: errors and cross correlations of the vector $\tvx$ must be weak. 
In this regime {\em the cosmic bias is always smaller than the
relative cosmic error}, except for accidental cancellations in
Eq.~(\ref{eq:cosmicbiasapp}) (in that case, the next order would be
needed in the expansion).  When the cosmic bias becomes large the
expansion in Eq.~(\ref{eq:expans}) breaks down; in this case,
numerical simulations show that the cosmic bias can be larger than the
relative cosmic error~\cite{HuGa99}.

\subsubsection{Cosmic Error and Cross-Correlations of Factorial Moments}

\label{sec:secfacmomtheo}

According to the above formalism, the knowledge of errors and
cross-corre\-la\-tions on a complete set of unbiased estimators, such
as the factorial moments, $F_k$, $k=1,...,\infty$, or count-in-cells
themselves, $P_N$, allows the calculation of the cosmic error (or
cross-correlations) on any counts-in-cells statistics.  The general
theoretical framework for computing the cosmic error on factorial
moments can be found in~\cite{SzCo96} and~\cite{SCB99}\footnote{See
the earlier work in~\cite{CBS95} for detailed calculations of the void
probability function cosmic error.}.  Here we review the main results.

First, it is important to notice that there is a source of error due
to the finiteness of the number of cells $C$ used in
Eqs.~(\ref{eq:estpn}) and (\ref{eq:estfk}).  This source of error,
which is estimated in~\cite{SzCo96}, can be rendered arbitrarily small
by taking very large number of sampling cells, $C$, or by using an
algorithm equivalent to infinite sampling, $C\rightarrow \infty$ as
proposed in~\cite{Szapudi98a}.  Measurements are often done using
$C\simeq V/v$, i.e. the number of cells necessary to cover the
sample, which {\em is not} a good idea.  Indeed, such small number of
sampling cells {\em does not, in general, extract all the
statistically significant information} from the catalog, except in
some particular regimes in the Poisson limit.  The best way to measure
count-in-cells statistics is thus to do {\em as massive oversampling
as possible}\footnote{This is because missing clusters cores, which
occupy a very small fraction of the volume, leads to underestimation
of higher-order moments.} and estimate the cosmic error independently,
as explained below.  Similarly, when measuring the two-point
correlation function using a Poisson sample $R$ to estimate $RR$ and
$DR$, in order to avoid adding noise to the measurements, the random
catalog $R$ should have as many objects as possible.  Having that in
mind, we shall assume from now that $C$ is very large.

The error generating function is defined as follows

\begin{equation}
{\cal E}(x,y)=\sum_{N,M} \left[ \langle \tP_N \tP_M \rangle - \langle
\tP_N \rangle \langle \tP_M \rangle \right],
\end{equation}

where the ensemble average $\langle \ldots \rangle$ denotes the
average over a large number of realizations of the catalog with same
geometry and same underlying statistics.  Then, the cosmic covariance
on factorial moments and count-in-cells reads

\begin{eqnarray}
\Delta_{k,l}\equiv {\rm Cov}(F_k,F_l) & = & \left.  \left(
\frac{\partial}{\partial x} \right)^k \left( \frac{\partial}{\partial
y} \right)^y {\cal E}(x+1,y+1)\right|_{x=y=0},\label{eq:errorini}\\
{\rm Cov}(P_N, P_M) & = & \left.  \left( \frac{\partial}{\partial x}
\right)^N \left( \frac{\partial}{\partial y} \right)^M {\cal
E}(x,y)\right|_{x=y=0}.
\end{eqnarray}

The error generating function can be written in terms of bivariate
distributions

\begin{equation}
{\cal E}(x,y)=\frac{1}{{\hat V}^2} \int_{\hat V} \d^{\dim} \vr_1 \d^{\dim}
\vr_2 \left[ {\cal P}(x,y) - {\cal P}(x) {\cal P}(y) \right]. 
\label{eq:errcosme}
\end{equation}

In this equation, ${\hat V}$ is the volume covered by cells included
in the catalog and ${\cal P}(x,y)$ is the generating function of
bicounts $P_{N,M}$ for cells separated by a distance $|\vr_1-\vr_2|$
(see also Sect.~\ref{sec:sec5bis} below):

\begin{equation}
{\cal P}(x,y) \equiv \sum_{N,M} x^N y^M P_{N,M}.  \label{eq:genebiv}
\end{equation}

The calculation of the function ${\cal E}(x,y)$, detailed in
Appendix~\ref{cosmicerrorfacmom}, is simplified by separating the
integral in Eq.~(\ref{eq:errcosme}) into two components, $E_{\rm
overlap}(x,y)$ and $E_{\rm disjoint}(x,y)$, according to whether cells
overlap or not.

At leading order in $v/V$, $\Delta_{k,l}$ has three contributions

\begin{equation}
\Delta_{k,l}=\Delta_{k,l}^{\rm F}+\Delta_{k,l}^{\rm
E}+\Delta_{k,l}^{\rm D}, \label{eq:cosmiccovtot}
\end{equation}

where $\Delta_{k,l}^{\rm F}$, $\Delta_{k,l}^{\rm E}$ and
$\Delta_{k,l}^{\rm D}$ are the finite volume, edge and discreteness
effect contributions, respectively.  From~\cite{SzCo96}
and~\cite{SCB99}, the first few terms in the three-dimensional case
are listed in Appendix~\ref{cosmicerrorfacmom}.

The finite-volume error comes from the disjoint cells contribution in
the error generating function.  The corresponding relative error, or
cross-correlation, $\Delta^{\rm F}_{k,l}/(F_k F_l)$ does not depend on
the number of objects in the catalog, and is proportional to the
integral of the two-point correlation function over the survey volume:

\begin{equation}
\xiav({\hat L}) \equiv \frac{1}{\hat V} \int_{r_{12}\geq 2R} \d^{\dim}
\vr_1 \d^{\dim} \vr_2 \xi(r_{12}).  \label{eq:xiavhatl}
\end{equation}

The edge effect term, $\Delta^{\rm E}_{k,l}/(F_k F_l)$, is the
contribution remaining from overlapping cells in the continuous limit,
$\barN \rightarrow \infty$.  It does not depend on the number of
objects in the catalog and is proportional to $\xiav v/V$.  A pure
Poisson sample does not have edge effect error at leading order in
$v/V$, in agreement with intuition.  The discreteness effect error,
$\Delta^{\rm D}_{k,l}/(F_k F_l)$, is the contribution from overlapping
cells which depends on ${\bar N}$ and thus disappears in the
continuous limit.  As discussed in the introduction of this chapter,
the separation between these three contributions is useful but
somewhat arbitrary.  For example, Eq.~(\ref{eq:xiavhatl}) actually
contains some edge effects through the constrain $r_{12} \geq 2R$, as
shown in Appendix~\ref{cosmicerrorfacmom}.

Furthermore, if next to leading order contributions in $v/V$ are
considered, the corresponding correction is proportional to the
contour of the survey, $\partial V$~\cite{Ripley88,CCDFS00}.  Each
contribution, $\Delta^{\rm X}_{k,l}/(F_k F_l)$, ${\rm X}={\rm F}$, E
or D contains a term proportional to $\partial V$.  This correction is
an edge correction, leading to terms such as edge-finite-volume and
edge-discreteness contributions in our nomenclature.

It is important to emphasize that the expressions given in
Appendix~\ref{cosmicerrorfacmom} {\em are of direct practical
use}\footnote{They have been implemented in the publically available
FORCE package~\cite{SCB99}.} for estimating errors on factorial
moments or on cumulants (Sect.~\ref{sec:cumerr}) using the theory of
propagation of errors explained above
(e.g.~\cite{HSB00,SBFMS00,SPLO01} for applications to actual
measurements in real galaxy catalogs).  Similarly as in
Eq.~(\ref{eq:cosmicxi}), a careful examination of these expressions
shows that prior knowledge of the shape of the two-point correlation
function $\xi$ [namely, $\xiav$ and $\xiav({\hat L})$] and
higher-order statistics, $S_p$ and $C_{p\,q}$ up to some value of $p$
and $q$ is necessary to compute $\Delta_{k,l}$.  To estimate cumulants
$\xiav$ and $S_p$, one can simply use the values directly measured in
the catalog or other existing estimates
(e.g.~\cite{Gaztanaga94,SMN96}), as well as existing fitting formulae
for $\xiav$ (\cite{HKLM91,PeDo94,JMW95,PeDo96}, see
Sect.~\ref{nlevtp}) and PT, EPT (\cite{CBBH97}, see
Sect.~\ref{sec:EPT}) or HEPT (\cite{ScFr99}, see
Sect.~\ref{sec:HEPT}) for $S_p$.  To compute $\xiav({\hat L})$ it is
necessary to make assumptions about the cosmological model.  The
cumulant correlators $C_{pq}$ can be estimated directly from the
catalog or from various models which further simplify the calculations
(e.g.~\cite{BeSc92,SzSz93a,SCB99}).  These models can be particular
cases of the hierarchical model, Eq.~(\ref{HM}), or can rely on PT
results (Sect.~\ref{sec:jpdf}) or extensions such as E$^2$PT
(Sect.~\ref{sec:EPT}).

\begin{figure}[h!]
\centering \centerline{\epsfxsize=14.  truecm
\epsfbox{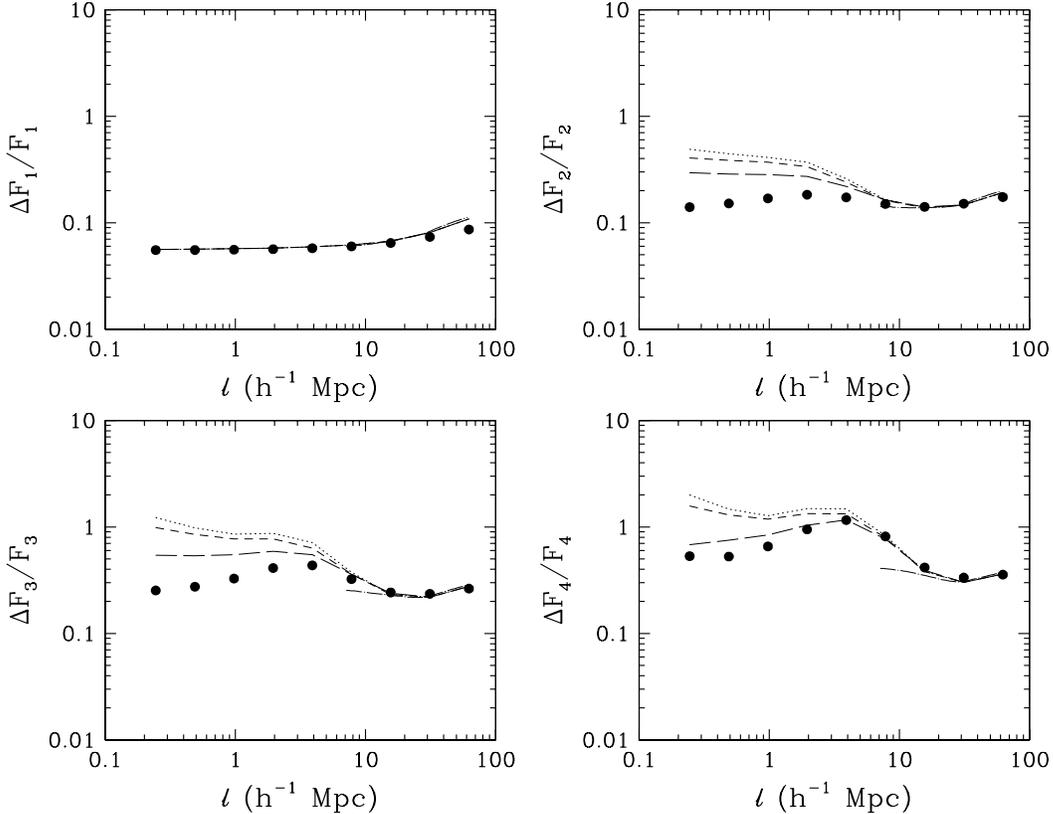}} \caption[]{The relative cosmic error on
factorial moments measured as a function of scale~\cite{CSJC99},
obtained from the dispersion over a large ensemble of subsamples
extracted from one of the Hubble volume
simulations~\cite{Evrardetal99}, as explained in the text.  The
dotted, dashed, long dashed, dot-long dashed curves correspond
respectively to theoretical predictions based on two particular cases
of the hierarchical model, namely SS and BeS, E$^2$PT and PT. The SS
model~\cite{SzSz93a} assumes $Q_{NM}=Q_{N+M}$ with the definition in
Eq.~(\ref{eq:defqnnew}).  The BeS model~\cite{BeSc92} is more
complicated, but obeys $Q_{NM}=Q_{N1}Q_{M1}$, as in the E$^2$PT
framework, described in Sect.~\ref{sec:EPT}.  The PT results are
shown only in the weakly nonlinear regime, $\xiav \la 1$.}
\label{fig:fig-chap7-1}
\end{figure}

Among the models tested, the best known so far is E$^2$PT as
illustrated by Figure~\ref{fig:fig-chap7-1}.  In this figure, taken
from~\cite{CSJC99}, the cosmic error on factorial moments is measured
from the dispersion over 4096 subsamples of size $L=125\ h^{-1}$ Mpc,
extracted from a $\tau$CDM simulation of size $2000\ h^{-1}$ Mpc
involving $1000^3$ particles~\cite{Evrardetal99}.  The accuracy of
theoretical predictions is quite good, especially at large, weakly
nonlinear scales.  At small scales, all the models tend to
overestimate the magnitude of the errors, including E$^2$PT, but the
disagreement between theory and measurements is at most a factor two
approximately.  This discrepancy suggests that details of the dynamics
still need to be understood in order to describe appropriately
multivariate distribution functions in the highly nonlinear regime.

\subsubsection{Cosmic Error and Cosmic Bias of Cumulants}

\label{sec:cumerr}

Using the results in Sects.~\ref{sec:secerrprop} and
\ref{sec:secfacmomtheo} it is possible to compute the cosmic bias and
the cosmic error on estimators (\ref{eq:estxi}) (\ref{eq:ests3}) and
(\ref{eq:ests4}) (see also~\cite{HuGa99}).  It would be too cumbersome
to put all the results here, but getting analytic expressions similar
to what was obtained for $\Delta_{k,l}$ is very easy with standard
mathematical packages.  For example, simple algebraic calculations
give for the cosmic bias

\begin{equation}
b_{\xiav}=\frac{F_2}{\xiav \barN^2} \left(
\frac{3\Delta_{11}}{\barN^2} - \frac{2\Delta_{12}}{\barN F_2} \right),
\end{equation}

\begin{equation}
b_{S_3} = b_{\xiav_3}- 3 b_{\xiav} - \frac{2\Delta_{23}}{F_2 F_3}
+\frac{3\Delta_{22}}{F_2^2},
\end{equation}

with

\begin{equation}
b_{\xiav_3}=\frac{F_3}{\xiav_3}{\barN^3} \left( \frac{6
\Delta_{11}}{\barN^2} -\frac{3 \Delta_{13}}{\barN F_3} \right) -3
\frac{F_2}{\xiav_3 F_1^2} \left( \frac{3 \Delta_{11}}{\barN^2}
-\frac{2\Delta_{12}}{\barN F_2} \right).
\end{equation}

Similarly, the cosmic errors read

\begin{equation}
\sigma_\xi^2 \simeq \frac{1}{\barN^6}\left(4 F_2^2 \Delta_{11} - 4
\barN F_2 \Delta_{21} + \barN^2 \Delta_{22} \right),
\end{equation}

\begin{eqnarray}
\sigma^2_{S_3} & \simeq & \frac{1}{\barN^{12} \xiav^6 S_3^2}\left[
\left( 2 \barN^3 F_2 - 6 \barN F_2^2 + 3 \barN^2 F_3 + F_2 F_3
\right)^2 \Delta_{11} \right.  \nonumber \\ & + & 2 \barN \left( -2
\barN^6 F_2 + 12 \barN^4 F_2^2 -18 \barN^2 F_2^3 - 3 \barN^5 F_3
\right.  \nonumber \\ & &+ \left.  4 \barN^3 F_2 F_3 +15 \barN F_2^2
F_3 - 6 \barN^2 F_3^2 -2 F_2 F_3^2 \right) \Delta_{12} \nonumber \\ &
+ & 2 \barN^3 \xiav \left( 2 \barN^3 F_2 - 6 \barN F_2^2 + 3 \barN^2
F_3 + F_2 F_3 \right) \Delta_{13} \nonumber \\ & + & \barN^2 \left(
\barN^3 - 3 \barN F_2 + 2 F_3 \right)^2 \Delta_{22} \nonumber \\ & + &
\left.  2 \barN^4 \xiav \left( \barN^3 - 3 \barN F_2 + 2 F_3 \right)
\Delta_{23} + \barN^6 \xiav^2 \Delta_{33} \right].
\end{eqnarray}

It is interesting to compare the results obtained for $\xiav$ to what
was derived for function $\xi(r)$.  For example, replacing
$\Delta_{kl}$ and $F_k$ with their value as functions of $\barN$ and
cumulants leads  to the following result for the cosmic
bias in the 3-D case~\cite{SCB99}

\begin{eqnarray}
b_{\xiav} & \simeq &\left( 0.04-\frac{1}{\xiav} \right)
\frac{v}{\barN V} +\left(16.5 - 7.6 S_3 - \frac{0.53}{\xiav} \right)
\frac{\xiav v}{V} \nonumber\\ && + \left( 3 - 2 C_{1\,2} -
\frac{1}{\xiav} \right) \xiav({\hat L}).
\end{eqnarray} 

In this equation, valid in the perturbative regime ($|b_{\xiav}| \ll
\sigma_{\xiav} \ll 1$) and at leading order in $v/V$ one can
recognize in the first, second and third terms the discreteness, edge,
and finite volume effect contributions, respectively.  As expected,
the last line is very similar to Eq.~(\ref{eq:cosmicbias1}).  Note
that the discreteness effect term is rather small and can be neglected
in most realistic situations, in agreement with
Eq.~(\ref{eq:cosmicbias1}).  An alternative calculation of $b_{\xiav}$
can be found in~\cite{HuGa99} with similar conclusions.

\begin{figure}
\centerline{\hbox{\psfig{figure=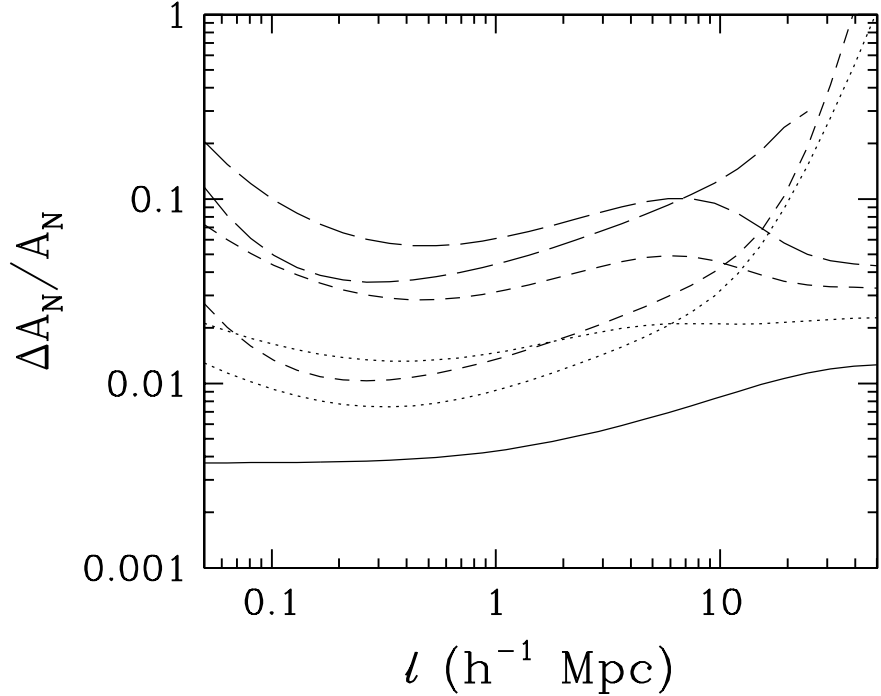,bbllx=48pt,bblly=112pt,bburx=521pt,bbury=488pt,width=12cm}}}
\caption[]{ Comparison of the cosmic errors for the factorial and
connected moments expected in the SDSS~\cite{SCB99}.  Standard CDM is
assumed for the two-point correlation function and E$^2$PT with
$n_{\rm eff} = -2.5$ for higher-order statistics.  Solid, dotted,
dash, and long dash lines correspond to orders $1$ through $4$,
respectively.  Of each pair of curves with the same line-types the one
turning up on large scales relates to the cumulant.  Note that the
perturbative approach used to compute the cosmic error on the
cumulants fails at large scales, explaining the right stopping point
of the long dash curve for $S_4$.  } \label{fig:cosmicSDSS}
\end{figure}

Figure~\ref{fig:cosmicSDSS} displays the cosmic error as a function of
scale for factorial moments and cumulants expected in the SDSS. It
illustrates how these different estimators perform and shows that the
relative error on the cumulants $\xiav$, $S_3$ and $S_4$ is expected
to be smaller 3, 5 and 15 percent, respectively in the scale range
$1-10\,h^{-1}$ Mpc~\cite{SCB99}.

\subsection{Multivariate Count-in-Cells}

\label{sec:sec5bis}

The generalization of count-in-cells to the multivariate case is quite
straightforward.  Here we focus on bivariate statistics, which were
used to compute the cosmic error on count-in-cells estimators in
Sect.~\ref{sec:secfacmomtheo}.

For a pair of cells at position $\vr_1$ and $\vr_2$ separated by
distance $r=|\vr_1-\vr_2|$, factorial moment correlators~\cite{SDES95}
are defined as

\begin{equation}
\facmomcor_{kl}(r_{12})\equiv \frac{ F_{kl} - F_k F_l }{ {\bar
N}^{k+l}},
\end{equation}

\begin{equation}
\facmomcor_{k0} \equiv \frac{ F_{k0} }{ {\bar N}^k } \equiv \frac{ F_k
}{ {\bar N}^k },
\end{equation}

where the joint factorial moment is given by

\begin{equation}
F_{kl}(r_{12})\equiv \langle (N)_k (N)_l \rangle.
\end{equation}

Similarly as factorial moments, $F_{kl}$ estimates joint moments of
the smoothed density field

\begin{equation}
F_{kl}(r_{12})= {\bar N}^{k+l} \langle [1+\delta(\vr_1)]^k
[1+\delta(\vr_2)]^l\rangle.
\end{equation}

The joint factorial moments and thus the factorial moment correlators
can be easily related to the quantities of physical interest, namely
the two-point density normalized cumulants -- also designed by
cumulant correlators~\cite{SzSz97}, $C_{p\,q}$ [Eq.~(\ref{fb:Spqdef})]. 
Indeed, as for the monovariate case, one can write

\begin{equation}
F_{kl}=\left.  \left( \frac{\partial}{\partial x} \right)^k \left(
\frac{\partial}{\partial y} \right)^l {\cal P}(x+1,y+1)
\right|_{x=y=0},
\end{equation}

\begin{equation}
{\cal M}({\bar N}x,{\bar N}y)=\exp\left[ {\cal C}(x,y) \right] ={\cal
P}(x+1,y+1),
\end{equation}

where ${\cal P}(x,y)$ is the generating function for bicounts defined
previously in Eq.~(\ref{eq:genebiv}), ${\cal M}(x,y)=\langle
\exp[x\delta(\vr_1)+y\delta(\vr_2)] \rangle$ is the moment generating
function (Sect.~\ref{sec:sec3.3.2}) and ${\cal C}(x,y)$ is the
two-point density cumulant generating function
[Eq.~(\ref{eq:cumulmulti})].  For example, the first few cumulant
correlators are~\cite{SzSz97}

\begin{eqnarray}
C_{1\,2}\ \xiav\ \xi & = & \facmomcor_{12} - 2 \xi, \\ C_{1\,3}\ \xiav^2\ \xi
& = & \facmomcor_{13} - 3 \facmomcor_{12} - 3 \facmomcor_{20} + 6 \xi,
\\ C_{2\,2}\ \xiav^2\ \xi & = & \facmomcor_{22} - 4 \facmomcor_{12} + 4
\xi - 2 \xi^2,
\end{eqnarray}

with $\xi \equiv \xi(r_{12})$.  We have used the approximation
$\facmomcor_{11} \simeq \xi$, valid when $r_{12} \gg R$.

An unbiased estimator for the joint factorial moment $F_{kl}$
analogous to Eq.~(\ref{eq:estfk}) is simply, for a set of $P$ pairs of
cells in the catalog separated by distance $r$ and thrown at random
(with random direction),

\begin{equation}
{\hat F}_{kl}^P(r) = \frac{1}{2P} \sum_{{\rm pairs}\ (i,j)}
\left[(N_i)_k (N_j)_l + (N_i)_l (N_j)_k \right].
\end{equation}

A possible (biased) estimator for the factorial moments correlators is
then, for the same set of cells,

\begin{equation}
\tfacmomcor_{kl}^P=\frac{{\hat F}_{kl}^P-{\hat F}_{k0}^P {\hat
F}_{l0}^P}{[{\hat F}_{10}^P]^{k+l}},
\end{equation}

with the definition

\begin{equation}
{\hat F}_{k0}^P \equiv \frac{1}{2P} \sum_{{\rm pairs}\ (i,j)}
\left[(N_i)_k + (N_i)_l \right].
\end{equation}

At this point, it is interesting to notice again that
${\tfacmomcor}_{11}$ can be used directly as an estimator of the
two-point correlation function, if the cell size $R$ is small compared
to the separation $r$ (e.g.~\cite{PeHa74,GrPe77}).  In that case, the
averages are done on sets of pairs of cells in a bin $\Theta$ as
defined in Sect.~\ref{sec:estxi2}.

Further generalization to higher-order multivariate statistics is
trivial.  For example, $\tfacmomcor_{111}$ can be used to estimate the
three-point correlation function (e.g.~\cite{GrPe77}),
$\tfacmomcor_{1111}$ to estimate the four-point correlation function
(e.g.~\cite{FrPe78}) and so on.

\subsection{Optimal Weighting}

\label{sec:sec6}

To optimize the measurements of $N$-point statistics, the data can be
given a varying spatial weight $\omega(\vr_1,\ldots,\vr_N)$ symmetric
in its arguments and properly normalized.  Furthermore, in realistic
redshift surveys, the average number density of galaxies changes with
distance $r$ from the observer:

\begin{equation}
{\bar n}_{\rm g} (r)={\bar n}_{\rm g} \phi(r),
\end{equation} 

where $\phi(r) \leq 1$ is the selection function.  Now, the estimators
defined so far are valid only for {\em statistically homogeneous}
catalogs, i.e. with constant ${\bar n}_{\rm g}(r)$.  One way to avoid
this problem is to use volume limited catalogs.  This method consists
in extracting from the parent catalog subsamples of depth $R_i$ such
that the apparent magnitude of objects in these catalogs at distance
$r=R_i$ from the observer would be larger than the magnitude limit. 
Such a selection criterion renders the number density of galaxies in
the subsamples independent of distance at the price of a significant
information loss\footnote{However, a number of volume-limited samples
can be constructed from the parent catalog to compensate for this.}. 
In order to be able to extract all the information from the catalog,
it is however possible to correct the estimators for the spatial
variation of ${\bar n}_{\rm g}(r)$.  Moreover, the signal to noise can
be further improved by appropriate choice of the weight function
$\omega$.

The generalization of Eq.~(\ref{eq:dprq}) reads

\begin{equation}
D^pR^q = \sum \frac{\omega(\vx_1,\ldots,\vx_p,\vy_1,\ldots,\vy_q)}{
\phi(\vx_1)\dots\phi(\vx_p)\phi(\vy_1)\dots\phi(\vy_q)}
\Theta(\vx_1,\ldots,\vx_p,\vy_1,\ldots,\vy_q).  \label{eq:estimasamp}
\end{equation}

(We assume that same selection effects are applied to the random
catalog $R$).  Note that the weight could be included in the bin
function $\Theta$, but we prefer to separate the idea of spatial
weighting from the idea of binning.  In principle, the binning can
change slightly the nature of the measured statistic $A$ in
$A_{\Theta} \neq A$.  Of course, up to now we have assumed that the
binned quantity is always close to the quantity of interest,
$A_{\Theta} \simeq A$, but this condition is not absolutely necessary:
the binning function $\Theta$ can be chosen arbitrarily and determined {\em
a priori}.  Then the statistic of interest becomes $A_{\Theta}$
instead of the original $A$.  For example, count-in-cells represent a
particular choice of the binning function.  On the other hand the
spatial weight should not bring any change, i.e., the weighted
quantity, should be, after ensemble average, equal to the real value
(or at least, very close to it): $\langle \tA_{\Theta,\omega} \rangle=
A_{\Theta}$.

The optimal weight by definition minimizes the cosmic error.  In what
follows, we assume that ${\bar n}_{\rm g}$ and $\phi(r)$ are
externally determined with very good accuracy.  As a result the cosmic
error for $N$-point statistics is given by
Eq.~(\ref{eq:cosmicerrorgen}), with the obvious correction to the
functional ${\cal S}_i$ with the weights and selection function.  The
optimal weight can then be found by solving an integral equation for
the function $\omega$~\cite{Hamilton93b,Hamilton97a,CSS98}.  There are
several methods to solve this equation, for example by pixelizing the
data, thus transforming the integral into a sum.  In this way, solving
the integral equation corresponds to inverting a matrix.  We shall
come back to that in end of this section and in
Sect.~\ref{s:QuadEst}.

Otherwise, it has been shown that within the following approximations,

\begin{enumerate}
\item the considered $N$-uplets occupy a region ${\cal R}$ small
enough compared to the size of the catalog that variations of function
$\phi$ in the vicinity of a $N$-uplet are negligible,
$\phi(\vr_1)\simeq\ldots\simeq \phi(\vr_N)$; 
\item edge effects are insignificant; 
\item the function $\omega$ depends only on position $\vr$ of the
region ${\cal R}$, i.e. the variations of $\omega$ within ${\cal R}$
are negligible;
\end{enumerate}

the function $\omega(\vr)$ that gives the optimal weight for the
two-point function (but it is likely to be the case for the higher
order functions) appears to be a functional of the selection function
only~\cite{Hamilton93b}.  

Within this simplifying
framework\footnote{Hamilton~\cite{Hamilton97a,Hamilton97b} developed a
general formalism for optimizing the measurement of the two-point
correlation function in real and Fourier space, relying on the
covariance matrix of the statistic $\langle \delta(\vr_i)
\delta(\vr_j) \rangle$, which would correspond to the binning function
$\Theta(\vr_1,\vr_2)= \delta_{\rm D}(\vr_1)\delta_{\rm D}(\vr_2)$.  He
proposed a way of computing the optimal sampling weight without
requiring these simplifying assumptions.}, the solution for the
optimal weight is very simple~\cite{Hamilton93b,CSS98}

\begin{equation}
\omega(r) \propto 1/\sigma^2(r), \label{eq:omegaopti}
\end{equation}

where $\sigma(r)$ is the relative cosmic error on the considered
statistics in a statistically homogeneous catalog with same geometry
and same underlying statistics as the studied one, but with a number
of objects such that its number density is ${\bar n}_{\rm g} \phi(r)$. 
This result actually applies as well to Fourier space (at least for
the power-spectrum~\cite{FKP94}) and to counts-in-cells
statistics~\cite{CSS98}.

To find the optimal weight, one has to make assumptions about the
higher-order statistics in order to compute the cosmic error, since
the latter depends on up to the $2k^{\rm th}$ order for estimators of
$k^{\rm th}$ order statistics.  To simplify the calculation of
$\sigma(r)$, the Gaussian limit is often assumed.  This is valid only in
the weakly nonlinear regime and leads to the following weight for the
two-point correlation function, commonly used in the
literature~\cite{LEPM92,Hamilton93b,MFES94,Efstathiou96,Hamilton97a}:

\begin{equation}
\omega(r) \propto \frac{1}{[1/{\bar n}_{\rm g}(r)+J(r)]^2}, \label{eq:weight1}
\end{equation}

where 

\begin{equation}
J(r)=\int_{r' \leq r} \d^{\dim}\vr' \xi(r').
\end{equation}

In Fourier space the result is~\cite{FKP94}

\begin{equation}
\omega(r) \propto \frac{1}{[1/{V{\bar n}_{\rm g}(r)}+P(k)]^2},
\label{eq:weight2}
\end{equation}

a result that can be easily guessed from Eq.~(\ref{eq:errpow}).  This
equation is valid for $\{k,\Delta k\} \gg 1/L$ where $L$ is the size
of the catalog in the smallest direction and $\Delta k$ is the width
of the considered bin.

Note that the function $\omega(r)$ is of pairwise nature.  It
corresponds to weighting the data with

\begin{equation}
n_{\rm g}(\vr) \rightarrow n_{\rm g}(\vr) \sqrt{\omega(r)}.
\end{equation}

Now, we turn to a more detailed discussion of optimal weighting in
count-in-cell statistics.  The problem of finding the optimal sampling
weight was studied in~\cite{CSS98}.  Similarly to
Eq.~(\ref{eq:estimasamp}), the weighted factorial moment estimator
reads

\begin{equation}
\tF_k^C=\frac{1}{C} \sum_{i=1}^C
\frac{(N_i)_k\,\omega(\vr_i)}{[\phi_{R}(r_i)]^k}, \label{eq:newsetfk}
\end{equation}

where $\phi_{R}(r)$ is the average of the selection function over a
cell.

To simplify the writing of the cosmic error as a function of the
sampling weight, the variations of the function $\omega$ and of the
selection function are assumed to be negligible within the cells,
which is equivalent to points (1) and (3) above.  Then the relative
cosmic error $\sigma_{F_k}[\omega,\phi]=(\Delta \tF_k/F_k)^2$ is

\begin{equation}
\sigma^2_{F_k}[\omega,\phi]=\sigma^2_{\rm F}[\omega]+\sigma^2_{\rm
E}[\omega] +\sigma^2_{\rm D}[\omega,\phi],
\end{equation}

where the finite volume, edge effect and discreteness contributions
read, respectively

\begin{eqnarray}
\sigma^2_{\rm F}[\omega] & = & \frac{\sigma^2_{\rm F}}{\xiav({\hat
L}){\hat V}} \int_{\hat V} \d^3\vr_1 \d^3\vr_2
\,\omega(\vr_1)\,\omega(\vr_2)\,\xi(r_{12}),\\ \sigma^2_{\rm
E}[\omega] &= & \frac{\sigma^2_{\rm E}}{\hat V} \int_{\hat V} \d^3\vr\,
\omega(\vr),\\ \sigma^2_{\rm D}[\omega,\phi] &=& \frac{1}{\hat V}
\int_{\hat V} \d^3\vr\, \omega^2(\vr)\, \sigma^2_{\rm D}(r).
\end{eqnarray}

In these equations, there are terms such as $\sigma^2_{\rm
F}=\sigma^2_{\rm F}[1]$ or $\sigma^2_{\rm E}=\sigma^2_{\rm E}[1]$. 
They correspond to the finite volume and edge effect errors in the
case of homogeneous sampling weight.  They do not depend on the number
density and are given by analytical expressions in
Appendix~\ref{cosmicerrorfacmom}.  The term $\sigma^2_{\rm D}(r)$ is
similar, but there is a supplementary $r$ dependence because the
average count ${\bar N}$ is proportional to the selection function
$\phi$.

Using Lagrange multipliers, it is easy to write the following integral
equation which determines the optimal weight~\cite{CSS98}

\begin{equation}
\frac{\sigma^2_{\rm F}}{\xiav({\hat L}){\hat V}} \int_{\hat V} \d^3\vu\,
\omega(\vu)\,\xi(|\vr-\vu|) + [\sigma^2_{\rm E}+\sigma^2_{\rm D}(r)]
\omega(\vr) +\lambda=0.
\end{equation}

The constant $\lambda$ is determined by appropriate normalization of
the weight function

\begin{equation}
\frac{1}{\hat V} \int_{\hat V} \d^3\vr\, \omega(\vr)=1.
\end{equation}

The solution of this integral equation can be found numerically. 
However, the approximation (\ref{eq:omegaopti}) was found to be
excellent, i.e. almost perfectly minimizes the cosmic
error~\cite{CSS98}.

Using the leading order theory of propagation of errors in
Sect.~\ref{sec:secerrprop}, it is easy to see that these
calculations apply as well to the variance and the cumulants, provided
that errors are small enough: in Eqs.~(\ref{eq:estxi}),
(\ref{eq:ests3}) and (\ref{eq:ests4}), $\tF_k$ would be computed with
Eq.~(\ref{eq:newsetfk}), using the sampling weight minimizing the
cosmic error of the cumulant of interest.

This result shows as well that for a statistically homogeneous
catalog, a weight unity $\omega=1$ is very close to optimal in most
practical cases for count-in-cell statistics.  This statement of
course is not necessarily true for $N$-point correlation functions,
particularly if the catalog presents a complicated geometry.  In that
case, the use of a weight might help to correct for edge effects at
large scales, although the LS estimator and its generalization to
higher order performs already well in this respect with an uniform
weight.  For traditional counts-in-cells estimators, the finite
extension of the cells prevents from correcting for edge effects. 
This is actually the main weakness of these statistics compared to the
$N$-point correlation functions, and often the latter are preferred to
the former, particularly when the geometry of the catalog is
complicated by the presence of numerous masks which reduce
considerably the range of scales available to counts-in-cells.

Finally it is worth noting the following point: the optimal weight is
actually difficult to compute, because it requires knowledge of
statistics of order $\l \leq 2k$ for an estimator of order $k$. 
Therefore, the Gaussian limit, given by Eqs.~(\ref{eq:weight1}) and
(\ref{eq:weight2}) for functions $\xi(r)$ and $P(k)$ respectively was
widely used in the literature.  However, this is rigorously valid only
in the weakly nonlinear regime.  where the shot noise error is likely
to be negligible, implying a simple, uniform weight to be nearly
optimal, unless the catalog is very diluted.  Discreteness errors are
less of a concern with modern surveys under construction, such as the
2dFGRS or the SDSS.

Furthermore, it was noticed in~\cite{CSS98} that the traditional
volume limited sample method does almost as good as a single optimized
measurement extracting all the information from the catalog, if the
depth of the subsample, $R_i$, is chosen such that for the scale
considered signal to noise is approximately maximal.  Of course,
estimating the cosmic error is still a problem, but the advantage of
the volume limited approach is that prior determination of the
selection function is not necessary, which simplifies considerably the
analysis.

\subsection{Cosmic Distribution Function and Cross-Correlations}

\label{sec:sec6bis}

\subsubsection{Cosmic Distribution Function and Likelihood}

For a set of (possibly biased) estimators, ${\tvf}=\{ \tf_k
\}_{k=1,K}$, let us define the covariance matrix as $C_{kl}={\rm
Cov}\{\tf_k, \tf_l \}$.  The extra-diagonal terms can be correlations
between a given estimator (e.g. of the power spectrum) at different
scales (as in Sect.~\ref{covapk}), between different estimators at the
same scale (e.g. factorial moments, see Sect.~\ref{sec:crossestsec}
below), or in general different estimators at different scales. 
Knowledge of these cross-correlations can in fact help to better constrain
theories with observations, because they bring more information
on the shape of the cosmic distribution function.

As mentioned in Sect.~\ref{basic}, the cosmic distribution function
$\Upsilon$ is the probability distribution for an estimator {\em given
a theory} (or class of theories parametrized in some convenient form),
i.e. $\Upsilon=\Upsilon(\tvf|{\rm theory})$ is the probability of
measuring $\tvf$ in a finite galaxy catalog given a theory.  Knowledge
of $\Upsilon(\tvf|{\rm theory})$ allows one to extract constraint on
cosmological parameters from the data through maximum likelihood
analysis, where the {\em likelihood function} is given by the cosmic
distribution function thought as a function of the parameters that
characterize the theory (with $\tvf$ replaced in terms of the observed
data).

In particular, if the cosmic distribution function $\Upsilon$ is
Gaussian, it is entirely determined once the covariance matrix $\vC$
is known:

\begin{equation}
\label{upsilonG} \Upsilon(\tvf|\vC,\vf,\vb)=\frac{1}{\sqrt{(2\pi)^K
|\vC|}} \exp\left[ -\frac{1}{2} \sum_{k,l} \delta \tf_k C_{kl}^{-1}
\delta \tf_l \right],
\end{equation}

where $\vC^{-1}$ and $|\vC|$ are respectively the inverse and the
determinant of the covariance matrix, $\vf$ is the true value of the
statistics in question ($\vf=\langle \tvf \rangle$ for unbiased
estimators) and $\vb$ a vector accounting for possible cosmic bias.
Both $\vC$ and $\vf$ (and $\vb$ if non-zero) are calculated from
theoretical predictions as a function of cosmological parameters.

It is very important to note that the Gaussian assumption for
$\Upsilon$ is in general different from assuming that the density
field is Gaussian unless the estimator $\tvf$ corresponds to the
density contrast\footnote{In this case $\Upsilon$ is proportional to
the density PDF.}.  For this reason, Eq.~(\ref{upsilonG}) is not
necessarily a good approximation for estimators that are not linear in
the density contrast even if the underlying statistic of the density
field is Gaussian.  We shall come back to this point in
Sect.~\ref{sec:cosmicdisshape}.

Why is it useful to take as $\vf$ non-linear functions of the density
contrast?  The problem is that the assumption of Gaussianity for the
density field itself is very restrictive to deal with galaxy
clustering: it does not include information on higher-order moments
which arise due to e.g. non-linear evolution, non-linear galaxy bias,
or primordial non-Gaussianity.  Since there is no general expression
for the multi-point PDF of the density field which describes its
non-Gaussian shape\footnote{The Edgeworth expansion,
Eq.~(\ref{fb:edge}), in principle provides a way to accomplish
this~\cite{Amendola96}.  In practice, however, its regime of validity
is very restricted.}, one must resort to a different approach.  The
key idea is that taking $\vf$ to be a statistic\footnote{These are
non-linear functions of the data, e.g. the power spectrum is
quadratic.} of the density field, it is possible to work in a totally
different regime.  Indeed, when the cosmic error is sufficiently
small, there must be many independent contributions to $\tvf$ so that,
by the central limit theorem, its cosmic distribution function should
approach Gaussianity\footnote{Note that, in contrast to the PDF of the
density field, this limit is usually approached {\em at small scales},
we shall discuss examples below in Sect.~\ref{sec:cosmicdisshape}.}. 
On the other hand, the cosmic error becomes large when probing
large-scales, where there are not many independent samples; in this
case, assumption of a Gaussian density field plus the nonlinear
transformation involved in $\tvf$ leads to a useful guess about the
asymptotic behavior of $\Upsilon$.  In practice, the specific shape of
$\Upsilon$ must be {\em computed} for a given set of theories, and the
limit of validity of the asymptotic forms discussed above should be
carefully checked, as discussed further in
Sect.~\ref{sec:cosmicdisshape}.

This remainder of this section is organized as follows.  In
Sect.~\ref{sec:crossestsec} we discuss about correlations between
different statistics.  As an example, we show how knowledge of the
number of objects in a galaxy catalog can be used to reduce the
error bar on the measurement of the two-point correlation function. 
Then, in Sect.~\ref{sec:cosmicdisshape}, we address the problem of
non-Gaussianity of the cosmic distribution function.

\subsubsection{Cross-Correlations Between Different Statistics}

\label{sec:crossestsec}

An important kind of cross-correlation is given by that between
statistics of different kind.  For example, the calculation leading to
Eq.~(\ref{eq:cosmicxi2}) is a {\em conditional} average with the
constraint that the average number density is equal to the observed
one:

\begin{eqnarray}
\left(\Delta {\hat \xi}| {\bar n}_{\rm g}\right)^2 & \equiv & \langle
{\hat \xi}^2 | {\bar n}_{\rm g} \rangle - \langle {\hat \xi} | {\bar
n}_{\rm g} \rangle^2 \\ &= & \frac{\int \xi^2 \Upsilon(\xi,{\bar
n}_{\rm g}) \d\xi} {\int \Upsilon(\xi,{\bar n}_{\rm g}) \d\xi}- \left[
\frac{\int \xi \Upsilon(\xi,{\bar n}_{\rm g}) \d\xi} {\int
\Upsilon(\xi,{\bar n}_{\rm g}) \d\xi} \right]^2.
\end{eqnarray}

The knowledge of this supplementary information decreases the expected
error on the measurement of $\xi(r)$ and provides better constraints
on the models.  The calculation of Bernstein leading to
Eq.~(\ref{eq:cosmicxi}) does not make use of the fact that ${\bar
n}_{\rm g}$ can be measured separately:

\begin{eqnarray}
\left( \Delta{\hat \xi}\right)^2 & = & \langle {\hat \xi}^2 \rangle -
\langle {\hat \xi} \rangle^2 \\ & = & \int \xi^2 \Upsilon(\xi,{\bar
n}_{\rm g}) \d\xi \d{\bar n}_{\rm g} -\left[\int \xi \Upsilon(\xi,{\bar
n}_{\rm g}) \d\xi \d{\bar n}_{\rm g} \right]^2
\end{eqnarray}

and therefore slightly overestimates the error on $\xi(r)$ as
emphasized in~\cite{LaSz93}.  For example, if the function $\Upsilon$
is Gaussian, we have

\begin{equation}
\left(\Delta {\hat \xi}| {\bar n}_{\rm g}\right)^2=\left( \Delta{\hat
\xi}\right)^2 \left[1 - \rho_{\xi,{\bar n}_{\rm g}}^2 \right],
\end{equation}

where the correlation coefficient $\rho_{AB}$ is defined for
estimators $\tA$ and $\tB$ as

\begin{equation}
\rho_{AB}\equiv \frac{\langle (\tA-\langle \tA \rangle)(\tB-\langle
\tB \rangle) \rangle}{\Delta \tA \Delta \tB}.
\end{equation}

{}From this simple result, we see that joint measurement of
(theoretically) {\em more correlated or anti-correlated} statistics
brings better constraints on the underlying theory.

In~\cite{SCB99} and as described in Sect.~\ref{sec:secfacmomtheo},
cross-correlations between factorial moments are computed analytically
at fixed scale.  From the theory of propagation of errors, it is
straightforward to compute cross-correlations between other
count-in-cells statistics of physical interest, such as average count
${\bar N}$, variance $\xiav$ and cumulants $S_p$.  Theoretical
calculations and measurements in numerical
simulations~\cite{SCB99,CSJC99} show that, for realistic galaxy
catalogs such as the SDSS, ${\bar N}$ and $\xiav$ are not, in general
strongly correlated, and similarly for correlations between ${\bar N}$
and higher-order statistics.  Interestingly, $\xiav$ and $S_3$ are not
very strongly correlated, but $S_3$ and $S_4$ are.  Actually, in
general and as expected, the degree of correlation between two
statistics of orders $k$ and $l$ decreases with $|k-l|$.

\subsubsection{Validity of the Gaussian Approximation} 

\label{sec:cosmicdisshape}

We now discuss the validity of the Gaussian approximation,
Eq.~(\ref{upsilonG}), for the cosmic distribution function.  To
illustrate the point, we take two examples, the first one about
count-in-cells statistics, the second one about the power-spectrum and
bispectrum.

\begin{figure}[h!]
\centering \centerline{\epsfxsize=14.  truecm
\epsfbox{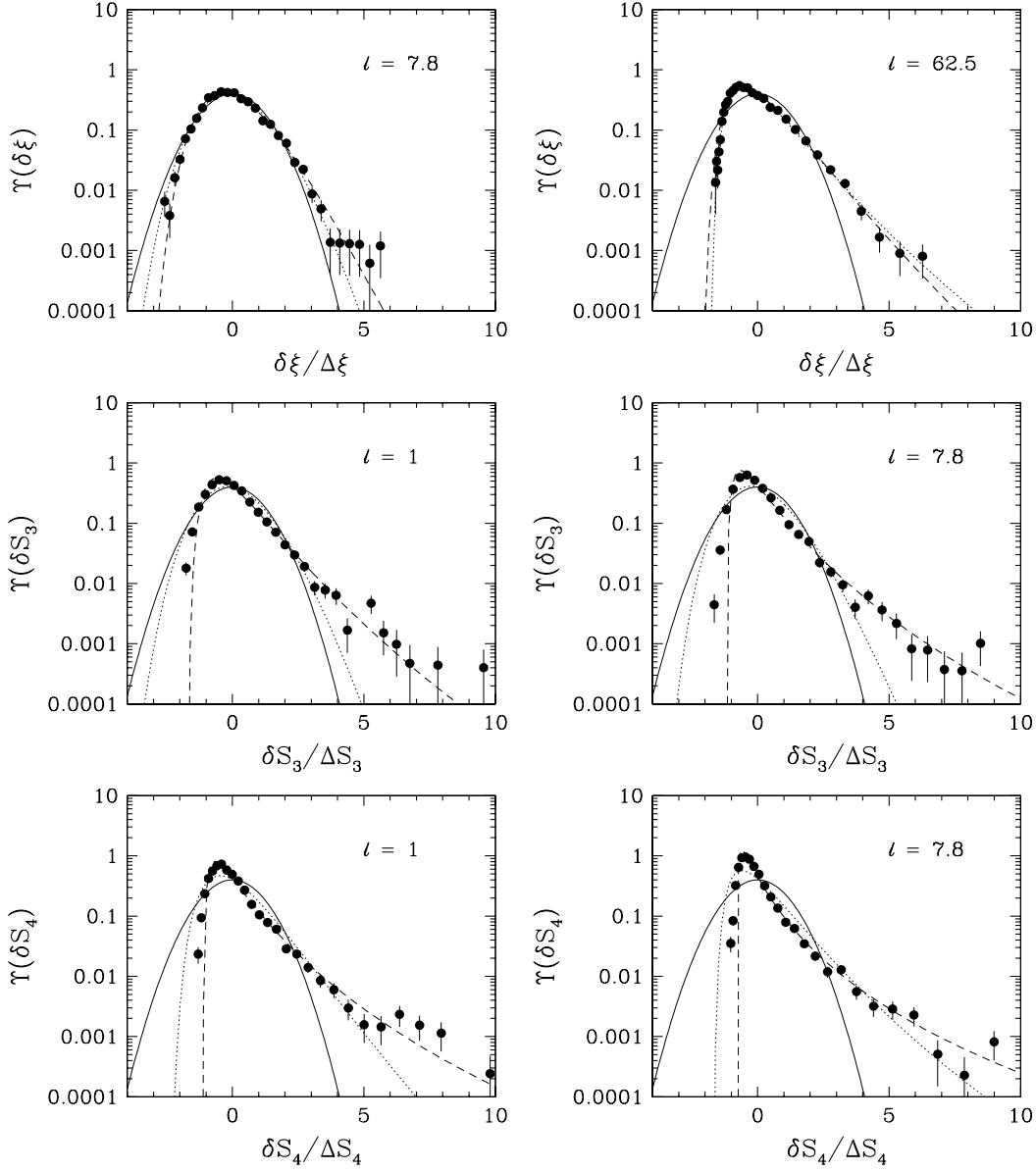}} \caption{The cosmic distribution
function of measurements $\Upsilon({\hat{\xiav}})$ (upper line of
panels), $\Upsilon({\hat S}_3)$ (middle line of panels) and
$\Upsilon({\hat S}_4)$ (lower line of panels) measured from a
distribution of subsamples extracted from a Hubble volume simulation
(see end of Sect.~\ref{sec:secfacmomtheo} for more details).  The
scale of the measurements, either $R=1,\ 7.8$ or $62.5h^{-1}$ Mpc, is
indicated on each panel.  The solid, dotted and dash curves correspond
to the Gaussian, lognormal and generalized lognormal
[Eq.~(\ref{eq:lognor2})] distributions, respectively.  With the choice
of the coordinate system, the magnitude of the cosmic error does not
appear directly, but is reflected indirectly by the amount of skewness
of the lognormal distribution.} \label{fig:fig-chap7-2}
\end{figure}

Exhaustive measurements in one of the Hubble volume
simulations~\cite{SCJC99} show that for count-in-cell statistics,
$\Upsilon(\tA)$ is approximately Gaussian if $\Delta \tA/A \la 0.2$. 
Therefore, at least for count-in-cells, Gaussianity is warranted only
if the errors are small enough.  When the cosmic errors become
significant, the cosmic distribution function becomes increasingly
skewed, developing a tail at large values of $\tA$~\cite{SCJC99}. 
This result applies to most counts-in-cells estimators (${\hat P}_N$,
${\hat F}_k$, ${\hat{\xiav}}$, ${\hat S}_p$).  One consequence is that
the most likely value is below the average, resulting in an {\em
effective cosmic bias}, even for unbiased statistics such as factorial
moments: typically, the measurement of a statistic $\tA$ in a finite
catalog is likely to underestimate the real value, except in some rare
case where it will overestimate it by a larger amount\footnote{This is
of course analogous to non-Gaussianity in the density PDF. Positive
skewness means that the most likely value is to {\em underestimate}
the mean, see Eq.~(\ref{s3demax}).  To compensate for this there is a
rare tail at large values compared to the mean, see e.g.
Fig.~\ref{SmoothEff}.}.  To take into account the asymmetry in the
shape, it was proposed in~\cite{SCJC99} to use a generalized version
of the lognormal distribution, which describes very well the shape of
function $\Upsilon(\tA)$ for a single statistic, as illustrated by
Fig.~\ref{fig:fig-chap7-2}:

\begin{eqnarray}
\Upsilon(\tA) & = & \frac{s}{\Delta A [s (\tA-A)/\Delta A+1] \sqrt{
2\pi \eta } } \nonumber \\ & & \times \exp\left( - \frac{ \{ \ln[s
(\tA-A)/\Delta A+1]+\eta/2 \}^2}{2 \eta} \right), \label{eq:lognor2}
\end{eqnarray}

\begin{equation}
\eta=\ln(1+s^2),
\end{equation}

where $s$ is an adjustable parameter.  It is fixed by the requirement
that the analytical function Eq.~(\ref{eq:lognor2}) have identical
average, variance, and skewness $S_3 = 3+s^2$, as the measured
$\Upsilon(\tA)$.

However, the generalization of Eq.~(\ref{eq:lognor2}) to multivariate
cosmic distribution functions is not easy, although feasible at least
in some restricted cases (e.g. see~\cite{Sheth95}).  An alternate
approach, would employ a multivariate Edgeworth
expansion~\cite{Amendola96}.

\begin{figure}[t]
\begin{center} \begin{tabular}{cc} {\epsfysize=6.5truecm
\epsfbox{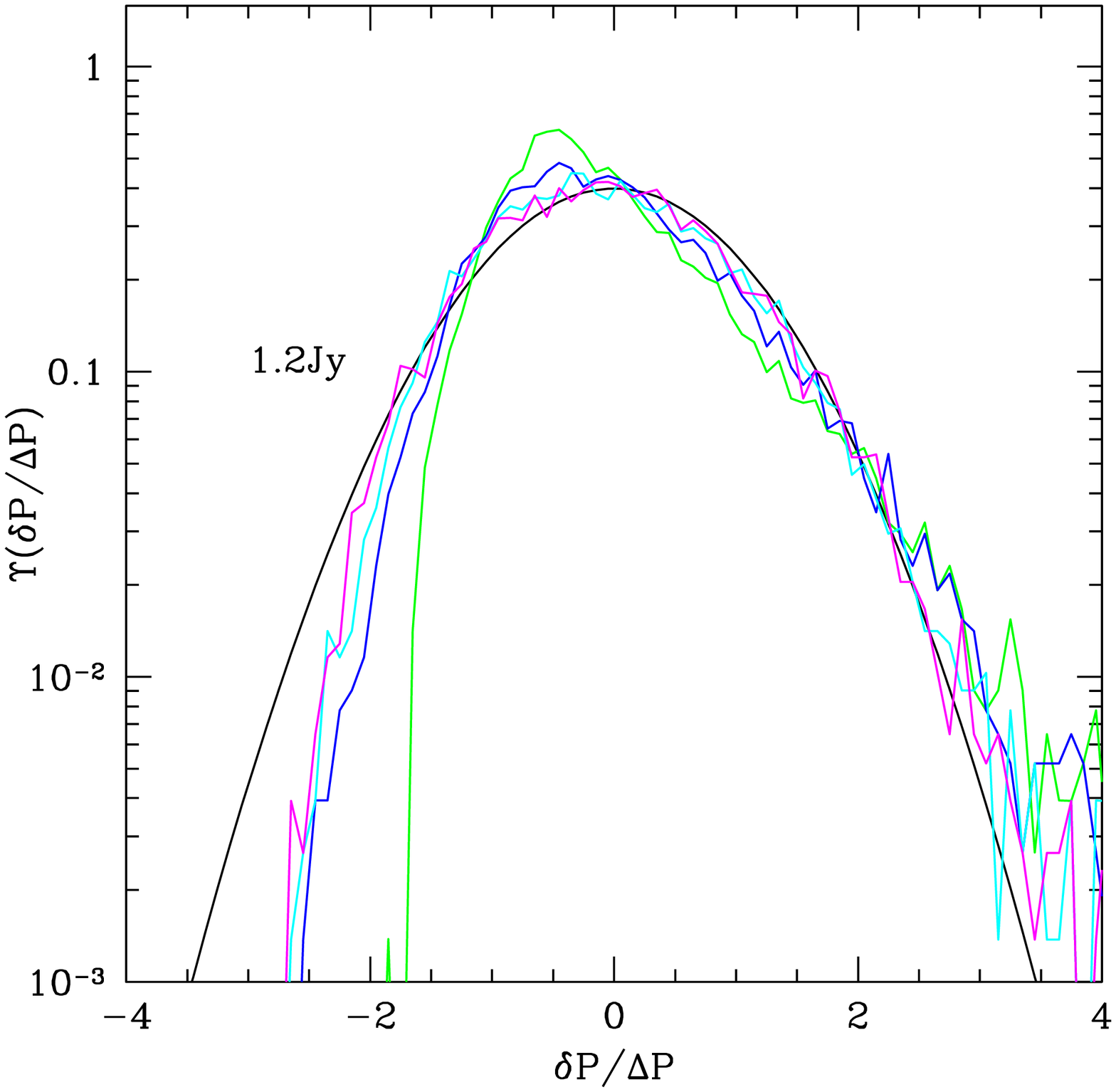}}& {\epsfysize=6.5truecm \epsfbox{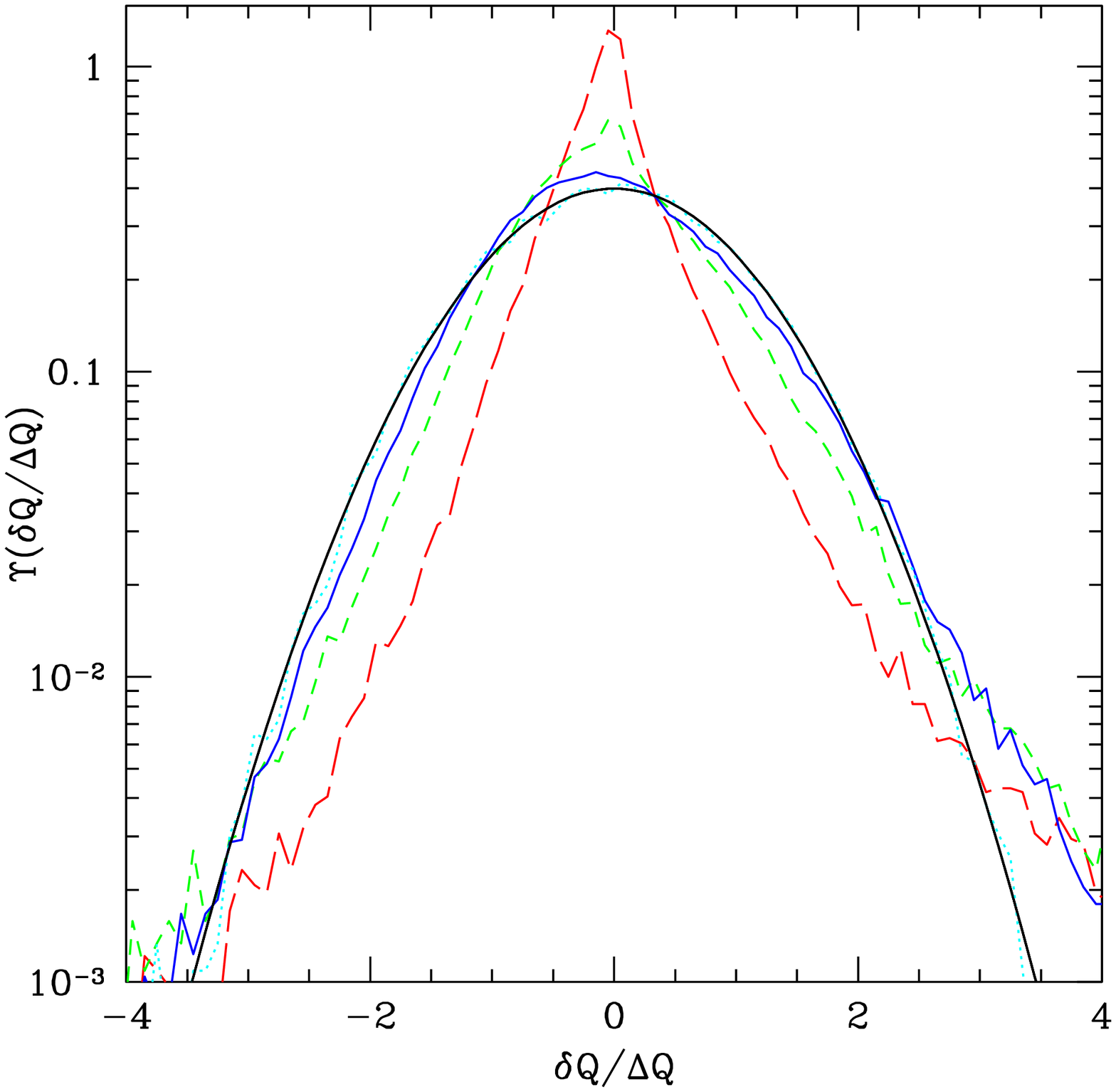}}
\end{tabular} \end{center} \caption{ {\em Left Panel:} Power spectrum
cosmic distribution function in a IRAS 1.2Jy-like survey as a function
of scale in logarithmic scale, smooth solid line denotes a Gaussian
distribution.  From least to most non-Gaussian, scales are
$k/k_f=1-10$, $k/k_f=11-20$, $k/k_f=21-30$, $k/k_f=31-40$, where
$k_f=0.005$ h/Mpc.  {\em Right Panel:} Cosmic distribution function of
$\delta Q/\Delta Q \equiv (Q-\bar{Q})/\Delta Q $ for different surveys
in models with Gaussian initial conditions: 2nd order Lagrangian PT
with $256^3$ objects in a volume of 600 Mpc/h radius (dotted), IRAS
1.2Jy (solid), IRAS 2Jy (dashed), IRAS QDOT (long-dashed).  The smooth
solid curve is a Gaussian distribution.} \label{UpsilonPkQ}
\end{figure}

Since the Gaussianity of the cosmic distribution function mainly
depends on the variance of the statistic under consideration, it is
expected that for surveys where errors are not negligible, Gaussianity
is not a good approximation.  Figure~\ref{UpsilonPkQ} illustrates this
for IRAS surveys in the case of the power spectrum and
bispectrum~\cite{Scoccimarro00}, as a function of normalized
variables, $\de A/\Delta A \equiv ({\hat A}-A)/\langle ({\hat
A}-A)^2\rangle^{1/2}$.  For the bispectrum, this choice of variable
makes the cosmic distribution function approximately independent of
scale and configuration.

The left panel of Fig.~\ref{UpsilonPkQ} shows the power spectrum
cosmic distribution function as a function of scale, from least to
most non-Gaussian, scales are $k/k_f=1-10$, $k/k_f=11-20$,
$k/k_f=21-30$, $k/k_f=31-40$, where $k_f=0.005$ h/Mpc.  As expected,
non-Gaussianity is significant at large scales, as there are only a
few independent modes (due to the finite volume of the survey), and
thus the power spectrum PDF is chi-squared distributed.  As smaller
scales are considered, averaging over more modes leads to a more
Gaussian distribution, although the convergence is slow since the
contributing modes are strongly correlated due to shot noise.

The right panel in Fig.~\ref{UpsilonPkQ} shows a similar plot for the
bispectrum.  In sparsely sampled surveys such as QDOT, deviation from
Gaussianity can be very significant.  In a large volume limited sample
of 600 Mpc/h radius with many galaxies (dotted curve), Gaussianity
becomes an excellent approximation, as expected.  The cosmic
distribution function for $\chi^2$ initial conditions was also
calculated in~\cite{Scoccimarro00}; in this case non-Gaussianity is
significant even for large volume surveys, and thus must be taken into
consideration in order to properly constrain primordial
non-Gaussianity~\cite{SFFF01,FFFS01}.

\subsection{Optimal Techniques for Gaussian Random Fields}

\label{sec:general}

Up to now, we have restricted our discussion to a particular subset of
estimators used commonly in the literature, which apply equally well
to two-point and higher-order statistics.  To give account of recent
developments, we now reinvestigate the search for optimal estimators
in the framework of Gaussian random fields.  That is, the cosmic
distribution function, with estimators $\tvf$ that {\em will} be taken
as density contrasts (measured in pixels or their equivalent in some
space of functions, such as spherical harmonics), will be assumed to
be Gaussian.  As discussed above, this approach is only justifiable to
obtain estimates of the power spectrum (or two-point correlation
function) at the largest scales, where Gaussianity becomes a good
approximation.

First we recall basic mathematical results about minimum variance and
maximum likelihood estimators (Sect.~\ref{sec:MLE}).  In
Sect.~\ref{s:QuadEst}, we discuss optimal weighting for two-point
statistics taking into account the full covariance matrix (compare to
Sect.~\ref{sec:sec6}), and in Sect.~\ref{s:UcEr} we briefly address
techniques for obtaining uncorrelated estimates of the power spectrum,
comparing with results discussed in previous sections when relevant. 
Finally we briefly describe the Karhunen-Lo\`eve transform, useful for
compressing large amounts of data expected in current and forthcoming
surveys (Sect.~\ref{sec:KL}).

\subsubsection{Maximum Likelihood Estimates}

\label{sec:MLE}

The basic results given here are well known in statistical
theory~\cite{SOA99,Wilks63}.  For more details and applications to
optimal measurements of the power spectrum in cosmological data sets
see e.g.~\cite{TTH97,Tegmark97,BJK98,Hamilton97a}.

Let's assume that we have at our disposal some data $\tvx$, say, a
vector of dimension $N$ with the cosmic distribution function
$\Upsilon(\tvx)$, which is Gaussian and can be expressed explicitly as
a function of $\tvx$ and a set of unknown parameters $\vf$, which we
aim to estimate, given our data.  When thought as a function of the
parameters $\vf$, $\Upsilon(\vf)$ is usually known as the {\em
likelihood function}\footnote{Therefore, the assumption of a Gaussian
density field means $\Upsilon(\tvx)$ as a function of $\tvx$ is
Gaussian, whereas in the limit that a large number uncorrelated data
contributes $\Upsilon(\vf)$ becomes Gaussian.  }.  The corresponding
estimators, ${\tvf}=(\tf_1,\cdots,\tf_K)$, $K \leq N$, are sought in
the space of functions of the data $\tvx$.  The problem of finding an
optimal estimator $\tvf$ can be formally approached at least in two
ways, the first one consisting in minimizing the cosmic error on
$\tvf$, the second one consisting in maximizing the likelihood.

We restrict ourselves to unbiased estimators, 
\begin{equation}
\langle\tvf \rangle\equiv\int\d^N\hat\vx\ \Upsilon(\tvx|\vf)\ \tvf(\tvx)
=\vf.  \label{eq:avens}
\end{equation}

The search for the first kind of optimal estimator, already discussed
in Sect.~\ref{sec:sec6}, consists in minimizing the cosmic error

\begin{equation}
\Delta^2 f_k = \langle (\tf_k - f_k)^2 \rangle,
\end{equation}

given the constraint (\ref{eq:avens}).  It is useful at this point to
assume that the likelihood function is sufficiently smooth and to
introduce the so-called Fisher information matrix

\begin{equation} \label{eq:fisher} F_{kl}\equiv \left\langle
\frac{\partial^2 [-\log \Upsilon(\vf)]}{\partial f_k \partial f_l}
\right\rangle= \left\langle \frac{\partial \log \Upsilon(\vf)}{\partial
f_k} \frac{\partial \log \Upsilon(\vf) }{\partial f_l} \right\rangle. 
\end{equation}

Let's assume that the matrices $\vF$, and the covariance matrix $\vC$
defined by $C_{kl}\equiv {\rm Cov}(f_k,f_l)=\langle \delta \tf_k
\delta \tf_l \rangle$ are positive definite. From the Cauchy-Schwarz
inequality one gets the so-called Cram\'er-Rao inequality

\begin{equation} (\Delta f_k)^2\ F_{kk}\geq 1, \end{equation}

so that the inverse of the Fisher matrix can be thought as the minimum
errors that one can achieve.  Through a change of variable this
inequality can be generalized in 
\begin{equation} (\va^{\rm t}\cdot\vC\cdot\va)\ 
({\vb^{\rm t}\cdot\vF\cdot\vb})
\geq\left(\va^{\rm t}\cdot\vb \right)^2, \label{eq:ineq1}
\end{equation}
where $\va$ and $\vb$ are two sets of constants. It implies

\begin{equation} |\vC| \geq \frac{1}{|\vF|}.  \label{eq:ineq2}
\end{equation}

An estimator $\tvf$ which obeys the equality in Eqs.~(\ref{eq:ineq1})
or (\ref{eq:ineq2}) is called {\em minimum variance bound} (MVB). 
This can happen if and only if the estimator $\tvf$ can be expressed
as a linear function of the derivative of log-likelihood function with
respect to the parameters:

\begin{equation} \left( \frac{\partial \log
\Upsilon}{\partial \vf} \right)^{\rm t}\cdot\vb = g(\vf)\ 
(\tvf-\vf)^{\rm t}\cdot\va, \label{eq:linearprop} 
\end{equation}

where the constant of proportionality $g(\vf)$ might depend on the
parameters but not on the data $\tvx$.  As a result, for an arbitrary
choice of the parameters $\vf$, minimum variance unbiased estimators
are not necessarily MVB.
\noindent The second way of seeking an optimal estimator consists in
maximizing directly the likelihood function in the space of
parameters, $\vf \rightarrow \tvf$.  The goal is to find $\tvf_{\rm
ML}$ such that

\begin{equation}
\Upsilon(\tvx)|_{\vf=\tvf_{\rm ML}(\tvx)} \geq \Upsilon(\tvx)|_{\vf}
\label{eq:MLest}
\end{equation}

for any possible value of $\vf$.  A practical, sufficient but not
necessary condition is given by the solution of the two sets of
equations

\begin{equation}
\frac{ \partial \log \Upsilon}{\partial \vf}=0
\end{equation}

\begin{equation}
\frac{ \partial^2 \log \Upsilon}{\partial f_k \partial f_l} <0.
\end{equation}

The solution of Eq.~(\ref{eq:MLest}), if it exists, does not lead
necessarily to an unbiased estimator nor a minimum variance estimator. 
But if by chance the obtained ML estimator is unbiased, then it
minimizes the cosmic error.  Moreover, if there is an MVB unbiased
estimator, it is given by the ML method.  Note that in the limit that
large number of uncorrelated data contributes, the cosmic distribution
function tends to a Gaussian and the ML estimator is asymptotically
unbiased and MVB. In that regime, the cosmic cross-correlation matrix
of the ML estimator is very well approximated by the inverse of the
Fisher information matrix

\begin{equation}
C_{kl}={\rm Cov}(f_k,f_l) = \langle \delta \tf_k \delta \tf_l \rangle
\simeq (\vF^{-1})_{k,l}. \label{CijFij}
\end{equation}

On the other hand, from the Gaussian assumption for $\Upsilon(\tvx)$,
it follows that the ML estimator for the power spectrum [${\hat
P}(k_\alpha) \equiv \tf_\alpha$] is the solution of

\be
\label{MLest} \tf_\alpha= \frac{1}{2} F_{\alpha \beta}^{-1}
\frac{\partial C_{ij}}{\partial f_\beta} [C^{-1}]_{ik} [C^{-1}]_{jl}
(\de_k \de_l -N_{kl})
\ee

(where $\de_k$ denotes the density contrast at $\vr_k$) for which the
estimate is equal to the prior, $\tvf=\vf$.  That is, in order to
obtain the ML estimator, one starts with some prior power spectrum
$\vf$, then finds the estimate $\tvf$, puts this back into the prior,
and iterates until convergence.  In Eq.~(\ref{MLest}), the Fisher
matrix is obtained from Eq.~(\ref{eq:fisher}),

\be F_{\alpha \beta}=\frac{1}{2} \frac{\partial C_{ij}}{\partial
f_\alpha} [C^{-1}]_{ik} [C^{-1}]_{jl} \frac{\partial C_{kl}}{\partial
f_\beta},
\label{fisherM}
\ee

the covariance matrix $C_{ij}= \xi_{ij}+N_{ij}$ contains a term due to
clustering (given by the two-point correlation function at separation
$|\vr_i-\vr_j|$, $\xi_{ij}$), and a shot noise term $N_{ij} \equiv
\nbar_i \de_D(\vr_i-\vr_j)$.  Applications of the ML estimator to  
measurements of the 2-D galaxy power spectrum was recently done for 
the APM~\cite{EfMo01} and EDSGC~\cite{HKN01} surveys (see 
Sect.~\ref{sec:ang2pt}).

\subsubsection{Quadratic Estimators}

\label{s:QuadEst}

In reality it is in general difficult to express explicitly the
likelihood function in terms of the parameters.  In addition, even if
we restrict to the case where the parameters are given by the power
spectrum as a function of scale as discussed in the previous section,
one must iterate numerically to obtain the ML estimates, and their
probability distribution also must be computed numerically in order to
provide error bars\footnote{However, see~\cite{BJK00} for an analytic
approximation in the case of the 2-D power spectrum using an offset
lognormal.}.  As a result, a useful approach is to seek an optimal
estimator, unbiased and having minimum variance, by restricting the
optimization to a subspace of estimators, as discussed in
Sect.~\ref{sec:sec6}.  Of course, this method is not restricted to the
assumption of Gaussianity, provided that the variance is calculated
including non-Gaussian contributions.  It turns out there is an
elegant solution to the problem~\cite{Hamilton97a,Hamilton00}, which
in its exact form is unfortunately difficult to implement in practice,
but it does illustrate the connection to the ML
estimate~(\ref{MLest}) in the Gaussian limit, and also provides a
generalization of the standard optimal weighting results,
Eqs.~(\ref{eq:weight1},\ref{eq:weight2}) to include non-Gaussian (and
non-diagonal) elements of the covariance matrix.

Since the power spectrum is by definition a quadratic quantity in the
overdensities, it is natural to restrict the search to quadratic
functions of the data.  In this framework, the unbiased
estimator\footnote{This is assuming that the mean density is perfectly
known.} of the power spectrum having minimum variance
reads~\cite{Hamilton97a,Hamilton00}

\be
\label{quadest} \tf_\alpha=  F_{\alpha \beta}^{-1}
\frac{\partial C_{ij}}{\partial f_\beta} [\tilde{C}^{-1}]_{ijkl}  
(\de_k \de_l -\hat{N}_{kl}), 
\ee

where the variance is given by Eq.~(\ref{CijFij}) and the Fisher
matrix by Eq.~(\ref{fisherM}) replacing ${1\over 2} [C^{-1}]_{ik}
[C^{-1}]_{jl}$ with $[\tilde{C}^{-1}]_{ijkl}$, where

\be\tilde{C}_{ijkl}= \langle (\de_{i}\de_{j} -\hat{N}_{ij}-\xi_{ij}) 
(\de_{k}\de_{l} -\hat{N}_{kl}-\xi_{kl}) \rangle
\ee

is the (shot noise subtracted) power spectrum covariance matrix. 
Here $\hat{N_{ij}}$ denotes the `actual' shot noise, meaning that the
self-pairs contributions to $\xi_{ij}$ are not included,
see~\cite{Hamilton00} for details.  In the Gaussian limit,
$[\tilde{C}^{-1}]_{ijkl} \rightarrow {1\over 2} [C^{-1}]_{ik}
[C^{-1}]_{jl}$ (symmetrized over indices $k$ and $l$) and the minimum
variance estimator, Eq.~(\ref{quadest}), reduces to ML estimator,
Eq.~(\ref{MLest}), assuming iteration to convergence is carried out as
discussed above.  If the iteration is not done, the estimator remains
quadratic in the data, and it corresponds to using Eq.~(\ref{MLest})
with a fixed prior; this should be already a good approximation to the
full ML estimator, otherwise it would indicate that the result depends
sensitively on the prior and thus there is not significant information
coming from the data.  The use of such quadratic estimators in the
Gaussian limit to measure the galaxy power spectrum is discussed in
detail in~\cite{THSVS98}, see also~\cite{Tegmark97,TTH97,BJK98}. 
Extension to minimum variance cubic estimators for the angular
bispectrum in the Gaussian limit is considered
in~\cite{Heavens98,GaMa00}.

Note that, the full minimum variance estimator involves inverting a
rank 4 matrix, a very demanding computational task, which however
simplifies significantly in the Gaussian limit where $\tilde{\vC}$
factorizes.  Another case in which the result becomes simpler is the
so-called FKP limit~\cite{FKP94}, where the selection function
$\nbar_{\rm g}(\vr)$ can be taken as locally constant, compared to the scale
under consideration.  This becomes a good approximation at scales much
smaller than the characteristic size of the survey, which for present
surveys is where non-Gaussian contributions become important, so it is
a useful approximation.  In this case the minimum variance pair
weighting for a pair $ij$ is only a function of the separation
$\alpha$ of the pair, not on their position or orientation, since
$\nbar_{i}$ and $\nbar_{j}$ are assumed to be constants locally.  As a
result, the power spectrum covariance matrix can be written in terms
of a two by two reduced covariance matrix, which although not 
diagonal due to non-Gaussian contributions, becomes so in the Gaussian 
limit, leading to the standard result Eq.~(\ref{eq:weight2}). We refer 
the reader to~\cite{Hamilton00} for more details.

\subsubsection{Uncorrelated Error Bars}

\label{s:UcEr}

Clearly, minimum variance estimates can be deceptive if correlations
between them are substantial.  Ideally one would like to obtain not
only an optimal estimator (with minimum error bars), but also
estimates which are uncorrelated (with diagonal covariance matrix),
like in the case of the power spectrum of a Gaussian field in the
infinite volume limit.  Once the optimal (or best possible) estimator
$\tvf$ is found, it is possible to work in a representation where the
cosmic covariance matrix $\vC$ becomes diagonal,

\begin{equation}
\vC\cdot\vPsi_j=\lambda_j \vPsi_j,
\end{equation}

where the eigenvectors $\vPsi_j$ form an orthonormal basis.  A new set
of estimators can be defined

\begin{equation}
\tvg \equiv \vPsi^{-1}\cdot \tvf,
\end{equation}

which are statistically orthogonal

\begin{equation}
\langle \delta \tg_i \delta \tg_j \rangle = \lambda_i \delta_{ij} =
\vPsi_i^{\rm t}\cdot\vC\cdot\vPsi_i\ \delta_{ij}.
\end{equation}

These new estimators can in principle be completely different from the
original set, but if by chance the diagonal terms of $\vC$ are
dominant, then we have $\tvg \simeq \tvf$.  In fact, if one takes the
example of the two-point correlation function (or higher order) in
case the galaxy number density is known, using the new estimator
$\tvg$ is equivalent to changing the binning function $\Theta$ defined
previously to a more complicated form.  Among those estimators which
are uncorrelated, it is however important to find the set $\tvg$ such
that the equivalent binning function is positive and compact in
Fourier space and $\langle \tvg \rangle \simeq \vf$, so that to keep
the interpretation of the power in this new representation as giving
the power centered about some well-defined
scale~\cite{Hamilton97b,Hamilton00}.

The above line of thoughts can in fact be pushed even further by
applying the so called ``pre-whitening'' technique to $\tvf$: if
$\tvf$ is decomposed in terms of signal plus noise, pre-whitening
basically consists in multiplying $\tvf$ by a function $h$ such that
the noise becomes white or constant.  If the noise is uncorrelated,
this method allows one to diagonalize simultaneously the covariance
matrix of the signal and the noise.  When non-Gaussian contributions
to the power spectrum covariance matrix are included, however, such a
diagonalization is not possible anymore.  However, in the FKP
approximation, as described in the previous section, it was shown that
an approximate diagonalization (where two of the contributions coming
from two- and four-point functions are exactly diagonal, whereas the
third coming from the three-point function is not) works extremely
well, at least when non-Gaussianity is modeled by the hierarchical
ansatz~\cite{Hamilton00}. The quantity whose covariance matrix has 
these properties corresponds to the so-called prewhitened power 
spectrum, which is easiest written in real space~\cite{Hamilton00}

\begin{equation}
{\hat \xi}(r) \rightarrow \frac{2 {\hat \xi}(r)}{1+[1+\xi(r)]^{1/2}}.
\end{equation}

Note that in the linear regime, ${\hat \xi}(k)$ reduces to the linear
power spectrum; however, unlike the non-linear power spectrum, ${\hat
\xi}(k)$ has almost diagonal cosmic covariance matrix even for
nonlinear modes.  More details on the theory and applications to
observations can be found in e.g.~\cite{Hamilton00,HaTe00a}
and~\cite{HaTe00b,PTH01,HTP00} respectively.

\subsubsection{Data Compression and the Karhunen-Lo\`eve Transform}

\label{sec:KL}

A problem to face is with modern surveys such as the 2dFGRS and SDSS, is
that the data set $\tvx$ becomes quite large for ``brute force''
application of estimation techniques.  Before statistical treatment of
the data as discussed in the previous sections, it might be necessary
to find a way to reduce their size, but keeping as much information as
possible.  The (discrete) Karhunen-Lo\`eve transform (KL) provides a
fairly simple method to do that (see e.g.~~\cite{VoSz96,TTH97} and
references therein for more technical details and
e.g.~\cite{PTH01,MSL00} for practical applications to observations). 
Basically, the idea is to work in the space of eigenvectors $\vPsi_j$
of the cross-correlation matrix $\vM\equiv \langle \delta \tvx\cdot
 \delta \tvx^{\rm t} \rangle$, i.e. to diagonalize the cosmic covariance
matrix of the data,

\begin{equation}
\vM\cdot\vPsi_j=\lambda_j\,\vPsi_j,
\end{equation}

where the matrix $\Psi$ is unitary, $\vPsi^{-1}=\vPsi^{\rm t}$.  A
new set of data, $\tvy$, can be defined

\begin{equation}
\tvy \equiv \vPsi^{\rm t}\cdot\tvx,
\label{eq:vye}
\end{equation}

which is statistically orthogonal

\begin{equation}
\langle \delta \ty_i \delta \ty_j \rangle=\lambda_i\ \delta_{ij}=
{}^{\rm t} \vPsi_i\cdot\vM\cdot\vPsi_j\ \delta_{ij}.
\end{equation}

The idea is to sort the new data from highest to lowest value of
$\lambda_i$.  Data compression will consist in ignoring data
$\ty_i$ with $\lambda_i$ lower than some threshold.

An interesting particular case of the KL transform is when the data
can be decomposed in signal plus noise uncorrelated with each
other~\cite{Bond95}:

\begin{equation}
\tvx=\tvs+\tvn.
\end{equation}

The signal and the noise covariance matrices read

\begin{equation}
\vS\equiv \langle \delta\tvs\cdot\delta\tvs^{\rm t} \rangle, \quad
\vN \equiv \langle \delta\tvn\cdot\delta\tvn^{\rm t} \rangle.
\end{equation}

Then, instead of diagonalizing the cosmic covariance matrix of the
data, one solves the generalized eigenvalue problem

\begin{equation}
\vS\cdot\vPsi_j=\lambda_j\ \vN\cdot\vPsi_j, \quad
\vPsi^{\rm t}_j\cdot\vN\cdot\vPsi_j=1.
\end{equation}

The new data vector given by Eq.~(\ref{eq:vye}) is statistically
orthogonal and verifies\footnote{In the approximation that the 
distribution of $\tvx$ is Gaussian, this also implies statistical 
independence.} 

\begin{equation}
\langle \delta \ty_i \delta \ty_j \rangle=(1+\lambda_i)\ \delta_{ij}.
\end{equation}

One can be easily convinced that this new transform is equivalent to a
KL transform applied on the ``prewhitened'' data, 
$(\vN^{\rm t})^{-1/2}\cdot\tvx$, where

\begin{equation}
\vN \equiv (\vN^{\rm t})^{1/2} \cdot \vN^{1/2}.
\end{equation}

The advantage of this rewriting is that the quantity $\lambda_i$ can
be now considered as a signal to noise ratio $1+\lambda_i=1+S/N$. 
Data compression on the prewhitened data makes now full physical
sense, even if the noise is inhomogeneous or correlated.

The KL compression is generally used as a first step to reduce the size
of the data set keeping as much information as possible, which can
then be processed by the methods of ML estimation or quadratic
estimation which otherwise would not be computationally feasible.  The
final results should be checked against the number of KL modes kept in
the analysis, to show that significant information has not been
discarded.  Note that in addition, since the methods generally used
after KL compression assume Gaussianity, one must check as well that
modes which probe the weakly non-linear regime are {\em not} included
in the analysis to avoid having undesired biases in the final results.

\subsection{Measurements in $N$-Body Simulations}

\label{sec:9}

Measurements of statistics in $N$-body simulations are of course
subject to the cosmic error problem, but can be contaminated by other
spurious effects related to limitations of the numerical approach used
to solve the equations of motion.  Transients, related to the way
initial conditions are usually set up were already discussed in
Sect.~\ref{s:transients}.  Here, we first consider the cosmic error
and the cosmic bias problems, which in practice are slightly different
from the case of galaxy catalogs.  Second, we briefly mention problems
due to $N$-body relaxation and short-range softening of the
gravitational force.

\subsubsection{Cosmic Error and Cosmic Bias in Simulations}

\label{sec:cosmicinsim}

Here we restrict to the case of $N$-body simulations of
self-gravitating collisionless dark matter.  Most of simulations are
done in a cubic box with periodic boundaries.  The first important
consequence is that the average number density of particles, ${\bar
n}_{\rm g}$, is perfectly determined.

The second consequence as mentioned earlier is that edge effects are
inexistent.  The only sources of errors are finite volume and shot
noise.  With the new generation of simulations, discreteness effects
are in general quite small except at small scales or if a sparse
synthetic catalog of ``galaxies'' is extracted from the dark matter
distribution.  Finite volume effects in simulations have been
extensively studied in~\cite{CBS94,CBS95,CBH96}.  For these effects to
be insignificant in measured moments or correlation functions of the
density distribution, the simulation box size $L$ has to be large
compared to the typical size of a large cluster, the correlation
length $R_0$.  Typically it is required that $R_0 \la L/20$.  Even if
this condition is fulfilled, the sampling scales (or separations) $R$
must be small fractions of the box size in order to achieve fair
measurements, typically $R \la L/10$.  Indeed, because of finite
volume effects, moments of the density distribution, cumulants and
$N$-point correlation functions tend to be systematically
underestimated, increasingly with scale.  This is a consequence of
cosmic bias  and {\em effective bias} due to the
skewness of the cosmic distribution function, as discussed in
Sect.~\ref{sec:sec6bis}.

The estimation of cosmic bias was addressed quantitatively at large
scales in~\cite{Seto99} using PT where it was found that although
moments can be affected by as much as 80\% at smoothing scales one
tenth of the size of the box (for $n=-2$), the skewness $S_{3}$ was
affected by at most 15\% at the same scale.  Finite volume effects for
velocity statistics are much more severe, as they are typically
dominated by long wavelength fluctuations, e.g.
see~\cite{JFPTCWCPEN98}.

The most obvious consequence of finite volume effects is the fact that
the high-density tail of the PDF develops a cutoff due to the finite
number of particles.  A method was proposed
in~\cite{CBS92,CBS94,CBS95} and exploited in other
works~\cite{CBH96,MBMS99} to correct the PDF for finite volume
effects, by smoothing and extending to infinity its large-$\delta$
tail.  Another way to bypass finite volume effects consists in doing
several simulations and taking the average value (see,
e.g.~\cite{JWACB95,GaBa95,BGE95}) of the moments or cumulants, with
the appropriate procedure for cumulants to avoid possible biases. 
This is however, by itself not necessarily sufficient, because in each
realization, large scale fluctuations are still missing due to the
periodic boundaries (e.g.~\cite{Seto99}).  In other words, doing a
number of random realizations of given size $L$ with periodic
boundaries is not equivalent to extracting subsamples of size $L$ from
a very large volume.  With many realizations one can reduce
arbitrarily the effect of the skewness of the distribution, but not
the influence of large-scale waves not present due to the finite
volume of the simulations.

\subsubsection{$N$-Body Relaxation and Force Softening}

\label{sec:relsof}

Due to the discrete nature of numerical simulations, there are some
dynamical effects due to interactions between small number of
particles.  To reduce these relaxation effects it is necessary to
bound forces at small interparticle separation, thus a softening
$\epsilon$ is introduced as discussed in Sect.~\ref{sec:numsim}. 
However, this softening does not guarantee the fluid limit.  The
latter is achieved locally only when the number of particles in a
softening volume $\epsilon^{\dim}$ is large.  Typically, the softening
parameter is of order the mean interparticle distance $\lambda$ in
low-resolution simulations, or of order $\lambda/20$ in high
resolution simulations (Sect.~\ref{sec:numsim}).  At early stages of
simulations, where the particles are almost homogeneously distributed,
relaxation effects are thus expected to be significant.  Later, when
the system reached a sufficient degree of nonlinearity, these effects
occur only in underdense regions\footnote{In fact, in these regions,
small but rare groups of particles experiencing strong collisions can
be found even at late stages of the simulations.}.  It is therefore
important to wait long enough so that the simulation has reached a
stage where typical nonlinear structures contain many particles.

Statistically, this is equivalent to say that the correlation length
should be much larger than the mean interparticle distance, $R_0 \gg
\lambda$~\cite{CBH96}.  This criterion is valid for most statistics
but there are exceptions.  For example, it was shown that the void
probability distribution function can be contaminated by the initial
pattern of particles (such as a grid) even at late
stages~\cite{CBS95}.  Indeed, underdense regions tend to expand and to
keep the main features of this initial pattern.  Another consequence
is that the local Poisson approximation is not valid if this initial
pattern presents significant correlations or anticorrelations (such as
a grid or a ``glass''~\cite{BGE95,White96}).

Finally, short-range softening of the forces itself can contaminate
the measurement of statistics at small scales.  With a careful choice
of the timestep (see, e.g.~\cite{EDWF85}) the effects of the softening
parameter are negligible for scales sufficiently large compared to
$\epsilon$, a practical criterion being that the considered scale $R$
verifies $R =\alpha \epsilon$ with $\alpha$ of order a few~\cite{CBH96}.

\clearpage 
\section{\bf Applications to Observations}

\subsection{The Problem of Galaxy Biasing}
\label{sec:bias}

Application to galaxy surveys of the results that have been obtained
for the clustering of dark matter is not trivial, because in principle
there is no guarantee that galaxies are faithful tracers of the dark
matter field.  In other words, the galaxy distribution may be a {\em
biased} realization of the underlying dark matter density field.

A simplified view of biasing often encountered in the literature is
that the two fields, galaxy and matter density fields, are simply
proportional to each other,

\begin{equation}
\delta_g(\vx)=b\,\delta(\vx).
\end{equation}

It implies in particular that the power spectra obey
$P_g(k)=b^2\,P(k)$.  As long as one considers two-point statistics
this might be a reasonable prescription; however, when one wants to
address non-Gaussian properties, this is no more sufficient: the
connection between dark matter fluctuations and galaxies, or clusters
of galaxies, should be given in more detail.

In principle, this relation should be obtained as a prediction of a
given cosmological model.  However, although significant progress has
been done recently to study galaxy formation from ``first principles''
via hydrodynamic numerical
simulations~\cite{CeOs00,KHW99,BCOS99,virgo99}, they still suffer from
limited dynamical range and rely on simplified descriptions of star
formation and supernova feedback, which are poorly understood.  This
fundamental problem implies that when dealing with galaxies, one must
usually include additional (non-cosmological) parameters to describe
the relation between galaxies and dark matter.  These parameters,
known generally as {\em bias parameters}, must be determined from the
data themselves.  In fact, the situation turns out to be more
complicated than that: since there is no generally accepted framework
for galaxy biasing yet, one needs to test the parameterization itself
against the data in addition to obtaining the best fit parameter set.

The complexity of galaxy biasing is reflected in the literature, where
many different approaches have emerged in the last decade or so.  In
addition to the hydrodynamic simulations, two other major lines of
investigations can be identified in studies of galaxy biasing.  The
simplest one, involves a phenomenological mapping from the dark matter
density field to galaxies, which is reviewed in the next section. 
Another approach, that has become popular in recent years, is to split
the problem of galaxy biasing into two different steps~\cite{WhRe78}. 
First, the formation and clustering of dark matter halos, which can be
modeled neglecting non-gravitational effects, this is the subject of
sections~\ref{hierbias} and~\ref{dhbias}.  This step is thought to be
sufficient to describe the spatial distribution of galaxy clusters. 
The second step, discussed in section~\ref{galbias}, is the
distribution of galaxies within halos, which is described by a number
of simplifying assumptions about the complex non-gravitational
physics.  It is generally believed that such processes are likely to
be very important in determining the properties of galaxies while
having little effects on the formation and clustering of dark matter
halos.

Note that observational constraints on biasing (from higher-order
correlations) are discussed in the next chapter (see
Sections~\ref{sec:biagau} and~\ref{sec:biagau2}).

\subsubsection{Some General Results}

\label{genres}

The first theoretical approach to galaxy biasing was put forward by
Kaiser~\cite{Kaiser84}, who showed that if rich galaxy clusters were
rare density peaks in a Gaussian random field, they will be more
strongly clustered than the mass, as observed~\cite{PeHa74,BaSo83}. 
These calculations were further extended in~\cite{PeHe85,BBKS86}.  In
particular, it was found that rare peaks were correlated in such a way
that

\begin{equation}
\mg\delta^2_{\rm peak}\md=b^2_{\rm peak}\ \mg\delta^2\md
\label{fb:peakbias}
\end{equation}

where $\delta_{\rm peak}$ is the local density contrast in the number
density of peaks with a {\em bias parameter}

\begin{equation}
b_{\rm peak}(\nu)={\nu\over\sigma}
\end{equation}

where $\sigma$ is the variance at the peak scale, and $\nu$ is the
intrinsic density contrast of the selected peaks in units of $\sigma$. 
These results led to studies of biasing in CDM numerical
simulations~\cite{DEFW85,WDEF87}, which indeed showed that massive
dark matter halos are more strongly clustered than the mass.  However,
numerical simulations also showed later that dark matter halos are not
always well identified with peaks in the linear density
field~\cite{KQG93}.

An alternative description of biasing which does not rely on the
initial density field, is the local Eulerian bias model.  In this
case, the assumption is that at scales $R$ large enough compared to
those where non-gravitational physics operates, the {\em smoothed}
(over scale $R$) galaxy density at a given point is a function of the
underlying {\em smoothed} density field at {\em the same point},

\begin{equation}
\hat{\delta_g}(\vx) = \mF[\hat{\delta}(\vx)], \ \ \ \ \ \hat{A}(\vx)
\equiv \int_{\vert\vx'\vert<R} \d^3\vx' A(\vx-\vx') W(\vx')
\label{eulocbias}
\end{equation}

where $W$ denotes some smoothing filter.  For large $R$, where
$\hat{\delta} \ll 1$, it is possible to perturbatively expand the
function $\mF$ in Taylor series and compute the galaxy correlation
hierarchy~\cite{FrGa93}.  Indeed, one can write

\begin{equation}
\hat{\delta_g}= \sum_{k=0}^\infty {b_k \over {k!}} \hat{\delta}^k,
\label{eq:taylor}
\end{equation}

where the linear term $b_1$ corresponds to the standard linear bias
factor.  In this large-scale limit, such a local transformation
preserves the hierarchical properties of the matter distribution,
although the values of the hierarchical amplitudes may change
arbitrarily.  In particular~\cite{FrGa93},

\begin{eqnarray} 
\sigma^2_g &=& b_1^2 \sigma^2 \nonumber \\
S_{g,3}&=&b_{1}^{-1}\left(S_3+3c_2\right) \nonumber \\
S_{g,4}&=&b_{1}^{-2}\left(S_4+12c_2S_3+4c_3+12c_2^2 \right) \nonumber \\
S_{g,5}&=&b_{1}^{-3}\Big[S_5
+20c_2S_4+15c_2S_3^2+\left(30c_3+120c_2^2\right)S_3 +5c_4+60c_2c_3+
\nonumber \\ &&+60c_2^3 \Big],  \label{eq:S_g}
\end{eqnarray}

where $c_k \equiv b_k/b_1$.  As pointed out in \cite{FrGa93}, this
framework encompasses the model of bias as a sharp threshold
clipping~\cite{Kaiser84,PoWi84,BBKS86,Szalay88}, where $\delta_g=1 $
for $\delta > \nu \sigma$ and $\delta_g = 0$ otherwise.  Although it
does not have a series representation around $\delta = 0$, such a
clipping applied to a Gaussian background produces a hierarchical
result with $S_{g,p} = p^{p-2}$ in the limit $\nu \gg 1$, $\sigma \ll
1$.  This is the same result as we obtain from \eq(\ref{eq:S_g}) for
an exponential biasing of a Gaussian matter distribution, $\delta_g =
\exp(\alpha \delta/\sigma)$, which is equivalent to the sharp
threshold when the threshold is large and fluctuations are
weak~\cite{BBKS86,Szalay88}.  The exponential bias function has an
expansion $\mF = \sum_k (\alpha \delta /\sigma)^k / k!$ and thus $b_k
= b_{1}^k$, independently of $\alpha$ and $\sigma$.  With $S_p= 0$, the
terms induced in \eq(\ref{eq:S_g}) by $b_k$ alone also give $S_{g,p} =
p^{p-2}$.

As a result of \eq(\ref{eq:S_g}), it is clear that for high order
correlations, $p>2$, a linear bias assumption cannot be a consistent
approximation even at very large scales, since non-linear biasing can
generate higher-order correlations.  To draw any conclusions from the
galaxy distribution about matter correlations of order $p$, properties
of biasing must be included to order $p-1$.

Let us make at this stage a general remark.  {}From Eq.~(\ref{eq:S_g})
it follows that in the simplest case, when the bias is linear, a value
$b_1>1$ reduces the $S_p$ parameters and it may suggest that this
changes how the distribution deviates from a Gaussian (e.g. the galaxy
field would be ``more Gaussian'' than the underlying density field,
given that $S_3$ is smaller).  However, this is obviously an incorrect
conclusion, a linear scaling of the density field cannot alter the
degree of non-Gaussianity.  The reason is that the actual measure of
non-Gaussianity is encoded not by the hierarchical amplitudes $S_p$
but rather by the dimensionless skewness $B_3=S_3\sigma$, kurtosis
$B_4=S_4 \sigma^2$, and so on, which remain invariant under linear
biasing.  These dimensionless quantities are indeed what characterize
the probability distribution function, as it clearly appears in an
Edgeworth expansion, Eq.~(\ref{fb:edge}).

Since Fourier transforms are effectively a smoothing operation,
similar results to those above hold for Fourier-space statistics at
low wavenumbers.  In this regime, the galaxy density power spectrum
$P_g(k)$ is given by

\begin{equation}
P_g(k)=b_1^2\,P(k), \label{P_g}
\end{equation}

and the galaxy (reduced) bispectrum obeys [recall Eq.~(\ref{q})]

\begin{equation}
Q_g(\vk_1,\vk_2,\vk_3)=\frac{1}{b_1} Q(\vk_1,\vk_2,\vk_3)+{b_2\over
b_1^2} \label{Q_g_eul}
\end{equation}

As discussed in Section~\ref{bispgrav}, $Q$ given by
Eq.~(\ref{qtree}), is very insensitive to cosmological parameters and
depends mostly on triangle configuration and the power spectrum
spectral index.  Since the latter is not affected by bias in the
large-scale limit, Eq.~(\ref{P_g}), it can be measured from the galaxy
power spectrum and used to predict $Q(\vk_1,\vk_2,\vk_3)$ as a
function of triangle configuration.  As first proposed
in~\cite{FrGa94,Fry94a}, a measurement of $Q_g$ as a function of
triangle shape can be used to determine $1/b_1$ and $b_2/b_1^2$.  So
far, this technique has only been applied to IRAS
galaxies~\cite{SFFF01,FFFS01}, as will be reviewed in the next chapter
(see Sect.~\ref{sec:q3z})\footnote{Similar relations to
Eq.~(\ref{eq:S_g}) and Eq.~(\ref{Q_g_eul}) can be obtained for
cumulant correlators, see~\cite{Szapudi98b}.}.

The results above suggest that local biasing does not change the {\em
shape} of the correlation function or power spectrum in the
large-scale limit, just scaling them by a constant factor $b_1^2$
independent of scale.  This derivation~\cite{FrGa93} assumes that the
smoothing scale is large enough so that $\hat{\delta} \ll 1$, but in
fact, it can be shown that this continues to hold in more general
situations.  For example, an arbitrary local transformation of a
{\em Gaussian} field, leads to a bias that cannot be an increasing function
of scale and that becomes constant in the large-scale limit,
irrespective of the amplitude of the rms
fluctuations~\cite{Coles93}\footnote{but this is an unrealistic
situation since Gaussianity breaks down when the rms fluctuations are
larger than unity.}.  However, it is easy to show that if the
underlying density field is hierarchical (in the sense that the
$C_{pq}$ parameters in Eq.~(\ref{fb:Spqdef}) are independent of
scale), a local mapping such as that in Eq.~(\ref{eulocbias}) does
lead to a bias independent of scale in the large-scale limit even if
$\hat{\delta} \gg 1$~\cite{BeSc92,ScWe98}.

Recent studies of galaxy
biasing~\cite{ScWe98,DeLa99,BCOS99,Matsubara99} have focused on the
fact that Eq.~(\ref{eulocbias}) assumes not only that the bias is local
but also deterministic; that is, the galaxy distribution is completely
determined by the underlying mass distribution.  In practice, however,
it is likely that galaxy formation depends on other variables besides
the density field, and that consequently the relation between
$\hat{\delta}_g(\vx)$ and $\hat{\delta}(\vx)$ is not deterministic but
rather stochastic,

\begin{equation}
\hat{\delta}_g(\vx) = \mF [\hat{\delta}(\vx)] +
\varepsilon_\delta(\vx), \label{eulocbias2}
\end{equation}

\noindent where the random field $\varepsilon_\delta(\vx)$ denotes the
scatter in the biasing relation {\em at a given} $\delta$ due to the
fact that $\hat{\delta}(\vx)$ does not completely determine
$\hat{\delta}_g(\vx)$.  Clearly for an arbitrary scatter, the effects
of $\varepsilon_\delta(\vx)$ on clustering statistics can be
arbitrarily strong.  However, under the assumption that the scatter is
local, in the sense that the correlation functions of
$\varepsilon_\delta(\vx)$ vanish sufficiently fast at large
separations (i.e. faster than the correlations in the density field),
the deterministic bias results hold for the two-point correlation
function in the large-scale limit~\cite{ScWe98}.  For the power
spectrum, on the other hand, in addition to a constant large-scale
bias, stochasticity leads to a constant offset (given by the rms
scatter) similar to Poisson fluctuations due to shot
noise~\cite{ScWe98,DeLa99}.

Another interesting aspect of stochasticity was studied
in~\cite{Matsubara99}, in connection with non-local biasing.  A simple
result can be obtained as follows.  Suppose that biasing is non-local
but linear, then we can write

\begin{equation}
\delta_g(\vx) = \int \delta(\vx') K(\vx-\vx') \d^3\vx', \label{nonloclin}
\end{equation}

where the kernel $K$ specifies how the galaxy field at position $\vx$
depends on the density field at arbitrary locations $\vx'$.  This
convolution of the density field leads to stochasticity in real space,
i.e. the cross-correlation coefficient $r$

\begin{equation}
r(s) \equiv \frac{\mg \delta(\vx) \delta_g(\vx') \md}{ \sqrt{ \xi_g(s)
\xi(s) }}, \label{rijreal}
\end{equation}

where $s \equiv |\vx-\vx'|$, is not necessarily unity.  However, due
to the convolution theorem, the cross-correlation coefficient in {\em
Fourier space} will be exactly unity, thus

\begin{equation}
\mg\delta_g(\vk)\delta(\vk')\md=\delta_D(\vk+\vk')\,b(k)P(k)
\label{fb:linbias}
\end{equation}

and

\begin{equation}
\mg\delta_g(\vk)\delta_g(\vk')\md=\delta_D(\vk+\vk')\,b^2(k)P(k),
\end{equation}

where the bias $b(k)$ is the Fourier transform of the kernel $K$.  The
study in~\cite{Matsubara99} showed on the other hand that the
real-space stochasticity (in the sense that $r<1$) at large scales was
weak for some class of models.  At small scales, however, significant
deviations from $r<1$ cannot be excluded, for example due to nonlinear
couplings in Eq.~(\ref{nonloclin}).  However, without specifying more
about the details of the biasing scheme, it is very difficult to go
much beyond these results.

Most of the general results discussed so far have been observed in
hydrodynamical simulations of galaxy formation.  For example, in
\cite{BCOS99} it has been obtained that at large scale ($R\ga 15$
Mpc/h) the bias parameter tends to be constant and the cross
correlation coefficient $r$ reaches unity for oldest galaxies.  The
authors stress that the bias shows a substantial scale dependence at
smaller scales, which they attribute to the dependence of galaxy
formation on the temperature of the gas (which governs its ability to
cool).  In addition, they observe a substantial amount of
stochasticity for young galaxies ($r\approx 0.5$), even at large
scales.  However, these results are in disagreement with observations
of the LCRS survey, where it was found that after correcting for
errors in the selection function the cross-correlation between early
and late-type galaxies is $r \approx 0.95$~\cite{Blanton00}.

Another assumption that enters into the local Eulerian biasing model
discussed above, is that the galaxy field depends on the underlying
density field at {\em the same time}.  In practice, it is expected
that to some extent the merging and tidal effects {\em histories}
affect the final light distribution.  This can lead to non-trivial
time evolution of biasing.  For instance, as shown in~\cite{Fry96}, if
galaxy formation was very active in the past but after some time it
becomes subdominant, then {\em in the absence of merging} the galaxy
density contrast is expected to follow the continuity equation,

\begin{equation}
a{\partial \delta_g\over\partial a}+\vu.\nabla\delta_g+(1+\delta_g)
\nabla.\vu=0 \label{fb:contdg}
\end{equation}

where $\vu$ is the peculiar velocity field of the dark matter field:
galaxies are simple test particles that follow the large-scale flows.
Formally this equation can be rewritten as

\begin{equation}
{\d\log(1+\delta_g)\over\d\tau}= {\d\log(1+\delta)\over\d\tau}
\label{solcont}
\end{equation}

where $\d/\d\tau$ is the convective derivative.  As a consequence the
galaxy density field is expected to resemble more and more the density
field in terms of correlation properties: both the bias parameters,
$b_{k}$, and the cross-correlation coefficient, $r$, are expected to
approach unity, galaxies ``de-bias'' when they just follow the
gravitational field~\cite{NuDa94,Fry96,TePe98}.  The higher-order
moments characterized by $S_p$ are also expected to get closer to
those for the dark matter field.  These calculations have been
illustrated in~\cite{Fry96,TKS99}.

One obvious limitation of these ``galaxy conserving'' schemes is the
assumption that there is no merging, which is expected to play a
central role in hierarchical structure formation.  In addition,
ongoing galaxy formation leads to galaxies formed at different
redshifts with different ``bias at birth''.  Indeed, models based on
the continuity equation predict a {\em slower} time evolution of bias
than observed in simulations~\cite{BCOST00,SLSDKW01}, i.e. galaxies
become unbiased {\em faster} than when these effects are neglected.

An interesting consequence of Eq.~(\ref{solcont}) has been unveiled
in~\cite{CLMP98} where they remark that the solution is

\begin{equation}
1+\delta_g(\vx,z)=[1+\delta_g^L(\vq)][1+\delta(\vx,z)]
\end{equation}

where the galaxy field at the {\em Lagrangian} position $\vq$ is
obtained from the linear density field at $\vq=\vx -\Psi(\vq,z)$ by
$\delta_g^L(\vq)= \sum b_k^L/k!  \delta^L(\vq)$.  That is, in this
model, the bias is assumed to be local in {\em Lagrangian} space
rather than Eulerian space.  In this particular case, unlike in peaks
biasing mentioned above, once the galaxy field is identified in the
initial conditions, its subsequent evolution is incorporated by
Lagrangian perturbation theory to account for displacement effects due
to the gravitational dynamics.  In this case, the tree-level
bispectrum amplitude becomes~\cite{CPK00}

\begin{eqnarray}
Q_g=\frac{1}{b_1} Q + \frac{b_2^L}{b_1^2} + \frac{4b_1^L}{7b_1^2}
\times \frac{\Delta Q_{12} P_g(k_1) P_g(k_2) + {\rm cyc.}}{P_g(k_1)
P_g(k_2) + {\rm cyc.}}, \label{fb:B_g}
\end{eqnarray}

where $\Delta Q_{12} \equiv 1- (\vk_1.\vk_2)^2/(k_1 k_2)^2$, and $b_1
\equiv 1+ b_0^L+b_1^L$.  Note that the last term in this expression
gives a different prediction than Eq.~(\ref{Q_g_eul}) for the
dependence of the galaxy bispectrum as a function of triangle
configuration that can be tested against observations;   application
to the PSCz survey bispectrum~\cite{FFFS01} suggests that the model in
Eq.~(\ref{Q_g_eul}) fits better the observations than
Eq.~(\ref{fb:B_g}).

Finally, we should also mention that a number of phenomenological
(more complicated) mappings from dark matter to galaxies have been
studied in detail in the literature~\cite{MPH98,CHWF98,NBW00,BNW01}.
The results are consistent with expectations based on the simpler
models discussed in this section.

\subsubsection{Halo Clustering in the Tree Hierarchical Model}

\label{hierbias}

As mentioned previously the validity of the prescription
(\ref{eulocbias}) is subject to the assumption that the mass density
contrast is small.  For biasing at small scales this cannot be a valid
assumption.  Insights into the functional relation between the halo
field and the matter field then demand for a precise modeling of the
matter fields.  The {\em tree hierarchical model},
Eq.~(\ref{fb:TreeHM}), has been shown to provide a solid ground to
undertake such an investigation~\cite{BeSc92,BeSc99}.  In these papers
the {\em connected part} of joint density distribution have been
computed for an arbitrary number of cells,
$p_c(\delta_1,\dots,\delta_p)$ and showed to be of the form,

\begin{equation}
p_c(\delta_1(\vx_1),\dots,\delta_p(\vx_p)) = \sum_{a=1}^{t_p}
Q_{p,a}(\delta_1,\dots,\delta_N) \sum_{\rm labelings}\ \prod_{\rm
edges}^{p-1} \xi_2(\vx_i,\vx_j), \label{fb:jointPc} 
\end{equation}

with

\begin{equation}
Q_{p,a}(\delta_1,\dots,\delta_p)=\Pi_i p(\delta_i)\nu_q(\delta_i)
\label{fb:jointQ} 
\end{equation}

where $\nu_q(\delta)$ is a function of the local density contrast that
depends on the number $q$ of lines it is connected to in the graph.
This form implies for instance that

\begin{equation}
p(\delta_1,\delta_2)=p(\delta_1)p(\delta_2)\left[1+\xi_2(\vx_1,\vx_2)
\,\nu_1(\delta_1)\,\nu_1(\delta_2)\right].
\end{equation}

At small scales, when the variance is large, the density contrast of
dark matter halos is much larger than unity, and should be reliably
given by a simple threshold condition, $\de_i \ga \de_{\rm thres}$.
Therefore the function $\nu_1$ describes the halo bias, and
higher-order connected (two-point) joint moments follow directly from
this {\em bias function} and the two-point correlation function of the
mass. In this framework a number of important properties and results
have been derived,

\begin{enumerate}

\item[(i)] the correlation functions of the halo population follow a
tree structure similar to the one of the matter field in the large
separation limit (e.g. when the distances between the halos are much
larger than their size);

\item[(ii)] the values of the vertices depend only on the internal
properties of the halos, namely on the reduced variable,

\begin{equation} 
x={\rho\over\overline{\rho}\sigma^2};
\label{fb:xdef}
\end{equation}

\item[(iii)] all vertices are growing functions of $x$ and have a
specific large $x$ asymptotic behavior,

\begin{eqnarray} \nu_1(x)&\equiv& b(x)\sim x\\ \nu_p(x)&\sim& b^p(x); 
\end{eqnarray}
\end{enumerate}

The large $x$ limit that has been found for the high-threshold
clipping limit is once again recovered, since we expect in such a
model that $S_{h,p}\to p^{p-2}$ when $x\to \infty$. Property (iii),
together with (ii), also holds for halos in the framework of the
Press-Schechter approach, as we shall see in the next section [see
discussion below Eq.~(\ref{epsilons})].

\begin{figure}
\centerline{ \epsfysize=7truecm\epsfbox{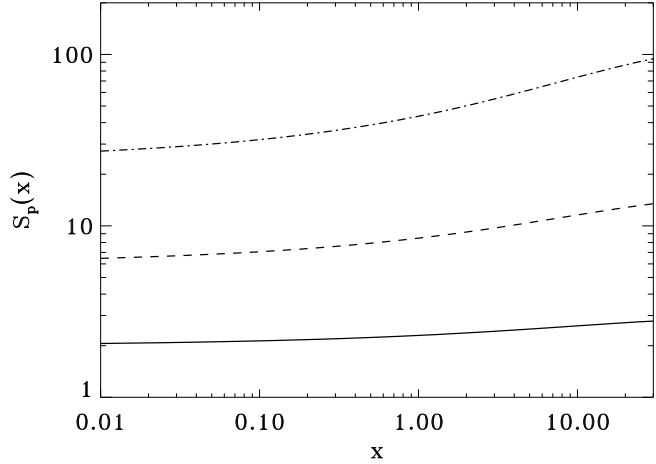} }
\caption{Example of a computation of the $S_3$, $S_4$ and $S_5$
parameters in the tree hierarchical model for dark matter halos
selected with a varying threshold in $x$, defined by Eq.
(\ref{fb:xdef}).  Calculations have been made with the vertex
generating function, $\zeta(\tau)=(1-\tau/\kappa)^{-\kappa}$ with
$\kappa=1.3$.  For large values of $x$ one explicitly sees the $S_p\to
p^{p-2}$ behavior expected in the high threshold limit.} 
\label{S3_S4_halos_HModel}
\end{figure}

In addition, it is possible to derive the functions $\nu_p(x)$ in
terms of the vertex generating function $\zeta(\tau)$.  These results
read,

\begin{equation}
\nu_p(x)=\int_{-\ii\infty}^{\ii\infty}\d y\,\varphi^{(p)}(y)\exp(xy)/
\int_{-\ii\infty}^{\ii\infty}\d y\,\varphi(y)\exp(xy)
\end{equation}

where the function $\varphi^{(p)}(y)$ can be expressed in terms of
$\zeta$ and its derivatives (see~\cite{BeSc99} for details).  In case
of the minimal tree model where all vertices are pure numbers, we have,

\begin{eqnarray}
\varphi(y)&=&y\zeta(\tau)+\tau^2/2,\ \ \tau/\zeta'(\tau)=-y;\\
\varphi^{(1)}(y)&=&\tau(y);\\ \varphi^{(2)}(y)&=&-{y\zeta''(\tau)\over
1+y\zeta''(\tau)};\\ \varphi^{(3)}(y)&=&-{y\zeta'''(\tau)\over
[1+y\zeta''(\tau)]^3};\\ \dots\nonumber
\end{eqnarray}

These results provide potentially a complete model for dark matter
halo biasing.  The explicit dependence of the skewness and kurtosis
parameters has been computed in these hierarchical models
in~\cite{BeSc99}, see Fig.~\ref{S3_S4_halos_HModel}.

\begin{figure}
\vspace{5 cm} \special{hscale=70 vscale=70 voffset=0 hoffset=20
psfile=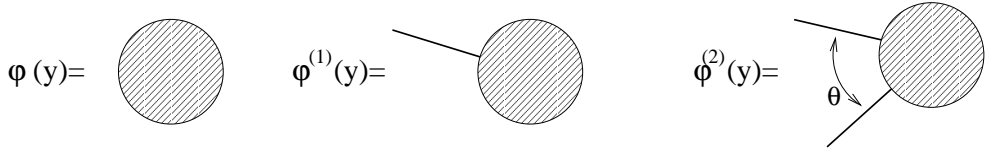} \caption{The functions $\varphi(y)$,
$\varphi^{(1)}(y)$ and $\varphi^{(2)}(y)$ are the generating functions
of trees with respectively 0, 1 and 2 external lines.  For orders
above 2 a possible angular dependence with the outgoing lines cannot
be excluded.} \label{nupdex}
\end{figure}

Although initially undertaken in the strongly nonlinear regime, these
results a priori extend to weakly nonlinear scales; that is, to scales
where halo separations are in the weakly nonlinear regime.  Indeed
only the tree structure, in a quite general sense (see
\cite{Bernardeau95,BeSc99} for details), is required to get these
results.  In this case the vertex $\nu_2(x)$ might bear a non-trivial
angular dependence originating from the expression of
$\varphi^{(2)}(y)$, see Fig.  \ref{nupdex}.  There is therefore a
priori no reason to recover the result in Eq.~(\ref{Q_g_eul}) for the
halo bispectrum.  The connection, if any, with simple relations such
as Eq.~(\ref{eulocbias}) is thus still to be understood.
Stochasticity emerging due to nonlinear effects is in particular
likely to limit the validity of Eq.~(\ref{eulocbias}).

\subsubsection{Halo Clustering in the Extended Press-Schechter Approach}

\label{dhbias}

The results obtained in the previous subsection correspond to the
correlations properties of dense halos detected in a snapshot of the
nonlinear density field.  This approach does not give any insights on
the merging history of the halos that is likely to be important for
the galaxy properties.  And because dark matter halos are highly
non-linear objects, their formation and evolution has traditionally
been studied using numerical simulations.

However, a number of analytical
models~\cite{MoWh96,MJW97,CLMP98,ShLe99}, based on the so-called
Press-Schechter (PS) formalism~\cite{PrSc74} and
extensions~\cite{BCKE91,Bower91,LaCo93,KaWh93}, revealed a good
description of the numerical simulation results.

The PS formalism aims at giving the comoving number density of halos
as a function of their mass $m$,

\begin{equation}
\frac{m^2 n(m)}{\bar{\rho}}= \sqrt{\frac{2 y^2}{\pi}} \exp
\Big(-\frac{y^2}{2}\Big) \frac{\d\ln y}{\d\ln m},
\end{equation}

where $\bar{\rho}$ denotes the average density of the universe, and $y
\equiv \delta_c/\sigma(m)$, with $\delta_c\approx 1.68$ the collapse
threshold given by the spherical collapse model and $\sigma^2(m)$ is
the variance of the linearly extrapolated density field smoothed at
scale $R=(3m/4\pi\bar{\rho})^{1/3}$.  The average number of halos in a
spherical region of comoving radius $R_0$ and over-density $\delta_0$
is

\begin{equation} {\cal N}(m|\delta_0)dm = \frac{m_0}{m}
f(\sigma,\delta_c|\sigma_0,\delta_0) \frac{\d \sigma^2}{\d m} \d m,
\end{equation}

where

\begin{equation}
f(\sigma,\delta_c|\sigma_0,\delta_0)=\frac{1}{\sqrt{2\pi}}
\frac{\delta_c-\delta_0}{(\sigma^2-\sigma^2_0)^{3/2}} \exp \Big[-
\frac{(\delta_c-\delta_0)^2}{2(\sigma^2-\sigma^2_0)} \Big]
\end{equation}

is the fraction of the mass in a region of initial radius $R_0$ and
linear over-density $\delta_0$ that is at present in halos of mass
$m$~\cite{BCKE91,Bower91}.  The Lagrangian halo density contrast is
then~\cite{MoWh96}

\begin{equation}
\delta_h^L(m|\delta_0)=\frac{{\cal N}(m|\delta_0)}{n(m)V_0}-1,
\label{delta_hl}
\end{equation}

where $V_0=4\pi R_0^3/3$.  When $R_0 \gg R$ so that $\sigma_0 \ll
\sigma$ and $|\delta_0|\ll \delta_c$, this gives

\begin{equation}
\delta_h^L(m|\delta_0)=\frac{y^2-1}{\delta_c} \delta_0. 
\label{delta_hl2}
\end{equation}

On the other hand, the Eulerian halo density contrast is~\cite{MoWh96}

\begin{equation}
\delta_h(m|\delta_0)=\frac{{\cal N}(m|\delta_0)}{n(m)V}-1,
\label{delta_h}
\end{equation}

where the volume $V=4\pi R^3/3$ is related to the initial volume by
$R_0=R(1+\delta)^{1/3}$ with $\delta(\delta_0)=\sum_{m=1}^{\infty}
\nu_m \delta_0^m$ given by the spherical collapse model.  When
considered as a function of $\delta$, Eq.~(\ref{delta_h}) gives a bias
relation similar to Eq.~(\ref{eq:taylor}) with bias
parameters~\cite{MJW97}

\begin{eqnarray}
b_1(m)&=&1+\epsilon_1,\ \ \ \ \ \ \ \ \ \
b_2(m)=2(1-\nu_2)\epsilon_1+\epsilon_2, \nonumber \\
b_3(m)&=&6(\nu_3-\nu_2)\epsilon_1+3(1-2\nu_2)\epsilon_2+\epsilon_3,
\label{biaspar}
\end{eqnarray}

with

\begin{equation}
\epsilon_1=\frac{y^2-1}{\delta_c},\ \ \ \ \
\epsilon_2=\frac{y^2(y^2-3)}{\delta_c^2},\ \ \ \ \
\epsilon_3=\frac{y^2(y^4-6y^2+3)}{\delta_c^3}.
\label{epsilons}
\end{equation}

This framework has been extended to give halo biasing beyond the
spherical collapse approximation, in particular~\cite{CLMP98} discuss
the use of the Zel'dovich approximation, the frozen-flow approximation
and second-order Eulerian PT. In addition,~\cite{SMT01} study the
effects of ellipsoidal collapse on both the mass function and the
biasing of dark matter halos.  They show that tidal effects change the
threshold condition for collapse to become a function of mass,
$\delta_c(m)$, and that the resulting halo bias and mass function are
in better agreement with numerical simulations than the PS ones.  In
particular, less massive halos are more strongly clustered than in PS
calculations as summarized by fitting formulae derived from N-body
simulations~\cite{Jing99,PCL99}, and low (high) mass halos are less
(more) abundant than predicted in PS~\cite{ShTo99,JFWCCECY01}.

The higher-order moments for dark matter halos can be calculated from
the expansion in Eqs.~(\ref{biaspar}) and (\ref{eq:S_g}), as first
done in~\cite{MJW97}.  For instance, in the rare peak limit $b_1\sim
y^2/\delta_c \gg 1$ and $b_2 \sim b_1^2$ so that the three-point
function obeys the hierarchical model with $Q_3=1$ (or equivalently
$S_3=3$).  This actually extends to any order to give $Q_N=1$, i.e.
$S_p=p^{p-2}$ in this limit~\cite{MJW97}.

The fact that dark matter halos are spatially exclusive induces
non-trivial features on their correlation functions at small scales,
which cannot be modeled simply as a biasing factor acting on the mass
correlation functions.  In particular, the variance becomes
significantly less than the Poisson value at small
scales~\cite{MoWh96}.  A detailed discussion of exclusion effects can
be found in~\cite{ShLe99}.

\subsubsection{Galaxy Clustering}

\label{galbias}

Since galaxy formation cannot yet be described from first principles,
a number of prescriptions based on reasonable recipes for
approximating the complicated physics have been proposed for
incorporating galaxy formation into numerical simulations of dark
matter gravitational clustering~\cite{KCDW99,SoPr99,CLBF00}.  These
``semi-analytic galaxy formation'' schemes can provide detailed
predictions for galaxy properties in hierarchical structure formation
models, which can then be compared with observations.

The basic assumption in the semi-analytic approach is that the
distribution of galaxies within halos can be described by a number of
simplifying assumptions regarding gas cooling and feedback effects
from supernova.  For the purposes of large-scale structure
predictions, the main outcome of this procedure is the number of
galaxies that populate a halo of a given mass, $N_{\rm gal}(m)$.
Typically, at large mass $\lexp N_{\rm gal}(m)\rexp \sim m^\alpha$
with $\alpha<1$, and below some cutoff mass $N_{\rm gal}(m)=0$.  The
physical basis for this behavior is that for large masses the gas
cooling time becomes larger than the Hubble time, so galaxy formation
is suppressed in large-mass halos (therefore $\lexp N_{\rm
gal}(m)\rexp $ increases less rapidly than the mass).  On the other
hand, in small-mass halos effects such as supernova winds can blow
away the gas from halos, also suppressing galaxy formation.

A useful analytical model has been recently developed, generally known
as ``the halo model'', which can be easily modified to provide a
description of galaxy clustering using knowledge of the $N_{\rm
gal}(m)$ relation and the clustering of dark matter halos described in
Sect.~\ref{dhbias}.  The starting point is a description of the dark
matter distribution in terms of halos with masses, profiles and
correlations consistent with those obtained in numerical simulations.
This is a particular realization of the formalism first worked out
in~\cite{ScBe91} for general distribution of seed masses, although
precursors which did not include halo-halo correlations were studied
long before~\cite{NeSc52,Peebles74c,McSi77}.

Let $u_m(\vr)$ be the profile of dark matter halos of mass $m$ (for
example, as given in~\cite{NFW97,MQGSL99}), normalized so that $\int
\d^3x' u_m(\vx-\vx') =1$, and $n(m)$ be the mass function, with $\int
n(m) m \d m=\bar{\rho}$ and $\bar{\rho}$ the mean background density.
The power spectrum in this model is written
as~\cite{ShJa97,PeSm00,Seljak00,MaFr00c,CoHu01,SSHJ01}

\begin{eqnarray}
\bar{\rho}^2 P(k)&=& (2\pi)^3 \int n(m) m^2 \d m |u_m(\vk)|^2 +
(2\pi)^6 \int u_{m_1}(k) n(m_1) m_1 \d m_1 \nonumber \\ & & \times
\int u_{m_2}(k) n(m_2) m_2 \d m_2 P(k;m_1,m_2), \label{pkm}
\end{eqnarray}

\noindent where $P(k;m_1,m_2)$ represents the power spectrum of halos
of mass $m_1$ and $m_2$.  The first term denotes the power spectrum
coming from pairs inside the same halo (``1-halo'' term), whereas the
second contribution comes from pairs in different halos (``2-halo''
term).  Similarly, the bispectrum is given by

\begin{eqnarray}
\label{Bhalo} \bar{\rho}^3 B_{123}&=& (2\pi)^3 \int n(m) m^3 \d m\
\Pi_{i=1}^3 u_m(\vk_i) + (2\pi)^6 \int u_{m_1}(k_1) n(m_1) m_1 \d m_1
\nonumber \\ & & \times \int u_{m_2}(k_2) u_{m_2}(k_3) n(m_2) m_2^2 \d
m_2 P(k_1;m_1,m_2) + {\rm cyc.}\nonumber \\ & & + (2\pi)^9 \Big(
\prod_{i=1}^3 \int u_{m_i}(k_i) n(m_i) m_i \d m_i \Big)
B_{123}(m_1,m_2,m_3) ,
\end{eqnarray}

\noindent where $B_{123}(m_1,m_2,m_3)$ denotes the bispectrum of halos
of mass $m_1,m_2,m_3$.  Again, contributions in Eq.~(\ref{Bhalo}) can
be classified according to the spatial location of triplets, from
``1-halo'' (first term) to ``3-halo'' (last term).  The halo-halo
correlations, encoded in $P(k;m_1,m_2)$, $B_{123}(m_1,m_2,m_3)$ and so
on, are described by non-linear PT plus the halo-biasing prescription
discussed in Sect.~\ref{dhbias}, Eq.~(\ref{biaspar}), plus
Eqs.~(\ref{eq:S_g}-\ref{Q_g_eul}) with mass correlation functions
obtained from perturbation theory.

To describe {\em galaxy} clustering, one needs to specify the
distribution (mean and the higher-order moments) of the number of
galaxies which can inhabit a halo of mass $m$.  This is an output of
the semi-analytic galaxy formation schemes,
e.g.~\cite{KCDW99,BCFBL00}, or some parameterization can be
implemented (see e.g.~\cite{SSHJ01,BeWe01,Benson01}) which is used
to fit the clustering statistics.  Assuming that galaxies follow the
dark matter profile, the galaxy power spectrum
reads~\cite{Seljak00,SSHJ01}

\begin{eqnarray}
\bar{n}_g^2 P_g(k) &=& (2\pi)^3 \int n(m) \lexp N_{\rm gal}^2(m) \rexp
\d m |u_m(\vk)|^2  \nonumber \\ & + & (2\pi)^6 \left[ \int u_{m}(k)
n(m) \d m b_1(m)  \lexp N_{\rm gal}(m)\rexp \right]^2  P_L(k),
\label{pkg}
\end{eqnarray}

\noindent and similarly for the bispectrum,  where the mean number
density of galaxies is

\begin{equation}
\bar{n}_g = \int n(m) \lexp N_{\rm gal}(m)\rexp \d m.
\end{equation}

\noindent Thus, knowledge of the number of galaxies per halo moments
$\lexp N_{\rm gal}^n(m) \rexp$ as a function of halo mass gives a
complete description of the galaxy clustering statistics within this
framework.  Note that in the large-scale limit, the galaxy bias
parameters reduce to [$u_{m}(k) \rightarrow 1$]

\begin{equation}
b_i \approx \frac{1}{\bar{n}_g} \int n(m) \d m\ b_i(m)\ \lexp N_{\rm
gal}(m) \rexp.  \label{beff}
\end{equation}

Therefore, in this prescription the large-scale bias parameters are
not independent, the whole hierarchy of $b_i$'s is a result of
Eqs.~(\ref{biaspar}) for $b_i(m)$ and the $\langle N_{\rm gal}(m)
\rangle$ relation, which can be described by only a few parameters. 
In addition, the higher-order moments $\langle N_{\rm gal}^n(m)
\rangle$ with $n>1$, determine the small-scale behavior of galaxy
correlations; however, relations can be obtained between these moments
and the mean which, if robust to details\footnote{The simplest of such
relations assumes Poisson statistics, where $\lexp N_{\rm gal}(N_{\rm
gal}-1)\ldots (N_{\rm gal}-j) \rexp =\lexp N_{\rm gal} \rexp^{j+1}$,
but it is known to fail for low-mass halos which have sub-Poisson
dispersions~\cite{KCDW99,BCFBL00}.  A simple fix assumes a binomial
distribution~\cite{SSHJ01}, with two free parameters that reproduce
the mean and second moment, and automatically predict the $n>2$
moments.  However, it is not known yet how well this model does
predict the $n>2$ moments.  Other prescriptions are given
in~\cite{BCFBL00,BeWe01,Benson01}; in particular, \cite{BeWe01} study
in detail the sensitivity of galaxy clustering to the underlying
distribution.}, means that the parametrization of the mean relation is
the main ingredient of galaxy biasing.  In this sense, this framework
promises to be a very powerful way of constraining galaxy biasing.

\begin{figure}
\centering
\centerline{\epsfxsize=8truecm\epsfysize=8truecm\epsfbox{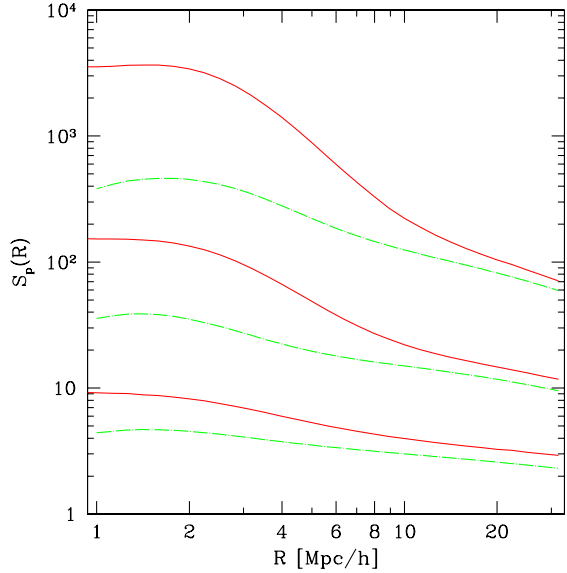}}
\caption{The $S_p$ parameters for $p=3,4,5$ (from bottom to top) for
dark matter (solid) and galaxies (dot-dashed) as a function of
smoothing scale $R$.  These predictions correspond to those of the
halo model, for galaxies they assume that $\lexp N_{\rm gal} \rexp =
(m/m_0)^{0.8}$ for $m>m_0=8\times 10^{11}$ $M_{\odot} h^{-1}$, $\lexp
N_{\rm gal} \rexp = (m/m_0)$ for $m_c<m<m_0$ and $\lexp N_{\rm gal}
\rexp =0$ for $m<m_c= 4\times 10^9$ $M_{\odot} h^{-1}$.}
\label{Sp_gal}
\end{figure}

The weighing introduced by the $\lexp N_{\rm gal}^n(m) \rexp$ on
clustering statistics has many desirable properties.  In particular,
the suppression of galaxy formation in high-mass halos leads to a
galaxy power spectrum that displays power-law-like
behavior\footnote{In addition, note that a power-law behavior
has also been obtained in numerical simulations by selecting
`galaxies' as halos of specific circular
velocities~\cite{CKKK99}.}~\cite{BCFBL00,Seljak00,PeSm00,SSHJ01} and
higher-order correlations show smaller amplitudes at small scales than
their dark matter counterparts~\cite{SSHJ01} (see Fig.~\ref{Sp_gal}),
as observed in galaxy catalogs.  A very important additional
consideration is that this high-mass suppression also leads to
velocity dispersion of galaxies in agreement with galaxy surveys such
as LCRS~\cite{JMB98}.

\subsection{Projection Effects}

\label{sec:projeff}
\def\lag{\langle }  \def\rag{\rangle}  \def\Om{\Omega_m}
\def\Ol{\Omega_{\Lambda}}  \def\drad{\chi}  \def\De{{\cal D}}
\def\deltap{\delta_{2D}}  \def\varphip{\varphi_{\rm proj}}
\def\varphic{\varphi_{\rm cyl}}

This subsection is devoted to the particular case of angular surveys.
These surveys constitute a large part of the available data and allow
to probe the statistical properties of the cosmic density field at
large scales, as we shall discuss in the next chapter, and furthermore
they do not suffer from redshift-space distortions.  Although they
do not really probe new aspects of gravitational dynamics, the
filtering scheme deserves a specific treatment.  It is also worth
noting here, as we shall briefly discuss in the next section, that
this filtering directly applies to weak lensing observations that are
now emerging, see e.g.~\cite{Mellier99} for a review.

In the following we first review the general aspects of projection
effects, and quickly turn to the widely used small-angle
approximation, where most applications have been done.  We then show
how the three dimensional (3D) hierarchical model projects into a two
dimensional (2D) hierarchy, where the 3D and 2D hierarchical
coefficients are simply related.  In Sects.~\ref{chpt} and
\ref{sec:q3projproj} we go beyond the hierarchical assumption to
present predictions for the projected density in PT.  Finally, in
Sect.~\ref{pdfpd}, we discuss the reconstruction of the one-point PDF
of the projected density.

\subsubsection{The Projected Density Contrast}
\label{pdc}

Let us describe the comoving position $\vx$ in terms of the radial
distance $\drad$ and angular distance $\De$ so that
$\vx=(\drad,\De\theta)$\footnote{See cosmology textbooks, e.g.
\cite{Peebles93}, or the pedagogical summary in~\cite{Hogg99} for a
detailed presentation of these aspects.}.  The radial distance is
defined by\footnote{Note that the $\Omega$ parameters refer here to 
those evaluated at redshift $z=0$.}

\begin{equation}  
\d\drad = \frac{c\,\d
z/H_0}{\sqrt{\Ol+(1-\Om-\Ol)(1+z)^2+\Om(1+z)^3}}, 
\label{fb:chidef}
\end{equation}

with $H_0$ Hubble's constant\footnote{Throughout this work we use
$H_0=100~ h~{\rm km/sec/Mpc}$.} and $c$ the speed of light, while the
angular distance is defined by,

\begin{equation}
\De(\drad) = \frac{c / H_0}{\sqrt{1-\Om-\Ol}} \sinh \left(
\sqrt{1-\Om-\Ol} \; \frac{H_0 \drad}{c} \right).  \label{fb:Dedef}
\end{equation}

In general, for angular surveys, the measured density contrast of
galaxy counts at angular direction $\theta$ is related to the 3D
density contrast through,

\begin{equation}
\delta_{2D}(\theta)=\int\d\drad\,\drad^2\,
\psi(\drad)\,\delta_{3D}(\drad,\De\,\theta)
\label{fb:defdeltap}
\end{equation}

where $\psi(\drad)$ is the selection function (normalized such that
$\int \d\drad\,\drad^2\,\psi(\drad) =1$); it is the normalized
probability that a point (galaxy) at a distance $\chi$ is included in
the catalog.

In practice the depth of the projection is finite due to the rapid
decrease of the selection function $\psi(\drad)$ with $\drad$ at
finite distance.  The selection function $\psi(\drad)$ for a sample
limited by apparent magnitudes between $m_1$ and $m_2$ is typically
given by,

\begin{eqnarray} 
\psi(\De) &=& \psi^* \int_{q_1(\De)}^{q_2(\De)} ~\d q\ \phi^*\,
q^\alpha\,e^{-q}, \quad q_i(\De) = 10^{-{5\over{2}}(M_i(\De)-M^*)},\
i=1,2 \label{lumi}
\end{eqnarray}

with

\begin{equation}
M_i(\De) = m_i-5 \log_{10} \De(1+z)-25,
\end{equation}

where $\psi^*$ is a normalization constant and $\phi(q)=\phi^*\,
q^\alpha\,e^{-q}$ is the luminosity function, i.e. the number density
of galaxies of a given luminosity.  $M^*$ and $\alpha$ might be
expressed as a function of redshift $z$ to account for k-corrections
and luminosity evolution.  When redshift information is available, one
can also rewrite the selection function in terms of the sample
redshift number counts $N(z)$ alone.

\subsubsection{The Small-Angle Approximation}
\label{sag}

The cumulants of the projected density can obviously be related to
those of the 3D density fields.  Formally they correspond to the ones
of the 3D field filtered by a conical-shaped window.  {}From
Eq.~(\ref{fb:defdeltap}) we obtain:

\begin{equation}
w_N(\theta_1,...,\theta_N) = \int \prod_{i=1}^{N} \d\drad_i
\drad_i^2\; \psi(\drad_i) \lag \delta(\drad_1, \De_1 \theta_1) \dots
\delta(\drad_N, \De_N {\theta}_N) \rag_c.  \label{fb:cum1}
\end{equation}

The computation of such quantities can be easily carried out in the
small angle approximation.  Such approximation is valid when the
transverse distances $\De\vert\vec\theta_i-\vec\theta\vert$ are much
smaller than the radial distances $\drad_{i}$.  In this case the
integral (\ref{fb:cum1}) is dominated by configurations where
$\drad_i-\drad_j\sim \De_i\vert\vec\theta_i-\vec\theta_j\vert \sim
\De_j\vert\vec\theta_i-\vec\theta_j\vert$.  This allows to make the
change of variables $\drad_i\to r_i$ with
$\drad_i=\drad_1+r_i\De_1(\theta_i-\theta_1)$.  Then, since the
correlation length (beyond which the multi-point correlation functions
are negligible) is much smaller than the Hubble scale $c/H(z)$ (where
$H(z)$ is the Hubble constant at redshift $z$) the integral over $r_i$
converges over a small distance of the order of
$\De_1\vert\vec\theta_i-\vec\theta_1\vert$ and the expression
(\ref{fb:cum1}) can be simplified to read

\begin{eqnarray}
w_N(\theta_1,...,\theta_N) &=& \int\d\drad_1\,\drad_1^{2N}\,\De^{N-1}
\; \psi(\drad_1)^N \nonumber\\ &&\hspace{-1.5cm} \times
\int_{-\infty}^{\infty} \prod_{i=2}^{N}\; (\theta_i-\theta_1)\d r_i \;
\xi_N\left[(\drad_1, \De_1 {\theta}_1), \dots, (\drad_N, \De_1
{\theta}_N) \right] \label{fb:cum2}
\end{eqnarray}

This equation constitutes the small-angle approximation for the
correlation functions.  If these behave as power-laws,
Eq.~(\ref{fb:cum2}) can be further simplified.  For instance, the
two-point function is then given by the {\em Limber
equation}~\cite{Limber53},

\begin{equation}
w_2(\theta)=\theta^{1-\gamma}\,r_0^{\gamma}\,\int\d\drad\,\drad^4\,
\De^{1-\gamma}\, \psi^2(\drad) \,\int_{-\infty}^{\infty}\d
r\,(1+r^2)^{-\gamma/2},
\label{eq:limber}
\end{equation}

if the 3D correlation function is $\xi_2(r)=(r/r_0)^{-\gamma}$.  The
fact that the last integral that appears in this expression
converges\footnote{It is given by $\int_{-\infty}^{\infty}\d
r\,(1+r^2)^{-\gamma/2}={{\sqrt{\pi }}\,\Gamma(\frac{-1 + \gamma}{2})}/
{\Gamma(\frac{\gamma}{2})}$, which converges for $\gamma>1$.}
justifies the use of the small-angle approximation.  It means that the
projected correlation functions are dominated by intrinsic 3D
structures, that is, the major contributions come from finite values
of $r_i$ which corresponds to points that are close together in 3D
space.

The small-angle approximation seems to be an excellent approximation
both for $w_2$ and for $w_3$ up to $\theta \simeq 2$ deg.  This can be
easily checked by numerical integration of a given model for $\xi_2$
and $\xi_3$, see e.g.~\cite{Peebles80,Bernardeau95,GaBe98}.

An equivalent way of looking at the small-angle approximation is to
write the corresponding relations in Fourier space.  The angular
two-point correlation function can be written in terms of the 3D power
spectrum as~\cite{Kaiser92},

\begin{equation}
w_2(\theta)=2\pi\,
\int\d\drad\,\drad^4\psi^2(\drad)\int\d^2\vk_{\perp}\,P(k_{\perp})\,
e^{\ii\De\vk_{\perp}\theta}.  \label{fb:w2Dcor}
\end{equation}

The expression (\ref{fb:w2Dcor}) shows that in Fourier space the small
angle approximation consists in neglecting the radial component of
$\vk$ (to be of the order of the inverse of the depth of the survey)
compared to $\vk_{\perp}$ (of the order of the inverse of the
transverse size of the survey).  Thus, in the small-angle
approximation, the power spectrum of the projected density field is,

\begin{equation}
P_{2D}(l)=2\pi\, \int\d\drad\,{\drad^4\psi^2(\drad)\over
\De^2}\,P\left({l\over \De}\right).  \label{fb:P2d}
\end{equation}

This can be easily generalized to higher-order correlations in Fourier
space,

\bea \langle \de_{2D}(\vl_{1}) \ldots \de_{2D}(\vl_{N}) \rangle_{c}
&=& (2\pi)^{N-1}\ \de_{\rm D} (\vl_{1}+\ldots+\vl_{N}) \nonumber \\ &
&  \times \int \d\drad\,{\drad^{2N}\psi^{N}(\drad)\over
\De^{2N-2}}\,P_{N}\left({\vl_{1}\over \De},\ldots,{\vl_{2}\over
\De}\right).  \label{PN2d} \eea

Note that the Fourier-space expression given above assumes in fact not
only the small-angle approximation, but also the flat-sky
approximation which neglects the curvature of the celestial sphere.
General expressions for the power spectrum and higher-order
correlations beyond the small-angle (and flat-sky) approximation can
be derived from Eq.~(\ref{fb:cum1}) by Legendre transforms, see
e.g.~\cite{FrTh99,VHM00}.

\subsubsection{Projection in the Hierarchical Model}
\label{sec:projhier}

The inversion of \eq(\ref{fb:cum2}), to relate $\xi_N$ in terms of
$w_N$ is still not trivial in general because the projection effects
mix different scales.  As in the case of the two-point correlation
function, i.e. Limber's equation, it is much easier to obtain a simple
relation between 3D and 2D statistics for models of $\xi_N$ that have
simple scale dependence.  In the {\em Hierarchical model} introduced
in Sect.~\ref{sec:HM},

\begin{equation}
\xi_N(\vr_1,...,\vr_N) = \sum_{a=1}^{t_N} Q_{N,a} \sum_{\rm
labelings}\ \prod_{\rm edges}^{N-1} \xi_2(\vr_A,\vr_B), 
\label{eq:hierarc}
\end{equation}

and, remarkably, it follows that the projected angular correlations
obey a similar relation:

\begin{equation} w_N(\theta_1,...,\theta_N) = \sum_{a=1}^{t_N} q_{N,a}
\sum_{\rm labelings}\ \prod_{\rm edges}^{N-1} w_2(\theta_A,\theta_B)
\label{hierw} \end{equation}

where $q_{N,a}$ is simply proportional to $Q_{N,a}$.  Moreover the
relation between $q_{N,a}$ and $Q_{N,a}$ depends only on the order $N$
and is independent on the particular tree topology.  To express $q_N$
in terms of $Q_N$ we can use a power-law model for the two-point
correlation: $\xi_2(r)=(r/r_0)^{-\gamma}$.  For small angles, we thus
have,

\begin{eqnarray} q_N &=& r_N Q_N, \label{limberqN} \\ r_N &=& {
I_1^{N-2} I_N \over{ I_2^{N-1}}} \quad {\rm with},\quad I_k =
\int_{0}^{\infty}\d\drad\,\drad^{2k}\,\De^{k-1}
\psi^k(\drad)\,\De^{-\gamma(k-1)}(1+z)^{-3(k-1)} \nonumber
\end{eqnarray}

where we have taken into account of redshift evolution of the
two-point correlation function in the non-linear regime assuming
stable clustering (see Sect.~\ref{stcl}), $\xi_2(r,z) =
\xi_2(r)\,(1+z)^{-3}$.  The integrals $I_k$ are just numerical values
that depend on the selection function and $\gamma$.  The values of
$\psi^*$ and $\phi^*$ in \eq(\ref{lumi}) are thus irrelevant for
$q_N$.  The only relevant parameters in the luminosity function are
$M^*$ and $\alpha$.

The resulting values of $r_N$ increase with $\gamma$ and $M^*$ and
decrease with $\alpha$, but do not change much within the
uncertainties in the shape of the luminosity function (see \S56
in~\cite{Peebles80}, and~\cite{Gaztanaga94}).  This is illustrated in
Table \ref{rjtable} where values of $r_N$ are plotted for different
parameters in the selection function.  In the analysis of the APM,
variations of $\gamma$ are only important for very large scales,
$\theta > 3^\circ$, where $\gamma$ changes from $1.8$ to $3$.  In this
case $r_N$ displays a considerable variation and \eq(\ref{limberqN})
is not a good approximation.

As an example we can consider the selection function given by the
characteristic ``bell shape'' in a magnitude limited sample:

\begin{equation}
\psi(r)\propto r^{-b}\ \exp[-r^2/{\calD}^2], \label{selec}
\end{equation}

where $\calD$ is related to the effective sample depth and $b$ is
typically a small number (e.g. for the APM $b\simeq 0.1$ and ${\calD}
\simeq 350$Mpc/h).  For this selection function and a power-law $P(k)
\propto k^n$ (e.g. $\gamma=-(n+3)$) we can calculate $r_3$ explicitly,

\begin{equation}
r_3 =\disp{ {8\over{3\sqrt{3}}} \left({\sqrt{27}\over 4}\right)^b\
{\Gamma[{3/2-b/2}]\Gamma[3/2-n-3/2\ b]\over\Gamma[3/2-n/2-b]^2} }
\left({3\over 2}\right)^n \label{r3}
\end{equation}

For $b=0$ and $n=0$ we find $r_3={8\over{3\sqrt{3}}} \simeq 1.54$,
while for $b=0$ and $n=-1$, closer to the APM case, $r_3 =
{2\pi\over{3\sqrt{3}}} \simeq 1.21$, comparable to the values given
in~\cite{Gaztanaga94}.

It is important to notice that although $r_N$ are unaffected by
changes in $\psi$, the overall normalization of $I_k$ can change
significantly.  Because of this, while the amplitude of $\xi_2$ is
uncertain by $40\%$ for $\Delta M^*=1.0$ and $\Delta\alpha=0.4$ the
corresponding uncertainty in $r_3$ is only $2\%$.  This is an
excellent motivation for using the hierarchical ratios $q_{N}$ as
measures of clustering.

\begin{table}

\centering
\caption{Projection factors for different slopes $\gamma$ and
parameters  $M^*_0$ and $\alpha_0$ in the luminosity function.}
\label{rjtable} \begin{tabular}{c c c c c c c c c c}
\hline  $\gamma$ & $M^*_0$ & $\alpha_0$ & $r_3$ & $r_4$ & $r_5$ &
$r_6$ & $r_7$ & $r_8$ & $r_9$ \\ \hline 1.7 & -19.8 & -1.0 & 1.19 &
1.52 & 2.00 & 2.71 & 3.72 & 5.17 & 7.25 \\ 1.7 & -19.3 & -1.2 & 1.21 &
1.57 & 2.12 & 2.93 & 4.13 & 5.88 & 8.44 \\ 1.7 & -20.3 & -0.8 & 1.18 &
1.48 & 1.93 & 2.56 & 3.46 & 4.73 & 6.51 \\ 1.8 & -19.8 & -1.0 & 1.20 &
1.55 & 2.08 & 2.85 & 3.98 & 5.62 & 8.00 \\ 3.0 & -19.8 & -1.0 & 1.54 &
2.85 & 5.78 & 12.4 & 27.8 & 63.9 & 150\\ \hline
\end{tabular}
\end{table}

Note that the above hierarchical prediction could only provide a good
approximation to clustering observations at small scales, where the
hierarchical model in Eq.~(\ref{eq:hierarc}) might be a good
approximation (see Sects.~\ref{sec:HM} and~\ref{sec:nptang}).  On
larger scales, accurate predictions require projection using the PT
hierarchy, which is different from Eq.~(\ref{eq:hierarc}), as the
N-point correlation functions have a significant shape dependence (see
Sect.~\ref{sec:TLPT}).  Despite this ambiguity on how to compare
angular observations to theoretical predictions, note that these two
approaches give results that agree within $20\%$ (e.g. see
Fig.~\ref{s3tl} below).

\subsubsection{The Correlation Hierarchy for the Projected Density}
\label{chpt}

We can define the area-averaged angular correlations $\wbar_p(\theta)$
in terms of the angular correlation functions $w_N(\te_1,...,\te_N)$:

\begin{equation}
\wbar_p(\theta) \equiv {1\over{A^p}} \int_A \d A_1\dots\d A_p~
w_p(\te_1,...,\te_p) = \lag\delta_{2D}^p(\theta)\rag_c, \label{wbar}
\end{equation}

where $A=2 \pi(1-\cos\theta)$ is the solid angle of the cone, $\d
A_p=\sin\theta_p\d\theta_p\d\varphi_p$ and $\delta_{2D}(\theta)$ is
the density contrast  inside the cone.  Thus $\wbar_p(\theta)$ only
depends on the size of the cone, $\theta$, analogous to smoothed
moments in  the 3D case.  The use of Eq.  (\ref{fb:cum2}) leads to,

\begin{eqnarray}
\wbar_p(\theta) &=&{1\over{A^p}} \int\d\drad_1 \drad_1^{3p-1}\;
\psi_1^p \; \int \prod_{i=1}^{p} {\d A_i} \nonumber\\ &&\hspace{-0cm}
\times \int_{-\infty}^{\infty} \prod_{i=2}^{p} \d r_i \;
\xi_p\left[(\drad_1, \De_1 {\theta}_1), \dots, (\drad_p, \De_1
{\theta}_p) \right].  \label{fb:cum2bis}
\end{eqnarray}

On can see that the cumulants of the projected density are thus
line-of-sight averages of the density cumulants in {\em cylindrical}
window function,

\begin{equation} \mg\delta_{{\rm proj},\theta}^p\md_c=
\int\d\drad\drad^{2p}\psi^p(\drad)\,\mg\delta^p_{\De\theta,\,{\rm
cyl}}\md_c\, L^{p-1}, \label{fb:proj1} \end{equation}

where $\delta^p_{\De\theta,\,{\rm cyl}}$ is the filtered 3D density
with a cylindrical filter of transverse size $\De \, \theta$ and depth
$L$.  For instance, written in terms of the power spectrum, the second
moment reads,

\begin{equation}
\wbar_2(\theta)=2\pi\,\int\d\drad\,\drad^4\,\psi^2(\drad)
\int\d^2\vk_{\perp}\,P(k_{\perp})\,W^2_{2D}(\De\theta\,k_{\perp})
\label{fb:w2Dexp}
\end{equation}

where $W_{2D}$ is the top-hat 2D window function, 

\begin{equation}
W_{2D}(l\,\theta)=2{J_1(l\theta)\over l\theta}.  \label{fb:W2Ddef}
\end{equation}

The relation (\ref{fb:proj1}) shows that the cumulant hierarchy is
preserved.  If we define the $s_{p}$ parameters in angular space,

\begin{equation}
s_p(\theta) \equiv { {\wbar_p(\theta)} \over
{~~[\wbar_2(\theta)]^{p-1}}},  \label{fb:sj}
\end{equation}

it follows that they are all finite and independent of $L$.

In the weakly nonlinear regime, we can compute exactly the hierarchy
for the projected density because the density cumulants for a
cylindrical window are those obtained for the 2D dynamics (see
Sect.~\ref{sec:sec2Ddyn}).  In case of a power-law spectrum the
$s_{p}$ are independent of the filtering scale.  The line-of-sight
integrations can then be performed explicitly\footnote{For CDM models
a semianalytic result can be obtained for the skewness,
see~\cite{Pollo97} for details.}.  Using Eq.~(\ref{fb:proj1}) and the
results of Sect.~\ref{sec:sec2Ddyn}, gives\footnote{It is important to
note that in Eq.  (\ref{fb:projectedsp}) the coefficients $S_p^{2D}$
need to be used and not those corresponding to 3D top-hat filtering as
suggested by the tree hierarchical model.}

\begin{eqnarray}
s_p &=& r_p\,S_p^{2D} \label{fb:projectedsp} \\ r_p &=& { I_1^{p-2}
I_p \over{ I_2^{p-1}}}, {\rm with}\quad I_k =
\int_{0}^{\infty}\d\drad\,\drad^{2k}\,
\psi^k(\drad)\,\De^{-(n+3)(k-1)}\,D_1^{2k-2}(z).  \label{fb:rpPT}
\end{eqnarray}

Note that the $r_p$ coefficients are very similar to those in the
nonlinear case except that the redshift evolution of the fluctuation
is taken here to be given by the linear growth rate.  This is actually
relevant only when the redshift under consideration is comparable to
unity.  

An interesting point is that it may seem inconsistent to use both
tree-level PT predictions and the small-angle approximation, as a
priori it is not clear whether their regimes of validity overlap.  As
shown in~\cite{GaBe98} for characteristic depths comparable to APM
there is at least a factor of five in scale where both approximations
are consistent, depending on the 3D power spectrum shape.  For deeper
surveys, of course, the consistency range is increased, so this is a
meaningful approach.

As expected, similar results hold for the hierarchy of correlation
functions in the weakly non-linear regime.  The results for the
angular three-point function and bispectrum have been studied with
most detail~\cite{FrTh99,FrGa99,BKJ00,VHM00}.  {}From
Eqs.~(\ref{fb:P2d}-\ref{PN2d}) and for power-law spectra, it follows
that the configuration dependence of the bispectrum is conserved by
projection, only the amplitude is changed by the projection factor
$r_{3}$, as in Eq.~(\ref{fb:rpPT})~\cite{GrPe77,FrTh99,FrGa99,BKJ00}.
However, as soon as the spectral index changes significantly on scales
comparable to those sampled by the selection function, this simple
result does not hold anymore~\cite{FrTh99}.  A number of additional
results regarding the shape dependence of projected correlations
include, i) a study of the dependence on configuration shape as a
function of depth~\cite{BJK00}, that also includes redshift-dependent
galaxy biasing; ii) the power of angular surveys to determine bias
parameters from the projected bispectrum in spherical
harmonics~\cite{VHM00}, and iii) comparisons of PT predictions and
numerical simulations in angular space~\cite{FrGa99}, as we summarize
in the next section.

\subsubsection{Comparison with Numerical Simulations}

\label{sec:q3projproj}

\begin{figure}
\centering \centerline{\epsfxsize=8truecm\epsfysize=8truecm\epsfbox{
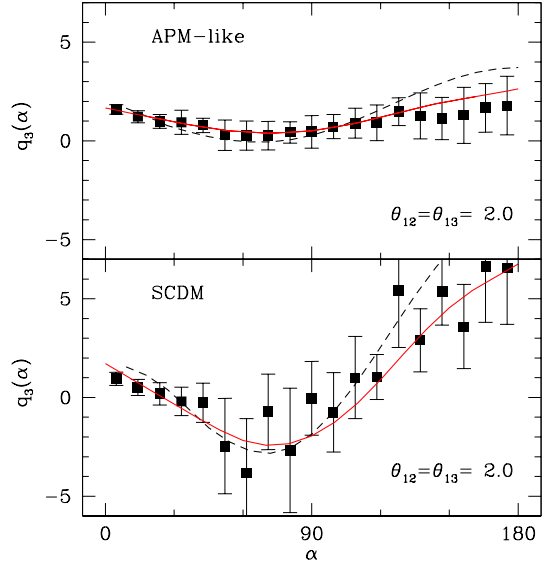}} \caption[junk]{Projected leading order PT predictions
(solid curves) and N-body results (points with sampling errors) for
the angular 3-point amplitude $q_3(\alpha)$ at fixed
$\theta_{12}=\theta_{13}=2$ deg for a survey with the APM selection
function.  N-body results correspond to the average and variance of 5
realizations of the APM-like model (top) and the SCDM model (bottom). 
The dashed lines show the corresponding PT predictions for
$r_{12}=r_{13}= 15$ Mpc/h projected with the hierarchical model.}
\label{q3scdmr2}
\end{figure}

We now illustrate the results described in the previous section and
compare their regime of validity against numerical simulations. 

Figure~\ref{q3scdmr2} shows the angular three-point correlation
function for APM-like and SCDM spectrum projected to the depth of the
APM survey, see~\cite{FrGa99} for more details.  As discussed before,
the configuration dependence of the three-point amplitude is quite
sensitive to the shape of the power spectrum.  Both the shape and
amplitude of $q_3(\alpha)$ predicted by PT (solid curves) are
reproduced by the N-body results (points) even on these moderately
small scales\footnote{At the mean depth of the APM, two degrees
corresponds to $\simeq 15$ h$^{-1}$ Mpc.}.  The error bars in the
simulation results are estimated from the variance between 5 maps from
different N-body realizations and have been scaled to 1-$\sigma$
uncertainties for a single observer.  The dashed lines correspond to
the results of the 3D $Q_3$ for $r_1=r_2=15 \Mpc$ multiplied by the
hierarchical projection factor in Eq.~(\ref{limberqN}), e.g. $q_3= Q_3
r_3$.  The model seems to work well for small $\alpha$, but there are
significant deviations for large $\alpha$, which illustrate that this
projection model does not work well, as discussed above.

In the weakly nonlinear regime the third moment of smoothed angular
fluctuations, defined in (\ref{fb:cum2bis}), can be explicitly written
in terms of the power spectrum using PT.  It is given by,

\begin{eqnarray}
\wbar_3 &=& {6 (2 \pi)^2} \int \d\drad\ \drad^6 \psi^3(\drad)
\displaystyle{\left[{6\over7}\left(\int k\d k
W_{2D}^2(k\,\De\,\theta)\,P(k)\right)^2 \right.} \\ \nonumber
&&\hspace{-1cm}+ \displaystyle{{1\over 2} \int k\,\d
k\,W^2_{2D}(k\,\De\,\theta)\,P(k)} \disp{\left.\int {k^2\ \d
k\,\De\theta}\,
W_{2D}(k\,\De\,\theta)\,W_{2D}'(k\,\De\,\theta)\,P(k)\right]}\nonumber
\label{wbar3pk}
\end{eqnarray}

where $W_{2D}'$ is the derivative of the top -hat window $W_{2D}$
defined in Eq.~(\ref{fb:W2Ddef}).  Therefore, in case of a power-law
spectrum $P(k) \sim k^n$, we have~\cite{Bernardeau95},

\begin{equation}
s_3= {r_3\ \left({36\over 7}-{3\over 2}\,(n+2)\right)}, \label{s_3}
\end{equation}

with $r_3$ given in general by Eq.~(\ref{fb:rpPT}).  The coefficient
$r_3$ is found in practice to be of order unity and to be very weakly
dependent on the adopted shape for the selection function.

It is worth to note that the hierarchical model in
Sect.~\ref{sec:projhier} yields a different prediction for $s_3$ than
the above tree-level value.  In the hierarchical case, $s_3 \simeq r_3
S_3$ (\cite{Gaztanaga94,Gaztanaga95}) with $S_3 =34/7-(n+3)$.  For
example, for $n \simeq -1$, the hierarchical model yields $s_3 \simeq
3.43$ while the tree level prediction yields: $s_3 \simeq 4.38$.  This
difference becomes smaller as we move towards larger $n$ (e.g. larger
scales), being zero at $n=4/7$, but it is significant for the range of
scales probed with current observations, even after taking
uncertainties into account.

\begin{figure}
\centering \centerline{\epsfxsize=8truecm\epsfysize=10truecm\epsfbox{
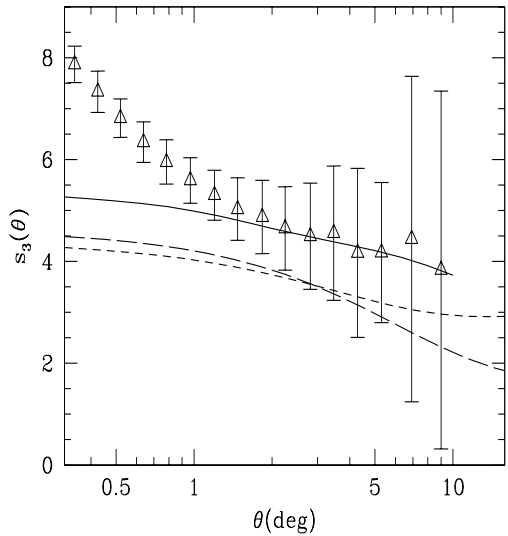}} \caption[junk]{Tree-level PT predictions for the APM-like
power spectrum (solid curves) and corresponding N-body results (points
with sampling errors) for the projected smoothed skewness
$s_3(\theta)$ as a function of the radius $\theta$ (in degrees) of the
cells in the sky.  The short and long dashed line show the
hierarchical prediction $s_3 \simeq r_3 S_3$, see text for details.}
\label{s3tl}
\end{figure}

Figure~\ref{s3tl} compares the predictions for the angular skewness
$s_3$ by tree-level PT (solid lines) for a power spectrum that matches
the APM catalogue and the APM measurements (triangles).  These
predictions correspond to a numerical integration of PT predictions in
Eq.~(\ref{wbar3pk})~\cite{GaBe98}.  The dashed lines show the
``naive'' hierarchical prediction $s_3 \simeq r_3 S_3$ at the angular
scale $\theta \simeq R/\calD$ given by the depth, $\calD$, of the
survey.  The long dashed line uses a fixed value of $r_3 =1.2$, while
the dashed line corresponds to $r_3= r_3(n)$ given by Eq.~(\ref{r3})
with $n=-(3+\gamma)$ given by the logarithmic slope of the variance of
the APM-like $P(k)$ at the angular scale $\theta \simeq R/\calD$.
These results are compared with the mean of 20 all sky simulations
described in~\cite{GaBe98} (error bars correspond to the variance in 20
observations).  As can be seen in the figure, the hierarchical model
gives a poor approximation, while the projected tree-level results
matches well the simulations for scales $\theta \ga 1$ deg, which
correspond to the weakly nonlinear regime where $\xi_2 \la 1$.  On
small scales the discrepancies between the tree-level results and the
simulations is due to 3D non-linear effects but also to projection: on
small scales the simulations follow the hierarchical model in
Eq.~(\ref{eq:hierarc}), rather than the PT predictions, and therefore
$s_3 \simeq r_3 S_3$ gives a good approximation, but $S_3$ should be
the non-linear 3D value (for example, as given by HEPT or EPT, see
Sects.~\ref{sec:HEPT} and \ref{sec:EPT}, respectively).  Further
comparisons with numerical simulations for $s_3$ and $s_4$ are
presented below in Fig.~\ref{s34apm} together with observational
results.

\subsubsection{Reconstructing the PDF of the Projected Density}
\label{pdfpd}

It is interesting to note that it is possible to write down a
functional relation between the cumulant generating function defined
in Eq.  (\ref{eq:varphieq}) for the projected density, $\varphip(y)$,
and the one corresponding to cylindrical filtered cumulants,
$\varphic(y)$~\cite{Valageas00b,MuJa00,BeVa00}.  This can be done from
the relation (\ref{fb:proj1}) which straightforwardly leads to,

\begin{equation}
\varphip(y)=\int {\d \drad\over \Xi_{\theta}(\drad)}\,
\varphic[y\,\drad^2\,\psi(\drad)\,\Xi_{\theta}(\drad)],
\label{fb:phiproj}
\end{equation}

with

\begin{equation}
\Xi_{\theta}(\drad)={\mg\delta^2_{\De\theta,\,{\rm cyl}}\md\over
\mg\delta_{2D}^2\md} \, L,
\end{equation}

which can be rewritten in terms of the matter fluctuation power
spectrum,

\begin{equation}
\Xi_{\theta}(\drad)={\int\d^2\vk\,P(k,z)\,W^2(k\,\De\,\theta)\over
\int\d\drad'\,\drad'^4\,\psi^2(\drad')\,\int\d^2\vk\,P(k,z')\,W^2(k\,\De'\,\theta)}.
\end{equation}

In this expression we have explicitly written the redshift dependence
of the power spectrum.  In the case of a power-law spectrum,

\begin{equation}
P(k,z)=P_0(z) \,\left({k\over k_0}\right)^n
\end{equation}

it takes a much simpler form given by,

\begin{equation}
\Xi_{\theta}(\drad)={P_0(z)\,\De^{-n-2}\over
\int\d\drad'\,\drad'^4\,\psi^2(\drad')\,P_0(z')\,\De'^{-n-2}}.
\end{equation}

Together with Eq.~(\ref{fb:phiproj}) this result provides the
necessary ingredients to reconstruct the one-point PDF of the
projected density with an inverse Laplace transform of $\varphip(y)$.
Note that projection effects alter the shape of the singularity in
$\varphi(y)$ though it preserves the large density exponential
cutoff~\cite{Valageas00b,BeVa00}.  So far this approach has only been
used in the literature to study the reconstruction of the one-point
PDF of the local convergence field in the context of weak lensing
observations~\cite{Valageas00b,MuJa00}.

We now turn to a brief summary of the basics of weak lensing and its
connections to projection effects.

\subsection{Weak Gravitational Lensing}

\def\mD{{\cal D}} 
\def\valpha{\vec{\alpha}}  
\def\vbeta{\vec{\beta}}
\def\omb{{\overline{\omega}}}

The first theoretical investigations on the possibility of mapping the
large-scale structure of the universe with weak gravitational lensing
dates back to the early
nineties~\cite{Blandford90,BSBV91,MiraldaEscude91,Kaiser92}.  It was
then shown that the number of background galaxies was large enough to
serve as tracers of the deformation field induced by the intervening
large scales structures.  In this context the observation of a
coherent shear pattern in the orientation of background galaxies is
interpreted as due to lensing effects caused by the mass concentration
along the line of sight.  The potential interest of such observations
has led to further theoretical investigations such as the
determination of the dependence of the results on cosmological
parameters~\cite{Villumsen96,BVWM97,JaSe97,VWBM99}, and to extensive
observational efforts.  The latter have recently led to the first
reliable detections of the so-called ``cosmic
shear''~\cite{vanWaerbekeetal00,BRE00,WTKDB00,KWL00}.

Although in nature totally different from galaxy counts, it is worth
pointing out that such observations eventually aim at mapping the
line-of-sight mass fluctuations so that techniques developed for
studying galaxy angular catalogues can be applied.  Here we briefly
introduce the physics of lensing with emphasis in connections to
angular clustering.  More comprehensive presentations can be found
in~\cite{BaSch99,Mellier99}.

\subsubsection{The Convergence Field as a Projected Mass Map}

The physical mechanism at play in weak lensing surveys is the
deflection of photon paths in gravitational potential fields.  The
deflection angle per unit distance, ${\delta \valpha/\delta s}$, can
be obtained from simple computations of the geodesic equation in the
weak field limit\footnote{See e.g.~\cite{MTW73,Sachs61} for a
comprehensive presentation of these calculations.}.  When the metric
fluctuations are purely scalar, the deflection angle reads

\begin{equation}
{\delta \valpha\over\delta s}=-2\,\vec{\nabla}_x{\phi\over c^2},
\end{equation}

where the spatial derivative is taken in a plane that is orthogonal to
the photon trajectory.

The direct consequence of this bending is a displacement of the
apparent position of the background objects.  This depends on the
distance of the source plane, $D_{OS}$, and on the distance between
the lens plane and the source plane $D_{LS}$.  It is given by,

\begin{equation}
\vec{\gamma}^S=\vec{\gamma}^I-{2\over c^2}\,{D_{LS}\over
D_{OS}\,D_{OL}} \,\vec{\nabla}_{\gamma}\left(\int \d s\
\phi(s,\gamma)\right)
\end{equation} 

where $\vec{\gamma}^I$ is the position in the image plane and
$\vec{\gamma}^S$ is the position in the source plane.  The gradient is
taken here with respect to the angular position (this is why a
$D_{OL}$ factor, distance to the lens plane, appears).  The total
deflection is obtained by an integration along the line-of-sight,
assuming the lens is thin compared to its angular distance.
Calculations are also usually done in the so-called Born approximation
for which the potential is computed along the unperturbed photon
trajectory.

The observable effect which is aimed at, however, is the induced
deformation of background objects.  Such an effect is due to the {\em
variations} of the displacement field with respect to the apparent
position.  These variations induce a change in both size and shape of
the background objects which are encoded in the amplification matrix,
$\mA$, describing the linear change between the source plane and the
image plane,

\begin{equation}
\mA=\left({\partial \gamma^I_i\over \partial \gamma^S_j}\right).
\end{equation}

Its inverse, $\mA^{-1}$, is actually directly calculable in terms of
the gravitational potential.  It is given by the derivatives of the
displacement with respect to the apparent position,

\begin{equation}
\mA^{-1}\equiv {\partial \gamma^S_i\over \partial \gamma^I_j}
=\delta_{ij}-2{D_{LS}\over D_{OS}\,D_{OL}}
\varphi_{,ij}.\label{fb:Amatdef}
\end{equation}

where $\varphi$ is the projected potential.  Usually its components
are written,

\begin{equation}
\mA^{-1}=\left( \begin{tabular}{cc} $1-\kappa-\gamma_1$&$-\gamma_2$\\
$-\gamma_2$&$1-\kappa+\gamma_1$ \end{tabular} \right),
\end{equation}

taking advantage of the fact that it is a symmetric matrix.  The
components of this matrix are expressed in terms of the convergence,
$\kappa$, (a scalar field) and the shear, $\gamma$ (a pseudo vector
field).

The key idea for weak lensing observations is then that collection of
tiny deformation of background galaxies can be used to measure the
local {\em shear} field from which the projected potential, and
therefore the convergence field, can be reconstructed~\cite{Kaiser92}. 
The latter has a simple cosmological interpretation: from the trace of
Eq.  (\ref{fb:Amatdef}) one obtains the convergence\footnote{~In these 
sections, $\Omega_{m}$ is understood to be at $z=0$.},

\begin{equation}
\kappa(\gamma)={3\over2}\Omega_m \int\d
z_s\,n(z_s)\int\d\chi\,{\mD(\chi_s,\chi)\,\mD(\chi)\over \mD(\chi_s)}
\,\delta(\chi,\gamma)\,(1+z)\label{fb:kappa},
\end{equation}

as the integrated line-of-sight density contrast.  In
Eq.~(\ref{fb:kappa}) $\chi$ is the distance along the line-of-sight
and $\mD$ are the angular distances.  In this relation sources are
assumed to be located at various redshifts with a distribution
$n(z_s)$ normalized to unity, and all the distances are expressed in
units of $c/H_0$.  The relation (\ref{fb:kappa}) is then entirely
dimensionless.  Note that in general the relation between the redshift
and the distances depends on cosmological parameters, see
Eq.~(\ref{fb:Dedef}).

\subsubsection{Statistical Properties}

To gain insight into the expected statistical properties of the
convergence field, it is important to keep in mind that in
Eq.~(\ref{fb:kappa}) the convergence $\kappa$ is not normalized as
would be the local projected density contrast.  The projected density
contrast is actually given by

\begin{equation}
\delta_{2D}(\gamma)={\kappa\over \omb}
\end{equation}

where $\omb$ is the mean lens efficiency,

\begin{equation}
\omb={3\over2}\Omega_m \int\d
z_s\,n(z_s)\int\d\chi\,{\mD(\chi_s,\chi)\,\mD(\chi)\over \mD(\chi_s)}
\,(1+z)\label{fb:omb}.
\end{equation}

This implies that the skewness of the convergence field is then given
by,

\begin{equation}
s_3^{\kappa}={s_3^{\rm proj}\over \omb}
\end{equation}

where $s_3^{\rm proj}$ is the skewness of the projected density
contrast given by Eq.~(\ref{s_3}).  As a consequence the skewness of
$\kappa$ is expected to display a strong $\Omega_{m}$
dependence.
This property has indeed been found in \cite{BVWM97} where it has been
shown using PT that

\begin{equation}
s_3^{\kappa}\approx 40\,\Omega_m^{-0.75}
\label{s3k}
\end{equation}

for sources at redshift unity\footnote{~For the same reasons that
$s_3^{\kappa}$ has a strong $\Omega_m$ dependence, it also depends
significantly on the source redshift distribution.}.  This result has
been subsequently extended to the nonlinear
regime~\cite{JaSe97,Hui99,MuCo00,MuJa01,vanWaerbekeetal00,CoHu01},
higher-order moments, the bispectrum~\cite{CoHu01}, and to more
complex quantities such as the shape of the one-point PDF of the
convergence field~\cite{Valageas00a,Valageas00b,MuJa00}.

\subsubsection{Next to Leading Order Effects}

Contrary to the previous cases, corrections to the previous leading
order PT results, e.g. Eq.~(\ref{s3k}), do not involve only
next-to-leading order terms due to the nonlinear dynamics but also
other couplings that appear specifically in the weak lensing context.
Let us list and comment the most significant of them: 

\begin{itemize}

\item[(i)] An exact integration of the lens equations leads to
lens-lens coupling and departures from the Born approximation.  This
induces extra couplings that have been found to be in all cases
negligible for a source population at redshift of about
unity~\cite{BVWM97,vWHSCB00}.

\item[(ii)] The source population clustering properties can also
induce non-trivial effects as described in~\cite{Bernardeau98a}.  This
is due to the fact that the source plane is by itself a random media
which introduces further couplings due to either intrinsic galaxy
number fluctuations or due to overlapping of lens and source
populations.  These effects have been found to be small if the
redshift distribution of the sources is narrow
enough~\cite{Bernardeau98a,HCTDMB00} which might indeed put severe
constraints on the observations.

\item[(iii)] The magnification effect (when $\kappa$ is large,
galaxies are enlarged and can thus be more efficiently detected) could
also induce extra couplings.  Although it is difficult to estimate the
extent of such an effect, it appears to have only modest effects on
the high-order statistical properties of the convergence
field~\cite{Hamana01}.

\end{itemize}

Finally it is important to note that the first reports of cosmic shear
detections have been challenged by suggestions that part of the signal
at small scales might be due to intrinsic galaxy shape
correlations~\cite{HRH00,CrMe00,CKB01}.  This is a point that should
be clarified by further investigations.

\subsubsection{Biasing from Weak Gravitational Lensing}

With the arrival of wide surveys dedicated to weak lensing
observations\footnote{~See for example,
http://terapix.iap.fr/Descart/}, a very powerful new window to
large-scale structure properties is being opened.  Weak lensing
observations can indeed be used not only to get statistical properties
of the matter density field, but also to map the mass distribution in
the Universe.  In particular it becomes possible to explore the
galaxy-mass local relation~\cite{vanWaerbeke98}.  Potentially, galaxy
formation models, biasing models, can be directly tested by these 
observations.  It is indeed possible to measure the correlation
coefficient $r_{\kappa}$ of the convergence field $\kappa$ with the
projected density contrast of the (foreground) galaxy $\delta_{g,2D}$,
\begin{equation}
r_{\kappa}=
{\langle\kappa\,
\delta_{g,2D}\rangle\over\sqrt{
\langle\kappa^2\rangle\langle\delta_{g,2D}^2\rangle}},
\end{equation}
a quantity which, within geometrical factors, is proportional to the
$r$ coefficient defined in Eq.  (\ref{rijreal}).  What has been
measured so far~\cite{HYG01} is however
${\langle\kappa\,\delta_{g,2D}\rangle/
\langle\delta_{g,2D}^2\rangle}$, that is, a quantity that roughly
scales like $\Omega_m\,r/b$.  Pioneering results suggest a scale
independent $r/b$ parameter of about unity for the favored
cosmological model ($\Omega_m=0.3$,
$\Omega_{\Lambda}=0.7$)~\cite{HYG01}.  Such observations are bound to
become common place in the coming years and will provide valuable
tests for galaxy formation models.

\subsection{Redshift Distortions}

\label{sec:reddis}

In order to probe the three-dimensional distribution of galaxies in
the Universe, galaxy redshifts are routinely used as an indicator of
radial distance from the observer, supplemented by the two-dimensional
angular position on the sky.  The Hubble expansion law tells us that
the recession velocity of an object is proportional to its distance
from us; however, the observed velocity has also a contribution from
peculiar velocities, which are generated due to the dynamics of
clustering and are unrelated to the Hubble expansion and thus
contaminate the distance information.  Therefore, the clustering
pattern in ``redshift space'' is somewhat different than the actual
real space distribution.  This is generically known as ``redshift
distortions''.

At large scales, the main effect of peculiar velocities is due to
galaxies infall into clusters.  Galaxies between us and the cluster
have their infall velocities added to the Hubble flow and thus appear
farther away in redshift space, whereas those galaxies falling into
the cluster from the far side have their peculiar velocities
subtracting from the Hubble flow, and thus appear closer to us than in
real space.  As a consequence of this, large-scale structures in
redshift space appear flattened or ``squashed'' along the line of
sight.  On the other hand, at small scales (smaller than the typical
cluster size) the main effect of peculiar velocities is due to the
velocity dispersion from virialization.  This causes an elongation
along the line of sight of structures in redshift space relative to
those in real space, the so-called ``finger of God'' effect (which
points to the observer's location).

\subsubsection{The Density Field in Redshift Space}
\label{des}

We now discuss the effects of redshift distortions on clustering
statistics such as the power spectrum, the bispectrum and higher-order
moments of the smoothed density field.  See the exhaustive
review~\cite{Hamilton98} for details on theoretical description of
linear redshift distortions and observational results.  In redshift
space, the radial coordinate $\vs$ of a galaxy is given by its
observed radial velocity, a combination of its Hubble flow plus
``distortions'' due to peculiar velocities.  Here we restrict to the
``plane-parallel'' approximation, so that the line of sight is taken
as a fixed direction, denoted by ${\hat z}$.  Plane-parallel
distortions maintain statistical homogeneity, so Fourier modes are
still the natural basis in redshift-space.  On the other hand,
statistical isotropy is now broken, because clustering along the line
of sight is different from that in the perpendicular directions.

However, when the radial character of redshift distortions is taken
into account, the picture changes.  Radial distortions respect
statistical isotropy (about the observer), but break statistical
homogeneity (since there is a preferred location, the observer's
position).  In this case Fourier modes are no longer special, in
particular, the power spectrum is no longer diagonal~\cite{ZaHo96}. 
Alternative approaches to Fourier modes have been suggested in the
literature~\cite{HeTa95,HaCu96,SML98}, here we review the simplest
case of plane-parallel distortions where most of the results have been
obtained.  We should note that this is not just of academic interest,
it has been checked with N-body simulations that results on monopole
averages of different statistics carry over to the radial case with
very small corrections~\cite{Scoccimarro00b}.

The mapping from real-space position ${\bf \vx}$ to redshift space in
the plane-parallel approximation is given by:

\begin{equation}
\vs=\vx - f \ \uz(\vx) {\hat z},
\end{equation}

where $f(\Omega_m) \approx \Omega_m^{0.6}$ is the logarithmic growth rate
of linear perturbations, and $\uu(\vx) \equiv - \vu(\vx)/({\cal H}
f)$, where $\vu(\vx)$ is the peculiar velocity field, and ${\cal
H}(\tau) \equiv (1/a)(da/d\tau)= Ha$ is the conformal Hubble parameter
(with FRW scale factor $a(\tau)$ and conformal time $\tau$).  The
density field in redshift space, $\ds(\vs)$, is obtained from the
real-space density field $\de(\vx)$ by requiring that the
redshift-space mapping conserves mass, i.e.

\begin{equation}
(1+\ds)\d^3\vs=(1+\de)\d^3\vx \label{mass}~.
\end{equation}

Using the fact that $\d^3\vs=J(\vx)\d^3\vx$, where $J(\vx)=|1-f \nabla_z
\uz(\vx)|$ is the {\em exact} Jacobian of the mapping in the
plane-parallel approximation, it yields

\begin{equation}
\ds(\vs)= \frac{\de(\vx)+1-J(\vx)}{J(\vx)}.  \label{dss}
\end{equation}

The zeros of the Jacobian describe caustics in redshift space, the
locus of points where the density field is apparently
infinite~\cite{McGill90}.  This surface is characterized in real space
by those points which are undergoing turn-around in the gravitational
collapse process, so their peculiar velocities exactly cancel the
differential Hubble flow.  In practice, caustics are smoothed out by
sub-clustering, see e.g. the discussion in~\cite{HKS00}.

An expression for density contrast in redshift space follows from
Eq.~(\ref{dss})~\cite{SCF99}

\begin{equation}
\ds(\vk) = \int \frac{\d^3\vx}{(2\pi)^3} {\rm e}^{-i \vk\cdot\vx} {\rm
e}^{i f k_z \uz(\vx)} \Big[ \de(\vx) + f \nabla_z \uz(\vx)
\Big]\label{d_s}~,
\end{equation}

where we assumed here that only points where $f \nabla_z \uz(\vx)<1$
contribute.  The only other approximation in this expression is the
use of the plane-parallel approximation, i.e. this is a fully
non-linear expression.  To obtain a perturbative expansion, we expand
the second exponential in power series,

\begin{eqnarray}
\ds(\vk)&=&\sum_{n=1}^{\infty} \int \d^3\vk_1 \ldots \d^3 \vk_n \dD_n \Big[
\de(\vk_1) + f \mu_1^2 \te(\vk_1) \Big] \frac{ (f \mu
k)^{n-1}}{(n-1)!} \nonumber \\ & & \times \frac{\mu_2}{k_2}\te(\vk_2)
\ldots \frac{\mu_n}{k_n}\te(\vk_n) \label{delta_s},
\end{eqnarray}

where $\dD_n \equiv \delta_{\rm D}(\vk - \vk_1 - \cdots - \vk_n)$, the
velocity divergence $\te(\vx) \equiv \nabla \cdot \uu (\vx)$, and
$\mu_i \equiv \vk_i \cdot {\hat z}/k_i$ is the cosine of the angle
between the line-of-sight and the wave-vector.  In linear PT, only the
$n=1$ term survives, and we recover the well-known formula due to
Kaiser~\cite{Kaiser87}

\begin{equation}
\ds(\vk)=\de(\vk) (1+f\mu^2) \label{delta_sl}.
\end{equation}

Equation (\ref{delta_s}) can be used to obtain the redshift-space
density field beyond linear theory.  In redshift space we can write

\begin{eqnarray}
\delta_s(\vk,\tau) &=& \sum_{n=1}^{\infty} D_1^n(\tau) \int \d^3\vk_1
\ldots \int \d^3\vk_n \dD_n \, Z_n (\vk_1, \ldots, \vk_n) \,
\delta_1(\vk_1) \cdots \delta_1(\vk_n), \nonumber \\ & &
\label{Zn}
\end{eqnarray}

where $D_1(\tau)$ is the density perturbation growth factor in linear
theory, and we have assumed that the n$^{\rm th}$-order growth factor
$D_n \propto D_1^n$, which is an excellent approximation (see
\cite{SCFFHM98}, Appendix B.3).  Since a local deterministic and
non-linear bias can be treated in an equal footing as non-linear
dynamics, it is possible to obtain the kernels $Z_n$ including biasing
and redshift-distortions.  {}From Eqs.~(\ref{eq:taylor}) and 
(\ref{delta_s})-(\ref{Zn}), the redshift-space kernels $Z_n$ for
the {\it galaxy} density field 
read~\cite{VHMM98,SCF99}\footnote{~Detailed expressions for the 
second-order solutions are given in~\cite{HBCJ95} including the 
(small) dependences on $\Omega_{m}$ for the unbiased case.}

\begin{eqnarray}
Z_1 (\vk) &=& (b_1+f \mu^2), \label{z1} \\ Z_2 (\vk_1,\vk_2) &=& b_1 F_2
(\vk_1,\vk_2) +f \mu^2 G_2 (\vk_1,\vk_2) \nonumber \\ & & + \frac{f
\mu k}{2} \Big[ \frac{\mu_1}{k_1} (b_1+f \mu_2 ^2) + \frac{\mu_2}{k_2}
(b_1+f \mu_1 ^2) \Big] + \frac{b_2}{2}, \label{z2} \\
Z_3(\vk_1,\vk_2,\vk_3) &=& b_1 F_3^{(s)} (\vk_1,\vk_2,\vk_3) +f \mu^2
G_3^{(s)} (\vk_1,\vk_2,\vk_3) \nonumber \\ & & + f \mu k \Big[b_1
F_2^{(s)} (\vk_1,\vk_2) +f \mu_{12}^2 G_2^{(s)} (\vk_1,\vk_2) \Big]
\frac{\mu_3}{k_3} \nonumber \\ & & + f \mu k (b_1+f \mu_1 ^2)
\frac{\mu_{23}}{k_{23}} G_2^{(s)}(\vk_2,\vk_3) \nonumber \\ & & +
\frac{(f \mu k)^2}{2} (b_1+f \mu_1 ^2) \frac{\mu_2}{k_2}
\frac{\mu_3}{k_3} + 3 b_2 F_2^{(s)} (\vk_1,\vk_2) + \frac{b_3}{6},
\label{z3}
\end{eqnarray}

where we denote $\mu \equiv \vk \cdot {\hat z}/k$, with $\vk \equiv
\vk_1 +\ldots +\vk_n$, and $\mu_i \equiv \vk_i \cdot {\hat z}/k_i$. 
As above, $F_2$ and $G_2$ denote the second-order kernels for the
real-space density and velocity-divergence fields, and similarly for
$F_3$ and $G_3$.  Note that the third order kernel $Z_3$ must still be
symmetrized over its arguments.  One can similarly obtain the PT
kernels $Z_n$ in redshift space to arbitrary higher order.

We note that there are {\em two} approximations involved in this
procedure: one is the mathematical step of going from Eq.~(\ref{d_s})
to Eq.~(\ref{delta_s}), which approximates the redshift-space mapping
with a power series; the other is the PT expansion itself (i.e., the
expansion of $\de(\vk)$ and $\te(\vk)$ in terms of linear fluctuations
$\de_1(\vk)$).  Therefore, one is not guaranteed that the resulting PT
in redshift space will work over the same range of scales as in real
space.  In fact, in general, {\em PT in redshift space breaks down at
larger scales than in real space}, because the redshift-space mapping
is only treated approximately, and it breaks down at larger scales
than does the perturbative dynamics.  In particular, a calculation of
the one-loop power spectrum in redshift space using
Eqs.~(\ref{z1}-\ref{z3}) does not give satisfactory results, because
expanding the exponential in Eq.~(\ref{d_s}) is a poor approximation. 
To extend the leading-order calculations, one must treat the
redshift-space mapping exactly and only approximate the dynamics using
PT~\cite{SCF99}.  To date, this program has only been carried out
using the Zel'dovich approximation \cite{FiNu96,TaHa96,HaCo98} and
second-order Lagrangian PT~\cite{Scoccimarro00}, as we shall discuss
below.

\subsubsection{The Redshift-Space Power Spectrum}
\label{pks}

The calculation of redshift-space statistics in Fourier space proceeds
along the same lines as in the un-redshifted case.  To leading
(linear) order, the redshift-space galaxy power spectrum
reads~\cite{Kaiser87}

\begin{equation}
P_s(\vk)= P_g(k)\ (1+\beta \mu^2)^2 = \sum_{\ell=0}^{\infty} a_\ell\
\pleg_\ell(\mu) \ P_g(k) \label{mult_l},
\end{equation}

where $P_g(k)\equiv b_1^2 P(k)$ is the real-space galaxy power spectrum,
$P(k)$ is the linear mass power spectrum, and $\beta\equiv f/b_1 \approx
\Omega_m^{0.6}/b_1$.  Here $\pleg_\ell(\mu)$ denotes the Legendre
polynomial of order $\ell$, and the multipole coefficients
are~\cite{Hamilton92,CFW94}

\begin{equation}
a_0 \equiv 1+ \frac{2}{3}\beta+\frac{1}{5}\beta^2, \ \ \ \ \ a_2
\equiv \frac{4}{3}\beta+\frac{4}{7}\beta^2, \ \ \ \ \ a_4 \equiv
\frac{8}{35}\beta^2 \label{PS_l};
\end{equation}

all other multipoles vanish.  Equation~(\ref{mult_l}) is the standard
tool for measuring $\Omega_m$ from redshift distortions of the power
spectrum in the linear regime; in particular, the
quadrupole-to-monopole ratio $R_{\rm P}\equiv a_2/a_0$ should be a
constant, independent of wavevector $k$, as $k \to 0$.  Note, however,
that in these expressions $\Omega_m$ appears only through the
parameter $\beta$, so there is a degeneracy between $\Omega_m$ and the
linear bias factor $b_1$.  Equation~(\ref{PS_l}) assumes deterministic
bias, for stochastic bias extensions see~\cite{Pen98,DeLa99}.

{}From equation (\ref{d_s}), we can write a simple expression for the
power spectrum in redshift space, $P_s(\vk)$:

\begin{eqnarray}
P_s(\vk) &=& \int \frac{\d^3 \vr}{(2 \pi)^3} {\rm e}^{-i \vk \cdot \vr}
\Big\langle {\rm e}^{i \lambda \Delta\uz} \Big[ \de(\vx) + f \nabla_z
\uz(\vx) \Big] \Big[ \de(\vx') + f \nabla_z' \uz(\vx') \Big]
\Big\rangle, \nonumber \\ & & \label{Ps}
\end{eqnarray}

where $\lambda \equiv fk\mu$, $\Delta\uz \equiv \uz(\vx)-\uz(\vx')$,
$\vr \equiv \vx-\vx'$.  This is a fully non-linear expression, no
approximation has been made except for the plane-parallel
approximation.  In fact, Eq.~(\ref{Ps}) is the Fourier analog of the
so-called ``streaming model''~\cite{Peebles80}, as modified
in~\cite{Fisher95} to take into account the density-velocity coupling.

The physical interpretation of this result is as follows.  The factors
in square brackets denote the amplification of the power spectrum in
redshift space due to infall (and they constitute the only
contribution in linear theory, giving Kaiser's~\cite{Kaiser87}
result).  This gives a positive contribution to the quadrupole ($l=2$)
and hexacadupole ($l=4$) anisotropies.  On the other hand, at small
scales, as $k$ increases the exponential factor starts to play a role,
decreasing the power due to oscillations coming from the pairwise
velocity along the line of sight.  This leads to a decrease in
monopole and quadrupole power with respect to the linear contribution;
in particular, the quadrupole changes sign.

In order to describe the non-linear behavior of the redshift-space
power spectrum, it has become popular to resort to a phenomenological
model to take into account the velocity dispersion
effects~\cite{PeDo94}.  In this case, the non-linear distortions of
the power spectrum in redshift-space are written in terms of the
linear squashing factor and a suitable damping factor due to the
pairwise-velocity distribution function

\begin{equation}
\label{Ppheno} P_s(\vk)= P_g(k)\ \frac{(1+\beta
\mu^2)^2}{[1+(k\mu\sigma_v)^2/2]^2} ~.
\end{equation}

Here $\sigma_v$ is a free parameter that characterizes the velocity
dispersion along the line-of-sight.  This Lorentzian form of the
damping factor is motivated by empirical results showing an
exponential one-particle\footnote{~Alternatively, if one assumes the
{\em two}-particle velocity distribution is exponential, the
suppression factor is the square root of that in Eq.~(\ref{Ppheno}),
with $\sigma_{v}$ the pairwise velocity dispersion along the line of
sight, see e.g.~\cite{BPH96}.  The observational results regarding
velocity distributions and their interpretation is briefly discussed
in Sect.~\ref{sec:2ptredsh}.} velocity distribution
function~\cite{PVGH94}; comparison with N-body simulations have shown
it to be a good approximation~\cite{CFW95}; however, these type of
phenomenological models tend to approach the linear PT result faster
than numerical simulations~\cite{HaCo98}.  In addition, although
$\sigma_v$ can be chosen to fit, say, the quadrupole-to-monopole ratio
at some range of scales, the predictions for the monopole or
quadrupole by themselves do not work as well as for their ratio.

Accuracy in describing the shape of the quadrupole to monopole ratio
as a function of scale is important, since this statistic gives a
direct determination of $\beta$ from clustering in redshift
surveys~\cite{Hamilton92,CFW94,CFW95,HaCo99}.  An alternative to
phenomenological models, is to obtain the redshift-space power
spectrum using approximations to the dynamics, as we now discuss.

In the case of the Zel'dovich approximation (ZA), it is possible to
obtain the redshift-space power spectrum as
follows~\cite{FiNu96,TaHa96}.  In the ZA, the density field obeys

\begin{equation}
1+\de(\vx)= \int \d^3\vq\ \de_D[\vx-\vq-\Psi(\vq)], \label{deltaZA}
\end{equation}

where $\Psi(\vq)$ is the displacement vector at Lagrangian position
$\vq$.  In the plane parallel approximation, one can treat redshift
distortions in the ZA by noting that it corresponds to amplifying the
displacement vector by $f \approx \Omega_m^{0.6}$ along the line of
sight; that is, the displacement vector in redshift space is
$\Psi_s(\vq)= \Psi(\vq)+ \hat{z}f( \Psi(\vq)\cdot \hat{z})$.  Fourier
transforming the corresponding expression to Eq.~(\ref{deltaZA}) in
redshift space, the power spectrum gives

\begin{equation}
P(\vk)= \int \d^3\vq\ \exp(i \vk \cdot \vq)\ \lexp \exp(i \vk \cdot
\Delta\Psi) \rexp, \label{PsZA}
\end{equation}

where $\Delta\Psi= \Psi(\vq_1)-\Psi(\vq_2)$ and $\vq=\vq_1-\vq_2$. 
For Gaussian initial conditions, the ZA displacement is a Gaussian
random field, so Eq.~(\ref{PsZA}) can be evaluated in terms of the
two-point correlator of $\Psi(\vq)$.  The results of these calculations
show that the ZA leads to a reasonable description of the quadrupole
to monopole ratio~\cite{FiNu96,TaHa96} provided that the zero-crossing
scale is fixed to agree with numerical simulations.  In general, the
ZA predicts a zero-crossing at wavenumbers larger than found in N-body
simulations~\cite{HaCo98}.  Furthermore, although the shape of the
quadrupole to monopole ratio resembles that in the simulations, the
monopole and quadrupole do not agree as well as their ratio.  This can
be improved by using second-order Lagrangian PT~\cite{ScSh01}, but the
calculation cannot be done analytically anymore, instead one has to
resort to numerical realizations of the redshift-space density field
in 2LPT.

\subsubsection{The Redshift-Space Bispectrum}
\label{sec:Qs}

Given the second-order PT kernel in redshift-space, the leading-order
(tree-level) galaxy bispectrum in redshift-space 
reads~\cite{HBCJ95,VHMM98,SCF99}

\begin{equation}
\label{Bs_def} B_s(\vk_1,\vk_2,\vk_3)= 2 Z_2 (\vk_1,\vk_2) \
Z_1(\vk_1)\ Z_1(\vk_2) \ P(k_1) \ P(k_2) + {\rm cyc.},
\end{equation}

which can be normalized by the power spectrum monopole to give the
reduced bispectrum in redshift space, $Q_s$,

\begin{equation}
\label{Qstree} Q_s(\vk_1,\vk_2,\vk_3) \equiv
\frac{B_s(\vk_1,\vk_2,\vk_3)}{a_0^2 \ (P_g(k_1) \ P_g(k_2) + {\rm
cyc.})},
\end{equation}

where ``cyc.''  denotes a sum over permutations of $\{k_1,k_2,k_3\}$. 
Note that $Q_s$ is independent of power spectrum normalization to
leading order in PT. Since, to leading order, $Q_s$ is a {\em
function} of triangle configuration which separately depends on
$\Omega_m$, $b$, and $b_2$, it allows one in principle to break the
degeneracy between $\Omega_m$ and $b$ present in measurement of the
power spectrum multipoles in redshift space~\cite{Fry94a,HBCJ95}. 
The additional dependence of (the monopole of) $Q_s$ on $\Omega_m$
brought by redshift-space distortions is small, typically about
$10\%$~\cite{HBCJ95}.  On the other hand, as expected, the quadrupole
of $Q_s$ shows a strong $\Omega_m$ dependence~\cite{SCF99}.

Decomposing into Legendre polynomials, $B_{s\ \rm eq}(\mu) =
\sum_{\ell=0}^\infty\ B_{s\ \rm eq}^{(\ell)}\ P_{\ell}(\mu)$, the
redshift-space reduced bispectrum for equilateral configurations
reads~\cite{SCF99}

\begin{eqnarray}
\label{Qs0} Q_{s\ {\rm eq}}^{(\ell=0)} &=& \frac{5\ (2520+
4410\,\gamma + 1890\,\beta + 2940\,\gamma\,\beta + 378\,\beta^2 +
441\,\gamma\,\beta^2)}{98\,b_{1}\, ( 15 + 10\,\beta + 3\,\beta^2 )^2}
\nonumber \\ & & + \frac{5\ (9\,\beta^3 + 1470\,b_{1}\,\beta +
882\,b_{1}\,\beta^2- 14\,b_{1}\,\beta^4 ) }{98\,b_{1}\, ( 15 + 10\,\beta +
3\,\beta^2 )^2},
\end{eqnarray}

where $\gamma \equiv b_{2}/b_{1}$.  This result shows that in redshift
space, $Q_{s,g} \neq (Q_{s} + \gamma)/b_{1}$ as in
Eq.~(\ref{Q_g_eul}), although it is not a bad
approximation~\cite{SCF99}.  In the absence of bias ($b_{1}=1$,
$\gamma =0$), Eq.~(\ref{Qs0}) yields

\begin{equation}
\label{Qs0b0} Q_{s\ {\rm eq}}^{(\ell=0)}= {{5\,\left( 2520 + 3360\,f +
1260\,{f^2} + 9\,{f^3} - 14\,{f^4}\right) }\over{98\,{{\left( 15 +
10\,f + 3\,{f^2} \right) }^2}}},
\end{equation}

which approaches the real-space result~\cite{Fry84b} $Q_{\rm eq}=4/7
=0.57$ in the limit $f \sim \Omega_m^{0.6} \rightarrow 0$.  On the other
hand, for $f=\Omega_m=1$, we have $Q_{s\ {\rm eq}}^{(0)}=0.464$: for
these configurations, the reduced bispectrum is suppressed by redshift
distortions.

\begin{figure}[t]
\begin{tabular}{cc} {\epsfysize=6.5truecm
\epsfbox{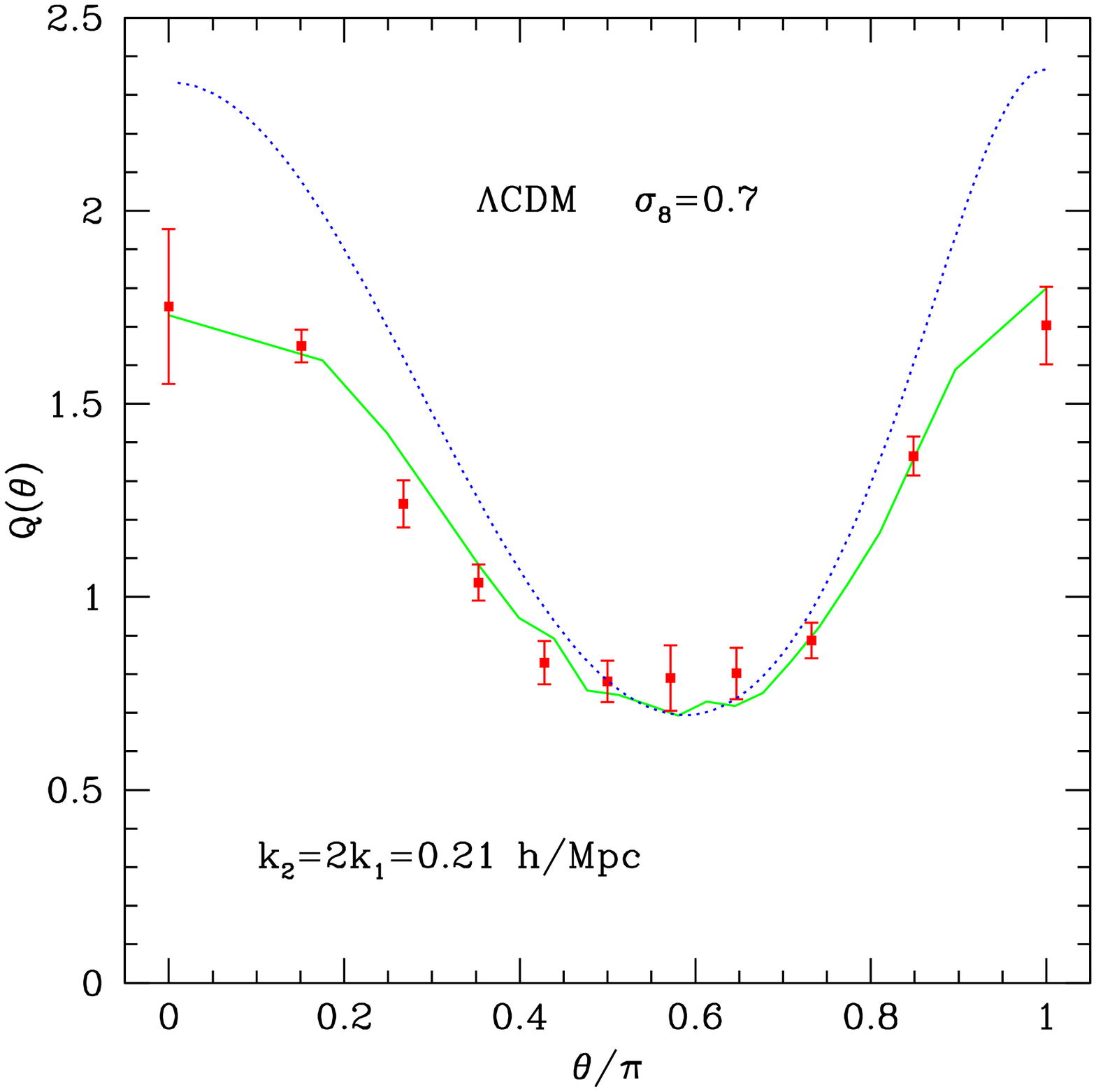}}& {\epsfysize=6.5truecm
\epsfbox{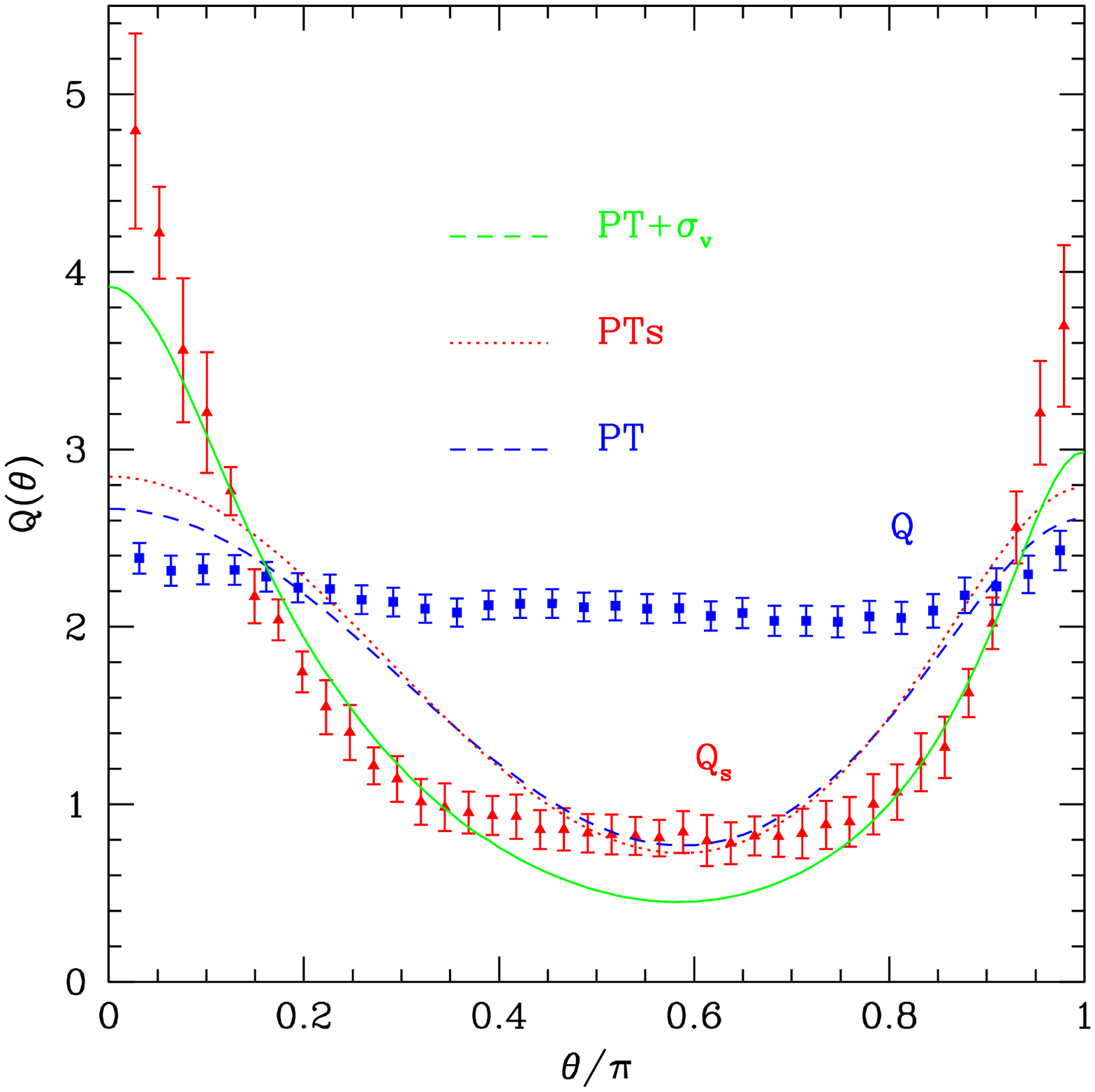}} \end{tabular} \caption{The left panel shows
the bispectrum in redshift space for configurations with $k_2=2k_1$ as
shown as a function of the angle $\theta$ between $\vk_1$ and $\vk_2$. 
The dotted line shows the predictions of second-order Eulerian PT,
whereas the solid lines correspond to 2LPT. Error bars correspond to
the average between 4 realizations.  The right panel shows the
bispectrum in redshift space for configurations with $k_2=2k_1=1.04$
h/Mpc, i.e. in the non-linear regime.  Square symbols denote $Q$ in
real space, whereas triangles denote the redshift-space bispectrum. 
Also shown are the predictions of PT in real space (dashed lines), PT
in redshift space (PTs, dotted line) and the phenomenological model
with $\sigma_v=5.5$, ( PT+$\sigma_v$, continuous line).  }
\label{bispz12}
\end{figure}

As discussed before for the power spectrum, leading-order calculations
in redshift space have a more restricted regime of validity than in
real space, due to the rather limited validity of the perturbative
expansion for the redshift-space mapping (instead of the perturbative
treatment of the dynamics).  The same situation holds for the
bispectrum.  The left panel in Fig.~\ref{bispz12} shows the reduced
bispectrum $Q$ as a function of angle $\theta$ between $\vk_1$ and
$\vk_2$ for configurations with $k_2=2k_1=0.21$h/Mpc.  The dotted line
shows the predictions of tree-level PT in redshift space,
Eq.~(\ref{Qstree}), whereas the symbols correspond to N-body
simulations of the $\Lambda$CDM model ($\Omega_m=0.3$,
$\Omega_{\Lambda}=0.7$, $\sigma_8=0.7$) with error bars obtained from
4 realizations.  The disagreement is most serious at colinear
configurations.  On the other hand, the solid lines obtained using
2LPT~\cite{Scoccimarro00} agree very well with the N-body
measurements.  Similarly good agreement is found for equilateral
configurations.  The key in the 2LPT predictions is that the
redshift-space mapping is done exactly (by displacing the particles
from real to redshift space in the numerical realizations of the 2LPT
density field), rather than expanded in power series.

At small scales, however, 2LPT breaks down and one must resort to some
kind of phenomenological model to account for the redshift distortions
induced by the velocity dispersion of clusters.  For the bispectrum,
this reads~\cite{SCF99}

\begin{equation}
\label{Bpheno} B_s(\vk_1,\vk_2,\vk_3)= \frac{B_s^{\rm
PT}(\vk_1,\vk_2,\vk_3)}{ \big[1+\alpha^2\ [(k_1\mu_1)^2 + (k_2\mu_2)^2
+ (k_3\mu_3)^2]^2 \sigma_v^2/2\big]^2},
\end{equation} 

where $B_s^{\rm PT}(\vk_1,\vk_2,\vk_3)$ is the tree-level
redshift-space bispectrum.  The assumption is that one can write the
triplet velocity dispersion along the line-of-sight in terms of the
pairwise velocity dispersion parameter $\sigma_v$ and a constant
$\alpha$ which reflects the configuration dependence of the triplet
velocity dispersion.  As noted above, $\sigma_v$ is determined from
simulations solely using the power spectrum ratio; the parameter
$\alpha$ is then fitted by comparison with the monopole-to-quadrupole
ratio of the equilateral bispectrum measured in the
simulations~\cite{SCF99}.  A somewhat different phenomenological model
can be found in~\cite{VHMM98}; in addition~\cite{Matsubara94} studies
using a similar phenomenological model the effects of redshift-space
distortions in the nonlinear regime for the three-point correlation,
assuming the validity of the hierarchical model in real space.

The right panel in Fig.~\ref{bispz12} shows the redshift-space bispectrum
at small scales, to show the effects of non-linear redshift
distortions.  The square symbols denote $Q$ is real space, which
approximately saturates to a constant independent of configuration. 
On the other hand, the redshift-space $Q$ shows a strong configuration
dependence, due to the anisotropy of structures in redshift space
caused by cluster velocity dispersion (fingers of God).  The
phenomenological model (with $\sigma_v=5.5$ and $\alpha=3$) in solid
lines does quite a good job at describing the shape dependence of
$Q_s$. 

Similar studies using numerical simulations have been carried out in
terms of the three-point correlation function, rather than the
bispectrum, to assess the validity of the hierarchical model in the
nonlinear regime in redshift space~\cite{MaSu94,SuMa94} and to compare
with redshift surveys at small scales~\cite{BBGKP95,GBGGHKP96,JiBo98}. 
They obtained analogous results to Fig.~\ref{bispz12} for the
suppression of $Q_{s}$ for equilateral configurations compared to $Q$
at small scales due to velocity dispersion.  However, studies of the
configuration dependence of $Q_{s}$ in the non-linear
regime~\cite{MaSu94,SuMa94,JiBo98} find no evidence of the
configuration dependence shown in the right panel in
Fig.~\ref{bispz12}.  This is surprising, as visual inspection of
numerical simulations shows clear signs of fingers of God; this
anisotropy should be reflected as a configuration dependence of
$Q_{s}$.  More numerical work is needed to resolve this issue.

\subsubsection{The Higher-Order Moments in Redshift Space}

In redshift space, the PT calculation of skewness and higher-order
cumulants cannot be done analytically, unlike the case of real space,
but can be done by a simple numerical integration for
$S_{3}$\cite{HBCJ95}\footnote{~Using a different approach,
\cite{WaTa01} recently derived a closed form for $S_{3}$ in redshift
space that does not agree with~\cite{HBCJ95}.  This apparent
disagreement merits further work.}.  The effects of redshift
distortions, however, are very small (of order $10\%$) for the
skewness and kurtosis. 

\begin{figure}
\centering
\centerline{\epsfxsize=8truecm\epsfysize=8truecm\epsfbox{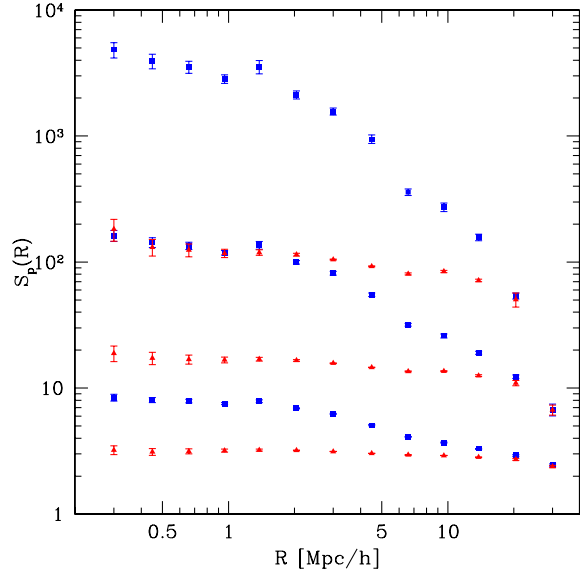}}
\caption{The $S_p$ parameters for $p=3,4,5$ (from bottom to top) in
redshift space for $\Lambda$CDM with $\sigma_8=0.9$ as a function of
smoothing scale $R$.  Square symbols denote measurements in real space
N-body simulations, whereas triangles correspond to redshift-space
values, assuming the plane-parallel approximation.}
\label{QzSpz}
\end{figure}

On the other hand, at small scales the effect of non-linear redshift
distortions is quite strong; since non-linear growth is suppressed in
redshift space due to cluster velocity dispersion, the skewness and
higher-order moments do not grow much as smaller scales are
probed~\cite{LIIS93,MaSu94,SuMa94,BBGKP95,ScGa01}.  Figure~\ref{QzSpz} shows
an example for the $S_p$ parameters for top-hat smoothing ($p=3,4,5$)
in the $\Lambda$CDM model; square symbols denote the real-space values
and triangles correspond to redshift-space quantities.  Note the close
agreement between real and redshift space at the largest scales, and
the suppression at small scales for the redshift space case.  The
latter looks almost scale independent; however, one must keep in mind
that correlation functions at small scales should be strongly
non-hierarchical, i.e. depend strongly on configuration as shown in
the right panel in Fig.~\ref{bispz12}.

\subsubsection{Cosmological Distortions}

Deep galaxy surveys can probe a large volume down to redshifts where
the effects of a cosmological constant, or more generally dark energy,
become appreciable.  A geometrical effect, as first suggested
in~\cite{AlPa79}, arises in galaxy clustering measures because the
assumption of an incorrect cosmology leads to an apparent anisotropy
of clustering statistics.  In particular, structures appear flattened
along the line of sight, and thus the power spectrum and correlation
functions develop anisotropy, similar to that caused by redshift
distortions~\cite{BPH96,MaSu96,PWRO98,deSt98,Matsubara00b}.  The
challenge to measure this effect is that redshift distortions are
generally larger than cosmological distortions, so a reliable measure
of cosmological distortions require an accurate treatment of redshift
distortions.  

Recent work along these lines~\cite{MaSz01}, using the approximation
of linear PT and that bias is linear, scale and time independent,
concludes that the best prospects for measuring cosmological
distortions in upcoming surveys is given by the LRG (luminous red
galaxy) sample of the SDSS. This sample of about 100,000 galaxies
seems to give a good balance between probing structure down to `high'
redshift ($z \sim 0.5$) and having a large enough number density so
that shot noise is not a limiting factor.  Analysis of redshift and
cosmological distortions gives a joint 3-$\sigma$ uncertainty on
$\Omega_{\Lambda}$ and $\Omega_{m}$ of about $15\%$, assuming
$\Omega_{\Lambda}=0.7$ and $\Omega_{m}=0.3$ as the fiducial model. 
Other surveys such as the quasar samples in 2dFGRS and SDSS, are
predicted to give less stringent constraints due to the sparse
sampling~\cite{MaSz01}.

Applications of cosmological distortions to the case of the
Lyman-$\alpha$ forest have been proposed in the
literature~\cite{HSB99,McME99}.  In this case, the distortions are
computed by comparing correlations along the line of sight to those by
cross-correlating line of sights of nearby quasars.  These studies
conclude that with only about 25 pairs of quasars at angular
separations of $\la 2'-3'$ it is possible to distinguish an open model
from a flat cosmological constant dominated model (with the same
$\Omega_m=0.3$) at the 4-$\sigma$ level.  These results, however,
assume a linear description of redshift distortions.  More recent
analysis using numerical simulations~\cite{McDonald01} suggests that
with $13 (\theta/1')^{2}$ pairs at separation less than $\theta$, and
including separations $<10'$, a measurement to $5\%$ can be made if
simulations can predict the redshift-space anisotropy with $5\%$
accuracy, or to $10\%$ if the anisotropy must be measured from the
data.

Finally, we should mention the effect of clustering evolution along
the line of sight, due to observation along the light cone.  Estimates
of this effect show that for wide surveys such as 2dFGRS and SDSS it
amounts to about 10\% in the power spectrum and higher-order
statistics, while it becomes significantly larger of course for deep
surveys, see e.g.~\cite{MSS97,HCS01}.

\clearpage 
\section{\bf Results from Galaxy Surveys}
\label{chapter9}

\subsection{Galaxies as Cosmological Tracers}

Following the discovery of galaxies as basic objects in our
universe~\cite{Sanford17,Hubble17,Hubble36}, it became clear that
their spatial distribution was not uniform but clustered in the sky,
e.g.~\cite{Zwicky38}.  In fact, the Local Supercluster was recognized
early on from two-dimensional maps of the galaxy
distribution~\cite{deVaucouleurs48}.  The first measurements ever of
the angular two-point correlation function $w(\theta)$, done in the
Lick survey~\cite{ToKi69}, established already one of the basic
results of galaxy clustering, that at small scales the angular
correlation function $w(\theta)$ has a power-law dependence in
$\theta$ [see Eq.~(\ref{w2pl}) below].

The first systematic study of galaxy clustering was carried out in the
1970's by Peebles and his collaborators.  In a truly groundbreaking
twelve-paper
series~\cite{Peebles73,HaPe73,PeHa74,Peebles74b,PeGr75,Peebles75,GrPe77,SePe77,SePe78,FrPe78,SePe79,FrPe80},
galaxies were seen for the first time as tracers of the large-scale
mass distribution in the gravitational instability
framework\footnote{~For an exhaustive review of this and earlier work
see~\cite{Fall79,Peebles80}.}.  These works confirmed (and extended)
the power-law behavior of the angular two-point function, established
its scaling with apparent magnitude, and measured for the first time
the angular power spectrum and the three- and four-point functions
which were found to follow the hierarchical scaling $w_{N} \sim
w_{2}^{N-1}$.  The theoretical interpretation of these observations
was done in the framework of galaxies that traced the mass
distribution in an Einstein de-Sitter universe\footnote{~In this case
self-similarity plus stable clustering leads to hierarchical scaling
in the highly non-linear regime, although it does not explain why
hierarchical amplitudes are independent of configuration, see
Sect.~\ref{snlr}.  These observations were partially motivated by work
on the BBGKY approach to the dynamics of gravitational
instability~\cite{Saslaw72} and also generated a significant
theoretical activity that led to much of the development of
hierarchical models.  For a recent historical account of these results
and a comparison with current views in the framework of biased galaxy
formation in CDM models see~\cite{Peebles01}.}.

These results, however, relied on visual inspections of poorly
calibrated photographic plates; i.e. with very crude magnitudes (e.g.,
Zwicky) or galaxy counts (e.g., Lick) estimated by eye, rather than by
some automatic machine.  These surveys were the result of adding many
different adjacent photographic plates and the uniformity of
calibration was a serious issue, since large-scale gradients can be
caused by varying exposure time, obscuration by our galaxy, and
atmospheric extinction.  These effects are difficult to disentangle
from real clustering, attempts were made to reduce them with smoothing
procedures, but this could also result in a removal of real
large-scale clustering.  More than 20 years after completion of Zwicky
and Lick surveys, there were major technological developments in
photographic emulsions, computers and automatic scanning machines,
such as the APM (Automatic Plate Measuring Machine, \cite{KBBI84}) and
COSMOS~\cite{MacGillivray84} micro-densitometers.  This allowed a
better calibration of wide field surveys, as measuring machines locate
sources on photographic plates and measure brightness, positions and
shape parameters for each
source~\cite{PFFM78,PEKBHH79,SFEM80,Hewett82,KoSz84,SSFM85,PrIn86,Sebok86}.

In the 1980's large number of redshifts and scanning machines gave
rise to a second generation of wide-field surveys, with a much better
calibration and a three-dimensional view of the universe\footnote{~For
a review of redshift surveys see
e.g.~\cite{Oort83,GiHa91,StWi95,Strauss99}.}.  The advent of CCD's
revolutionized imaging in astronomy and soon made photographic plate
techniques obsolete for large scale structure studies.
Nowadays, photometric surveys are done with
large CCD cameras involving millions of pixels and can sample
comparable number of galaxies.  Furthermore, it is possible with
massive multi-fiber or multi-slit spectroscopic techniques to build
large redshift surveys of our nearby universe such as the
LCRS~\cite{SLOTLKS96} the 2dFGRS (e.g.~see~\cite{CDMSNC01}) or the SDSS
(e.g.~see~\cite{York00}) as well as of the universe at higher
redshifts such as in the VIRMOS (e.g.~see~\cite{Lefevre01}) and DEIMOS
surveys (e.g.~see~\cite{DaNe01}).

This significant improvement in the quality of surveys and their
sampled volume allowed more accurate statistical tests
and therefore constrain better theories of large-scale structure. 
Stringent constraints from upper limits to the CMB anisotropy
(e.g.~\cite{UsWi84}), plus theoretical inputs from the production of
light elements (e.g.~\cite{YTSSO84}) and the generation of
fluctuations from inflation in the early
universe~\cite{Starobinsky82,Hawking82,GuPi82,BST83} led to the
development of CDM models~\cite{Peebles82,BFPR84} where most of the
matter in the universe is not in the form of baryons.  The
three-dimensional mapping of large scale structures in redshift
surveys showed a surprising degree of
coherence~\cite{KOSS81,HDLT83,DGH86} which when compared with
theoretical predictions of the standard CDM model (e.g.~\cite{DEFW85})
led to the framework of biased galaxy formation, where galaxies are
not faithful tracers of the underlying dark matter distribution
(Sect.~\ref{sec:bias}).  Subsequent observational challenge from the
angular two-point function in the APM survey~\cite{MESL90} and counts
in cells in the IRAS survey~\cite{EKSLREF90,SFRLE91} led to the demise
of standard CDM models in favor of CDM models with more large-scale
power, with galaxies still playing the role of (mildly) biased tracers
of the mass distribution (e.g.~\cite{DEFW92}).

The access to the third dimension also allowed analyses of peculiar velocity
statistics through redshift distortions~\cite{Kaiser87,Hamilton92}
(Sect.~\ref{sec:reddis}, see~\cite{Hamilton98} for a recent review) and
measurements of higher-order correlations became more reliable with the
hierarchical scaling (Sect.~\ref{sec:HM}), $\xi_N \sim Q_N \xi_2^{N-1}$, being
established by numerous measurements in 3D
catalogs~\cite{BaFr91,Gaztanaga92,BSDFYH93,FrGa94b}.  However, it was not
until recently that surveys reached large enough scales to test the weakly
non-linear regime and therefore predictions of PT against
observations~\cite{FrGa94,Gaztanaga94,GaFr94,Gaztanaga95,FrGa99,SFFF01,FFFS01}.
This is an important step forward, as higher-order statistics encode precious
additional information that can be used to break degeneracies present in
measurements of two-point statistics, constrain how well galaxies trace the
mass distribution, and study the statistics of primordial fluctuations.  It is
the purpose of the present chapter to review the observational efforts along
these lines.

In this Chapter, we discuss the various results obtained from
measurements in galaxy catalogs for traditional statistics such as
$N$-point correlation functions in real and Fourier space and
counts-in-cells cumulants (thus leaving out many results on the shape
of the CPDF itself, including the void probability function).  We do
not attempt to provide a comprehensive review of all relevant
observations but rather concentrate on a subsample of them.  The
choice reflects the connections to PT and thus there is a strong
emphasis on higher-order statistics.  In particular, we do not discuss
about cosmic velocity fields, except when redshift distortions are a
concern.  Also, we do not discuss the spatial distribution of clusters
of galaxies since the statistical significance of measurements of
higher-order statistics is still somewhat marginal.

The remainder of this chapter is mainly divided into two large
sections, one concerning angular surveys (Sect.~\ref{sec:angsur}), the
other one concerning redshift surveys (Sect.~\ref{sec:redsur}). 
Finally, Sect.~\ref{sec:outlook} reviews ongoing and future surveys.

\subsection{Results from Angular Galaxy Surveys}
\label{sec:angsur}

\subsubsection{Angular Catalogs}

\label{sec:angcat}

\begin{table}
\caption{Angular Catalogs.  The first 5 entries correspond to ``old''
catalogs (1961-1974) based on counts or magnitude/diameters estimates
by eye and with poor calibration.  The survey Area $\Omega$ is given
in steradians, the depth (mean luminosity distance)
and effective size $ D_E \equiv
(\Omega/4\pi)^{1/3} 2 D$ are in Mpc/$h$.  The sign $\sim$ reflects the
fact that different sub-samples have different values for that
quantity.} 
\bigskip
\def\ss{\rule[0ex]{0mm}{3.5ex}\rule[-1.75ex]{0mm}{3.5ex}\\*}
\def\df{\dotfill} \def\hf{\hfill}
\begin{tabular}{c c c c c c c} \hline \rule[-0.6ex]{0mm}{3.2ex}
Name & Area $\Omega$ & magnitudes & Depth $D$ & $D_E $ & \# gal/ster &
Ref \ss \hline Zwicky & $1.8$ ster & $m_z <15$ & 70 & 73 & $\sim 7000$ &
\cite{Zwicky61}\ss Lick & $ 3.3 $ ster & $m <19$ & 220 & 280 & $\sim
10^5$ & \cite{ShWi67} \ss Jagellonian & $0.01 $ ster & $m <21$ & 400 &
74 & $\sim 10^6$ & \cite{RDFBS73}\ss ESO/Uppsala & $\sim 1.8 $ ster &
$d_l > 1'$ & 60 & 63 & $\sim 2000$ & \cite{HLSW74} \ss UGC & $\sim 1.8
$ ster & $d_l > 1'$ & 70 & 74 & $\sim 2000$ & \cite{Nilson73} \ss
\hline 
APM & $1.3 $ ster & $b_J=17-20$ & 400 & 380 & $\sim 10^6$ &
\cite{MESL90} \ss EDSGC & $0.3 $ ster & $b_J=17-20$ & 400 & 230 & $\sim
10^6$ & \cite{CNL92} \ss IRAS 1.2Jy & $ 9.5 $ ster & $f_{60\mu m} > 1.2
Jy$ & 80 & 145 & 480 & \cite{FHSDYS95} \ss 
DeepRange & $0.005$ ster &
$I_{AB} < 22.5$ & 2000 & 150 & $\sim 10^8$ & \cite{PLSO98} \ss
SDSS & $\simeq 3 $ ster &
$  r'< 22$ & 1000 & 1300 & $\sim 10^7$ & \cite{York00} \ss
\hline
\end{tabular} \label{angularcat}
\end{table}

We begin our discussion of angular clustering with a brief description
of results from the older generation of catalogs that sets the stage
for the more recent results, and then go into a more detailed
description of the current state of the subject. 
Table~\ref{angularcat} lists the main angular catalogs that have been
extensively analyzed. We show the characteristic parameters of 
the samples used in the relevant clustering analyses. The 
information is organized as follows.  The
second column gives the total area, $\Omega$, of the catalog while the
fourth column shows its mean depth, $D$ (associated with the limiting
magnitude in the third column).  The fifth column gives the volume in
terms of a characteristic length, $D_E$.  The sixth column gives the
surface density, $n_g$, which also relates to the mean depth.  The
three numbers, $\Omega$, $D$ and $D_E$ control volume (area) and edge
effects discussed in Chapter~\ref{sec:chapter7}.  In particular
samples with similar volumes can have quite different sampling biases
due to edge effects, because of differences in the shape (angular
extent) of the survey.  The galaxy number density, $n_g$, relates to
discreteness errors (Chapter~\ref{sec:chapter7}), which of course are
more significant when the total number of objects in the catalog is
small.  Finally, let us note that some of these catalogs were
constructed with different photometric filters (typically blue).

\medskip

The original \ul{Zwicky catalog} (\cite{Zwicky61} 1961-1968) contains galaxies
to magnitude $m<15.7$. In the most angular clustering analyses
only galaxies brighter than $m=14.5-15$ (with $\sim 2000 gal/sr$)
and only in the North galactic cap ($\Omega \sim 1.8 ster$) were used.  
The mean depth is estimated to be about 50-80 Mpc/$h$.  The base
sample used for redshift surveys (see Sect.~\ref{sec:redsur}) 
is a wide survey ($\Omega \simeq 2.7$) with about $20000$ galaxy
positions ($m<15.5$) taken from photographic plates with 
different calibrations.  There
have been several studies of systematic errors in Zwicky photometry, showing
an important magnitude scale error (see \cite{GaDa00} and references therein),
however, it is not clear how seriously this affected the clustering
properties.

\medskip

The \ul{Lick catalog} (\cite{ShWi67} 1967) consists of 1246 plates of
6x6 square degrees.
Counts were done by eye.  In the analyses presented by Peebles and
collaborators, only 467 plates with $|b_{\rm II}|>40$ degrees were
used.  These plates have overlapping regions which were used to reduce
the counts to a uniform limiting magnitude.  Calibration was based on
matching the surface density of counts, $\langle n \rangle$, which is
much less reliable than calibration based on comparing positions and
magnitudes of individual sources.  Errors on count estimates were
assumed to be independent from cell to cell and to increase the
variance by an additive factor proportional to $\langle n \rangle$. 
In~\cite{GrPe77} large-scale gradients in the counts were removed by
applying a ``smoothing factor'' which led to some controversy
concerning the significance of the
analysis~\cite{GKL84,DKG86,GrPe86a,GrPe86b}.

\medskip

The \ul{Jagellonian field} (\cite{RDFBS73} 1973) consists of a $6
\times 6$ square degrees area with galaxy counts in cells of $3.75'
\times 3.75'$, e.g.~in a $98 \times 98$ grid (higher resolution and
deeper than in the Lick catalog).  There was no attempt to correct for
the lack of uniform optics and plate exposure across the large field
of view (e.g.~vignetting effects).  Although this is quite a deep
survey, its angular extent is small and it is clear that it should
suffer significantly from the volume and edge effects described in
Chapter~\ref{sec:chapter7}.

\medskip

The \ul{ESO/Uppsala} \cite{HLSW74} and \ul{UGC} \cite{Nilson73}
catalogs are based on several hundreds of copies of large
(ESO/Palomar) Schmidt plates.  Galaxies were found with a limiting
visual diameter of about $1'$.  There is evidence for the selection
function to depend on declination, which has to be taken into account
while inverting the angular correlations (e.g.~see \cite{JMB91}). 
Compensation for this effect is likely to produce large scale
artifacts, especially because the sample is relatively small.

\medskip

The \ul{APM galaxy catalog} (\cite{MESL90} 1990) is based on 185 UK
IIIA-J Schmidt photographic plates, each corresponding to $6 \times 6$
square degrees on the sky to $b_J=20.5$ and mean depth of $400$
Mpc/$h$ (a factor of two deeper than the Lick catalog) for $b <-40$
degrees and $\delta<-20$ degrees.  These fields were scanned by the
APM machine \cite{KBBI84}.  Galaxy and star magnitudes and positions
in the overlapping regions (of 1 degree per plate) were used to match
all plates to a single calibration/exposure.  Because there are
calibration errors for individual galaxies and positions in a plate, a
more careful analysis of vignetting and variable exposure within a
plate could be done (as compared to just using the counts).  The
resulting matching errors can be used to perform a study of the biases
induced in the clustering analysis.  In the results shown here, an
equal-area projection pixel map was used with a resolution of $3.5'
\times 3.5'$ cells.

\medskip

The \ul{EDSGC Survey} (\cite{CNL92} 1992) consists of 60 UK IIIA-J
Schmidt photographic plates corresponding to $6 \times 6$ square
degrees on the sky to $b_J=20.5$ and mean depth of $400$ Mpc/$h$.  In
fact, the raw photographic plates are the same in both the APM and
EDSGC catalogs, but the later only includes scans of a fraction (1/3)
of the APM plates, in the central part.  The EDSGC database was
constructed from COSMOS scans~\cite{MacGillivray84}, with different
calibration and software analyses.  Therefore these two catalogs can
be considered as fairly independent realizations of the systematic
errors.

\medskip

The \ul{IRAS 1.2 Jy} (\cite{SHDYFT92} 1990) is a redshift subsample of
the IRAS Point Source Catalog~\cite{CBC87} and is included here
because it has also been used to measure angular clustering.  This
catalog belongs to a newer generation of wide field surveys, where
magnitudes and positions of objects have been obtained by automatic
measurements.  The CfA~\cite{HDLT83} and SSRS~\cite{DPDMST91} redshift
catalogs have also been used to study angular clustering.  More
details on redshift samples will be given in Sect.~\ref{sec:zcat}.

\medskip

The \ul{DeepRange Catalog} (\cite{PLSO98} 1998) consists of 256
overlapping CCD images of 16 arc minutes on a side, including 1 arc
minute overlap to allow the relative calibration of the entire survey. 
Images were taken to $I_{\rm AB} < 24$ with a total area extending
over a contiguous $4 \times 4$ square degrees region.  The median redshift for
the deeper slices used in the clustering analysis, $I_{\rm AB} =
22-22.5$ is $z \simeq 0.75$ which corresponds to a depth of
approximately $2000 \Mpc$.  The $I_{\rm AB} = 17-18$ slice has $z
\simeq 0.15$, i.e.~a similar depth to the APM catalog.  Note the large
surface density of this survey.  Although this is quite a deep survey
its angular extent is rather small and it suffers from the volume and
edge effects described in Chapter~\ref{sec:chapter7}, especially at
the brighter end.

\medskip

The \ul{Sloan Digital Sky Survey} (SDSS, eg see \cite{York00})
was still under construction when this review was written
and only preliminary results are known at this stage.
These results are discussed in a separate 
section, see \ref{sec:outlook} for more details.

\medskip

Smaller, but otherwise quite similar in design to DeepRange, 
wide mosaic optical
catalogs have been used to study higher-order correlations.  For
example, the INT-WFC~\cite{RoEa99} with $\sim 70000$ galaxies to
$R<23.5$ over two separated fields of 1.01 and and 0.74 square
degrees.  There are a number of such surveys currently under analyses
or in preparation, such as the FIRST radio source survey~\cite{BWH95},
the NOAO Deep Wide-Field Survey~\cite{JaDe99}, the Canada-France Deep
Fields~\cite{CLBFLCM01}, VIRMOS~\cite{Lefevre01}, DEIMOS~\cite{DaNe01}
or the NRAO VLA Sky Survey~\cite{CCGYPTB98}.

Most of the catalogs described above have magnitude information,
allowing one to study subsets at different limiting magnitudes or
depth.  This can be used for instance to test Limber equation
[Eq.~(\ref{eq:limber})] and the homogeneity of the
sample~\cite{GrPe77,MESL90}.  Even with the new generation of better
calibrated surveys, there has been some concerns about variable
sensitivity inside individual plates in the APM and EDSGC
catalogs~\cite{Djorgovski98} and some questions regarding large-scale
gradients in the APM survey have been raised~\cite{FHS92}.  Later
analysis checked the APM calibration against external CCD measurements
over 13000 galaxies from the Las Campanas Deep Redshift Survey showing
an rms error in the range 0.04-0.05 magnitudes~\cite{MES96}.  These
studies concluded that atmospheric extinction and obscuration by dust
in our galaxy have negligible effect on the clustering and also gave
convincing evidence for the lack of systematics errors.

\subsubsection{The Angular Correlation Function and Power Spectrum}

\label{sec:ang2pt}

The  angular two-point correlation function in early surveys was
estimated from the Zwicky catalog, Jagellonian field and Lick survey
in~\cite{ToKi69,PeHa74,Peebles75,GrPe77,GoTu79}.  For catalogs with
pixel maps (counts in some small cells), such as Lick and Jagellonian,
the estimators used were basically factorial moment correlators as
described in Sect.~\ref{sec:sec5bis}, whereas for catalogs with
individual galaxy positions (such as Zwicky) the estimators were based
on pair counts as discussed in Sect.~\ref{sec:estxi2}.

The angular two-point function was found to be consistent between the
Zwicky, Lick and Jagellonian samples.  For a wide range of angular
separations, the estimates were well fitted by a power-law:

\be
w_2(\theta) \simeq \theta^{1-\gamma}, ~~~\gamma \simeq 1.77 \pm 0.04
\label{w2pl}
\ee

The resulting 3D two-point function, after using Limber's equation
[Eq.~(\ref{eq:limber})] for the deprojection of a power-law model,
gives consistent results for all catalogs with:

\be \xi_2(r)
\simeq \left({r_0\over{r}}\right)^\gamma, ~~~\gamma \simeq 1.8,~~~ r_0
\simeq 5 \pm 0.1 \Mpc \label{xi2r0}
\ee

for scales between $0.05 \Mpc < r < 9 \Mpc$~\cite{Peebles75,GrPe77}. 
On the largest scales, corresponding to $r \ga 10 \Mpc$, the results
are quite uncertain because correlations are small and calibration
errors become relevant.  The results in~\cite{GrPe77} suggested a
break in $\xi_2(r)$ for $r \ga 10 \Mpc$.  The position of this break,
however, depends on the smoothing corrections applied to the Lick
catalog (which is the one probing the largest scales) on angles
$\theta \ga 3$ degrees~\cite{GrPe86a,GrPe86b}.

Several other groups have measured small numbers of Schmidt and 4-m
plates to produce galaxy surveys of a few hundred square degrees down
to $b_J \simeq 20$ and a few square degrees down to $b_J \simeq
23$~\cite{SFEM80,Hewett82,KoSz84,SSFM85,PrIn86,Sebok86}.  Most of
these studies also show a power-law behavior with consistent values
and a sharp break at large scales, the location of the latter
depending on the size of the catalog\footnote{~More recent studies
using CCD cameras, find that the power-law form of the small-scale
angular correlation function remains in deep samples with amplitude
decreasing with fainter
magnitudes~\cite{CJB93,RSMF93,HuLi96,PLSO98,CLBFLCM01}, with
indications of a less steep power-law at the faint end, $I_{AB} \ga
23$~(e.g.~\cite{InPr95,PLSO98,CLBFLCM01}).  }.  This sharp break,
expected in CDM models, is at least in part caused by finite volume
effects, i.e. the integral constraint discussed in e.g.
Sect.~\ref{sec:secest}\footnote{~It is worth pointing out that the
cosmic bias caused by the small volume, the boundary or shot-noise in
the sample typically yields lower amplitudes of $w_2$ for the smaller
(nearby) samples.  This has been noticed by several authors
(e.g.~\cite{DMSDY88}, Fig.~3 in~\cite{JMB91}) and sometimes
interpreted as a real effect (see also Fig.~8 in~\cite{HuGa99}).}. 
Thus most of these analyses show uncertain estimations for $w_2$ in
the weakly non-linear regime, which is also the case for the
ESO/Uppsala and UGC catalogs~\cite{JMB91}.

The APM catalog has enough area and depth to probe large scales in the
weakly non-linear regime.  The first measurements of the angular
two-point correlation function~\cite{MESL90} led to the discovery of
``extra'' large-scale power (corresponding to shape
parameter\footnote{~The shape parameter when the contribution of
baryons is neglected ($\Omega_{b} \ll \Omega_{m}$) reads $\Gamma
\approx \Omega_{m} h$, see e.g.~\cite{EBW92,BoEf84,BBKS86}.  However,
for currently favored cosmological parameters it is more accurate to
use $\Gamma = \exp[-\Omega_b(1+\sqrt{2h}/\Omega_m)] \times\Omega_m
h$~\cite{Sugiyama95}.} $\Gamma \sim 0.2$), significantly more than in
the standard CDM model ($\Gamma=0.5$).  This result has been confirmed
by measurement of $w_{2}(\theta)$ in the EDSGC catalog~\cite{CNL92},
and subsequent analyses of the inferred 3D power spectrum from
inversion of the APM angular correlation function~\cite{BaEf93} and
angular power spectrum~\cite{BaEf94} and inversion from
$w_{2}(\theta)$ to the 3D two-point function~\cite{Baugh96} (see
Sect.~\ref{sec:w2pk} for a brief discussion of inversion procedures). 
Both APM and EDSGC find more power than the Lick catalog on scales
$\theta \ga 2$ degrees, suggesting that the Lick data were
overcorrected for possible large scale
gradients~\cite{MESL90,MESL90b,MESL90c,MES96}.

\begin{figure}
\centering \centerline{\epsfxsize=9.truecm \epsfbox{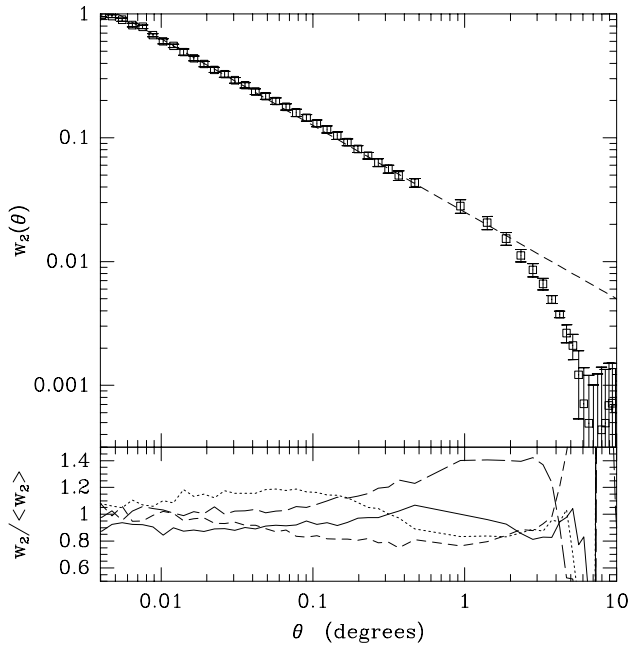}}
\caption{ The two-point angular correlation function $w_2(\theta)$
(squares with error-bars), estimated from counts-in-cells and
pair-counts in the APM map compared with a power-law $w_2 \sim
\theta^{-0.7}$ (dashed line).  Errors are from the dispersion in 4
disjoint subsamples within the APM. The lower panel shows the ratio of
the values in each zone to the average value in the whole sample.}
\label{w2apmz4}
\end{figure}

Figure~\ref{w2apmz4} shows the two-point angular correlation function
$w_2(\theta)$ estimated for $\theta> 1$ degree from counts in the
pixel maps (i.e. the factorial moment correlator $W_{11}$, see
Sect.~\ref{sec:sec5bis}) and at smaller scales from galaxy pair counts
(using the $DD/DR-1$ estimator, see Sect.~\ref{sec:estxi2}).  A fit of
the two-point angular correlation with a power-law $w_2 \simeq A
\theta^{1-\gamma} $, for scales $\theta < 2$ degrees gives $A \simeq
2.7 \times 10^{-2}$ and $\gamma \simeq 1.7$ (shown as a dashed line). 
After inverting the Limber equation, the corresponding 3D two-point
correlation function is in good agreement with Eq.~(\ref{xi2r0}),
with a slightly flatter slope $\gamma \simeq 1.7$. 
The uncertainty in the value of the correlation length $r_{0}$ is
controlled mainly by the accuracy in the knowledge of the selection
function in Eq.~(\ref{eq:limber}) and by the cosmic errors that we
discuss below.

The APM data show good match between several disjoint magnitude slices
when scaled according to the Limber equation to the same depth (see
Fig.~25 in \cite{MES96}).  The agreement is good up to very large
scales $\theta D \ga 40 \Mpc$; this indicates that the APM catalog can
be used to explore the weakly non-linear regime.  Similar conclusions
apply to the EDSGC catalog (see~\cite{CNL92}), which is compared in
terms of $w_{2}(\theta)$ to APM in~\cite{MES96} (see
also~\cite{HuGa99}): both catalogs agree well for $0.1 < \theta < 0.5$
degrees.  At larger angular scales, the EDSGC results differ from APM,
essentially because of finite volume and edge effects due to its
smaller area.  More worrisome is that at smaller scales, $\theta<0.1$
degrees there are also discrepancies (presumably related to deblending
of galaxies in high-density regions, see~\cite{SzGa98}) which can be
quite significant for higher-order moments as we shall discuss in 
Sect.~\ref{sec:cicang}.

The errors shown in Fig.~\ref{w2apmz4} are obtained from the scatter
among four disjoint subsamples in the APM, which is often an
overestimate of the true cosmic errors at large scales (see end of
Sect.~\ref{sec:secintegra}).  However, as discussed at length in
Chapter~6, error bars give only a partial view of the real
uncertainties (especially in the case of spatial statistics), since
measurements at different scales are strongly correlated.  This is
illustrated in the bottom panel of Fig.~\ref{w2apmz4}, where the
variations of the measured $w_2$ from subsample to subsample are
coherent (and quite significant at the largest scales where edge
effects become important).  As a result, the values of $w_2$ change
mostly in amplitude and to a lesser extent in slope from zone to zone. 
These cross-correlations are not negligible and must be taken into
account to properly infer cosmological information since the
measurements at different scales are not statistically independent. 
Only very recently the effect of the covariance between estimates at
different scales was included in the analyses of
APM~\cite{EiZa01,EfMo01} and EDSGC~\cite{HKN01} angular clustering, by
focusing on large-scales and using the Gaussian approximation to the
covariance matrix, similar to Eq.~(\ref{covawtheta}).  We discuss
these results in the next section.

Finally, note that the nearly perfect power-law behavior of the
angular correlation function imposes non-trivial constraints on models
of galaxy clustering.  Since in CDM models the dark matter two-point
correlation function is not a power-law, this implies that the bias
between the galaxy and mass distribution must be scale dependent in a
non-trivial way.  The current view (see discussion in
Sect.~\ref{galbias}) is that this happens because the number of
galaxies available in a given dark matter halo scales with the mass of
the halo as a power-law with index smaller than unity.  In these
scenarios, the fact that the galaxy two-point function follows a
power-law is thus a coincidence.  Given the accuracy of the power-law
behavior (see Fig.~\ref{w2apmz4}) this situation is certainly
puzzling, it seems unlikely that such a cancellation can take place to
such an accuracy\footnote{~However, one must keep in mind that features
in the spatial correlation function can be significantly washed out
due to projection, as first emphasized in~\cite{FaTr77}.}.  On the
other hand, these models predict at small-scales that galaxy velocity
dispersions and $S_{p}$ parameters are significantly smaller than for
the dark matter, as observed.  We shall come back to discuss this in
more detail below.

\begin{figure}
\centering \centerline{\epsfxsize=9.truecm \epsfbox{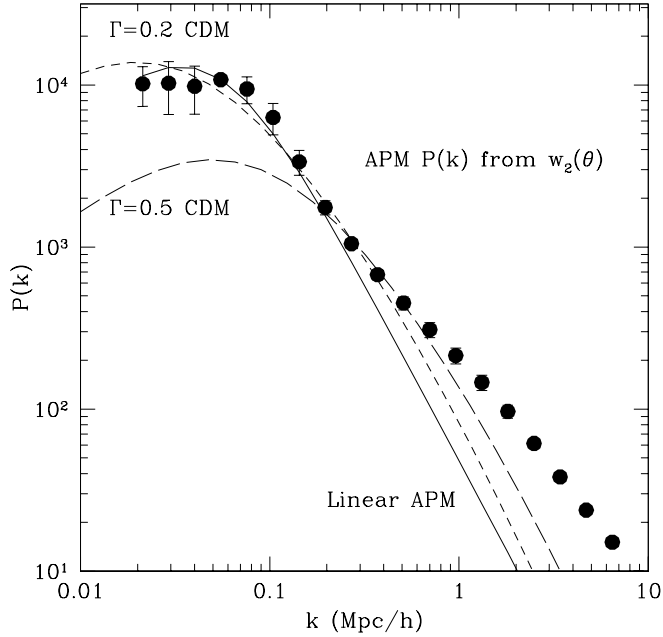}}
\caption{The APM 3D power spectrum reconstructed from $w_2(\theta)$. 
The continuous line shows a linear $P(k)$ reconstruction.  The short
and long dashed lines show linear CDM models with $\Gamma=0.2$ and
$\Gamma=0.5$, normalized to the data at $k \simeq 0.3 h/$Mpc.}
\label{pkw2apm}
\end{figure}

\subsubsection{Inversion from Angular to 3D Clustering}

\label{sec:w2pk}

The cosmological information contained in the angular correlation
function of galaxies can be extracted in basically two different ways. 
One is to just project theoretical predictions and compare to
observations in angular space.  It also is useful to carry out the
alternative route of an inversion procedure from Eq.~(\ref{fb:w2Dcor})
to recover the 3D power spectrum, and compare to theoretical
predictions in the more familiar 3D space.  This has the advantage
that it is possible to carry out parameter estimation on the scales
not affected by non-linear evolution\footnote{~In angular space, this
distinction is harder to make due to projection, particularly for the
two-point correlation function.  For example, for APM, $w(\theta)$ at
$\theta=1,2,3,5$ degrees has contributions from 3D Fourier modes up to
$k=1,0.4,0.3,0.2$ h/Mpc, respectively~\cite{DoGa00}.}.  To successfully
apply this method, however, one must be able to propagate
uncertainties from angular space to 3D space in a reliable way. 
Recent work has developed techniques that make this possible.

To go from the angular correlation function to the 3D power spectrum
(or two-point function) requires the inversion of an integral equation
with a nearly singular kernel, since undoing the projection is
unstable to features in the 3D correlations that get smoothed out due
to projection.  The inverse relation between $\xi_2(r)$ and
$w_{2}(\theta)$ can be written down formally using Mellin
transforms~\cite{FaTr77,Parry77}, however in practice this result is
difficult to implement since it involves differentiation of noisy
quantities.  Most inversions from
$w(\theta)$~\cite{BaEf93,Baugh96,GaBa98} and the angular power
spectrum~\cite{BaEf94} in the APM survey used an iterative
deconvolution procedure suggested by Lucy~\cite{Lucy74} to solve
integral equations.  However, although Lucy's method can provide a
stable inversion, it does not provide a covariance matrix of the
recovered 3D power spectrum.  Error bars on the reconstructed 3D power
spectrum have been estimated by computing the scatter in the spectra
recovered from four different zones of the APM
survey~\cite{BaEf93,BaEf94}; this can only be considered as a crude 
estimate and cannot be used to constrain cosmological parameters 
in terms of rigorous confidence intervals.

A number of methods have emerged in the last couple of years to
overcome these limitations.  These techniques involve some way of
constraining the smoothness of the 3D power spectrum to suppress
features in it that lead to minimal effects on the angular clustering
and thus make the inversion process unstable.  A method using a
Bayesian prior on the smoothness of the 3D power spectrum was proposed
in~\cite{DoGa00}.  An improved method, based on SVD
decomposition~\cite{EiZa01}, identifies and discards those modes that
lead to instability.  Both methods give the covariance matrix for the
estimates of the 3D power spectrum given a covariance matrix of the
angular correlations, which can be done beyond the Gaussian
approximation.  The resulting 3D covariance matrix shows significant
anti-correlations between neighboring bins~\cite{DoGa00,EiZa01}; this
is expected since oscillatory features in the power spectrum are
washed out by projection and thus are not well constrained from
angular clustering data.  Another technique based on maximum
likelihood methods for performing the inversion is presented
in~\cite{EfMo01} (see e.g. discussion in Sect.~\ref{sec:general}). 
This has the advantage of being optimal for Gaussian fluctuations, on
the other hand, the assumption of Gaussianity means that errors and
their covariances are underestimated at scales affected by non-linear
evolution where non-Gaussianity becomes important.  Including the
covariance matrix of angular correlations showed that constraints on
the recovered large-scale 3D power spectrum of APM galaxies become
less stringent by a factor of two~\cite{EiZa01,EfMo01} compared to
some of the previous analyses that assumed a diagonal covariance matrix.

Figure~\ref{pkw2apm} displays the APM 3D power spectrum $P(k)$
reconstructed from the angular two-point correlation
function~\cite{BaEf93,GaBa98} inverting Limber's Eq.~(\ref{fb:w2Dcor})
using Lucy's method.  The errorbars are obtained from the dispersion
on $w_2(\theta)$ over four zones as shown in Figure~\ref{w2apmz4} and
should thus be considered as a crude estimate, especially at large
scales (see~\cite{EfMo01} for comparison of errors in different
inversion methods).  The solid curve corresponds to a reconstruction
of the linear part of the spectrum, which can be fitted by:

\be
P_{\rm linear}^{\rm APM}(k) \simeq 7\times 10^5 {k
\over{\left[1+(k/0.05)^2\right]^{1.6}}} \label{linearAPM}
\ee

for $k<0.6 h/$Mpc, and $\Omega_{m}=1$~\cite{BaGa96}.  This
linearization has been obtained assuming no bias between APM galaxies
and dark matter\footnote{~Unfortunately, as shown in~\cite{BaGa96},
this assumption is inconsistent at small scales: the higher-order
moments predicted by evolving the linear spectrum in
Eq.~(\ref{linearAPM}) are in strong disagreement with the APM
measurements at scales $R \la 10$ Mpc/h (see Fig.~\ref{s34apm} below),
indicating that galaxy biasing is operating at non-linear scales.  On
the other hand, the large-scale correlations ($R>10$ Mpc/h) are
consistent with no significant biasing, see Sect.~\ref{sec:biagau}.},
following the linearization first done in~\cite{HKLM91} and extended
in~\cite{PeDo94} based on the mapping from the linear to non-linear
power spectrum (see e.g. Sect.~\ref{nlevtp} for a discussion). 
Equation~(\ref{linearAPM}) has been obtained by running N-body
simulations and agrees well with the mapping prescription
of~\cite{JMW95}.  Note how non-linear effects become important at
$k>0.1 h/$Mpc\footnote{~In fact, it has been demonstrated
in~\cite{FoGa98a} that the one-loop PT predictions presented in
Sect.~\ref{sec:1Lpk} work very well for this spectrum on scales where
the fit in Eq.~(\ref{linearAPM}) is valid, $k<0.6$ h/Mpc.}.

As can be seen from Fig.~\ref{pkw2apm}, a comparison to CDM models on
linear scales ($k<0.3$ h/Mpc) favors low values of power-spectrum
shape parameter $\Gamma$, showing more power on these scales than the
standard CDM model with $\Gamma=0.5$.  Indeed, the most recent
analyses including the effects of the covariance matrix discussed
above concludes using the deprojected data for $k \leq 0.19$ h/Mpc
that $0.05 \leq \Gamma \leq 0.38$ to 95\%
confidence~\cite{EfMo01}\footnote{~In addition, it was shown that
galactic extinction, as traced by the maps in~\cite{SFD98}, had little
effect on the power spectrum over the APM area with $\delta <
-20^{o}$.}.  Similar results have been obtained from a similar recent
likelihood analysis of the EDSGC survey angular power
spectrum~\cite{HKN01}.  Figure~\ref{pkw2apm} suggests that on very
large scales ($k<0.05$ h/Mpc), the APM data might show an indication
of a break in the power spectrum~\cite{GaBa98}.  From the figure, it
might seem as if this is a 3-sigma detection, but as mentioned above
different points are not independent.  Analytical studies, using
different approximations to account for the covariance matrix between
different band powers, indicate that this might be only a 1-$\sigma$
result~\cite{EiZa01,EfMo01}\footnote{~However, the initial suggestion
by ~\cite{GaBa98} for a break in the APM was confirmed with realistic
numerical simulations which show that a mock galaxy catalog as big as
the APM can be use to recover such a break when placed at different
scales (see Fig.  11-12 in \cite{GaBa98}).  The level of significance
for this detection was not studied, so these apparently discrepant
analyses require further investigation.}.

The above results on the shape parameter of the power spectrum have
been confirmed by analyses of redshift catalogues as will be discussed
in Sect.~\ref{sec:2ptredsh}, and will soon be refined by measurements
in large ongoing surveys such as the 2dFGRS or the SDSS
(Sect.~\ref{sec:outlook}).

On smaller scales, a detailed study ~\cite{GaJu01} of the
reconstructed 3D 2-point correlation function in the APM
~\cite{Baugh96} shows an inflection point in the shape of $\xi_2(r)$
at the transition to the non-linear scale $r \simeq r_0 \sim 5$ Mpc/h,
very much as expected from gravitational instability (see
Sect.~\ref{stcl}).

\begin{table}
\centering
\caption{The angular three and four-point amplitudes $3 q_3$ and
$16 q_4 \equiv 12 r_a + 4 r_b$, at physical scales (in Mpc/h)
specified in the third column by ${\cal D} \theta$. The last five 
entries correspond to the newer generation of galaxy catalogs (see 
Table~\ref{angularcat}). Error bars should be considered only as rough 
estimates, see text for discussion.} 
\begin{tabular}{ccccccc} \hline
\rule[-0.6ex]{0mm}{3.2ex} {$3 q_3$} & {$16 q_4$} & { ${\cal D}
\theta$} & {Sample} & Year & {Ref.} & Estimator \\* \hline
\rule[-0.6ex]{0mm}{3.2ex} $1.9 \pm 0.3$ & --- & 0.4-1.2 
& Jagellonian& 1975 & \cite{Peebles75} & cumulant corr.  \\*
\rule[-0.6ex]{0mm}{3.2ex} $3.5 \pm 0.4$ & --- & 0.1-4 
& Zwicky (-Coma)& 1975 & \cite{PeGr75} & multiplet counts \\*
\rule[-0.6ex]{0mm}{3.2ex} $5.3 \pm 0.9$ & --- & 0.1-4 
& Zwicky & 1977& \cite{GrPe77} & ''\\* 
\rule[-0.6ex]{0mm}{3.2ex} --- & $100 \pm 18$ &0.1-2 
& Zwicky & 1978 & \cite{FrPe78} & ''\\*
\rule[-0.6ex]{0mm}{3.2ex} $4.7 \pm 0.7$ & --- & 0.3-10 
&Lick & 1977 &\cite{GrPe77} & cumulant corr.  \\* 
\rule[-0.6ex]{0mm}{3.2ex} --- &$77 \pm 7$ & 0.5-4 
& Lick & 1978 & \cite{FrPe78} & '' \\*
\rule[-0.6ex]{0mm}{3.2ex} $4.8 \pm 0.1$ & $40 \pm 3$ & 0.3-5 
& Lick &1992 & \cite{SSB92} & '' \\* 
\rule[-0.6ex]{0mm}{3.2ex} $\simeq 3$ &--- & 0.3-5 ($k$) 
& Lick & 1982 & \cite{FrSe82} & bispectrum \\*
\rule[-0.6ex]{0mm}{3.2ex} $2.7 \pm 0.1$ & --- & 0.2-2 
& ESO-Uppsala &1991 & \cite{JMB91} & multiplet counts \\* 
\rule[-0.6ex]{0mm}{3.2ex}$5.4 \pm 0.1$ & --- & 0.2-2 
& UGC & 1991 & \cite{JMB91} & '' \\ *
\hline
\rule[-0.6ex]{0mm}{3.2ex}$3.8 \pm 0.3$ & $35 \pm 10$ & 4-20 
& IRAS 1.2Jy & 1992 & \cite{MSS92b}& cumulant corr.  \\* 
\rule[-0.6ex]{0mm}{3.2ex} $3.5 \pm 0.1 $ & $31\pm 1$ & 0.5-50 
& APM (17-20) & 1995 & \cite{SDES95} & '' \\*
\rule[-0.6ex]{0mm}{3.2ex} $3.9 \pm 0.6$ & --- & 4 
& APM & 1999 &\cite{FrGa99} & '' \\* 
\rule[-0.6ex]{0mm}{3.2ex} $2-6$ & --- & 4-30 
& APM & 1999 &\cite{FrGa99} & '' \\* 
\rule[-0.6ex]{0mm}{3.2ex} $1.5-3$ & --- & 0.2-3 
& LCRS & 1998 &\cite{JiBo98} & multiplet counts \\* 
\rule[-0.6ex]{0mm}{3.2ex} $ 8-3$ & ---  & 0.5-3 
& DeepRange & 2000 & \cite{SPLO01} & ''  \\* 
\rule[-0.6ex]{0mm}{3.2ex} $ 2-1$ & --- & 3-6 
& DeepRange & 2000 & \cite{SPLO01} & ''  \\*
\rule[-0.6ex]{0mm}{3.2ex} $5-1$ & --- & 0.5-20 
& SDSS & 2001 &\cite{Gaztanaga01,Gaztanaga01b} & '' \\* 
 \hline
\end{tabular}
\label{q3q4}
\end{table}

\subsubsection{Three-Point Statistics and Higher Order}

\label{sec:nptang}

Angular surveys provide at present the best observational constraints
on higher-order correlation functions in the non-linear regime.  In
most cases, however, a detailed exploration of the different
configurations available in three-point and higher-order correlations
has not been given, due to limitations in signal to noise\footnote{~In
addition, even with the currently available computational power and
fast algorithms relying on e.g. $K$D-tree
techniques~\cite{MCGGGKNSSSW01}, measuring directly higher-order
correlation functions can be very computationally intensive.}.  This
will have to await the next generation of photometric surveys (e.g.
SDSS~\cite{York00} and DPOSS~\cite{Djorgovski98}). 

Table~\ref{q3q4} summarizes the measurements achieved in various
surveys.  As can be seen in third column of Table~\ref{q3q4}, the
limited size of surveys means that most of the measurements only
probed the nonlinear regime, except those done in the IRAS and APM
catalogs.  The first measurements of the three-point angular
correlation function $w_{3}$ in the Jagellonian
field~\cite{Peebles75}, Lick and Zwicky surveys~\cite{GrPe77}
established that at small scales the {\em hierarchical model} (see
Sect.~\ref{sec:HM}) gives a good description of the data,

\be
w_{3}(\te_{1},\te_{2},\te_{3})= q_{3} \Big[ 
w_{2}(\te_{1})w_{2}(\te_{2}) + w_{2}(\te_{2})w_{2}(\te_{3})+ 
w_{2}(\te_{3})w_{2}(\te_{1}) \Big],
\label{HA3ang}
\ee 

where $q_{3}$ is a constant of order unity with little dependence on
scale or configuration (within the large error bars) at the range of
scales probed.  In addition, the four-point function was found to be
consistent in the Lick and Zwicky catalogs with the hierarchical
relation,

\bea w_4(1,2,3,4)&=&r_a\ \Big[ w_2(1,2)\, w_2(2,3)\, w_2(3,4) + {\rm
cyc.\ (12~terms)}\ \Big] \nonumber \\
& & +r_b\ \Big[ w_2(1,2)\, w_2(1,3)\, w_2(1,4) + {\rm cyc.\ (4~terms)}\
\Big],  \label{HA4ang}
\eea

where $w_{2}(i,j)\equiv w_{2}(\te_{ij})$ with $\te_{ij}$ being the
angular separation between points $i$ and $j$.  The amplitudes $r_{a}$
and $r_{b}$ correspond to the different topologies of the two type of
tree diagrams connecting the four points (see e.g. Fig.~\ref{fig4_2}
and discussion in Sect.~\ref{sec:HM}), the so-called snake ($r_{a}$,
first diagram in Fig.~\ref{fig4_2}) and star diagrams ($r_{b}$, second
diagram in Fig.~\ref{fig4_2}).  The overall amplitude of the
four-point function is thus $16 q_{4}\equiv 12r_{a}+4r_{b}$, which we
quote in Table~\ref{q3q4}, together with the three-point amplitude
$3q_{3}$.  These are useful to compare with the angular skewness and
kurtosis in Table~\ref{s3s4} discussed in Sect.~\ref{sec:cicang}
because in the hierarchical model $s_{N}\simeq N^{N-2}q_{N}$ to very
good accuracy\footnote{~And similarly in the 3D case, 
see~\cite{BSS94,Gaztanaga94}
for accurate estimates of the small corrections to this relation.}. 
In addition, as discussed in Sect.~\ref{sec:projhier}, the $q_{N}$
coefficients are very weakly dependent on details of the survey such
as the selection function and its uncertainties, so it is meaningful
to compare $q_{N}$ from different galaxy surveys.

In the first and second column of Table~\ref{q3q4}, in addition to the
numerical values of $q_3$ and $q_4$, we quote as well the error on the
estimate calculated by the authors.  Except when noted otherwise,
errorbars were obtained from the dispersion in different zones of the
catalog.  Since typically the number of zones used is small (about
four in most cases), the estimated errors are very
uncertain\footnote{~However, as discussed in end of
Sect.~\ref{sec:secintegra} for the two-point correlation function,
when the number of subsamples is large, this method tends to
overestimate the real cosmic errors.}.  In addition, this method
obviously cannot estimate the cosmic variance, which can be a
substantial contribution for surveys with small area.  Many, if not
most, of the differences between the various numerical values given in
Table~\ref{q3q4} can be explained by statistical fluctuations and
systematics due to the finiteness of the catalogs~\cite{HuGa99} (see
Chapter~6 for a detailed discussion of these issues), as we know
briefly discuss.

The results of $q_{3}$ in the Zwicky catalog do not seem to be very
reliable since the value found in~\cite{PeGr75} changed by more that
$50\%$ due to the omission of only 14 galaxies in the Coma cluster
(see~\cite{GrPe77}).  Similar effects have been found in other samples
(e.g. ESO-Uppsala~\cite{JMB91}).  This sensitivity reflects that the
sample is not large enough to provide a fair estimate of higher-order
statistics.  Likewise, the rather low value for $q_3$ found in the
Jagellonian sample is likely strongly affected by finite volume
effects due to the small area covered.  Similarly, the values obtained
from the projected LCRS by~\cite{JiBo98} could be partially
contaminated by edge effects due to the particular geometry of the
catalog (6 strips of $1.5 \times 80$ degrees) and perhaps also by
sampling biases due to inhomogeneous sampling around high density
regions\footnote{~Due to the fixed number of fibers per field and
``fiber collisions''.  Using random catalog generation~\cite{JiBo98}
checked that these effects appeared to be insignificant.}. 

Work has been done as well to study the dependence of $q_3$ on
morphological type, but dividing the data in smaller subsamples tends
to produce stronger statistical biases.  In the ESO-Uppsala and UGC
catalogs, \cite{JMB91} found that spirals have significantly smaller
values of $q_3$.  This could be interpreted through the well-known
density-morphology relation~\cite{Dressler80,PoGe84}: spirals avoid
rich clusters and groups, an effect that could be more important at
smaller scales (this is illustrated to some extent in
Fig.~\ref{Sp_gal}).  The results for the full sample in the
ESO-Uppsala and UGC catalogs showed good agreement with the
hierarchical scaling (note however that error bars quoted in this case
are just due to the dispersion in the fit to the hierarchical model
rather than reflecting sample variance).

The measurements of the three-point correlation function in the Lick
survey did not show any strong evidence for a dependence of $q_{3}$ on
the shape of the triangle, although a marginal trend was found that
colinear triangles had a higher $q_{3}$ than isosceles~\cite{GrPe77}. 
The three-point statistics was analyzed in terms of the bispectrum
in~\cite{FrSe82}, who found the same amplitude for $q_{3}$ than in
real space, but some indications of a scale dependence beyond the
hierarchical scaling, with $q_{3}$ increasing as a function of
wavenumber $k$ with a peak corresponding to the angular scale
($2.5^{o}$) of the break in $w_{2}(\theta)$, and then decreasing again
at large $k$.  A later re-analysis of the large-scale Lick
bispectrum~\cite{Fry94a} showed a marginal indication of dependence on
configuration shape, too small compared to the one expected in
tree-level PT, and thus in principle an indication of a large galaxy
bias [see Eq.~(\ref{Q_g_eul})].  However, the scales involved were not
safely into the weakly non-linear regime and thus this result is
likely the effect of non-linear evolution rather than a large galaxy
bias~\cite{SCFFHM98}.

The four-point function measurements in the Lick survey were not able
to test the relation in Eq.~(\ref{HA4ang}) in much detail, but
assuming Eq.~(\ref{HA4ang}) measurements for some specific
configurations (such as squares and lines) gave a constraint on the
amplitudes $r_{a}$ and $r_{b}$ which were then translated into a
constraint on 3D amplitudes by deprojection (see
Sect.~\ref{sec:projhier}), resulting in $R_{a}=2.5 \pm 0.6$ and
$R_{b}=4.3 \pm 1.2$~\cite{FrPe78}.  These results on the Lick survey
were considerably extended in~\cite{SSB92} to higher-order $q_{N}$'s
up to $N=8$ in the context of the {\em degenerate} hierarchical
model\footnote{~In this case all amplitudes corresponding to different
tree topologies are assumed to have the same amplitude $q_{N}$, and
thus $R_{a}=R_{b}$, etc., see Sect.~\ref{sec:HM}.}, by using two-point
moment correlators\footnote{~In the same spirit, it is worth noticing
that four-point correlation function estimates for particular
configurations can be obtained through measurements of the dispersion
of the two-point correlation function over subsamples (or cells)
extracted from the catalog (see~\cite{BoSh80,Fry83a}): this is a
natural consequence of the theory of cosmic errors on $w_2$ detailed
in Sect.~\ref{sec:secintegra}.  This method has the potential defect
of being sensitive to possible artificial large scale gradients in the
catalog.}.  This confirmed the hierarchical scaling $w_{N} \sim q_{N}
w_{2}^{N-1}$ up to $N=8$, at least for these configurations, with
$q_{N} \approx 1-2$\footnote{~Note however that the errors quoted by
the authors come from a fitting procedure, not sampling variance.  For
$N>6$ correlations are consistent with zero when using the sampling
variance among twelve zones.}.  

The same technique was applied to the IRAS 1.2~Jy survey
in~\cite{MSS92b}, verifying the hierarchical scaling for $N=3,4$ but
with $q_{N}$'s with $N>4$ being consistent with zero, and also to the
APM survey~\cite{SDES95} which showed non-zero amplitudes up to $N=6$,
with a trend of increasing $q_{N}$ as a function of $N$, i.e.
$q_{N}=1.2,2,5.3,10$ for $N=3,4,5,6$, unlike the case of the Lick
catalog.  The APM survey was later re-analyzed in terms of cumulant
correlators [e.g. see Eq.~(\ref{fb:Spqdef})] in~\cite{SzSz97}, showing
hierarchical scaling for $N=4,5$ to within a factor of
two\footnote{~The scales probed in this case, $0.8^{o}<\te<4.5^{o}$,
are in the transition to the nonlinear regime (1 degree corresponds to
about 7 Mpc/h at the APM depth), so it is not expected to show
hierarchical scaling.  On the other hand, galaxy biasing might help
make correlations look more hierarchical, as illustrated in
Fig.~\ref{Sp_gal} by the suppression in the growth of $S_{p}$
parameters as small scales are probed.}.  In addition, it showed that
at scales $\te \ga 3.5$ degrees the factorization property predicted
by PT, Eq.~(\ref{eq:cpqfac}), starts to hold.  By measuring $\lexp
\de_{1}^{3} \de_{2} \rexpc$ and $\lexp \de_{1}^{2} \de_{2}^{2} \rexpc$
and assuming the hierarchical model as in Eq.~(\ref{HA4ang}) it was
possible to constrain (after deprojection) $R_{a} \simeq 0.8$ and
$R_{b} \simeq 3.7$, in reasonable agreement with the Lick
results~\cite{FrPe78} mentioned in the previous paragraph.  These
imply an average $q_{4} \simeq 2.2 $.

The analysis of the three-point function in the DeepRange
survey~\cite{SPLO01} shows a general agreement with the hierarchical
model with large errors in $q_{3}$, with a consistent decrease as a
function of depth.  Indeed, a fit to the hierarchical model,
Eq.~(\ref{HA3ang}) gives $q_{3}=1.76,1.39,2.80,1.00,0.34,0.57$ for
I-band magnitudes $I=17-18,18-19,19-20,20-21,21-22,22-22.5$,
respectively.  This trend is also present in the count-in-cells
measurements and, if confirmed in other surveys, have interesting
implications for the evolution of galaxy bias (see
Fig.~\ref{s3z_SPLO}).  Note that in this work errors were estimated
using the FORCE code~\cite{SzCo96,CSS98,SCB99}, which is based on the
full theory of cosmic errors as described in
Chapter~\ref{sec:chapter7}.

\begin{figure}
\centering \centerline{\epsfxsize=8truecm \epsfbox{ 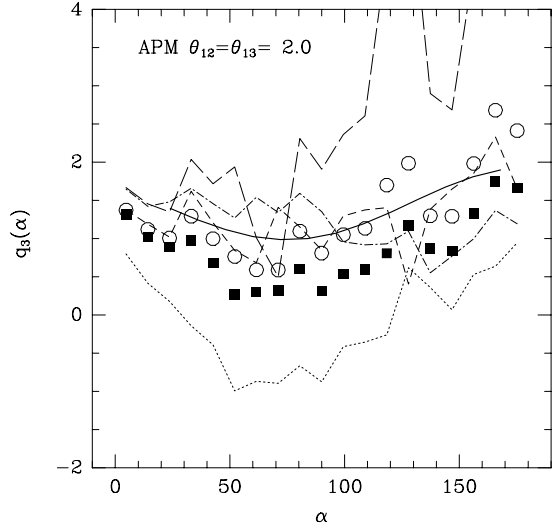}}
\caption[junk]{The angular three-point amplitude $q_3(\alpha)$ from PT
prediction (thick continuous line) compared with the APM measurements
at $\theta_{12}=\theta_{13}=2^{o}$: closed squares and open circles
correspond to the full APM map and to the mean of 4 disjoint zones. 
Other curves show results for each of the zones (from \cite{FrGa99}).}
\label{q3apmr2}
\end{figure}

Some of the analyses above probed the weakly nonlinear regime, where
the $q_N$'s are expected to show a characteristic angular dependence
predicted by PT, even after projection from 3D to angular
space~\cite{FrTh99,FrGa99,BKJ00}.  Measurements of $q_3$ in the Lick
catalog showed a marginal indication that colinear configurations are
preferred compared to isosceles triangles~\cite{GrPe77,Fry94a} (but
see~\cite{FrSe82}).  Projecting the three-point function in redshift
space from the LCRS survey, \cite{JiBo98} found a marginal enhancement
for colinear triangles, but the scales probed ($r \la 12$ Mpc/h) are
not safely in the weakly non-linear regime.

\begin{figure}
\centering \centerline{\epsfxsize=15truecm\epsfysize=12truecm\epsfbox{
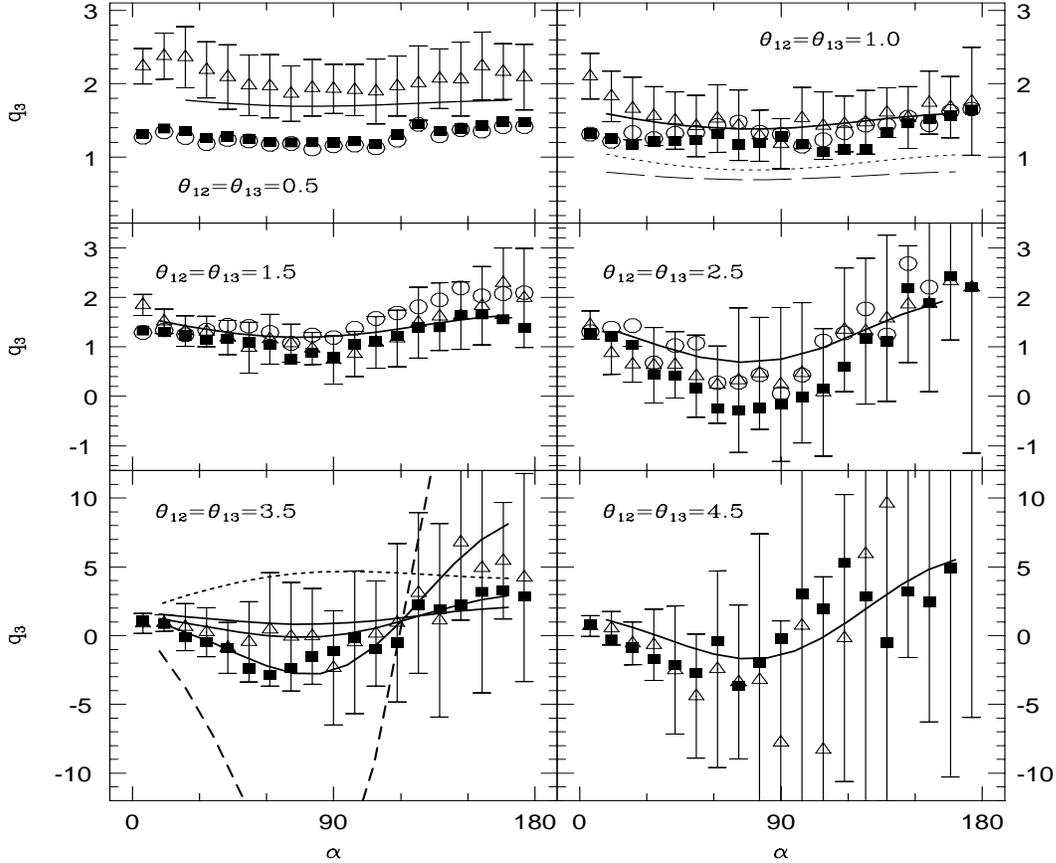 }} \caption[junk]{The projected three-point amplitude $q_3$
in PT (solid curves) and N-body results (open triangles with
errorbars) for the APM-like power spectrum are compared with $q_3$
measured in the APM survey (closed squares and open circles, with same
meanings as in Fig.  1).  Each panel shows the amplitude at different
$\theta_{12}=\theta_{13}$.  In upper right panel, dotted and dashed
curves correspond to PT predictions with $b_1=1, b_2=-0.5$ and $b_1=2,
b_2=0$, respectively.  In the lower left panel, upper and lower solid
curves conservatively bracket the uncertainties in the inferred
APM-like power spectrum, long-dashed curve corresponds to SCDM, and
the dotted curve shows the leading-order prediction for the $\chi^2$
non-Gaussian model.} \label{q3panel}
\end{figure}

For angular catalogs, the APM survey presents the best available
sample to check the angular dependence of $q_3$ predicted by
PT~\cite{FrGa99}.  Figures~\ref{q3apmr2} and \ref{q3panel} show the
measurements of $q_3(\alpha)$ in the APM survey at
$\theta_{12}=\theta_{13}=0.5-4.5$ degrees estimated by counting pairs
and triplets of cells of a given angular configuration, see
Sect.~\ref{sec:sec5bis}.  Closed squares correspond to estimations in
the full APM map, while open circles are the mean of $q_3$ estimated
in 4 disjoint zones.  The value of $3 q_3 \simeq 3.9 \pm 0.6$ at
$\alpha \simeq 0$, shown in Table \ref{q3q4}, is in agreement with the
cumulant correlators estimated (with $4 \times 4$ bigger pixels)
in~\cite{SzSz99}.  Furthermore, the average over $\alpha$ is
comparable to the values of $s_3/3$ in Table~\ref{q3q4} and in
particular to the APM and EDSGC
estimations~\cite{Gaztanaga94,MSS92b,SzGa98}.

Figure~\ref{q3apmr2} shows the results for individual zones in the APM
(same as the ones in Fig.~\ref{w2apmz4}) for all triangles with
$\theta_{12}=\theta_{13}=2$ degrees.  These estimations of $q_3$ are
subject to larger finite-volume effects, because each zone is only 1/4
the size of the full APM\footnote{~The fact that the average over the
four zones (open circles) is not equal to the measurement in the full
APM map is a manifestation of estimator bias~\cite{HuGa99,SCB99}.}. 
As in Fig.~\ref{w2apmz4}, there is a strong covariance among the
estimations in different zones, which results in a large uncertainty
for the overall amplitude $q_3$.  
 Because the zones cover a range of galactic latitude, a number
of the systematic errors in the APM catalog (star-galaxy separation,
obscuration by the galaxy, plate matching errors) might be expected to
vary from zone to zone.  No evidence for such systematic variation is
found in $q_3$: the scatter in individual zone values are compatible
with the sampling variance observed in N-body
simulations~\cite{FrGa99}.  On larger scales, $\theta \ga 3$ degrees,
the individual zone amplitudes exhibit large variance, and in addition
boundary effects come into play.  As seen in Fig.~\ref{q3panel} at
these scales $q_3$ is consistent with zero within the errors.

The APM results are compared with the values of $q_3$ predicted by PT
with the linear APM-like spectrum in Eq.(\ref{linearAPM}) (solid
curves) and with measurements in N-body simulations (open triangles
with errorbars) with Gaussian initial conditions corresponding to the
same initial spectrum.  Since the APM-like model has, by construction,
the same $w(\theta)$ as the real APM map, it is assumed that the
sampling errors are similar in the APM and in the simulations.  This
might not be true on the largest scales, where systematics in both the
APM survey and the simulations (periodic boundaries) are more
important.

At scales $\theta \ga 1$ degree, the agreement between the APM-like
model and the APM survey is quite good; this corresponds roughly to
physical scales $r \ga 7$ h$^{-1}$ Mpc, not far from the non-linear
scale ($r_0 \simeq 5$, where $\xi_2 \simeq 1$).  
Also note that the $q_3$ predicted in
the SCDM model (dashed curve in lower-left panel of
Fig.~\ref{q3panel}) clearly disagrees with the APM data; this
conclusion is independent of the power spectrum normalization and it
is therefore complementary to the evidence presented by two-point
statistics~\cite{MESL90,EKSLREF90} (see discussion in
Sects.~\ref{sec:ang2pt} and~\ref{sec:w2pk}).  At smaller angles,
$\theta \la 1$ deg, $q_3$ in the simulations is larger than in either
the real APM or PT (top-left panel in Fig.~\ref{q3panel}).  The
discrepancy between simulations and PT on these relatively small
scales is due to non-linear evolution.  The reason for the discrepancy
with the real APM is probably an indication of galaxy biasing at small
scales: this will affect the inference of the linear power spectrum
from the data~\cite{BaGa96} and also suppress higher-order correlations
compared to the dark matter~\cite{SSHJ01} (see e.g. Fig.~\ref{Sp_gal}
and discussion in Sect.~\ref{dhbias}).

\begin{table}
\centering
\caption[junk]{The reduced skewness and kurtosis from counts-in-cells
in angular space.  In most cases, only the mean values over a range of
scales were published.  In cases where measurements of the individual
$s_p$ for each smoothing scale are reported in the literature, we
quote the actual range and the corresponding range of scales.  Error bars
should be considered only as rough estimates, see text for
discussion.} 
\begin{tabular}{cccccc} \hline
\rule[-0.6ex]{0mm}{3.2ex} {$s_3$} & {$s_4$} & { ${\cal D} \theta$} &
{Sample} & Year & {Ref.} \\* \hline
\rule[-0.6ex]{0mm}{3.2ex} $2.9 \pm 0.9$ & $12 \pm 4$ & 1-8 & 
Zwicky &1984 & \cite{SBL84} \\* 
\rule[-0.6ex]{0mm}{3.2ex} $2.4 \pm 0.4$& $9.5 \pm 2.4$ & 2-20 & 
CfA & 1994 & \cite{FrGa94}  \\*
\rule[-0.6ex]{0mm}{3.2ex} $2.2 \pm 0.3$ & $ 8 \pm 3 $ & 2-20 & 
SSRS &1994 & \cite{FrGa94}  \\* 
\rule[-0.6ex]{0mm}{3.2ex}$ 2.5 \pm 0.4$ & $11 \pm 3$ & 2-20 & 
IRAS 1.9Jy & 1994 & \cite{FrGa94}\\* 
\rule[-0.6ex]{0mm}{3.2ex} $3.8 \pm 0.1 $ & $33\pm 4$ & 7-30 & 
APM (17-20) & 1994 & \cite{Gaztanaga94}  \\*
\rule[-0.6ex]{0mm}{3.2ex} $5.0 \pm 0.1 $ & $59 \pm 3$ & 0.3-2 & 
APM (17-20) & 1994 & \cite{Gaztanaga94}  \\* 
\rule[-0.6ex]{0mm}{3.2ex} $7-4$ & $170 - 40$ & 0.1-14 & 
EDSGC& 1996 & \cite{SMN96}  \\* 
\rule[-0.6ex]{0mm}{3.2ex}$3.0 \pm 0.3 $ & $20 \pm 5$ & 0.1 & 
APM (17-20) & 1998 & \cite{SzGa98} \\* 
\rule[-0.6ex]{0mm}{3.2ex} $ 6- 2$ & $120- 10$& 0.1-6 & 
DeepRange & 2000 & \cite{SPLO01} \\*
\rule[-0.6ex]{0mm}{3.2ex} $5-2$ & 100-20 & 0.5-20 
& SDSS & 2001 &\cite{Gaztanaga01,Gaztanaga01b,SFSS01} \\* 
 \hline

\end{tabular}
\label{s3s4}
\end{table}

\subsubsection{Skewness, Kurtosis and Higher-Order Cumulants}

\label{sec:cicang}

Table~\ref{s3s4} shows the results for the skewness ($s_3$) and
kurtosis ($s_4$) in several of the angular catalogs described in
Sect.~\ref{sec:angcat}.

The analysis of the Zwicky sample by~\cite{SBL84} used moments of
counts in cells to estimate the hierarchical amplitudes $q_{N}$,
assuming the degenerate hierarchical model in Sect.~\ref{sec:HM}. 
Because counts in cells were used, the measurement is closer to $s_3$
than to $q_3$.  As noted in Sect.~\ref{sec:nptang}, the Zwicky catalog
has been shown to be sensitive to a few galaxies in the Coma cluster,
a signature that the survey is not large enough to be a fair sample
for the estimation of higher-order moments.  Indeed, in~\cite{SBL84}
it was found that the mean over a four-subsample split changed from
the values in Table~\ref{s3s4} to $s_3=4.2 \pm 0.9$ and $s_4=-7 \pm
12$, a manifestation of the estimation biases discussed in Chapter~6.

In~\cite{FrGa94} angular positions from volume limited subsamples of
redshift catalogs (CfA, SSRS and IRAS 1.9 Jy) were used to estimate
the angular moments\footnote{~The values in Table~\ref{s3s4}, from
Table 8 in~\cite{FrGa94}, have been multiplied by $r_3 \simeq 1.2$ and
$r_4 \simeq 1.5$ for a direct comparison in angular space.}.  Note for
example how the values for $s_{3}$ and $s_{4}$ in the CfA survey from
these smaller samples are lower than in the parent Zwicky sample. 
This suggests again that there are significant systematic
finite-volume effects~\cite{Gaztanaga94,SzCo96,HuGa99,SCB99}.

\begin{figure}
\centering \centerline{\epsfysize=10.truecm 
\epsfbox{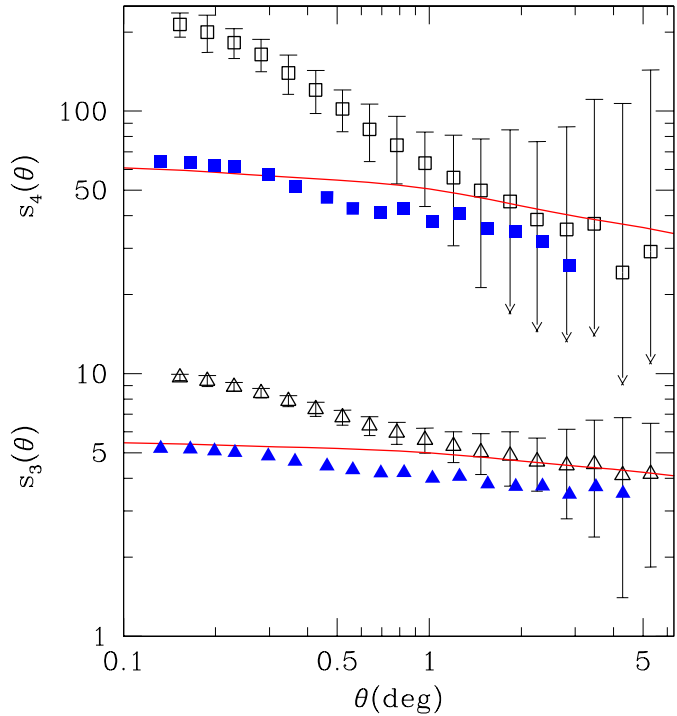}} \caption[junk]{The angular skewness, $s_3$, and
kurtosis, $s_4$, from the APM catalog (filled triangles and squares)
as compared with PT results (continuous line) and APM-like all-sky
N-body simulations (open triangles and squares).} \label{s34apm}
\end{figure}

Figure~\ref{s34apm} shows $s_3$ (filled triangles) and 
$s_4$ (filled squares) measured in the APM
survey~\cite{Gaztanaga94}.  The open figures with errorbars
correspond to the mean of 20 N-body all-sky simulations presented in ~\cite{GaBe98}
with the linear ``APM-like'' power spectrum in
Eq.~(\ref{linearAPM}), with 1-$\sigma$ error-bars scaled to the size of the
APM\footnote{~These errors should be considered more realistic
than those given in the fifth and sixth entry in Table~\ref{s3s4},
which were derived by combining results at different angular scales
assuming they are uncorrelated~\cite{Gaztanaga94}. These errorbars 
also correspond roughly to a 2-$\sigma$ confidence in
a single all-sky map: they are twice as large as the ones in Fig.\ref{s3tl}.}.  
The continuous
line show the tree-level PT results of~\cite{Bernardeau95} numerically
integrated for the APM-like power spectrum, as described
in~\cite{GaBe98}\footnote{~See e.g. Eq.~(\ref{wbar3pk}) and
Sect.~\ref{chpt} for a discussion of projection in the weakly
non-linear regime.}.  The uncertainties in the shape of the power
spectrum and the evolution of clustering of APM galaxies are
comparable or smaller than the simulation error
bars~\cite{Gaztanaga95}.

As can be seen in Fig.~\ref{s34apm}, APM measurements are somewhat
below the PT predictions or $N$-body results at $\theta \ga 1$
degrees~\cite{Bernardeau95}, indicating possibly a slight bias for APM
galaxies. But note that this difference is not very significant given the
errors and the fact that there is a strong covariance and a significant
negative bias on these scales (see section 4.1 in ~\cite{GaBe98}).
At smaller angles, $\theta \la 1$ degree, the N-body results are
clearly higher than either PT (due to non-linear evolution) or the
real APM results (see also top-left panel in Fig.~\ref{q3panel} for the
corresponding result for the three-point function).  The latter is
likely due to galaxy biasing operating at small scales~\cite{BaGa96},
as discussed in the last section.

Estimation of higher-order moments from the EDSGC~\cite{SMN96} up to
$p=8$ are in good agreement within the errors with APM on scales
$\theta \ga 0.1$ degrees.  On smaller scales, $\theta \la 0.1$
degrees, the EDSGC estimates are significantly larger than the APM
values, indicating systematic problems in the deblending of crowded
fields~\cite{SzGa98}\footnote{~Measurements in this paper were done
with an infinite oversampling technique~\cite{Szapudi98a}.  In
general, results without significant oversampling could underestimate
$S_{p}$ (see also \cite{HuGa99}) but this does not explain the
difference with the APM analysis, where the oversampling was
adequate.}.  The DeepRange results~\cite{SPLO01} for the corresponding
APM slice ($I_{AB}=17-18$) give values of $S_3$ and $S_4$ which are
intermediate between the APM and the EDSGC. This is also the case for
the R INT-WFC catalog~\cite{RoEa99}.  At larger scales, on the other
hand, they both give slightly smaller results.  This is not a very
significant deviation but might indicate that the DeepRange survey is
not large enough at this bright end and it therefore suffers from the
same biases that are apparent when the APM $S_p$ estimations are split
in its $6\times 6$ square degree plates.  For the fainter slices the
DeepRange results are less subject to volume effects and seem to
indicate smaller values of $S_3$ and $S_4$~\cite{SPLO01} as a function
of depth (see Fig.~\ref{s3z_SPLO} below).  Finally, we note also that
the skewness has been estimated for radio sources in the FIRST
survey~\cite{MMLW99} (see also~\cite{CrKa98} for measurements of the
angular correlation function), giving values $s_{3}=1-9$ for a depth
corresponding to 1-50 Mpc/h, approximately.

\subsubsection{Constraints on Biasing and Primordial Non-Gaussianity}

\label{sec:biagau}

Galaxy biasing and primordial non-Gaussianity can leave significant
imprints in the structure of the correlation hierarchy, as discussed
in detail in Sect.~\ref{sec:bias} and Sects.~\ref{ngic}
and~\ref{ngic2}, respectively.  These effects are best understood at
large scales, where PT applies and simple arguments such as local
galaxy biasing (see e.g. Sect.~\ref{genres}) are expected to hold. 
The APM survey is at present the largest angular survey probing scales
in the weakly non-linear regime, thus most constraints on biasing and
primordial non-Gaussianity from angular clustering have been derived
from it.  For constraints derived from galaxy redshift surveys see
Sect.~\ref{sec:biagau2}.

The lower-left panel in Fig.~\ref{q3panel} shows the linear prediction
(dotted lines), corresponding to the projection of
Eq.~(\ref{zetaI})~\cite{Peebles99b}, for $\chi^2$ initial conditions
(see Sect.~\ref{ngic}) with the APM-like initial
spectrum~\cite{FrGa99}.  Although the error bars are large and highly
correlated, the projected three-point function for this model is
substantially larger than that of the APM measurements and the
corresponding Gaussian model for intermediate $\alpha$.  This may seem
only a qualitative comparison, since as discussed in Sect.~\ref{ngic},
non-linear corrections for this model are very significant even at
large scales.  However, non-linear corrections lead to even more
disagreement with the data: although the shape dependence resembles
that of the Gaussian case, the amplitude of $q_{3}$ when non-linear
corrections are included becomes even larger than the linear result,
especially at colinear configurations (see Fig.~\ref{figiso}).

This is also in agreement with~\cite{GaMa96}, who used the deprojected
$S_p$ from the APM survey~\cite{Gaztanaga94} to constrain non-Gaussian
initial conditions from texture topological defects~\cite{TuSp91}
which, as in the case of the $\chi^{2}$ model, also have dimensional
scaling $\xi_{N} \sim B_{N} \xi_{2}^{N/2}$, with $B_{3} \approx B_{4}
\approx 0.5$ (see Fig.~\ref{S3text}).  In this case it was found that
N-body simulations of texture-type initial conditions lead to a
significant rise at large scales in the $S_{p}$ parameters not seen in
the APM data, even when including linear and non-linear (local) bias
to match the amplitude of $S_{p}$ at some scale.

Constraints on a non-local biasing model from the APM $S_p$ parameters
were considered in~\cite{GaFr94}.  The model of cooperative galaxy
formation~\cite{BCFW93}, where galaxy formation is enhanced by the
presence of nearby galaxies, was suggested to produce a
scale-dependent bias to create additional large-scale power in the
standard CDM model and thus match the APM angular correlation
function.  However, the effect of this scale-dependence bias is to
imprint a significant scale dependence on the $S_p$ parameters that is
ruled out by the APM measurements (see also Fig.~\ref{Qk_pps}
below).

The upper right panel in Fig.~\ref{q3panel} shows the PT predictions
for the APM-like initial power spectrum, Eq.~(\ref{linearAPM}), with
linear bias parameter $b_1=2$ (dashed curve) and a non-linear (local)
bias model [see e.g. Eq.~(\ref{eq:taylor}] with $b_1=1$, $b_2=-0.5$
(dotted curve).  Even if the errors are $100\%$ correlated, these
models are in disagreement with the APM data.  A more quantitative
statement cannot be made about constraints of bias parameters from the
APM higher-order moments since a detailed analysis of the covariance
matrix is required.  However, for linear bias the measurements imply
that APM galaxies are unbiased to within $20-30\%$~\cite{FrGa99}. 
These constraints agree well with the biasing constraints obtained
from the inflection point of the reconstructed $\xi_2(r)$ in the APM
~\cite{GaJu01}.  On the other hand, consideration of non-linear
biasing can open up a wider range of acceptable linear bias
parameters~\cite{FrGa94,GaFr94,VHM00}.

\begin{figure}
\centering \centerline{\epsfysize=10.truecm 
\epsfbox{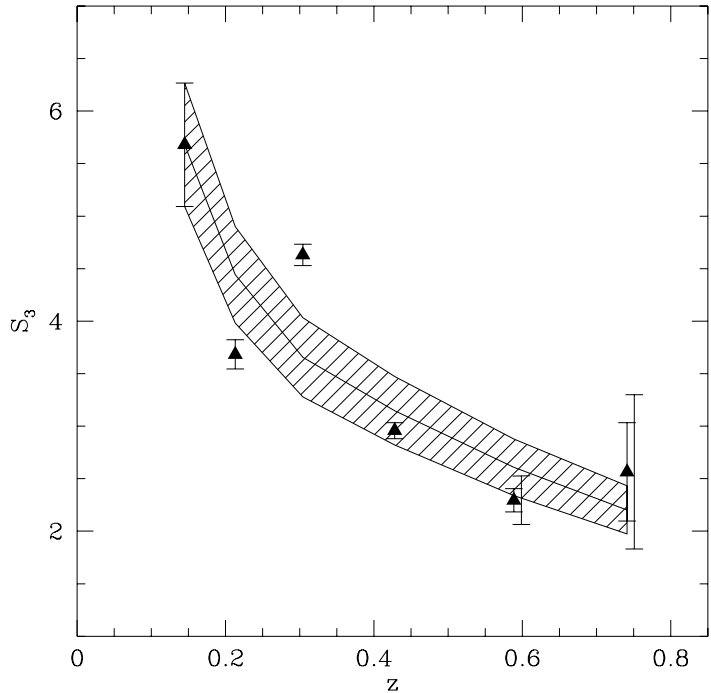}} \caption[junk]{The solid symbols display $s_3$
measured at $0.04^\circ$ for 6 magnitude slices ($I_{AB}=17-18, 18-19,
19-20, 20-21, 21-22, 22-22.5$, corresponding to increasing mean
redshift) of the DeepRange catalog (from~\cite{SPLO01}).  Each value
of $s_3$ is plotted at the median $z$ of the slice.  The shaded band
shows the predictions of a model of galaxy bias evolution, see text
for details.  The right-shifted error bars for the two faintest
measurements include errors due to star/galaxy
separation~\cite{SPLO01}.}
\label{s3z_SPLO}
\end{figure}

An alternative to wide surveys which probe the weakly non-linear
regime at recent times, deep galaxy surveys can probe the redshift
evolution and also reach weakly non-linear scales at high redshift. 
Although presently this is not possible due to the small size of
current deep surveys, it will become so in the near future.  An early
application along these lines is in Fig.~\ref{s3z_SPLO}, which shows
the redshift evolution of $s_3$ for measurements of~\cite{SPLO01} in
the DeepRange catalog at a fixed angular scale of $0.04^\circ$.  This
corresponds to about $0.3 \Mpc$ at $z \simeq 0.15$ and $1.5 \Mpc$ at
$z \simeq 0.75$, so the scales involved are in the non-linear
regime\footnote{~Although a fixed angular scale does not correspond to
a fixed spatial scale as a function of $z$, the comparison is
meaningful because the measured $s_{3}(\te)$ is scale independent
(hierarchical) to a good approximation.}.

The redshift evolution in Fig.~\ref{s3z_SPLO} is just the opposite of
that expected in generic (dimensional) non-Gaussian models, where the
skewness $s_{3}$ should increase with redshift (see e.g. discussion in
Sect.~\ref{ngic2}).  However, since these scales are in the non-linear
regime the predictions based on PT cannot be safely used, and galaxy
biasing can behave in a more complicated way.  In any case, the trend
shown in Fig.~\ref{s3z_SPLO} can be matched by a model, shown in a
shaded band, where $S_3(z)= S_3(0)\, (1+z)^{-0.5}$~\cite{SPLO01}, which
may indicate that galaxy bias is increasing with redshift, as expected
in standard scenarios of galaxy formation (see discussion in
Sect.~\ref{sec:bias}), and contrary to the evolution expected from
strongly non-Gaussian initial conditions.  A more quantitative
constraint will have to await the completion of future deep surveys
that can probe the weakly non-linear regime.

\subsection{Results from Redshift Galaxy Surveys}

\label{sec:redsur}

\subsubsection{Redshift Catalogs}

\label{sec:zcat}

Redshift surveys map the three-dimensional distribution of galaxies in
a large volume, and are thus ideally suited to use higher-order
statistics to probe galaxy biasing and primordial non-Gaussianity. 
Table~\ref{tab:zcat} shows a list of the main wide-field redshift
catalogs .  For a more general review on redshift catalogs
see~\cite{Oort83,GiHa91,StWi95,Strauss99}.

\begin{table}
\centering \bigskip
\def\ss{\rule[0ex]{0mm}{3.5ex}\rule[-1.75ex]{0mm}{3.5ex}\\*}
\def\df{\dotfill} \def\hf{\hfill} \caption{Optical and infrared (last
four) redshift catalogs.  The survey area $\Omega$ is given in
stereo-radians, the depth and effective size $ {\cal D}_E \equiv
(\Omega/4\pi)^{1/3} 2 {\cal D}$ are in Mpc/h.  } 
\begin{tabular}{c c c c c c c } \hline
\rule[-0.6ex]{0mm}{3.2ex} Name & Area $\Omega$ & magnitudes & 
Depth ${\cal D}$ & $ {\cal D}_E $& \# gal/ster & Ref \ss \hline 
CfA & $ 1.8 $ ster & $m_Z < 14.5$ & 50 & 52& $\sim 1000$ & \cite{HDLT83} \ss 
SSRS & $ 1.8 $ ster & $D(0) > 0.1$ &50 & 52 & $\sim 1000$ & \cite{DPDMST91} \ss 
PPS & $\sim 1$ ster & $m >15.5-15$ & 80 & 70 & $\sim 3000$ & \cite{GiHa91} \ss 
LCRS & $ 0.02 $ ster & $R< 17.8 $ & 300 & 70 & $1.3\times 10^{6}$ & \cite{SLOTLKS96} \ss
Stromlo-APM & $ 1.3 $ ster & $ b_J <17.15$ & 150 & 140 & 1400 & \cite{LPME96} \ss
Durham/UKST & $ 0.45 $ ster & $b_J < 17 $ & 140 & 90 & 5500 & \cite{RSPBWOCF98} \ss
2dFGRS      & $0.6 $ ster   & $b_J < 19.5$ & 300 & 220 & $\sim 2.5 \times 10^5$ & \cite{Norberg012dF} \ss 
SDSS        & $\simeq 3 $ ster &  $r'< 18$ & 275 & 341 & $\sim 10^6$ & \cite{Zehavi01SDSS} \ss
\hline
QDOT & $ 10 $ ster & $f_{60\mu m} > 0.6 Jy$ &90 & 170 & 245 &\cite{EKSLREF90} \ss 
IRAS 1.9Jy & $ 9.5 $ ster &$f_{60\mu m} > 1.9 Jy$ & 60 & 110 & 220 & \cite{SHDYFT92} \ss 
IRAS 1.2Jy & $ 9.5 $ ster & $f_{60\mu m} > 1.2 Jy$ & 80 & 145 & 480 &\cite{FHSDYS95} \ss
PSCz & $ 10.5 $ ster & $f_{60\mu m} > 0.6 Jy$ & 100 & 188 & 1470 &\cite{SSMKORMETWFCH00} \ss 
\hline
\end{tabular}  \label{tab:zcat}
\end{table}

Redshift surveys require a predefined sample of targets to obtain
redshifts, therefore they are often defined from angular surveys where
galaxies are detected photometrically.  Below we shortly discuss the
main characteristics of the surveys in Table~\ref{tab:zcat}, for a
brief description of the photometric parent catalogs see
Sect.~\ref{sec:angcat}.

The Center for Astrophysics survey (hereafter \ul{CfA}, \cite{HDLT83})
and the Perseus-Pisces redshift Survey (\ul{PPS}, \cite{GiHa91}) are
both based on the Zwicky catalog.  The CfA survey, perhaps the most
analyzed redshift survey in the literature, consists of 2417 galaxies
with Zwicky magnitudes less than 14.5, covering over $2.67$ ster 
($1.8$ ster in the North Galactic cap) with a
median redshift corresponding to 3300 km/sec.  The PPS survey,
centered around the Perseus-Pisces supercluster, contains over 3000
galaxies.  The Southern Sky Redshift Survey (hereafter \ul{SSRS},
\cite{DPDMST91}) is based on the ESO/Uppsala angular sample, and
contains about 2000 galaxies.  These surveys suffer from the same
calibration problems as their parent catalogs, but with redshift
information they were aimed to represent a fair sample of the
universe. Recent  extensions of these surveys to
deeper magnitudes ($m<15.5$, $2000$ redshift, ${\cal D} \simeq 80 Mpc/h$) 
are denoted by CfA2 and SSRS2 and have been merged into the Updated
Zwicky Catalog (UZC, \cite{FKGPBMTE99}).

The \ul{LCRS} \cite{SLOTLKS96}, consists of redshifts selected from a
well calibrated CCD survey of 6 narrow $1.5\times 80$ degrees strips
in the sky.  Although this survey is much deeper and better calibrated
than any of the previous ones, it is also potentially subject to
important selection and boundary effects: narrow slices,
density-dependent sampling (because of a constant number of fibers per
field) and the exclusion of galaxies closer than $55''$.  All these
effects tend to underweight clusters and, even if properly corrected,
could introduce important sampling biases in higher-order
statistics\footnote{~For example, it is impossible to recover any lost
configuration dependence of correlation functions in the non-linear
regime by a correction procedure, since the correcting weight for lost
galaxies would have to decide whether they were aligned or
isotropically distributed.}.

The \ul{Stromlo-APM} redshift survey (\cite{LPME96}) consists of 1790
galaxies with $b_J<17.15$ selected randomly at a rate of 1 in 20 from
APM scans in the south Galactic cap.  The \ul{Durham/UKST} galaxy
redshift survey (\cite{RSPBWOCF98}) consists of 2500 galaxy redshifts
to a limiting apparent magnitude of $b_J=17$, covering a 1500 sq deg
area around the south galactic Pole.  The galaxies in this survey were
selected from the EDSGC and were sampled, in order of apparent
magnitude, at a rate of one galaxy in every three.

The \ul{IRAS Point Source Redshift Catalog} (hereafter \ul{PSCz},
\cite{SSMKORMETWFCH00}) is based on the IRAS Point Source Catalog (see
\cite{CBC87}), with several small additions applied to achieve the
best possible uniformity over the sky.  The survey objective was to
get a redshift for every IRAS galaxy with 60 micron flux $f_{60}>0.6$
Jy, over as much of the sky as possible.  Sky coverage is about $84\%$
with 15411 galaxies.  Earlier subsamples of PSCz include the updated
\ul{QDOT} catalog~\cite{EKSLREF90}, the \ul{IRAS
1.9Jy.}~\cite{SHDYFT92} and the \ul{IRAS 1.2Jy.}~\cite{FHSDYS95}
redshift surveys.  The QDOT survey chooses at random one in six
galaxies from PSCz, leading to 1824 galaxies with galactic latitude
$|b|>10^o$.  The other subsamples are shallower but denser than QDOT;
the 2Jy catalog, complete to a flux limit $f_{60}>2$Jy.  contains 2072
galaxies, whereas the 1.2Jy.  catalog, with $f_{60}>1.2$ contains 4545
galaxies.  IRAS galaxies are mostly biased towards spiral galaxies
which tend to undersample rich clusters.  Thus IRAS galaxies are both
sparser and a biased sample of the whole galaxy population.

\medskip

The \ul{Sloan Digital Sky Survey} (SDSS, see e.g.~\cite{York00}) and the
two degree field \ul{2dF Galaxy Redshift Survey} (2dFGRS, see
\cite{CDMSNC01}) were still under construction when this review was
written and only preliminary results are known at this stage.  These
results are discussed in section \ref{sec:outlook}.

\medskip

Other recent redshift surveys for which there is not yet measurements
of higher-order statistics include the Canada-France Redshift
survey~\cite{LLCHT95}, the Century survey~\cite{GKWTFMHSF97}, the ESO
Slice Project~\cite{ESO98}, the Updated Zwicky
Catalog~\cite{FKGPBMTE99} and the CNOC2 Field galaxy
survey~\cite{CNOC299}.

\subsubsection{Two-Point Statistics}

\label{sec:2ptredsh}

We now briefly discuss results on two-point statistics from redshift
surveys, with emphasis on the power spectrum.  We first address
optical surveys and then infrared surveys.

The analysis of the redshift-space correlation function in the CfA
survey~\cite{DaPe83} found that, after integration over the parallel
direction to project out redshift distortions, the resulting two-point
function agreed with that derived from inversion in angular catalogs,
Eq.~(\ref{xi2r0}), with $\gamma \simeq 1.77$ and $r_{0}=5.4\pm 0.3$
Mpc/h, for projected separations $r_{p}<10$ Mpc/h.  At larger scales,
the redshift-space correlation function estimates become steeper and
there was marginal evidence for a zero crossing at scales larger than
about 20 Mpc/h.\footnote{~The measured redshift 2-point function
will be found to be flatter than the real space one, with more power on
large scales and less power on smaller scales, as expected from 
theory (see Sect.\ref{sec:reddis}), with evidence for a larger correlation
length in redshift space, $s_0>r_0$, in all CfA, SSRS and IRAS catalogues 
~\cite{FrGa94b}.}.
Modeling the redshift-space correlation function as a
convolution of the real-space one with an exponential pairwise
velocity distribution function\footnote{~An exponential form was first
suggested in~\cite{Peebles76b} and has since been supported by
observations, see e.g.~\cite{LSB98} for a recent method applied the
LCRS survey.  The interpretation of this technique, however, rests on
the assumption of a scale-independent velocity dispersion, which seems
consistent in LCRS~\cite{JMB98}, but may not necessarily be true in
general, see e.g.~\cite{HaTe00b,JBS01} for the PSCz survey. 
Theoretically, exponential distributions arise from summing over
Gaussian distributions, both in the weakly and highly non-linear
regimes, see~\cite{JFS98} and~\cite{Sheth96a,DiGe96} respectively. 
These results are also supported by N-body
simulations~\cite{FDSYH94b,ZQSW94}.} with velocity dispersion
$\sigma_{v}$, \cite{DaPe83} obtained that $\sigma_{v}=340 \pm 40$
km/sec at $r_{p}=1$ Mpc/h, well below the predictions of CDM models.

These results were extended a decade later with the analysis of the
power spectrum in the extension of the CfA survey to $m_{Z}<15.5$. 
In~\cite{VPGH92} it was shown that, in agreement with previous results
from the APM survey~\cite{MESL90} and IRAS
galaxies~\cite{EKSLREF90,SFRLE91}, the standard CDM model was
inconsistent with the large-scale power spectrum at the 99\%
confidence level.  In addition, \cite{PVGH94} studied the relation
between the real space and redshift space power spectrum in CDM
simulations, using the Eq.~(\ref{Ppheno}), and showed that agreement
between the small-scale power spectrum and $\Gamma=0.2$ CDM models
required a velocity dispersion parameter $\sigma_{v}\approx 450$
km/sec, somewhat larger than the value obtained by modeling the
two-point function in~\cite{DaPe83}.  A joint analysis of the CfA/PPS
power spectrum gave a best fit CDM shape parameter $\Gamma=0.34\pm
0.1$~\cite{BaFr91}.  Similarly, a joint analysis of the CfA/SSRS
samples in~\cite{dVGHP94} showed a power spectrum consistent with CDM
models with $\Gamma \approx 0.2$ and bias within 20\% of unity when
normalized to COBE~\cite{COBE92a,COBE92b} CMB fluctuations at the
largest scales.  A recent analysis~\cite{PTH01} of the redshift-space
large-scale ($k \la 0.3$ h/Mpc) power spectrum of the Updated Zwicky
catalog~\cite{FKGPBMTE99}, which includes CfA2 and SSRS, was done
using the quadratic estimator and decorrelation techniques (see
Sects.~\ref{s:QuadEst}-\ref{sec:KL}).  The measurements in different
subsamples are well fit by a $\Lambda$CDM model with normalization
$b_{1} \sigma_{8}=1.2-1.4$.

The analysis of the LCRS redshift space power spectrum was done
in~\cite{LKSLOTS96}, where they used Lucy's method~\cite{Lucy74} to
deconvolve the effects of the window of the survey, which are
significant given the nearly two-dimensional geometry.  They obtained
results which were consistent with previous analyses of the CfA2 and
SSRS surveys.  An alternative approach was carried out
in~\cite{LSLKOT96}, where they estimated the two-dimensional power
spectrum, which was found to have a ``bump'' at $k = 0.067$ Mpc/h with
amplitude a factor of $\approx 1.8$ larger than the smooth best fit
$\Gamma=0.3$ CDM model.  This is reminiscent of similar features seen
in narrow deep ``pencil beams'' redshift surveys, e.g.
\cite{BEKS90}\footnote{~See e.g.~\cite{KaPe91,PaGo91} and the recent
analysis in~\cite{YCWEMCJFPEPT01} for a discussion of the statistical
significance of these features.}.  A recent linear analysis of the
LCRS survey~\cite{MSL00} using the KL transform methods (see e.g.
Sect.~\ref{sec:KL}), parameterized the power spectrum in redshift space
by a smooth CDM model, and obtained a shape parameter $\Gamma=0.16 \pm
0.10$, and a normalization $b_{1} \sigma_{8}=0.79\pm 0.08$.

The two-point correlation function of LCRS galaxies was measured
in~\cite{TOKLSLSMGE97,JMB98}, and integrated along the line of sight
to give the projected correlation function in real space, which was
found to agree with Eq.~(\ref{xi2r0}), with $\gamma \simeq 1.86 \pm
0.04$ and $r_{0}=5.06\pm 0.12$ Mpc/h~\cite{JMB98}.  After modeling the
pairwise velocity distribution function by an exponential with
dispersion and mean (infall) velocity, the inferred pairwise velocity
at 1Mpc/h was found to be $\sigma_{v}=570\pm 80$ km/sec, substantially
higher compared to other surveys.  In fact, another analysis of the
LCRS survey in~\cite{LSB98} found a pairwise velocity dispersion of
$\sigma_{v}=363 \pm 44$ km/sec, closer to previous estimates.  In this
case, the deconvolution of the small-scale redshift distortions was
done by a Fourier transform technique, assuming a constant velocity
dispersion and no infall [i.e. negligible $\vu_{12}$, see
Eq.~(\ref{v12})].  At least part of this disagreement can be traced to
the effects of infall, as shown in~\cite{JiBo98b}.  For other recent
methods and applications to determining the small-scale pairwise
velocity dispersion and infall see e.g.~\cite{DMW97}
and~\cite{JFFJD00}, respectively.

Results from the power spectrum of the Stromlo-APM
survey~\cite{LEMP96,TaEf96}, the Durham/UKST survey~\cite{HBSR99} and
the ESO Slice Project~\cite{CBMZG01} are in agreement with previous
results from optically selected galaxies, and show an amplification
compared to the power spectrum of IRAS galaxies implying a relative
bias factor $b_{\rm opt}/b_{\rm iras} \approx 1.2-1.3$. This is 
reasonable, since IRAS galaxies are selected in the infrared and are 
mostly spiral galaxies which, from the observed morphology-density 
relation~\cite{Dressler80,PoGe84}, tend to avoid clusters. We shall 
come back to this point when discussing higher-order statistics.

The first measurements of counts-in-cells in the QDOT
survey~\cite{EKSLREF90,SFRLE91} showed that IRAS galaxies were more
highly clustered at scales of 30-40 Mpc/h compared to the predictions
of the standard CDM model, in agreement with the angular correlation
function from APM~\cite{MESL90}.  The QDOT power spectrum was later
measured in~\cite{FKP94} using minimum variance weighting, giving
redshift-space values $\sigma_{8}=0.87 \pm 0.07$ and $\Gamma=0.19\pm
0.06$.  Measurement of the power spectrum of the 1.2Jy
survey~\cite{FDSYH93} confirmed and extended this result, although it
showed somewhat less power at large scales than QDOT\footnote{~It was
later shown that the QDOT measurements were sensitive to a small
number of galaxies in the Hercules
supercluster~\cite{Efstathiou95,TaEf95}, which was over-represented in
the QDOT sample presumably due to a statistical fluctuation in the
random numbers used to construct the survey.}.  The measurement of the
two-point function in redshift space for the 1.2Jy
survey~\cite{FDSYH94,FDSYH94b} implied a real space correlation
function as in Eq.~(\ref{xi2r0}), with $\gamma \simeq 1.66 $ and
$r_{0}=3.76$ Mpc/h for scales $r \la 20$ Mpc/h, consistent with the
fact that IRAS galaxies are less clustered than optically selected
galaxies.  In addition, the inferred velocity dispersion at 1Mpc/h was
$\sigma_{v}=317^{+40}_{-49}$ km/sec.

\begin{figure} 
\centering{\epsfxsize=12.truecm \epsfbox{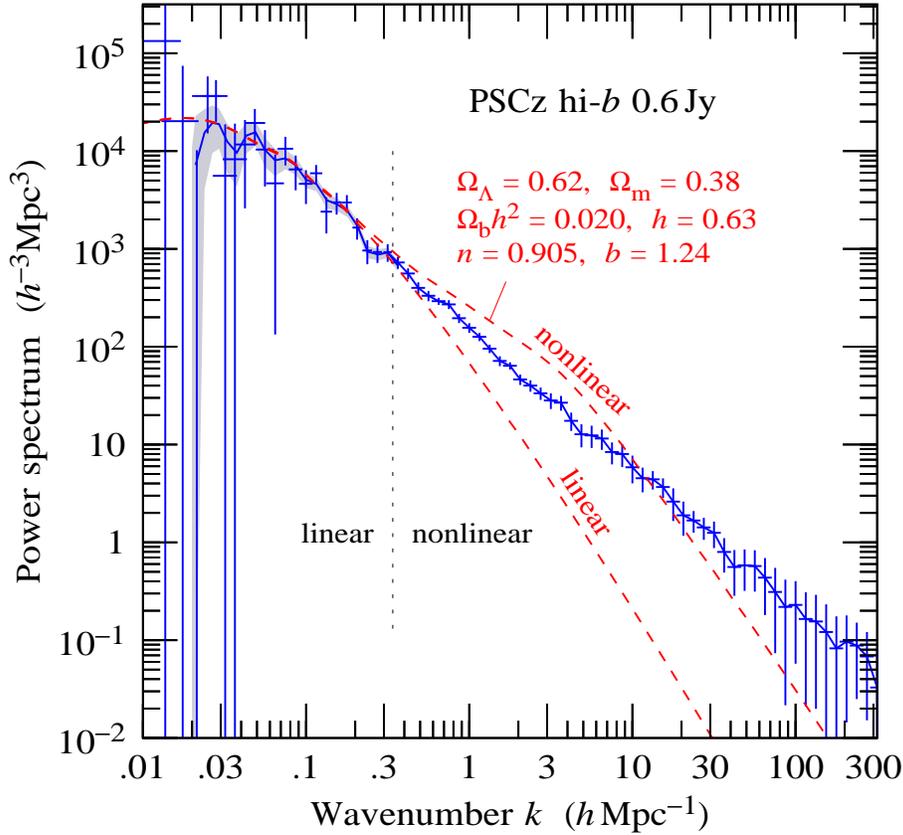}} 
\caption{The {\em real space} power spectrum of PSCz galaxies.  To the
left of the vertical line is the linear measurement of~\cite{HTP00}
(points with uncorrelated errorbars~\cite{HaTe00a}), while to the
right is the nonlinear measurement of~\cite{HaTe00b} (points with {\em
correlated} error bars). The dashed line corresponds to the flat 
$\Lambda$CDM concordance model power spectrum from~\cite{TZH01} with 
parameters as indicated, nonlinearly evolved according to the 
prescription in~\cite{PeDo96}. (from~\cite{HaTe00b})}
\label{PkPSCz}
\end{figure}

Measurements in the PSCz survey are currently the most accurate
estimation of clustering of IRAS galaxies.  At large scales, the power
spectrum is intermediate between that of QDOT and 1.2Jy surveys,
whereas at smaller scales it decreases slightly more
steeply~\cite{STEFKMMORSW99}.  The shape of the large-scale power
spectrum is consistent with $\Gamma=0.2$ CDM models, although it does
not strongly rule out other
models~\cite{STEFKMMORSW99,TBTHESFKMMORSW99}.  A comparison with the
Stromlo-APM survey shows a relative bias parameter of $b_{\rm
Stromlo}/b_{\rm PSCz} \approx 1.3$ and a correlation coefficient
between optical and IRAS galaxies of $R \geq 0.72$ at the 95\%
confidence limit on scales of the order of 20
Mpc/h~\cite{SSTEFKMMORSW99}.  These results were considerably extended
in~\cite{HaTe00b} to obtain the power spectrum in {\em real space} by
measuring the redshift-space power perpendicular to the line of sight
and parameterizing the dependence on non-perpendicular modes to
increase signal to noise.  The resulting power spectrum is reproduced
in Fig.~\ref{PkPSCz}.  It shows a nearly power-law behavior to the
smallest scales measured, with no indication of an inflexion at the
non-linear scale, and no sign of turnover at the transition to the
stable clustering regime.  Compared to the best fit CDM model
(obtained from a joint analysis with CMB fluctuations in~\cite{TZH01}
and shown as a dashed line), the PSCz requires a significant
scale-dependent bias. 

Finally, we briefly mention results on the parameter $\beta \approx
\Omega^{0.6}/b_{1}$ from measurements of the anisotropy of the power
spectrum in redshift space\footnote{~For an exhaustive review of these
results up to mid 1997 see~\cite{Hamilton98}.  } (see
Sect.~\ref{pks}).  These measurements are complicated by the fact that
surveys are not yet large enough to see a clear transition into the
linear regime predictions, Eq.~(\ref{mult_l}).  In addition, different
methods seem to give somewhat different answers~\cite{Hamilton98};
however, the average and standard deviation of reported values
are~\cite{Hamilton98} $\beta_{\rm opt}=0.52\pm 0.26$ and $\beta_{\rm
iras}=0.77 \pm 0.22$ for optically selected and IRAS galaxies,
respectively, which is roughly consistent with the relative bias
between these two populations.  On the other hand, the most recent
results from the PSCz survey find $\beta=0.39 \pm 0.12$~\cite{TBHT00},
and $\beta=0.41^{+0.13}_{-0.12}$~\cite{HTP00}.  Constraints from the
most recent optically selected surveys are considerably noisier, e.g.
Stromlo-APM does not even exclude $\beta \sim 1$~\cite{LEMP96,TaEf96},
and LCRS is consistent with no distortions at all,
$\beta=0.30\pm0.39$~\cite{MSL00}.  Resolution of these issues will
have to await results from the full-volume 2dFGRS and SDSS surveys
(see also Sect.~\ref{sec:outlook}).

\subsubsection{Three-Point Statistics}

\label{sec:q3z}

Determination of three-point statistics from redshift surveys has been
carried out mostly in the non-linear regime for optically selected 
surveys, and mostly in the weakly non-linear regime for IRAS surveys. 
Table~\ref{tab:q3z} shows different estimates of the three-point
function (top list) and the bispectrum (bottom list). 

As discussed before, the CfA sample covers a small volume to be a fair
estimate of higher-order correlations.  Even more so, estimates in the
Durham-AAT and KOSS samples are subject to large estimator biases as
they have only a few hundred redshifts.  Nonetheless, these results
roughly agree with each other, although the values of $Q_{3}$ are seen
to fluctuate significantly.  Note that the values in
Table~\ref{tab:q3z} are not directly comparable to those inferred from
deprojection of angular catalogs (Table~\ref{q3q4}) as they are
affected by redshift distortions (see e.g. Fig.~\ref{bispz12}).

The LCRS survey provides the best estimate to date of the three-point
function at small scales~\cite{JiBo98}.  Estimation of $Q_{3}$ in
redshift space and in projected space (by integrating along the line
of sight) showed values lower by a factor of about 2 than $\Lambda$CDM
simulations where clusters have been underweighted by $m^{-0.08}$,
essentially equivalent to assuming that the number of galaxies as a
function of dark matter halo mass $m$ scales as $N_{\rm gal}(m) \sim
m^{0.9}$ in the notation of Sect.~\ref{dhbias}.  The authors conclude
that the hierarchical model is not a good description of the data,
since they see some residual (small) scale and configuration
dependence.  However, as discussed at the end of Sect.~\ref{sec:Qs},
one {\em does not} expect the hierarchical model to be a good
description for correlations in redshift-space since velocity
dispersion creates ``fingers of god'' along the observer's
direction~\cite{SCF99}.  The fact that these are clearly seen by
visual inspection of the galaxy distribution ought to show up in a
clear shape dependence of the three-point function: colinear
configurations should be significantly amplified (see
Fig.~\ref{bispz12}).  Surprisingly, this is not seen in the LCRS
measurements~\cite{JiBo98}.

\begin{table}
\centering 
\caption[junk]{Some measurements of $Q_3$ in redshift catalogs.  In
most cases, only the mean values over a range of scales were
published.  In cases where measurements of the individual values for
each scale are reported in the literature, we quote the actual range
of estimates over the corresponding range of scales.  The top half of
the table is in configuration space, the bottom part in Fourier space. 
Scales are in Mpc/h and h/Mpc, respectively.  When possible, we give
estimates for equilateral (eq) and colinear (col) configurations. 
Error bars should be considered only as rough estimates, see text for
discussion.}
\begin{tabular}{ccccc}
\hline   \rule[-0.6ex]{0mm}{3.2ex} 
{$3~Q_3$} &  {Scales} & {Sample} & Year & {Ref} \\* \hline
$2.4 \pm 0.2$ &  --- & CfA & 1980 &\cite{Peebles80} (eq.[57.9]) \\* 
$2.04 \pm 0.15$ &  --- &\cite{Rood82} & 1981 & \cite{Peebles80} \\* 
$ 2.4 \pm 0.3$ &  1-2 & CfA & 1984 & \cite{EfJe84} \\* 
$ 1.8 \pm 0.2$ &  1-3 &Durham-AAT & 1983 & \cite{BESEP83} \\* 
$ 3.9 \pm 0.9$ &  1-2 & KOSS & 1983 & \cite{BESEP83} \\* 
$ 1.5-4.5$ &  1-8 & LCRS & 1998 & \cite{JiBo98} \\*
\hline
$Q_{\rm eq} \simeq 0.5$ &  0.1-1.6  & CfA/PPS & 1991&\cite{BaFr91} \\* 
$Q_{3} \approx 1$ &  0.05-0.2  & QDOT & 2001 &\cite{SFFF01} \\* 
$Q_{\rm eq} \simeq 0.2$; $Q_{\rm col} \simeq  0.6$ &  0.05-0.2  & IRAS 1.9Jy & 2001 &\cite{SFFF01} \\* 
$Q_{\rm eq} \simeq 0.4$; $Q_{\rm col} \simeq  0.8$ &  0.05-0.2  & IRAS 1.2Jy & 2001 &\cite{SFFF01} \\* 
$Q_{\rm eq} \simeq 0.4$; $Q_{\rm col} \simeq  1.4$ &  0.05-0.4  & PSCz & 2001 &\cite{FFFS01} \\* 
\hline
\end{tabular} \label{tab:q3z}
\end{table}

\begin{figure} 
\centering{\epsfxsize=12.truecm \epsfbox{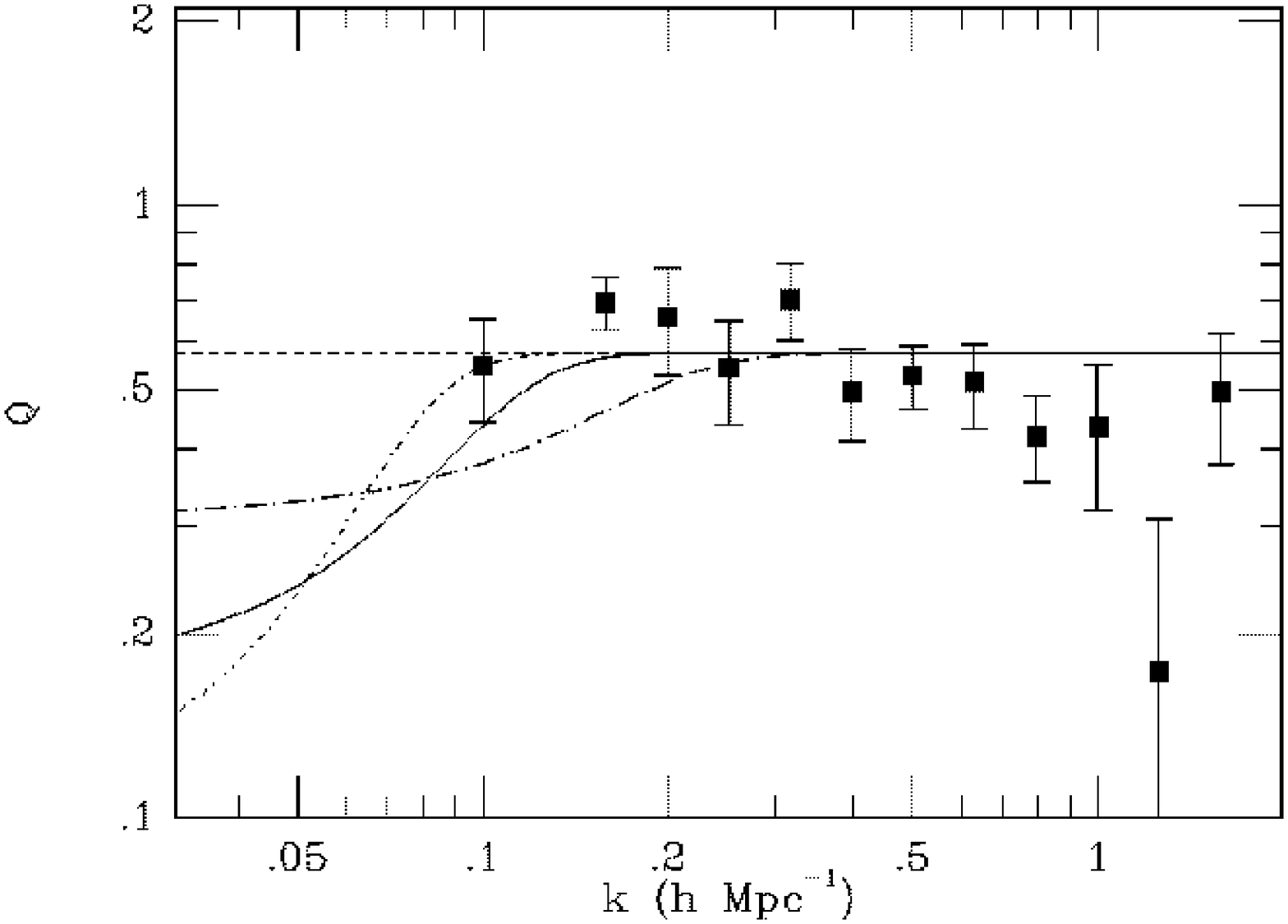}} \caption{ The
redshift-space reduced bispectrum $Q_{\rm eq}$ for equilateral
triangles as a function for scale $k$, for CfA/PPS galaxies (from
\cite{BaFr91}).  The dashed line shows the PT prediction: $Q_{\rm eq}
=4/7$.  The other lines show predictions for the cooperative galaxy
formation models, see Sect.~\ref{sec:biagau2}.}
\label{Qk_pps}
\end{figure}

Measurements of the bispectrum (for equilateral configurations) in
redshift space were first carried out for the CfA survey and a sample
of redshifts in the Pisces-Perseus super-cluster~\cite{BaFr91}.  This
was the first measurement that reached partially into the weakly
non-linear regime and compared the bispectrum for equilateral
configurations with PT predictions, $Q_{\rm eq}=4/7$.  As shown in
Fig.~\ref{Qk_pps} the agreement with PT predictions is very good, even
into the non-linear regime\footnote{~This is due to accidental
cancellations in redshift space.  At larger $k$'s, in the absence of
redshift distortions, $Q_{\rm eq}(k)$ increases, see e.g.
Fig.~\ref{fig_1LBeq_sf}.  However, velocity dispersion suppresses this
rise, resulting in approximately the same value as in PT~\cite{SCF99}. 
The same is not true for colinear configurations, see
Fig.~\ref{bispz12}.}.  The errors bars in each bin indicate the
variance among different subsamples, 3 from the CfA and 3 from the
Perseus-Pisces surveys .  This result was interpreted as a support for
gravitational instability from Gaussian initial conditions and in
disagreement with models of threshold bias~\cite{BBKS86,JeSz86}, which
predicted $Q_{3} \sim 1$.  The results in Fig.~\ref{Qk_pps} were later
used in~\cite{FrGa94} to constrain models of non-local bias that had
been proposed to give galaxies extra large-scale power in the standard
CDM scenario (see Sect.~\ref{sec:biagau2} for a discussion).  In
addition, \cite{BaFr91} measured the trispectrum for randomly
generated tetrahedral configurations, showing a marginal detection
with hierarchical scaling consistent with $Q_{4} \sim 1$.

\begin{figure}[t]
\centering \mbox{\epsfxsize=12.cm \epsffile{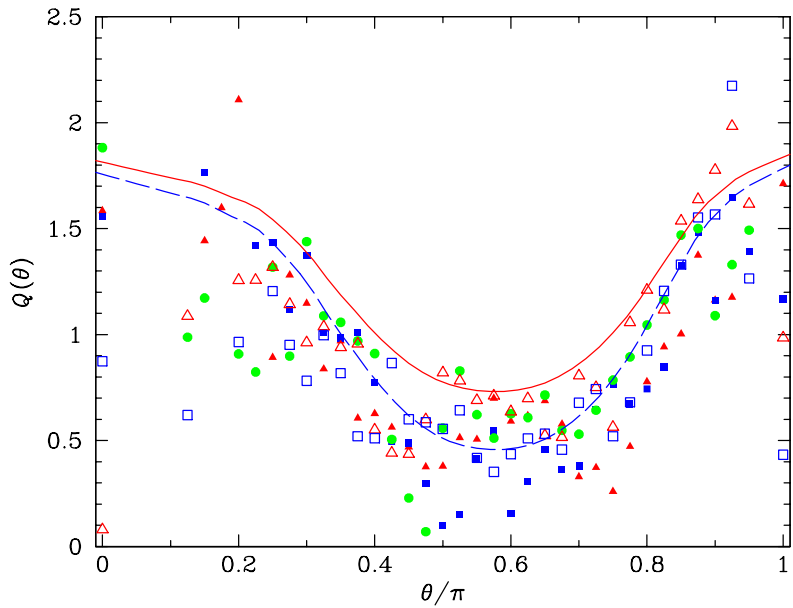}}
\caption{The bispectrum $Q_3$ vs.~$\theta$ for the PSC$z$ catalog for
triangles with $ 0.2 \leq k_1 \leq 0.4$ h/Mpc and with two sides of
ratio $k_2/k_1=0.4-0.6$ separated by angle $\theta$.  The solid curve
shows $Q_3$ in redshift space averaged over many 2LPT realizations of
the $\Lambda$CDM model.  Symbols show results from the PSC$z$ survey
for bands in $k_1$: filled triangles, $k_1=0.20$--$0.24$h/Mpc; filled
squares 0.24--0.28; filled circles 0.28-0.32; open circles 0.32--0.36;
and open squares 0.36--0.42.  The dashed curve shows the 2LPT
prediction for $\Lambda$CDM with the best-fit bias parameters $1/b =
1.20$, $b_2/b^2=-0.42$.  Taken from~\protect\cite{FFFS01}.  }
\label{bispPSCz_theta}
\end{figure}

Detailed measurements of the bispectrum in the weakly non-linear
regime were not done until a decade later, with the analyses of the
IRAS surveys~\cite{SFFF01,FFFS01}, which probe a large enough volume
of roughly spherical shape.  In~\cite{SFFF01}, measurements were done
for the QDOT, 1.9Jy and 1.2Jy surveys.  In order to constrain galaxy
bias and primordial non-Gaussianity, a likelihood method that takes
into account the covariance matrix of the bispectrum for different
triangles and the non-Gaussian shape of the likelihood function (see
e.g. Fig.~\ref{UpsilonPkQ}) was used, developed
in~\cite{Scoccimarro00b}.  This is essential to recover accurate
estimates of errors on bias parameters and primordial non-Gaussianity
without systematic estimator biases due to the finite volume of the
survey\footnote{~A likelihood analysis for analysis of the bispectrum
was first proposed in~\cite{MVH97}, based on the Gaussian
approximation for the likelihood function and a second-order Eulerian
PT calculation of the covariance matrix.  Extensions to redshift space
are given in~\cite{VHMM98}.}.  The results from QDOT were marginal,
due to the very sparse sampling (one galaxy every six) $Q_{3}$ was
only shown to be of order unity without any discernible dependence on
configuration.  The results from 1.9Jy and 1.2Jy showed a systematic
shape dependence similar to that predicted by gravitational
instability.  

These results were considerably extended with the analysis of the PSCz
bispectrum~\cite{FFFS01}.  Figure~\ref{bispPSCz_theta} shows the PSCz
reduced bispectrum $Q_3$ as a function of the angle $\theta $ between
$\vk_{1}$ and $\vk_{2}$ for triangles with $k_1/k_2 \approx 2$ and
different scales as described in the figure caption~\cite{FFFS01}. 
The configuration dependence predicted by gravitational
instability~\cite{Fry84b,HBCJ95} (solid lines for an unbiased
distribution, predicted by 2LPT, see e.g. Fig.~\ref{bispz12}) is
clearly seen in the data.  This is also the case for all triangles,
not just those shown in Fig.~\ref{bispPSCz_theta}, see Fig.~1
in~\cite{FFFS01}.  

Implications of these results for galaxy biasing and primordial
non-Gaussianity are discussed in~Sect.~\ref{sec:biagau2}.

\subsubsection{Skewness, Kurtosis and Higher-Order Cumulants}

\begin{table}
\centering 
\caption[junk]{Some measurements of $S_3$ and $S_4$ in redshift
catalogs.  In most cases, only the mean values over a range of scales
were published.  In cases where measurements of the individual values
for each scale are reported in the literature, we quote the actual
range of estimates over the corresponding range of scales.  In most
cases error bars should be considered only as rough estimates, see
text for discussion.}
\begin{tabular}{cccccc}
\hline   
\rule[-0.6ex]{0mm}{3.2ex}{$S_3$} & {$S_4$} & {Scales} & {Sample} & Year & {Ref} \\* \hline
$2\pm 1 - 6 \pm 4$ & --- & $5 - 20$ & QDOT & 1991 & \cite{SFRLE91} \\*
$1.5 \pm 0.5$ & $4.4 \pm 3.7$ & 0.1-50 & IRAS 1.2Jy & 1992 &\cite{BDS92},\cite{BSDFYH93} \\* 
$1.9 \pm 0.1$ & $4.1 \pm0.6$ & 2-22 & CfA & 1992 & \cite{Gaztanaga92} \\* 
$2.0 \pm 0.1$ & $5.0\pm 0.9$ & 2-22 & SSRS & 1992 & \cite{Gaztanaga92} \\* 
$2.1 \pm 0.3$ & $7.5 \pm 2.1$ & 3-10 & IRAS 1.9Jy & 1994 & \cite{FrGa94}\\* 
$2.4 \pm 0.3$ &$13 \pm 2$ & 2-10 & PPS & 1996 & \cite{GBGGHKP96} \\* 
$2.8 \pm 0.1$ & $6.9 \pm 0.7$ & 8-32 & IRAS 1.2Jy & 1998 & \cite{KiSt98} \\* 
$1.8 \pm 0.1$ & $5.5 \pm 1$ & 1-10 & SSRS2 & 1999 & \cite{BCDMBS99} \\* 
$1.9 \pm 0.6$ &$7 \pm 4$ & 1-30 & PSCz & 2000 & \cite{SBFMS00} \\* 
$1.82 \pm 0.21$ &$\sim 3$ & 12.6 & Durham/UKST & 2000 & \cite{HSB00} \\* 
$2.24 \pm 0.29$ & $\sim 8$ & 18.2 & Stromlo/APM & 2000 & \cite{HSB00}\\* 
\hline
\end{tabular} \label{tab:s3s4z}
\end{table}

Table \ref{tab:s3s4z} shows different estimates for
$S_3=\xibar_3/\xibar_2^2$ and $S_4=\xibar_4/\xibar_2^3$, the ratios of
the cumulants $\xibar_N$ obtained by counts-in-cells.  The shape of
the cells correspond to top-hat spheres, unless stated otherwise.

The QDOT results by~\cite{SFRLE91} were obtained from counts-in-cells
with a Gaussian window.  The errors, from a minimum variance scheme,
are quite large but they suggest a hierarchical scaling $\xibar_3
\simeq \xibar_2^2$, with a value of $S_{3}$ consistent with gravity
from Gaussian initial conditions, as argued in~\cite{CoFr91}.

\begin{figure}
\vspace{10cm} \special{hscale=35 vscale=35 voffset=-50 hoffset=0
psfile=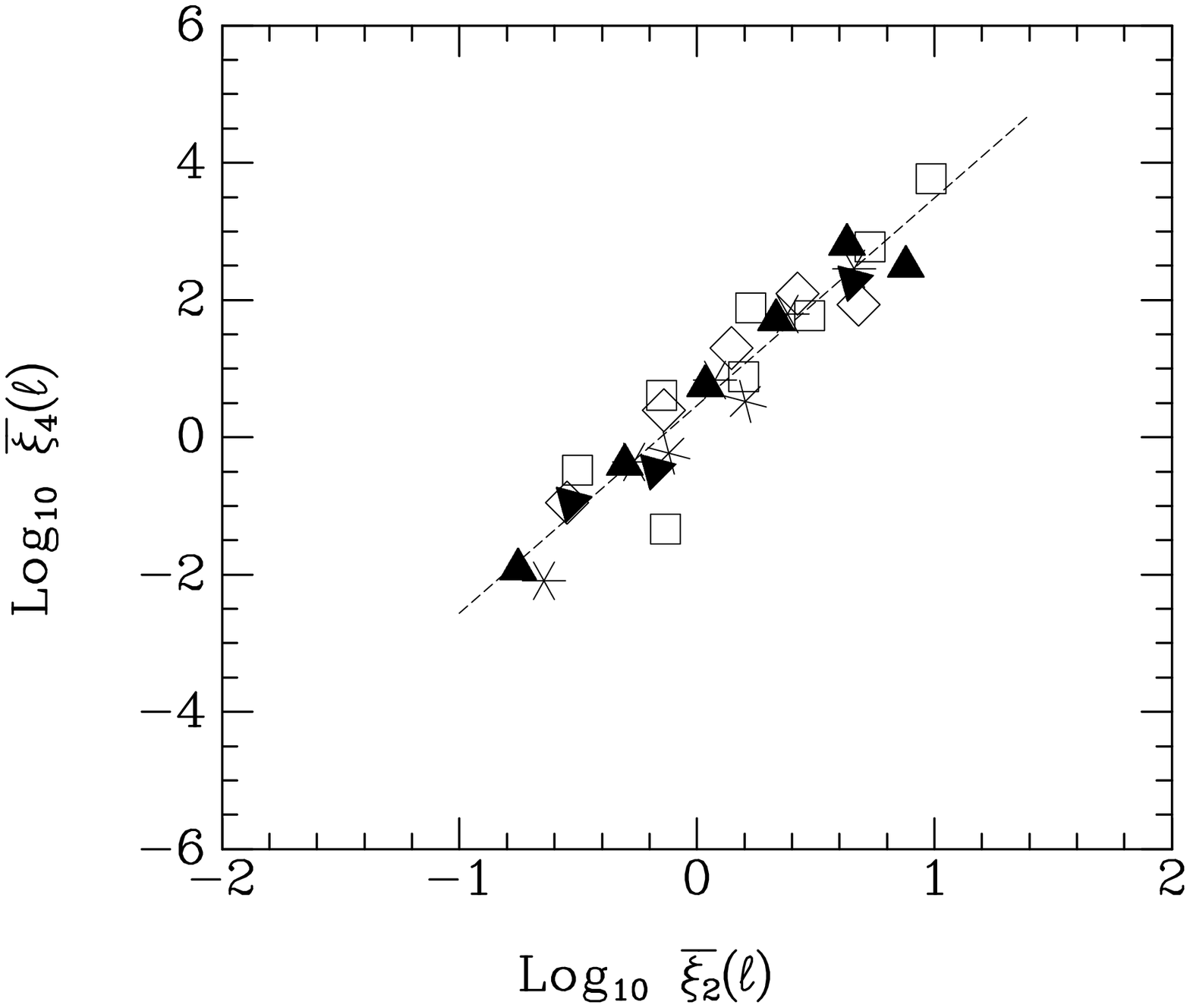} \special{hscale=35 vscale=35 voffset=100 hoffset=0
psfile=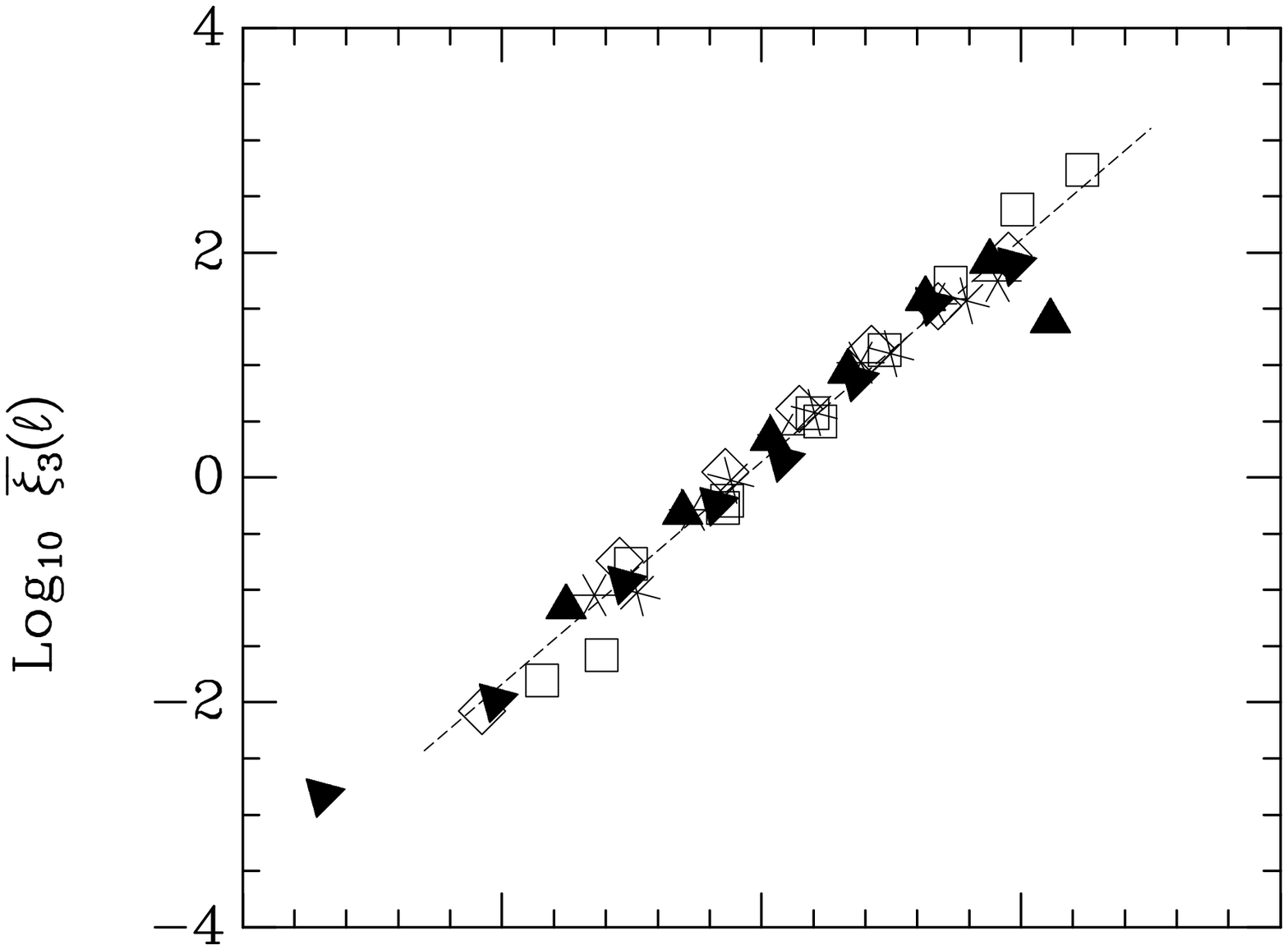} \special{hscale=80 vscale=80 voffset=-150
hoffset=60 psfile=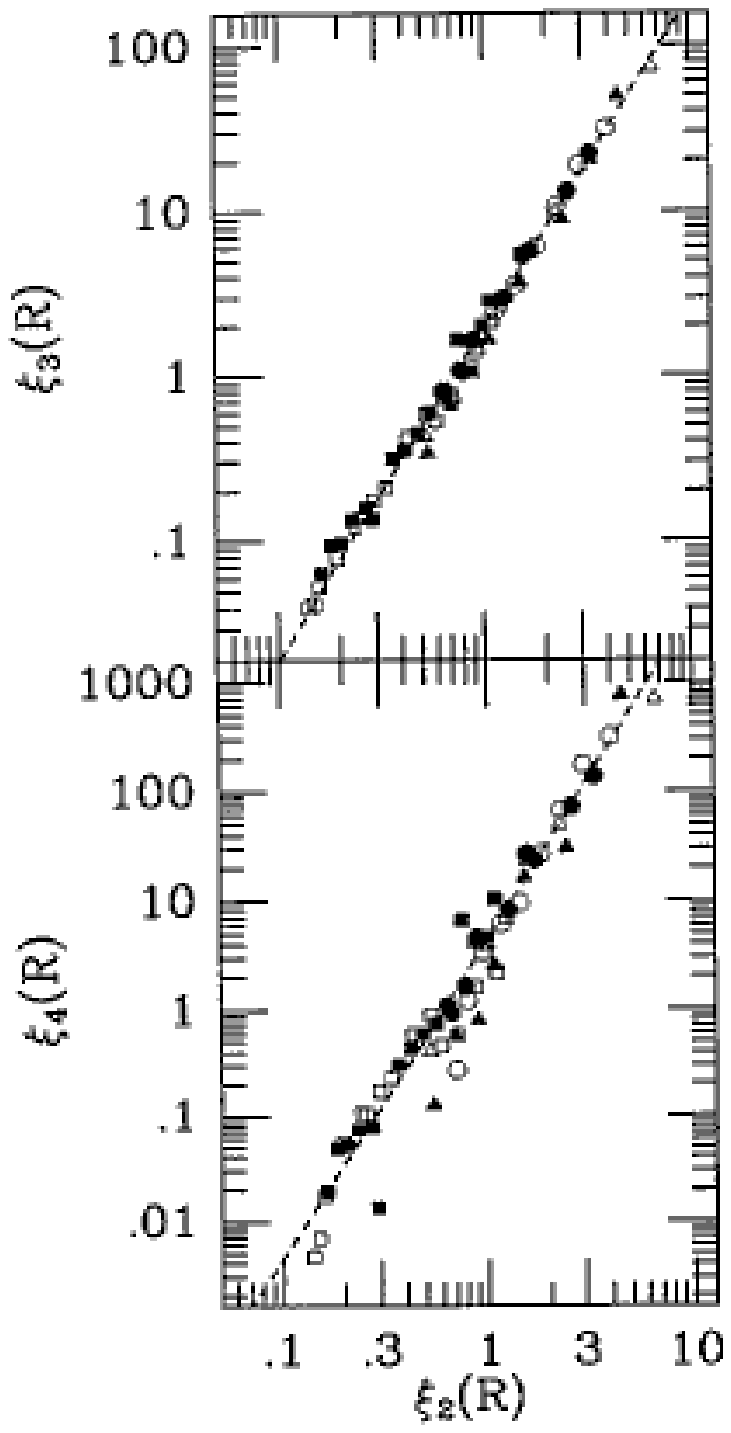}
\caption{ Values of $\xibar_3(R)$ and $\xibar_4(R)$, as a function of
$\xibar_2(R)$ in the IRAS (left, from \cite{BSDFYH93}) and in the CfA
and SSRS (right, from \cite{Gaztanaga92}) redshift catalogs.  The
lines show the best fit amplitude for the hierarchical scaling 
$\xibar_N=S_N ~\xibar_2^{N-1}$.} \label{fig:s34z}
\end{figure}

Figure~\ref{fig:s34z} displays the 1.2 Jy IRAS results (\cite{BDS92},
\cite{BSDFYH93}, left panel) and CfA-SSRS results (\cite{Gaztanaga92},
right panel).  There is a convincing evidence for the hierarchical
scaling in $\xibar_3$ and $\xibar_4$ (denoted by straight lines) but
the resulting $S_3$ and $S_4$ amplitudes are probably affected by
sampling biases (see discussion below).  Note that the scaling is
preserved well into the non-linear regime, this is in agreement with
expectations from N-body simulations which show that in redshift space
the growth of $S_{p}$ parameters towards the non-linear regime is
suppressed by velocity dispersion from virialized regions
(\cite{LIIS93,MaSu94}, see e.g. Fig.~~\ref{QzSpz}).

In their analysis of higher-order moments in the CfA, SSRS, and IRAS
1.9 Jy catalogs, \cite{FrGa94} studied the sensitivity of $S_{p}$ to
redshift distortions, by calculating moments in spherical cells and
conical cells.  The latter were argued to be less sensitive to the
redshift space mapping that acts along the line of
sight\footnote{~This is certainly true in the limit of large radial
distances.  At finite size structures will still look less
concentrated in conical cells than in real space due to velocity
dispersion.  Note that the conical geometry may introduce
a change in $\xi_N$ since not all $N$-point configurations
are weighted equally. }.  They find that although cumulants 
$\xibar_p$ are sensitive to the change in
cell geometry, the $S_{p}$ parameters were not.

On the other hand, \cite{GBGGHKP96} estimated the third and fourth
order cumulants using moments of counts centered in
galaxies~\cite{BBGKP95} in the PPS. After a somewhat ad-hoc correction
for virial fingers to recover ``real space'' quantities, they find a
variation of $S_3$ and $S_4$ with scale, compatible with a non
negligible cubic term, e.g. $\xibar_3 \sim S_3~\xibar_2^2+
C_3~\xibar_2^3$.  Since the scale where the cubic term becomes
important is found to be about 5Mpc/h, this is perfectly consistent
with gravitational clustering: at these scales loop corrections are
expected to increase (the real-space) $S_3$ and $S_4$, see e.g.
Figs.~\ref{numSC} and~\ref{QzSpz}.

An alternative method to moments of count-in-cells was proposed
in~\cite{KiSt98}, who parameterized the count PDF by an Edgeworth
expansion (see Sect.~\ref{sec:edgeworth}) convolved with a Poisson
distribution to take into account discreteness effects.  This method is
only applicable at large enough scales (and small enough $\de/\sigma$)
so that the Edgeworth expansion holds, however convolution with a
Poisson distribution helps to regularize the resulting PDF (i.e. it is
positive definite\footnote{~However, for future applications to
surveys not as sparse as the IRAS galaxy distribution, such as 2dFGRS
and SDSS, this will not be the case.}).  The advantage of this method
is that one can obtain the $S_{p}$ from a likelihood analysis of the
shape of the PDF near its maximum, rather than relying on the tails of
the distribution which are sensitive to rare clusters, as in the
moments method\footnote{~The peak of the PDF is however sensitive to
the largest voids in the sample (see e.g. Fig.~\ref{SmoothEff}), which
can influence the most likely value of $\de$ and thus the $S_{p}$
derived from such method.}.  One disadvantage is that error estimation
in this framework is more complicated, although in principle not
insurmountable.  Results from N-body simulations show this method to
be more reliable at large scales~\cite{KiSt98} than the standard
approach.  Application to the 1.2Jy survey~\cite{KiSt98} resulted in
values for $S_3$ and $S_4$ significantly higher than in previous work
using moments~\cite{BSDFYH93}, see Table~\ref{tab:s3s4z}.

Measurements of the higher-order moments in the SSRS2 survey were
obtained in~\cite{BCDMBS99}.  Results for $S_3$ and $S_4$ were shown
to be consistent with hierarchical at all scales probed (the error
bars quoted in Table~\ref{tab:s3s4z} were found be averaging over all
scales assuming uncorrelated measurements).  A study of the errors in
numerical simulations showed that bootstrap resampling errors were
underestimates by factor of order two.  A re-analysis of the data
using the Edgeworth method of~\cite{KiSt98} showed that $S_3$ changed
upward by a factor of about two to $S_{3} \sim 3$, similar to the
change seen in the IRAS 1.2Jy survey.

\begin{figure}[t]
\centering \mbox{\epsfxsize=9.cm \epsffile{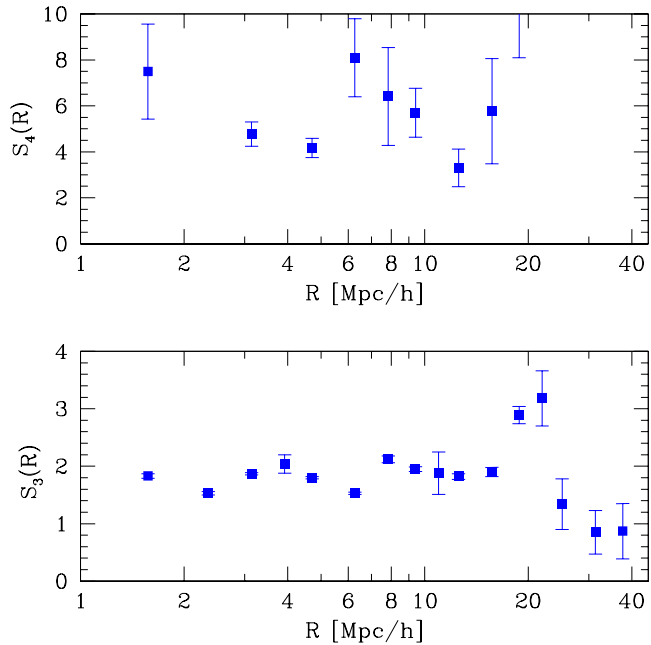}}
\caption{The redshift-space skewness $S_{3}$ and kurtosis $S_{4}$ as a
function of smoothing scale in the PSCz survey~\protect\cite{FFFS01}.}
\label{S3S4PSCz}
\end{figure}

A recent analysis of the PSCz survey~\cite{SBFMS00}, which should be
affected much less than previous IRAS surveys by finite volume
effects, was carried out by using minimum variance estimates of
moments of counts-in-cells in volume limited subsamples (see
Sect.~\ref{sec:sec6}).  The values of $S_3$ and $S_4$ found, shown in
Fig.~\ref{S3S4PSCz}, are consistent within the errors\footnote{~One
should take into account that errors in previous analyses have been
underestimated.  The more realistic errors in~\cite{SBFMS00} were
obtained using the FORCE code~\cite{SzCo96,CSS98,SCB99}, which is
based on the full theory of cosmic errors as described in
Chapter~\ref{sec:chapter7}.  } with that of previous IRAS results,
including those found by deprojection from angular
counts~\cite{MSS92b,BSDFYH93,FrGa94b} and also (for $S_{3}$) in
agreement with the amplitude obtained from measurements of the
bispectrum~\cite{FFFS01} (see Fig.~\ref{bispPSCz_theta}).  They also
found that the measurements of $S_3$ and $S_4$ agreed very well with
the predictions of the semi-analytic galaxy formation model
in~\cite{BCFBL00}, based on models of spiral galaxies in the framework
of $\Lambda$CDM models.

A similar analysis technique was used in the Stromlo-APM and
Durham/UKST surveys~\cite{HSB00}.  In this case measurements of the
skewness are in agreement with those found in shallower redshift
surveys (CfA, IRAS 1.2Jy, SSRS2) but with larger (but more realistic)
errors.  Comparison with deprojected values for $S_{3}$ and $S_{4}$
obtained from the parents catalogs APM~\cite{Gaztanaga94} and
EDSGC~\cite{SMN96} shows a systematic trend where redshift surveys
give systematically smaller values than angular surveys.  The most
significant contribution to this apparent discrepancy is likely to be
redshift distortions: as shown in Fig.~\ref{QzSpz} for scales $R \la
20$ Mpc/h the $S_{p}$ parameters are suppressed in redshift
space\footnote{~This is for dark matter, however at these scales bias
should not make a qualitative difference.  Furthermore, deviations in
galaxy surveys are seen at similar scales~\cite{HSB00}.}.  At scales
larger than 20 Mpc/h results from the redshift and parent angular
surveys should agree, since redshift distortions do not affect the
$S_{p}$ significantly~\cite{HBCJ95}.  In this regime, the results from
APM/EDSGC surveys seem systematically higher, although no more than
$1\sigma$ given the large error bars.  In this case other systematic
effects might be taking place.  Deprojection from angular surveys
using the hierarchical model rather than the configuration dependence
predicted by PT can cause an overestimation of the 3D $S_{p}$ that can
be as much as 20\% for $S_{3}$ (see e.g. Fig.~\ref{s3tl}).  In
addition, finite volume effects~\cite{SzCo96,HuGa99,SCB99} as
discussed in Chapter~6 can lead to underestimation of $S_{p}$ from
redshift surveys that are typically sampling a smaller
volume\footnote{~These effects are thought to be dominant for smaller
surveys such as CfA/SSRS, see~\cite{HuGa99} for a detailed
discussion.}.

\subsubsection{Constraints on Biasing and Primordial Non-Gaussianity}
\label{sec:biagau2}

We now review implications of the above results for biasing and
primordial non-Gaussianity, concentrating on higher-order statistics. 
Effects of primordial non-Gaussianity on the power spectrum have been
considered in~\cite{FKP94,StPe96,STEFKMMORSW99}.  The results
presented here are complementary to recent studies of the impact of
primordial non-Gaussian models in other aspects of large-scale
structure such as the abundance of massive
clusters~\cite{RGS99,KST99,Willick00,PGBB99}. 

Results on the redshift-space bispectrum in the CfA/PPS
sample~\cite{BaFr91} (see Fig.~\ref{Qk_pps}) and the skewness of
CfA/SSRS surveys~\cite{Gaztanaga92} were used in~\cite{FrGa94} to put
constraints on the non-local (scale-dependent) bias in the cooperative
galaxy formation (CGF) model~\cite{BCFW93} proposed to generate enough
large-scale power in the context of otherwise-standard CDM. This model
corresponds to a (density-dependent) threshold bias model where
galaxies form in regions satisfying $\de > \nu \sigma- \kappa
\de(R_{s})$, where $\kappa$ is the strength of cooperative effects and
$R_{s}$ describes the ``scale of influence'' of non-locality. 
Figure~\ref{Qk_pps} shows the predictions of CGF models for
$(\kappa,R_{s})=(0.84,10 \Mpc)$ (dot-long-dashed), $(2.29,20 \Mpc)$
(solid) , and $(4.48,30 \Mpc)$ (dot-short-dashed), all of which have
similar large-scale power to a $\Gamma=0.2$ CDM model.  Because of the
scale dependence induced by the CGF models, additional linear bias is
required to suppress these features, which in turns implies non-zero
non-linear bias to maintain agreement with $Q_{3} \sim 0.5$ and also
would be in disagreement with the normalization implied by the
CMB~\cite{COBE92a}.  In addition, this would make the agreement with
the simple prediction of PT from Gaussian initial conditions purely
accidental.  Similar results follow from the analysis of the skewness
$S_{3}$, see~\cite{FrGa94}.

As discussed in Sect.~\ref{sec:q3z}, the detection of the
configuration dependence of the bispectrum in IRAS surveys (see e.g.
Fig.~\ref{bispPSCz_theta}) gives a tool to constrain galaxy bias,
primordial non-Gaussianity, and break degeneracies present in
two-point statistics.  Using a maximum likelihood method that takes
into account the non-Gaussianity of the cosmic distribution function
and the covariance matrix of the bispectrum~\cite{Scoccimarro00b}, the
constraints on local bias parameters from IRAS surveys assuming
Gaussian initial conditions\footnote{~In addition, these constraints
assume a fixed linear power spectrum shape given by $\Gamma=0.21$, in
agreement with power spectrum measurements. 
See~\cite{Scoccimarro00b,SFFF01} for sensitivity of bias parameters on
the assumed power spectrum shape.  The dependence of the bispectrum on
the assumed $\Omega_{m}$ is negligible, as first pointed out
in~\cite{HBCJ95}.} read~\cite{SFFF01,FFFS01}

\begin{eqnarray}
\frac{1}{b_1} &=& 1.32^{+0.36}_{-0.58},\ \ \ \ \
\frac{b_2}{b_1^2}=-0.57^{+0.45}_{-0.30},\ \ \ \ \ {\rm (2Jy.)} \\
\frac{1}{b_1} &=& 1.15^{+0.39}_{-0.39},\ \ \ \ \
\frac{b_2}{b_1^2}=-0.50^{+0.31}_{-0.51},\ \ \ \ \ {\rm (1.2Jy.)} \\
\frac{1}{b_1} &=& 1.20^{+0.18}_{-0.19},\ \ \ \ \
\frac{b_2}{b_1^2}=-0.42^{+0.19}_{-0.19},\ \ \ \ \ {\rm (PSCz)}
\end{eqnarray}

with the best fit model shown as a dashed line in
Fig.~\ref{bispPSCz_theta} for the PSCz case.  These results for the
linear bias of IRAS galaxies, when coupled with measurements of the
power spectrum redshift distortions, which determine $\beta =
\Omega^{0.6}_m/b_1 \simeq 0.4 \pm 0.12$ for the PSC$z$ survey
\cite{HTP00,TBHT00}, allows the break of the degeneracy between linear
bias and $\Omega_m$, giving $\Omega_m=0.16 \pm 0.1$.

If bias is local in Lagrangian, rather than Eulerian, space the
bispectrum shape depends differently on bias parameters~\cite{CPK00},
see Eq.~(\ref{fb:B_g}).  Physically this corresponds to galaxies that
form depending exclusively on the initial density field, and then
evolved by gravity.  Eulerian bias, on the other hand, corresponds to
the other extreme limit where galaxies form depending exclusively on
the present (non-linear) density field.  Both limits are undoubtedly
simplistic, but analysis of the bispectrum in the PSCz survey suggest
that the Eulerian bias model is more likely than the Lagrangian
one~\cite{FFFS01}.

The bispectrum results can also be used to constraint non-Gaussian
initial conditions.  In this case one must also take into account the
possibility of galaxy biasing, which is more complicated since the
usual formula for Gaussian initial conditions, Eq.~(\ref{Q_g_eul}), is
not valid anymore, but it is calculable in terms of the primordial
statistics~\cite{Scoccimarro00}.  Using a $\chi^2$ model as an example
of dimensional scaling models (where $\xi_{N} \sim \xi_{2}^{N/2}$, see
Sect.~\ref{ssec:chisq}), it was shown that the IRAS 1.2Jy bispectrum
is inconsistent with the amplitude and scaling of this type of initial
conditions at the $95\%$ level~\cite{SFFF01}.

The PSC$z$ bispectrum provides stronger constraints upon non-Gaussian
initial conditions.  In~\cite{FFFS01} $\chi_N^2$ statistics were
considered as a general example of dimensional scaling models.  For
$N=1$, this corresponds to the predictions of some inflationary models
with isocurvature perturbations~\cite{Peebles99a,AMM97,LiMu97}; as $N
\to \infty$ the model becomes effectively Gaussian, and for a fixed
power spectrum (taken to fit that of PSC$z$) the primordial bispectrum
obeys $Q^I \propto N^{-1/2}$~\cite{Scoccimarro00}.  From the PSC$z$
data, it follows that $ N > 49 (22)$ at 68\% (95\%) CL. Since the
primordial dimensionless skewness is $ B_3=2.46$ for a $\chi_1^2$
field~\cite{Peebles99b}, the PSC$z$ bispectrum constrains $ B_3 < 0.35
(0.52)$.  These results are independent of (local) biasing, and they
are obtained by marginalizing over bias parameters~\cite{FFFS01}.

\subsection{Recent Results from 2dFGRS and SDSS}

\label{sec:outlook}

Looking at the overall picture, clustering statistics have been
measured in a wide range of observational data.  The catalogs listed
in Tables~\ref{angularcat} and~\ref{tab:zcat} cover angular surface
densities that are almost six orders of magnitude apart, solid angles
ranging over more than three orders of magnitude, depths that go from
$50$ to $2000 \Mpc$, and volumes ranging over three orders of
magnitude.  They also involve quite different systematics, from
photographic plates to satellite missions and different observational
filters.  Despite these large differences, and after carefully
correcting for systematic effects, all data on higher-order statistics
in the weakly non-linear regime seems in good agreement with
gravitational instability from Gaussian initial conditions.  This
provides a remarkable step forward in our understanding of structure
formation and points to gravity as the basic mechanism to build cosmic
structure from small primordial fluctuations generated in the early
universe.

Needless to say, the observational results reviewed here, although
providing a consistent picture, have significant limitations.  The
magnitude of statistical and systematic errors is still rather large
and the range of scales available in the weakly non-linear regime is
quite restricted.  In the next few years this is expected to change
significantly, with the completion of the new generation of wide field
surveys such as 2dFGRS and SDSS. Here we provide a brief summary of
the results that have been recently reported in the literature from
these preliminary samples.

The 2dFGRS has recently publically released their first versions of
galaxy and quasars catalogs, containing 100,000~\cite{CDMSNC01} and
10,000 redshifts~\cite{CSBSLML01}, respectively.  The completed survey
is expected to reach 250,000 galaxies and 25,000 quasars.  The parent
source catalog is an extended and revised version of the APM
survey~\cite{MES96}, with galaxies with magnitudes $b_{J}<19.45$. For 
a  review of the recent results see~\cite{Peacock01}.  

A measurement of the redshift-space two-point correlation function was
presented in~\cite{PCN01} from analysis of 141,402 galaxies.  Using a
phenomenological model similar to that in Eq.~(\ref{Ppheno}) with
input real-space power spectrum obtained by deprojection from the APM
survey~\cite{BaEf94}, they obtain a velocity dispersion parameter
$\sigma_{v} =385$ km/sec and, after marginalizing over
$\sigma_{v}$, a best fit estimate of $\beta=0.43 \pm 0.07$. These 
results are obtained by considering only the two-point function data 
for $8 \Mpc < r< 25 \Mpc$. 

A preliminary analysis of the redshift-space power spectrum is
presented in~\cite{Percival01}, based on a sample of 147,024 galaxies. 
After taking into account the window of the survey, and assuming
linear perturbation theory at scales $0.02 \la k \la 0.15$ h/Mpc, they
obtain that models containing baryons oscillations are marginally
($\sim 2 \sigma$) preferred over featureless spectra.  Assuming scale
invariance for the primordial power spectrum, their analysis gives
$\Omega_{m} h=0.20 \pm 0.03$ and a baryon fraction
$\Omega_{b}/\Omega_{m}=0.15 \pm 0.07$, in good agreement with recent
determinations from measurements of the CMB power
spectrum~\cite{BOOMERANG,DASI}.  The most recent analysis~\cite{THX01}
of the publically released 100,000 galaxy sample using KL eigenmodes 
finds however no significant detection of baryonic wiggles, although 
their results are consistent with the previous analyses using a 
larger sample, but less sophisticated techniques.

Using a series of volume-limited samples, \cite{NBHMPCF01} present a
measurement of the projected correlation function by integrating the
redshift-space two-point function along the line of sight.  The result
is well described by a power-law in pair separation over the range
$0.1 \Mpc <r<10 \Mpc$, with $r_{0}=4.9 \pm 0.3 \Mpc$ and
$\gamma=1.71\pm 0.06$, see Eq.~(\ref{xi2r0}).  Measurements for
different samples spanning a factor of 40 in luminosity show a
remarkable little variation in the power-law slope, with all
correlation functions being almost parallel with amplitudes spanning a
factor of about three.

These results have been confirmed by recent measurements in a
preliminary sample of the SDSS survey~\cite{Zehavi01SDSS} containing
29,300 galaxy redshifts.  They find a scale-independent luminosity
bias for scales $r<10 \Mpc$, with different subsamples having nearly
parallel projected correlation functions with power-law slope $\gamma
\sim 1.8$.  For the whole sample, the correlation length is $r_{0}=6.1
\pm 0.2 \Mpc$ and the power-law slope $\gamma=1.75\pm 0.03$, for
scales $0.1 \Mpc <r<16 \Mpc$.  The inferred velocity dispersion is
$\sigma_{v} \simeq 600 \pm 100$ km/sec, nearly independent of scale 
for projected separations $0.15 \Mpc < r_{p}< 5 \Mpc$.

A series of papers have recently analyzed angular clustering of over a
million galaxies in a rectangular stripe of $2.5^{\circ}\times
90^{\circ}$ from early SDSS data.  The analysis of systematic effects
and statistical uncertainties is presented in~\cite{Scranton01SDSS},
where the angular correlation function is calculated and the impact of
several potential systematic errors are evaluated, from star/galaxy
separation to the effects of seeing variations and CCD systematics,
finding all of them to be under control.  The Limber scaling test is
performed and showed to make angular correlation functions
corresponding to all four magnitude bins agree when scaled to the same
depth\footnote{~These correspond to $r*=18-19,19-20,20-21,21-22$, with
median redshifts $\bar{z} \sim
0.17,0.25,0.35,0.46$~\cite{Dodelson01SDSS}.}.  Analysis of statistical
errors includes calculation of covariance matrices for $w_{2}(\theta)$
in the four slices using 200 realizations of mock catalogs constructed
using the {\tt PTHalos} code~\cite{ScSh01} and also using the
subsampling and jackknife methods.

Analysis of the angular correlation function is presented
in~\cite{Connolly01SDSS}, which is found to be consistent with results
from previous surveys (see also~\cite{Gaztanaga01}).  On scales
between 1 degree and 1 arcminute, the correlation functions are well
described by a power-law with exponent of about -0.7, in agreement
with Eq.~(\ref{w2pl}).  The amplitude of the correlation function
within this angular interval decreases with fainter magnitudes in
accord with previous galaxy surveys.  There is a characteristic break
in the correlation functions on scales close to 1-2 degrees, showing a
somewhat smaller amplitude at large scales (for the corresponding
magnitude slice) than the APM correlation function.  On small scales,
less than an arcminute, the SDSS correlation function does not appear
to be consistent with the same power-law fitted to the larger angular
scales.  This result should however be regarded as preliminary due to
the still limited amount of data (only $1.6 \%$ of the final size of
the SDSS photometric sample) and the uncertainties in modeling the
covariance matrix of $w_{2}(\theta)$ obtained from the mock catalogs
described above.

The angular power spectrum $P_{2D}(l)$ is obtained
in~\cite{Tegmark01SDSS} for large angular scales corresponding to
multiple moments $\ell \la 600$.  The data in all four magnitude bins
is shown to be consistent with a simple $\Lambda$CDM ``concordance''
model with non-linear evolution (particularly evident for the
brightest galaxies) and linear bias factors of order unity.  The
results were obtained using KL compression, quadratic estimators and
presented in terms of uncorrelated band powers
(Sect.~\ref{sec:general}).  These results, together with those of the
angular correlation function~\cite{Scranton01SDSS,Connolly01SDSS}, are
used in~\cite{Dodelson01SDSS} to perform an inversion to obtain the 3D
power spectrum, using a variant of the SVD decomposition method
of~\cite{EiZa01}\footnote{~See Sect.~\ref{sec:w2pk} for a brief
discussion of inversion procedures and results.} with the
corresponding covariance matrix computed from the mock catalogs.  The
resulting 3D power spectrum estimates from both inversions agree with
each other and with previous estimates from the APM survey for $0.03
{\rm h/Mpc} < k< 1 {\rm h/Mpc}$.  These results are shown to agree
with an alternative method presented in~\cite{Szalay01SDSS}, where the
projected galaxy distribution is expanded in KL eigenmodes and the 3D
power spectrum parameters recovered are $\Gamma=0.188\pm 0.04$ and
$b_{1}\sigma_{8}=0.915\pm 0.06$.

Preliminary results for the higher order correlations in the SDSS have
been presented in~\cite{Gaztanaga01,Gaztanaga01b,SFSS01}, including
$s_3$, $s_4$, $q_3$ and $c_{12}$ statistics.  In all cases a very
good agreement with previous surveys was found.  In particular, at the
bright end the agreement with the APM results is quite remarkable
despite the important differences in survey design and calibration. 
These results confirm the need for non-trivial biasing at small 
scales, as discussed in Sects.~\ref{sec:nptang}-\ref{sec:cicang} (see 
also Fig.~\ref{s34apm}).

\clearpage 
\section{\bf Summary \& Conclusions}

As illustrated throughout this work, PT provides a valuable tool to
understand and calculate predictions for the evolution of large scale
structure in the universe.  The last decade has witnessed a
substantial activity in this area, with strong interplay with
numerical simulations of structure formation and observations of
clustering of galaxies and, more recently, weak gravitational lensing. 
As galaxy surveys become larger probing more volume in the weakly
non-linear regime, new applications of PT are likely to flourish to
provide new ways of learning about cosmology, the origin of primordial
fluctuations, and the relation between galaxies and dark matter.

The general framework of these calculations is well established and
calculations have been pursued for a number of observational
situations, whether it is for the statistical properties of the local
density contrast, the velocity divergence, for the projected density
contrast, redshift measurements or for more elaborate statistics such
as joint density cumulants. All these results provide robust
frameworks for understanding the observations or for reliable error
computations. There are however a number of outstanding issues that
remain to be addressed in order to improve our understanding of
gravitational instability at large scales, \\

\begin{itemize}

\item[--] Most of the calculations have been done assuming Gaussian
initial conditions, except for some specific cases such as $\chi^{2}$
models.  Although present observations are consistent with Gaussian
initial conditions, deriving quantitative constraints on primordial
non-Gaussianity requires some knowledge or useful parametrization of
non-Gaussian initial conditions and how they evolve by gravity.

\item[--] Predictions of PT for velocity field statistics are still in
a rudimentary state compared to the case of the density field. 
Upcoming velocity surveys will start probing scales where PT
predictions can be used.  In addition, robust methods for calculating
redshift distortions including the non-linear effects due to the
redshift-space mapping are needed to fully extract information from
the next generation of galaxy redshift surveys.

\item[--] Another observational context in which a PT approach can be
very valuable is the Lyman-$\alpha$ forest observed in quasar spectra. 
The statistical properties of these systems should be accessible to
perturbative methods since most of the absorption lines correspond to
modest density contrasts (from 1 to 10).  This is a very promising
field for observational cosmology.

\item[--] Accurate constraints on cosmological parameters from galaxy
surveys require precise models of the joint likelihood of low and
higher-order statistics including their covariance matrices.  To date
this has only been investigated in detail numerically, or analytically
in some restricted cases. \\ 

\end{itemize}

In addition, as we probe the transition to the non-linear regime, there
are a few technical issues that need more investigation, \\

\begin{itemize}

\item[--] Most results have been obtained in the tree-level
approximation, for which systematic calculations can be done and the
emergence of non-Gaussianity can be characterized in an elegant way. 
There is no such systematic framework for loop corrections, and only a
few general results are known in this case.  Furthermore, loop
corrections are found to be divergent for power-law spectra with index
$n \geq -1$, the interpretation of which is still not clear.  Although
this issue is irrelevant for realistic spectra such as CDM, its
resolution may shed some light into the physics of the non-linear
regime.

\item[--] The SC collapse prescription (Sect.~\ref{sec:SCmodel}) leads
to a good description of $S_{p}$ parameters in the transition to the
non-linear regime when compared to N-body simulations and exact
one-loop corrections when known.  Is it possible to improve on this
approximation, or make it more rigorous in any well-controlled way
while maintaining its simplicity?

\item[--] The development of HEPT (Sect.~\ref{sec:HEPT}) and EPT
(Sect.~\ref{sec:EPT}) suggests that there is a deep connection between
gravitational clustering at large and small scales.  Is this really
so, or is it just an accident?  Why do strongly non-linear clustering
amplitudes seem to be so directly related to initial conditions?  \\
\end{itemize}

{}From the observational point of view, the next few years promise to
be extremely exciting, with the completion of 2dFGRS and SDSS and deep
surveys that will trace the evolution of large-scale structure towards
high redshift\footnote{~See e.g.~\cite{CDS01} for a recent assessment
of how well upcoming deep surveys will determine correlation
functions.}.  Observations of the so-called Lyman break
galaxies~\cite{SADGPK98} are should soon provide a precious probe of
the high-redshift universe, in particular regarding the evolution of
galaxy bias~\cite{ASGDPK98,PoGi01,BWS02}.  Furthermore, weak lensing
observations will provide measurements of the projected mass density
that can be directly compared with theoretical predictions.  In
addition, CMB satellites and high-resolution experiments will probe
scales that overlap with galaxy surveys and thus provide a consistency
check on the framework of the growth of structure.

Outstanding observational issues abound, most of them perhaps related
to the way galaxies form and evolve.  One of the most pressing ones,
as discussed many times in Chapter~8, is probably to have a convincing
explanation of why correlation functions scale as power-law's at non-linear
scales. The scaling in Figs.~\ref{w2apmz4} and~\ref{PkPSCz} is certainly
remarkable and preliminary results from 2dFGRS~\cite{NBHMPCF01} and
SDSS~\cite{Zehavi01SDSS,Connolly01SDSS} seem already to confirm and
extend these results.  In the CDM framework, however, this simple
behavior is thought to be the result of accidental cancellation of the
dark matter non-power-law form by scale-dependent bias due to the way
dark matter halos are populated by galaxies (see discussion in
Sect.~\ref{dhbias}).  Although this may seem rather adhoc, this model
has, on the other hand, many observable consequences.  The same
weighting that makes the two-point function depend as a power-law of
separation~\cite{Seljak00,PeSm00,SSHJ01} suppresses the velocity
dispersion and mean streaming of galaxies~\cite{SHDS01,SDHS01} as
observed, see e.g.~\cite{JMB98}.  In addition, this weighting affects
higher-order statistics in the non-linear regime, suppressing them in
comparison with their dark matter counterparts~\cite{SSHJ01} (see
Fig.~\ref{Sp_gal}) as observed, see e.g. Fig.~\ref{s34apm} for a
comparison between dark matter and $S_{p}$ in the APM survey.

There are also complementary indications that galaxies do not trace
the underlying dark matter distribution at small scales from
measurements of higher-order statistics.  As discussed in
Sect.~\ref{sec:w2pk}, reconstruction of the linear power spectrum from
galaxy surveys leads to significant disagreement of higher-order
moments if no biasing is imposed at small scales, as shown in
Fig.~\ref{s34apm} for APM galaxies.  A promising way to confirm that
the underlying higher-order statistics of the dark matter are much
higher than those of galaxies at small scales is by measuring
higher-order moments in weak gravitational lensing.  This will likely
be done in the near future, as weak lensing surveys are already
beginning to probe the relation between dark matter and
galaxies~\cite{HYG01}.

In any case, statistical analysis of future observations are going to
decide whether the small-scale behavior of correlations is dictated by
biasing or if a new framework is needed to understand galaxy
clustering at non-linear scales.  What seems clear, whatever the
outcome, is that the techniques described here will be a valuable tool
to achieve that goal.

\vskip 2pc

This project was possible thanks to the hospitality of several
institutions that supported frequent visits.  We thank CSIC, IAP, IAS,
IEEC, SPhT, and also CITA during the initial stages of this work.  We
also benefited greatly from discussions with F. Bouchet, J. Frieman,
J. Fry, R. Juszkiewicz and I. Szapudi that help set the structure of
this review.  We thank Marc Kamionkowski for many helpful comments
about a previous version of this work.  This project has made
extensive use of NASA's Astrophysics Data System Abstract Service.

\clearpage 
\appendix

\section{\bf The Spherical Collapse Dynamics}
\label{AppendixSC}

The spherical collapse dynamics can be obtained from the Friedmann equations
of the expansion  factor in different cosmologies.  It  amounts to solve the
motion  equation for  the radius  $R$ of  a shell  collapsing under  its own
gravity,
\begin{equation}
{\d^2 R\over \d t^2}=-G\,{M(<R)\over R^2}
\end{equation}
where $M(<R)$  is the mass encompassed  in a radius  $R$.  The corresponding
density contrast can be defined as,
\begin{equation}
\delta_{\rm sc}(t)={M(<R)\over\rhob\,4\pi\,R^3/3}-1.
\end{equation}
Explicit solutions are known for open or closed universes without cosmological
constant. The complete derivation of them can be found in
\cite{Peebles80} where the density contrast is expressed as a function
of time $t$. We present the results here in a slightly
different way by expressing the nonlinear density contrast 
as a function of the linear density contrast, $\epsilon$ ($\equiv
D_+(t)\delta_{\rm init}$)~\cite{Bernardeau92b}. 

For an open universe the background evolution is described by
parameter $\psi_0$ 
so that the current value of the density parameter is given by,
\begin{equation}
\Omega_0={2\over 1+\cosh\psi_0}
\end{equation}
Similarly the density fluctuation is characterized by  a
parameter $\theta$. There is a minimal value of the linear density
contrast below which the density fluctuation is still below critical 
and does not collapse. This is given by,
\begin{equation}
\epsilon_{\rm min}={9\over 2}{\sinh\psi_0\left(\sinh\psi_0-\psi_0\right)
\over\left(\cosh\psi_0-1\right)^2}
\end{equation}
As a result, if the linear density contrast $\epsilon>\epsilon_{\rm min}$,
the evolution of the perturbation density is given by,
\begin{equation}
\delta_{\rm sc}(\epsilon)=\left({\cosh\psi_0-1\over-\cos\theta+1}\right)^3
\left({-\sin\theta+\theta\over\sinh\psi_0-\psi_0}\right)^2-1
\end{equation}
with
\begin{equation}
\epsilon=\epsilon_{\rm min}\left[\left({-\sin\theta+\theta
\over\sinh\psi_0-\psi_0}\right)^{2/3}+1\right].
\end{equation}
If $\epsilon<\epsilon_{\rm min}$ we have,
\begin{equation}
\delta_{\rm sc}(\epsilon)=\left({\cosh\psi_0-1\over\cosh\theta-1}\right)^3
\left({\sinh\theta-\theta\over\sinh\psi_0-\psi_0}\right)^2-1
\end{equation}
with
\begin{equation}
\epsilon=-\epsilon_{\rm min}\left[\left({\sinh\theta-\theta
\over\sinh\psi_0-\psi_0}\right)^{2/3}-1\right].
\end{equation}

The Einstein de Sitter case is recovered when $\psi_0\to 0$. It
implies that $\epsilon_{\rm min}\to 0$. In this case the solution 
reads, for $\epsilon<0$,
\begin{eqnarray}
\delta_{\rm sc}(\epsilon)&=&{9\over 2}{(\sinh\theta-\theta)^2\over
(\cosh\theta-1)^3}-1\\
\epsilon&=&-{3\over 5}\left[{3\over
4}\left(\sinh\theta-\theta\right)\right]^{2/3}
\end{eqnarray}
and for $\epsilon>0$,
\begin{eqnarray}
\delta_{\rm sc}(\epsilon)&=&{9\over 2}{(\theta-\sin\theta)^2\over
(1-\cos\theta)^3}-1\\
\epsilon&=&{3\over 5}\left[{3\over
4}\left(\theta-\sin\theta\right)\right]^{2/3}.
\end{eqnarray}

In the limit $\Omega_0\to 0$ we have $\psi_0\to\infty$. It implies
that $\epsilon_{\rm min}\to 3/2$. Moreover $\epsilon$ is finite when $\theta$
is close to $\psi_0$ so that,
\begin{equation}
\epsilon={3\over 2}\left({\exp\theta\over\exp\psi_0}\right)-1,\ \ \ \ 
\delta_{\rm sc}={\exp\psi_0\over\exp\theta}
\end{equation}
which gives,
\begin{equation}
\delta_{\rm sc}(\epsilon)={1\over (1-2\epsilon/3)^{3/2}}
\end{equation}

The case of a closed universe is obtained by the change of variable $\psi_0\to
\ii\psi_0$.

\section{\bf Tree Summations}
\label{LegTrans}

In this appendix we present methods for performing tree summations.
These calculations have been developed initially in 
different contexts (such as polymer physics, see e.g.~\cite{DCJ87}).
In cosmology, these computations techniques have been introduced 
in~\cite{Schaeffer84} and presented in details in a more complex
situation in~\cite{BeSc92}.

\subsection{For One Field}
\label{LegTrans1}

The issue we address is the computation of  the sum of all tree
diagrams (in a specific sense given in the following) connecting an
arbitrary  number of points.  More specifically we define $\varphi(y)$
as (minus)  the sum of all diagrams with the weight $(-y)^n$ for
diagrams of $n$ points.

For computing the contribution of each order  
the rule is to build all possible minimal connection (that means
$n-1$ connections for $n$ points) and to affect the value $\nu_p$
to points connected to $p$ neighbors.
The value of each
diagram is then given by the product of the vertices $\nu_p$ 
it is composed of.

The function $\varphi(y)$ then  corresponds to the cumulant
generating function,
\begin{equation}
\varphi(y,\nu_1,\nu_2,\ldots)= -\sum_{n=2}^{\infty}(-y)^n \sum_{{\rm
trees\ connecting\ }n{\rm\ points}}\left( \lower2.ex\hbox{$\;
\buildrel{\Pi} \over {{\rm vertices}}\;$}
\nu_p\right).
\end{equation}
At the end of the calculation the 
value of $\nu_1$ will be unity, but for the time being we assume it
is a free parameter. Then $\varphi$ is a function of
$y$ and of the vertices  $\nu_p$. We can then define $\tau$ as 
\begin{equation}
-\tau={1\over -y}{\partial (-\varphi)\over\partial \nu_1}.
\end{equation}
Like $\varphi$, $(-\tau)$ is a function of $y$ and of the vertices
$\nu_p$.  This corresponds to all the diagrams for which {\em one}
external line (connected to a $\nu_1$ vertex) has been marked
away. This is the  sum of so called diagrams with one free external
line. It is possible to write down an implicit equation for $\tau$,
\begin{equation}
-\tau=-y\,\left(\nu_1-\nu_2\,\tau+\nu_3\,{\tau^2\over 2}+\ldots+
\nu_p\,{(-\tau)^{p-1}\over (p-1)!}+\ldots\right).
\label{tauexp}
\end{equation}
This equation expresses the fact that $\tau$ can be reconstructed
in a recursive way (see Fig. \ref{taugraph}). 
Note the factor $(p-1)!$ which corresponds to the symmetry factor. 
If one defines the generating function of the vertices,
\begin{equation}
\zeta(\tau)=\sum_{p=1}^{\infty}\nu_p\,{(-\tau)^p\over p!}
\end{equation}
then we have,
\begin{equation}
\tau=-y{\partial \zeta\over\partial \tau}.
\end{equation}

To complete the calculation we need to introduce the Legendre transform
$\mL(\tau,\nu_2,\ldots)$ defined as
\begin{equation}
\mL(y,\tau,\nu_2,\ldots)=\varphi+y\,\nu_1\,\tau.
\end{equation}
It is important to note that $\mL$ is viewed as a function
of $\tau$ and not of $\nu_1$. We then have the remarkable property
due to the Legendre transform,
\begin{equation}
{\partial \mL\over\partial\tau}={\partial \varphi\over \partial \nu_1}
{\partial \nu_1\over\partial\tau}+y\,\tau\,{\partial \nu_1\over\partial\tau}
+y\,\nu_1=y\,\nu_1
\label{fb:dLsdtau}
\end{equation}
{}From Eq. (\ref{tauexp}) we have,
\be
y\,\nu_1=\tau-y\,\sum_{p=2}^{\infty}\nu_p\,{(-\tau)^{p-1}\over (p-1)!}
\ee
which, after integrating relation (\ref{fb:dLsdtau}), implies that,
\be
\mL=c+{\tau^2\over 2}+y\,\sum_{p=2}^{\infty}\nu_p\,{(-\tau)^p\over p!}
=c+{\tau^2\over 2}+y\zeta(\tau)+y\,\nu_1\,\tau,
\ee
which leads to (the integration constant $c=0$ is such that
$\varphi(y)\sim -y^2$ at leading order in $y$),
\begin{equation}
\varphi(y)=y\zeta(\tau)-{1\over 2}y\tau\zeta'(\tau).
\end{equation}
This equation, with Eq. (\ref{tauexp}), gives the tree generating
function expressed as a function of the vertex generating function
$\zeta$.

\subsection{For Two Fields}
\label{LegTrans2}

We can extend the previous results to joint tree summations. 
It corresponds to either 2 different fields taken
at the same position (as the density and the velocity divergence for
instance), or to two fields taken at different locations. 

We want to construct the joint generating function,
$\varphi(y_1,y_2)$, of the joint cumulants,
\be
\varphi(y_1,y_2)=-\sum_{n,m,n+m\ge2} C_{nm}{(-y_1)^n\over n!}
{(-y_2)^m\over m!}
\ee
where $C_{nm}$ is the value of each cumulant. In this case for each
diagram there are $n$ vertices of type 1, and $m$ of type 2. 
They take respectively the value $\nu_p$ and $\mu_q$ if they
are connected respectively to $p$ or $q$ neighbors. Obviously
if the two fields are identical the two series identify. 
Moreover in order to account for cell separation, a weight $\xi$ 
is put for each line connecting points of different nature.

The generating function $\varphi$ is then a function of 
$y_1,y_2,\xi,\nu_1,\dots,\mu_1,\dots$. 
One can define the two functions $\tau_1$ and $\tau_2$ by, 
\be
\tau_1={1\over -y_1}{\partial(-\varphi)\over \partial \nu_1},\ \ \ 
\tau_2={1\over -y_2}{\partial(-\varphi)\over \partial \mu_1}.
\ee
It is easy to see that the functions $\tau_1$ and $\tau_2$ are given 
respectively by,
\ba
\tau_1&=&y_1\sum_{p=1}^{\infty}\nu_p{(-\tau_1)^{p-1}\over (p-1)!}+
\xi\,y_2\sum_{p=1}^{\infty}\mu_p{(-\tau_2)^{p-1}\over (p-1)!},\\\label{tauexp1}
\tau_2&=&\xi\,y_1\sum_{p=1}^{\infty}\nu_p{(-\tau_1)^{p-1}\over (p-1)!}+
y_2\sum_{p=1}^{\infty}\mu_p{(-\tau_2)^{p-1}\over (p-1)!}.\label{tauexp2}
\ea
This expresses the fact that there is a joint recursion 
between the two functions. A factor $\xi$ is introduced whenever
a vertex of a given type is connected to vertex of the other
type.

Defining the Legendre transform as 
$\mL=\varphi+y_1\tau_1\nu_1+y_2\tau_2\mu_1$, one obtains,
\be
{\partial\mL\over\partial\tau_1}=y_1\nu_1,\ \ \ 
 {\partial\mL\over\partial\tau_2}=y_2\mu_1.
\ee
One should then solve the linear system for $\nu_1$ and 
$\mu_1$ given by Eqs. (\ref{tauexp1}, \ref{tauexp2}). 
One eventually gets for $\varphi$,
\be
\varphi=y_1\zeta_1(\tau_1)+y_2\zeta_2(\tau_2)+{1\over
2(1-\xi^2)}\left(\tau_1^2-2\xi\tau_1\tau_2+\tau_2^2\right),
\ee
where $\zeta_1$ and $\zeta_2$ are respectively the generating functions
of $\nu_p$ and $\mu_p$.
This result can be rewritten in a more elegant form,
\be
\varphi(y_1,y_2)=y_1\zeta_1(\tau_1)+y_2\zeta_2(\tau_2)-{1\over2}
y_1\tau_1\zeta_1'(\tau_1)-{1\over2} y_2\tau_2\zeta_2'(\tau_2).
\ee
If $\xi$ is unity, for instance for the computation of 
the joint density distribution of $(\delta,\theta)$, we have,
\be
\tau=\tau_1=\tau_2=-y_1\zeta'(\tau)-y_2\zeta'(\tau_2).
\ee

\subsection{The Large Separation Limit}

The other case of interest is when $\xi$ is small (which means that the
correlation function at the cell separation is much smaller
that the average correlation function at the cell size).

It is then possible to expand $\varphi(y_1,y_2)$ at leading
order in $\xi$. The results reads,
\be
\varphi(y_1,y_2)=
\varphi_1(y_1)+\varphi_2(y_2)-\tau_1^{(0)}(y_1)\xi\tau_2^{(0)}(y_2)
\ee
where $\tau_1^{(0)}$ and $\tau_2^{(0)}$ are the respectively 
the functions $\tau_1$ and $\tau_2$ computed when $\xi=0$.

\section{\bf Geometrical Properties of Top-Hat Window Functions}
\label{tophatgeom}

\def\vl{{\overline l}}

In this section we recall the properties of top-hat window
function. The derivations are presented in a systematic way 
for any dimension of space $D$.
The window function $W_{D}$ in Fourier space is given by
\begin{equation}
W_D(k)=2^{D/2}\,\Gamma\left({D/2}+1\right)
\,{J_{D/2}(k)\over k^{D/2}}.
\end{equation}
We are interested in computing the angle integrals
of $W_D(\vert \vl_1-\vl_2\vert)$ times a geometrical function
that can be expressed in terms of Legendre polynomials.
In particular we want to compute $
\int\d^D\Omega\,W_D(\vert \vl_1-\vl_2\vert)
\left[1-{(\vl_1.\vl_2)^2/(l_1^2\,l_2^2)}\right]
$ and $\int\d^D\Omega\,W_D(\vert \vl_1-\vl_2\vert)
\left[1+\vl_1.\vl_2/l_1^2\right]$. In general the only angle
the intervene in the angular integral, $\d^D\Omega$, is the relative
angle $\varphi$ so that, ${\d^D\Omega/\Omega_{\rm tot.}}$
reduces to ${\Gamma(D/2)/(\sqrt{\pi}\,\Gamma[(D-1)/2])}
\sin(\varphi)^{D-2}\d\varphi$, $0\le\varphi\le\pi$.

In order to complete these calculations,
we need the summation theorem (GR, 8.532) for Bessel function,
\begin{equation}
{J_{\nu}(\vert \vl_1-\vl_2\vert)\over \vert \vl_1+\vl_2\vert^{\nu}}
=2^{\nu}\Gamma(\nu)\sum_{k=0}^{\infty}\left(\nu
+k\right)\,{J_{\nu+k}(l_1)\over l_1^{\nu}}\,{J_{\nu+k}(l_2)\over l_2^{\nu}}\,
C_k^{\nu}(\cos\varphi)
\end{equation}
where $C_k^{\nu}$ are Gegenbauer polynomials. 
Note that in case of $\nu=0$ the previous equations reads,
\begin{equation}
J_0(\vert \vl_1-\vl_2\vert)=J_0(l_1)\,J_0(l_2)+
2\sum_{k=1}^{\infty}J_{k}(l_1)\,J_{k}(l_2)\,\cos(k\varphi).
\end{equation}
In the following, the only property of interest for the Gegenbauer
polynomials is (GR, 7.323)
\begin{eqnarray}
\int_0^{\pi}C_k^{\nu}(\cos\varphi)\,\sin^{2\nu}(\varphi)\d\varphi&=&0,\ \ \ 
{\rm for}\ k\ge1\\ 
\int_0^{\pi}C_0^{\nu}(\cos\varphi)\,\sin^{2\nu}(\varphi)\d\varphi&=&
{\pi\,\Gamma(2\nu+1)\over 2^{2\nu}\,\Gamma^2(1+\nu)}
\end{eqnarray}

As a result we have,
\begin{eqnarray}
&&{\Gamma(D/2)\over \sqrt{\pi}\,\Gamma[(D-1)/2]}
\int_0^{\pi}\sin(\varphi)^{D-2}\d\varphi\,W_D(\vert \vl_1-\vl_2\vert)
\left[1-{(\vl_1.\vl_2)^2\over l_1^2\,l_2^2}\right]=\nonumber\\
&&\ \ \ 
{2^D\,\Gamma^2(D/2)\,\Gamma(D/2+1)\over \sqrt{\pi}\,\Gamma[(D-1)/2]}
\sum_{k=0}^{\infty}\left({D\over2}+k\right)\,
{J_{\nu+k}(l_1)\over l_1^{\nu}}\,{J_{\nu+k}(l_2)\over l_2^{\nu}}\nonumber\\
&&\ \ \ \times
\int\sin(\varphi)^{D}\d\varphi\,C_k^{D/2}(\cos\varphi).
\end{eqnarray}
The only non-vanishing term of this summation is the one corresponding
to $k=0$. We finally have,
\begin{equation}
\int{\d^D\Omega\over \Omega_{\rm tot.}}
\,W_D(\vert \vl_1-\vl_2\vert)
\left[1-{(\vl_1.\vl_2)^2\over l_1^2\,l_2^2}\right]=
\left(1-{1\over D}\right)\,W_D(l_1)\,W_D(l_2).
\end{equation}
This result writes as a kind  of commutation
rule: the filtering can be applied to the wave vectors separately
provided the angular kernel is properly averaged.

The second relation can be obtained from the observations that 
\ba
l{\d\over \d l}\left[J_{D/2-1}(l)\over l^{D/2-1}\right]
&=&-{J_{D/2}(l)\over l^{D/2}},\\
{J_{D/2-1}(l)\over l^{D/2-1}}&=&l{\d\over \d l}
\left[J_{D/2}(l)\over l^{D/2}\right]+d{J_{D/2}(l)\over l^{D/2}}. 
\ea
The summation theorem 
applied to $J_{D/2-1}(\vert \vl_1+\vl_2\vert)/
\vert \vl_1+\vl_2\vert^{D/2-1}$, leads to,
\ba
&&\int{\d^D\Omega\over \Omega_{\rm tot.}}
{J_{D/2-1}(\vert \vl_1+\vl_2\vert)\over
\vert \vl_1+\vl_2\vert^{D/2-1}}
=\nonumber\\
&&\ \ \ \ \ \ 
{{2\,\left( D-2  \right) \,{\sqrt{\pi }}\,{\rm \Gamma}(-2 + D)}\over
{{2^{D/2}}\,{\rm \Gamma}({(-1 + D)/ 2})}}
{J_{D/2-1}(l_1)\over l_1^{D/2-1}}
{J_{D/2-1}(l_2)\over l_2^{D/2-1}}.
\ea
Taking the derivative of this equality with respect to $l_1$ 
leads to,
\ba
&&\int{\d^D\Omega\over \Omega_{\rm tot.}}
\,W_D(\vert \vl_1-\vl_2\vert)
\left[1-{\vl_1.\vl_2\over l_1^2}\right]=\nonumber\\
&&\ \ \ \ \ 
W_D(l_1)\left[W_D(l_2)+
{l_2\over D}\,{\d \over\d l_2}W_D(l_2)\right].
\ea

\section{\bf One-Loop Calculations: Dimensional Regularization}
\label{DimReg}

To obtain the behavior of the one-loop $p$-point spectra for $n <-1$,
one can use dimensional regularization (see e.g. \cite{Collins84}) to
simplify considerably the calculations.  Since we are interested in
the limit where the ultraviolet cutoff $k_c \rightarrow \infty$, all
the integrals run from $0$ to $\infty$, and divergences are regulated
by changing the dimensionality $d$ of space: we set $d = 3 +
\varepsilon$ and expand in $\varepsilon \ll 1$. For example, for
one-loop bispectrum calculations, we need the following one-loop
three-point integral:

\begin{equation}
J(\nu_1,\nu_2,\nu_3) \equiv 
\int  \frac{\d^d {\bf q}}{(q^2)^{\nu_1} [({\bf k_{1}}-{\bf q})^2]^{\nu_2}
[({\bf k_{2}}-{\bf q})^2]^{\nu_3}} 
\label{J}.
\end{equation}

\noindent When one of the indices vanishes, e.g. $\nu_{3}=0$, this
reduces to the standard formula for dimensional-regularized two-point
integrals~\cite{Smirnov91}

\begin{equation}
 J(\nu_1,\nu_2,0) =
 \frac{\Gamma (d/2 -\nu_1)
\Gamma (d/2 -\nu_2)
\Gamma (\nu_1 + \nu_2 - d/2)}{\Gamma (\nu_1) \Gamma (\nu_2) \Gamma (d-
\nu_1 - \nu_2)} \ \pi^{d/2}  \ k_{1}^{d-2 \nu_1 -2 \nu_2}
\label{I},
\end{equation}

\noindent which is useful for one-loop power spectrum calculations.
The integral $J(\nu_1,\nu_2,\nu_3)$ appears in triangle diagrams for
massless particles in quantum field theory, and can be evaluated for
arbitrary values of its parameters in terms of hypergeometric
functions of two variables. The result is~\cite{Davydychev92}

 \begin{eqnarray}
 	J(\nu_1,\nu_2,\nu_3) & = & \frac{\pi^{d/2} k_{1}^{d-2 \nu_{123}
 	}}{\Gamma (\nu_1) \Gamma (\nu_2)  \Gamma (\nu_3) \Gamma (d-
    \nu_{123})} \times \ \Bigg( \Gamma (\nu_3) 
    \Gamma (\nu_{123} - d/2) 
 	\nonumber   \\
 	 &  & \times {\rm F}_{4} (\nu_3,\nu_{123} - d/2;1 +\nu_{23}- 
 	 d/2, 1 + \nu_{13}- d/2 ; x , y)
 	\nonumber  \\
 	 &  & \times \Gamma (d/2-\nu_{13}) \Gamma (d/2-\nu_{23}) +
 	 y^{d/2-\nu_{13}} \Gamma (\nu_2) \Gamma (d/2-\nu_1)
 	\nonumber  \\
 	 &  & \times {\rm F}_{4} (\nu_2,d/2 - \nu_1;1 +\nu_{23}- 
 	 d/2, 1 -\nu_{13}+ d/2 ; x , y)
 	\nonumber  \\ 
 	 &  & \times \Gamma (\nu_{13}-d/2) \Gamma (d/2-\nu_{23})
 	 + x^{d/2-\nu_{23}} \Gamma (\nu_1) \Gamma (d/2-\nu_2)
 	\nonumber  \\
 	 &  & \times {\rm F}_{4} (\nu_1,d/2 - \nu_2;1 -\nu_{23}+ 
 	 d/2, 1 + \nu_{13}- d/2 ; x , y)
 	\nonumber  \\ 
 	 &  & \times \Gamma (d/2-\nu_{13}) \Gamma (\nu_{23}-d/2) 
 	 + x^{d/2-\nu_{23}} y^{d/2-\nu_{13}} \Gamma (d/2-\nu_3)
 	\nonumber  \\ 
 	 &  & \times 
 	 {\rm F}_{4} (d-\nu_{123},d/2 - \nu_3;1 -\nu_{23}+ 
 	 d/2, 1 -\nu_{13}+ d/2 ; x , y)
 	\nonumber  \\ 
 	 &  & \times \Gamma (d-\nu_{123}) \Gamma (\nu_{23}-d/2)
 	 \Gamma (\nu_{13}-d/2) \Bigg)
 	\label{Jresult},
 \end{eqnarray}

\noindent where $\nu_{123} \equiv \nu_{1}+\nu_{2}+\nu_{3}$, $\nu_{ij}
\equiv \nu_{i}+ \nu_{j}$, $x \equiv ({\bf k}_{2}-{\bf
k}_{1})^{2}/k_{1}^{2}$, $y \equiv k_{2}^{2}/k_{1}^{2}$, and ${\rm
F}_{4}$ is Apell's hypergeometric function of two variables, with the
series expansion:

\begin{equation}
	{\rm F}_{4} (a,b;c,d;x,y) = \sum_{i=0}^{\infty} \sum_{j=0}^{\infty}
     \frac{x^{i}y^{j}}{i!\ j!} \ \frac{ (a)_{i+j} (b)_{i+j}}{(c)_{i}
     (d)_{j}}
	\label{F4exp},
\end{equation}

\noindent where $(a)_{i} \equiv \Gamma (a+i) / \Gamma (a)$ denotes the
Pochhammer symbol.  When the spectral index is $n=-2$, the
hypergeometric functions reduce to polynomials in their variables due
to the following useful property for $-a$ a positive integer:

\begin{equation}
	{\rm F}_{4} (a,b;c,d;x,y) = \sum_{i=0}^{-a} \sum_{j=0}^{-a-i}
     \frac{x^{j}y^{i}}{j!\ i!} \ \frac{  (b)_{i+j}}{(c)_{i}
     (d)_{j}} \ \frac{(-1)^{i+j} (-a)!}{(-a-i-j)!}
	\label{F4pol}.
\end{equation}

When using expressions such as Eq.~(\ref{Jresult}), divergences appear
as poles in the gamma functions; these can be handled by the following
expansion ($n=0,1,2, \ldots$ and $\varepsilon \rightarrow 0$):

\begin{eqnarray}
	\Gamma(-n + \varepsilon) &=& \frac{(-1)^n}{n!} \left[
	\frac{1}{\varepsilon} + \psi (n+1) + \frac{\varepsilon}{2}
	\left( \frac{\pi^2}{3} + \psi^2 (n+1) - \psi' (n+1)
	\right)\right],
	\nonumber \\ & &  \label{gammapoles}
\end{eqnarray}

\noindent plus terms of order $\varepsilon^2$ and higher. Here $\psi
(x) \equiv d \ln \Gamma (x) /dx$ and

\begin{eqnarray}
	\psi (n+1) & = & 1 + \frac{1}{2} + \ldots +  \frac{1}{n} - \gamma_e
	\label{psi},  \\
	\psi' (n+1) & = & \frac{\pi^2}{6} - \sum_{k=1}^{n} \frac{1}{k^2}
	\label{psiprime},
\end{eqnarray}

\noindent with $\psi (1) = - \gamma_e = -0.577216 \ldots$
 and $\psi' (1) = \pi^2 /6$.

\section{\bf PDF Construction from Cumulant Generating Function}
\label{app:pdedelta}

In this section we present the mathematical relation
between the cumulant generating function defined in section \ref{fb:Stoch}
and the one-point probability distribution function of the local
density, and more generally the counts in cells probabilities.

In this presentation we follow the calculations (and most
of notations) developed in \cite{BaSc89a}. 

\subsection{Counts-in-Cells and Generating Functions}

Let us consider a cell of volume $V$ placed at random in the
field. We note $P(N)$ the probability that this cell contains $N$
particles. One can define the probability distribution function
$\mP(\lambda)$  with,
\be
\mP(\lambda)=\sum_{N=0}^{\infty}\,\lambda^N\,P(N).
\ee
By construction the counts in cells probabilities $P(N)$ are obtained
by a Taylor expansion of $\mP(\lambda)$ around $\lambda=0$,
\be
P(N)={1\over N!}{\d^n\over\d\lambda^n}\mP(\lambda=0).
\ee
Remarkably the (factorial) 
moments of this distribution are obtained by a Taylor
expansion of $\mP(\lambda)$ around $\lambda=1$,
\ba
\mP(1)&=&1\nonumber\\
{\d\over\d\lambda}\mP(1)&=&\bn\nonumber\\
{\d^2\over\d\lambda^2}\mP(1)&=&\mg N(N-1)\md\nonumber\\
\dots&&\nonumber\\
{\d^p\over\d\lambda^p}\mP(1)&=&\mg N(N-1)\dots(N-p+1) \md.
\ea
If the field is an underlying Poisson distribution
of a continuous field, then the factorial moments,
$\mg N(N-1)\dots(N-p+1) \md$, are equal to $\bn^p\,M_p$
where $M_p$ is the $p^{th}$ moment of the local density distribution.
$\mP(\lambda)$  can therefore be written in terms of the moment
generating function (see sec. \ref{sec:sec3.3.2}), $
\mP(\lambda)=\mM[(\lambda-1) \bn]$,
which in turns can be written in terms of the cumulant generating
function, $\mC(\lambda-1)$,
\be
\mP(\lambda)=\exp\left(\mC[(\lambda-1) \bn]\right).
\ee
When the cumulant generating function is written in terms of the $S_p$
generating function, the counts in cells read,
\be
P(N)=\oint{1\over 2\pi \ii}{\d \lambda\over \lambda^{N+1}}
\exp\left[-{\bn\xiav(1-\lambda)+
\varphi(\bn\xiav(1-\lambda))\over\xiav}\right]
\ee
where the integral is made in the complex plane around the
singularity $\lambda=0$. One can change the variable to use
$y=\bn\xiav(1-\lambda)$, so that,
\be
P(N)={-1\over \bn\xiav}
\oint{\d y\over 2\pi \ii}\left(1-{y\over \bn\xiav}\right)^{-(N+1)}
\exp\left[-{y+\varphi(y)\over\xiav}\right]
\ee

\subsection{The Continuous Limit}

The contributing values for $y$ are finite, so that, in the continuous
limit $\lambda$ should be close to unity. As a result one can write
\be
\left(1-{y\over\bn\xiav}\right)^{N+1}=\exp\left[-(N+1)\log
\left(1-{y\over\bn\xiav}\right)\right]=
\exp\left[-{N\over\bn\xiav}y\right]
\ee
It implies that
\be
P(\rho)\d\rho=-{\d\rho\over \xiav}\int_{-\ii\infty}^{+\ii\infty}
{\d y\over 2\pi \ii}\exp\left[-{y+\varphi(y)\over \xiav}+{y\rho\over
\xiav}\right].
\label{fb:InvLapTr}
\ee
This is the inverse Laplace transform.

It is important to note that the counts in cells $P(N)$ can be
recovered by a Poisson convolution of the continuous distribution. A
Poisson distribution is given by,
\be
P_{\rm Poisson}(N,\bn)={\bn^N\over N!}e^{-\bn}=
\oint{1\over 2\pi \ii}{\d\lambda\over\lambda^{N+1}}
\exp\left(-\bn(1-\lambda)\right)
\ee
Then 
\ba
&&\int\d\rho\,P(\rho)\,P_{\rm Poisson}(N,\bn\rho)=\nonumber\\
&&-\int{\d\rho\over \xiav}\int_{-\ii\infty}^{+\ii\infty}
{\d y\over 2\pi \ii}\oint{\d\lambda\over 2\pi \ii}
{1\over\lambda^{N+1}}
\exp\left[-{y+\varphi(y)\over \xiav}+{y\rho\over
\xiav}-\bn\rho(1-\lambda)\right]
\ea
The integration over $\rho$ leads to $\delta_{\rm
Dirac}(y-\bn\xiav(1-\lambda))$, which finally implies,
\be
\int\d\rho\,P(\rho)\,P_{\rm Poisson}(N,\bn\rho)=P(N).
\ee
This is not surprising since we assumed from the very beginning that
any discrete field would be the Poisson realization of a continuous
field.

\subsection{Approximate Forms for $P(\rho)$ when $\xiav\ll1$}

In this paragraph, we review the various approximations that have
been used for $P(\rho)$. It obviously depends on the regime we are
interested in, that on the amplitude of the density fluctuations
$\xiav$.

When $\xiav$ is small, it is possible to apply a saddle point
approximation. This point is defined by
\be
\rho_s={\d\varphi(y_s)\over\d y}.
\ee
It leads to
\be
P(\rho)={1\over\sqrt{-2\pi\xiav\,\varphi''(y_s)}}
\exp\left[-{1\over\xiav}(y_s+\varphi(y_s)-y_s\varphi'(y_s))\right].
\label{fb:pdfsaddle}
\ee In case $\varphi(y)$ is a obtained through a tree summation, as
for the weakly non-linear regime, one finally gets the formula
(\ref{fb:pdeltasaddle}).

Obviously such a result makes sense  only if $\varphi''(y)$ is
negative. Because of the presence of a singular point on the real
axis this will not be always the case. In practice it will be true
only for values of the density smaller than a critical value,
$\rho_c$. These values are given in table \ref{fb:tabSing}
for the results obtained in the quasi-linear regime.

For $\rho>\rho_c$ the shape the saddle point position is pushed
towards the singularity.  The behavior of the PDF will then be
dominated by the behavior of $\varphi(y)$ around this point.  Let us
write generally $\varphi(y)$ as, \be
\varphi(y)=\varphi_s+r_s(y-y_s)+\dots-a_s(y-y_s)^{\omega_s} \ee where
the expansion around the singular point has been decomposed into its
regular part, $\varphi_s+r_s(y-y_s)+\dots$ and singular part
$a_s(y-y_s)^{\omega_s}$, where $w_s$ is a non-integer value ($w_s=3/2$
in the quasi-linear theory).  In \ref{fb:InvLapTr} the integration
path for $y$ will be pushed towards the negative part of the real axis
($y<y_s$).  It can thus be described by the real variable $u$ varying
from 0 to $\infty$ with, \be y=y_s+u\,e^{\pm \ii\pi} \ee where the
sign is changing according to whether $y$ is above or under the real
axis.  Expanding the singular part in the exponential one gets, 
\be
P(\rho)= {-a_s\over \xiav^2}\int_0^{\infty}\d u\,u^{\omega_s}
{e^{\pm \ii\pi(\omega_s-1)}\over 2 \pi \ii} \exp\left(-{\rho-r_s\over
\xiav}u\right) \ee which gives, \be P(\rho)={a_s\over
\Gamma(-\omega_s)\xiav^2}
\left({\rho-r_s\over\xiav}\right)^{-\omega_s-1}
\exp\left(-{\varphi_s\over \xiav}-\vert
y_s\vert{\rho\over\xiav}\right)
\label{fb:pdfcutoff}
\ee
taking advantage of the relation,
$\Gamma(\omega_s+1)\Gamma(-\omega_s)=-\sin(\pi\omega_s)/\pi$.
For the parameters describing the quasi-linear theory one gets the
relation (\ref{fb:pdeltatail}).

\subsection{Approximate Forms for $P(\rho)$ when $\xiav\gg1$}

Two scaling domains have been found (see \cite{BaSc89a} for a
comprehensive presentation of the scaling laws). One corresponds
to the rather dense regions. It corresponds to cases where
$\varphi(y)$ is always finite in \ref{fb:InvLapTr}. For large values
of $\xiav$ it is therefore possible to write,
\be
P(\rho)={-1\over\xiav^2}\int_{-\ii\infty}^{+\ii\infty}
{\d y\over 2\pi \ii}\varphi(y)\exp(xy)
\ \ {\rm with}\ \ x=\rho/\xiav.
\ee
One can see that the PDF is a function of $x$ only. Roughly speaking,
in this integral, $y\sim 1/x$ so that the validity domain of this 
expression is limited to cases where $\varphi(1/x)\ll \xiav$. It will
be limited to a regime where 
\be
x\gg (\xiav/a)^{1/(1-\omega)},
\label{fb:pdfplaw}
\ee
if $\varphi(y)$ behaves like $a\,y^{1-\omega}$ at large $y$.

If $x$ is small, in a regime where $\varphi(y)$ can be approximated by
its power law asymptotic shape, the PDF eventually reads,
\be
P(\rho)={1\over\xiav^2}{a(1-\omega)\over \Gamma(\omega)}
x^{2-\omega}.
\ee

For large values of $x$, one recovers the exponential cut-off
found in the previous regime, (\ref{fb:pdfcutoff}), with further
simplifications since $\xiav\gg1$,
\be
P(\rho)={a_s\over \Gamma(-\omega_s)\xiav^2}
\left({\rho\over\xiav}\right)^{-\omega_s-1}
\exp\left(-\vert y_s\vert \rho/\xiav\right)
\label{fb:pdfcutoff2}
\ee

The second scaling regime corresponds to the underdense regions. They
are described by the asymptotic form of $\varphi(y)$, which implies,
\be
P(\rho)={-1\over \xiav}\int_{-\ii\infty}^{+\ii\infty}
{\d y\over 2\pi\ii}\exp\left(-a{y^{1-\omega}\over\xiav}+
{\rho y\over\xiav}\right)
\ee
A simple change of variable, $t^{1-\omega}=y^{1-\omega}\xiav/a$,
shows that it can be written,
\be
P(\rho)={-1\over \xiav}\left(a\over\xiav\right)^{-1/(1-\omega)}
\int_{-\ii\infty}^{+\ii\infty}
{\d t\over 2\pi\ii}\exp\left(-t^{1-\omega}+z\,t\right)
\ee
with 
\be
z={\rho\over\rho_v},\ \ \rho_v=\xiav\,\left(a\over\xiav\right)^{1/(1-\omega)},
\ee
which can be written,
\be
P(\rho)={1\over \pi\rho_v}\int_0^{\infty}
\d u\sin[u^{1-\omega} \sin\pi u]\,e^{-zu+u^{1-\omega}\cos\pi u}.
\ee
For large values of $z$, the power law behavior of (\ref{fb:pdfplaw})
is recovered, and the two regimes overlap. 

Small values of $z$ however describe the small density cut-off.  The
expression of the PDF can be obtained by a saddle point approximation,
and it appears to be a particular case of the results obtained in Eq. 
(\ref{fb:pdfsaddle}).  Note that the shape of this cut-off depends
only on $\omega$, 

\be P(\rho)={1\over\rho_v} {(1-\omega)^{1\over
2\omega}\over\sqrt{2\pi\omega}}\,
z^{-{1\over2}-{1\over2\omega}}\,\exp\left[-\omega(1-\omega)^{1-\omega\over\omega}\,z^{1-\omega\over\omega}\right].
 \ee

\subsection{Numerical Computation of the Laplace Inverse Transform}

The starting point of the numerical computation of the local density
PDF from the cumulant generating function is equation (\ref{fb:InvLapTr}). In
case the cumulant generating function can be obtained from a vertex
generating function $\mG$, the latter is the natural variable to use. 
The technical difficulty is actually to choose the path to follow in
the $y$ or $\mG$ complex plane. The original path for $y$ runs
from ${-\ii\infty}$ to ${+\ii\infty}$ along the imaginary axis. But as
the functions $\tau(y)$ or $\varphi(y)$ are not analytic over the
complex plane (there is at least one singularity on the local axis for
$y=y_s<0$) the crossing point of the path with the real axis cannot be
moved to the left side of $y_s$ (otherwise the PDF would simply
vanish!). Actually the crossing point of the path for the numerical
integration is the position of the saddle point, $y_{\rm saddle}$
defined by,
\begin{equation}
0=\rho-1-\left.{\d \varphi(y)\over \d y}\right\vert_{y=y_{\rm saddle}}.
\end{equation}
This equation has a solution as long as $\rho<1+\delta_c$ and it is then
at a point $y_{\rm saddle}>y_s$ (see Section \ref{fb:TheDensityPDF}).
In case of $\rho>1+\delta_c$ the crossing point of the integration path is
then simply chosen to be $y=y_s$. The integration path is subsequently
built in such way that $\rho y -1-\varphi(y)$ is kept real and negative
to avoid unnecessary oscillation of the function to integrate. In
practice the path is built step by step with an adaptive integration
scheme~\cite{Bernardeau94a,CBBH97}.

\section{\bf Cosmic Errors: Expressions for the Factorial Moments}
\label{cosmicerrorfacmom}

In this appendix, we first explain how the cosmic error on the factorial
moments of count-in-cells is calculated. We then list the
corresponding
analytic expression for the cosmic covariance matrix up to third
order in the three-dimensional case.

\subsection{Method} 
{}From now, to simplify we assume that
the cells are spherical (or circular, in two dimensions), but the
results are valid in practice with the obvious appropriate corrections
for any compact cell.

The local Poisson assumption allows us to neglect correlations
inside the union $C_{\cup}$ of volume $v_{\cup}$ of two overlapping
cells and the non-spherical contribution of $C_{\cup}$.  As a result,
the generating function for bicounts in overlapping cells
reads~\cite{SzCo96}
\begin{equation}
   {\cal P}_{\rm over}(x,y)={\cal P}_{\cup}\left[q(x+y)+p x y\right].
\end{equation}
The generating function ${\cal P}_{\cup}(x)$ is the same as ${\cal
P}(x)$ but for a cell of volume $v_{\cup}$, and
\begin{equation}
  p=[1-f_{\dim}(r/R)]/[1+f_{\dim}(r/R)], \quad
  q=f_{\dim}(r/R)/[1+f_{\dim}(r/R)],
\end{equation}
where $f_{\dim}(r/R)$ represents the excess of volume (or area) of
$v_{\cup}$ compared to $v_R$,
\begin{equation}
   v_{\cup}=v_R[1+f_{\dim}(r/R)],
\end{equation}
and $r$ is the separation between the two cells.  We have
$f_3(\psi)=(3/4)\psi-(1/16)\psi^3$, and and $f_2(\psi)=1-(1/\pi)\left[
2 {\rm arccos} (\psi/2)-\sqrt{1-\psi^2/2} \right]$ in three and two
dimensions, respectively.

The generating function for disjoint cells
is Taylor expanded
\begin{equation}
  {\cal P}_{\rm disjoint}(x,y)\simeq {\cal P}(x) {\cal P}(y)\left[ 1 +
  {\cal R}(x,y)\right]+{\cal O}(\xi/\xiav^2),
\end{equation}
with
\begin{equation}
  {\cal R}(x,y)=\xi \sum_{M=1,N=1}^{\infty} (x-1)^M (y-1)^N
  \frac{S_{NM}}{N!\,M!}\barN^{N+M}\xiav^{N+M-2}.
\end{equation}

It is then easy to calculate cross-correlations on factorial moments,
$\Delta_{k,l}$, by computing the double integral in
Eq.~(\ref{eq:errcosme}) after applying partial derivatives in
Eq.~(\ref{eq:errorini}), with the further assumption that the
two-point correlation function is well approximated by a power-law of
index $-\gamma\simeq -1.8$ for $r\leq 2R$.\footnote{~The results do not
depend significantly on the value of $\gamma$~\cite{SzCo96}.}

\subsection{Analytic Results}
The cosmic errors for the factorial moments as discussed in 
Sect.~\ref{sec:secfacmomtheo}, Eq.~(\ref{eq:cosmiccovtot}),
are now detailed here, up to
third order (in the three-dimensional case):
\begin{eqnarray}
 \Delta_{11}^{\rm F} & = & {{\bn}^2}\,\xiav({\hat L}),\\
 \Delta_{11}^{\rm E} & = & 5.508\bn^2\xiav\frac{v}{V},  \\
 \Delta_{11}^{\rm D} & = & \bn \frac{v}{V}, \\
 \Delta_{22}^{\rm F} & = & 4 {{\bn}^4}\,\xiav({\hat L})\,
  \left( 1 + 2\,\xiav\,Q_{12} + 
    {{\xiav}^2}\,Q_{22} \right),\\
   \Delta_{22}^{\rm E} & = & 
    {{{ \bn}}^4}\,{\xiav} \frac{v}{V} \left( 17.05\, 
    + 3.417\,{{{ \xiav}}} 
    + 45.67\,{{{ \xiav}}}\,Q_3
    + 42.24\,{{{ \xiav}}^2}\,Q_4 \right) 
    \\
   \Delta_{22}^{\rm D} & = & 
     {{{ \bn}}^2} 
    \frac{v}{V}\left( 0.648\,
    + 4\,{{{ \bn}}}
    + 0.502\,{ \xiav}
    + 8.871\,{{{ \bn}}}\,{ \xiav}
    +  6.598\,{{{ \bn}}^2}\,{{{ \xiav}}^2}\,Q_3 \right), \\
\Delta_{33}^{\rm F}& = & 9{{\bn}^6}\,\xiav({\hat L})\,
  \left( 1 + 2\,\xiav + {{\xiav}^2} + 
    4\,\xiav\,Q_{12} + 
    4\,{{\xiav}^2}\,Q_{12} + 
    6\,{{\xiav}^2}\,Q_{13} \right. \nonumber \\ & & \left. + 
    6\,{{\xiav}^3}\,Q_{13}  + 
   4\,{{\xiav}^2}\,Q_{22} + 
    12\,{{\xiav}^3}\,Q_{23} + 
    9\,{{\xiav}^4}\,Q_{33} \right).\\
\Delta_{33}^{\rm E}& = &
    {{{\bn}}^6}\,{\xiav}\, \frac{v}{V}\left( 
   34.62 + 
   99.26\,{{{\xiav}}} + 
   39.60\,{{{\xiav}}^2} + 
   180.3\,{{{\xiav}}}\,Q_3 + 
   331.1\,{{{\xiav}}^2}\,Q_3 + 
\right. \nonumber \\
   & & 
    93.50\,{{{\xiav}}^3}\,{{Q_3}^2} +  
   633.5\,{{{\xiav}}^2}\,Q_4 +  
   441.3\,{{{\xiav}}^3}\,Q_4 +   \nonumber \\
   & & \left.
   1379\,{{{\xiav}}^3}\,Q_5 + 
   1668\,{{{\xiav}}^4}\,Q_6 \right), \\
\Delta_{33}^{\rm D}& = &
   \,{{{\bn}}^3}\,\frac{v}{V}, \left( 
   0.879 + 
   5.829\,{{{\bn}}} + 
   9.\,{{{\bn}}^2} + 
   2.116\,{\xiav} +
   27.13\,{{{\bn}}}\,{\xiav} + \right. \nonumber \\
   & &
   66.53\,{{{\bn}}^2}\,{\xiav} +  
   10.59\,{{{\bn}}}\,{{{\xiav}}^2} + 
   74.23\,{{{\bn}}^2}\,{{{\xiav}}^2} + 
   1.709\,{{{\xiav}}^2}\,Q_3 + \nonumber \\ & &
   42.37\,{{{\bn}}}\,{{{\xiav}}^2}\,Q_3 + 
   148.5\,{{{\bn}}^2}\,{{{\xiav}}^2}\,Q_3 + 
   111.2\,{{{\bn}}^2}\,{{{\xiav}}^3}\,Q_3 +  \nonumber \\ & & \left.
   44.40\,{{{\bn}}}\,{{{\xiav}}^3}\,Q_4 + 
   296.4\,{{{\bn}}^2}\,{{{\xiav}}^3}\,Q_4 + 
   349.3\,{{{\bn}}^2}\,{{{\xiav}}^4}\,Q_5  
   \right).
\end{eqnarray}
The cosmic cross-correlations read
\begin{eqnarray}
 \Delta_{12}^{\rm F} & = & 2{{\bn}^3}\,\xiav({\hat L})\,
  \left( 1 + \xiav\,Q_{12}
    \right),\\
 \Delta_{12}^{\rm E} & = & {{\bn}^3} \xiav \frac{v}{V}
  \left( 8.525+ 
  11.42\,{\xiav}\,Q_3 \right),\\
\Delta_{12}^{\rm D} & = & {{\bn}^2} \frac{v}{V} \left(2.0 + 
  1.478\,\xiav \right),\\
 \Delta_{13}^{\rm F} & = & 3 {{\bn}^4}\,\xiav({\hat L})\,
  \left( 1 + \xiav + 
    2\,\xiav\,Q_{12} + 
    3\,{{\xiav}^2}\,Q_{13} \right),\\
\Delta_{13}^{\rm E} & = & {{\bn}^4} \xiav \frac{v}{V}
  \left( 9.05 + 
  11.42\,{{\xiav}} + 
  21.67\,{{\xiav}}\,Q_3 + 
  42.24\,{{\xiav}^2}\,Q_4 \right),\\
\Delta_{13}^{\rm D} & = & {{\bn}^3} \frac{v}{V}
  \left( 3.0 + 
  6.653\,\xiav + 
  4.949\,{{\xiav}^2}\,Q_3 \right),\\
 \Delta_{23}^{\rm F} & = & 6 {{\bn}^5}\,\xiav({\hat L})\,
  \left( 1 + \xiav + 
    3\,\xiav\,Q_{12} + 
    3\,{{\xiav}^2}\,Q_{13} + 
    {{\xiav}^2}\,Q_{12} + \right. \nonumber \\ & & \left.
    2\,{{\xiav}^2}\,Q_{22} + 
    3\,{{\xiav}^3}\,Q_{23} \right),\\
 \Delta_{23}^{\rm E} & = & {{\bn}^5} \xiav \frac{v}{V}
  \left( 23.08 + 
  33.09\,{{\xiav}} + 
  90.17\,{{\xiav}}\,Q_3 + 
  55.19\,{{\xiav}^2}\,Q_3 + \right. \nonumber \\ & & \left.
  211.2\,{{\xiav}^2}\,Q_4 + 
  229.9\,{{\xiav}^3}\,Q_5
  \right),\\
\Delta_{23}^{\rm D} & = & {{\bn}^3}\frac{v}{V}
 \left( 1.943 + 6.\,{{\bn}} + 
  4.522\,\xiav + 
  26.61\,{{\bn}}\,\xiav + 
  9.898\,{{\bn}}\,{{\xiav}^2} + \right. \nonumber \\ & & \left.
  3.531\,{{\xiav}^2}\,Q_3 + 
  39.59\,{{\bn}}\,{{\xiav}^2}\,Q_3 + 
  39.53\,{{\bn}}\,{{\xiav}^3}\,Q_4 
 \right).
\end{eqnarray}
Note that the finite volume effect terms $\Delta_{pq}^{\rm F}$ 
would be the same in the 2-D case.
In the above equations, $\xiav({\hat L}$ is the integral of
the two-point correlation function over the survey volume 
[Eq.~(\ref{eq:xiavhatl})] and 
\begin{equation}
   Q_N\equiv \frac{S_N}{N^{N-2}}, \quad Q_{NM}\equiv
   \frac{C_{NM}}{N^{N-1}M^{M-1}}.
   \label{eq:defqnnew}
\end{equation}
Note that these $Q_N$ and $Q_{NM}$ are slightly different from what
was defined in Eqs.~(\ref{Qn}) and
(\ref{HM}). They are also often used in the literature instead of
$S_p$ or $C_{p\,q}$.

An accurate approximation for $\xiav({\hat L})$
is~\cite{CSJC99,CCDFS00}
\begin{equation}
  \xiav({\hat L}) \simeq 
  \frac{1}{{\hat V}^2} \int_{\hat V} \d^{\dim}r_1 \d^{\dim}r_2
  \xi(r_{12})-\frac{1}{\hat V}\int_{r\leq 2R} d^{\dim}r \xi(r).
\end{equation}
This actually means that, rigorously, the finite volume error as we
defined it here actually contains an edge effect term.  For practical
calculations, however, the following approximation generally works
quite well
\begin{equation}
  \xiav({\hat L}) \simeq \xiav(L),
\end{equation}
where $\xiav(L)$ was defined in Eq.~(\ref{eq:xiavLdef}).

\clearpage 

\end{document}